%% file: deav3.tex
\documentclass[10pt,a4paper]{memoir}
\usepackage{latexsym}
\usepackage{epsf}
\usepackage{amssymb}
\usepackage{color}
\usepackage{amsmath} 
\usepackage[dvips]{graphicx}
\usepackage{subfig}

\settrimmedsize{297mm}{210mm}{*}
\settypeblocksize{634pt}{448.13pt}{*}
\setulmargins{4cm}{*}{*}
\setlrmargins{*}{*}{1.5}
\setmarginnotes{17pt}{51pt}{\onelineskip}
\setheadfoot{\onelineskip}{2\onelineskip}
\setheaderspaces{*}{2\onelineskip}{*}
\checkandfixthelayout

\makeatletter \@addtoreset{equation}{section} \makeatother

\begin{document}

\vspace{15cm}
\begin{center}


\begin{titlingpage}

\begin{center}


\textsc{\LARGE Departamento de F\'{i}sica Fundamental (UNED)}\\[1.5cm]

\vspace{3cm}

\textsc{\Large Master's thesis (Complex Systems)}\\[0.5cm]


\hspace{\fill}\rule{\linewidth}{.7pt}\hspace{\fill}\\
\vskip 0.7cm
{ \huge \bfseries Coupled Ising models and interdependent discrete choices under social influence in homogeneous populations}\\[0.4cm]

\hspace{\fill}\rule{\linewidth}{.7pt}\hspace{\fill}\\

\vskip 0.5cm
\textsc{\Large Ana Fern\'andez del R\'io}\\
\emph{Supervisor:} \textsc{Elka Korutcheva}

\vfill

{\large December 2010}

\end{center}

\end{titlingpage}

\begin{abstract}
\small 
The use of statistical physics to study problems of social sciences is motivated and its current state of the art briefly reviewed, in particular for the case of discrete choice making. The coupling of two binary choices is studied in some detail, using an Ising model for each of the decision variables (the {\itshape opinion or choice moments or spins}, socioeconomic equivalents to the magnetic moments or spins). Toy models for two different types of coupling are studied analytically and numerically in the mean field (infinite range) approximation. This  is equivalent to considering a social influence effect proportional to the fraction of adopters or average magnetisation. In the nonlocal case, the two spin variables are coupled through a Weiss mean field type term. In a socioeconomic context, this can be  useful when studying individuals of two different groups, making the same decision under social influence of their own group, when their outcome is affected by the fraction of adopters of the other group. In the local case, the two spin variables are coupled only through each individual. This accounts to considering individuals of a single group each making two different choices which affect each other. In both cases, only constant (intra- and inter-) couplings and external fields are considered, i.e., only completely homogeneous populations. Most of the results presented are for the zero field case, i.e. no externalities or private utilities. Phase diagrams and their interpretation in a socioeconomic context are discussed and compared to the uncoupled case. The two systems share many common features including the existence of both first and second order  phase transitions, metastability and hysteresis. To conclude, some general remarks, pointing out the limitations of these models and suggesting further improvements are given.  
\end{abstract}
\end{center}
\newpage
\tableofcontents
\pagestyle{Ruled}
\chapterstyle{ell}

\include{secs/intro}

\include{secs/socphy}

\include{secs/isiopi}
\include{secs/2p1cho}
\include{secs/1p2cho}
\include{secs/conc}

\chapter*{Acknowledgements}

I want to very specially thank my supervisor Elka Korutcheva for her (much appreciated) unconditional support and for her invaluable guidance since we began this project. The results presented on chapters \ref{cha:2p1cho} and \ref{cha:1p2cho} are the result of work done in collaboration with Javier de la Rubia and her, so they deserve at least as much credit for them. We would like to thank Prof. N. Tonchev for discussions on the mean field analysis of both models and for indicating previous work of Galam, Salinas and Shapir on the local model of which we were not aware of. Both his comments and the reference provided proofed to be very helpful. I would also like to thank Daniel Guinea for his sociological insight and for giving me the chance to work on real data (results hopefully coming soon). Finally, I wish to thank Alberto Ramos for some insightful discussions and for pointing out some useful references.

\bibliographystyle{hplain}
\bibliography{deav3}



\end{document}

%% file: secs/intro.tex
\chapter{Introduction}
\label{cha:intro}

The general direction of this work is the study of interdependent discrete choice making under social influence. \emph{Interdependent} because it considers different choices which are related to and affect each other. \emph{Under social influence} because we will be considering populations of individuals which are subject to social interactions with other individuals that have an effect in their decision making process. Rather than studying any particular socioeconomic case of interest, the goal is to develop (or in some sense collect) a general theoretical framework, methodology and set of appropriate tools to study these type of problems in the spirit of \emph{social physics} (which will be soon explained in more detail). We will be borrowing many ideas and results from the study of condensed matter through the use of statistical physics and can thus be considered within the scope of \emph{statistical physics of choice/opinion dynamics}. 

This type of approach has been increasingly popular in the literature in the last years (see next chapter for more details and references) in the treatment of many problems traditionally way out of the realm of physics.  Whether this enterprise of building a more {\itshape physics like} body of knowledge for the social sciences will be fruitful or proof of any real use is a question to be elucidated over the years to come, but has already given some interesting and novel insight on some areas. These subtle though crucial matters on the motivations and validity of this approach will be briefly discussed in the next chapter.

In particular, this work will be studying the coupling of two binary choices (opinions) in a radically simple scenario of social structure, social influence and coupling between the decisions. Two examples are considered. Two different groups  making a choice about the same subject in which the outcome for the individuals in one of the groups will depend on the choices of the other group (as well as on social pressure to conform to its own group) in what we will call the \emph{non local model}. It can be used in problems such as the study of  the public opinion of a certain issue in two neighbouring countries or of two interrelated social groups deciding whether to join or not a specific community or to take a certain position. In the service or product demand interpretation side it may be used to study for example the adoption of a certain technology by the different members/companies of two related sectors. In the \emph{local model}, which is the second example considered, there is a single group of individuals in which each is making two decisions (subject to social influence) which are interdependent in that they either tend to reinforce or exclude each other. Examples of interest range from the study of the public opinion on two different related issues to results in simultaneous elections for different institutions, female participation in the labour market vs forming a family, continuing studies vs teenage pregnancy, etc. From an economic perspective, it can be used to study the demand interplay for two similar/different products of the same/different brand.

A very wide variety of problems from the social sciences, ranging from sociology to economy to political science, can in principle be considered under this approach. The new results presented in this work are, however, for radically simple models and should be considered as toy models aimed at understanding the basic qualitative features arising from the coupling between the decisions. The key simplification that possibly renders them useless for their application to explain any problem of real interest in the social sciences context is the fact that they allow for no variation whatsoever within the individuals from a group: they all interact with all of the others in the same way and they all have the same natural propensity to make a positive choice. Further more, most of the specific results presented are for the zero field case, i.e., equal {\itshape a priori} probability of deciding for or against for all individuals (the choice making outcome is determined through social/decision interactions only). They are however a necessary preliminary step: if we want to understand what qualitative differences arise when characterising the individuals' attitude towards the decision making we need to understand in as much detail as possible the model with constant parameters. This heterogeneity can be introduced using probability distributions for the coupling parameters and externalities rather than fixed parameters. This has been a widely explored subject in condensed matter (spin glasses, random field theories). See \cite{Cohen1985, Goldenfeld1992, Mezard1987, Parisi1992, Young1997} for complete reviews on the subject.

Further more, in the end we are studying two coupled Ising models under two different inter spin interaction setups. This is arguably a problem of theoretical interest in itself as well as having other possible applications besides their use in social sciences, for example in the study of plastic phase transitions (and except where stated otherwise, the results are to the best of our knowledge original). Example on previous work done on similar coupled systems can be found in \cite{Galam1988, Galam1989, Galam1990, Galam1995, Imry1975, Korutcheva1988, Simon2000}.

Besides the introduction of {\itshape disorder} or heterogeneity in the population there are many simple and direct possible extensions of this work which may prove interesting, from using different methodologies (Monte Carlo simulations, renormalization group theory), to understanding other social interaction schemes (nearest neighbours or others, complex networks). These will be pointed out throughout the text when relevant.  
 
In the next chapter we will introduce the cause for Social Physics and briefly review its current status and successes. In chapter \ref{cha:isiopi} we will focus on its applications to the study of discrete choice theory, present the problem of interdependent choices and describe our approach in its study. In chapters \ref{cha:2p1cho} and \ref{cha:1p2cho} we discuss and compare analytical and numerical results for the nonlocal and local models respectively. Our main goal will be to describe and understand both models' zero field phase diagrams in a decision making context. We finally present our main conclusions and future lines of work in chapter \ref{cha:conc}.

%% file: secs/socphy.tex
\chapter{Socioeconomic Physics}
\label{cha:socphys}
What exactly do we mean when we say we are seeking to do a \emph{science of society}? In using the word physics besides economy or sociology what is meant is obviously not an attempt to deduce or explain these disciplines from first physical principles in the sense it is used for example in physical chemistry or biophysics (although it can include in some way the search of some {\itshape social first or natural principles}). More often it refers to the more or less direct application of models, tools or results from different fields in physics (in particular statistical physics as we will see) to the explanation of socioeconomic phenomena when it seems reasonable (and of course the opinion of what is or not reasonable is subject to a lot of discussion). 

At a more  philosophical  or epistemological level, it is fundamentally a matter of approach and methodology. It involves thinking about problems the way physicists are used to. Building mathematical models to try and understand a certain phenomenon (generally involving idealisations or simplifying hypothesis of the real system of interest to make the model tractable) and then comparing its predictions about reality with experiment. Then going back again and again to relax some of the initial hypothesis, add some complexity or make some changes to the initial model and checking the conclusions drawn against real data until one feels reassured that it is a completely satisfactory model of the phenomenon under study. One of the keys of the method is that it is empirical, resulting in models or theories trying to be increasingly accurate in explaining reality. The other lays in its mathematical characterisation, which allows for flawless logical deductive reasoning and provides the tools and level of abstraction necessary to build a consistent and interrelated body of knowledge. 

Of course this means nothing more (and nothing less) than using the scientific method and so in no justice can it be attributed to physicists alone. It is common to all pure and applied sciences as well as to many lines of research throughout the history of social sciences. In fact, as we shall see, the development of social and physical or mathematical  disciplines has not been at all unrelated (the term \emph{Social Physics}, for example, was actually coined by the French philosopher Auguste Comte (1798 $-$ 1857), considered by many one of the founders of sociology). 

The current use of the word physics in this context is related to the relatively new {\itshape trend} of growing research in the social sciences context coming from physicists and mathematicians (as well as from trained social scientists with increasing interest and adherence to this approach, and in the best of the cases, from  both working together). It has been a question of noticing that the underpinning framework and tools developed by physicists over the last century, and specially the last decades, to study problems completely unrelated to that of social interactions, could be well suited for the study of some problems in the social sciences field. The key is in the profound revolution undergone by physics from a \emph{mechanistic} and \emph{deterministic} view to an \emph{statistical mechanical} one during the first half of the $XX^{th}$ century. This paradigmatic change (together with what we can call the {\itshape computer revolution}) was in the end what brought about many new research lines (specially during the last decades of last century and to our days) that are now sometimes generically referred to as the study of \emph{complex systems}. This {\itshape new} discipline has undergone a spectacular evolution in the last decades and has found increasing applications in {\itshape foreign} fields such as genetics, behavioural and evolutionary biology, data analysis, neuroscience\ldots, and more recently, the social sciences.    

Another factor that propitiates the emergence of a more scientific approach to the study of society in our days is the availability of huge quantities of data
and the computational means and human expertise to analyse it. The scientific method obliges us to confront our theories with experiment. A remarkable difference between using it as a guiding principle in the social sciences and pure science is that in the latter there is normally (but not always, cosmology may serve as example) the possibility to repeat controlled experiments to verify our hypothesis, and indeed the science of designing better experiments has been of key importance in constructing or current physical understanding of the world. In the socioeconomic context, we seldom get the chance to repeat an experiment under controlled circumstances and very often we can only consider to have a single realization of the experiment.

This makes the study and analysis of available data a challenge in itself. Appropriately collecting and understanding socioeconomic data is a complicated task. The growth of the quantitative social sciences is also stimulating the development of this new approach and the interchange of knowledge between traditionally (at least in the last century) disconnected fields. In fact, quantitative economy, sociology or political science build models out of the data and so the quest is basically the same. If we do wish to make a difference between both approaches, we can consider quantitative social sciences as collecting and extracting the actual trends found in real data related to the problems under study. Socioeconomic physics builds models out of some hypothesis and studies what trends emerge from them. In the end of course, both have to converge as the aim is to find theoretical models out of some defined natural principles or assumptions that do (at least partially) explain the actual general trends watched in reality. This would help us not only understand what is happening, but also why it is, and so possibly give us some intuition on how to change or manage the problem under consideration. Again, let us note that the frontier between both disciplines is a rather artificial  one as the goal is basically the same for both and the tools in many cases interchangeable. In an analogy to theoretical and experimental physics, the difference is sometimes rather diffuse (specially in the {\itshape infancy} of a new area of study).

Let us briefly review the main historical milestones in this approach to the study of social sciences and its current status of development before discussing in some more detail why this kind of {\itshape statistical physics of social dynamics} can prove to be valuable.

\section{Brief historical perspective}
We have considered useful to include this section because it comes to a surprise to a lot of people that this approach towards the understanding of society is in fact not new at all, and that the interconnectedness between the development of physical and social sciences is greater than generally acknowledged. It by no means intends to be a detailed account but merely point out some important names and milestones mentioned by Philip Ball in his account on the subject (chapters 1 and 3 of \cite{Ball2004} and \cite{Ball2002}), to which the reader can refer for more details and full references. 

Although we can trace the first attempts to define something like natural principles of society back to the Greek, the English philosopher Thomas Hobbes (1588-1679) can be fairly said to be where the story begins. In {\itshape De cive} ({\itshape On the citizen}, 1642) and {\itshape Leviathan} (1651) he sought to logically deduce the best form of government out of some natural first principles for individuals composing the society.

Hobbes was a man of the XVII$^{th}$ century, the times of Descartes, Galileo and Newton and the starting point of the Age of Enlightment. He had an extraordinary humanistic formation and the turbulence of his times (with Civil War and Cromwell's rule in England, the Thirty Years War in Europe, the emergence of a middle class and new political ideas) helps understand why he decided to take such a quest. He served as a secretary for Francis Bacon, had a lot of interest in geometry and strong connections to the circle of French mechanistic philosophers, included Marin Mersenne (1558$-$1655) and Pierre Gassendi (1592$-$1655), colleagues of Descartes. He even travelled to Florence to meet Galileo in 1636. Hobbes was a strict determinist, he took a reductionist approach  and had a completely mechanistic view of human behaviour (which comes as no surprise, he was a man of his age) in a way that seems either frightening, senseless or naive nowadays. His basic assumptions were also greatly influenced by his time (that of a perpetual motion of individuals in the seek of power over the others) and so was his conclusion (that for the sake of preservation individuals will be willing to transfer all power to some authority, and that the better option to provide stability was absolute monarchy). It is the methodology that he uses that  makes his work remarkable. There is however no sign of anything resembling math in his work.

The English economist, philosopher, scientist and founding member of the Royal Society William Petty (1623$-$1687) was one of Hobbes disciples and the first advocate of using quantitative measures (data) of society in order to uncover the fundamental laws (in a Newtonian sense) which he supposed to rule our society, in what he called \emph{political arithmetic}. It was his friend John Graunt (business man and Fellow of the Royal Academy) one of the first to start collecting demographic data (basically births and deaths). Looking for trends in this type of data sets became increasingly popular (the astronomer Edmund Halley amongst the interested). By the middle of the XVIII$^{th}$ century the new field already had the name of \emph{Statistics}, although it was considered to have nothing to do with mathematics but rather with tabulating social data.

And then along came the French mathematician Marie Jean Antoine Nicolas de Caritat de Condorcet (1743$-$1794). Under Jean Le Rond D\'{}Alambert's influence,  he began to consider the mathematical theory of probability (first developed in the study of gambling) in its applications to the study of these collections of data to understand different human processes. In {\itshape Essai sur l\'{}application de l\'{}analyse a la probabilit\'e des d\'ecisions rendues \`a la pluralit\'e des voix} ({\itshape Essay on the Application of Analysis to the Probability of Majority Decisions}, 1785), he describes several important results related to voting, such as  Condorcet's jury theorem (which states that if each member of a voting group is more likely than not to make a correct decision, the probability that the highest vote of the group is the correct decision increases as the number of members of the group increases) and Condorcet's paradox (which shows that majority preferences become intransitive with three or more options). He also states his confidence that collecting large numbers of data and developing the appropriate mathematical tools is all that is needed to reduce the science of human affairs to (in his words) ''{\itshape laws similar to those which Newton discovered with the aid of calculus}''. In his last piece of work, {\itshape Equise d\'{}un tableau historique des progr\`es de l\'{}esprit humain} ({\itshape Sketch for a Historical Picture of the Progress of the Human Spirit}, 1793), he describes an incredibly optimistic science based utopic society to which he thought we were headed (which is ever more striking when you consider the fact that he wrote it while fugitive of Robespierre before being captured and condemned to the guillotine). Other XVIII$^{th}$ century remarkable characters interested in some way in this scientific approach towards the understanding of human behaviour are the Baron de Montesquieu, David Hume, Fran\c{c}ois Quesnay and Adam Smith.

By the 1820s, the French mathematician and astronomer Pierre-Simon Laplace (1749$-$1827) and his pupil Sim\'eon Denis Poisson (1781$-$1840) had already found the Gaussian curve in the distribution of astronomical measurement errors and  connected error rates to probability theory. Laplace had also already used the same curve to fit social data. The French mathematician Joseph Fourier (1768$-$1830), at the time director of the {\itshape Bureau de Statistique} of the {\itshape D\'{e}partement de la Seine}, had also applied it on demographic data. It was then when the Belgian astronomer Adolphe Quetelet (1796$-$1874) came to the Royal Observatory in Paris to learn from Laplace and Poisson and became deeply impressed by these ideas. He popularised Laplace's data and was to develop what he called \emph{social mechanics}, an statistics based approach to the understanding of social processes which was to influence Jeremy Bentham, John Stuart Mill and Karl Marx amongst others. The English Henry Thomas Buckle (1821$-$1862) was a follower of Auguste Comte's positivism and a great enthusiast of Quetelet's work. He argued that the statistical regularities observed in the data transcended human intervention and policy making and that this was evidence of  underlying guiding principles (however obscure and difficult to grasp) to all individual actions. This led him to make the case for a science of history in {\itshape History of Civilisation in England} which compiles a large number of regularities in demographic data. Immanuel Kant, Thomas Malthus and Auguste Comte were other XIX$^{th}$ century thinkers involved more or less directly with this scientific approach towards the understanding of human affairs.

Meanwhile physicists were in the quest of explaining the new field of thermodynamics, ({\itshape the study of heat flow} which had appeared after the Industrial Revolution when studying how to make engines more efficient) in terms of Newtonian mechanics.  Quetelet and Buckle were to also have their influence in James Clerk Maxwell (1831$-$1879) when he introduced the Gaussian curve for the velocities of particles in developing the kinetic theory of gases. This began the statistical mechanics revolution in physics.  He eloquently acknowledges this when he writes ''{\itshape those uniformities which we observe in our experiments with quantities of matter containing millions of millions of molecules are uniformities of the same kind as those explained by Laplace and wondered at by Buckle arising from the slumping together of multitudes of causes each of which is by no means uniform with the others}''. Ludwig Boltzmann (1844$-$1906), who showed that any group of randomly moving particles in a gas must converge to Maxwell's distribution, also knew of Buckle's work and drew analogies between particles in a gas and interacting humans. It seems the introduction of statistics in physics was made easier by its previous apparent {\itshape successes} in the scientific study of human societies.

The rest of the story is better known. During the late XIX$^{th}$ and most of the XX$^{th}$ century, quantitative social sciences and physical sciences developed mostly in an independent way. Physics underwent a series of well known paradigmatic revolutions besides (but not unrelated from) the introduction of statistical methods in the construction of theories (and not only in the assessment of experimental results): quantum mechanics and relativity. Quantum mechanics  deepened the philosophical sense of the uncertainty mechanical statistics had already introduced, and probabilities became a fundamental part of the physical world instead of merely encoding our limited amount of knowledge. For decades, physicists set out to explain the physical principles underlying chemistry and then to further explore the basic constituents of matter. Special relativity generalises Galileo's principle of invariance to Maxwell's theory of electromagnetism and breaks to pieces the Newtonian sense of absolute distance and time (replacing them with a constant speed of light). General relativity extends these ideas to Newton's theory of gravity. Quantum field theory was developed and used to build theories both quantum and relativistic for electromagnetic and nuclear strong and weak interactions. Physics of the XX$^{th}$ century has been in justice that of the search of unification of the more fundamental laws of matter and its interactions. While a quantum theory of gravity is still due, we have nowadays an outstandingly unified and accurate comprehension of how the subatomic world and fundamental interactions work, which has certainly brought us closer than ever to understanding our Universe.

Meanwhile, there was also being a lot of work around more prosaic uses of all these new ideas. Knowing how quarks interact and more about the origin of the Universe or black holes was all very well. Having abstract theories about idealised gases that could be traced back to first physical principles remarkable and very useful. But many immediately perceived the great potential of the new statistical mechanical framework for understanding more {\itshape everyday life}, non-ideal systems and so for more practical applications. Understanding the basic principles underlying the existence of ferromagnetism is needed, but it is not enough to know at exactly what temperature will an specific piece of a certain alloy become magnetised. Effective models for different phenomena where developed and improved as new mathematical tools were available and the theoretical understanding of the subject grew. Technological improvements gave us better ways to test our theories and to experimentally uncover new properties of matter to be explained. The study of phase transitions and critical phenomena became a subject of its own. Relations with and applications to other theoretical disciplines such us information theory, game theory, graph theory, nonlinear dynamics and chaos or topology and geometry are being fruitfully explored to our days and have greatly enriched the statistical physics body of knowledge. The development of computers that could be afforded not only by national laboratories and the subsequent improvement of algorithms (particularly Monte Carlo methods) and computing capacity made accessible problems that were formerly intractable.  By the end of last century, what we have referred to as the study of complex systems was a well established discipline, encompassing applications outreaching to non physical areas such as biology or neuroscience. 

During all this time, quantitative social sciences continued to evolve and were also transformed by the computer revolution which made it possible to {\itshape easily} analyse huge amounts of data. There was also some effort on a more theoretical, first principle, physical approach, specially from economists (as one would hope, sociologists have been traditionally more reluctant to neglect psychological and cultural factors as this approach seemed to convey) although they generally did not consider social interactions. Neoclassic microeconomic theory is a good example. Also worth noting is the introduction by the sociologist Vilfredo Pareto in 1897 of power laws to explain wealth distribution. The American  George Kingsley Zipf was probably the more enthusiast of social scientists about a scientific theory of sociology (which he based on a Hamiltonian {\itshape least effort} type principle) during the first part of the XX$th$ century. In his works he collected data and showed they could be fit to a power law for a wide variety of areas: demographic data, travel statistics, marriage data, income distributions, war casualties, language properties\ldots He believed power distributions were the key to explaining societies as the Gaussian curve was to the physical world. Some of these ideas have been lately revived in relation with critical phenomena and their possible applications in explaining human affairs.

At some point the link between some of the models from social quantitative sciences and some statistical mechanical ones and the hint that its tools could be applied to a wide variety of problems brought the two worlds again together. It has only been in the last few years that \emph{socioeconomic physics} has really become an incipient line of research for people coming both from mathematical or physical and social sciences backgrounds.

We will now have a look at the different current lines of research in this area, which will give the reader an idea of the variety of problems being considered and the stage of development of the subject.

\section{State of the art}

In this section we will try to give a loose picture of what is going on and what kind of problems are under attack. It does not intend to be exhaustive but descriptive. References \cite{Castellano2009, Martino2006} are excellent technical reviews and \cite{Ball2004a} gives a nice general picture. Ball's book on the subject for the general public \cite{Ball2004} is a comprehensive and detailed account on its history and current status. The case of discrete choice will not be considered explicitly as it will be explored in detail in next chapter. It does however arise repeatedly in relation to many of the general problems studied. Discrete choice scenarios can be used to model a wide diversity of processes.

\subsection{Economic markets}

It is worth starting by discussing in some detail the role of mathematical and physical models in economic theory. It somewhat has a different history of thinking from that outlined in last section, and is for sure the field within social sciences with an older and more respected tradition in this sense. There have been vast amounts of work produced both during last century (in a more classical framework) and more recently in an statistical mechanical approach sometimes called \emph{econophysics}. Reference \cite{Mantegna1999} gives a complete outlook on the subject. Other recent interesting reviews are \cite{Burda2003, Feigenbaum2003, Stanley2008, Stauffer2000}.

It was Adam Smith in the XVIII$^{th}$ century that first introduced a theory on how markets work, based upon Hobbes' political model but substituting the search of power by that of wealth (with competition acting as a compensating force). In fact, in economic theory, there has been traditionally no concerns in assuming fundamental, immutable laws governing the economy. It can be considered to have remained mainstream since the early XIX$^{th}$ century to our days. Sophisticated mathematical models have been developed and used by economists without encountering the {\itshape suspicion} they have in other social sciences. It has been a traditional, mainstream part of economic research for over two centuries. Another intriguing trait which makes quantitative economy rather different from other quantitative social sciences is that they have, on a general basis, developed theoretical models with little attention to real data until quite recently.

Adam Smith's market was one in equilibrium, and we shall see this concept has persisted in economic theory. It was Karl Marx who first argued for inevitable cyclic crisis in a capitalist system in his {\itshape scientific} economic theory. Marx's is still a self-regulating market (there is a negative feedback), but not in a state of equilibrium. Conventional economic theory has been, however, that of an equilibrium market driven by external forces. Others, such as Keynes, have supported that there were endogenous forces driving the markets. But it  has been traditionally a minority, particularly in what concerns mathematical microeconomic models devised to explain it.

The French physicist Louis Bachelier (PhD student to Henri Poincar\'e) was the first to introduce random fluctuations in explaining the prices of stocks and shares, assuming thus that they fluctuate following Gaussian statistics \cite{Bachelier1900}. His work went unnoticed at the time (it was {\itshape too early} for it to be appreciated by economists or physicists), but it is the basic idea underlying most of conventional market models. But random walk models for prices fail to explain real data. There is a specially marked deviation for large fluctuations. Real price fluctuations follow distributions with {\itshape fat tails}. Mathematician Benoit Mandelbrot introduced L\'evy flights (random walks with occasional big leaps) which introduce the fat tail effect (although does not explain the origin of the high relative frequency of large fluctuations) \cite{Mandelbrot1963}. This description has gradually been accepted by the academic world while traders have usually gone for the simpler, Gaussian models for practical reasons. In fact, data sets in the 1990s already indicated that even L\'evy flight models were not enough to satisfactorily explain data either (they overestimate large fluctuations as opposed to Gaussian distributions that severely underestimate them). Price fluctuations seem to be \emph{scale-free} (they follow power law distributions).

Conventional economic theory, although based on mathematical models and first principle approaches as supported in this work, has remained {\itshape captive} of the idea of a market in equilibrium and Gaussian statistics, which at this point seems evidently disproved. Part of the problem has been the lack of emphasis in confronting theories with experiment, although in most cases the community was well aware of the problems of the models when trying to explain real data. Different attempts of improving the models and introducing more realistic dynamics have been made with different degrees of success or {\itshape usability}.

The introduction of the statistical mechanical frameworks to its study is directed at enriching neoclassical microeconomic theory with more realistic features, especially in the key ingredient that was traditionally ignored: interactions. These models not only account for the large fluctuations observed in real data (the fat tailed distributions), they explain their origin (which is precisely in the interactive nature of the system that makes the global picture no longer a {\itshape blown up} version of individual behaviour). Many of these developments go {\itshape hand in hand} with the introduction of social interaction terms in classical economic discrete choice theory (where choices could be, for example, sell or buy) which is precisely the matter of discussion of this work. For more details on this approach, see next chapter. Here we will briefly comment  two examples of interesting work along other (different but related) lines.

Kirman was one of the first to propose a trading model in which agents can be divided into  \emph{fundamentalists} (who will buy and sell according to the dogma that prices reflect fundamentals) and the \emph{chartists} (which search for trends in empirical data and buy/sell accordingly). Interesting effects which can be seen in these models (and in reality) include \emph{clustering} (groups of traders who buy and sell primarily amongst themselves) and \emph{herding} (tendency to mimic others). Thomas Lux and Michele Marchesi \cite{Lux1999} have studied such a model in which chartists are divided into optimists (who tend to buy) and pessimists (who tend to sell). Agents can change from optimist to pessimist and from chartist to fundamentalist (and {\itshape vice versa}): if another attitude seems to be more profitable, agents will tend to adjust to it. They will also tend to imitate the rest, so the larger, say, the group of optimist chartists is, the more probable than a  fundamentalist or a pessimist may become one of them. Prices are determined by what the agents are doing according to demand and supply as well as by the change in fundamental values, which is still the main driving force of the system. In their model, Lux and Marchesi consider normally distributed fundamental fluctuations. This model gives a realistic macroeconomic picture in which interactions between traders transform a Gaussian variation of fundamentals to a completely different distribution in real price fluctuations.

Another interesting project is \emph{Sugarscape}. Devised by Epstein and Axtell to serve as testing ground for social theories,  it gives a really complex and sophisticated interacting-agents environment. It consists of a toroidal lattice that can be populated by agents who follow the impulse to seek for sugar, which can be also distributed along the grid in a determined manner. Agents will move towards sugar and consume it. All sorts of interaction rules between agents can be defined, including trade which is the case of interest here. Simulations in this environment  suggest that trade, while generating wealth (in that regions with sustained trading relations are capable of supporting larger populations), also tend to skew its distribution amongst the population, as do scenarios where agents have limited or different access to information.

\subsection{Opinion dynamics}

This is the study of how agreement or majorities evolve. The whole of discrete choice theory can be considered under the study of opinion dynamics (when opinions can be characterised discretely), so we will have much to say about this subject in the following chapters. Here we will briefly review some simple models widely explored in the literature. These are in many cases strongly related to the Ising model and thus to the general approach of this work. These models are reviewed in more detail in \cite{Castellano2009}. Note that they are focused on the dynamics of opinion evolution (how opinions spread) while in our study we will be focusing on finding stable states of systems in statistical equilibrium.

The \emph{voter model} is a very simple model that was first introduced in the context of species competition but can be used to mimic the dynamic evolution of the propagation of agreement about binary opinions. A regular lattice of agents with spin or opinion $s_{i}=\pm 1$ is made interact as follows. At each step of time, an agent $i$ is selected at random, one of its neighbours $j$ is then also randomly chosen and agent $i$ is set to conform to it $s_{i}=s_{j}$. This model has drawn a lot of interest because it is one of the few non equilibrium stochastic processes that can be solved exactly in any dimension. It is equivalent to zero temperature Glauber dynamics in one dimensional lattices and to a model of random walkers that coalesce upon encounter. This dynamic always makes the system tend to a greater uniformity. For infinite systems, for one or two dimensional lattices complete consensus is always arrived at, while for three dimensional (or higher dimensional) systems domains of opposite opinions coexist indefinitely.  Finite systems of any dimension asymptotically reach total consensus (in a time depending on the size of the system). Different variations of this model have been studied in the literature: the noisy voter model (which accounts for nonzero temperature), multi-type voter model (associated to multiple choices or Potts rather than Ising variables), introduction of quenched disorder (statistical fluctuations fixed in time), AB-model (in which agents must pass through state AB to get from opinion A to opinion B), introduction of memory\ldots There has also been a lot of work on their definition on complex networks and the implications for different topologies.

Another {\itshape popular} model for binary opinion dynamics introduced by Galam to describe public debates is the \emph{majority rule model}. In this setup, we still consider  spin agents with $s_{i}=\pm 1$. To make them interact, we select a group of $r$ agents (with $r$ selected at each step from a given distribution) and set all agents of the group to the opinion of the majority (with a bias towards one of the opinions needed if $r$ can be even). There is a threshold on the initial fraction of agents with $s_{i}=1$ that will determine which of the opinions prevails. When studied on a regular lattice with neighbours conforming the groups of interaction (with a site picked at random on every step of evolution), for dimension higher than one and when the initial magnetisation is zero, slowly evolving metastable states appear and two natural timescales arise depending on their existence or not. Variants of this model studied in the literature include their dynamics when defined on complex networks of different topologies, models where agents can move in space, introduction of inflexible agents (that never change opinion) or of multi-state opinion. The \emph{majority-vote model} makes strong contact with the models to be studied in next chapter. At each step a spin is selected at random and updated to the opinion of the majority of its neighbours with probability $p$ and to that of the minority with probability $q=1-p$ (zero \emph{noise parameter} $q=0$ is equivalent to the zero temperature Glauber dynamics of the Ising model). These systems show second order phase transitions from an ordered to a disordered state at a value $q_{c}$.

In \emph{social impact theory}, individuals are characterised by a binary opinion and by two parameters that estimate their persuasiveness (capability to make someone change his/her opinion) and supportiveness (capability to make someone keep his/her opinion) and that are drawn from probability distributions (see \cite{Holyst2000} for a short review). Impact of each individual over the others is also dependent on the distance between agents. Random fields can also be included to represent other sources of influence on the opinion of each individual. In the zero field case,  these models lead to a tendency in the population to favour one of the opinions (most spins aligned in the same direction), but with stable domains of spins in the minority opinion state still persisting. If an external random field is considered, these minority domains become metastable. Sophisticated versions of this model have also been studied in the context of opinion dynamics, with modifications to include learning of the agents, different types of agents or the effects of a strong leader. Another fancy extension is a model that considers agents that are Brownian particles in motion with some internal energy that allows them to move and that interact through a scalar opinion field.

In the \emph{Sznajd model}  a one dimensional spin chain is considered where a pair of neighbouring spins are chosen randomly. If they both share the same opinion, they convince both of their remaining neighbours. Depending on the particular model considered, there is an additional interaction rule imposing that if the chosen pair disagrees, each agent convinces the other neighbour (this in fact can be seen to be equivalent to voter dynamics). The only steady states of this system are those of all spins in the same direction or half pointing in each direction. These models have been extended to higher dimensions, synchronous updating (where frustration can occur and complete consensus is prevented), its combination with other convincing strategies, in complex networks of different topologies\ldots If we introduce  more than two possible states of opinion we can use \emph{bounded confidence} (only agents similar enough interact) conditions and in such setup, for more than three options, most final states will include at least two of them. The Sznajd model has been used in marketing models and in commercial competition scenarios. It has also been used to explain voting behaviour (see \cite{Bernardes2002}, which will be mentioned again in next chapter).

\emph{Continuous opinion dynamics} is also a topic of interest (\cite{Lorenz2007} is a recent review). Bounded confidence assumptions can be made. The general system is that of a complex network with agents at its nodes each with a continuous opinion variable $x_{i}\in [0,1]$. Agents are made interact only if the difference in their opinion is bellow a certain determined tolerance in which case both agents draw their position closer by another set quantity. In the case of the \emph{Deffuant model}, at each time step an agent and one of its neighbours are selected at random, which is a suitable approximation for large populations in which people meet in small groups. In the \emph{Hegselmann-Krause model}, the randomly chosen agent at every time step interacts with all of its neighbours (with opinions similar enough to its own),  which should be better to describe consensus dynamics through formal meetings involving a larger number of individuals. In both cases the final configuration is a collection of Dirac's deltas. The number and size of the clusters depends on the specific values of the tolerance and the amount by which interacting agents adjust their opinions to that of their neighbours. These models have also been studied on different topologies, on discrete rather than continuous opinions, with stochastic tolerances, introducing external perturbations\ldots

Nowadays computational analysis and Monte Carlo algorithms allow to simulate the evolution of an opinion dynamical system with nearly as many particularities as one would wish to include, and so a wide variety of more complex dynamics than exposed here has been considered. Most recent approaches start to include \emph{noise} (understood as statistical fluctuations or disorder) either in the form of \emph{annealed} (evolving in time with the temperature) or in other \emph{quenched} (fixed in time) manner (or both). The general framework we will expose in next chapter allows to reformulate most of the models presented and naturally allow for the introduction of both annealed and quenched disorder.

\subsection{Cultural dynamics}

Although some other settings have been occasionally considered, most physical modelling on the dissemination and acquisition of cultural traits has been related in one way or the other to \emph{Axelrod's model} of cultural dynamics \cite{Axelrod1997, Castellano2000}. 

The basic version of the model consists of a vectorial extension of an opinion dynamics model of the type discussed in the previous section. It characterises each \emph{culture} with a list of \emph{features}, each of which can have a value (\emph{trait}) out of a finite set of choices. These represent language, religion, gastronomy\ldots or whatever we want to use as defining a culture. Agents (that can represent in this case different cultures) are set on a two dimensional regular lattice. At each time step, two neighbours are chosen at random and, with a probability proportional to the \emph{cultural overlap} (number of features that have the same value), update the value of a randomly selected feature to that of their neighbour. It turns out the number of different cultures in the final state  grows with an increasing number of traits per feature (which decreases the probability of interaction) and decreases if the number of features per culture are raised (higher probability to interact). It first increases and then decreases when raising the lattice size (small grids cannot support diversity because of spatial limitations, large grids make the evolution last longer thus favouring assimilation). In fact the change between the single predominant culture picture and that of different cultural islands shows many characteristics of a second order phase transition. At the critical point, the size distribution of the different cultural regions follows a power law form. 

Extensions of these models that have been considered in the literature include complex networks of different topologies, random noise, scenarios in which interaction between agents depends on the trait of the majority of its neighbourhood (which makes the model more robust to the introduction of noise) and the effects of external fields.

\subsection{Social diffusion}

Other interesting models concerning the spread of opinion or of certain behaviours are those with an \emph{epidemics} perspective. These models have deep relations to the ones already reviewed in this section and to those that will be the centre of our discussion on discrete choice on next chapter.

In \cite{Campbell}, Campbell and Ormerod propose a simple model for the spread of crime. Agents are divided into three groups: individuals not susceptible to crime under any circumstances, criminals and potential criminals. Susceptibility of the honest but subject to temptation depends on the relative size of the other two groups, making them more prone to crime if they are in a high crime background. The probability of an individual becoming a criminal is also linked to the level of social deprivation and the toughness of the possible punishment. The phase diagram on those parameters is studied. They find large regions of the parameter space where there are two possible equilibria, one related to a low crime incidence and one to a high one. The phase diagram in fact looks very much like that of van der Waal's model of gases. Multiple states are then possibly associated to first order transitions and can be metastable. In any case, in some regions of the parameter space, small changes in social conditions or crime punishment policies may provoke abrupt changes in the level of criminality. In the rest, however, even large changes on these will have little effect.

The same authors use a similar model to study the prevalence of marriage in a society depending on social attitudes (favouring it or considering it old fashioned) and economic incentives (such as tax discounts)\cite{Ormerod1998}, showing the same characteristics.

These simple models already illustrate how an abrupt change on society can be driven by gradual change in {\itshape external factors} and how specific socioeconomic circumstances can determine how effective different policies may be. Many of these features will come up again in next chapter when discussing discrete choice theory with social interactions.

\subsection{Crowd and traffic dynamics}

A lot of fruitful work has been done on understanding actual dynamics involving humans, both as pedestrians (see \cite{Helbing2000} for an example) and in vehicles (traffic flow, \cite{Helbing1999} is an interesting example).  We will not get into many details but briefly comment on the usual approach. Work along these lines can help design safer and in general more appropriate spaces for large numbers of people and to better manage traffic congestion.

Models of crowd dynamics generally assume that each individual has a particular destination and a particular preferred speed and that he will keep both unless he needs to avoid colliding. These simple models already give rise to nice emergent features, such as walkers tending to traverse corridors organised in right and left streams. If the agents are trying to get out from a room, a phase transition is observed as we raise the value of desired average speed to a state in which the rate of exit drops (panic situations, see \cite{Helbing2000}). Some other work focus on how trails are formed across empty open spaces when there are restrictions on the access/exit points.

Traffic models make the basic assumptions pedestrian models do, but the additional restriction of one dimensional motion makes it easier to model. In general, these seem to point at the existence of a critical traffic density over which jams set abruptly. An additional {\itshape traffic phase} has been proposed (synchronised flow, dense traffic flowing rather smoothly), that provides an image similar to van der Waals theory of the solid (traffic jam or no-flow phase), liquid (synchronised flow phase) and gas (free-flow phase) states of matter. The effect of {\itshape inhomogeneities} such as junctions or traffic lights in the system's dynamics has also be studied with very interesting results \cite{Helbing1999} and {\itshape traffic phase diagrams} constructed.

\subsection{Group growth and alliances}
\label{subsec:groupgrowth}

A common trait in human behaviour is a tendency to form groups and alliances. These (objectively or subjectively) provide protection and assurance, and in many cases the only way to engage in certain enterprises which need of a collective  summation of efforts.  These can be studied in many different contexts. Here, we will briefly describe three interesting examples. 

In \cite{Ma2009} a model is proposed for the {\itshape recruitment} of non-profit making and benevolent-based organisations. This chosen setup is also particularly close to the ones we will be discussing on next chapter. In this model, agents can choose to join the organisation and cooperate, join it and \emph{free-ride} (take advantage of the benefits of belonging to the group without cooperating) or not join the group. Each agent has an estimated value of the community. The community has an additional value proportional to the fraction of the population already engaged in it and an extra added value for the fraction of these which are cooperators. So both cooperators and free-riders make the group look more attractive to others, but the more the larger the number of cooperators is. A fixed cost is assumed for all cooperating individuals. Each free-rider will have an (individual dependent) cost that depends on his own perception of the {\itshape wrongness} of his actions (moral burden) and which is also proportional to the fraction of the population that is cooperating (the more everybody is cooperating, the worse the social image of {\itshape slackers}). The relationship between the level of cooperation amongst the population and the size of the community are studied. The dynamics of such systems are explored both in parallel and random sequential updating. These systems present two types of equilibria. There are fixed points (which in this case are Nash equilibria) with a mixed population of cooperators and free-riders. And there are also cyclic solutions with periodic variations in the level of cooperation and size group. This model provides two simple mechanisms for such organisations to persist in the presence of free-riders.    

Another interesting example is Axtell's agent modelling approach to the growth of businesses \cite{Axtell1999}. Data strongly suggests power law distributions for both  size (as in number of employees) and growth rates (see \cite{Stanley1996} for a study on the growth rates of all publicly traded US manufacturing companies between 1975 and 1991). In Axtell's model, agent's utilities depend on the amount of work the agent is doing (which generates money) but also on how much time they get for leisure. Heterogeneous populations are considered (different agents have different preferences about money and time). Each agent has information only about him and his {\itshape neighbours} or {\itshape friends}. The model has also built into it the condition that agents get back more for the same time cost if they join efforts. Each agent gets to make a decision, in his perceived best interests (maximising its utility), at randomly chosen time intervals. It can decide to stay as it is, join a friend's company or start their own firm.

The most remarkable thing about this model is that it does not specifically impose \emph{increasing returns} ({\itshape the bigger the better}) which is normally a premise of most models (and the standard, {\itshape un-mathematical} theory) of firm growth. It can be expected from the hypothesis of being better off joining efforts. But it can not be guaranteed because agents adapt their level of effort to the given situation. When a slacker joins a big firm of hard workers, he can lower his amount of effort and still everybody's returns won't be significantly affected (but will decrease).

The system studied presents no Nash equilibria, there is a constant process of firms appearing, showing different growth rates, and disappearing. It shows the familiar {\itshape booms} and {\itshape busts} we sometimes see in reality. In fact it generates power law distributions both for sizes and growth rates that can be used to fit the data of about twenty million US firms is 1997 \cite{Axtell1999}. It has been (to our knowledge), the only microeconomic model that correctly predicts the scale free behaviour observed in real data.

In Axtell's model, business growth is attained through its ability to attract and retain hard working employees and are condemned when they become dominated by slackers that will try to get off doing as less as they possibly can. This is of course a very rudimentary model which disregards many significant factors. For example, there is no mechanism to fire agents from a firm. The fact that it does fit the data so well, can be used to argue in favour of the importance of {\itshape looking after} your good workers (which is increasingly popular in some management sectors). It suggests this may be the key factor underlying success (or survival) rather than the ability to maximise profit. More specifically, firm death is self-induced in this model and unrelated to external factors, yet it still gives a correct picture of reality. Other interesting considerations can be made with respect to agent's average utility in different sized companies. For example, {\itshape workaholics} are usually better off in relatively small companies which are less likely to be predated by slackers, still  utility is in general higher than the average for employees of very big firms.

Another very interesting feature of this model is that its appropriateness to fit real data can be used to illustrate  a sort of \emph{universality} in modelling social processes. It seems the only basic ingredient needed for the power laws to emerge is \emph{purposiveness} (utility maximising). If agents make choices at random or adjust their efforts at random these do not appear. They do still appear (with a different slope), if we vary other features of the model, such as neighbourhood size (knowledge of the market) or more evidently favouring increasing returns of scale. This {\itshape lack of importance of the details} is what is meant by universality and it could explain how such an {\itshape unrealistic} model can make correct predictions about real data.

The last example we will mention is the \emph{landscape model} devised by Axelrod and Bennet to explain how alliances are formed \cite{Axelrod1993,Axelrod1995}. For the problem of interest, a list is made with the relevant characteristics that will be considered to affect the alliance making process. The degree of {\itshape attraction/repulsion} between each two agents will depend on these factors. The energy of this system is then studied, under the constraint of only two alliances been formed, for every configuration,  and Nash equilibria of the system determined. Axelrod and Bennet have used this model to explain two completely different scenarios: company alliances regarding Unix standardisation in the 1980s and the formation of political alliances previous to the Second World War.

In the case of the standardisation of Unix, agents are companies that are attracted to each other depending on their size. There is also a source of repulsion based upon how much products and markets overlap between each two firms. There are nine firms in total and the state of minimum energy of the model is very close to the real outcome (note there are 256 different configurations).

For pre-war Europe, seventeen countries were analysed in 1936, considering as relevant factors economy, demography, ethnicity, religion, territorial disputes, ideology and past history. Under the two alliances restriction, this means 65536 possible configurations. The one of minimum energy is  very close to the real outcome, with only Poland and Portugal misplaced. The energy landscape shows another local minimum (metastable state) in which all countries are united against the Soviet Union, Yugoslavia and Greece. Remarkably, when this model is recalculated for 1939, the {\itshape anti-soviet} metastable state is not present anymore. It can be understood as history going through a kind of \emph{spinodal point} (where metastable states appear/disappear) after which the anti-soviet alliance is no longer possible.

\subsection{Other research lines}

The number of research lines including in some way this physical approach to the understanding of human nature is large and increasing and it seems impossible to give a reasonable account of all without extending for too long. Let us simply mention some of the remaining before we close this section.

\emph{Cooperation} has drawn a lot of interest and active research. It involves the study of what conditions are necessary for cooperation to evolve and persist in some stable form, generally using evolutionary game theory dynamics. See \cite{Axelrod1981} for a {\itshape classical} paper. 

Another very interesting topic is what we can call \emph{language dynamics}. Its concerns are all processes related to the emergence, evolution, interaction and extinction of languages. In some approaches, communication strategies are considered innate and genetically transmitted. Successful communication gives individuals selective advantage and thus evolutionary game theory is usually used in this setup. Other lines of research consider language in itself as a complex dynamical system that self-organises and evolves in time. {\itshape Language complex networks}, for example, have shown very interesting features that give insight on language structure and acquisition. Work has also been done considering a community of language users as a complex dynamical system that collectively develops a communication system (based on interactive communication problem solving).

The study of \emph{complex networks} should deserve, not only a subsection of its own but a whole chapter. They are omnipresent in all of social physics in that we can generally define any problem on a network of a given topology and use it to model the exact relations each individual has. And the study of their topology and dynamics is on its own a matter of both theoretical and sociological interest. Models with \emph{co-evolution} of the network topology and of the individuals at its nodes are also recently being considered. A more thorough review should be in order but is out of the scope of this work. See \cite{Newman2000, Newman2003} for introductory reviews.

\section{Why bother?}

Is there really any point in trying to use mathematical models to explain human behaviour? Can statistical mechanics provide us with a framework in which to actually learn something useful about societies? We hope that at this point, the reader will at least be persuaded that there is actually a lot of interesting research (though maybe not too much on a practical side yet) going on. In this section we will try to argue on more general or philosophical grounds why we should expect statistical mechanics to provide the adequate tools for the investigation of such problems. Why it makes sense to study this type of problems in this setup.

Statistical mechanical methods can be employed when we need to treat the behaviour of a system concerning whose condition we have some knowledge but not enough for a complete specification. We will be rather studying the average properties of a representative ensemble of systems, i.e., the average behaviour of a collection of systems of the same structure as the one we are interested in but distributed over a range of different possible states. If we consider the conceptual space spanned by all the degrees of freedom of all particles, each specific system will be represented by a point and the ensemble can be characterised by the density of these points (systems) in this space. The postulates of statistical mechanics state that all systems compatible with our knowledge of the system of interest should be considered and all with identical probabilities (and equal a priori phases in the case of quantum mechanical statistics). Have a look at \cite{Tolman1938} for an excellent text book on the principles of statistical mechanics and the statistical description of thermodynamic principles. 

The important point here is that statistical mechanics, more than a physical theory, is a theoretical framework in which to study the expected average behaviour of a system about which we have limited information or about which there is uncertainty. It is the natural way to study collective trends emerging from (many times simple) interactions between large numbers of individual entities (which yields impossible to have a complete specification of all of the individual's characteristics). It gives us the right tools to study systems where the behaviour is more complicated than that arising simply from {\itshape adding} all individual effects, where interaction counts. Indeed, complicated emergent unexpected trends can be found in many systems interacting under very simple rules. Statistical physics provides  tools specifically suited for the study of such \emph{self-organised}, emerging properties. 

Trends do definitely emerge in human societies. Despite how complex individual's psychology may be, in many situations, human behaviour tends to be collectively predictable at least to some extent. And social norm is important although different individuals may have a different perception about it or may be less susceptible to it. Hardly no one decides to go to work wearing a tie around their head (although considerably more tend to find it appropriate after the Christmas party has been going on for a while). It is easy to predict with a considerable rate of success that a judge will not be wearing a white wig to court, as it was in the past to predict the contrary. The fact that trends appear when analysing human behaviour is related to considering large numbers (that {\itshape smooth out} individual characteristics) and to the fact that on a wide variety of problems we really only have a very limited amount of options. No matter how complicated or intricate  individual's {\itshape inner worlds} or psychological traits are or how different their social or cultural backgrounds are, in a yes or no referendum, all have to settle for yes or no (or for abstention if you wish). There is no need for assuming limited \emph{free will}. It is the fact that it can only manifest through a yes or a no, together with considering large populations, what makes patterns emerge.    

Of course in many cases we are drawing direct analogies between humans and identical particles subject to immutable laws governing their action. So skepticism about how good this approach can really be is more than fair. However, physicists have been studying interacting systems of all sorts for decades and it thus seems reasonable that there is \emph{something} that can be learnt from that experience that can be of use for  at least some problems concerning the social sciences. Besides, models can be adapted or complicated and still the tools of statistical mechanics be appropriate for their study. And toy models do have the virtue of providing us with insight of what qualitative characteristics emerge under different circumstances and which can be expected to be {\itshape inherited} by more sophisticated models. For example, Schelling's model of racial segregation \cite{Dall'Asta2008, Gauvin2009, Schelling1971} may not give a particularly good fit for real data on neighbourhood racial segregation (as is expected taking to account its simplifying assumptions), but it does point out an underlying simple mechanism by which these patterns can emerge even when the population's tolerance towards integration is high. Paraphrasing Einstein, a good model will be the simplest possible one, but not simpler. In the social sciences context, its usefulness and applicability as well as its capacity to explain real data or uncover basic qualitative features underlying the process, is what should make a model {\itshape good enough}. 
 
What are the {\itshape symptoms} of an interactive statistical mechanical system and are they present in socioeconomic human affairs? Interactions introduce strong nonlinearities and can produce highly unpredictable behaviour even on deterministic hypothesis. Complex systems sometimes undergo \emph{discontinuous or first order phase transitions}. These are abrupt changes in the collective state of a system due to \emph{nucleation} (small clusters of a distinct thermodynamic state or a qualitatively different social situation). Under certain conditions, statistical fluctuations make groups of the particles or individuals shift to a new state (the clusters we have referred to) and they then {\itshape drag} most of the rest behind. As they are associated with metastability, they give the possibility for multiple equilibria and \emph{irreversible} changes in the state of a system by unmaking the changes (\emph{hysteresis}). These type of situations can be found in human societies, where a general trend can change rapidly after some \emph{critical mass} has been attained (some examples have been briefly reviewed in last subsection).

Another type of change that an statistical mechanical system can undergo are \emph{continuous or second order phase transitions} (sometimes also called critical transitions). Although these are also associated to statistical fluctuations and multiple equilibria, the mechanism that drives the change in this case is different from that of first order phase transitions. They are related to  the system arriving at a \emph{critical point} where it must {\itshape decide} between two possible ways to go (an additional symmetry appears). Which equilibria will be favoured will be determined through statistical fluctuations. The qualitative state (phase determined by an order parameter) of the system changes in a drastic way at a critical point as the order parameter goes continuously to zero. At the critical point, different quantities describing the system (heat capacity, magnetic susceptibility \ldots) show different types of divergences characterised by critical exponents. These describe (scale free) power laws as those considered by Zipf and Pareto and that are present in many social data collections, from wealth distribution to casualties in war to scientific citations (as we hope the last subsections have helped to convey). The underlying mechanism for these distributions for some socioeconomic problems could be understood as the system operating near a critical point. There are some dynamical systems whose evolution {\itshape prevents} them from escaping criticality, through different (power law distributed in size) collapses (fluctuations) that only bring the system again to the verge of collapse. This type of out of equilibrium behaviour is called \emph{self organised criticality} and could also be understood as the underlying explanation of some of the distributions that appear in studying human behaviour. Another interesting feature about critical phenomena is their \emph{universality} (as discussed in subsection \ref{subsec:groupgrowth}). Critical exponents are the same for large groups of systems (which can be divided into universality classes). This can be used to reinforce our general argument when using statistical physics to study social dynamics that sometimes {\itshape details do not matter}. See \cite{Domb, Ma1976, Stanley1971} for  detailed accounts on the modern theory of phase transitions and critical phenomena. Very recent and interesting texts are \cite{Sethna2006} in the context of complex systems and \cite{Pettini2007} in that of geometry and topology.

Most of the skepticism (our outright rejection) about mathematical models or scientific theories about society is related to the notion that they represent a mechanistic and deterministic picture of how humans act. In essence they are related to what they imply about \emph{free will}. Any model of human processes does indeed involve a great degree of simplification (that is needed if we want our models to {\itshape give back more information than is put into them}). But statistical mechanics does leave room for all that complexity. It is exactly the right tool when dealing with such uncertainties and {\itshape disorder} can be coded into statistical mechanical models in many ways (see section \ref{sec:isiopisin} for a discussion on \emph{socioeconomic temperature}, for example). In fact, in many cases the use of this framework is aimed precisely at enriching previous socioeconomic models in this direction (specially regarding market economics and discrete choices). As well as providing a natural way to introduce interactions between individuals and study emergent properties, which in a very fundamental way, could even be considered to be what defines society as such.

Of course we have to take every caution about these models and the interpretations that arise. The objective is not to find a universal equation that encodes all human behaviour or anything of the like. Physics of society will not give answers to all questions but rather a consistent framework that, in some contexts, may provide us with a deeper understanding of the nature of human societies. It is however daunting that such virulent criticism should be focused sometimes on this type of approach when nowadays such rigour is seldom asked even from the more high level decision makers. It is strikingly easy (simply turn the TV on) to find political and economical analysts and {\itshape experts} making categorical affirmations about reality based on nothing more that their own {\itshape ideology} (which many times resembles more a desire of reality to conform to their wishes than a structured framework of reasoning). Yet many times these are accepted as {\itshape dogma} by political elites while dismissing genuinely honest attempts (both based on theoretical models and on empirical data) to understand the facts for being {\itshape too simple}. The really good part of using the scientific method is that, if done correctly and honestly, the worst that can happen is that we find out we were wrong and that we have been loosing our time. In these days, when a systematic way of encoding our knowledge about human affairs is probably in more urge than ever, it seems worth taking the risk. The main value of socioeconomic physics should be that of challenging preconceptions  and  generating new hypothesis about how human society works.

%% file: secs/isiopi.tex
\chapter{Discrete choices and the Ising model}
\label{cha:isiopi}

The formal study of discrete choices has been a subject of interest to economists (and more scarcely and recently to political scientists and sociologists) for the last decades. The theoretical background was set up by pioneers such as the economist Nobel Laureates Daniel McFadden \cite{McFadden1974, McFadden1984} or Thomas Schelling \cite{Schelling1971, Schelling1973}. Most of the work produced is intended at the end of the day for data analysis and its application to identify the specific {\itshape weights} of several explanatory variables. The formal underlying approach is always that of agents maximising an (individual) utility function with a deterministic and a stochastic part. Traditionally, direct interactions between agents affecting the outcome were not considered (some of Schelling's and Becker's work are remarkable early exceptions \cite{Becker1981, Schelling1973}, see \cite{Dall'Asta2008,Gauvin2009} for statistical mechanical descriptions of Schelling's model of racial segregation)\footnote{The different language conventions between the social sciences and physics can raise some confusion. In the traditional discrete choice literature, when there is a term accounting for social interactions, this has sometimes been referred to as interdependent or with externalities. Please note the words are not used in the same sense in this work, where interdependence refers to coupling over different choices (rather than on the same choice of different individuals) and externalities to terms unrelated to choice or social interaction, i.e., external fields in the physical sense or \emph{private utilities} in the socioeconomic literature.}. Even if this particular framework was in general started to be applied to problems from sociology later than in economy, it was in the context of sociological problems that social interactions were first introduced.

The use of more or less sophisticated models from condensed matter in the study of discrete choices can be motivated on two grounds. On the one side, drawing an analogy between ferromagnetic models and a collective decision making process is quite straightforward. In the case of binary choices for example, we can consider the magnetic moment or spin of a particle as representing the choice of an agent, option one for spin up, option two for spin down. Social interaction is then introduced naturally in the model as the particles' magnetic moments interacting and tending to align. Depending on the exact model of ferromagnetism used we can account for different interaction setups, either each agent interacting with a determined subset of the rest or each agent with all of the rest. The latter is equivalent to using mean field approximation. Sociologically, it involves social pressure on individuals is through their (accurate) perception of what the average behaviour of the group is. Fields can be introduced to mimic factors besides the desire to conform to the group. As we have mentioned, randomness  can be introduced (both in the field or the coupling) to encode the particularities of individuals towards the choice through probability distributions characterising the group, models which have also been already extensively studied in physics \cite{Cohen1985, Goldenfeld1992, Mezard1987, Parisi1992, Young1997}. 

On the other hand, it were social scientists who first introduced interactions in the traditional utility approach. They were the first to notice the similarities and to start using the tools provided by statistical mechanics in their problems. As we will see, some models from condensed matter have a direct translation to models independently studied in the social sciences scenario. 

The intimate relation between both approaches (which we will soon discuss) opens two doors. Socioeconomic utility defined problems which have no equivalence with a well known (or even defined) model can still benefit from the tools of statistical mechanics (see for example \cite{Nadal1998} for a physicist's approach to a non interaction framework for buyer behaviour where the appropriateness of the logit choice function is argued upon on terms of a {\itshape maximum entropy} principle). Models already studied extensively in a condensed matter context  can be {\itshape translated} to social language and see if they make any sense. In our work we will be basically using the simplest of all models, the constant Ising model (of which all other models can be considered more or less sophisticated modifications) and see what it has to tell us about groups of people making interdependent decisions. 

In this chapter we explain in some detail the relation between socioeconomic utility and physical ferromagnetism, motivate the introduction of coupling over more than one choice type and define the general model of interest for the study of two coupled binary choices. We finally sum up well known results and their interpretation in a socioeconomic context for the single Ising model which we will be using as a basic {\itshape building block} for our coupled models.  

\section{Models from condensed matter and the socioeconomic utility scenario}
\label{sec:isiopisin}

We will focus on the binary choice case. The extension of the socioeconomic utility approach described bellow to an finite set of options ($>2$), while posing some technical challenges, is quite direct. On the statistical mechanical side, we can also find models which could be understood to encode such behaviour such as the Heisenberg or Potts model for ferromagnetism. The  study of multiple choices is of great interest in many problems from the social sciences (voting in a multiple party scenario or in a two party system with strong abstention,  choosing from different means of transport, on which shop of the mall to make your shopping, which brand to choose, which neighbourhood to live in, which labour union to join, which professional option to take, how many children to have\ldots). However, many of the problems are (or can be reformulated in terms) of binary options, ranging from the incidence of \emph{social pathologies} (whether to abide the law or not, to become a teenage mother or not, to drop from school\ldots) to economic demand (whether to buy or not a certain good, companies electing from one of two production technologies, free vs owned software\ldots) to political science (whether to vote or not, to vote yes or no, to support or not any particular movement or idea\ldots) to opinion dynamics (whether to agree or not on any proposition) to other social decisions (to live in the city or the suburbs, to have children or not, to marry or not, to go to university or not\ldots).

Any binary group choice problem can be generically described by the maximisation problem of the individual's utility function V:

\begin{equation}
max_{s_{i}\in\{-1,1\}}\;V(s_{i}, h_{i}, E_{i}(\vec{s}), \epsilon_{i}(s_{i}))
\end{equation}

\noindent where $s_{i}$ is the choice made by the individual $i$ (with alternatives coded -1 and 1) which are the components of vector $\vec{s}$. $E_{i}$ represents the i agent's  subjective belief on the choices of the rest of the agents, $h_{i}$  characterises the agent's personal attributes or preferences  and $\epsilon_{i}$ an (individual and choice dependent) random shock.

To go any further, the assumption of an additive decomposition of the utility function  is made

\begin{equation}
V(s_{i}, h_{i}, E_{i}(\vec{s}), \epsilon_{i}(s_{i}))= U(s_{i}, h_{i},E_{i}(\vec{s}))+ \epsilon_{i}(s_{i})
\end{equation}

\noindent where U is the deterministic utility and $\epsilon_{i}(s_{i})$ is the random utility. 

When interactions are not considered (neglecting dependence on $E_{i}(\vec{s})$), a typical approach to the problem is to consider $U(s_{i},h_{i})=h_{i}s_{i}$ yielding the maximisation problem

\begin{equation}
\label{eq:unoint}
max_{s_{i}\in\{-1,1\}}\;h_{i}s_{i}+\epsilon_{i}(s_{i})
\end{equation}

\noindent where $h_{i}$ measures the deterministic difference in the payoff of the two choices for agent i and is sometimes referred to \emph{idiosyncratic willingness to buy or adopt} (IWA). As it will son be clear, this is equivalent to the Weiss model without spin interaction where $h_{i}$ is an external field.

At this point different assumptions can be made for the probability distributions of $h_{i}$ and $\epsilon_{i}$ depending on the particular problem, a common one being a logistic distribution for the difference of the random payoff terms $\epsilon_{i}(-1)-\epsilon_{i}(1)$.

Economists such as Durlauf, Brock and Blume developed a theoretical framework for the study of such problems with social interactions in the 1990s and noted their similarities and direct relations to statistical mechanical models \cite{Blume1993, Brock2001, Durlauf1997, Durlauf1999}. What follows intends to give a comprehensive summary of the basic ideas and the reader may refer to these references for more details.

Introducing a new term in the utility function: 

\begin{equation}
V(s_{i}, h_{i}, E_{i}(\vec{s}), \epsilon_{i}(s_{i}))= u(s_{i}, h_{i})+S(s_{i},E_{i}(\vec{s}))+ \epsilon_{i}(s_{i})
\end{equation}

\noindent where $u$ is the private deterministic utility, $S$ the social deterministic utility and $\epsilon_{i}(s_{i})$ the private random utility. We can thus rewrite \eqref{eq:unoint} as

\begin{equation}
max_{s_{i}\in\{-1,1\}}\;h_{i}s_{i}-E_{i}\left(\sum_{j\neq i}\frac{J_{ij}}{2}(s_{i}-s_{j})^{2}\right)+\epsilon_{i}(s_{i})
\end{equation}

\noindent where $J_{ij}$ codifies how much agents i and j want to conform to each other in their opinions ($J_{ij}>0$). When {\itshape translated} into a ferromagnetism model, the first term corresponds to an external field, the second to spin interaction terms and the last is random noise.

Under the logistic assumption described above for independently distributed random shocks

\begin{equation}
P(\epsilon_{i}(-1)-\epsilon_{i}(1)\leq z)=\frac{1}{1+e^{-\beta_{i}z}}
\end{equation}

\noindent with positive $\beta_{i}$, we have a complete enough specification to define conditional probabilities of an individual's choice (conditioned on his characteristics and expectations). We can rewrite the social utility as $S(s_{i},J_{ij}, E_{i}(s_{j}))=\sum_{j\neq i}J_{ij}(s_{i}E_{i}(s_{j})-1)$ ($s_{i}^{2}=1$) and taking into account that $P(s_{i}|h_{i},E_{i}(s_{j})\; \forall j\neq i)=P(V(s_{i}=1)>P(V(s_{i}=-1)))$, we arrive at the expression:

\begin{equation}
P(s_{i}|h_{i},E_{i}(s_{j})\; \forall j\neq i) = e^{(\beta_{i}h_{i}s_{i}+\sum_{j\neq i}\beta_{i}J_{ij}s_{i}E_{i}(s_{j}))}
\end{equation}

We can use this to calculate the probability of any given opinion configuration of the system $\vec{s}$. Individual choices are independent once you condition them to all $h_{i}$, $E_{i}(s_{j})$ and so this is only equivalent to the product of the probabilities for all agents. The system can be characterised by the expectation value of the choice or average magnetisation (which is related to the fraction of the population with $s_{i}=1$) $s=\frac{1}{N}\sum_{i}s_{i}$ where $N$ is the number of individuals. We can calculate it as the sum of possible values (1 and $-1$) times their probabilities (and normalise to total probability one). This yields the set of $N$ equations

\begin{equation}
\label{eq:eqsi}
E(s_{i}) = \tanh\left(\beta_{i}\left(h_{i}s_{i}+\sum_{j\neq i}J_{ij}s_{i}E_{i}(s_{j})\right)\right)
\end{equation}

The parameter $\beta_{i}$ can be set to a constant value $\beta$ for all individuals through a redefinition of the rest of parameters. If $E_{i}(s_{j}) = s$, $J_{ij}=\frac{J_{0}}{N}$ and $h_{i}=h$ for all agents, the average choice of the population can thus be expressed as the equation of state of the Curie$-$Weiss (mean field) Ising model: 

\begin{equation}
\label{eq:eqs}
s=\tanh(\beta(h+J_{0}s))
\end{equation}.

It is thus obvious that we can understand each individual's utility $V$ as the analogous of the particle's energy  (with opposite sign, maximising utility will be equivalent to minimising free energy). Deterministic private utility is related to external fields in a ferromagnetic setup, deterministic social utility to spin interaction terms and private random utility is a random noise (which a direct relation with temperature in the canonical ensemble as will be discussed bellow).

Equivalence between the subjective expectations and the mathematical expectation $E_{i}(s_{j}) = s$ is what in the socioeconomic literature is referred to as \emph{rational expectations} of the agents. They all have an accurate understanding of what is going on with the rest. This can be a good approximation for many situations where there is plenty of public information available for all individuals  or when the group of interaction is limited to a small enough number of other agents. Besides, at least part of any distortion to this rationality about the others' beliefs can be built into the $J_{ij}$ or $h_{i}$. In any case the effects of this kind of irrationality related to the knowledge available to the individuals is of a lot of interest (for example in trade market decisions, or in including some media effect shifting the group's perception). An example of work on these lines can be found in \cite{Collet2010}. 

In assuming constant $J_{ij}$  for all $i$ and $j$ we are considering that each individual only cares about what he thinks is the expected choice of the group, i.e., he is only influenced by his perception of the average degree of acceptance (or the fraction of the group that is going for it). The introduction of randomness in this coupling is certainly of interest \cite{Durlauf1997, Weisbuch2000} but constant interactions seem a good starting point. Specially so when you consider much more variation across the group in their private than in their social utility (which seems a natural assumption in many cases). This makes it equivalent to a mean field model, as all agents are interacting with all other agents (and with all of them in the same way). We can therefore also modify this setup in making each individual interact with a subset of the others (only a finite number of $J_{ij}\neq 0$ for each agent $i$), which can be translated to non mean field solutions of non infinite range models in the statistical mechanical language (for example nearest neighbours) \cite{Durlauf1997}. These are interesting to model processes where individuals are affected mostly by their relevant connections (be that friends, business competitors, people in the individual's neighbourhood, any social connections, individuals of the same gender\ldots depending on the problem under study). However, the mean field approach can be argued to be a better approximation for many problems of interest in the social sciences (much more than for any physical model of magnetism anyway) and they are much more {\itshape easy} to study (which is why they were developed for ferromagnetism in the first place). They are a nice way of introducing a general tendency to conform to the group and the study of condensed matter has shown us that they can be quite reliable as an approximation of non mean field models as long as you stay far enough from critical regions.  Other interesting scenarios are those in which the individual's {\itshape sampling neighbourhood} can evolve (see for example \cite{Epstein2001} for an example in which the strength of a social norm increasing is seen as agents tending to {\itshape think} less about what to do in that they limit the scope of their sampling in the population). Reference \cite{Sznajd-Weron2000} studies binary opinion evolution in the Sznajd model, an Ising chain where interaction is with neighbouring pairs in a simple way, and the equilibrium states will be of total consensus or half of the population either way (the focus of this work being on the dynamical evolution towards these equilibria). The certainly very interesting case of the study of these type of problems in complex networks are found, for example in \cite{Wu2004} or in \cite{Bernardes2002}, where such networks are used to explain Brazilian election results.

The IWAs $h_{i}$ may seem therefore a natural way to encode most of the population's heterogeneity (as done traditionally in models not contemplating interactions) and so making it constant over all the group is probably the more drastic simplification of all. It is however, as has already been argued for the coupled case that will be studied in this work, a necessary starting point. Work relaxing this hypothesis can be found in \cite{Brock2001, Collet2010, Galam1997}. The zero field equivalent is thus one considering no deterministic private utilities for the agents at all. Loosely speaking, these models describe \emph{fashions} and \emph{traditions}, understood as choices that will {\itshape a priori} have no cost or benefit for the individual and so will be determined exclusively by the social cost or payoff they provide through imitation. Note however that we must be careful with this interpretation as there is no way of encoding cultural or sociological inclinations which are present in most traits we generally consider as fashions or traditions. Studying the zero field case is however interesting as it is enough to determine the regions where the social utility will make a difference even when the deterministic private utilities are {\itshape turned on} (see section \ref{sec:sinisi}).

Several general conclusions can be made about the effects of introducing interactions, as there are qualitative features emerging in this kind of setup for all models as compared to their non interactive equivalents. In an statistical mechanical sense, there are both first order and second order phase transitions as well as metastability and hysteresis. In a discrete choice context these kind of models yield (see section \ref{sec:sinisi} for these explained in more detail for the constant parameter Ising model):

\begin{itemize}

\item \emph{Microeconomic specification of the model that may not uniquely determine its macroeconomic properties}, i.e., there is more than one stable state (physically equivalent or metastable) for some range of values of the parameters (weak private deterministic utilities). 

\item \emph{Regions where social utility counts and regions where it does not}. The critical values will be separating phases where only private deterministic utility is determining the outcome from those where social influence (spontaneous magnetisations) can have an important effect in the decision making process.

\end{itemize}

Note that we are giving $\beta$ the statistical mechanical role of inverse of the temperature $\beta = \frac{1}{K_{B}T}$. Socioeconomic temperature measures the degree of statistical uncertainty about the individual's decisions. This can be seen as codifying personal characteristics affecting the opinion that are not being taken into account in the deterministic private utility. We can also give it a deeper meaning about the human nature not being predictable in essence, about \emph{free will}. If we assume that the deterministic utility contains all information about what should be a useful decision, then the temperature codes the probability of individuals to behave {\itshape irrationally}, to not choose what in principle is more convenient for them. This makes room in our models for a more complex human psychology than the detractors of a science of society generally acknowledge (or are even aware of).  

The use of a logistic distribution with constant $\beta$ is equivalent to studying our statistical physics models in the canonical ensemble. This is appropriate for the study of isolated, conservative systems, i.e., systems in thermodynamic equilibrium. Note that the subject of our study will be therefore the study of states of equilibrium for these systems rather than their evolution. The tools we will be using are specially designed for the treatment of statistical equilibrium and anyway the proposed systems are generally not appropriate for the study of the real evolution of the social system. They do not intend to reproduce the real time social evolution but rather investigate the states of equilibrium where a system representing the decision making process will end. If we however suppose that the evolution is slow and so from a state of equilibrium to another slightly different (which may turn out not to be a valid assumption for many problems), each moment in time will be represented by a functionally equivalent but slightly different Hamiltonian. We can then see the evolution of the system as an evolution in the parameter space. Dynamical approaches can be found for example in \cite{Ma2009, Sznajd-Weron2000, Weisbuch2000, Wu2004}. Another limitation of this kind of approach is that physical models generally consider symmetric interactions, which may not be a good approximation for many interesting problems is economy, sociology or political science. 

Note that the utility approach does not need to appeal to large population sizes. As any statistical model, any affirmation on expectation values will have a higher degree of certainty the larger the sample, but there is no real need to invoke a large population limit. We will however be using the thermodynamic limit extensively in this work and so the study of finite size effects in this context is also an interesting road to follow.

\section{Coupled choices and coupled order parameters}

On many occasions, individuals can be considered to be simultaneously making more than one decision that affects each other. We can understand this to be the case for some socioeconomic problems of interest. Examples could range from the interdependence of social pathologies (e.g. dropping school vs getting into crime), economic decisions (e.g. buying a TV of  brand A vs buying a DVD reader of brand B), political opinions (e.g. voting party A to the Congress vs voting party A to Senate in simultaneous elections) or other social traits (e.g. females entering the labour market vs having children).

There are also problems where we are interested in a single choice a group is making but we consider the decision making process of the group affects and is affected by the same decision process going on on another group. Examples can again be related to social pathologies (e.g. dropping school in two neighbouring districts), economic decisions (e.g. architects and engineers choosing to use a certain design software), political (e.g. public opinion in two neighbouring countries) or of other type (e.g. females without children joining labour market vs females with child care responsibilities joining the labour market).

In both cases we can consider to be studying the coupling of two discrete choice spin variables. In the first case described, the decision process is modelled using a different variable for each type of decision and a term coupling both together. In the second case, there is a different variable for each group and these are also coupled. Coupled systems of this type have been studied in  statistical mechanical contexts,  see for example \cite{Galam1988, Galam1989, Galam1990, Galam1995, Imry1975, Korutcheva1988, Simon2000}.   

Of course, the extension to more than two coupled models (more than two decisions being simultaneously made) is of interest. Extension of this work in this sense would be straightforward but probably pose technical challenges. How to generalise our results to a finite number of choices is to be explored.

A spin system having magnetic properties that can be translated to those of the fraction of adopters (or demand in an economic context) of a system of two coupled binary choices and that accounts for an heterogeneous population  may be therefore described by the general Hamiltonian:

\begin{equation}
\label{eq:hamgen}
H = -\frac{1}{N}\sum_{(i,j)}\left(J^{s}_{ij}s_{i}s_{j}-J^{t}_{ij}t_{i}t_{j}\right)-\frac{1}{N}\sum_{(i,j)_{k}}k_{ij}s_{i}t_{j}-\sum_{i}\left(h^{s}_{i}s_{i}+h^{t}_{i}t_{i}\right)
\end{equation} 

The intra couplings $J^{s}_{ij}$, $J^{t}_{ij}$ are positive and quantify the strength of social pressure for each choice and the summation is considered over all pairs $(i,j)=1\leq i<j \leq N$. When $J^{s,t}_{ij}=J_{s,t}$ are constant over all the population (homogeneously socially influenced population), terms on the intra couplings will give the dynamics of two infinite range constant Ising models. As discussed for the single choice case, the introduction of quenched disorder (heterogeneity) in the intra couplings would account for a certain distribution of how sensitive the individual is to social pressure on that certain issue amongst the population ($J^{s,t}_{ij}$ as independent identically distributed random variables for all equivalent pairs of spin locations). A similar model is studied using field theory and Monte Carlo simulations in \cite{Simon2000}.  

 As we have seen in the previous subsection for the case of a single spin model, infinite range can be  considered a good approximation for some social influence mediated decision making settings, but we could also choose to study the system in a nearest neighbour scheme ($J^{s,t}_{ij}\neq0$  only for $(i,j)$ nearest neighbours, see for example \cite{ Simon2000}) or other in different dimensions or even on complex networks \cite{Palchykov2009} as discussed for the single discrete choice setting.

The inter coupling $k_{ij}$ measures the degree of interrelation between both choices. It will be positive when both decisions tend to reinforce each other and negative when they tend to exclude each other. Different terms from the one used could also be considered \cite{Galam1988, Galam1989, Galam1990}.

Different coupling schemes are interesting in different situations. Analogously to the intra case discussed above, using the \emph{ non local } (infinite range) interaction for the inter coupling $\sum_{(i,j)_{k}}s_{i}t_{j}=\sum_{1\leq i<j \leq N}s_{i}t_{j}$ is equivalent to coupling both variables through their expectation value. It is the type of system studied in \cite{Korutcheva1988}. This will in general be a good setup to study for example the coupling of the same choice in two different (large enough) populations. See section \ref{sec:nlosoc} for a further discussion of the use of these models in a social sciences context.

The \emph{local inter coupling interaction } setup  $\sum_{(i,j)_{k}}s_{i}t_{j}=N\sum_{i}s_{i}t_{i}$ with $k_{ij}=k_{i}\delta_{ij}$ is of much interest as it has the natural interpretation of two choices which all members of a (large enough) group have to make in which  the interaction of both choices is only through each individual. This type of setup has been studied in the context of plastic phase transitions for example in \cite{Galam1995}. See section \ref{sec:locsoc} for more details on its socioeconomic implications.   

Note that in both cases the interaction is exactly reciprocal (symmetric) in the sense that the costs or payoffs due to the alignment or not between both variables are the same for the two groups/choices. 

Constant inter couplings $k_{ij}=k$ represent homogeneous populations with respect to the choice interaction. Social heterogeneity can also be introduced through quenched disorder (in the non local case) or random local coupling (in the local case) in the inter coupling interaction. These type of models have been studied in some cases \cite{Galam1995} and their application in the coupled discrete choice scenario may prove interesting.

The last term of equation \eqref{eq:hamgen} accounts for all externalities\footnote{As explained for the single Ising model in the previous subsection, when we use the terms {\itshape externalities} or {\itshape external field} we are using condensed matter language. These terms will in general not be exogenous to the population. There can be some effects that can be considered external in some way to the group (such as prices, socioeconomic level, government policies, etc.) but these terms will also account for any personal preferences of the agents regarding both decisions and so we must not misinterpret the use of the word external as indicating they have nothing to do with the social group under consideration. They are however exogenous to both social and choice interaction.}. As discussed for the single model, when the external fields are constant we are considering homogeneous populations in all external and personal factors (except for social influence and choice interaction) and is thus probably the more dramatic simplification of those considered regarding application to socioeconomic problems. Taking into account each individual's particularities accounts to using external random fields $h^{s}_{i}$, $h^{s}_{i}$ or \emph{idiosyncratic willingness to adopt} (hereafter IWA). In a demand context, we can set $h^{s}_{i}=b^{s}_{i} +p_{s}$, $h^{t}_{i}=b^{t}_{i} +p_{t}$ with $b^{s}_{i}$, $b^{t}_{i}$ the \emph{idiosyncratic willingness to buy} (IWB) and $p_{s}$, $p_{t}$ the prices. It is useful to consider them as independently drawn from  particular probability distributions encoding the group's attitude towards each particular choice in the setup studied. 

As we have seen in the previous section, the simplest socioeconomic utility approach that does not consider social interaction  is equivalent to studying only the random external field terms. It thus seems a  natural approach to first consider constant (intra and inter) coupling constants and to introduce all heterogeneity through the externalities to which the system is subject. This system will be studied in \cite{R'io2010}. This model is still homogeneous in all features concerning social interactions in the decision making process. All individuals have the same inherent desire of resembling their peers and the same perception of the group's average. The interaction between both choices is identical for all of them too.

The (coupled) order parameters of system \eqref{eq:hamgen} will be each variable's average choice or magnetisation defined by:
\begin{align}
s &= \frac{1}{N}\sum_{i}s_{i} = 2\mu_{s}-1\\
t &= \frac{1}{N}\sum_{i}t_{i}= 2\mu_{t}-1
\end{align}

\noindent with $\mu_{s}$, $\mu_{t}$ the \emph{fractions of adopters} (fraction of the population with $s_{i}=1$ and $t_{i}=1$ respectively). 

Depending on whether none, one or both order parameters are zero we can distinguish three types of phases:
\begin{enumerate}
\item \emph {Paramagnetic phase ($s=0$, $t=0$):} There is complete disorder in both variables and we can expect half of the population deciding in favour and half against for both decisions at all times.
\item \emph{ Ferromagnetic phases ($s\neq0$, $t\neq0$):} There is order in both variables, i.e., there is a tendency to uniformity in the population for both decisions. We can distinguish four cases depending on the sign of both magnetisations.
\item \emph{ Mixed phases ($s=0$, $t\neq 0$ or $s\neq 0$, $t= 0$):} Although both phases are interrelated, there will be order in one of the variables but not in the other. Again, we can distinguish four different phases depending on which of the order parameters is zero and the sign of the non zero one.
\end{enumerate} 

The system with constant $J_{s}$, $J_{t}$ and $k$ with random external fields is being studied in detail in \cite{R'io2010} (working paper) and is a natural extension to two coupled decisions of that studied in \cite{Gordon2007} (at finite as well as zero temperature). In this work we present a preliminary study of the two inter coupling interaction scenarios introduced above in the constant external field case (delta distributions for $p(h^{s}_{i})$ and $p(h^{t}_{t})$). In fact, although constant external fields are taken into consideration for the derivation of some expressions and in some general discussions, we will be focusing in understanding the zero external field case (or the contribution to the demand due solely to social/choice interaction). That is, we will be describing the phase diagram of a system of coupled choices in completely homogeneous populations in the absence of private deterministic utilities. Note that in the coupled case the zero field case is no longer limited to choices affected purely by social influence (loosely described as {\itshape fashions} or {\itshape traditions} in the previous subsection) as the interaction between both choices can be translated into actual losses or benefits for the individuals transcending social influence (see chapters \ref{cha:2p1cho}, \ref{cha:1p2cho}). The details of the non zero field will be considered as a particular example of the coupled random field Ising models in \cite{R'io2010}.

\section{The Ising model}
\label{sec:sinisi}

Let us briefly review well known results for the single Ising model in the mean field approximation. This will help us in the understanding of which specific features are brought about by the presence of the coupling between both decisions and will allow us to present concepts and tools from statistical mechanics that we will be using later. This model was solved exactly in the 1930s and in this section we will only collect well known results for convenience (see for example \cite{Baxter1982} for the solution of the model and \cite{Brush1967} for a historical review of the Ising model). 

\subsection{The model}

The Hamiltonian of a single infinite range Ising model with constant external field for N spin 1 variables $s_{i}$ is give by

\begin{equation}
\label{eq:isingsingleham}
H = -\frac{1}{N}\left(\sum_{(i,j)}J_{ij}s_{i}s_{j}\right) - \sum_{i}h_{i}s_{i} 
= -\frac{1}{2N}\left(J_{0}\sum_{i\neq j}s_{i}s_{j}\right) - h\sum_{i}s_{i} 
\end{equation}

\noindent where the sums on $i$ are over all N agents, sums on $(i,j)$ over all possible $\frac{N(N-1)}{2}$ different pairs of agents ($1 \leq i < j \leq N$) and sums on $i \neq j$ over all pairs ($1 \leq i \leq N, 1 \leq j \leq N, i\neq j$). Agent's $i$ decision is given by $s_{i}$ ($s_{i}=+1$, decide in favour; $s_{i}=-1$, decide against), $J_{ij}$ is the coupling between agent $i$'s
 decision  and agent $j$'s decision and $h_{i}=h$ a constant external field. We will be considering identical coupling between every two agents ($J_{ij}=J_{0}$ for all $i,j$). We can understand the need for the $\frac{1}{N}$ factor as for keeping the effective field that each agent will be subject to finite in the thermodynamic limit ($N \to \infty$). Note that both terms scale as $\sim N$ making the Hamiltonian extensive.

Defining the average magnetisation of the system $s = \frac{1}{N}\sum_{i=1}^{N}s_{i}$ and using mean field approximation yields:

\begin{equation}
H \approx \frac{1}{2}Js^{2}-\left(\frac{1}{N} Js+h \right)\sum_{i}s_{i}
\end{equation} 

\noindent where $J=\sum_{j(\neq i)}J_{ij}=(N-1)J_{0}$. This notation is useful as it makes it easy to extend to the case where not all $J_{ij}$ are identical but J is (coupling between every two agents is not always the same but following the same schema for all of them). Note the scaling is still in $N$ (though hidden in J dependence on it in the first term). When approaching the thermodynamic limit (and thus $(N-1) \to N$), we can write this Hamiltonian in terms of the constant coupling between every two agents $J_{0}$ as:

\begin{equation}
H \approx \frac{N}{2}J_{0}s^{2}-\left( J_{0}s+h \right)\sum_{i}s_{i}
\end{equation}

We are now ready to calculate the partition function for the representative canonical ensemble. We will be  considering every possible spin configuration ($2^{N}$ in total), each of which has probabilities that will be weighed out by the Boltzmann distribution according to their energy for the given parameters in the canonical ensemble. There will be $\frac{N!}{a!(N-a)!}$ configurations with $a$ spins up.

\begin{equation}
Z = Tr e^{-\beta H} = e^{-\frac{\beta}{2}Js^{2}}\left( 2\cosh\left(\beta(\frac{J}{N}s+h)\right)\right)^{N}
\end{equation}

The free energy for the system can be written as

\begin{equation}
F = \frac{1}{2}Js^{2}-{N}{\beta}\ln\left(2\cosh\left(\beta(\frac{J}{N}s+h)\right)\right) 
\end{equation}
 
\noindent and the free energy density

\begin{equation}
\label{eq:fising}
f = \frac{1}{2N}Js^{2}-\frac{1}{\beta}\ln\left(2\cosh\left(\beta(\frac{J}{N}s+h\right)\right)
\end{equation}

\noindent In the thermodynamic limit \eqref{eq:fising} can be written in terms of $J_{0}$  as

\begin{equation}
\label{eq:singlef}
f = \frac{1}{2}J_{0}s^{2}-\frac{1}{\beta}\ln\left(2\cosh\left(\beta\left(J_{0}s+h\right)\right)\right)
\end{equation}

The average magnetisation $s$ is the order parameter defining the phase of the system. For $s=0$, the system is in its \emph{paramagnetic} phase, there is complete disorder with all individual contributions to the magnetisation cancelling out and continuous flipping of individual agents' spin in statistical equilibrium. In the binary opinion dynamics context it involves a non preferred option situation where half of the population decides for and half against. 

We will be much more interested in the \emph{ferromagnetic} phase with $s = \pm m \neq 0$ and $0 < m \leqslant 1$ (or \emph{anti-ferromagnetic phase} if $J<0$ which is of little interest in our context), where all agents tend to align their spin in the same direction and an emergent order (or spontaneous magnetisation) arises, i.e., the interaction between agents (or social influence on the decision process) will make the population tend towards uniformity in their decision making. 

\subsection{Equation of state and its solutions}

Stable equilibrium solutions for the expected value of the  magnetisation of our system in the mean field regime will be those minimising the free energy \eqref{eq:fising} (which is equivalent to maximising agents' utilities). Candidates for relative minima and maxima will be given by $\frac{\partial f}{\partial s} = 0$. For these values, the sufficient condition for them to be minima (stable equilibria) will be $\frac{\partial^{2}f}{\partial s^{2}} > 0$ when evaluated at the critical point. When we calculate the derivatives for \eqref{eq:fising} we get:

\begin{align}
 \frac{\partial f}{\partial s} &= J_{0}s-J_{0}\tanh\lbrack\beta(J_{0}s+h)\rbrack \\
\frac{\partial^{2}f}{\partial s^{2}} &= J_{0} - \beta J_{0}\,^{2}\frac{1}{\cosh^{2}(\beta(J_{0}s+h))}
\end{align}

Free energy local maxima and minima will be therefore achieved, for $J_{0} \neq 0$, at the solutions of the system's equation of state, first arrived at by Bragg and Williams in 1934:

\begin{equation}
\label{eq:isingstateq}
s = \tanh\lbrack\beta(J_{0}s+h)\rbrack
\end{equation}

Solutions to this equation of state will give the system's free energy critical points. We can consider points where local minima are achieved as stable solutions of the equation of state (and the rest, corresponding to local maxima and inflection points, non stable solutions). 

The first obvious remarks we can make when looking at this equation are that the paramagnetic phase can only exist at a finite temperature if the external field $h=0$ and that in the absence of stochastic fluctuations $T=0$ ($\beta \to \infty$) $s \to \pm 1$ (complete consensus in the population's decision). On the other hand, for sufficiently large temperatures ($\beta \to 0$) the paramagnetic is the only possible phase (independently of the value of the external field $h$). 

Let us begin by studying the situation when $h=0$. In this case equation \eqref{eq:isingstateq} simplifies to

\begin{equation}
\label{eq:isingstateqsinh}
s = \tanh(\beta J_{0}s)
\end{equation}

This is a prototypical example of a second order or continuous phase transition in $\beta J_{0}$ that we can study linearising for $s \ll 1$  equation \eqref{eq:isingstateqsinh} yielding 

\begin{equation}
s = \beta J_{0}s + O(s^{3})
\end{equation}

\noindent and thus the critical values are $T_{c}= \frac{J_{0}}{k_{B}}$ for constant $J_{0}$ or $J_{0c}= \frac{T}{k_{B}}$ for constant T.

\begin{figure}
\centering
\subfloat[]{\includegraphics[width=0.5\textwidth]{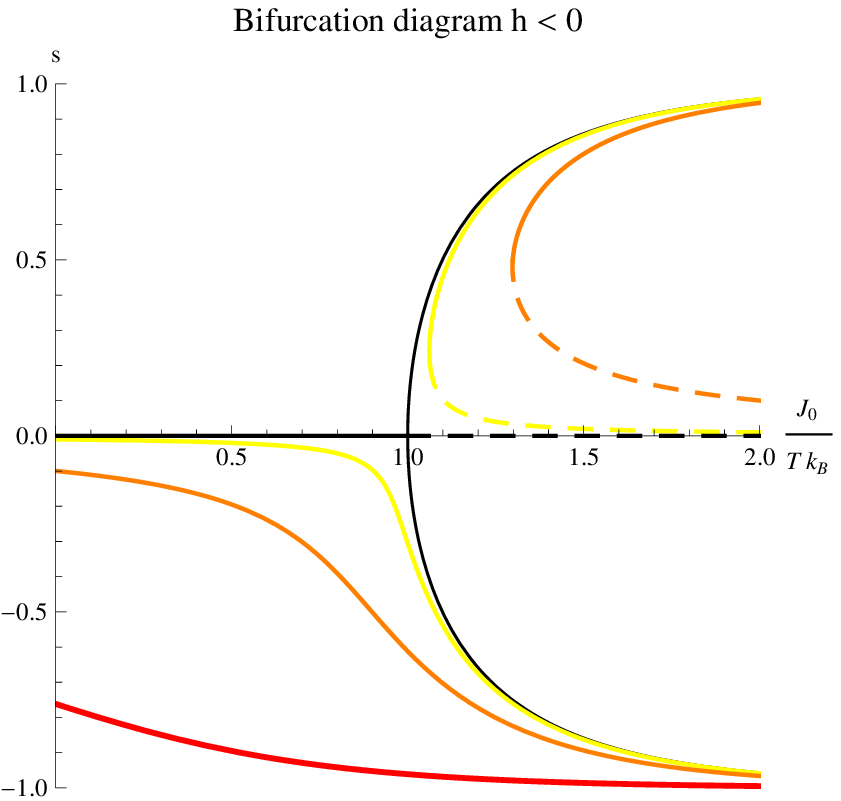}}
\subfloat[]{\includegraphics[width=0.5\textwidth]{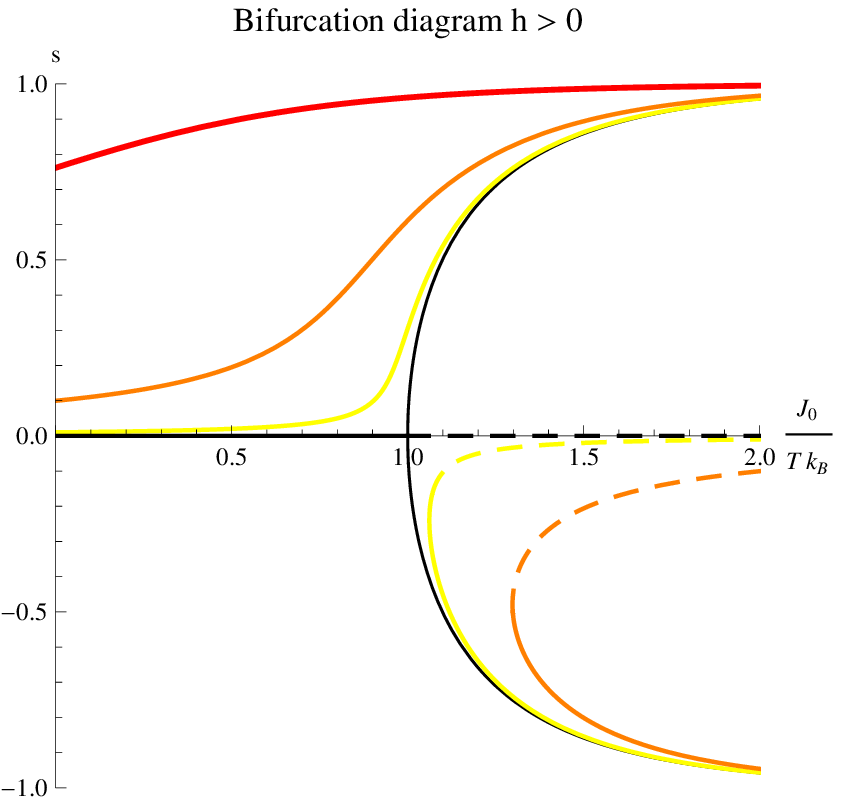}}
\caption{ Average magnetisation is plotted against $\beta J_{0}$ for (a) Different negative values of $h$ (b) Different positive values of $h$. Black line represent he zero field case, yellow $h=0.01$, orange $h=0.1$ and red $h=1$. Dashed lines represent non stable critical points.}
\label{fig:bifdiagJT}
\end{figure}

The situation is as depicted by the black lines in figure \ref{fig:bifdiagJT}. For a given coupling $J_{0}$, for temperatures above the critical one ($J_{0}\beta < 1$) the only possible stable equilibrium is the paramagnetic phase ($s=0$). Bellow the critical temperature ($J_{0}\beta > 1$) the paramagnetic phase becomes unstable and we will have two possible (physically identical) values for the magnetisation $s=\pm m$. The situation is analogous for fixed temperatures but now we will have paramagnetic phase for coupling (social influence) bellow the critical coupling.

\begin{figure}
\centering
\subfloat[]{\includegraphics[width=0.33\textwidth]{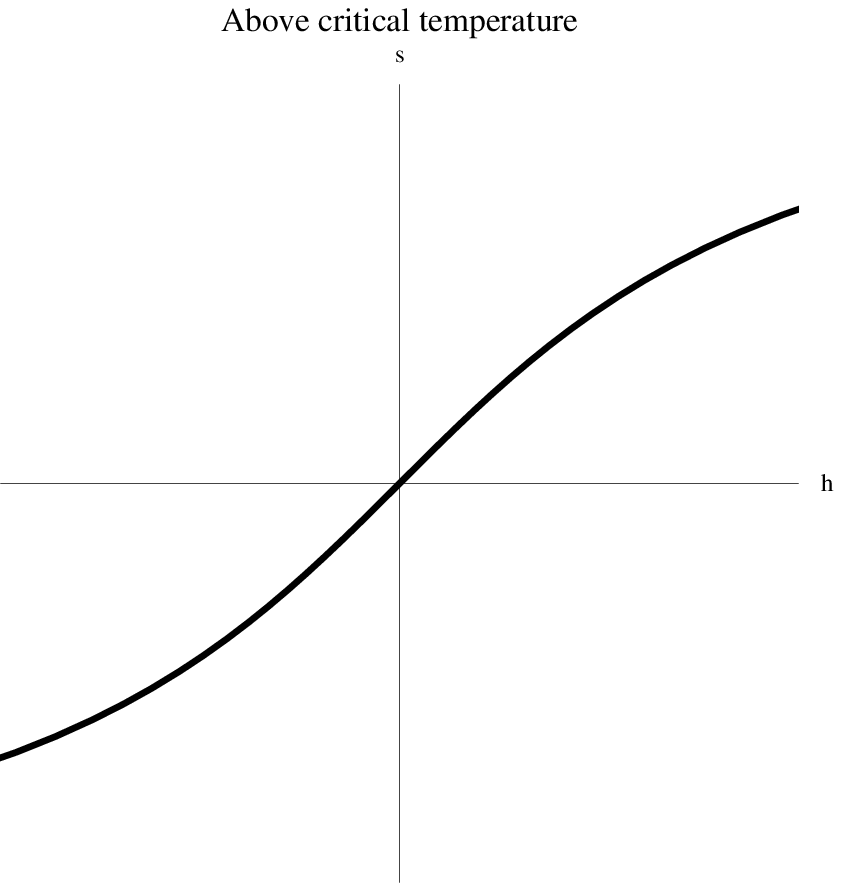}}
\subfloat[]{\includegraphics[width=0.33\textwidth]{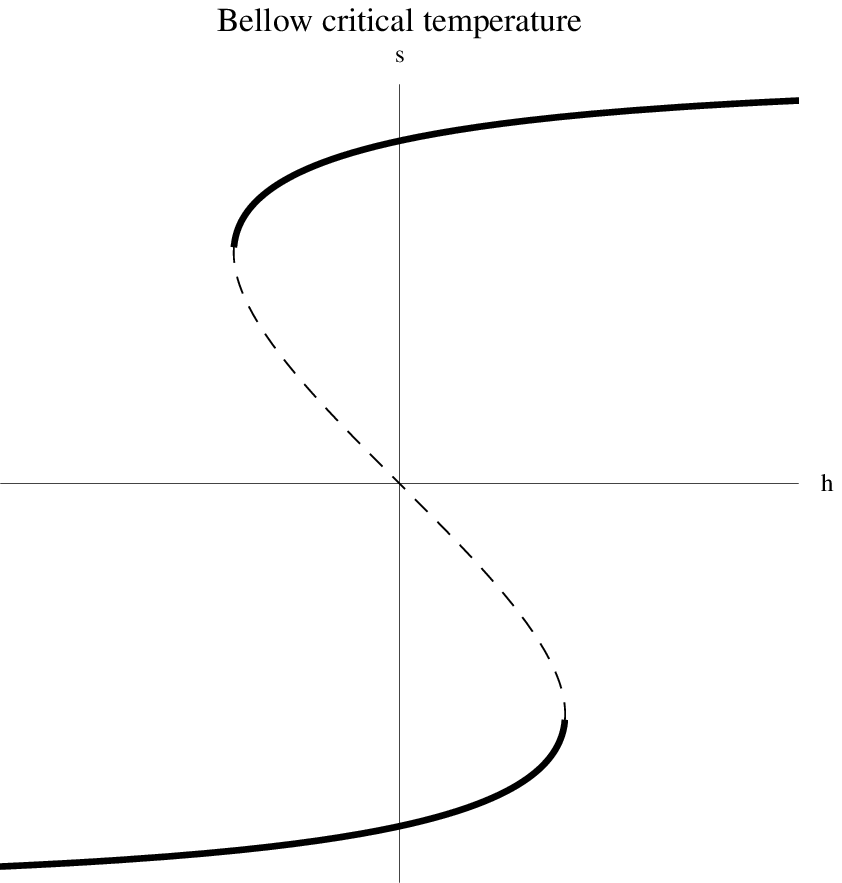}}
\subfloat[]{\includegraphics[width=0.33\textwidth]{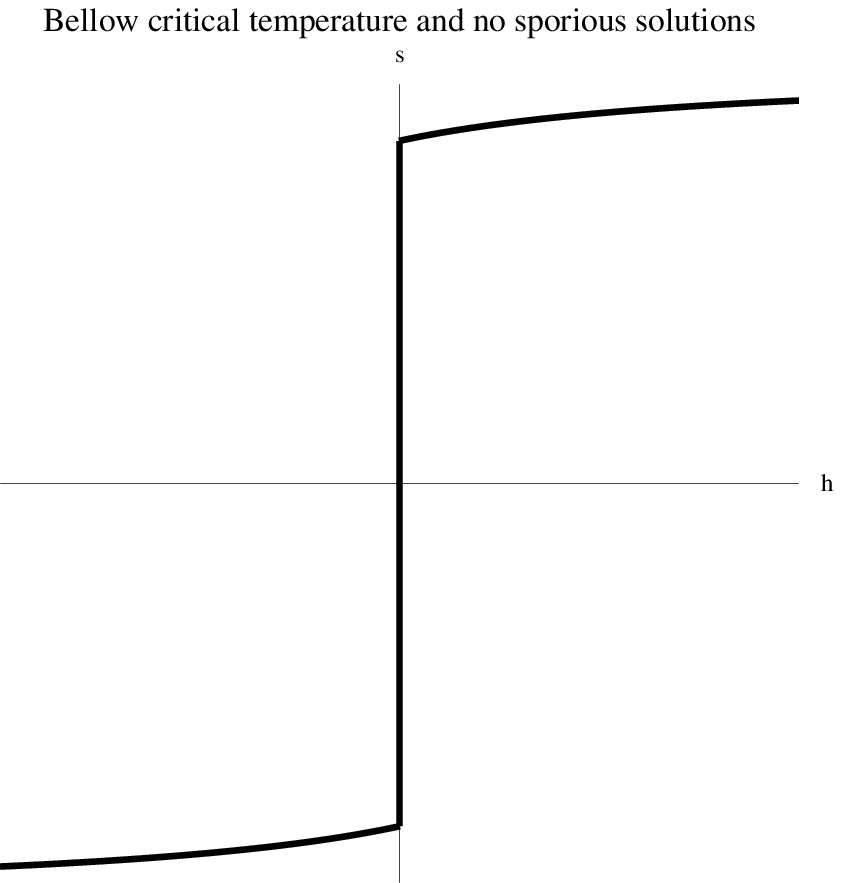}}
\caption{ Average magnetisation is drawn against $h$ for (a) $J_{0}<k_{B}T$  and (b) $J_{0}>k_{B}T$ at a fixed finite temperature and coupling. (c) Shows the result of considering part of the solutions of (b) metastable or spurious}
\label{fig:bifdiagh}
\end{figure}

Now, what happens if we {\itshape turn on} the external field h at finite temperature T? With a nonzero constant field above critical temperature the paramagnetic phase disappears. There will now be only one local minimum at a positive or negative value depending on the direction of the field. This change is smooth in h and there is no change in the stability scheme of the system. In this sense the onset of magnetisation is not through a phase transition as the paramagnetic phase in this case should be understood as the  $s=0$ value corresponding to the zero value for the external field rather than as a  qualitatively different phase in the sense of emerging processes. The graph of magnetisation against h is given in  \ref{fig:bifdiagh} (a).

If we were already in the ferromagnetic phase prior to the field onset under critical temperature, the average magnetisation will now shift in the direction of the field although the situation is more complex than in the one explained in the paragraph above. Now the stability picture for our system is changing qualitatively as we vary h and we have two different regimes: either there is only one critical point (where there is a local minimum) and negative for negative fields and positive for positive ones or there are three critical points (different from zero), one positive, one negative (but no longer symmetric with respect to the origin) where the free energy has local minima and another {\itshape in the middle} where it has a local maximum. In other words, if the field applied is small enough the multiple minima regime (for which the interaction between agents is responsible) is not broken but shifted in the direction of the field and deformed (the values of f at them will not be the same as was the case in the former situation). These values $h_{-}$ and $h_{+}$ (which will depend on the explicit vale of the coupling and temperature) enclose a region in the parameter space where there are metastable states (these are sometimes referred to as spinodal points), of a negative magnetisation in the presence of positive field (or positive if the field is negative), allowing for the possibility of hysteresis. Note that now the paramagnetic phase looses its condition of local maximum (which is also shifted by the field) and is no longer a solution of the equation of state at all.

In fact both magnetisations in this multiple equilibria regime are not exactly on equal terms (and that is why there are three and not two graphs in the figure). To begin with, the higher (in absolute value) stable magnetisation will be an attractor for a bigger region of the $s-h$ plane than the lower one. Secondly, we will only see this effect when reversing the direction of the field and for very particular values of the average magnetisation of the system (which is considered to be in statistical equilibrium already when the field is changed). But further more the general statistical mechanical approach will be making us disregard all the positive (negative) field and negative (positive) magnetisations as spurious solutions (metastable states that will sooner or later decay to the main one) because they are much less probable than the other stable equilibria in most cases.

The partition function will have contributions from every possible spin configuration, but some of these will be more important than others depending on the number of possible configurations with that average magnetisation and the exponential term giving more weight to lower energy states. In our particular case, multiple solutions exist when the distribution of these contributions as a function of the average magnetisation of that given configuration has two local maxima (and a local minima in between). One of them is however much sharply peaked (corresponding to the higher absolute value solution) and more so as N becomes larger and thus disregarding the contributions of the rest of terms becomes safe and we can consider  \ref{fig:bifdiagh} (c) to better represent the situation in the thermodynamic limit (see for example chapter 3 in \cite{Baxter1982}). Basically, the free energy \eqref{eq:singlef} may have multiple local minima in the multiple equilibria regime, but only when there is no external field are they both global minima. When $h$ is non zero, only one of them will be the absolute minimum and thus the system's {\itshape ground state} and the expected average magnetisation value in a system of such characteristics.

We must understand however that the relative importance of these extra \emph{metastable} solutions in the low temperature, low field regime depends both on the actual shape of $f$ and the relative importance of both local minima (which is related to $N$ besides the parameter values as we have seen) and on our initial knowledge of our system of interest. If the secondary minimum is {\itshape deep enough} and our system is in a previous known state of equilibrium with a given magnetisation, changing the value of $h$ may make $f$ reach its new local secondary  minimum and the system may stay there (for a relatively long time) for low enough temperatures (stochastic fluctuations are small enough to make it unlikely that they will shift the system far enough from this metastable value to reach the region of attraction of the absolute minimum free energy magnetisation). We can therefore consider the use of figure \ref{fig:bifdiagh} (c) when studying the magnetisation of a system of which we don't have any previous information (besides parameters) and (b) when we have more precise information of the state the system is in or, for example when we want to consider what will happen when slowly changing the value of the parameters drifting the state from a state of thermodynamic equilibrium to a slightly different one. 

In all cases,  above the critical temperature there is no spontaneous magnetisation due to self organisation and so s changes smoothly and continuously with h. Bellow it, besides the magnetisation due to the action of the external field there will be spontaneous magnetisation given by

\begin{equation}
\label{eq:sponmag}
s_{0}= \lim_{h \to 0^{+}} s(h,T)= \tanh(\beta J_{0}m_{0})
\end{equation}

\noindent due to the interaction between agents. This spontaneous magnetisation $m_{0}$ behaves exactly in the same way as the total average magnetisation in the no external field case. Note it becomes negligible at very high temperatures or fields. 

Figure \ref{fig:bifdiagJT} shows plots for the average magnetisation vs $\frac{J_{0}}{K_{B}T}$ for different values of $h$ that complete the picture of the non zero external field case (positive filed in (a) and negative in (b)). They both show the characteristics we have been discussing: the paramagnetic phase will only  be a solution (and then it will be a stable one) when $\beta J_{0} < 1$. In this regime there will always be only one stable magnetisation (with same sign as the field). When $\beta J_{0} > 1$ there is always spontaneous magnetisation. For sufficiently intense fields the case will be similar as that discussed above with one stable equilibria. For a region of small intensity field (between $h_{-}$ and $h_{+}$ that will depend on the values of coupling and temperature) there is a region where two stable equilibria coexist, one negative and one positive with different absolute values, the larger one being the one with the same sign as the field. We must be cautious regarding these multiple equilibria as in the thermodynamic limit the largest magnetisation (in absolute value) completely dominates the partition function.

\subsection{Phase diagram}

We now have all the necessary information to describe the phase diagram of the system. It is in fact so simple it is not even worth showing in figures. Let us begin by considering the case of zero external field. The phase diagram  $(J_{0},k_{B}T)$ cross section is then divided into two regions by the line of slope 1 across the origin representing a continuous second order phase transition. For $K_{B}T>J_{0}$ the system is in the paramagnetic phase. For $K_{B}T>J_{0}$ in the ferromagnetic phase (where two physically identical stable magnetisations $s = \pm m$ exist).

When $h \neq 0$, the rest of $h = constant$ sections are much less interesting if we are only considering the phase of the system, as the paramagnetic phase is never a solution in this case so the $k_{B}T-J_{0}$ phase diagram is a completely ferromagnetic cross section. Nevertheless, it is worth mentioning in this case there will only be one possible stable magnetisation not two physically identical ones of opposite sign (with sign depending on the sign of constant h) if we do not consider multiple equilibria.

The other two possible sections when considering the three parameters are the $h-K_{B}T$ and $h-J_{0}$ sections. The ferromagnetic phase predominates in both cases and only for some values of $h=0$ is the paramagnetic phase  stable, namely,  $J_{0}<K_{B}T_{c}$ for the $h-J_{0}$ section, and $K_{B}T_{c}>J_{0}$ for the other. This {\itshape paramagnetic segment} thus end in a critical point. For temperatures bellow it (couplings above it), there is a jump in $h=0$ in the sign of the magnetisation (first order transitions) while for the rest of values the change in the magnetisation is smooth with $h$. Regions for positive $h$ will have $s>0$ and regions for negative $h$ will have $s<0$ (not considering metastable solutions).

\subsection{Socioeconomic interpretation}

The phase diagram of the model gives rise to interesting socioeconomic interpretations. As compared to a non interactive scheme, the more relevant feature is that there are values of the parameters (IWA, strength of the social pressure, importance of the statistical fluctuations) for which there is a multiple equilibria regime (physically identical or metastable depending on the case). 

In the absence of private utilities (trends or traditions, zero IWA), there is a critical temperature (higher the stronger the social coupling is) at which a second order phase transition takes place. For temperatures (statistical fluctuations) bellow it social influence on its own will make a preferred option emerge in the collective choice making. There are two equally probable equilibrium states of high and low demand or acceptance in the group. Second order phase transitions are associated to these type of symmetry breaking, involving high nonlinearity in the response of the system (small variations in the parameters can even change the number of equilibria). 

When $h\neq 0$, the critical temperature is still the temperature for which the social utility becomes relevant. The choice of the population will be polarised in the direction indicated by private utility. At low enough values of $h$ though, the multiple equilibria scenario persists for small temperatures. The state of average choice contradicting private deterministic utility is however now metastable and thus less probable. There is therefore the possibility for \emph{hysteresis} (irreversible change in the state of the system when changing a parameter as it {\itshape decays} to the real ground state). First order transitions of this type are associated to nucleation processes. A metastable state represents a collectively reinforced choice that persists even when everybody would be better off changing their minds. As some individuals do start changing their minds (this is where the temperature comes in) the others will tend to imitate them and rapidly collapse to the more convenient state.

As first noted by Durlauf and Brock \cite{Durlauf1997}, this allows us to explain the ''{\itshape complementary nature of the roles of economic fundamentals and social norms in explaining the degree of social pathologies in different neighbours}''. It naturally integrates the two classical approaches to explain the prevalence of some social pathologies (let us say, for example, becoming a criminal) in certain neighbourhoods. Some, normally on the liberal side of the spectrum, have argued that the underlying reason was the lack of opportunities ({\itshape economic fundamentals}, small private utility returns discouraging crime). Others, usually in the more conservative spectrum, relayed on a sort of {\itshape culture of crime}. This is understood as a strong coupling regime in a group already in a state of propensity to crime. A {\itshape social norm} about crime that influences the individuals beyond economic or social benefits they may gain from  changing their minds (thus giving policies aimed at increasing these private utilities limited effect). The Ising model describes a situation where the average behaviour will be governed by private utilities unless these are weak enough, where congenital social problems can overcome the individual's sense of {\itshape what is best for him/her} in a self perpetuating socially undesirable situation. It does give reason to be optimistic: if enough efforts are put into raising the individuals' utilities above the spinodal point, the model predicts an spectacular and drastic change (phase transition) between a high crime incidence situation to a low one.

To study economic demand contexts, one can replace the IWA by $h_{i}=b_{i}-p$ with $p$ fixed price and for a given parameter configuration study how the demand depends on the price. In the regions of metastability, these will be multi-valued functions. In \cite{Gordon2007} the random field Ising model is studied in this context at zero temperature. As for the case studied in this section, multiple equilibria appear (the number of them depending on the number of modes of the particular probability distribution chosen for the $b_{i}$s). Under some conditions it is in fact equivalent to the standard Ising model we have studied (for an unimodal distribution). Note that in the random field Ising model the source of disorder is fixed in time (\emph{quenched}) as opposed to the randomness associated to the temperature (\emph{annealed}).


%% file: secs/2p1cho.tex
\chapter{Nonlocal coupling: single choice coupled over two populations}
\label{cha:2p1cho}
Let us consider two groups of size $N_{s}$ and $N_{t}$, all agents making a binary choice such that $s_{i}=\pm 1$ (for the s type agents) and $t_{i}=\pm 1$ (for t type agents). Each individual's probability to make a decision is affected by the perception of what the rest of its group is doing, for which the average magnetisation (or fraction of adopters) of the group is assumed to be a well approximation for all members. This social conditioning may be of different strength for each of the groups but is the same for all members of the same group. 

The choice made by each individual is also affected by what the other group is doing. In this case, the interdependence between both groups may be of social influence character (as the intra-couplings described above), of a different character (concerning economic or social gains/costs of a group depending on the choice of the members of the other group when these are reciprocal), or of both. An individual from one of the groups will have terms in its utility function accounting for its gains and losses due to the interaction between both groups, and we will consider the case where these terms depend only on the average magnetisation  (or fraction of adopters) of the other group. In cases where the interaction between both groups is due to social pressure, this means that all the individuals of each group have the same (correct) perception of the degree of acceptance of the other group too. When considering interactions involving other tradeoffs not associated to social recognition or distinction, it is only appropriate to use it when it is a good approximation to consider each individual's gains/losses related to the interaction as proportional to the average magnetisation of the other group (and identical for all individuals).  

As opposed to the intra-group social pressure, the inter-group interaction can take both signs, either reinforcing ($k>0$) or discouraging ($k<0$) alignment between both groups in their choice making.

Besides being subject to social interaction with the members of their group and the other, the probability of making a particular choice for each individual can also be affected by external factors or subjective preferences, but these must be constant within each group (although may vary from one group to the other).   
 
The system we just described can be analysed considering two types of spin variables described by Weiss mean field approximation of the Ising model (with constant external field) plus an additional Weiss mean field type interaction between both kinds of spins. In the thermodynamic limit ($N_{s}\to \infty$, $N_{s}\to \infty$) this is equivalent to considering infinite range (intra-and inter-group) interactions for which Hamiltonian \eqref{eq:hamgen} takes the form:       

\begin{equation}
\label{eq:ham2p1cho1}
H = \sum_{(i,j)}\left(-\frac{J_{s}}{N_{s}}s_{i}s_{j}-\frac{J_{t}}{N_{t}}t_{i}t_{j}-\frac{k(N_{s}+N_{t})}{2N_{s}N_{t}}s_{i}t_{j}\right)-\sum_{i}\left(h_{s}s_{i}+h_{t}t_{i}\right)
\end{equation} 

\noindent where summations over $(i,j)$  are $1\leq i<j \leq N_{s}$ for the first term, $1\leq i<j \leq N_{t}$ for the second term and $1\leq i\leq N_{s}$, $1\leq j\leq N_{t}$  for the mixed term. Summations over $i$ are $1 \leq i \leq N_{s}$ for the fourth term and $1 \leq i \leq N_{t}$ for the last term.  

When both groups are of the same size, this is the system studied in \cite{Korutcheva1988} and this section is an extension and review in the light of its application in opinion dynamics of the work presented there. 

Note that while we are choosing the interpretation one choice, two populations, equation \eqref{eq:ham2p1cho1} can also be considered to describe a single choice and population where social interaction can take three different values depending on the pair of individuals and where individuals, can be subject to one of two possible externalities. This model is thus not coupled in strict sense and can be understood as the introduction of minimal heterogeneity through a binary categorisation of the individuals, with the particularity that now some of the social interactions can have the opposite  {\itshape antisocial} (antiferromagnetic) effect (corresponding to negative values of the inter-coupling in \eqref{eq:ham2p1cho1}). 

From this perspective it seems natural to divide the population into a larger number of smaller groups which interact between them and this could be a valid coarse grain type approach to study the effects of the different subgroup composition in the choice making scenario (basic socioeconomic communities or opinion groups of a country's  society for example). Individuals of a given subgroup are characterised by the strength of the social influence to which they are subject from each other group (including its own), so they will tend to imitate members of some groups more than members of others, some groups may have no social influence at all on others and there can even be a negative {\itshape antisocial} relation between groups. At this point, it is a good idea to remember that social influence needs to be reciprocal between the groups in this setting, as it somewhat limits the richness of scenarios that can be considered in the social sciences context. There is an obvious limitation to this approach in that subgroups need to be big enough to yield the use of statistical mechanics and thermodynamic limit appropriate and this bounds the number of subgroups that we can safely use and so the amount of randomness or heterogeneity that can be considered using this approach. Any effects due to group size (relative to the rest) will not be captured by this approach either.

Hamiltonian \eqref{eq:ham2p1cho1} for $N=N_{s}=N_{t}$ also describes a group where each individual makes two choices each of which depends on the acceptance of both choices in the rest of the population. While interesting, we have focused on the interpretation described in detail because we feel it is more relevant for most common problems tackled in the socioeconomic literature.

In what remains of the section we will describe our results. In section \ref{sec:nlmod} we compute the system's free energy in the mean field approximation and in section \ref{sec:eqsta} we derive the equations of state and conditions for stability. We then present an analysis on the dependence on each parameter of our numerically calculated solutions to the average magnetisation vector $(s,t)$ in section \ref{sec:nlnumsol}, construct the system's phase diagram in section \ref{sec:nlophadia} and analyse its sociological interpretations in section \ref{sec:nlosoc}.

\section{The model}
\label{sec:nlmod}
For equally sized groups (which will always be the case in the thermodynamic limit), and using mean field theory (on all intra- or inter-coupling terms), we can rewrite \eqref{eq:ham2p1cho1} as
 
\begin{equation}
\label{eq:ham2p1cho2}
H = \frac{NJ_{s}}{2}s^{2}+\frac{NJ_{t}}{2}t^{2}+Nkwm 
 -\left(J_{s}s+kt+h_{s}\right)\sum_{i}s_{i} -\left(J_{t}t+ks+h_{t}\right)\sum_{i}t_{i} 
\end{equation} 

\noindent The corresponding partition function for the representative canonical ensemble ($Z=Tr e^{-\beta H}$ where $Tr$ indicates sum over all possible spin configurations) can be expressed  
\begin{equation}
Z = e^{-\beta(\frac{N}{2}s^{2}J_{s}+\frac{N}{2}t^{2}J_{t}+Nkst)} 
\lbrack 2\cosh\left(\beta\left(J_{s}s+kt+h_{s}\right)\right)2\cosh\left(\beta\left(J_{t}t+ks+h_{t}\right)\right) \rbrack^{N}
\end{equation}

\noindent where $\beta = \frac{1}{K_{B}T}$, $K_{B}$ is Boltzmann's constant and T the temperature (which in this case accounts for statistical fluctuations).

Finally the system's free energy density ($f=F/N$ with $F$ the free energy $F=K_{B}T\log(Z)$) will be given in the mean field approximation by
\begin{equation}
f = \frac{1}{2}J_{s}s^{2}+\frac{1}{2}J_{t}t^{2}+kst-\frac{1}{\beta}\ln\left(2\cosh\left(\beta\left(J_{s}s+kt+h_{s}\right)\right)\right)  
 -\frac{1}{\beta}\ln\left(2\cosh\left(\beta\left(J_{t}t+ks+h_{t}\right)\right)\right)
\label{eq:freeE}
\end{equation}

\noindent Stable states of the system will be those minimising the free energy.

\section{Equations of state: solutions and stability}
\label{sec:eqsta}
Critical points of the free energy will be determined through the free energy first derivatives:

\begin{align}
\frac{ \partial f}{ \partial s} & = J_{s}s+kt-J_{s}\tanh \left(\beta\left(J_{s}s+kt+h_{s}\right)\right)-k\tanh\left(\beta\left(J_{t}t+ks+h_{t}\right)\right) \\
\frac{ \partial f}{ \partial t} &= J_{t}t+kw-J_{t}\tanh \left(\beta\left(J_{t}t+ks+h_{t}\right)\right)-k\tanh\left(\beta\left(J_{s}s+kt+h_{s}\right)\right)
\end{align}
 
\noindent that when set to zero and after simple algebraic manipulation yield the system:

\begin{equation}
\begin{array}{l}
a\left(s-\tanh\left(\beta\left(J_{s}s+kt+h_{s}\right)\right)\right) = 0  \\
a\left(t-\tanh\left(\beta\left(J_{t}t+ks+h_{t}\right)\right)\right) = 0
\end{array}
\end{equation}

\noindent with:

\begin{equation}
a = J_{s}J_{t}-k^{2}
\end{equation}

We have thus two differentiated cases to consider:
\begin{enumerate}
\item{\emph{Degenerate case ($a=0$):} Substituting $J_{t}=J_{s}/k^{2}$ in the original system both turn out to give the same equation of state:

\begin{equation}
J_{s}s+kt-J_{s}\tanh\left(\beta\left(J_{s}s+kt+h_{s}\right)\right) 
-k\tanh\left(\beta\left(\frac{k^{2}}{J_{s}}t+ks+h_{t}\right)\right) = 0 
\label{eq:eqstadeg}
\end{equation}
}
\item{\emph{Non degenerate case ($a \neq 0$):} Critical points will be given by solutions to the system of equations of state:

\begin{equation}
\begin{array}{l}
s = \tanh\lbrack\beta\left(J_{s}s+kt+h_{s}\right)\rbrack\\
t = \tanh\lbrack\beta\left(J_{t}t+ks+h_{t}\right)\rbrack
\end{array}
\label{eq:eqstandeg}
\end{equation}
We will be focusing our attention to this case.}
\end{enumerate}

The stability of the solutions to the equations of state of the system will be determined by the Hessian evaluated at the critical point. Its components will be given by the free energy's \eqref{eq:freeE} second derivatives:

\begin{align}
\frac{ \partial^{2}f}{\partial s^{2}} &=  J_{s}-\beta J_{s}\,^{2}\frac{1}{\cosh^{2}\lbrack \beta \left(J_{s}s+kt+h_{s}\right)\rbrack} 
-\beta k^{2}\frac{1}{\cosh^{2}\lbrack \beta \left(J_{t}t+ks+h_{t}\right)\rbrack}\label {eq:fss}\\
\frac{ \partial^{2}f}{\partial t^{2}} &=  J_{t}-\beta J_{t}\,^{2}\frac{1}{\cosh^{2}\lbrack \beta \left(J_{t}t+ks+h_{t}\right)\rbrack}
-\beta k^{2}\frac{1}{\cosh^{2}\lbrack \beta \left(J_{s}s+kt+h_{s}\right)\rbrack}\\
\frac{ \partial^{2}f}{\partial s \partial t} &= k -\beta k J_{s}\frac{1}{\cosh^{2}\lbrack \beta \left(J_{s}s+kt+h_{s}\right)\rbrack}
-\beta k J_{t}\frac{1}{\cosh^{2}\lbrack \beta \left(J_{t}t+ks+h_{t}\right)\rbrack}
\end{align}

Critical points will be minima (stable solutions) for positive definite Hessian, maxima (unstable solutions) for negative definite Hessian and saddle points (stable on one of the directions and unstable on the other) for indefinite Hessian. If the Hessian is singular, its analysis will give no information on the stability of the solutions.

In our case, the Hessian's determinant can be written as
\begin{equation}
\det(\mathcal{H})=a\left(1-\beta J_{s}\gamma_{s}-\beta J_{t}\gamma_{t}\right) +\beta^{2}a^{2}\gamma_{s}\gamma_{t}
\label{eq:dethess}
\end{equation}

\noindent and the free energy's second derivative \eqref{eq:fss} as
\begin{equation}
\frac{ \partial^{2}f}{\partial s^{2}} = J_{s}-\beta J_{s}\,^{2}\gamma_{s}-\beta k^{2}\gamma_{t} 
\label{eq:fss2}
\end{equation}

\noindent where $\gamma_{s}=\frac{1}{\cosh^{2}\lbrack \beta \left(J_{s}s+kt+h_{s}\right)\rbrack}$ and $\gamma_{t}=\frac{1}{\cosh^{2}\lbrack \beta \left(J_{t}t+ks+h_{t}\right)\rbrack}$.

So solutions $(m_{s},m_{t})$ of the system of equations of state are saddle points if  $det(\mathcal{H}(m_{s},m_{t}))<0$. If $det(\mathcal{H}(m_{s},m_{t}))>0$ they are minima for $\displaystyle{\partial^{2}f}/{\partial s^{2}} > 0$ and maxima for $\displaystyle{\partial^{2}f}/{\partial s^{2}} < 0$\footnote{Alternatively we can substitute the condition in $\displaystyle{\partial^{2}f}/{\partial s^{2}} > 0$ by the equivalent ones  $\displaystyle{\partial^{2}f}/{\partial t^{2}} > 0$ or $\displaystyle{\partial^{2}f}/{\partial s^{2}}+ \displaystyle{\partial^{2}f}/{\partial t^{2}}  > 0$ (positive trace) as they all involve definite positive Hessian if its determinant is positive.}.

\subsection{Non degenerate zero field case}
We may start by making some general remarks upon simple inspection of the system of equations of state \eqref{eq:eqstandeg} when $h_{s}=h_{t}=0$.  The paramagnetic phase $(s,t)=(0,0)$ is always a critical point of the free energy (at finite temperature) and it will be the only one for very high temperatures ($\beta \to 0$). 

For $T=0$ ($\beta \to \infty$) the only solution is a completely ordered ferromagnetic phase $(s,t)=(\pm 1,\pm 1)$. In fact, for positive $k$ $(1,1)$ and $(-1,-1)$ are always solutions of \eqref{eq:eqstandeg} and so is the case of $(-1,1)$ and $(1,-1)$ for negative $k$. When $k<J_{s}$ and $k<J_{t}$ all four possible solutions are present regardless the sign of $k$. 

At finite temperature, for any ferromagnetic solution $(m_{s},m_{t})$, $(-m_{s},-m_{t})$ is also a solution, so these always appear in pairs. Further more, the system \eqref{eq:eqstandeg} is invariant under the change $k \to -k$ and $(s,t)\to(-s,t)$ (or $(s,t)\to(s,-t)$), so the absolute value of the average magnetisations which are critical points of the free energy is the same for $k$ and $-k$, with opposed relative signs between both. 

Mixed phases can only exist at $k=0$ and only the type where the non zero magnetisation is the one associated to the highest intra-coupling. They will exist in the region where one of the (uncoupled in this case) Ising models is in the paramagnetic phase while the other is still in the ferromagnetic phase. That is, for $J_{s}>J_{t}$, between $J_{uc}^{t}=K_{B}T_{uc}$ and $J_{uc}^{s}=K_{B}T_{uc}^{s}$. These will also appear in pairs ${(m_{s},0), (-m_{s},0)}$ (or ${(0,m_{t}), (0,-m_{t})}$ if $J_{t}>J_{s}$) as they are subject to the same symmetries as ferromagnetic solutions.

Linearising equations \eqref{eq:eqstandeg} for $|s| \ll 1$ and $|t| \ll 1$  (at finite non zero temperature) yields

\begin{equation}
\begin{array}{l}
s = \beta\left(J_{s}s+kt\right) + O(s^{3},t^{3}, s^{2}t, st^{2})\\
t = \beta\left(J_{t}t+ks\right) + O(s^{3},t^{3}, s^{2}t, st^{2})
\end{array}
\end{equation}

\noindent and further simplification of this system leads us to the expression
\begin{equation}
l(\beta)=a\beta^{2}-(J_{s}+J_{t})\beta + 1 = 0
\label{eq:Tceq}
\end{equation}.

Roots of $l(\beta)$ as defined in equation \eqref{eq:Tceq} give two values of $\beta$ where the behaviour of the solutions to the system \eqref{eq:eqstandeg} will change qualitatively:

\begin{equation}
\label{eq:Tc}
\beta =\frac{J_{s}+J_{t} \pm \sqrt{(J_{s}+J_{t})^{2}-4a}}{2a}=\frac{J_{s}+J_{t} \pm \sqrt{(J_{s}-J_{t})^{2}+4k^{2}}}{2a}
\end{equation}

Depending on the sign of $a$ \eqref{eq:Tc} will yield either one or two physically relevant (positive) values for the temperature and we can thus differentiate two different behaviours according to the relative inter- and intra-coupling strengths.  In the \emph{strong coupling regime ($k^{2}>J_{s}J_{t}$)} there is only one such physical value for the temperature $\beta_{c}$ satisfying equation \eqref{eq:Tceq} and in the \emph{weak coupling regime ($k^{2}<J_{s}J_{t}$)} there will be two $\beta_{c}<\beta_{b}$ ($K_{B}T_{c}=\frac{1}{\beta_{c}}>K_{B}T_{b}=\frac{1}{\beta_{b}}$). We are using the notation $\beta_{b} =\frac{J_{s}+J_{t} + \sqrt{(J_{s}+J_{t})^{2}-4a}}{2a}$ and $\beta_{c} =\frac{J_{s}+J_{t} - \sqrt{(J_{s}+J_{t})^{2}-4a}}{2a}$. 

We can now turn to study the stability of the solutions using \eqref{eq:dethess} and \eqref{eq:fss2}. From simple inspection of these equations we can see that 
they still have the same symmetries as the equations of state \eqref{eq:eqstandeg} in the zero field case. Namely, they are invariant under the transformation $(s,t)\to(-s,-t)$ and $(s,t)\to(-s,t)\;k\to-k$, and so each pair $(m_{s},m_{t}),\, (-m_{s},-m_{t})$ of ferromagnetic solutions will always have the same stability. The same applies to pairs of solutions related through $(m_{s},m_{t})\to(-m_{s},m_{t})$ (or $(m_{s},m_{t})\to(m_{s},-m_{t})$) for the same value and opposite sign of $k$.

As for mixed phases ($\gamma_{s}\neq1$ or $\gamma_{t}\neq1$), little can be said  about their stability with this simple first inspection as they exist neither at $\beta \to 0$ nor at $\beta \to \infty$.

For ferromagnetic phases ($\gamma_{s}\neq1$ and $\gamma_{t}\neq1$) at $T = 0$ ($\beta \to \infty\;\gamma_{s},\,\gamma_{s} \to 0$) the sign of the determinant will be that of $a$ and the second derivative \eqref{eq:fss2} is always positive, so existing critical points of $f$ of this type (either ${(1,1),(-1,-1)}$ or ${(-1,1),(1,-1)}$ depending on the sign of $k$ in the strong coupling regime and ${(1,1),(-1,-1),(-1,1),(1,-1)}$ in the weak coupling regime) will be minima when $a>0$ and saddle points when $a<0$. A nonzero $k$ necessarily breaks the symmetry present in the free energy at $k=0$ (where all four solutions are on equal grounds) privileging two depending on the sign of $k$. For low enough $k$ ($a<0$) however,  the other two solutions remain as metastable states.

At very high temperatures ($\beta \to 0$) we recover the same behaviour for the only existing solution, the paramagnetic phase. For the paramagnetic phase $\gamma_{s,t}=1$ and we can further simplify equations \eqref{eq:dethess} and \eqref{eq:fss2} to

\begin{align}
\det(\mathcal{H})&=a\left(1-\beta J_{s}-\beta J_{t}\right) +\beta^{2}a^{2} = al(\beta) \label{eq:nldethesspara}\\
\frac{ \partial^{2}f}{\partial s^{2}} &= J_{s}+\beta\left( J_{s}\,^{2}- k^{2}\right) \label{eq:fss3}
\end{align}

\noindent which allows us to study analytically the stability of $(s,t)=(0,0)$ in more detail as previously done in \cite{Korutcheva1988}.

When studying the values of the temperature where the sign of the Hessian determinant changes, we recover equation \eqref{eq:Tceq} thus making clear the meaning of the roots of $l(\beta)$ ($\beta_{b}$, $\beta_{c}$)  given by \eqref{eq:Tc} as values at which the stability of the paramagnetic phase changes. 

In the strong coupling regime, for $T>T_{c}$ the paramagnetic phase will have negative determinant and for $T>T_{c}$ positive one. In the weak coupling regime, for  $T>T_{c}>T_{b}$  or $T<T_{b}<T_{c}$ the determinant is positive and for $T_{b}<T<T_{c}$ negative.

As for the sign of the second derivative \eqref{eq:fss3}, it will be positive only when

\begin{equation}
\beta < \frac{J_{s}}{J_{s}\,^{2}+k^{2}}
\end{equation}

\noindent Solving for $J_{s}$ we can get the equivalent conditions

\begin{align}
\beta^{2} &< \frac{1}{4k^{2}} \\
J_{s}\,^{-}&< J_{s}<J_{s}\,^{+}
\end{align}

\noindent with

\begin{equation}
J_{s}\,^{\pm} = \frac{1 \pm \sqrt{1-4\beta^{2}k^{2}}}{2\beta}
\end{equation}.

This means that when in the weak coupling regime, the paramagnetic phase will be stable for $T>T_{c}$, saddle point when $T_{b}<T<T_{c}$ and unstable for $T<T_{b}<T_{c}$. Temperature $T_{c}$ therefore gives a critical point at which there is a second order phase transition (see the next two sections for more details). For temperatures bellow it any stable solution will be ferromagnetic and so there can not be a change in the magnetic phase at $T_{b}$ (that is associated as we will see to the onset of new saddle type ferromagnetic solutions).

In the strong coupling regime, for temperatures above $T_{c}$  the paramagnetic phase is a saddle point and bellow it a maximum of the free energy. This analysis suggests no phase transition at $T_{c}$, as the paramagnetic phase is never stable regardless the value of the temperature. This fact, together with there being no stable solutions for $T=0$, is already suggesting there will be no stable solutions at all, making the discussion of phase transitions is irrelevant. 

In the strong coupling regime, there will always be  global minima of the free energy density at $(s,t)=\{(1,1), (-1,-1)\}$ or $(s,t)=\{(1,-1),(-1,1)\}$ depending on the sign of k and the temperature and so the system will show a tendency to evolve towards higher absolute value magnetisations. It is however prevented from reaching a state of stable equilibrium. Physically, frustration prevents the system from evolving towards a minimum of the free energy density. Sociologically, if the inter-coupling is too strong agents {\itshape can\'{}t make up their minds}. 

In the weak coupling regime, there is always a local (and global) minimum for the paramagnetic phase at high temperatures and two ferromagnetic stable solutions (with relative signs depending on the sign of $k$) for low temperatures. At low enough temperatures, two more local (metastable) minima appear. Physically, particles tend to align their spins according to the coupling setting, giving ferromagnetic phases, unless the temperature is high enough to yield complete disorder. At low enough temperatures, if the system was in a previous state of thermodynamic equilibrium and there is a slow change in the parameters driving it to the current situation, the system may be trapped in one of the metastable states for a while (giving rise to hysteresis). In the discrete choice scenario, it means that in the absence of externalities and if statistical fluctuations are small enough, the system will evolve to a state of equilibrium where there are four possible solutions. If $k>0$ these are given by either two high or low fractions adopters and if $k<0$ by one state of high fraction of adopters for one choice  and low for the other. 

When $k = 0$, at low temperatures, there are four minima where the free energy has the same value. As we move to higher temperatures these merge into two minima along the $t=0$ axis (if $J_{s}>J_{t}$), corresponding to mixed phases when the critical temperature for the $t$ system is arrived at ($K_{B}T_{uc}^{t}=J_{t}$), and these merge into a single minimum at the origin (paramagnetic phase) for temperatures above the $s$ system's critical temperature ($K_{B}T_{uc}^{s}=J_{s}$). The introduction of a nonzero $k$ breaks the symmetry and so for the regions where four minima still exist two of them have a higher value of the free energy (metastable states). The existence of metastable states is associated to a first order phase transition at $k=0$.

\section{Numerical analysis for the zero field case}
\label{sec:nlnumsol}
We have used the Newton-Raphson algorithm to numerically solve the system of equations \eqref{eq:eqstandeg} for fixed $J_{s}=1$ and different values of the rest of the parameters (couplings thus measured in $J_{s}$ units). For all numerical solutions used, a tolerance\footnote{Iterations are carried out until the absolute value of each of the components of the function is smaller than the given tolerance.} of $5\cdot 10^{-6}$ concerning convergence is considered with a minimum of 100 and a maximum of 1000 iterations per solution used. Whenever convergence is not attained or problems are encountered (non-invertible Jacobian, for example) in the process, the solutions are discarded. In an attempt to exhaust the solution space, the algorithm is applied to a variety of initial values $(s_{0},t_{0}) \in I\times I$ with $\times$ indicating the Cartesian product and $I=\{-1.0,-0.9,-0.8,-0.7,-0.6,-0.5,-0.4,-0.3,-0.2,-0.1,-0.01,0.0,0.01,0.1,0.2,0.3,0.4,0.5,0.6,0.7,\\ 0.8,0.9,1.0\}$. Python libraries developed for the computation and analysis of these is available at {\text https://github.com/anafrio/}. Results are summarised bellow. In all plots, green is used for $s$ and blue for $t$. Dark points are used for stable solutions and lighter asp ($\times$, for saddle points) or cross ($+$, for maxima) for non-stable solutions.

The type of graphs we will be analysing plot both average magnetisations (jointly in the same graph) against some parameter. We must therefore be careful to take into account the symmetries of the system for a correct interpretation. Ferromagnetic solutions will always appear as symmetric positive and negative magnetisation branches. When $k$ is positive, this corresponds to having either both (for $s$ and $t$) positive or negative branches (both groups have either a high or low demand). When $k$ is negative, the pair of solutions will consist in the $s$ positive with $t$ negative branch and $s$ negative $t$ positive (one group with high demand the other with low one). 

\subsection{Dependence on temperature}
There is a markedly different behaviour in the different coupling regimes as expected from our previous analysis and the different qualitative behaviour of $l(\beta)$'s roots (equation \eqref{eq:Tceq}).

\begin{figure}
\centering
\includegraphics[width=\textwidth]{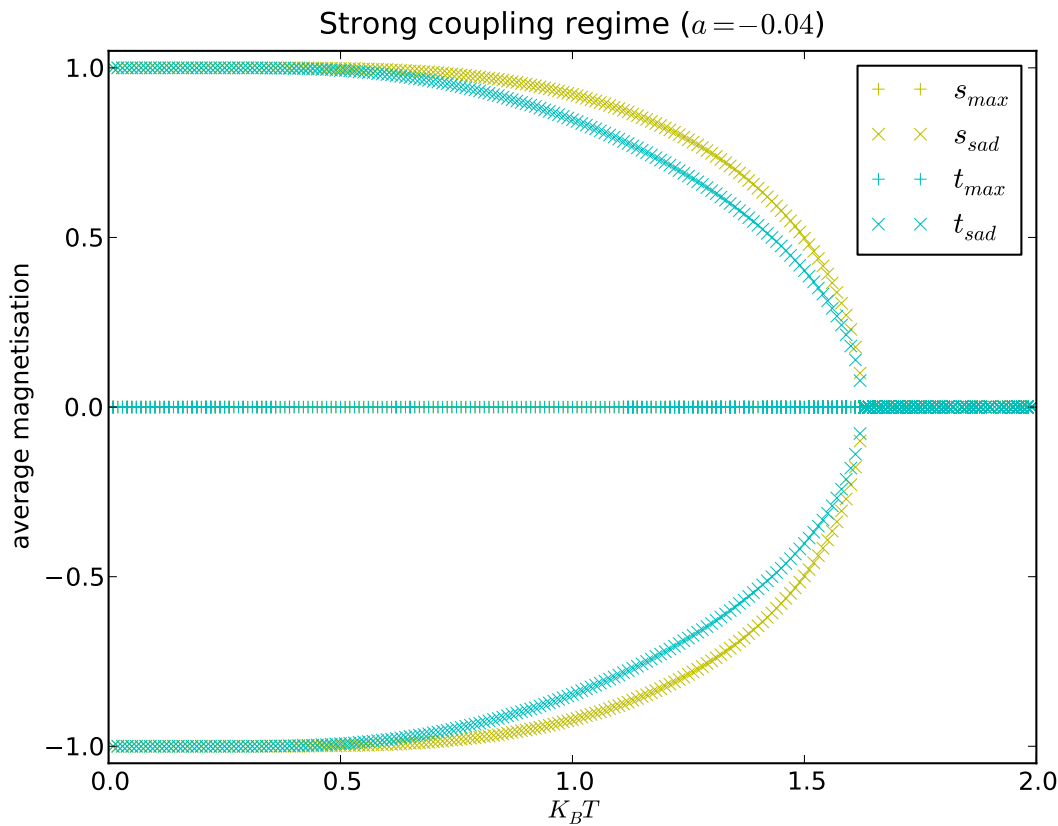}
\caption{Dependence on temperature of the numerically calculated average magnetisations in the strong coupling regime: $J_{s}=1$, $J_{t}= 0.6$ , $k = \pm 0.8$ ($K_{B}T_{c} = 1.62$). Different solutions are plotted for temperatures  between 0.01 and 2 every 0.01 ($K_{B}T$). Magnetisations are plotted in light green for $s$ and light blue for $t$ using asps ($\times$) for saddle point solutions and crosses ($+$) for maxima of $f$.}
\label{fig:nlanaTsc}
\end{figure}

In the strong coupling regime, no stable solutions are found regardless the temperature. For temperatures above $T_{c}$ given by \eqref{eq:Tc}, the only existing solution is the paramagnetic one and it is a saddle point of the free energy. For temperatures bellow it, as we already saw, the paramagnetic phase continues to be a solution, but its stability changes as it now becomes a maximum. Two more ferromagnetic (saddle type) solutions appear at this point: $(m_{s},m_{t})$ and $(-m_{s},-m_{t})$ for positive $k$ and $(m_{s},-m_{t})$ and $(-m_{s},m_{t})$ for negative k ($m_{s},m_{t}>0$). Figure \ref{fig:nlanaTsc}  depicts this situation for $J_{s}=1$, $J_{t}=0.6$, $k=\pm 0.8$ ($a=-0.04$)  with \eqref{eq:Tc} $K_{B}T_{c} = 1.62$.

\begin{figure}
\centering
\includegraphics[width=\textwidth]{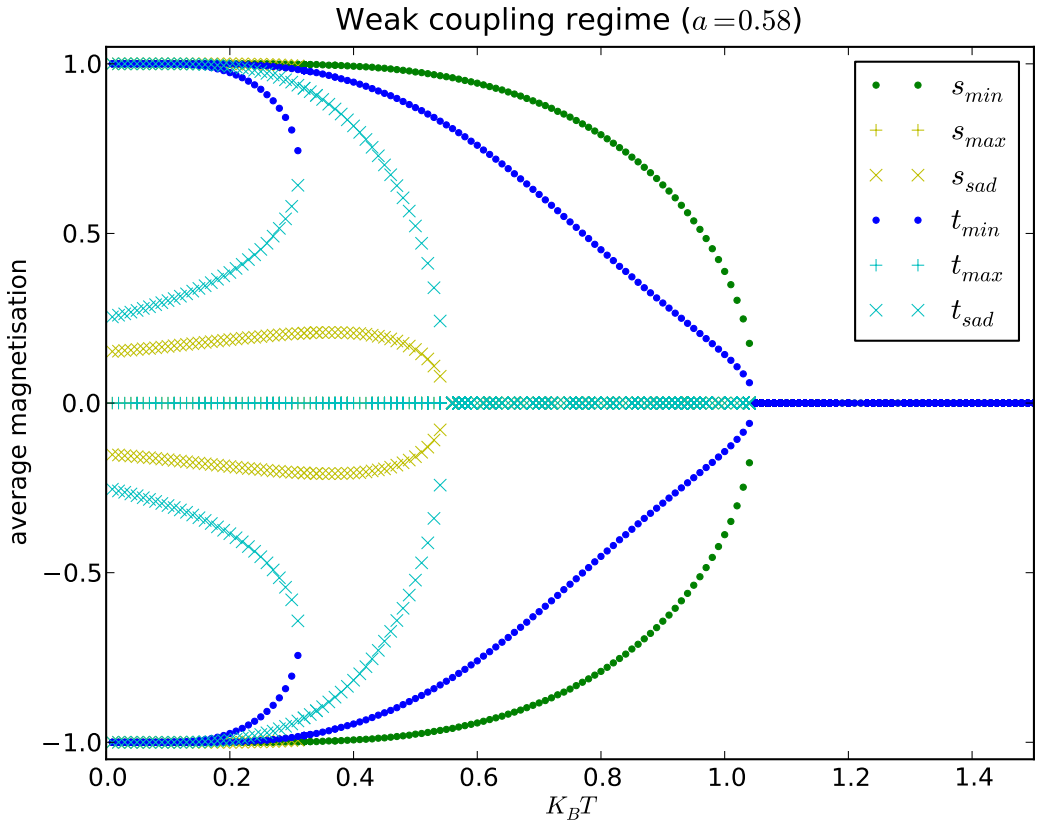}
\caption{Dependence on temperature of the numerically calculated average magnetisations for the weak coupling regime: $J_{s}=1$, $J_{t}= 0.6$ , $k = \pm 0.15$ ($K_{B}T_{b} = 0.55$, $K_{B}T_{c} = 1.05$). Different solutions are plotted for temperatures  between 0.01 and 1.5 every 0.01 ($K_{B}T$). Magnetisations are plotted in green for $s$ and blue for $t$. Dark points are used for stable solutions and lighter asp ($\times$, for saddle points) or cross ($+$, for maxima) for non stable solutions. }
\label{fig:nlanaTwc}
\end{figure}

In the weak coupling regime there are stable solutions for all values of the temperature. For temperatures above both $T_{b}$ and $T_{c}$ (given by \eqref{eq:Tc}) the only existing solution is again the paramagnetic one but now it is a minimum of the free energy. Once bellow $T_{c}$ (the only real critical temperature in terms of magnetic phase transitions), and while above $T_{b}$, two ferromagnetic stable solutions, $(m_{s},m_{t})$ and $(-m_{s},-m_{t})$ for positive $k$ and $(m_{s},-m_{t})$ and $(-m_{s},m_{t})$ for negative k ($m_{s},m_{t}>0$) appear, and the paramagnetic phase becomes a saddle point. At $T_{b}$, the paramagnetic solution becomes a maximum, ferromagnetic solutions remain stable and two new saddle point solutions (with both magnetisations having the same sign when k is negative and different sign when it is positive) appear. At a still lower temperature $T_{a}$, there is another qualitative change. Two new saddle points and two more minima appear. In this case the new stable solutions have opposite relative sign between both magnetisations as the main previously existing ones and are local minima where the free energy value is higher than for the latter and so are metastable. The saddle point ones have the same sign behaviour as the previously existing ones. Figure \ref{fig:nlanaTwc}  shows this for $J_{s}=1$, $J_{t}=0.6$, $k=\pm 0.15$ ($a=0.58$, $K_{B}T_{c} = 1.05$ and $K_{B}T_{b} = 0.55$). These type of points, as we have seen, are usually called \emph{spinodal points} and so in this setup $T_{a}$ can be regarded as the spinodal temperature.

There is thus a phase transition at $T=T_{c}$ in this case. As has already been mentioned, it is in fact a second order transition as indicated by the smooth change in the average magnetisations towards zero as the critical temperature (or critical point in any of the parameters as will be shown in the next subsections) is approached.  

\begin{figure}
\centering
\subfloat[]{\includegraphics[width=0.33\textwidth]{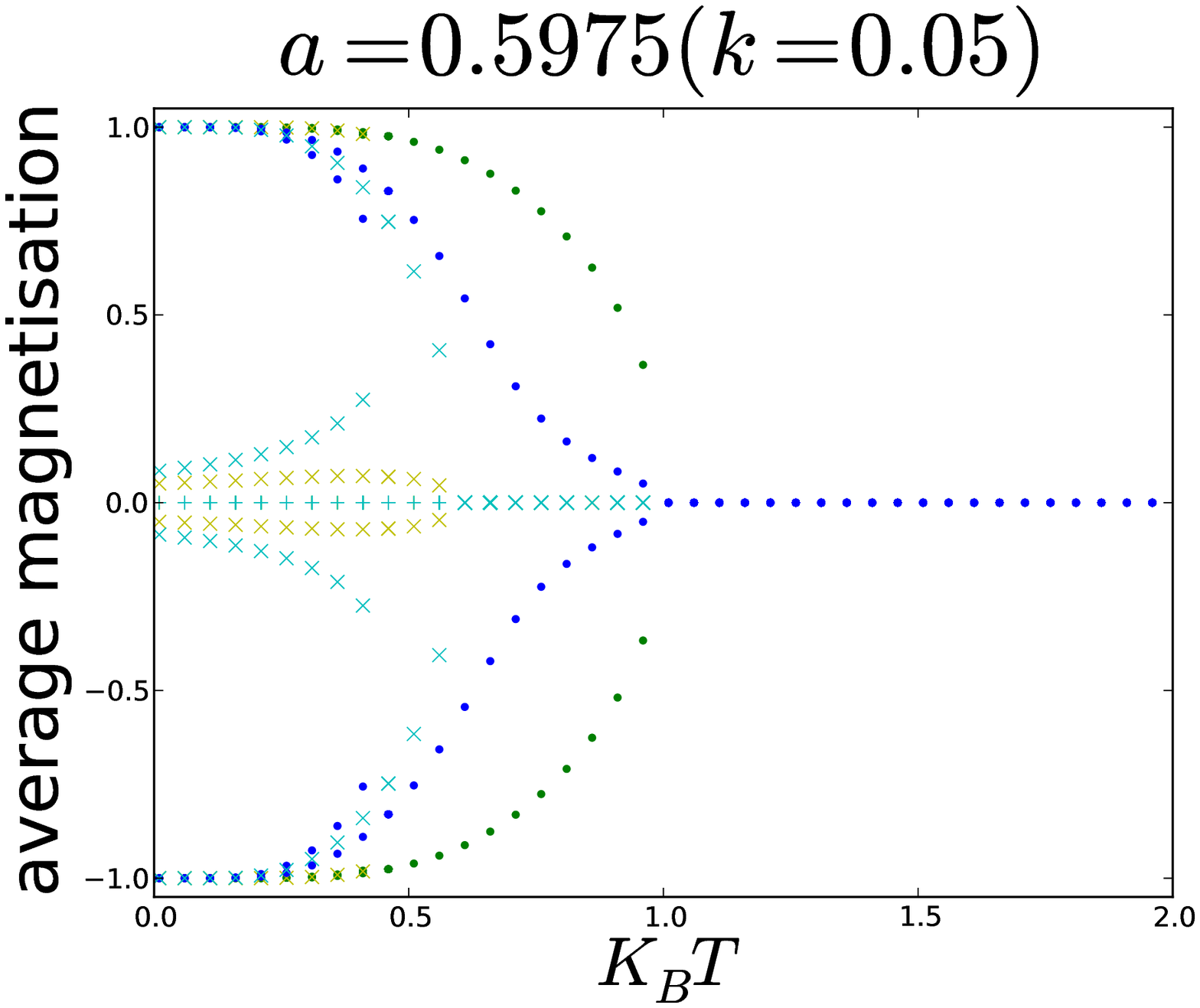}}
\subfloat[]{\includegraphics[width=0.33\textwidth]{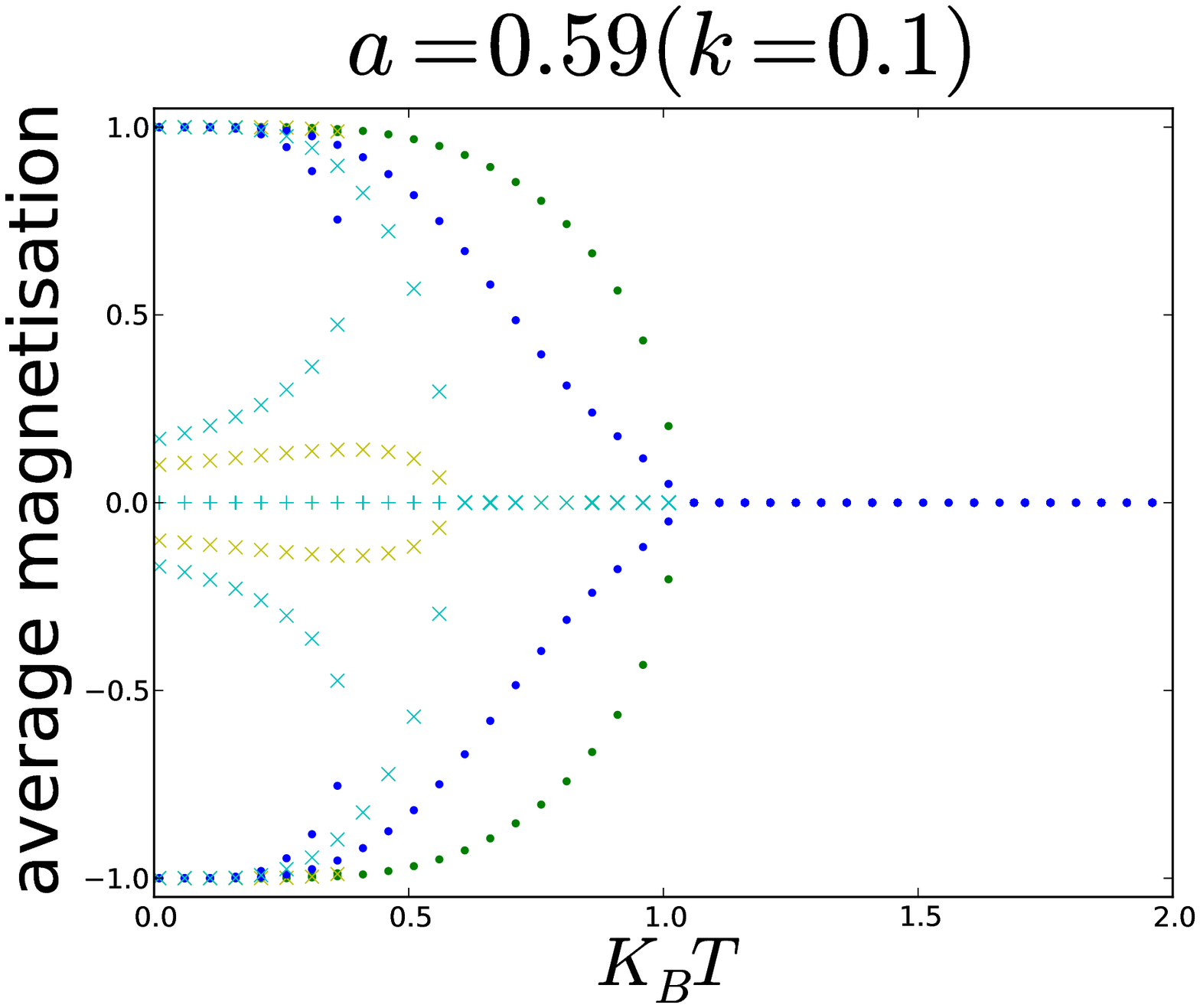}}
\subfloat[]{\includegraphics[width=0.33\textwidth]{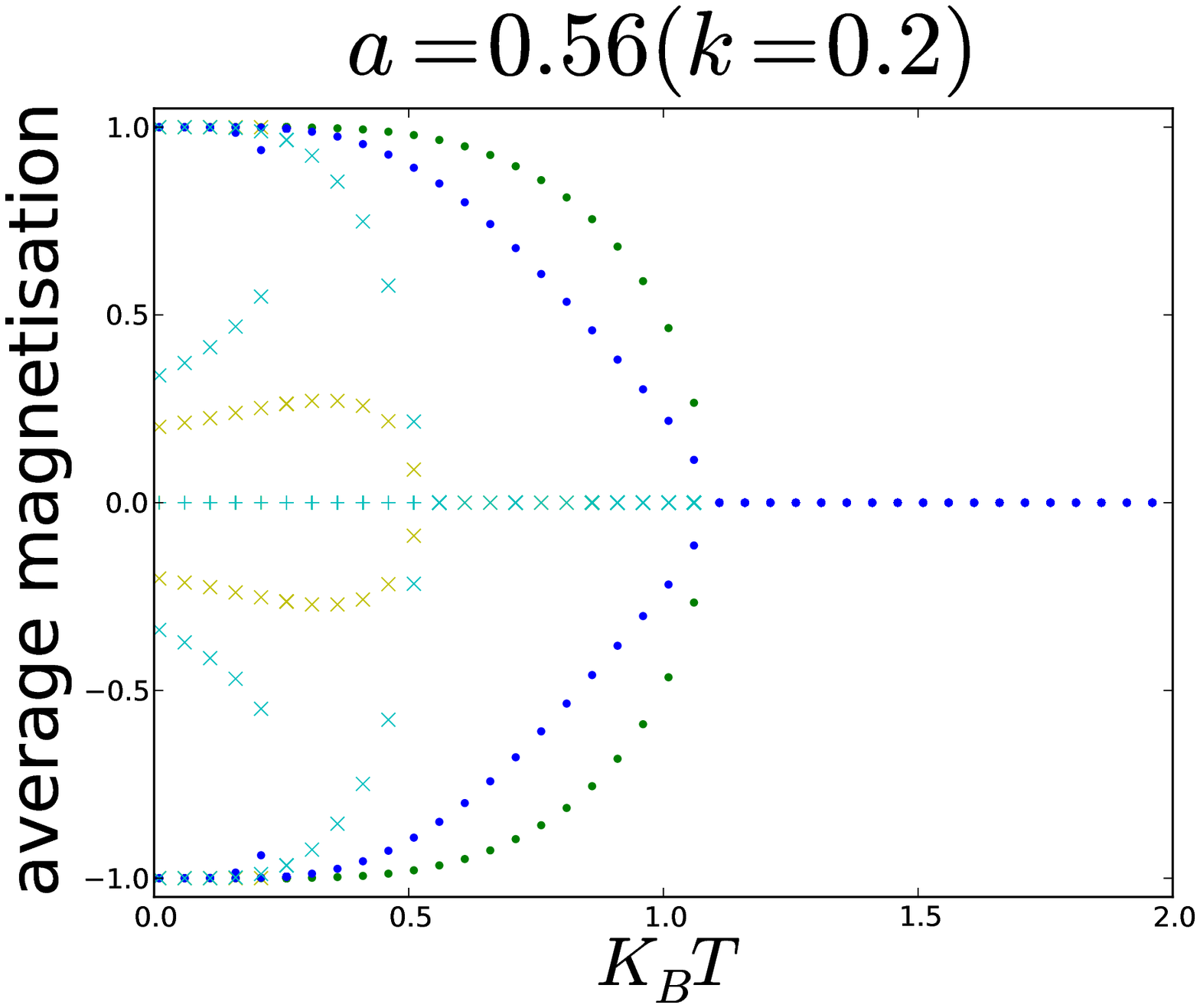}}\\
\subfloat[]{\includegraphics[width=0.33\textwidth]{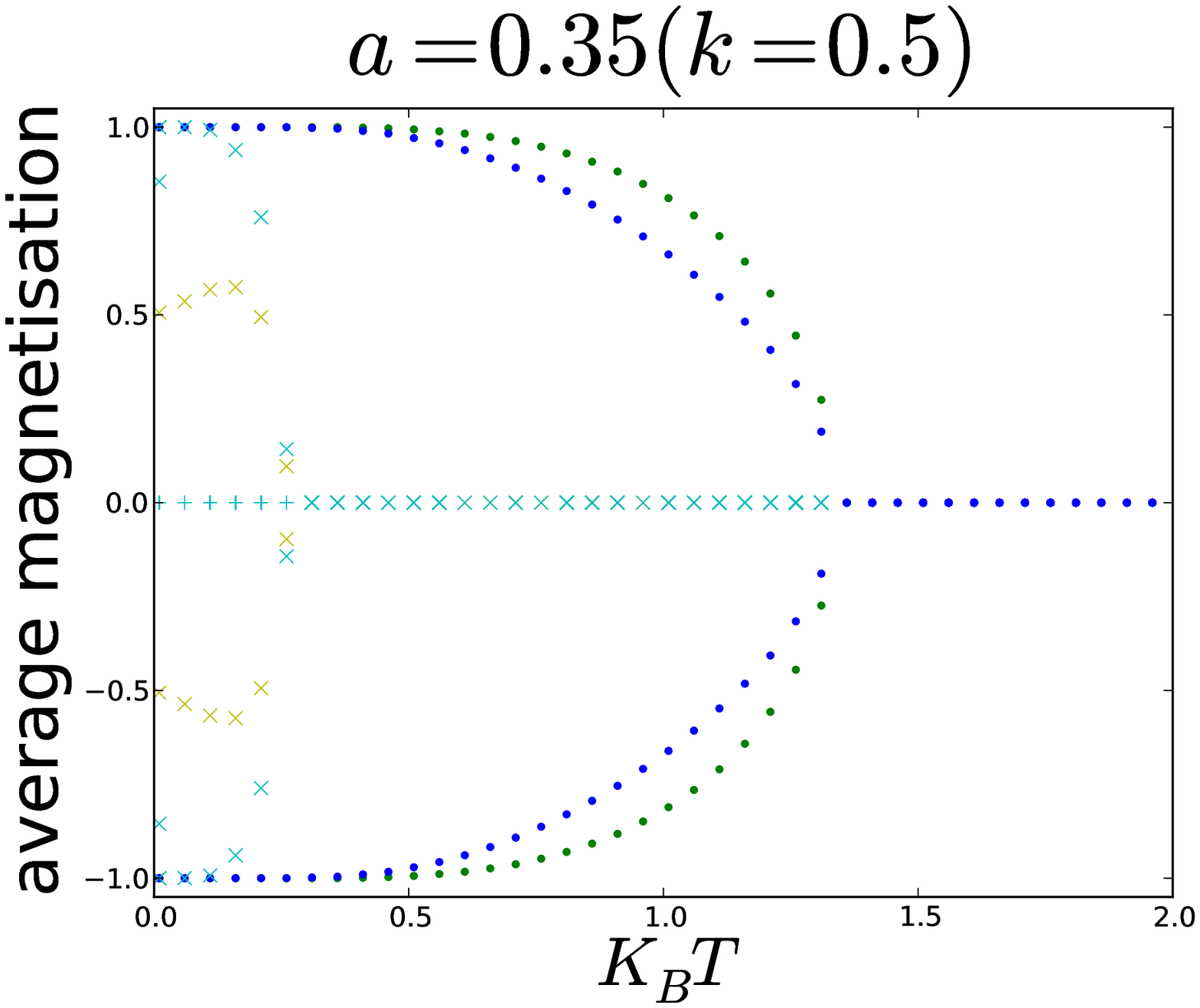}}
\subfloat[]{\includegraphics[width=0.33\textwidth]{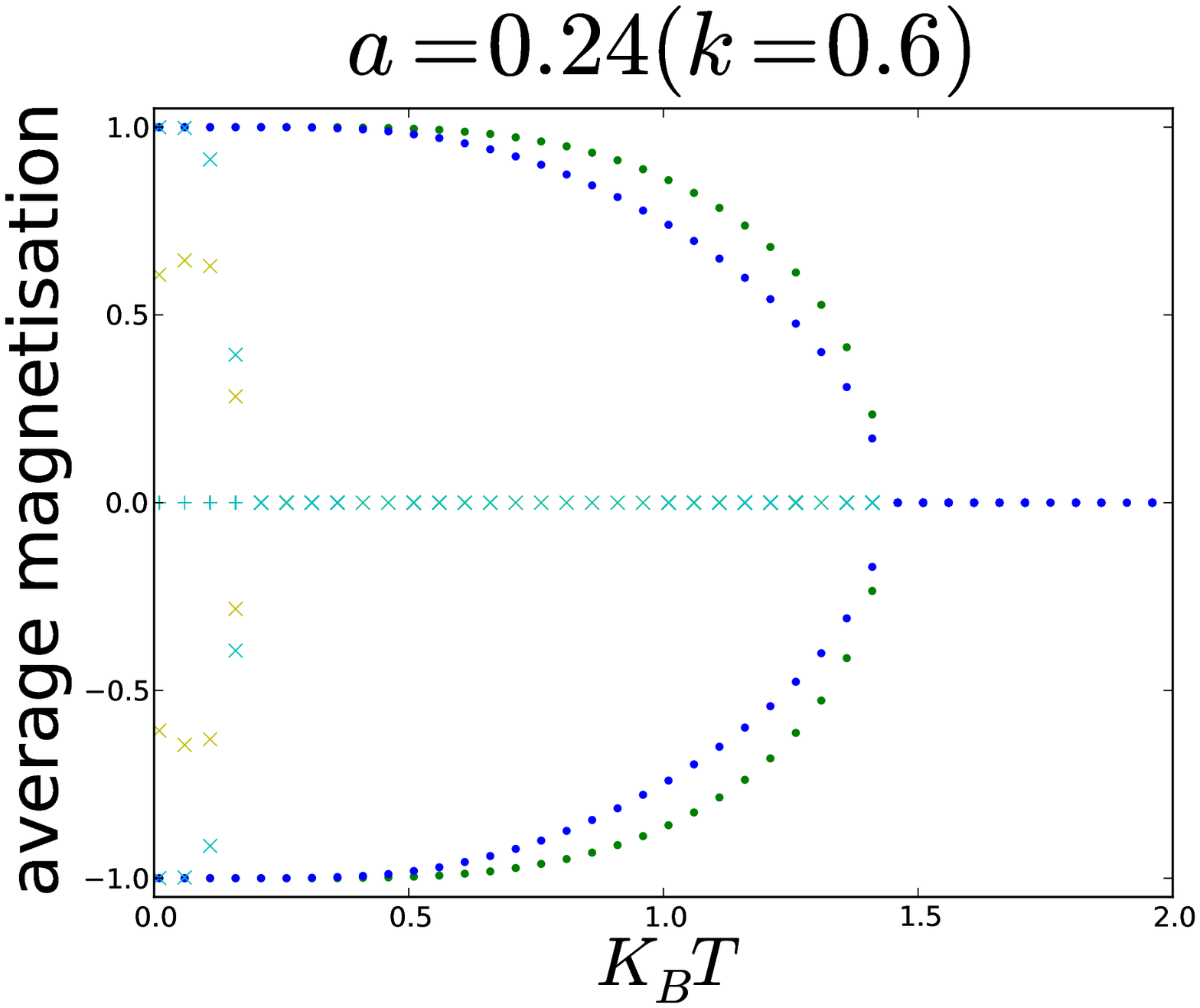}}
\subfloat[]{\includegraphics[width=0.33\textwidth]{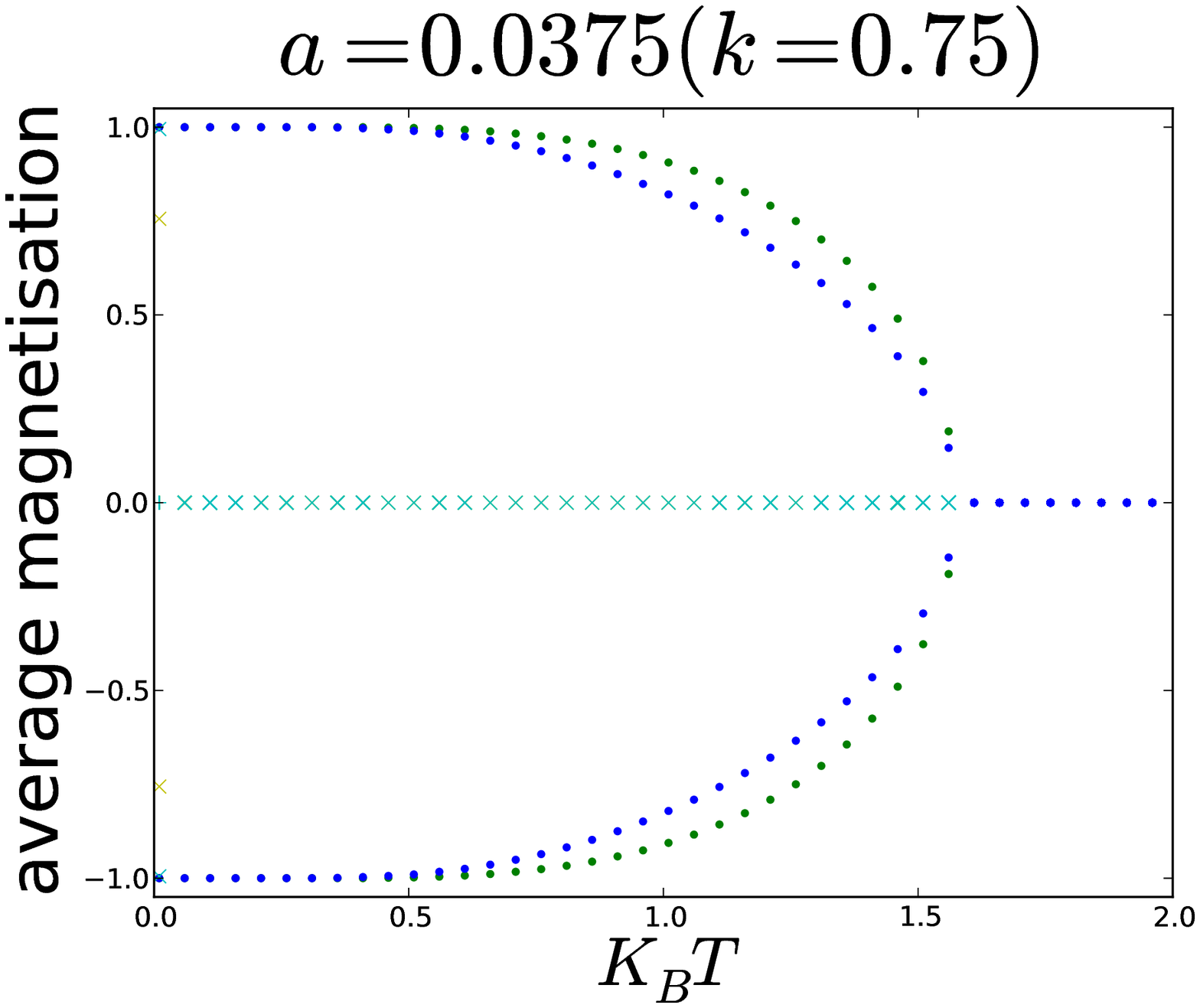}}\\
\subfloat[]{\includegraphics[width=0.33\textwidth]{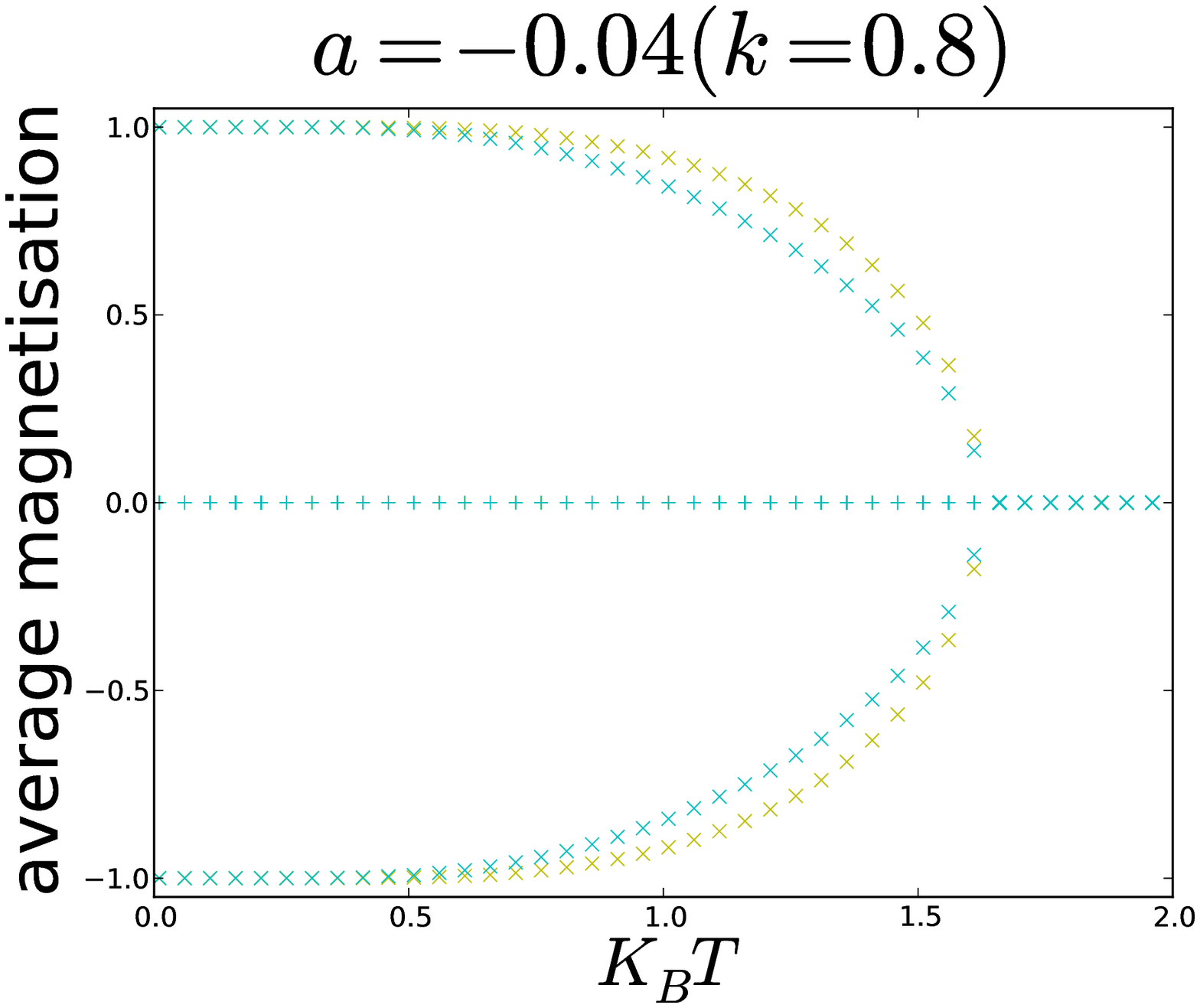}}
\subfloat[]{\includegraphics[width=0.33\textwidth]{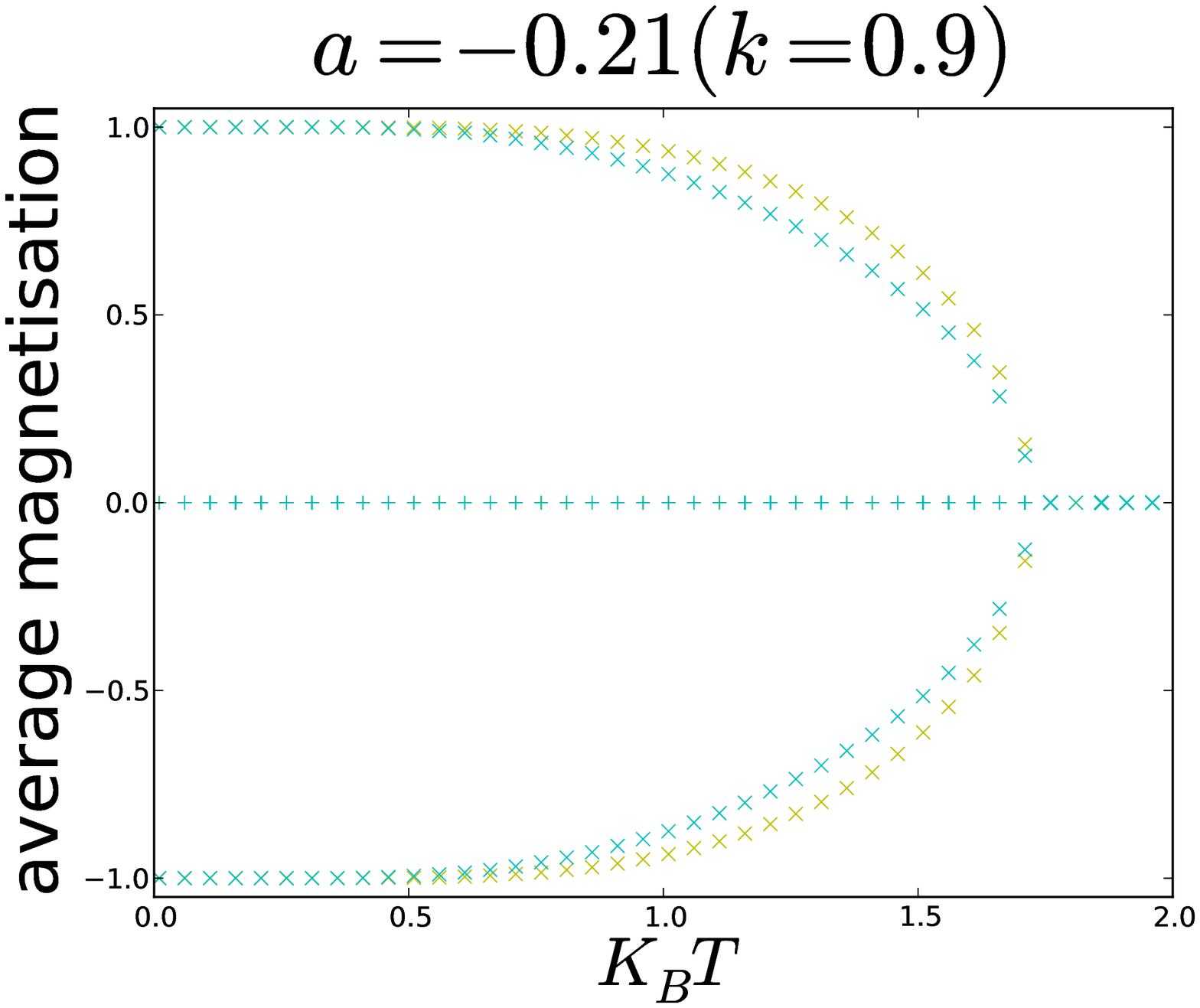}}
\subfloat[]{\includegraphics[width=0.33\textwidth]{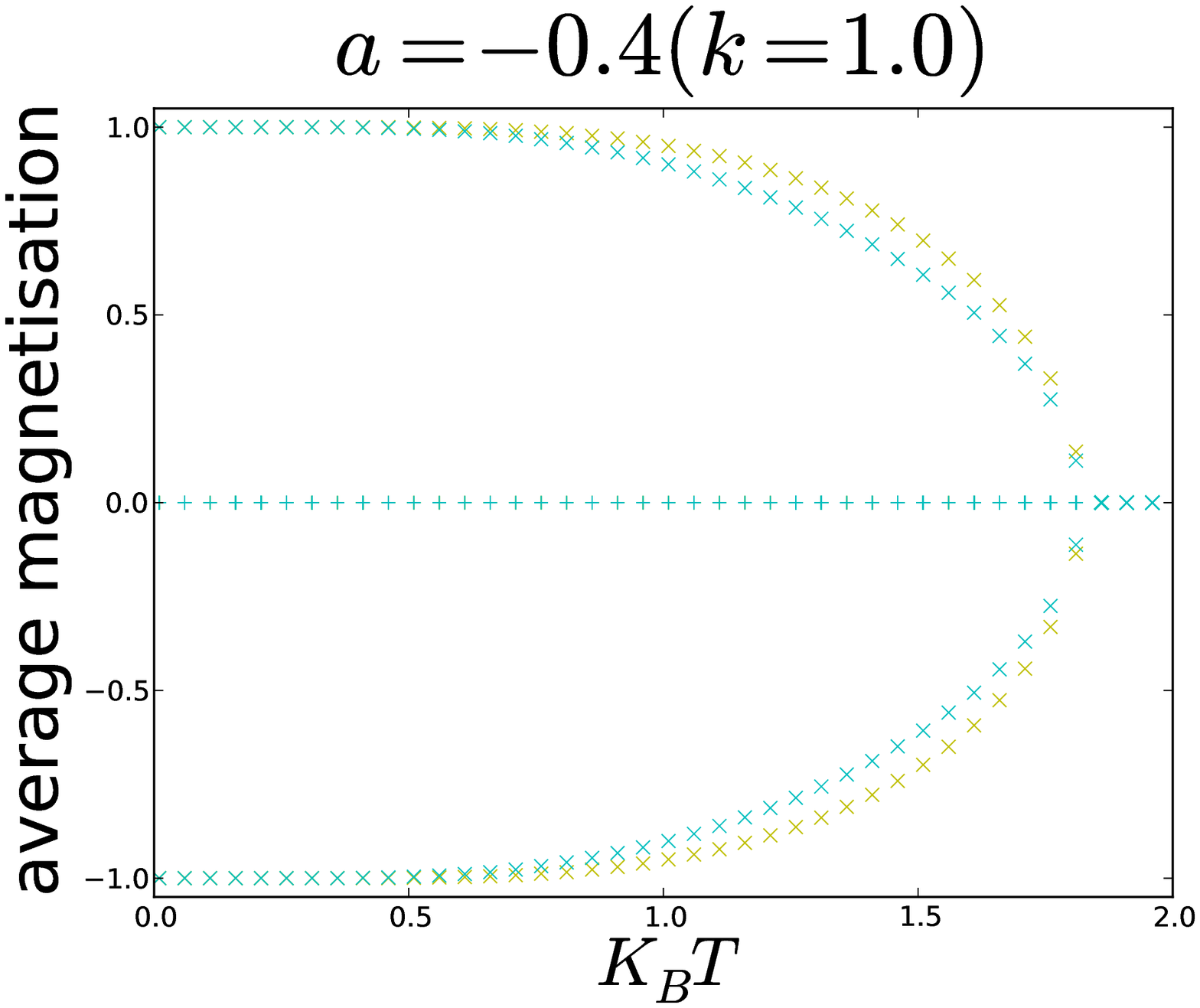}}
\caption{Dependence on temperature of the numerically calculated average magnetisations $(s,t)$ for different values of the inter-coupling $k$. $J_{s}=1$ and $J_{t}= 0.6$  for all plots. (a) $k = 0.05$, (b) $k = 0.1$, (c) $k = 0.2$, (d) $k = 0.5$, (e) $k = 0.6$, (f) $k = 0.75$, (g) $k = 0.8$, (h) $k = 0.9$ and (i) $k = 1$. In all cases, different solutions are plotted for temperatures  between 0.01 and 2 every 0.05 ($K_{B}T$). Magnetisations are plotted in green for $s$ and blue for $t$. Dark points are used for stable solutions and lighter asp ($\times$, for saddle points) or cross ($+$, for maxima) for non stable solutions.}
\label{fig:nlanakT}
\end{figure}

Figure \ref{fig:nlanakT} depicts how the dependence on the temperature varies when we change $k$ leaving $J_{t}$ fixed ($J_{t}=0.6$). Starting from a value of the inter-coupling $k$ (figure \ref{fig:nlanakT} a) small enough to be in the weak coupling with metastable states (characterised by $T_{a}$, $T_{b}$, $T_{c}$) as we move to higher absolute values of $k$ (figure \ref{fig:nlanakT} b, c) without leaving the weak coupling regime, $T_{c}$ moves to higher values and $T_{a}$ and $T_{b}$ to smaller ones. The absolute values of stable $s$ and $t$ at every point become more and more similar and the region where they are both practically one larger. For a strong enough inter-coupling (figure \ref{fig:nlanakT} d, e), the metastable states disappear (no $T_{a}$) while there are still additional saddle point solutions that also disappear when we move to even stronger values of $k$ (figure \ref{fig:nlanakT} f, no $T_{b}$). When we move into $k$ values in the strong coupling regime (figure \ref{fig:nlanakT} g), all stable solutions become saddle points and saddle points become maxima. Now the only characteristic temperature is $T_{c}$ (which does not represent a phase transition temperature in strict sense anymore) that moves to higher values as we move to stronger and stronger couplings (figure \ref{fig:nlanakT} h, i). 

\begin{figure}
\centering
\subfloat[]{\includegraphics[width=0.33\textwidth]{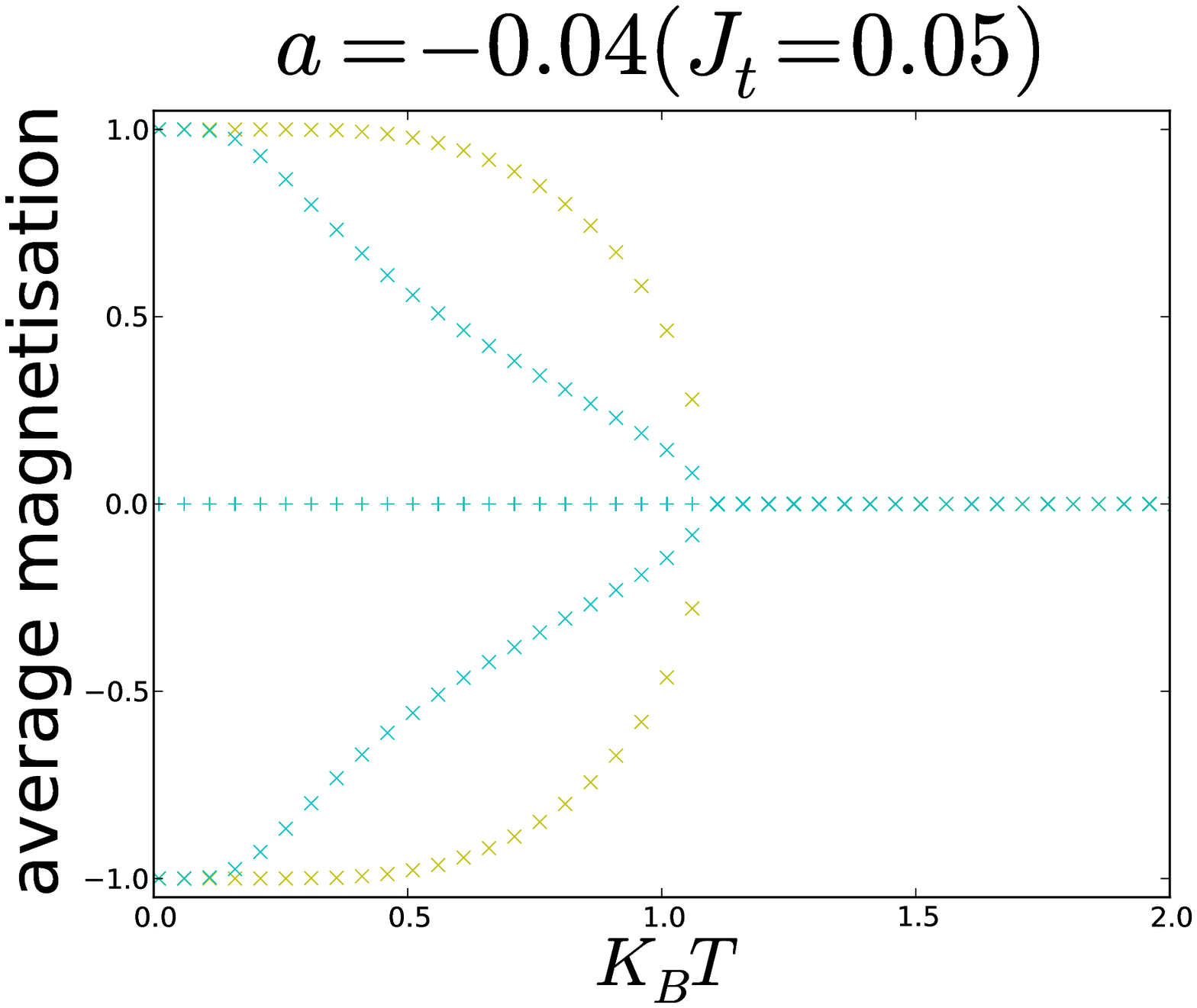}}
\subfloat[]{\includegraphics[width=0.33\textwidth]{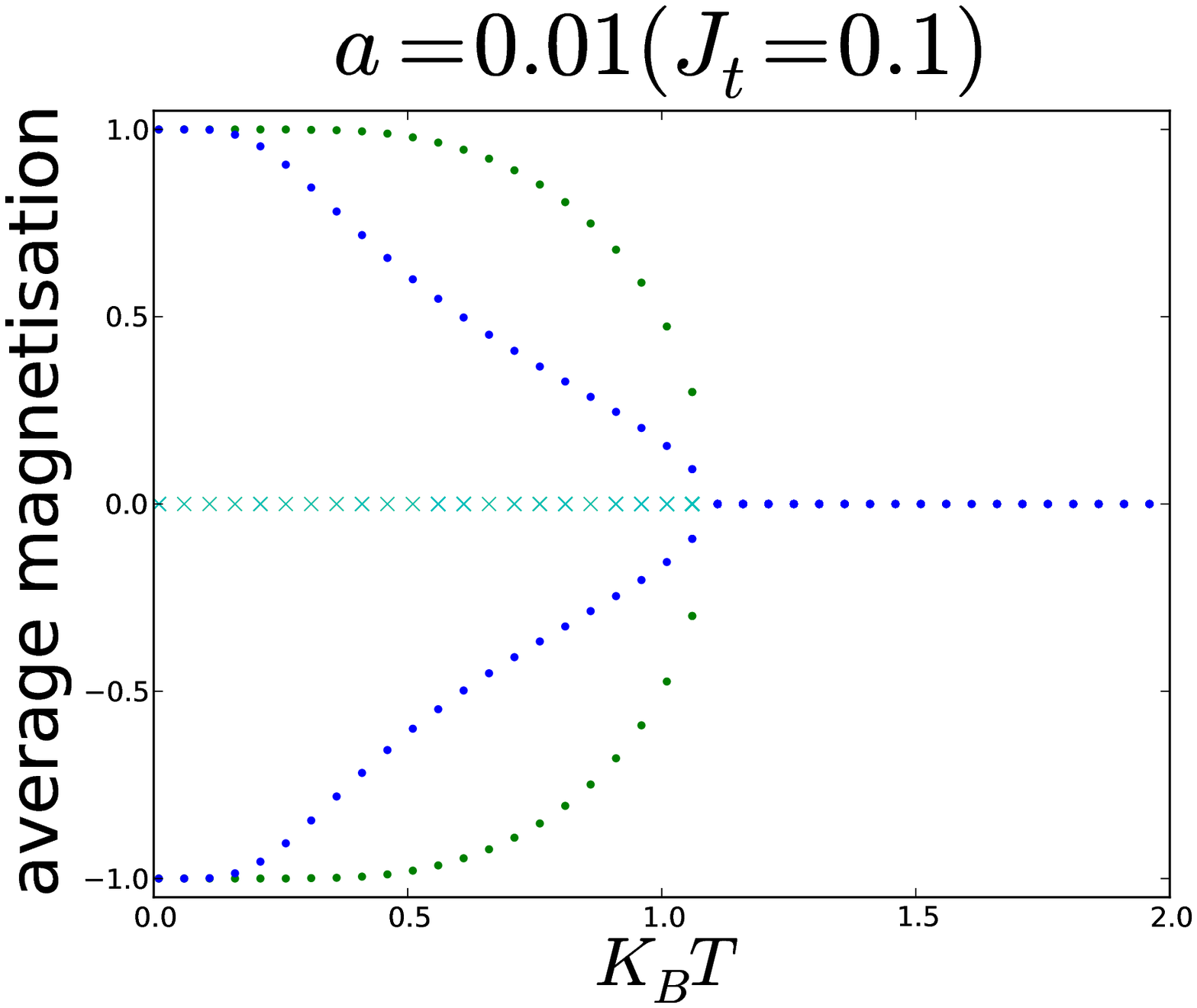}}
\subfloat[]{\includegraphics[width=0.33\textwidth]{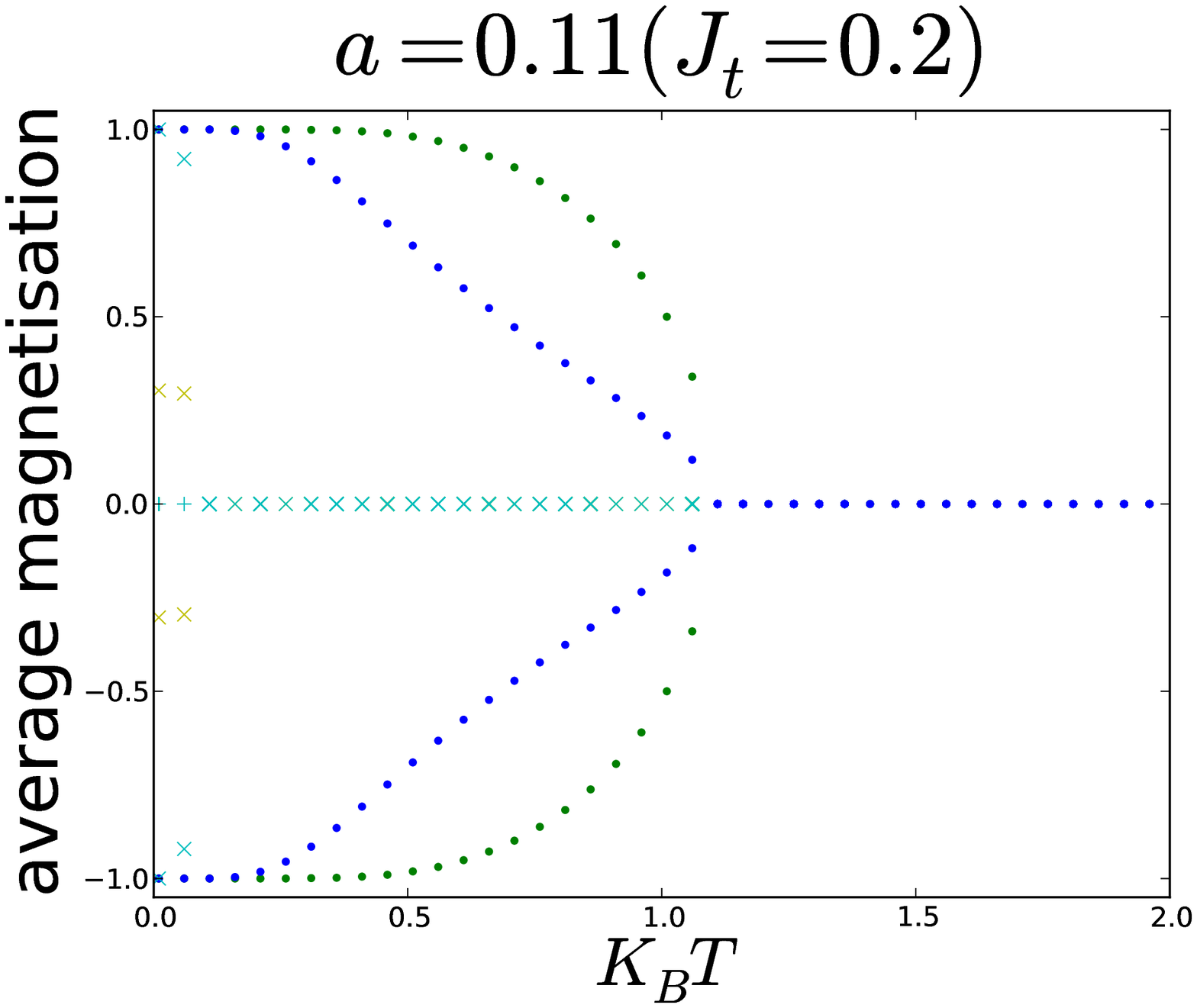}}\\
\subfloat[]{\includegraphics[width=0.33\textwidth]{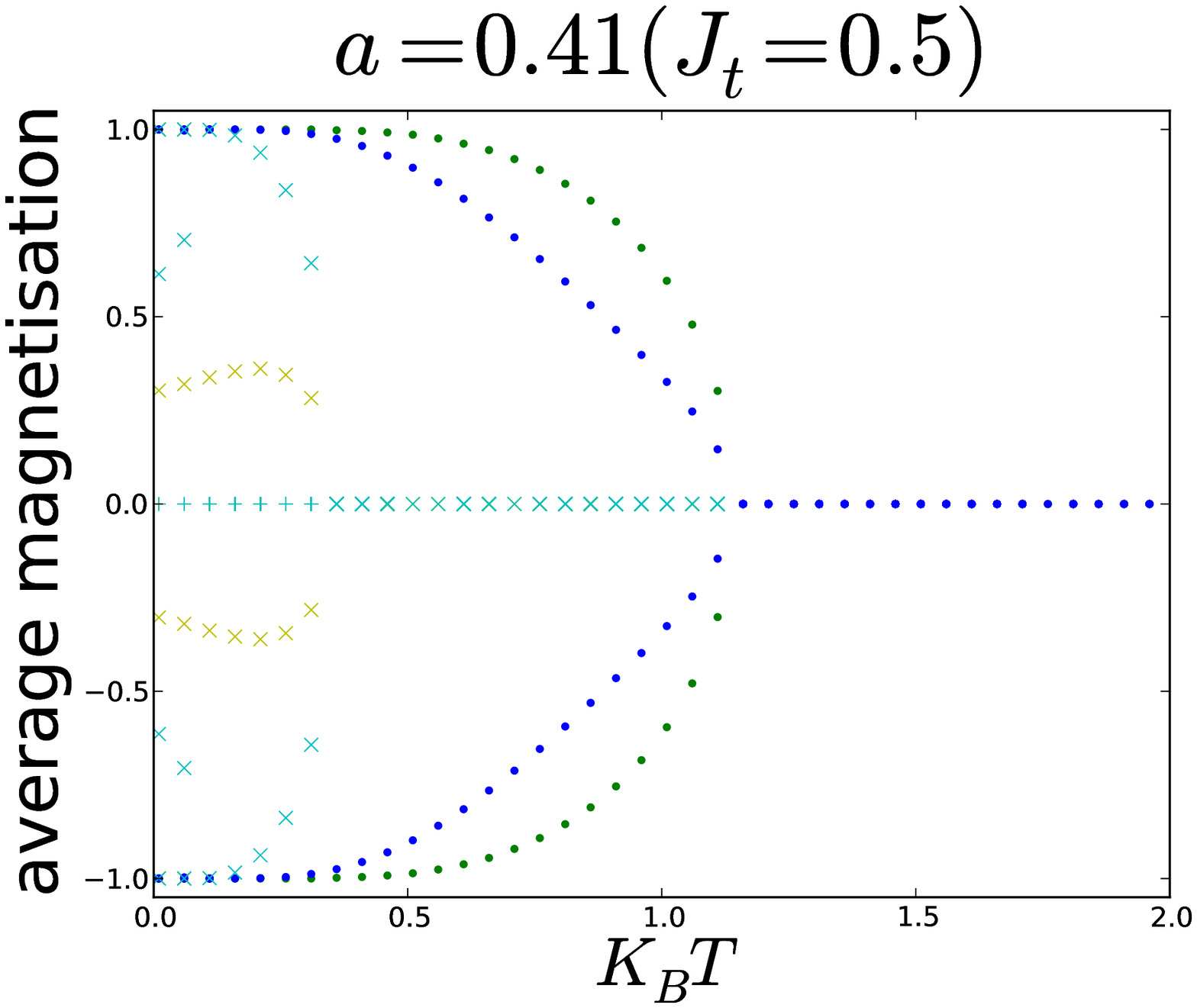}}
\subfloat[]{\includegraphics[width=0.33\textwidth]{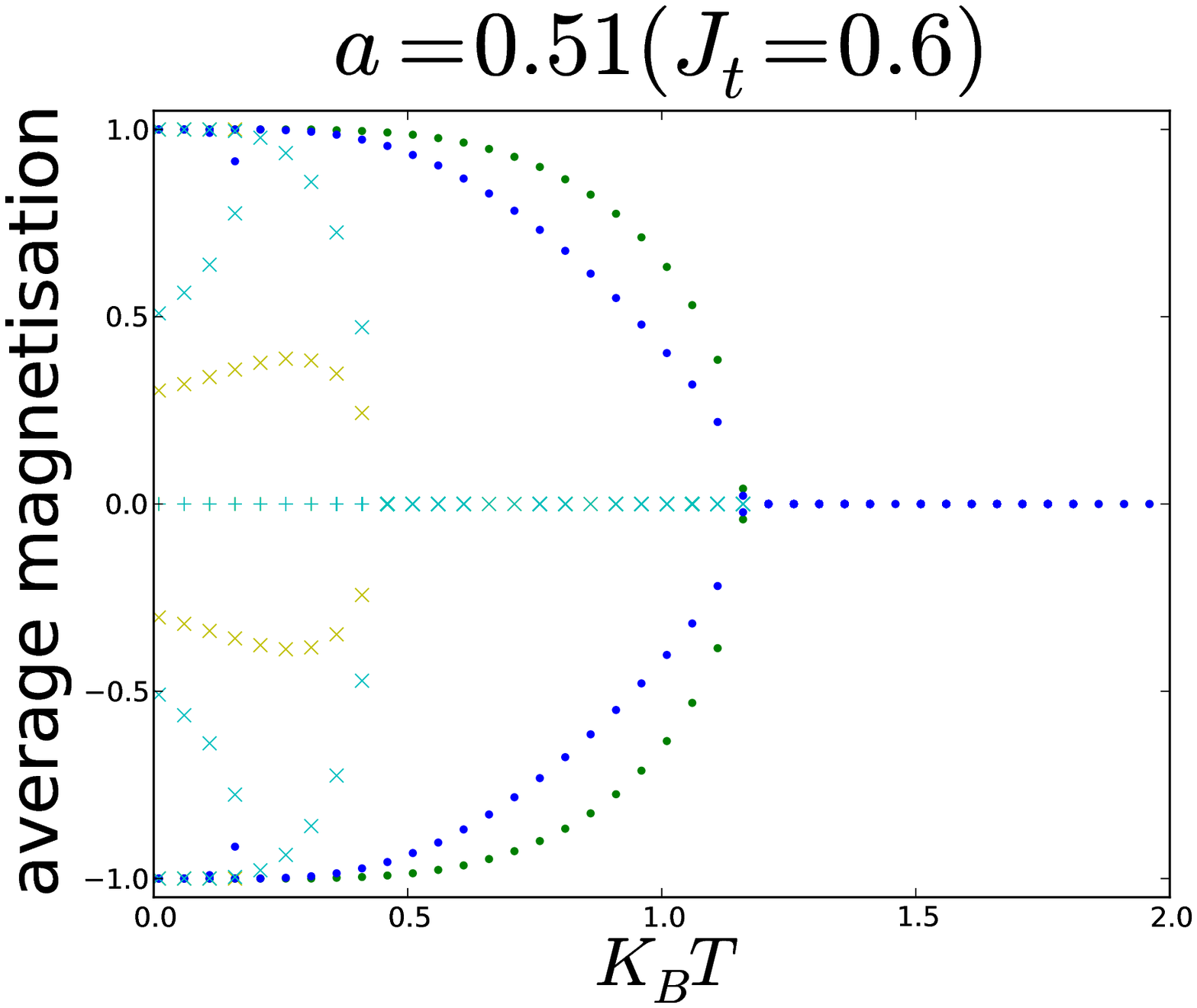}}
\subfloat[]{\includegraphics[width=0.33\textwidth]{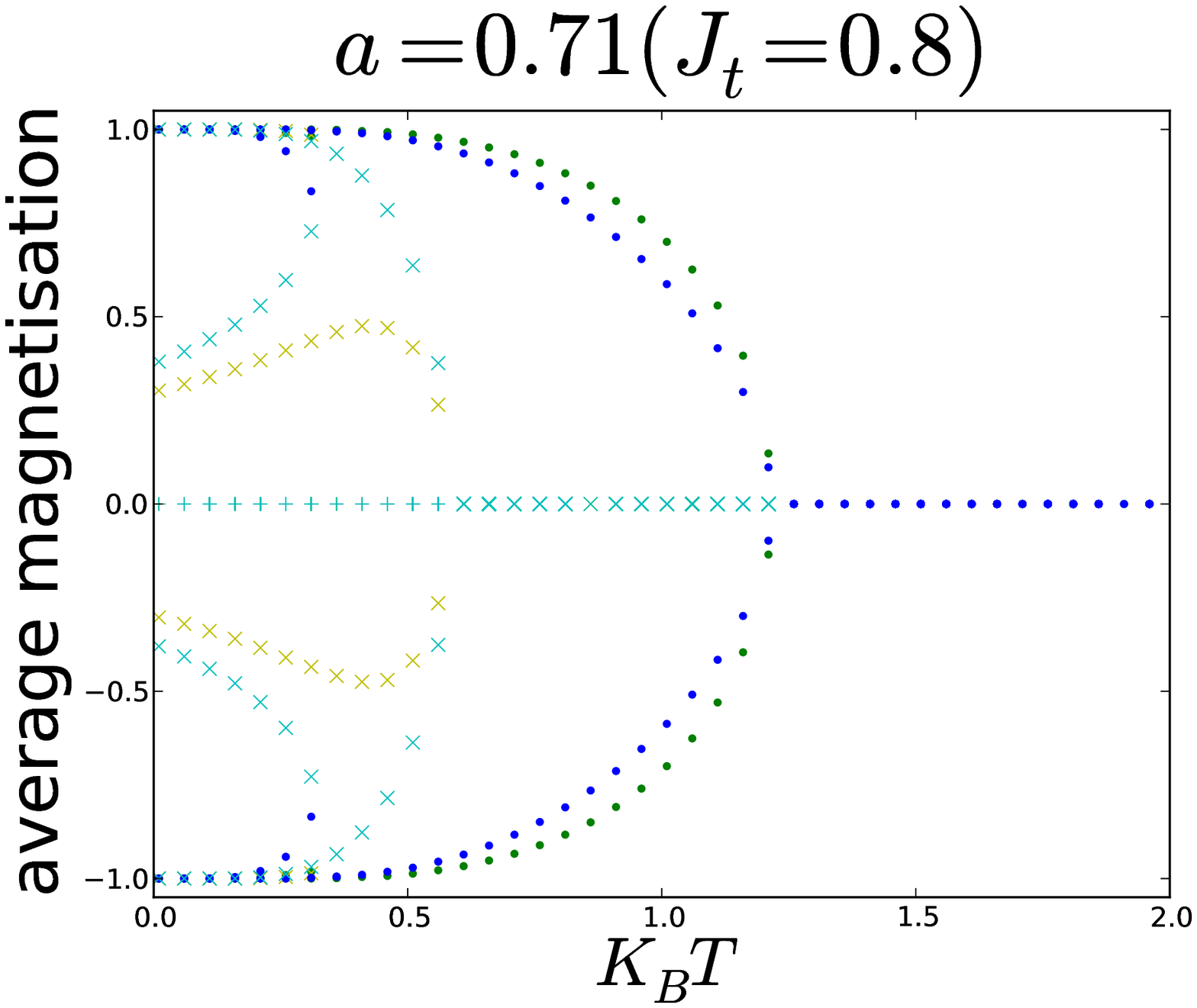}}\\
\subfloat[]{\includegraphics[width=0.33\textwidth]{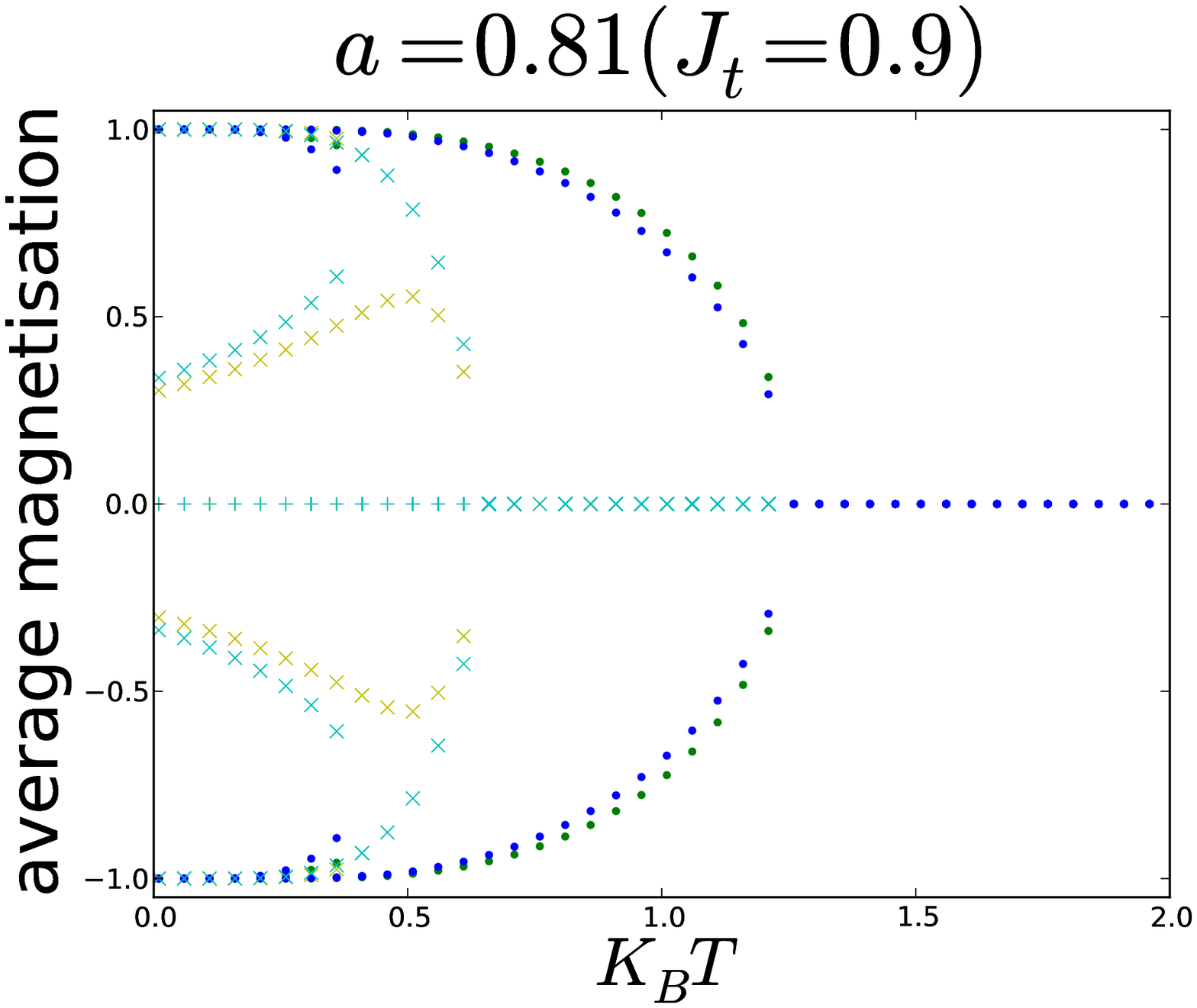}}
\subfloat[]{\includegraphics[width=0.33\textwidth]{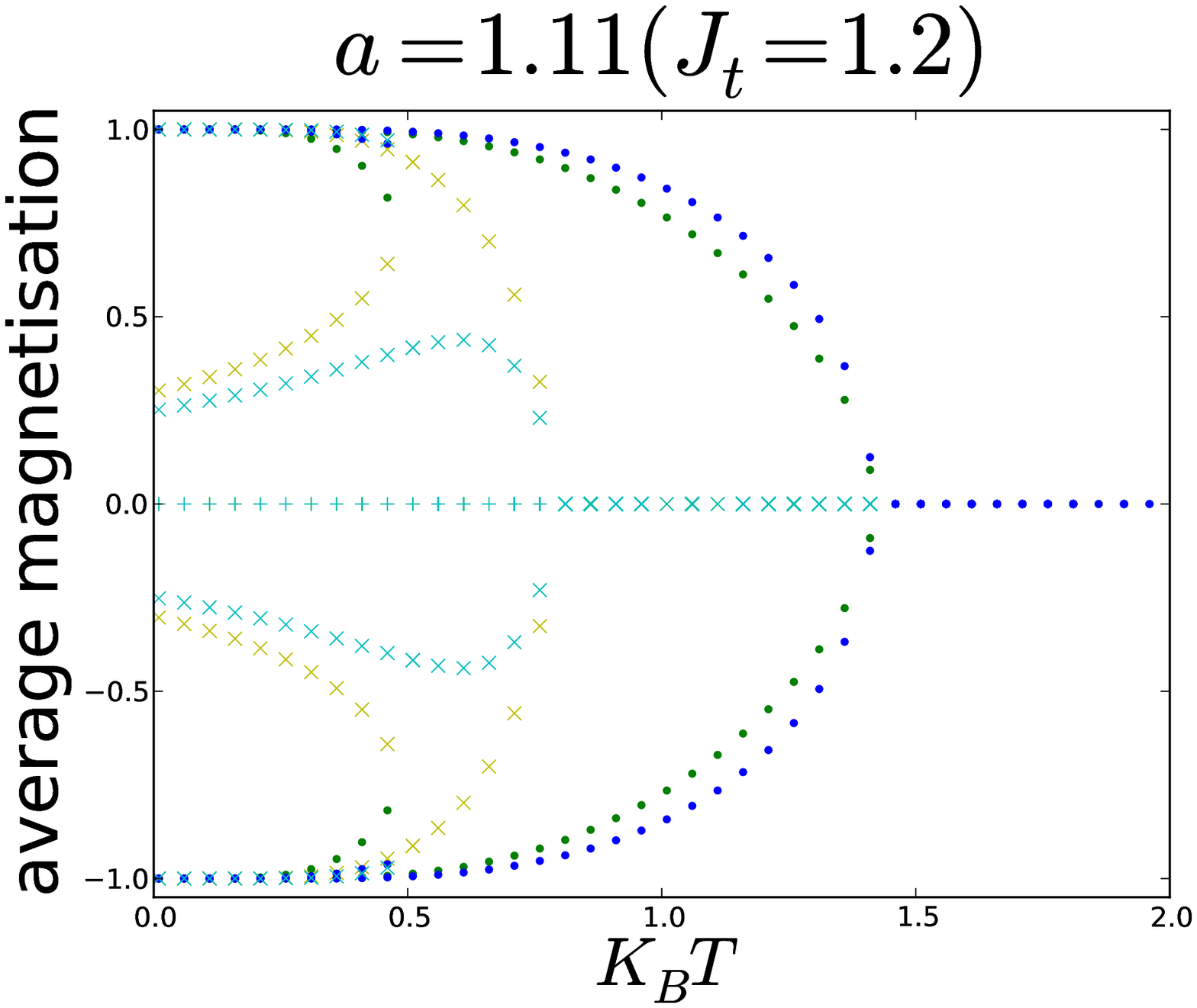}}
\subfloat[]{\includegraphics[width=0.33\textwidth]{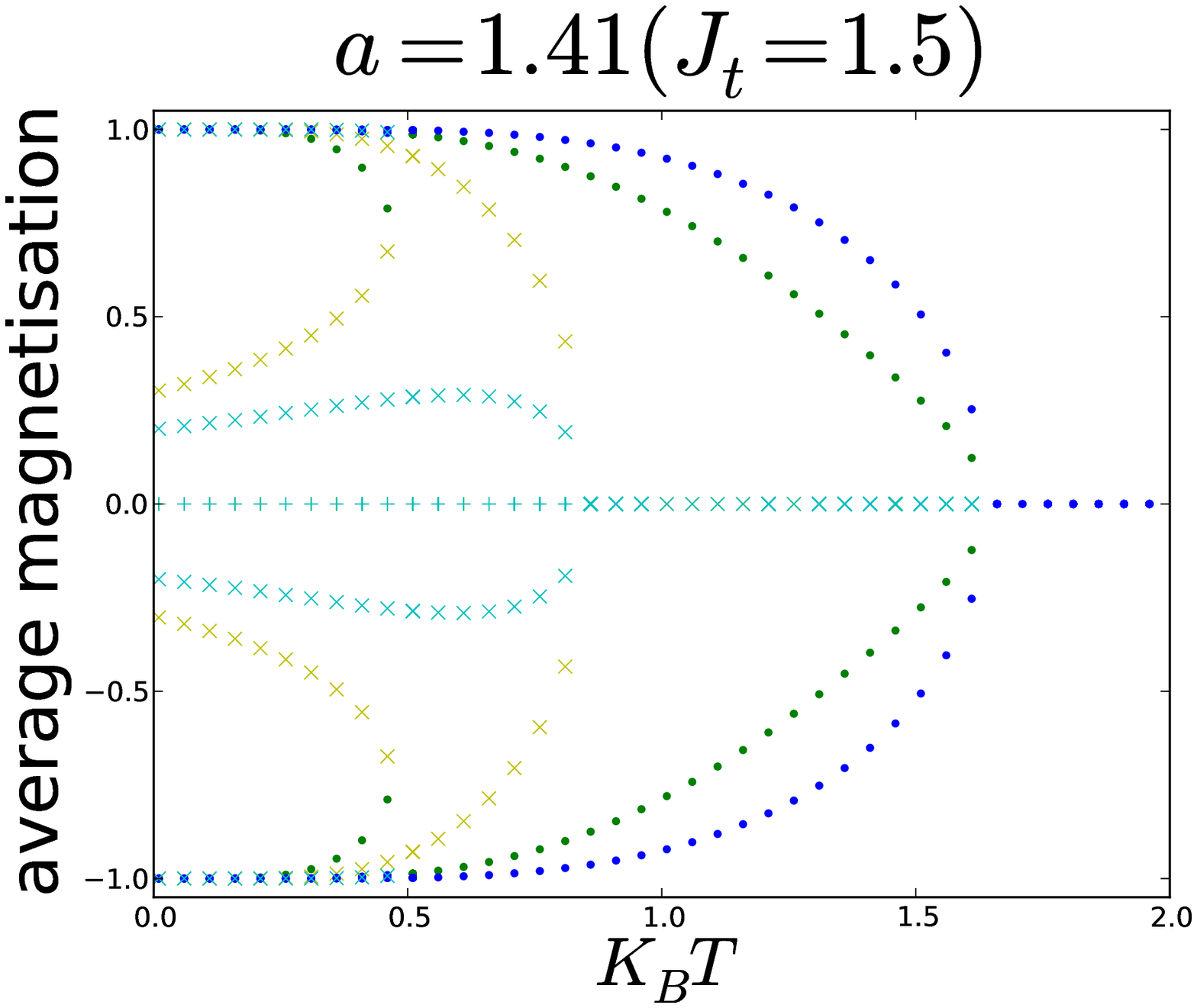}}
\caption{Dependence on temperature of the numerically calculated average magnetisations $(s,t)$ for different values of the inter-coupling $J_{t}$. $J_{s}=1$ and $k=\pm 0.3$  for all plots. (a) $J_{t} = 0.05$, (b) $J_{t} = 0.1$, (c) $J_{t} =0.2 $, (d) $J_{t} = 0.5$, (e) $J_{t} =0.6 $, (f) $J_{t} = 0.8$, (g) $J_{t} =0.9$, (h) $J_{t} =1.2 $ and (i) $J_{t} = 1.5$. In all cases, different solutions are plotted for temperatures  between 0.01 and 2 every 0.05 ($K_{B}T$). Magnetisations are plotted in green for $s$ and blue for $t$. Dark points are used for stable solutions and lighter asp ($\times$, for saddle points) or cross ($+$, for maxima) for non stable solutions.}
\label{fig:nlanaJT}
\end{figure}

We can also study how the dependence on the temperature varies as we change $J_{t}$ for fixed $k$ ($k=\pm 0.3$). This situation is depicted in figure \ref{fig:nlanaJT}. Starting from small values of $J_{t}$ (strong coupling regime) with no stable solutions characterised by $T_{c}$ (figure \ref{fig:nlanaJT} a), as we increase $J_{t}$, $T_{c}$ moves to higher values and eventually when the weak coupling regime is reached (figure \ref{fig:nlanaJT} b), ferromagnetic solutions become stable bellow $T_{c}$ and paramagnetic solutions above $T_{c}$. As we continue to increase $J_{t}$, first additional saddle points (at $T_{b}$) appear (figure \ref{fig:nlanaJT} c). For higher $J_{t}$, both $T_{b}$ and $T_{a}$ become larger (figure \ref{fig:nlanaJT} d). At some point, two more saddle solutions, together with metastable solutions, appear for $T<T_{a}$ (figure \ref{fig:nlanaJT} e). If we continue to increase $J_{t}$, $T_{a}$, $T_{b}$ and $T_{c}$ move to higher values (figure \ref{fig:nlanaJT} f,g) but then after a given temperature ($K_{B}T$ around 1 in this case), $K_{B}T_{a}$ will remain constant at approximately 0.5.  Throughout all the process the region where average magnetisations are practically one in absolute value becomes larger as $J_{t}$ increases. The more similar $J_{s}$ and $J_{t}$ are (the closer $J_{t}$ is to one in this analysis), the more similar do absolute values of $s$ and $t$ in each ferromagnetic solution  become. Notice also that obviously for $J_{t}<J_{s}=1$, $s$ ferromagnetic branches are higher in absolute value than $t$ branches, and that this situation reverses when $J_{t}$ becomes larger than $J_{s}=1$.

\subsection{Dependence on inter-coupling}

If we consider the dependence of the function $l$ defined in equation \eqref{eq:Tceq} on the inter-coupling $k$, we can rewrite the function's roots as given by

\begin{equation}
k_{c}=\pm \sqrt{J_{s}J_{t}-\frac{1}{\beta}(J_{s}+J_{t})+\frac{1}{\beta^{2}}}
\end{equation}

\noindent There will  be two additional values (to the two provided by $k_{d}^{2}=J_{s}J_{t}$)  $k_{c}$ (of equal absolute value and opposite sign) at which the sign of the determinant of the Hessian for the paramagnetic phase \eqref{eq:nldethesspara} changes (saddle to minimum/maximum), unless $K_{B}T_{c0t}=J_{t}<K_{B}T<J_{s}=K_{B}T_{c0s}$ (for $J_{t}<J_{s}$)\footnote{Note that at the uncoupled critical values $K_{B}T_{uc}^{t}=J_{t}$ and $J_{s}=K_{B}T_{uc}^{s}$,  $k_{c}=k_{d}$.}. The sign of the second derivative for the paramagnetic phase \eqref{eq:fss3} will change at $k^{2}=J_{s}(J_{s}+K_{B}T)$.

We thus recover the critical temperatures for the uncoupled case with an interesting interpretation. For a constant temperature above both uncoupled critical temperatures $T_{uc}^{s}$ and $T_{uc}^{t}$, if $| k_{c} |<\sqrt{J_{s}J_{t}}$ the paramagnetic phase will be stable for $k^{2} <k_{c}^{2}$, saddle point for $k_{c}^{2}<k^{2} <J_{s}J_{t}$ and unstable for $k^{2}>J_{s}J_{t}$. There is thus a second order phase transition in $k_{c}$ in this case. If $| k_{c} | > \sqrt{J_{s}J_{t}}$ the paramagnetic phase will be stable for $k^{2} <J_{s}J_{t}$, saddle point for $J_{s}J_{t}<k^{2} <k_{c}^{2}$ and unstable for $k^{2}>k_{c}^{2}$ (in this case there is no phase transition as there are no stable solutions at all for $k^{2}>J_{s}J_{t}$). For temperatures bellow both uncoupled critical temperatures, if $| k_{c} |<\sqrt{J_{s}J_{t}}$ the paramagnetic phase will be a maximum for $k^{2} <k_{c}^{2}$, saddle point for $k_{c}^{2}<k^{2} <J_{s}J_{t}$, and unstable again for $k^{2}>J_{s}J_{t}$. If $| k_{c} | > \sqrt{J_{s}J_{t}}$ the paramagnetic phase will be unstable for $k^{2} <J_{s}J_{t}$, saddle point for $J_{s}J_{t}<k^{2} <k_{c}^{2}$, and unstable for $k^{2}>k_{c}^{2}$. There is thus no phase transition at low temperatures and the change is related to the apparition of new saddle point ferromagnetic solutions that as we have seen announce the onset of the metastable states. At intermediate temperatures ($J_{t}<K_{B}T<J_{s}$ when $J_{t}<J_{s}$), the paramagnetic phase will be a saddle point for $k^{2} <J_{s}J_{t}$ and maximum for $k^{2} >J_{s}J_{t}$ (again no phase transition). There are no metastable states in this last case. 

Regarding the dependence on $k$, we will be showing it graphically in detail for low and high temperature (understood as $K_{B}T<J_{t}<J_{s}$ and $K_{B}T>J_{s}>J_{t}$ respectively) behaviour. The intermediate case ($J_{t}<K_{B}T<J_{s}$) can be seen as the limiting case of low temperature behaviour when there are neither metastable nor saddle point ferromagnetic solutions. 

In all cases there will only be stable solutions for $k^{2}<J_{s}J_{t}$. Magnetisations appear as symmetric with respect to the $k=0$ axis, although the relative sign between both magnetisations in every pair of solutions will be different depending on the sign of $k$. For fixed $J_{s}$ and $J_{t}$, above a certain temperature, there can be stable paramagnetic or ferromagnetic solutions depending on the value of $k$. Bellow this temperature however, the paramagnetic phase no longer exists as a minimum, and for low enough temperatures new metastable ferromagnetic solutions appear.

\begin{figure}
\centering
\includegraphics[width=\textwidth]{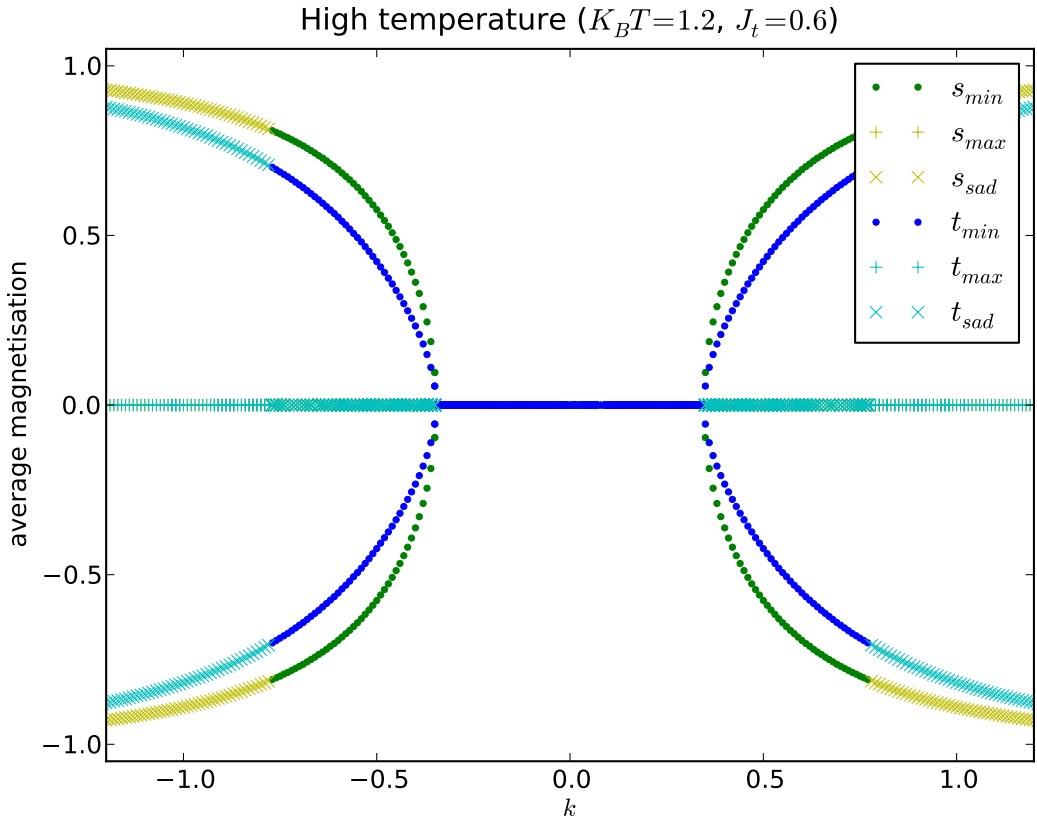}
\caption{Dependence on inter-coupling of the numerically calculated average magnetisations for  $J_{s}=1$, $J_{t}= 0.6$ , $K_{B}T = 1.2$ ($k_{c}=\pm 0.35$). Degenerate case (limiting value between both coupling regimes) for $|k|=\sqrt{J_{s}J_{t}}=0.77$. Different solutions are plotted for $k$  between -1.2 and 1.2 every 0.01. Magnetisations are plotted in green for $s$ and blue for $t$. Dark points are used for stable solutions and lighter asp ($\times$, for saddle points) or cross ($+$, for maxima) for non stable solutions.}
\label{fig:nlanak1}
\end{figure}

The situation at high temperatures is depicted in figure \ref{fig:nlanak1} for $J_{t}=0.6$ and $K_{B}T=1.2$ ($k_{c}=\pm 0.35$). For low positive values of $k$ (bellow $k_{c}$), only the paramagnetic phase is a solution and it is stable. At $k_{c}$ there is a second order phase transition, and for values above it (but still in the weak coupling regime), the paramagnetic solution becomes saddle point while a pair of stable ferromagnetic solutions $(m_{s},m_{t})$, $(-m_{s},-m_{t})$ ($m_{s},m_{t}>0$) appear. For values of $k$ in the strong coupling regime ($k>0.77$ in this case), these become saddle point and the paramagnetic phase a maximum. For negative values of $k$ the situation is analogous, but the pair of ferromagnetic solutions have relative signs $(m_{s},-m_{t})$, $(-m_{s},m_{t})$ ($m_{s},m_{t}>0$).

\begin{figure}
\centering
\includegraphics[width=\textwidth]{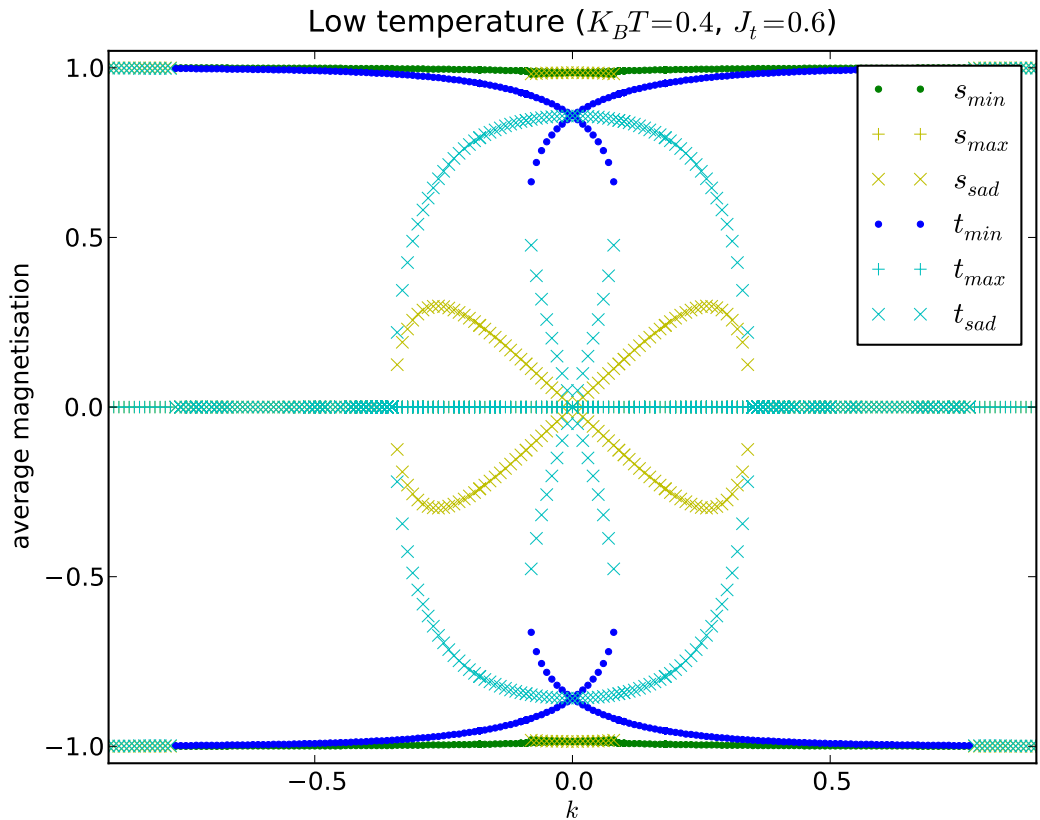}
\caption{Dependence on inter-coupling of the numerically calculated average magnetisations for  $J_{s}=1$, $J_{t}= 0.6$ , $K_{B}T = 0.4$ ($k_{c}=\pm 0.35$). Degenerate case (limiting value between both coupling regimes) for $|k|=\sqrt{J_{s}J_{t}}=0.77$. Different solutions are plotted for $k$  between -0.9 and 0.9 every 0.01. Magnetisations are plotted in green for $s$ and blue for $t$. Dark points are used for stable solutions and lighter asp ($\times$, for saddle points) or cross ($+$, for maxima) for non stable solutions.}
\label{fig:nlanak2}
\end{figure}

An example of low temperature behaviour with metastable states is plotted in figure \ref{fig:nlanak2} ($J_{t}=0.6$, $K_{B}T=0.4$  and $k_{c}=\pm 0.35$). For low positive values of $k$ ($k<k_{c}$), the paramagnetic solution is a maximum. Up to a value $k_{a}$ (spinodal inter-coupling) there are two pairs of ferromagnetic stable solutions, the main one having both $s$ and $t$ positive or negative and very near one in absolute value, and the metastable one with a lower value of $t$ (as $J_{t}<J_{s}$ in this case) and opposite relative signs between $s$ and $t$. There are also two pairs of saddle point solutions with relative $s$ to $t$ sign opposite to that of the metastable pair. Above $k_{a}$ the metastable solutions (and its associated saddle type pair) disappear. At $k=k_{c}$ the paramagnetic phase becomes a saddle point of the free energy density and ferromagnetic saddle points disappear. When $k$ is high enough to be in the strong coupling regime ($k>0.77$), it becomes a maximum and the previously stable main ferromagnetic branch becomes saddle point. For negative values of $k$ the situation is analogous, but pairs of ferromagnetic solutions have opposite relative signs.

\begin{figure}
\centering
\subfloat[]{\includegraphics[width=0.33\textwidth]{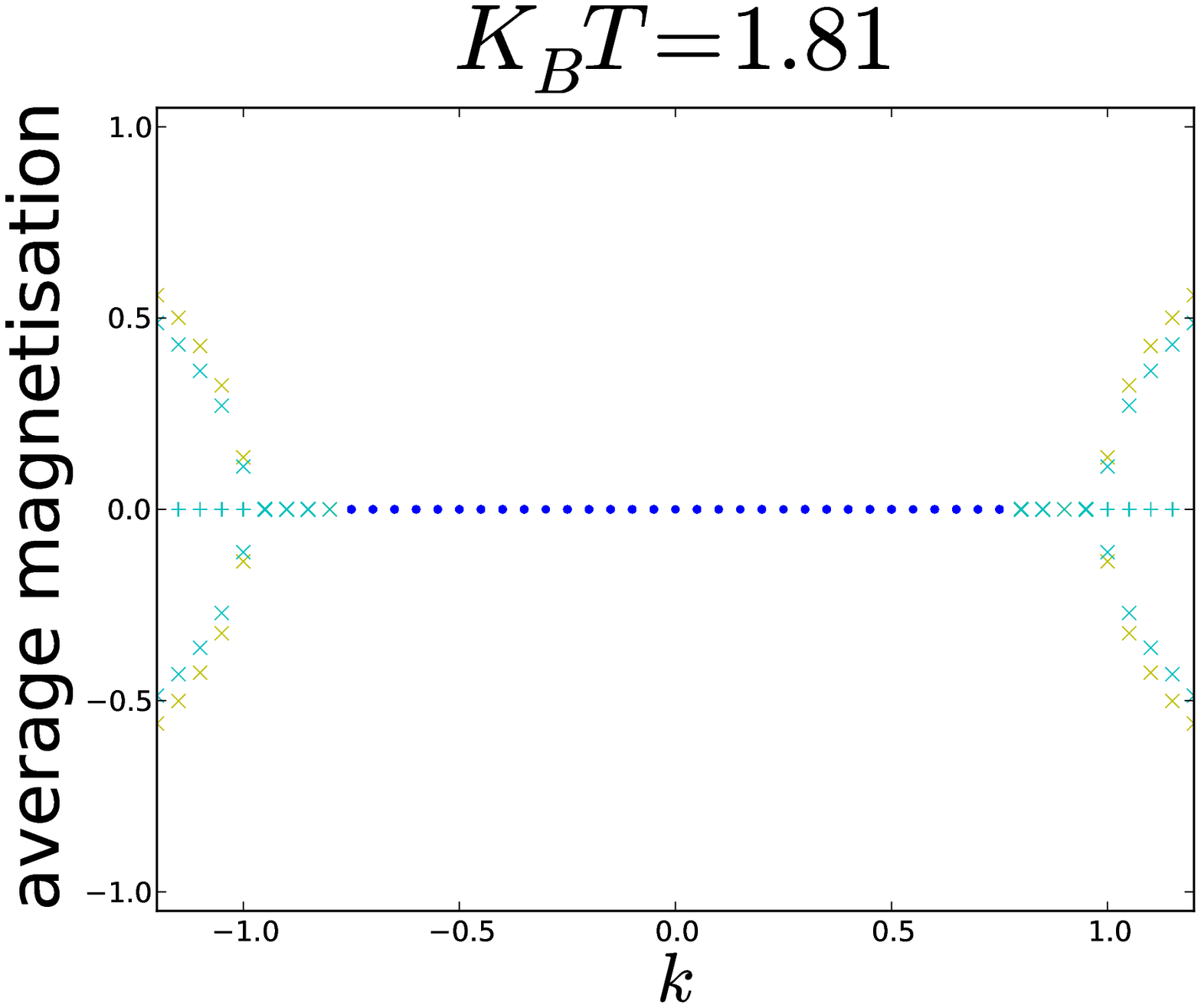}}
\subfloat[]{\includegraphics[width=0.33\textwidth]{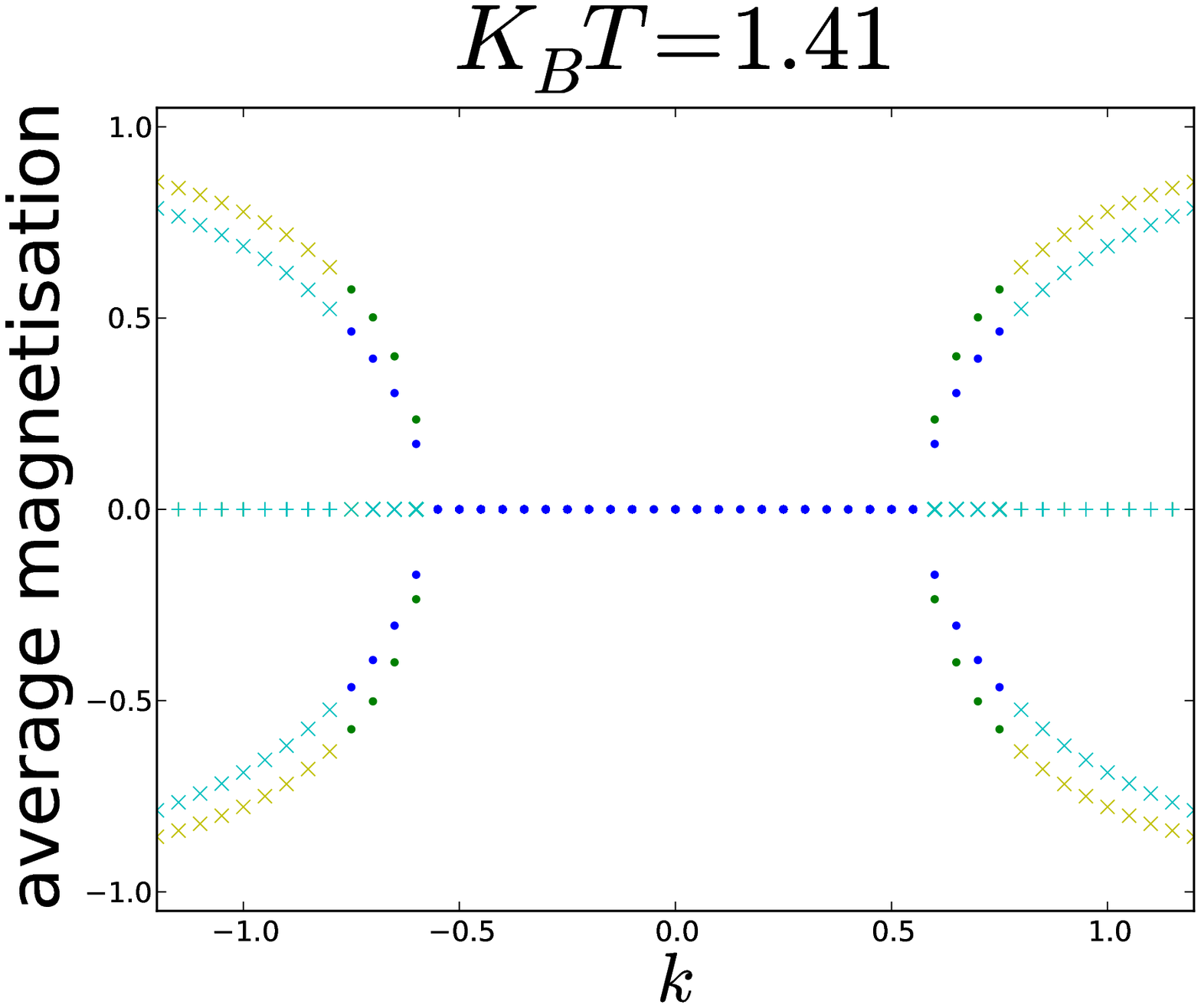}}
\subfloat[]{\includegraphics[width=0.33\textwidth]{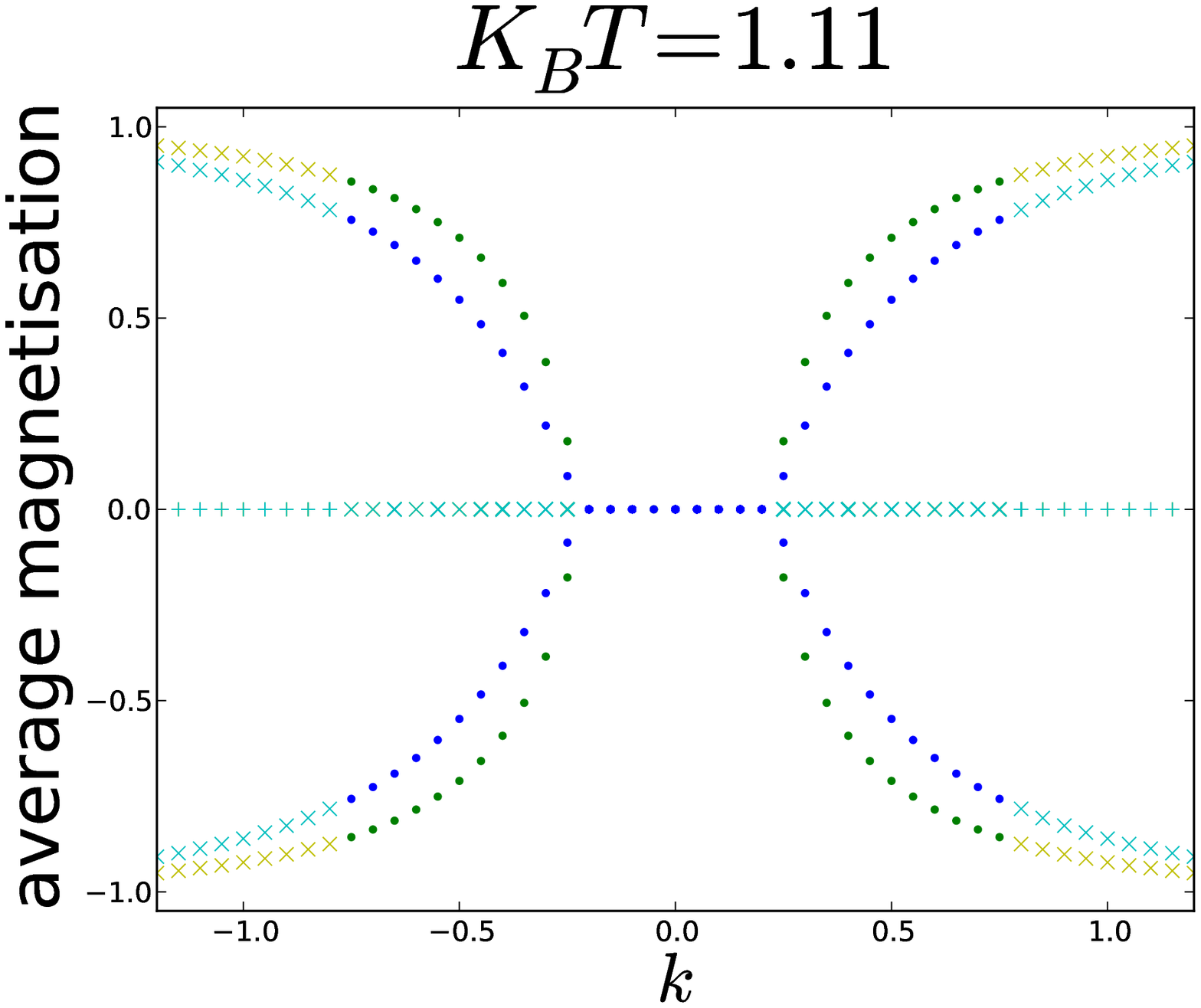}}\\
\subfloat[]{\includegraphics[width=0.33\textwidth]{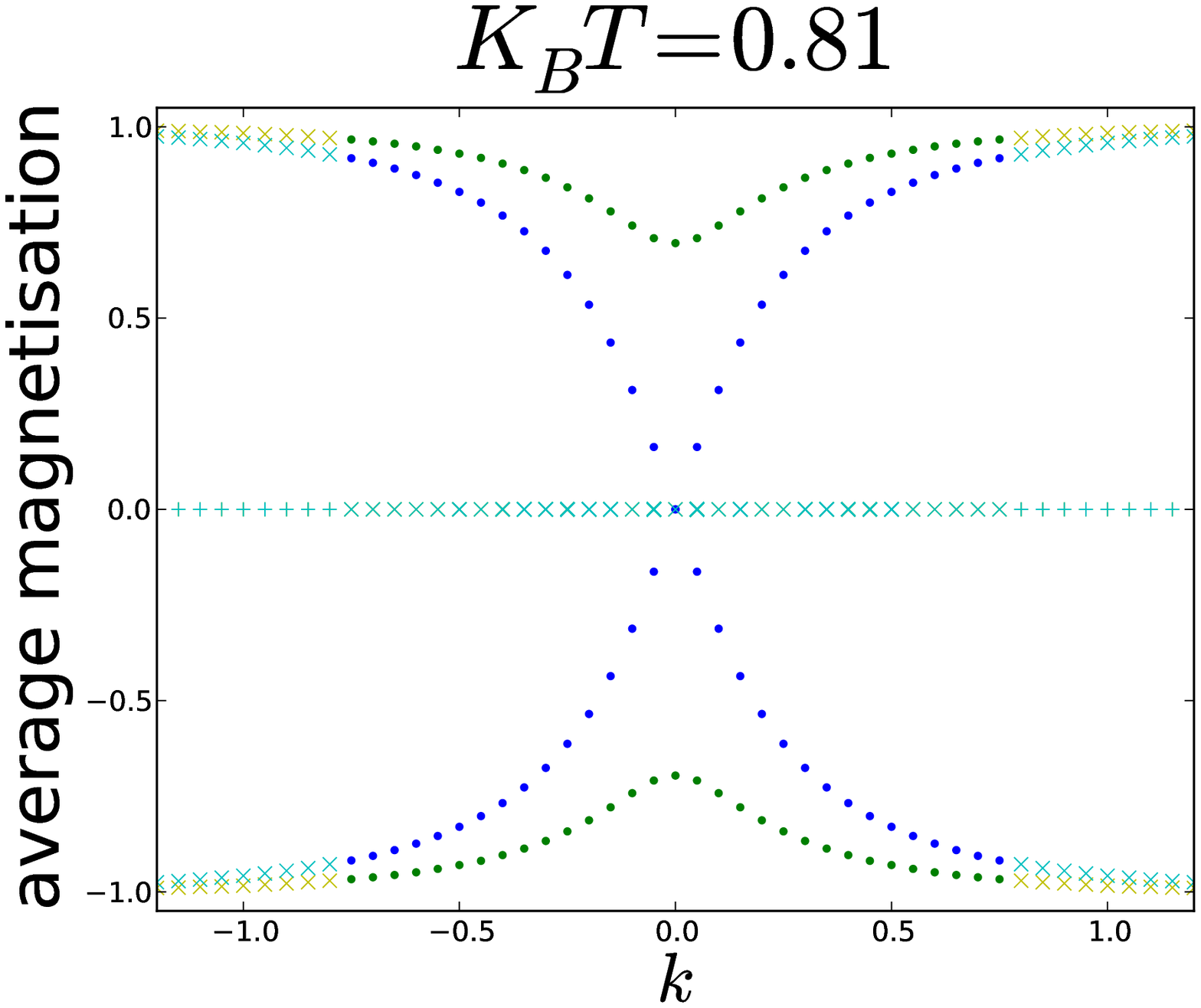}}
\subfloat[]{\includegraphics[width=0.33\textwidth]{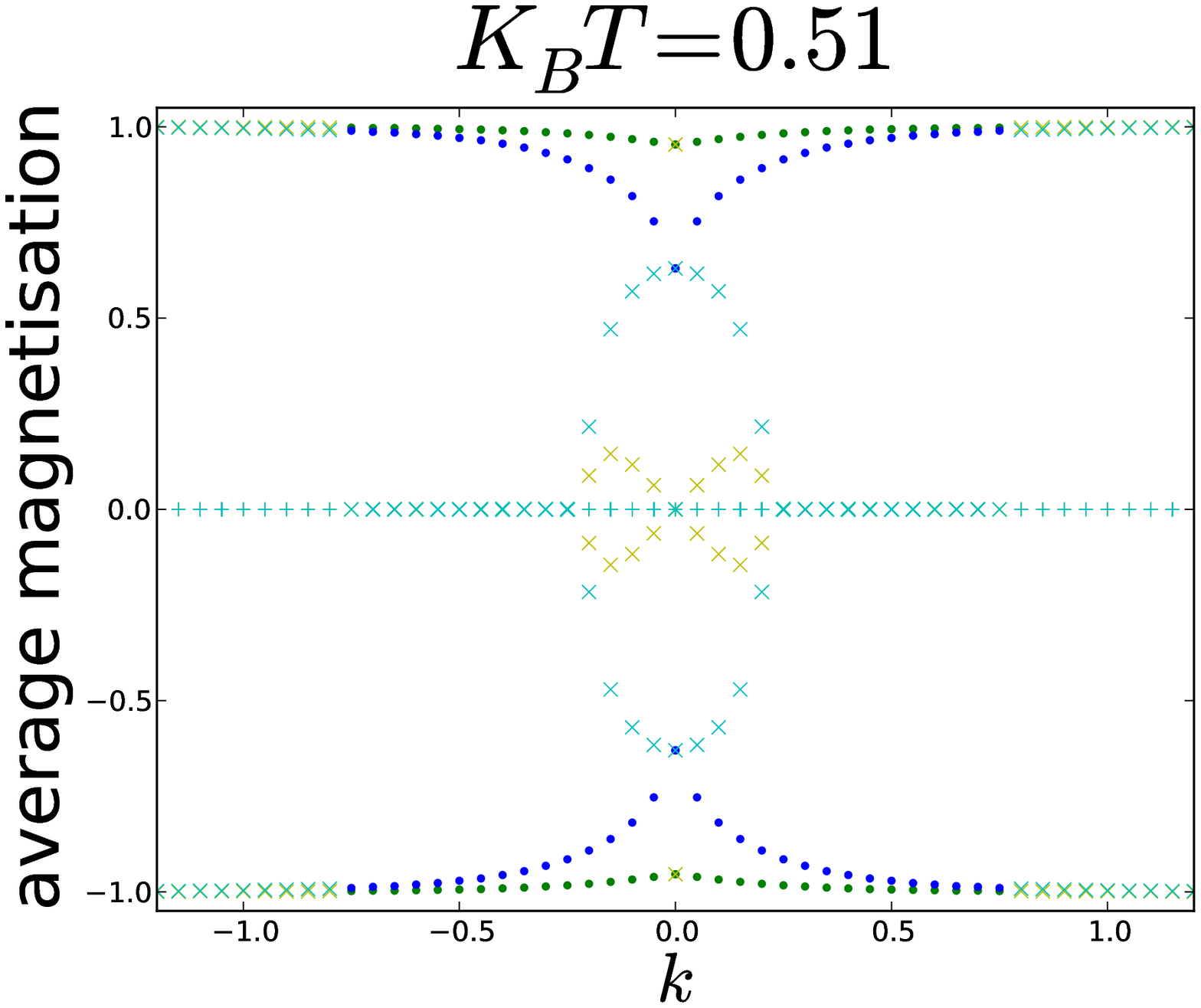}}
\subfloat[]{\includegraphics[width=0.33\textwidth]{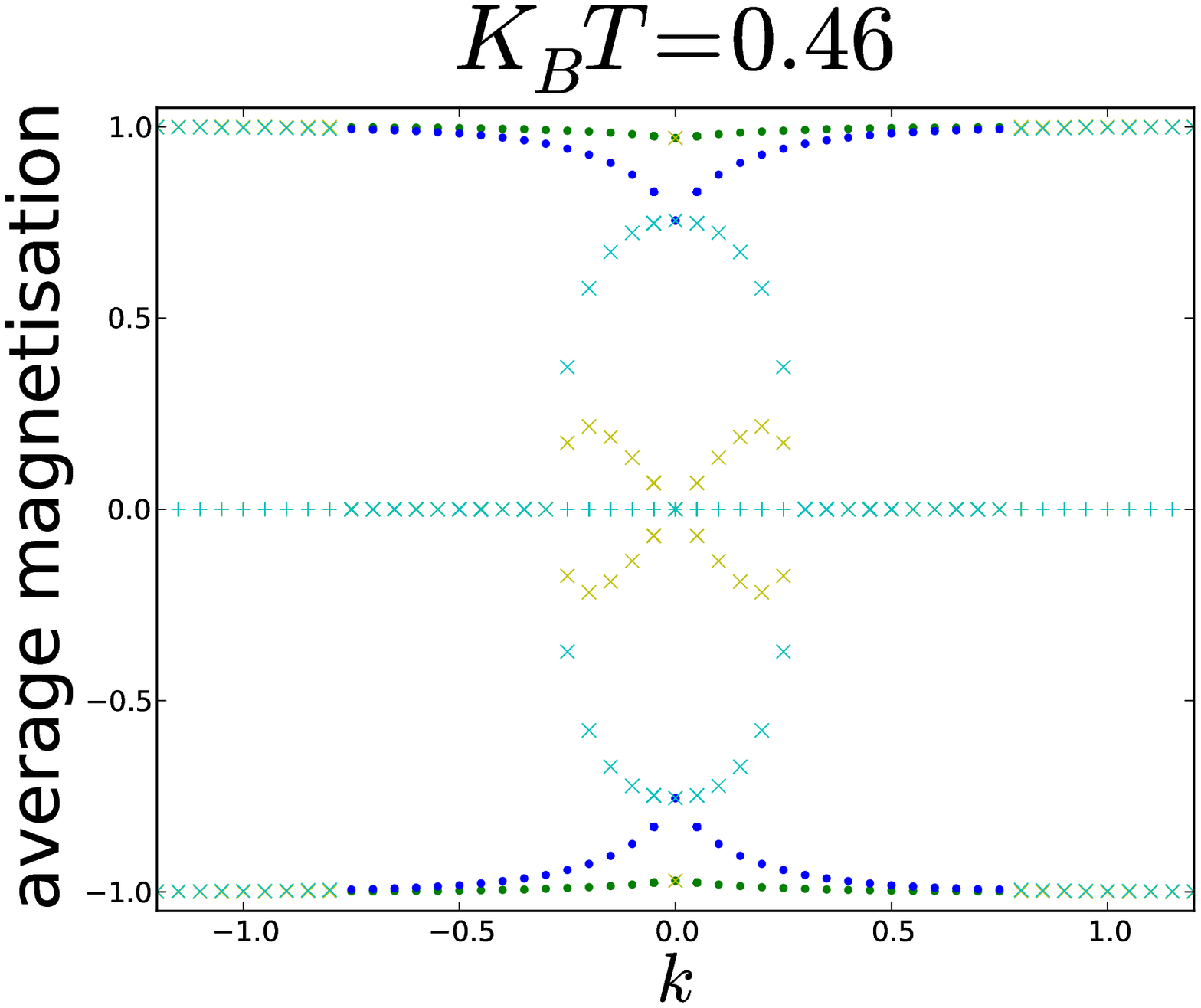}}\\
\subfloat[]{\includegraphics[width=0.33\textwidth]{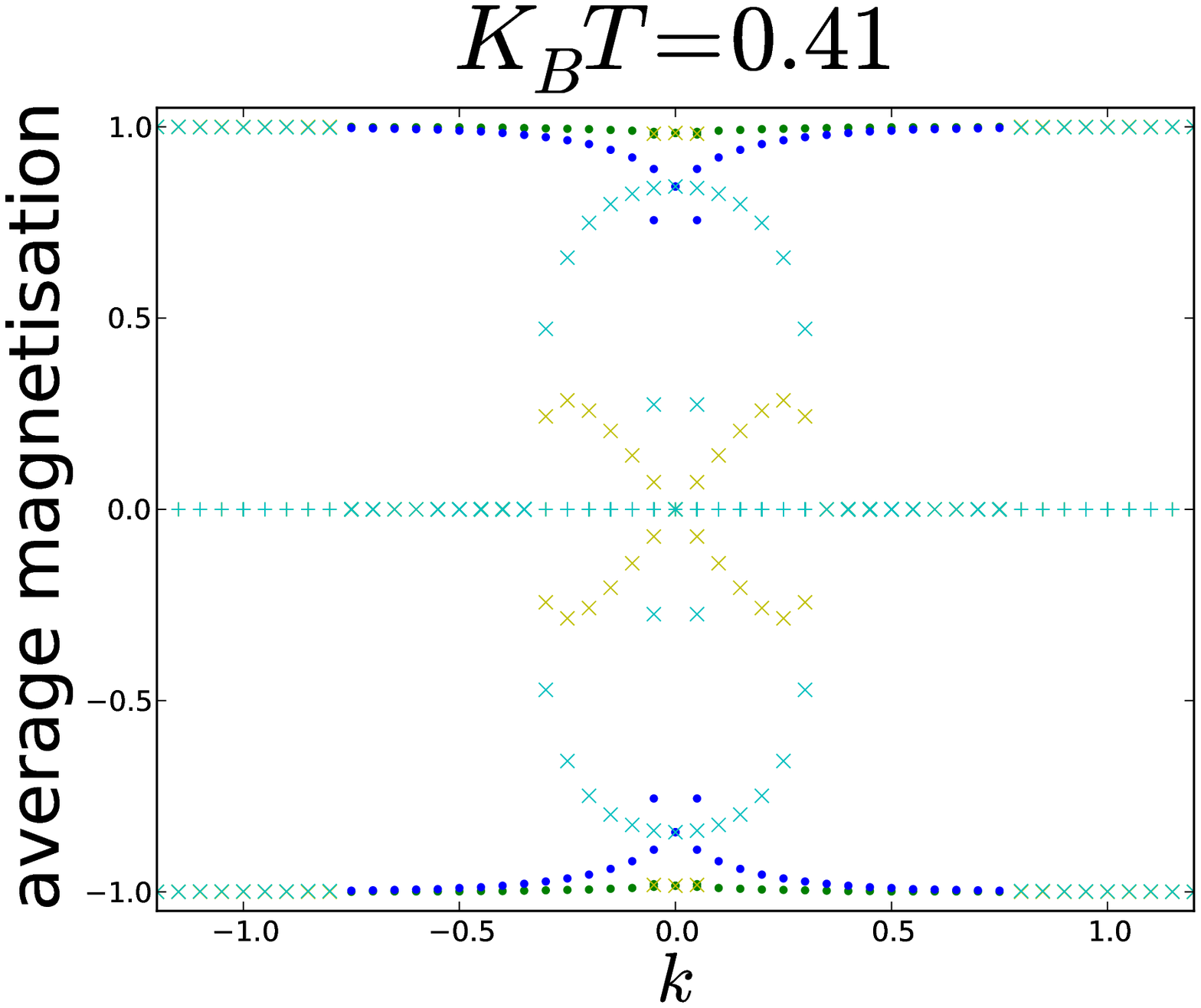}}
\subfloat[]{\includegraphics[width=0.33\textwidth]{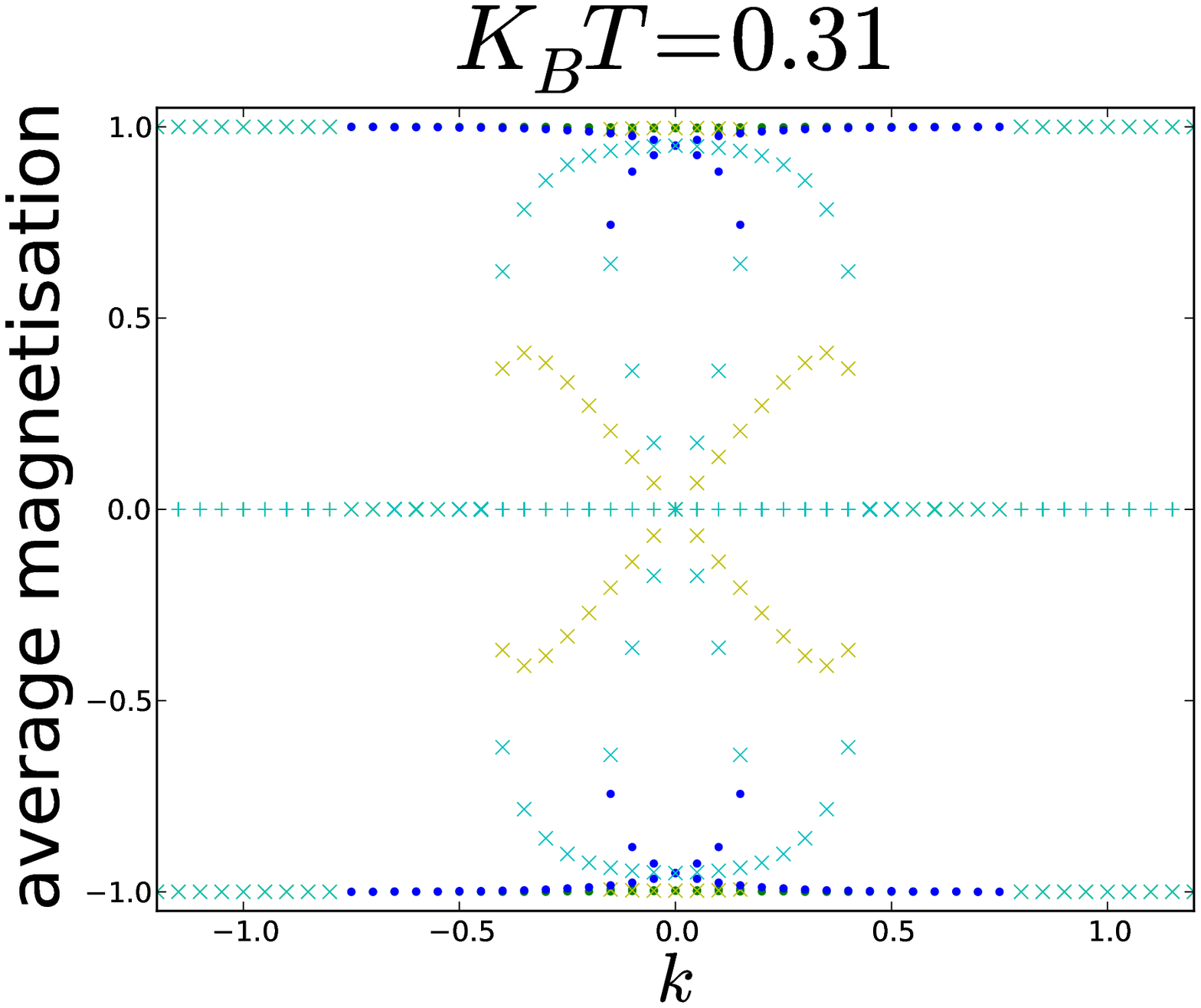}}
\subfloat[]{\includegraphics[width=0.33\textwidth]{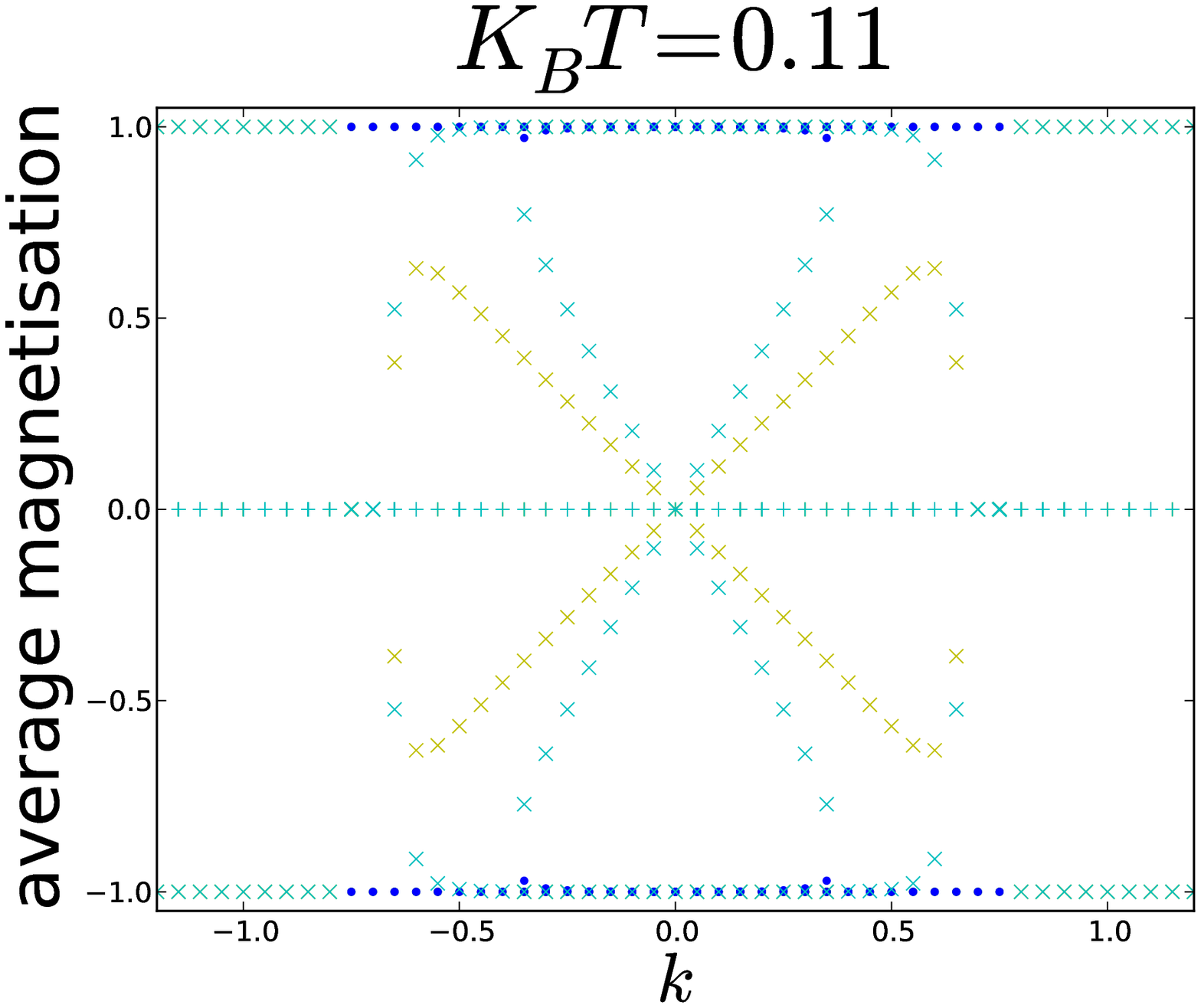}}
\caption{Dependence on inter-coupling of the numerically calculated average magnetisations $(s,t)$ for different values of the temperature. $J_{s}=1$ and $J_{t}= 0.6$  for all plots. (a) $K_{B}T = 1.81$, (b) $K_{B}T = 1.41$, (c) $K_{B}T = 1.11$, (d) $K_{B}T = 0.81$, (e) $K_{B}T = 0.51$, (f) $K_{B}T = 0.46$, (g) $K_{B}T = 0.41$, (h) $K_{B}T = 0.31$ and (i) $K_{B}T = 0.11$. In all cases, different solutions are plotted for $k$ between -1.2 and 1.2 every 0.05 ($K_{B}T$). Magnetisations are plotted in green for $s$ and blue for $t$. Dark points are used for stable solutions and lighter asp ($\times$, for saddle points) or cross ($+$, for maxima) for non stable solutions.}
\label{fig:nlanaTk}
\end{figure}

Figure \ref{fig:nlanaTk} shows how the dependence with $k$ varies as we lower the temperature for fixed intra-couplings ($J_{t}=0.6$). For temperatures even higher than the one used in figure \ref{fig:nlanak1}, $k_{c}$ may be higher than the value of $k$ at which the coupling becomes strong ($|k|=|k_{d}|=0.77$ in this example), and so the only stable solution is the paramagnetic phase (figure \ref{fig:nlanaTk} a). As we lower the temperature, $|k_{c}|$ becomes smaller, and once it falls in the weak coupling regime (figure \ref{fig:nlanaTk} b), we recover the qualitative behaviour depicted in figure \ref{fig:nlanak1} with both paramagnetic and ferromagnetic stable solutions in the weak coupling case depending on the value of $k$. As we continue to move to lower temperatures, $|k_{c}|$ becomes smaller and the paramagnetic phase is stable for smaller regions  (figure \ref{fig:nlanaTk} c) until it disappears (figure \ref{fig:nlanaTk} d). Throughout this process, the smaller the temperature, the more different are $s$ and $t$ in absolute value (at low $k$), and the larger the region where they are both practically one (at high $k$). Note that at this point (and for $T_{uc}^{t}<T<T_{uc}^{s}$), a pair of mixed phase solutions (${(m_{s},0),(-m_{s},0)}$) will be the only stable solutions at $k=0$ (and nonzero temperature). For small enough temperatures, the characteristic value $k_{c}$ reappears, but now no longer associated to a phase transition but rather to the point  under which a saddle point ferromagnetic pair of additional solutions appear (figure \ref{fig:nlanaTk} e). As we keep moving towards smaller temperatures,  $|k_{c}|$ now  moves to higher values (figure \ref{fig:nlanaTk} f), and absolute values of $s$ and $t$ become again more and more similar as the region where they are both practically one increases. Finally, for small enough values (figure \ref{fig:nlanaTk} g), two spinodal values $k_{a}$ (of equal absolute value and opposite sign) at which the metastable pair of ferromagnetic solutions appear (together with new saddle point ferromagnetic ones). When we move towards even lower temperatures, $|k_{a}|$  and $|k_{c}|$ become larger,  and so does the region with metastable solutions (figure \ref{fig:nlanaTk} h, i). The metastable solutions will be more and more similar to the main ones (and so very similar to $\pm 1$) in absolute value (and with opposite relative signs) the lower the temperature. 

\begin{figure}
\centering
\subfloat[]{\includegraphics[width=0.33\textwidth]{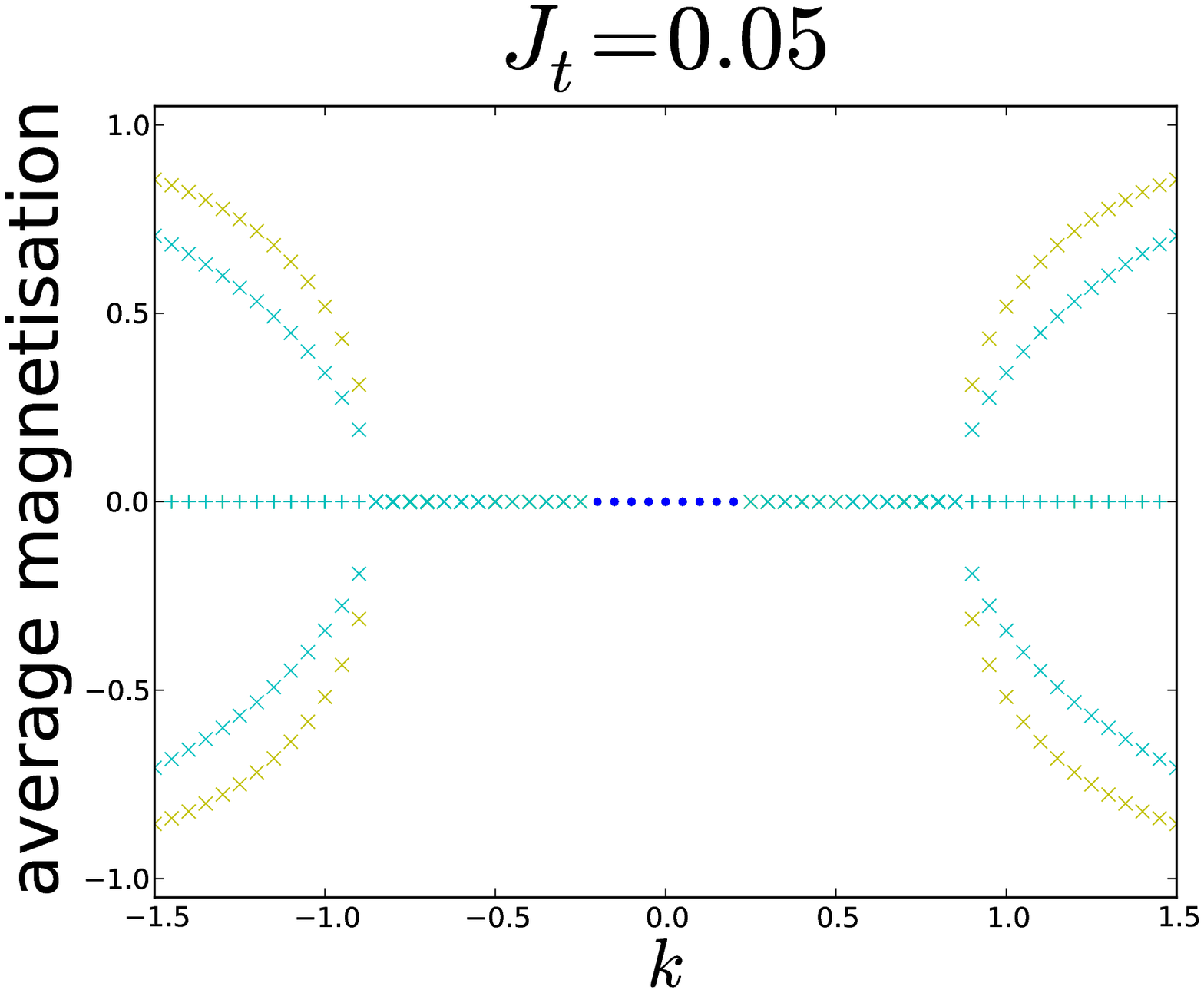}}
\subfloat[]{\includegraphics[width=0.33\textwidth]{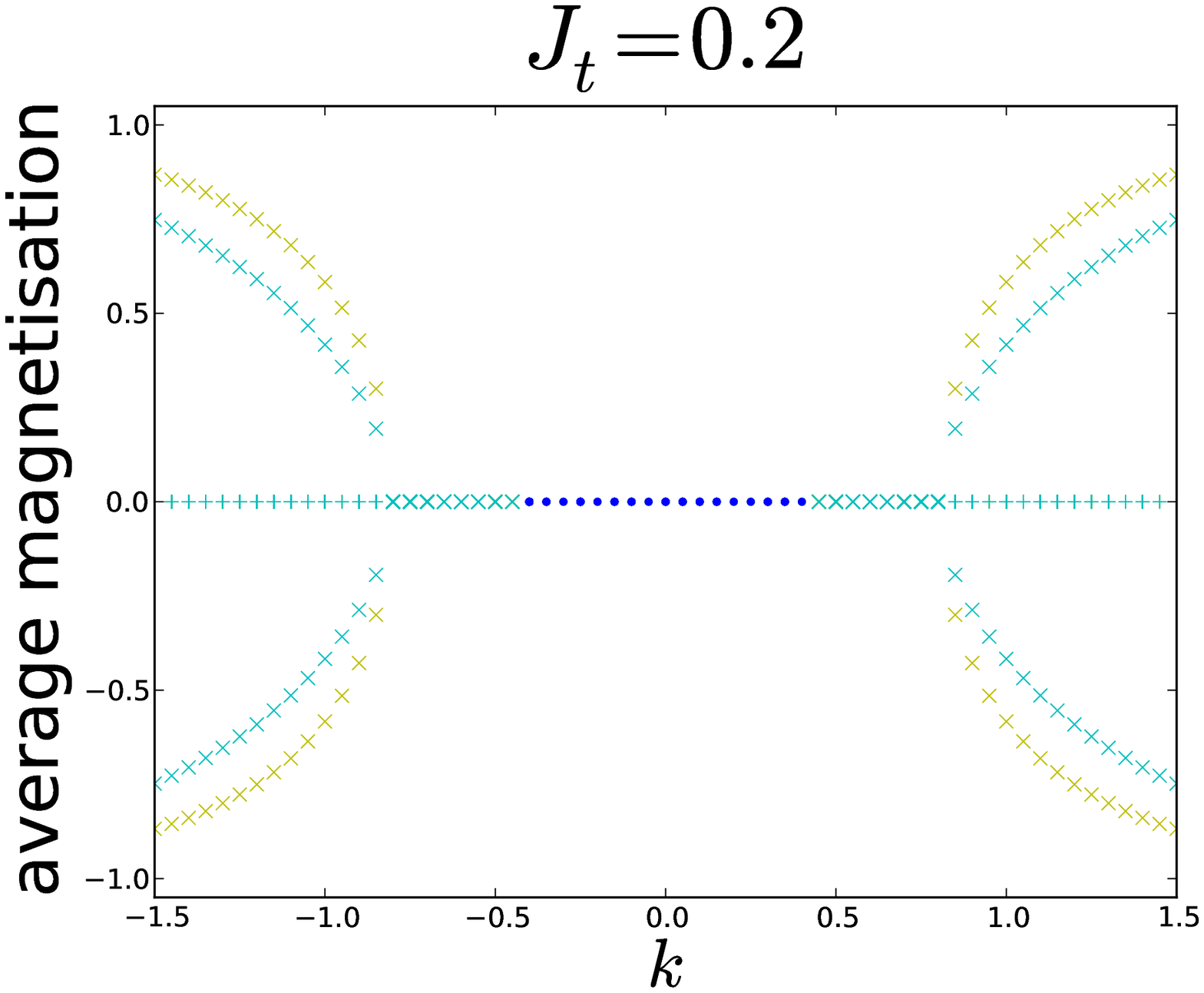}}
\subfloat[]{\includegraphics[width=0.33\textwidth]{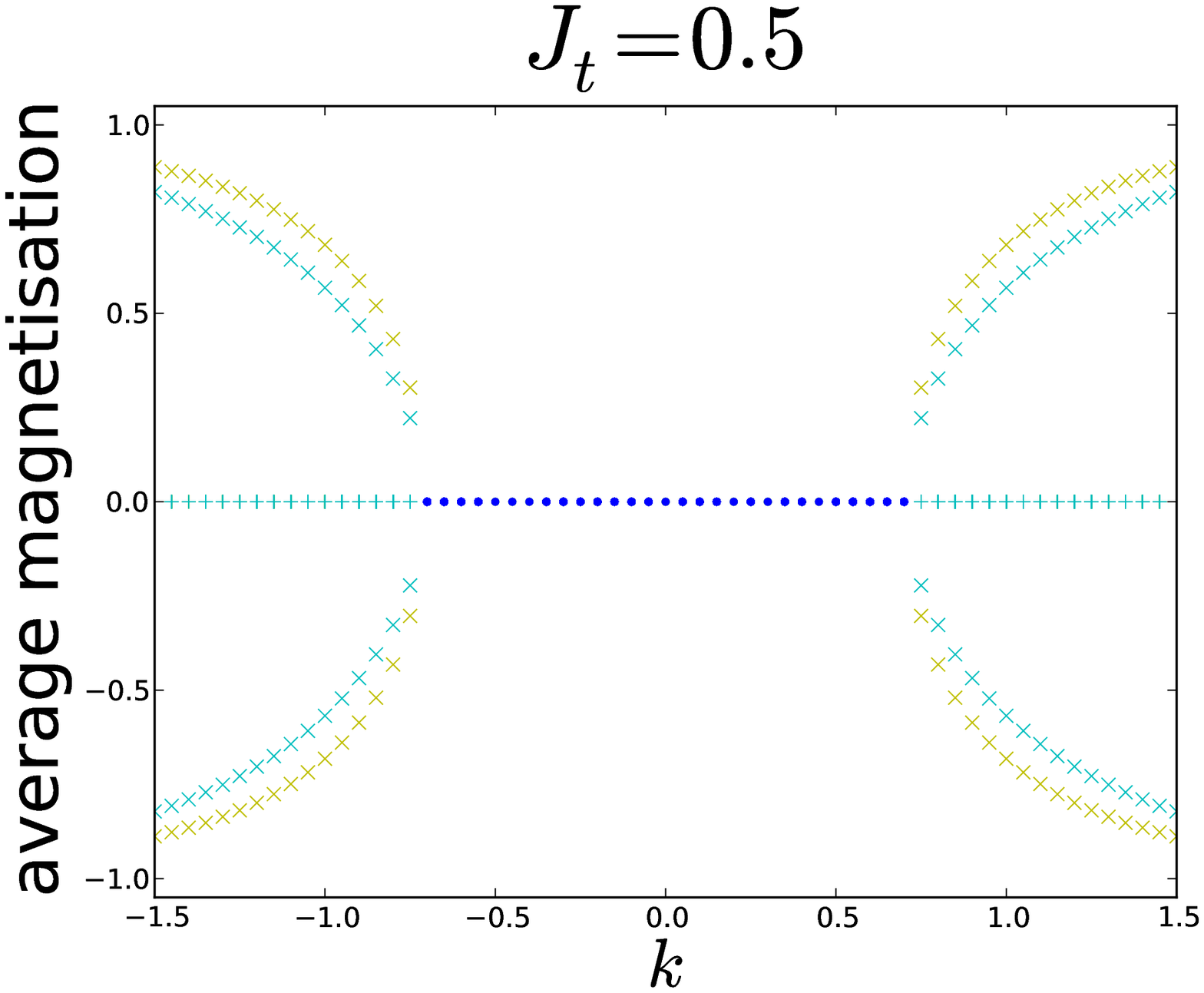}}\\
\subfloat[]{\includegraphics[width=0.33\textwidth]{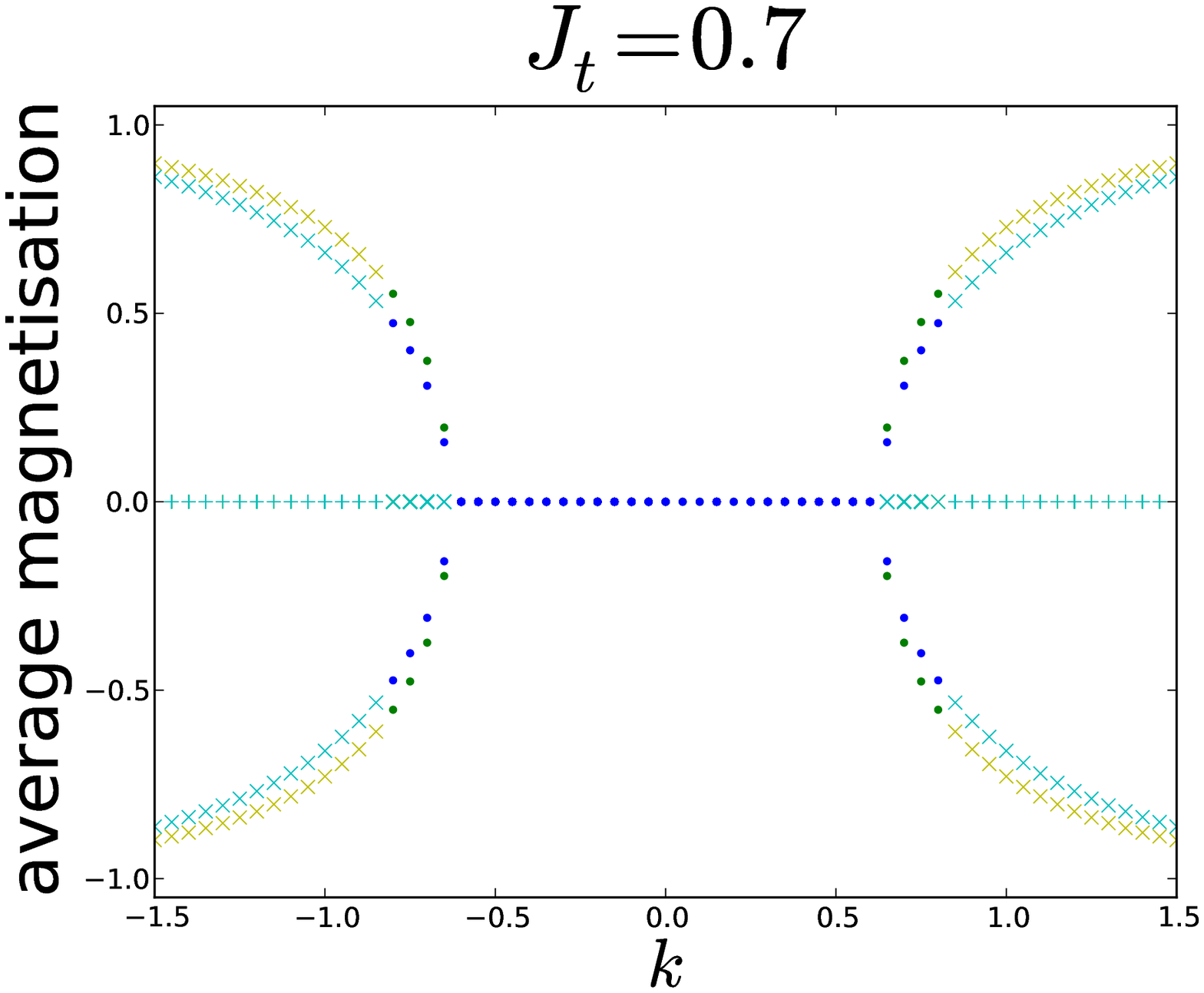}}
\subfloat[]{\includegraphics[width=0.33\textwidth]{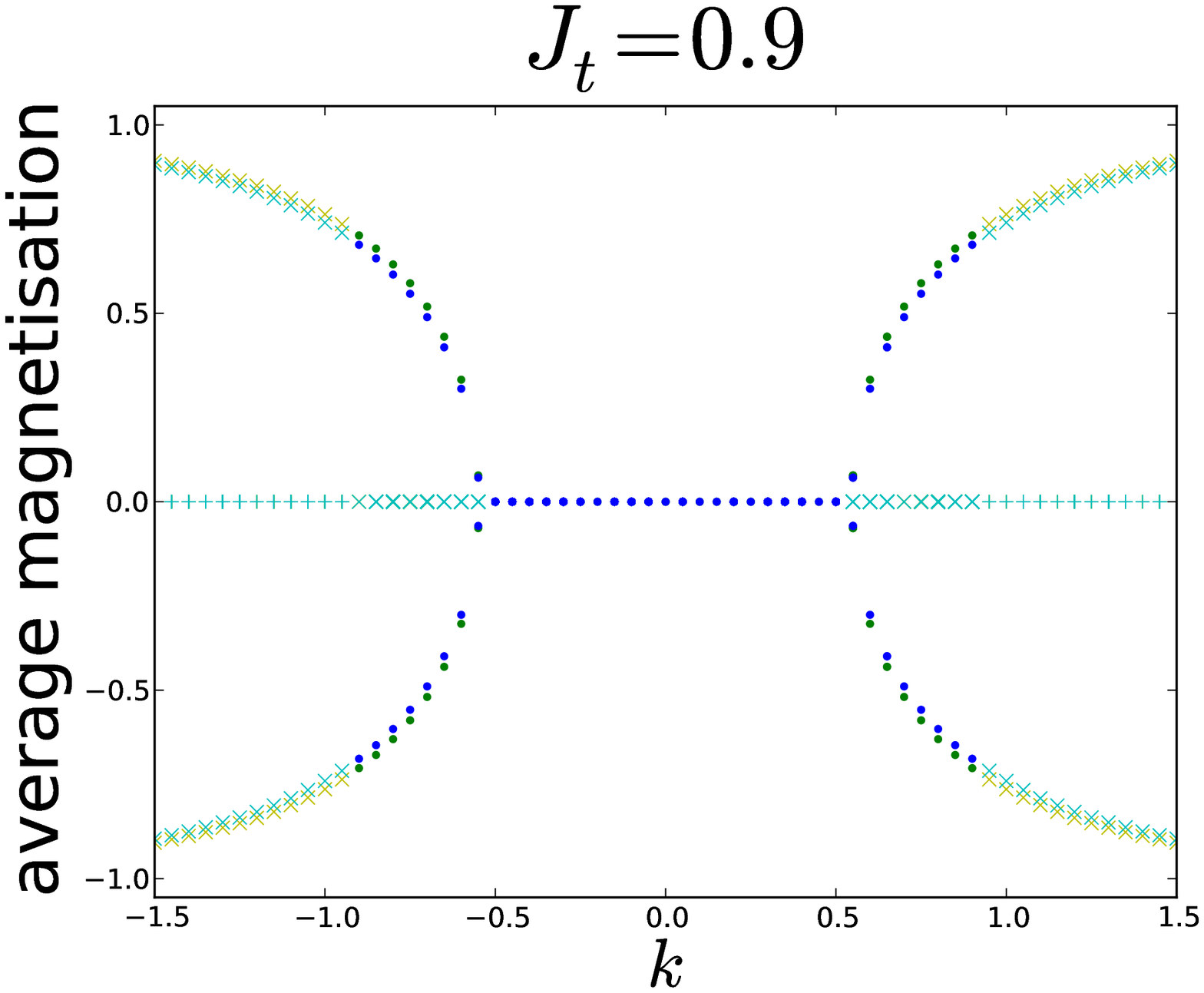}}
\subfloat[]{\includegraphics[width=0.33\textwidth]{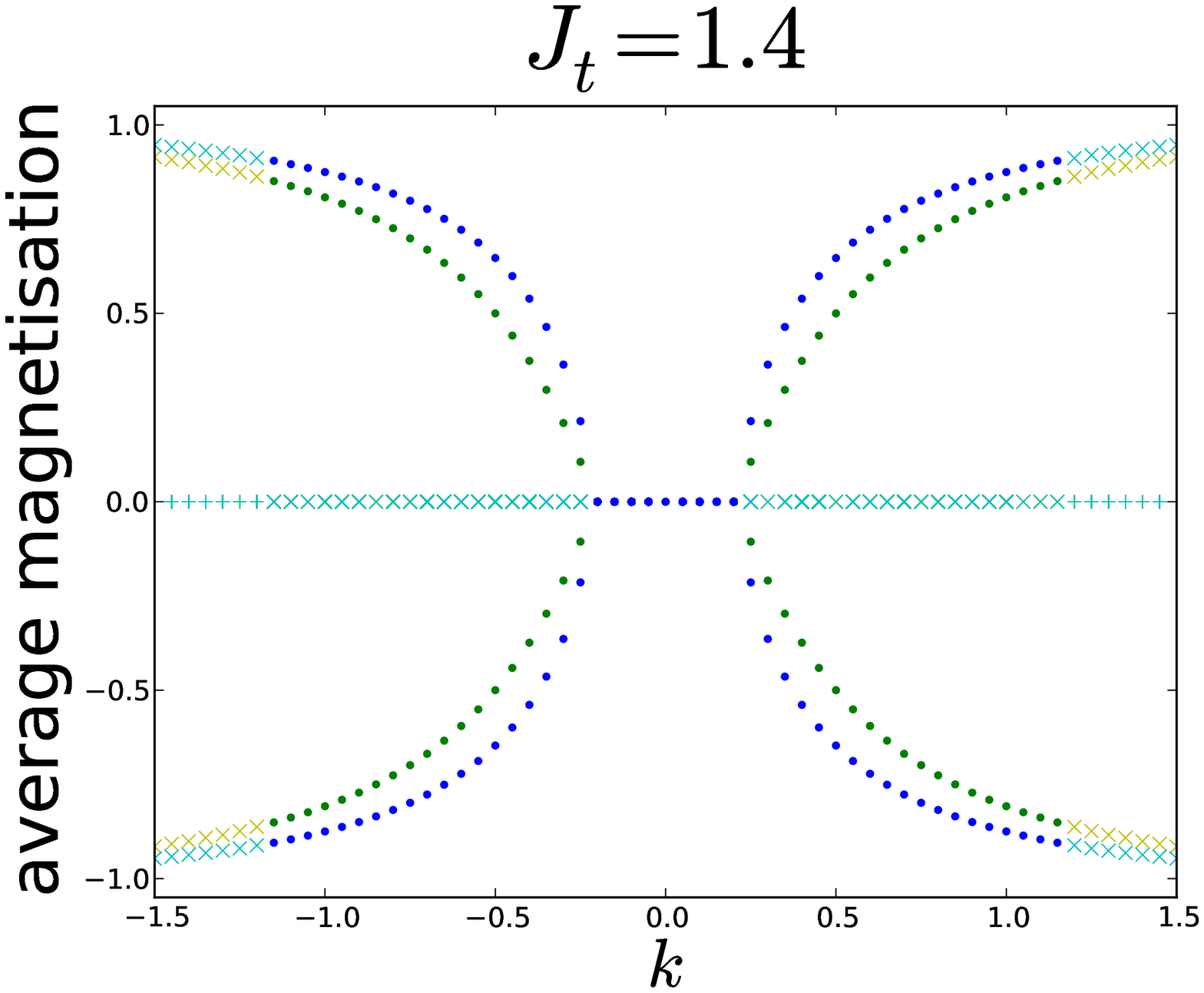}}\\
\subfloat[]{\includegraphics[width=0.33\textwidth]{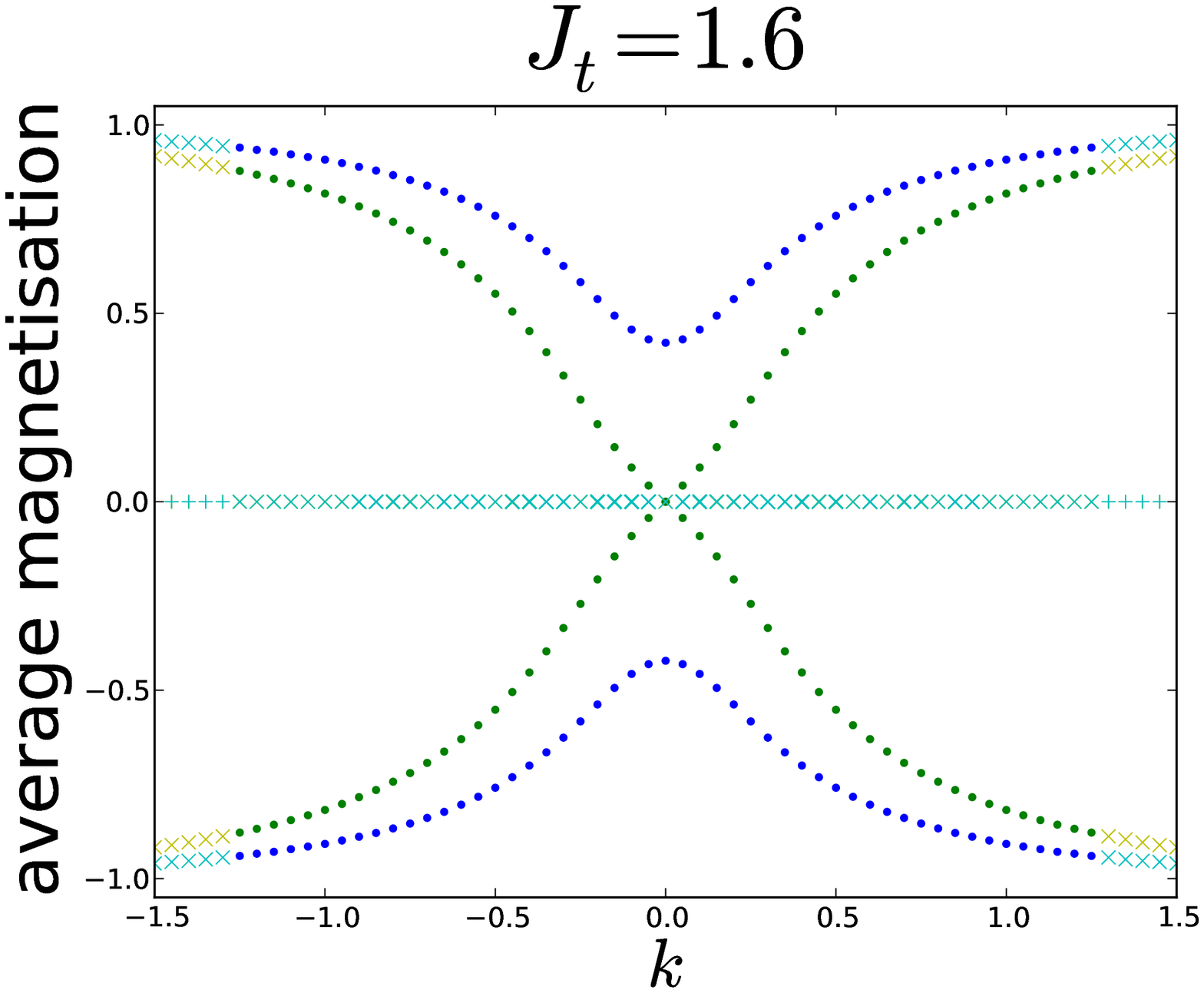}}
\subfloat[]{\includegraphics[width=0.33\textwidth]{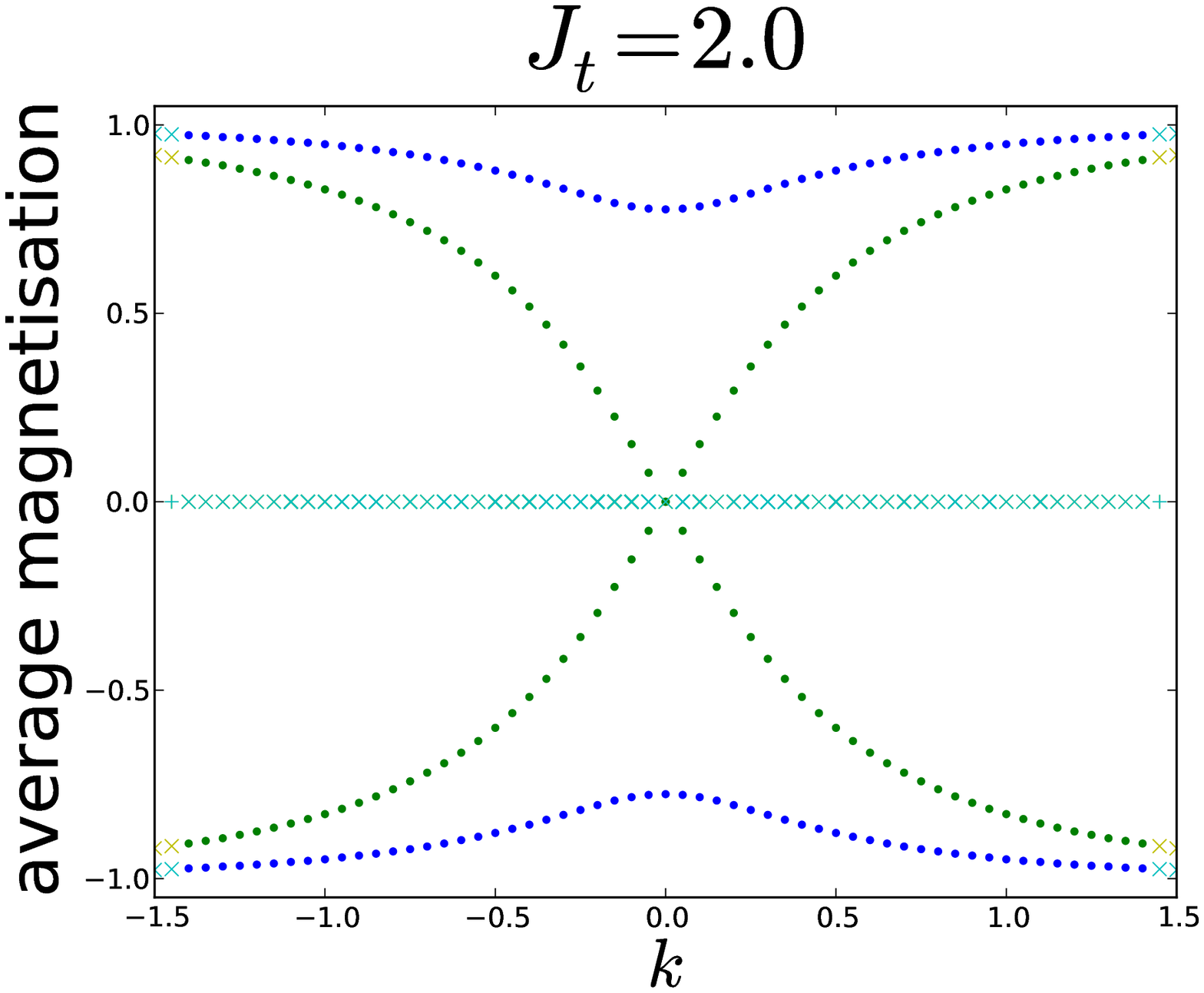}}
\subfloat[]{\includegraphics[width=0.33\textwidth]{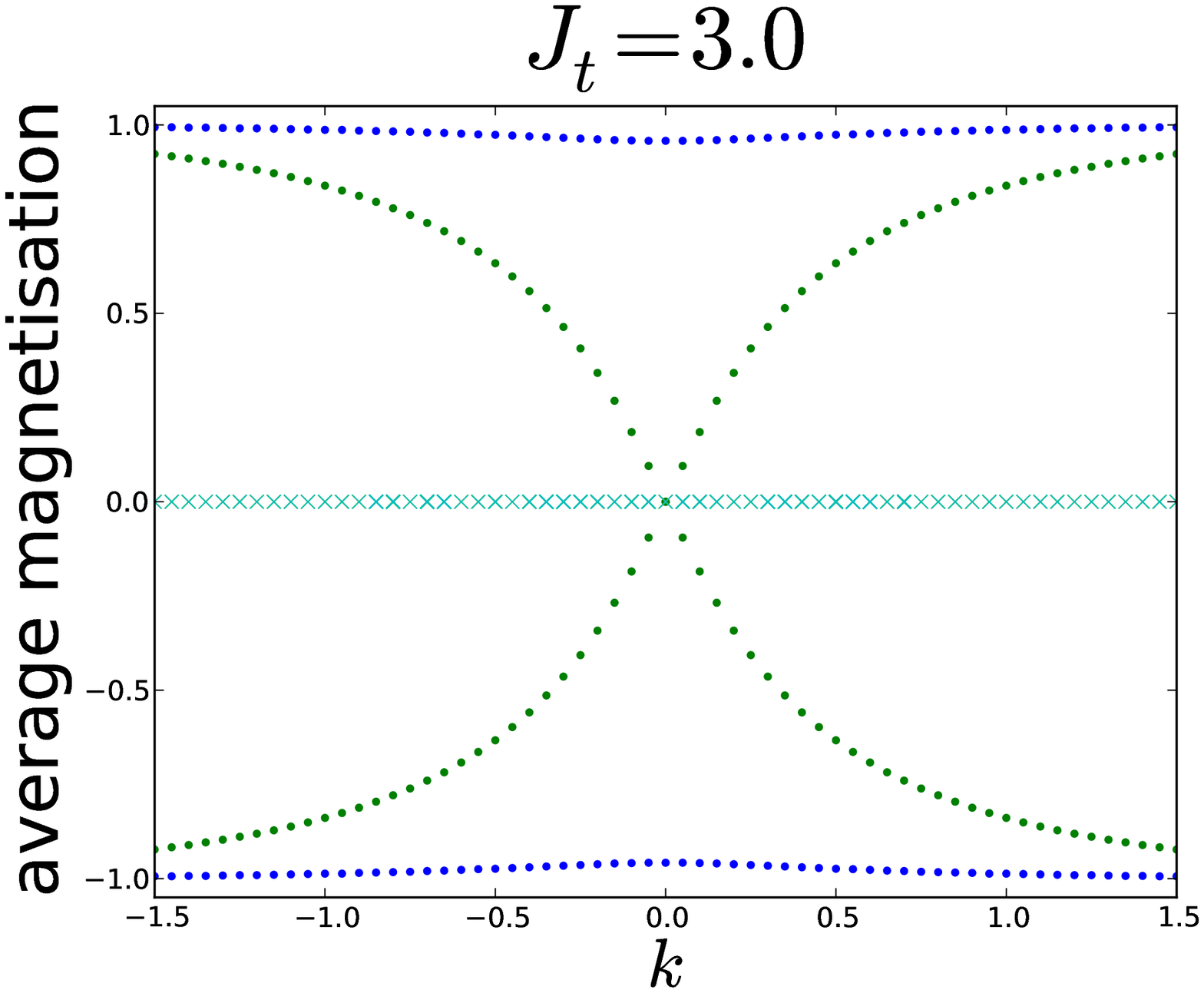}}
\caption{Dependence on inter-coupling of the numerically calculated average magnetisations $(s,t)$ for different values of $J_{t}$ at high temperature. $J_{s}=1$ and $K_{B}T= 1.5$  for all plots. (a) $J_{t} =0.05 $, (b) $J_{t} =0.2 $, (c) $J_{t} =0.5$, (d) $J_{t} =0.7$, (e) $J_{t} = 0.9$, (f) $J_{t} = 1.4$, (g) $J_{t} = 1.6$, (h) $J_{t} = 2$ and (i) $J_{t} = 3$. In all cases, different solutions are plotted for $k$ between -1.5 and 1.5 every 0.05 ($K_{B}T$). Magnetisations are plotted in green for $s$ and blue for $t$. Dark points are used for stable solutions and lighter asp ($\times$, for saddle points) or cross ($+$, for maxima) for non stable solutions.}
\label{fig:nlanaJk1}
\end{figure}

\begin{figure}
\centering
\subfloat[]{\includegraphics[width=0.33\textwidth]{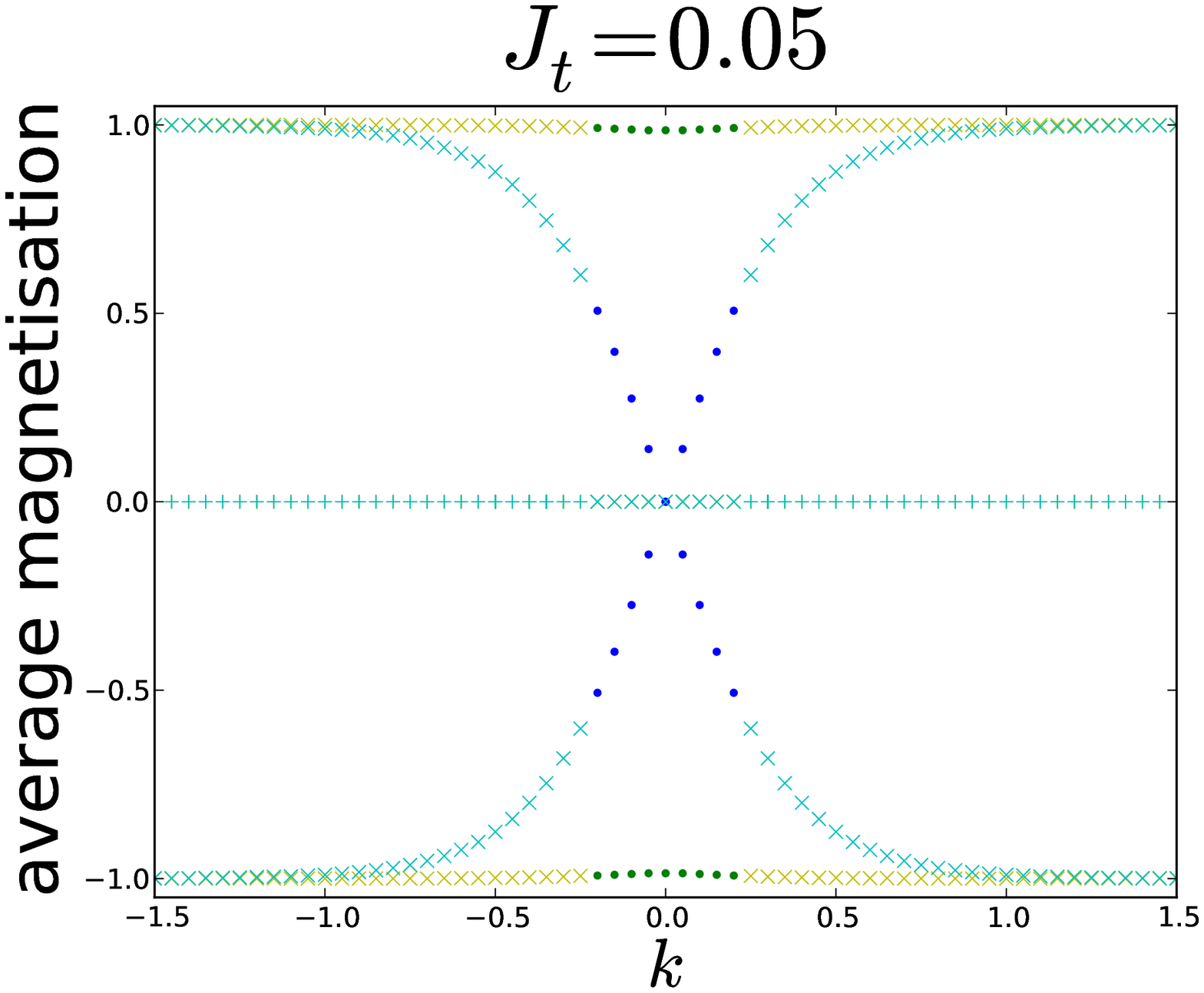}}
\subfloat[]{\includegraphics[width=0.33\textwidth]{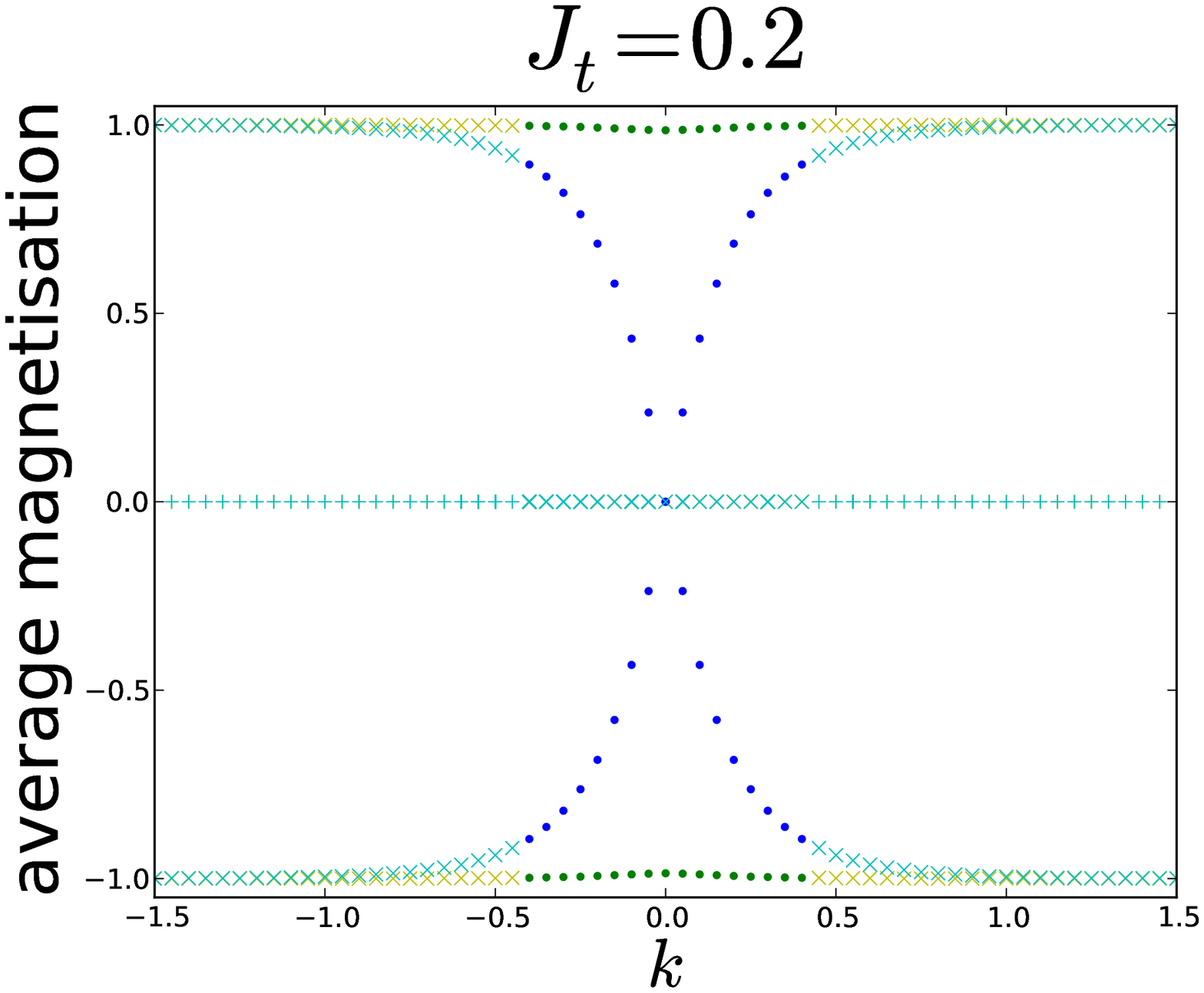}}
\subfloat[]{\includegraphics[width=0.33\textwidth]{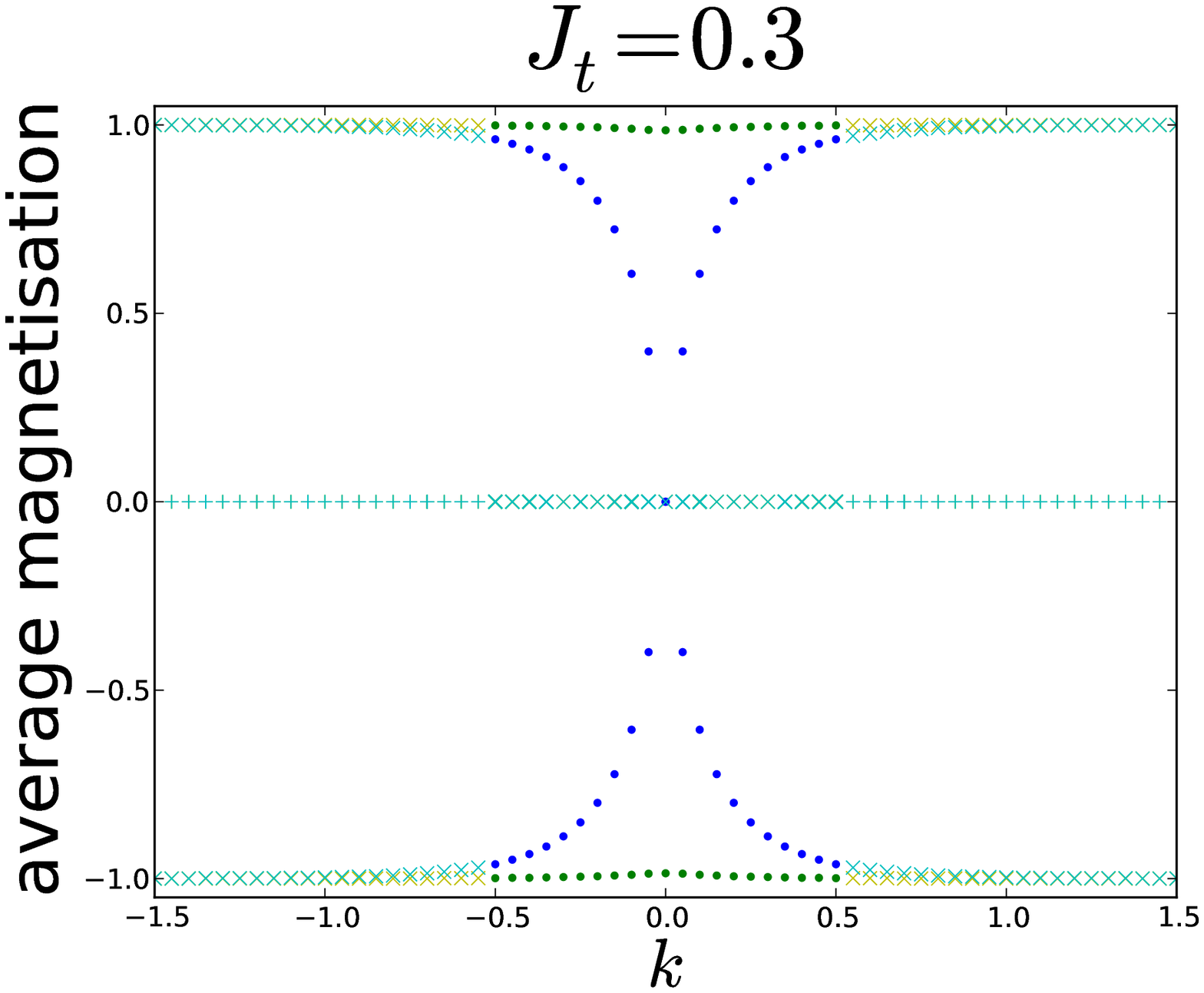}}\\
\subfloat[]{\includegraphics[width=0.33\textwidth]{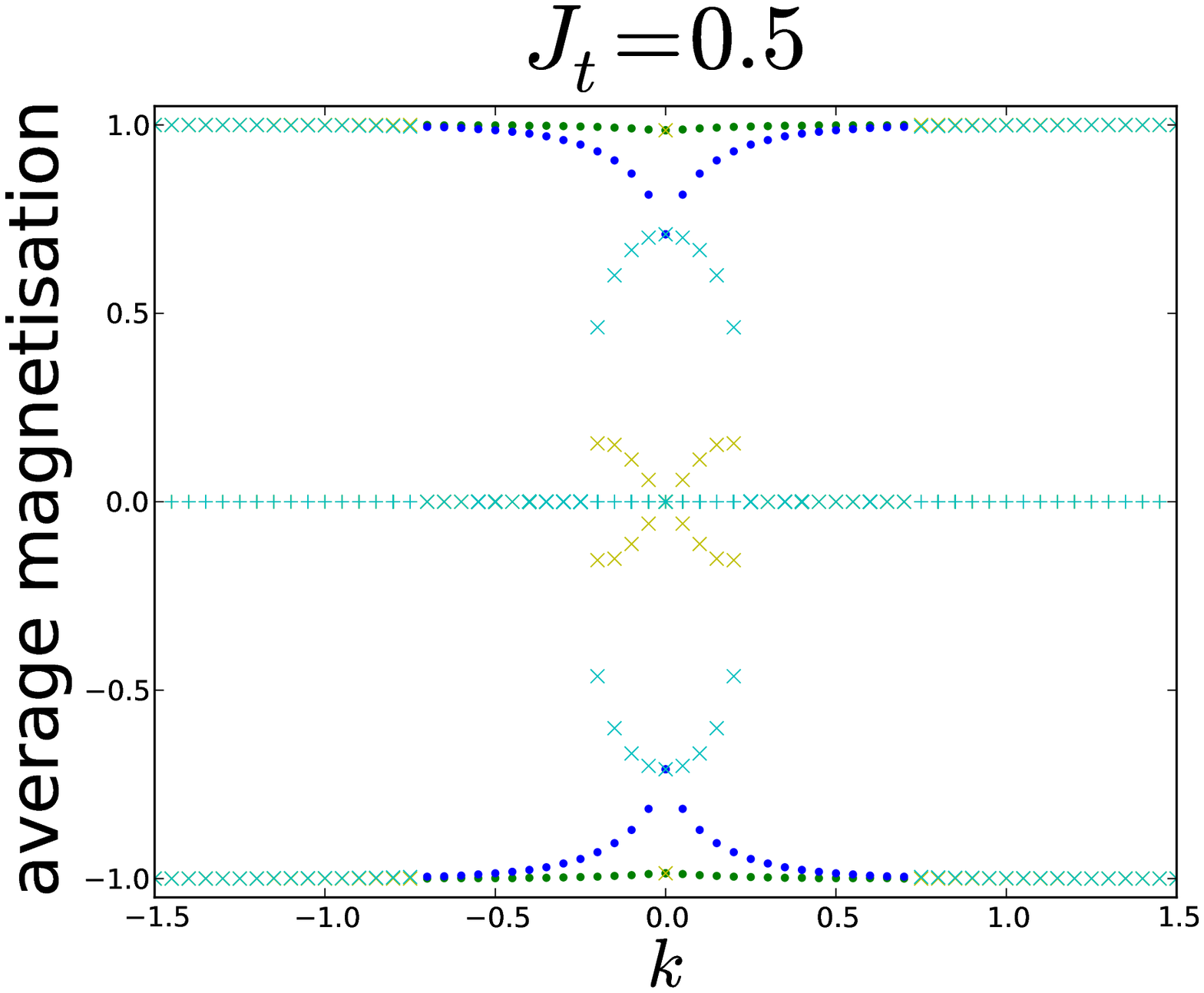}}
\subfloat[]{\includegraphics[width=0.33\textwidth]{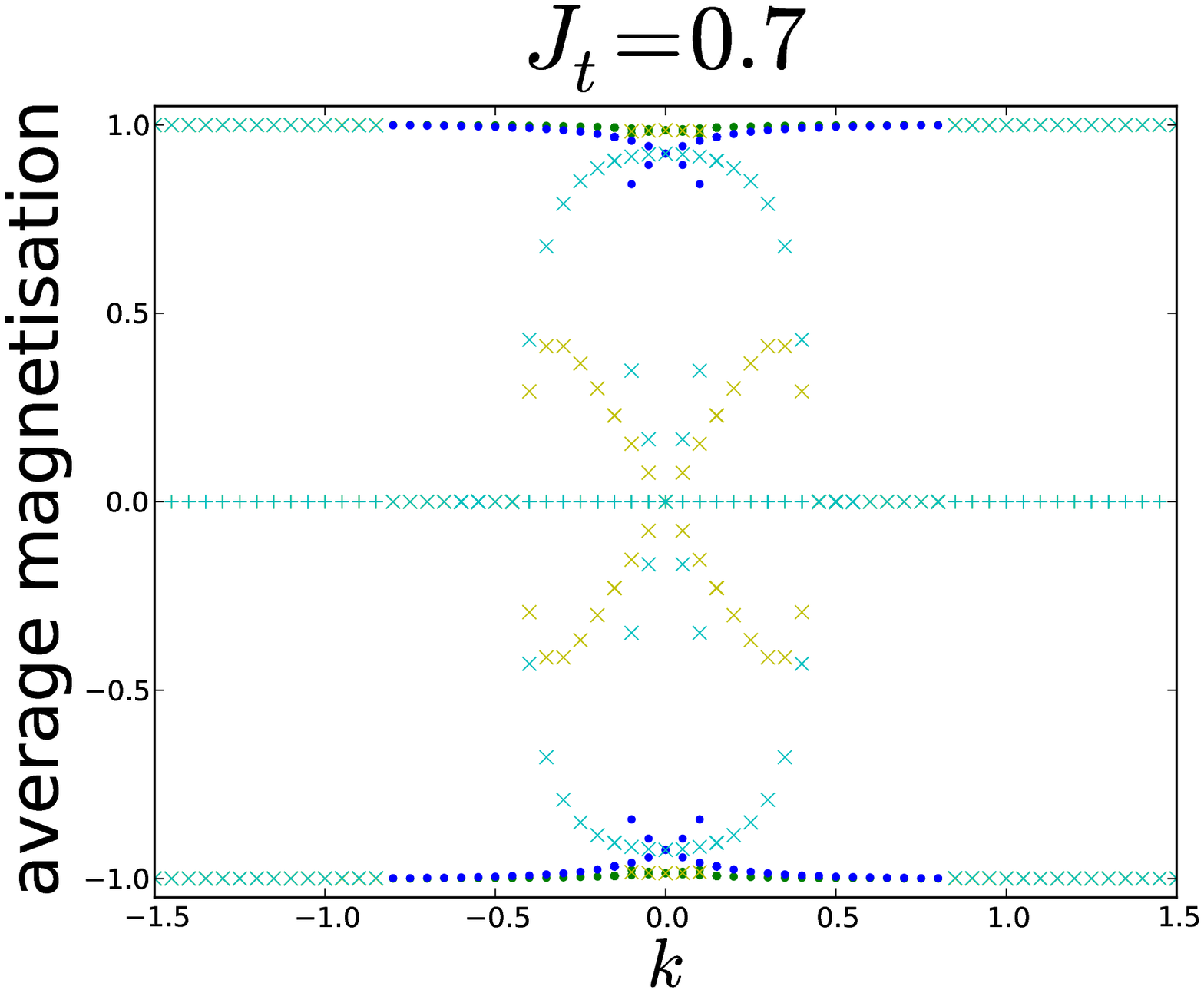}}
\subfloat[]{\includegraphics[width=0.33\textwidth]{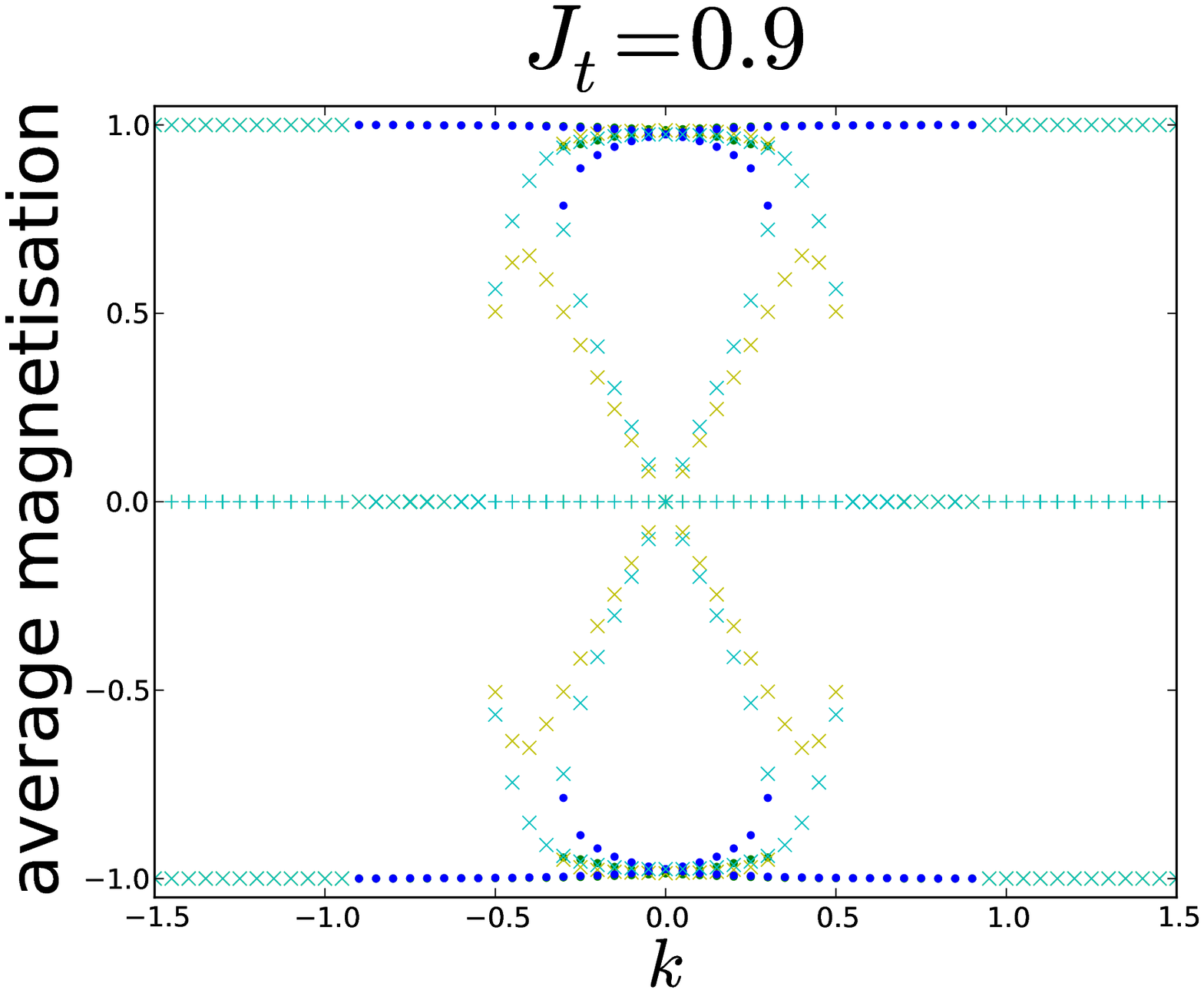}}\\
\subfloat[]{\includegraphics[width=0.33\textwidth]{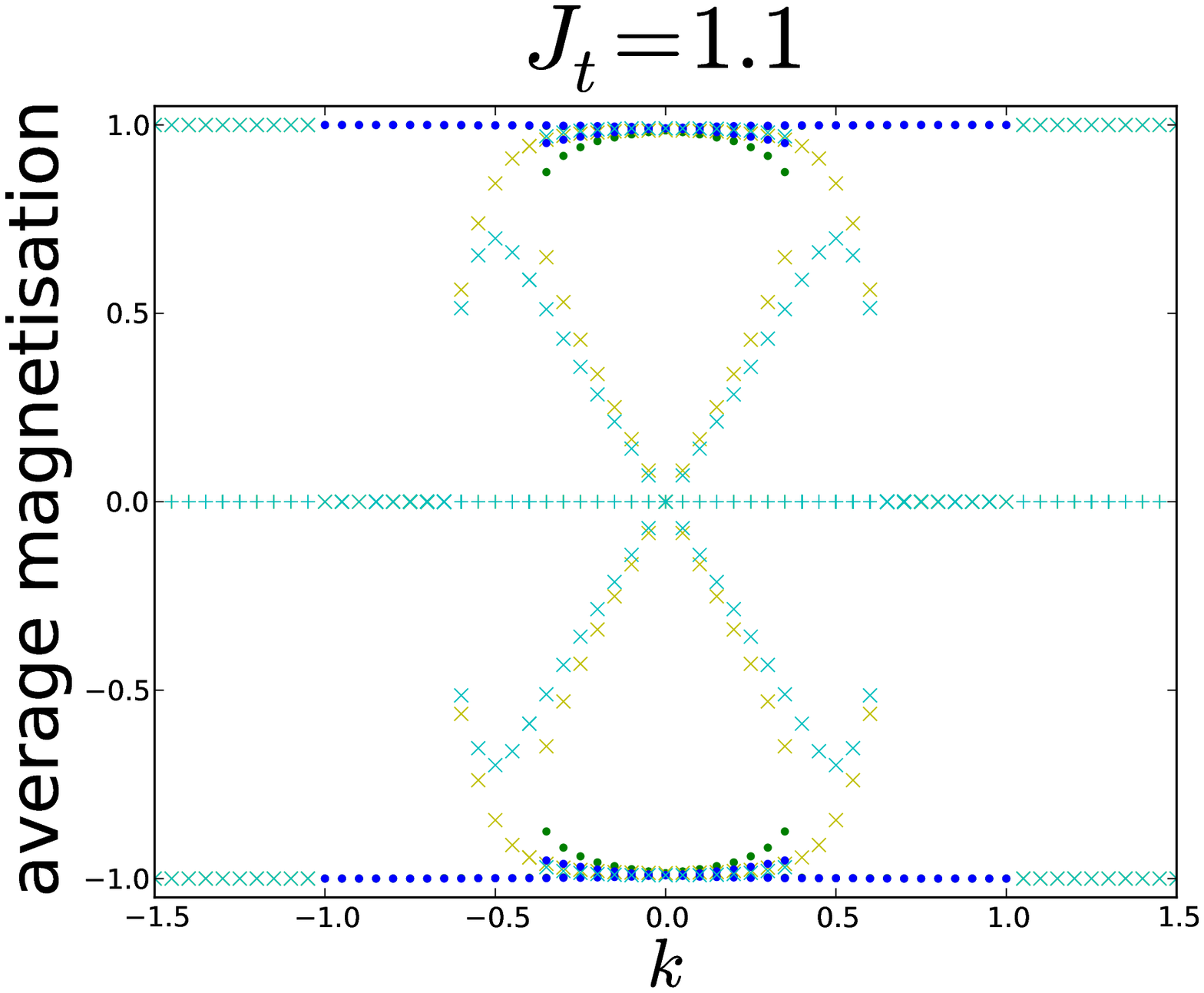}}
\subfloat[]{\includegraphics[width=0.33\textwidth]{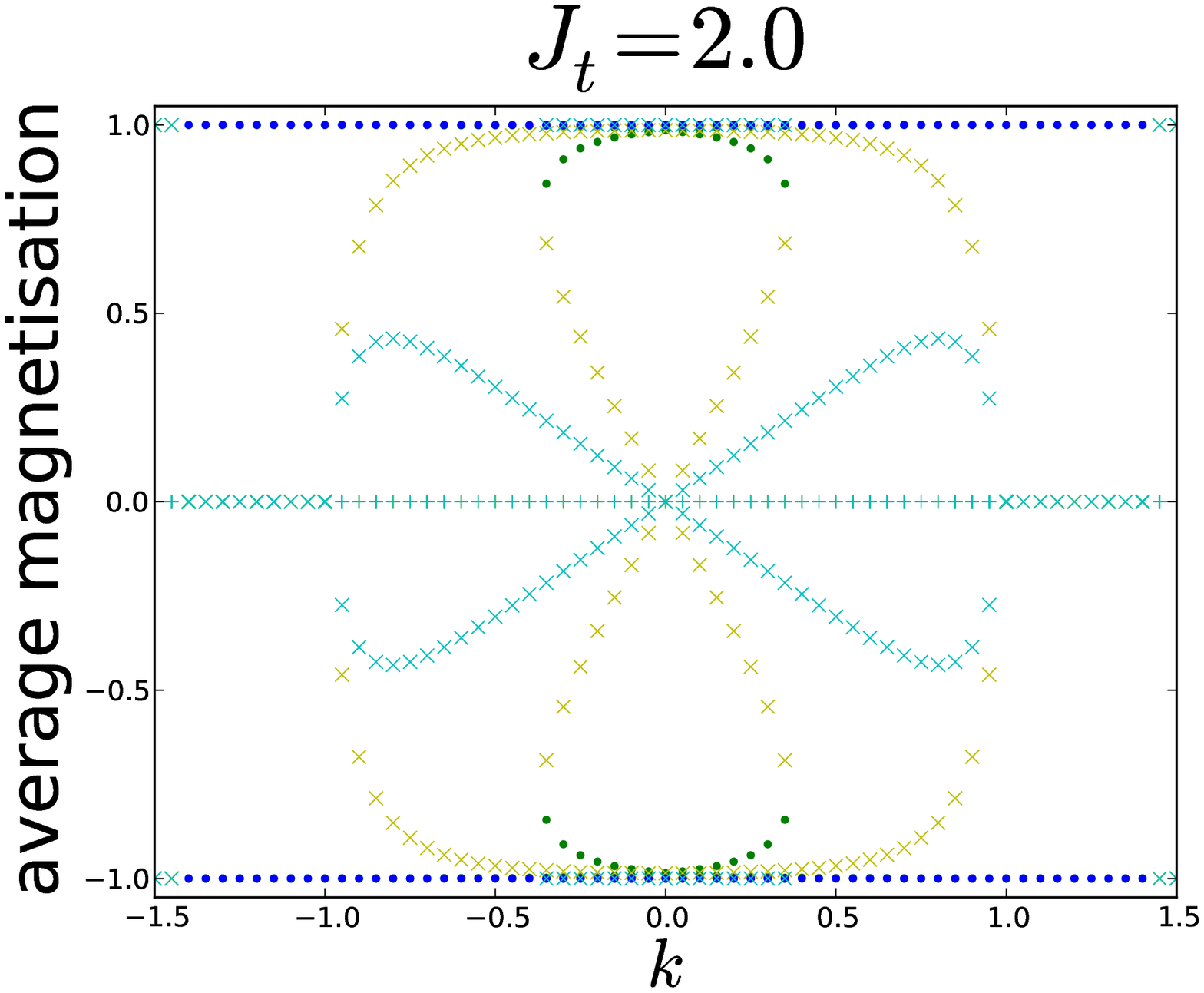}}
\subfloat[]{\includegraphics[width=0.33\textwidth]{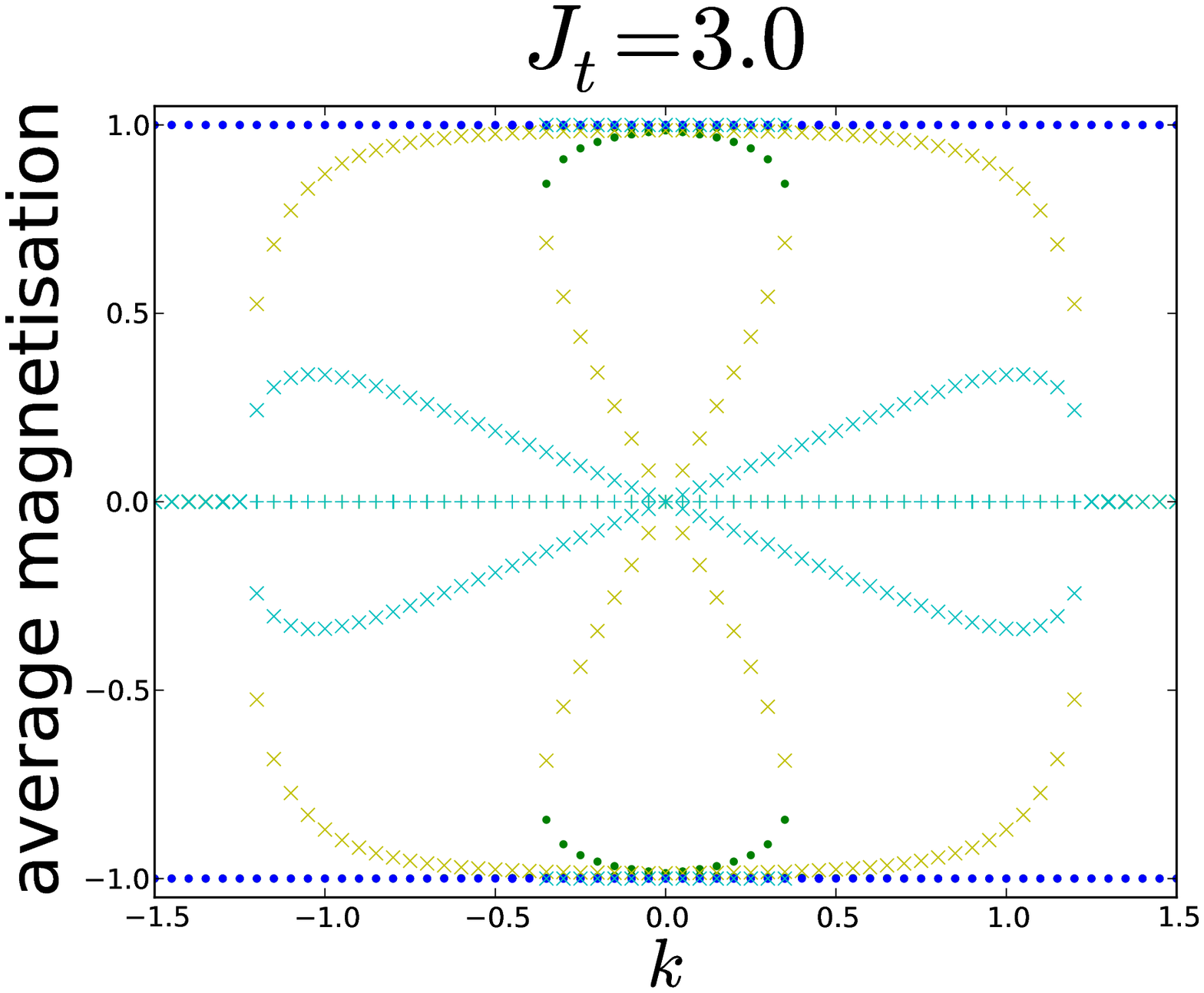}}
\caption{Dependence on inter-coupling of the numerically calculated average magnetisations $(s,t)$ for different values of the $J_{t}$ at low temperature. $J_{s}=1$ and $K_{B}T= 0.4$  for all plots. (a) $J_{t} = 0.05$, (b) $J_{t} = 0.2$, (c) $J_{t} =0.3$, (d) $J_{t} =0.5$, (e) $J_{t} = 0.7$, (f) $J_{t} = 0.9$, (g) $J_{t} =1.1 $, (h) $J_{t} = 2$ and (i) $J_{t} = 3$. In all cases, different solutions are plotted for $k$ between -1.5 and 1.5 every 0.05 ($K_{B}T$). Magnetisations are plotted in green for $s$ and blue for $t$. Dark points are used for stable solutions and lighter asp ($\times$, for saddle points) or cross ($+$, for maxima) for non stable solutions.}
\label{fig:nlanaJk2}
\end{figure}

If we vary $J_{t}$ for fixed $T$, the dependence on $k$ will change as shown in figure \ref{fig:nlanaJk1} at high temperatures and figure \ref{fig:nlanaJk2} at low temperatures. It is convenient to make this distinction because the $k-J_{t}$ sections of the phase diagram will have qualitatively different behaviour (regarding both the existence of paramagnetic regions and of metastable solutions) depending on the value of the temperature (see next section for more details).

As depicted in figure \ref{fig:nlanaJk1} ($K_{B}T=1.5$), at high temperature and low $J_{t}$ the situation is qualitatively similar to that of high temperatures discussed above. While $|k_{c}|>\sqrt{J_{s}J_{t}}$, only the paramagnetic phase is stable for a short range of values of $k$ (figure \ref{fig:nlanaJk1} a). This region increases as $J_{t}$ increases and $|k_{c}|$ moves to higher values (figure \ref{fig:nlanaJk1} b, c), and for  $|k_{c}|<\sqrt{J_{s}J_{t}}$ (figure \ref{fig:nlanaJk1} d), both paramagnetic and ferromagnetic stable solutions exist. As we continue to increase $J_{t}$, $|k_{c}|$ keeps getting smaller and the region where stable solutions exist larger. Thus the paramagnetic phase will be stable for smaller and smaller regions as the ferromagnetic region of stability grows (figure \ref{fig:nlanaJk1} e, f). For high enough $J_{t}$ the paramagnetic phase disappears completely as a stable solution (figure \ref{fig:nlanaJk1} g). For $J_{t}> K_{B}T=1.5$, the only change in stability will be that associated to the change of coupling regime. Only the ferromagnetic phase is stable, except for $k=0$ where we recover mixed phases. As $J_{t}$ continues to increase, the range of values where $t$ is practically one will move to lower values of $|k|$, while $s$ is much less affected, so the difference between both of them in absolute value a low $k$ becomes greater (figure \ref{fig:nlanaJk1} h, i). In general, the more similar $J_{s}$ and $J_{t}$ are (the more similar $J_{t}$ is to one in this case), the more similar are $s$ and $t$ in absolute value even at low $k$. At $J_{t}=J_{s}=1$, there is another qualitative change as $|t|>|s|$ for higher values of $J_{t}$. There are no metastable states in this setup regardless the values of $k$ and $J_{t}$.

Figure \ref{fig:nlanaJk2} shows an example of low temperature behaviour of the dependence on $k$ when varying $J_{t}$ for fixed $T$ ($K_{B}T=0.4$). As in the high temperature case, increasing the value of $J_{t}$ will make the region where stable solutions exist larger. In this case, there are never paramagnetic stable solutions. At low values of $J_{t}$, in the stable region, $|s|$ will be practically one for all values of $k$, while $|t|$ can be much smaller at low values of $k$ (the only stable region). For $k=0$, stable solutions will be mixed phases (a pair with $s\neq 0$ and of equal absolute $s$ and opposite signs). As $J_{t}$ increases, $|t|$ will become more similar to $|s|$ (and thus to one), and the stability region increases (figure \ref{fig:nlanaJk2} a, b, c). For $J_{t}>K_{B}T=0.4$, two values $k_{c}$ appear such that there will be ferromagnetic saddle solutions for $|k|<|k_{c}|$ and the mixed phase at $k=0$ disappears (figure \ref{fig:nlanaJk2} d). For still higher intra-couplings, spinodal points $k_{a}$ and metastable states appear, with both $|k_{c}|$ and $|k_{a}|$ higher the higher $J_{t}$ is (figure \ref{fig:nlanaJk2} e, f). At this point, both ferromagnetic main branches are very close to one in absolute value for all values of $k$, and only metastable states can have significant lower average magnetisation $t$. For $J_{t}>J_{s}=1$, this behaviour is reversed, and it is now $|s|$ in the metastable branch that can be significantly lower (figure \ref{fig:nlanaJk2} g). At this point, $|k_{c}|$ will continue to move to higher values as we increase $J_{t}$, but not so $|k_{a}|$ that will remain constant (figure \ref{fig:nlanaJk2} h, i).

\subsection{Dependence on intra-couplings}

Regarding dependence of the function $l$ defined in equation \eqref{eq:Tceq} on the intra-coupling $J_{t}$, we can rewrite the function's roots as given by

\begin{equation}
J_{t}^{c}=\frac{\beta(k^{2}\beta+J_{s})-1}{\beta(J_{s}\beta-1)}
\end{equation}.

The numerator will be negative for $0<\beta<\frac{-J_{s}+\sqrt{J_{s}^{2}+4k^{2}}}{2k^{2}}$ ($K_{B}T>\frac{2k^{2}}{-J_{s}+\sqrt{J_{s}^{2}+4k^{2}}}$) and positive for higher $\beta$ (lower temperature). The denominator of the expression is negative for $0<\beta<\frac{1}{J_{s}}$ ($K_{B}T>J_{s}$) and positive for higher $\beta$ (lower temperature). For $\beta=\frac{-J_{s}+\sqrt{J_{s}^{2}+4k^{2}}}{2k^{2}}$ $J_{t}^{c}=0$ and for $K_{B}T=J_{s}$ there is no value $J_{t}^{c}$. Depending on the values chosen for $k$ and $T$, there will be therefore either none (physically relevant)
 or one value of $J_{t}$ where the stability of the paramagnetic phase changes from saddle to minimum/maximum (change in the sign of \eqref{eq:nldethesspara}), besides of that given by the change in the coupling regime $J_{t}=\frac{k^{2}}{J_{s}}$. As for the sign of the second derivative \eqref{eq:fss3}, it is negative for $K_{B}T<\frac{k^{2}-J_{s}^{2}}{J_{s}}$ and positive for $K_{B}T>\frac{k^{2}-J_{s}^{2}}{J_{s}}$.

Figure \ref{fig:locanaJ1} shows an example on what we will call high temperature behaviour ($K_{B}T>J_{s}$) for $k=0.3$ and $K_{B}T=1.5$ ($J_{t}^{c}=1.32$). The change between both coupling regimes takes place at $J_{t}=0.09$. For $J_{t}<0.09$ there are no stable solutions. The paramagnetic solution is a saddle point that becomes stable for $J_{t}>J_{t}^{d}=0.09$. At $J_{t}=J_{t}^{c}=1.32$, there is a second order phase transition, the paramagnetic phase looses its stability and becomes again a saddle point, while the two main ferromagnetic pair of solutions appear as minima of the free energy.

\begin{figure}
\centering
\includegraphics[width=\textwidth]{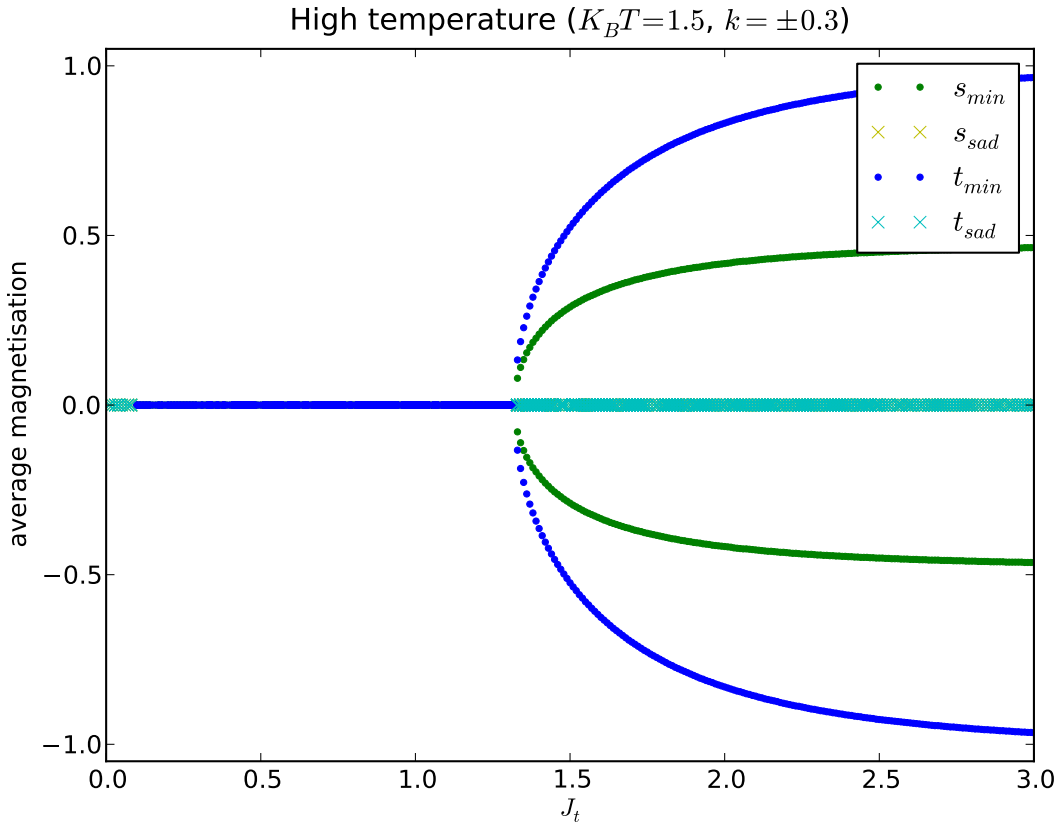}
\caption{Dependence on the intra-coupling $J_{t}$ of the numerically calculated average magnetisations at low temperature ($J_{s}=1$, $k= \pm 0.3$ , $K_{B}T = 1.5$, $J_{t}^{c}=1.32$). Degenerate case (limiting value between both coupling regimes) for $J_{t}=\frac{k^{2}}{J_{s}}=0.09$. Different solutions are plotted for $J_{t}$  between 0 and 3 every 0.01. Magnetisations are plotted in green for $s$ and blue for $t$. Dark points are used for stable solutions and lighter asps ($\times$) for saddle points, non stable solutions.}
\label{fig:nloanaJ1}
\end{figure}

\begin{figure}
\centering
\includegraphics[width=\textwidth]{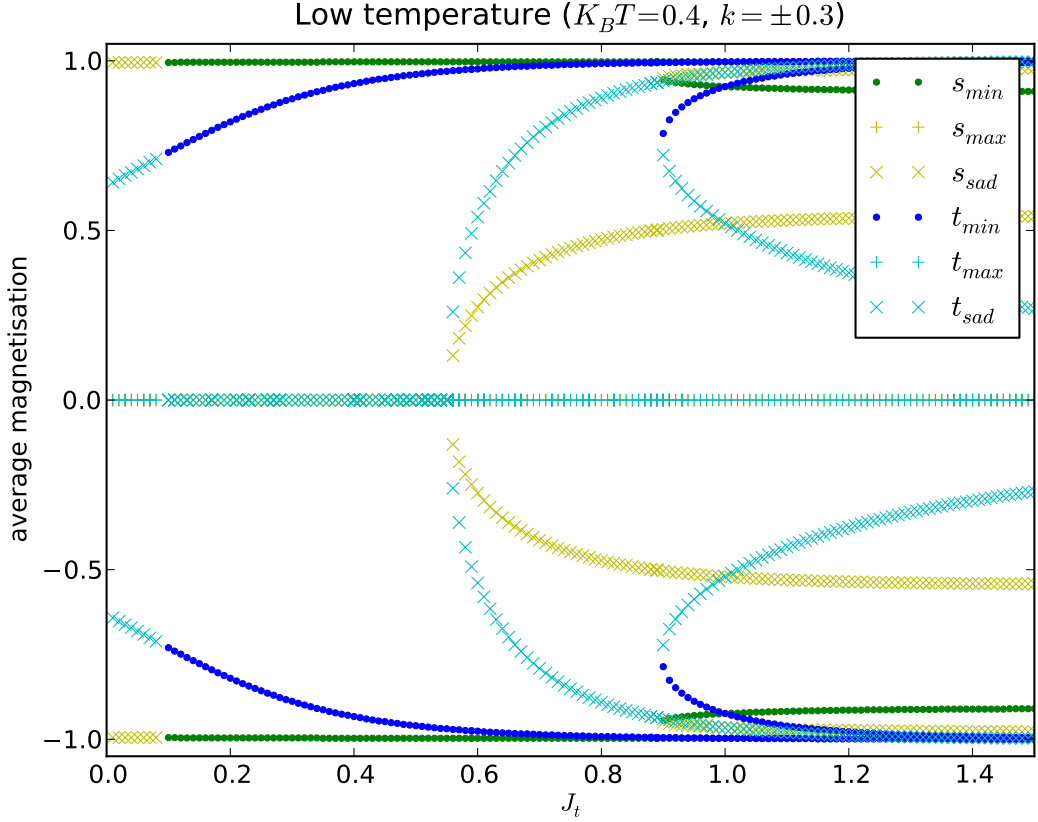}
\caption{Dependence on the intra-coupling $J_{t}$ of the numerically calculated average magnetisations at low temperature ($J_{s}=1$, $k= \pm 0.3$ , $K_{B}T = 0.4$, $J_{t}^{c}=0.55$). Degenerate case (limiting value between both coupling regimes) for $J_{t}=\frac{k^{2}}{J_{s}}=0.09$. Different solutions are plotted for $J_{t}$  between 0 and 1.5 every 0.01. Magnetisations are plotted in green for $s$ and blue for $t$. Dark points are used for stable solutions and lighter asp ($\times$, for saddle points) or cross ($+$, for maxima) for non stable solutions. }
\label{fig:nloanaJ2}
\end{figure}

In figure \ref{fig:nloanaJ2} the situation at low temperatures (($K_{B}T<J_{s}$)) is illustrated ($k=0.3$, $K_{B}T=0.4$, $J_{t}^{c}=0.55$)). Change between both coupling regimes still takes place at $J_{t}=0.09$. In this case, there are no phase transitions, and $J_{t}^{c}$ is associated to the onset of saddle point ferromagnetic solutions. For $J_{t}<0.09$ there are no stable solutions. The paramagnetic solution is a maximum and the ferromagnetic main pair of solutions are saddle points. For $J_{t}>0.09$ the stability regime begins, $(0,0)$ becomes a saddle point while the ferromagnetic main branches become minima. At $J_{t}^{c}=0.55$ two new saddle point ferromagnetic branches appear and at a certain (spinodal) value $J_{t}^{a}$, the metastable ferromagnetic states appear. 

\begin{figure}
\centering
\subfloat[]{\includegraphics[width=0.33\textwidth]{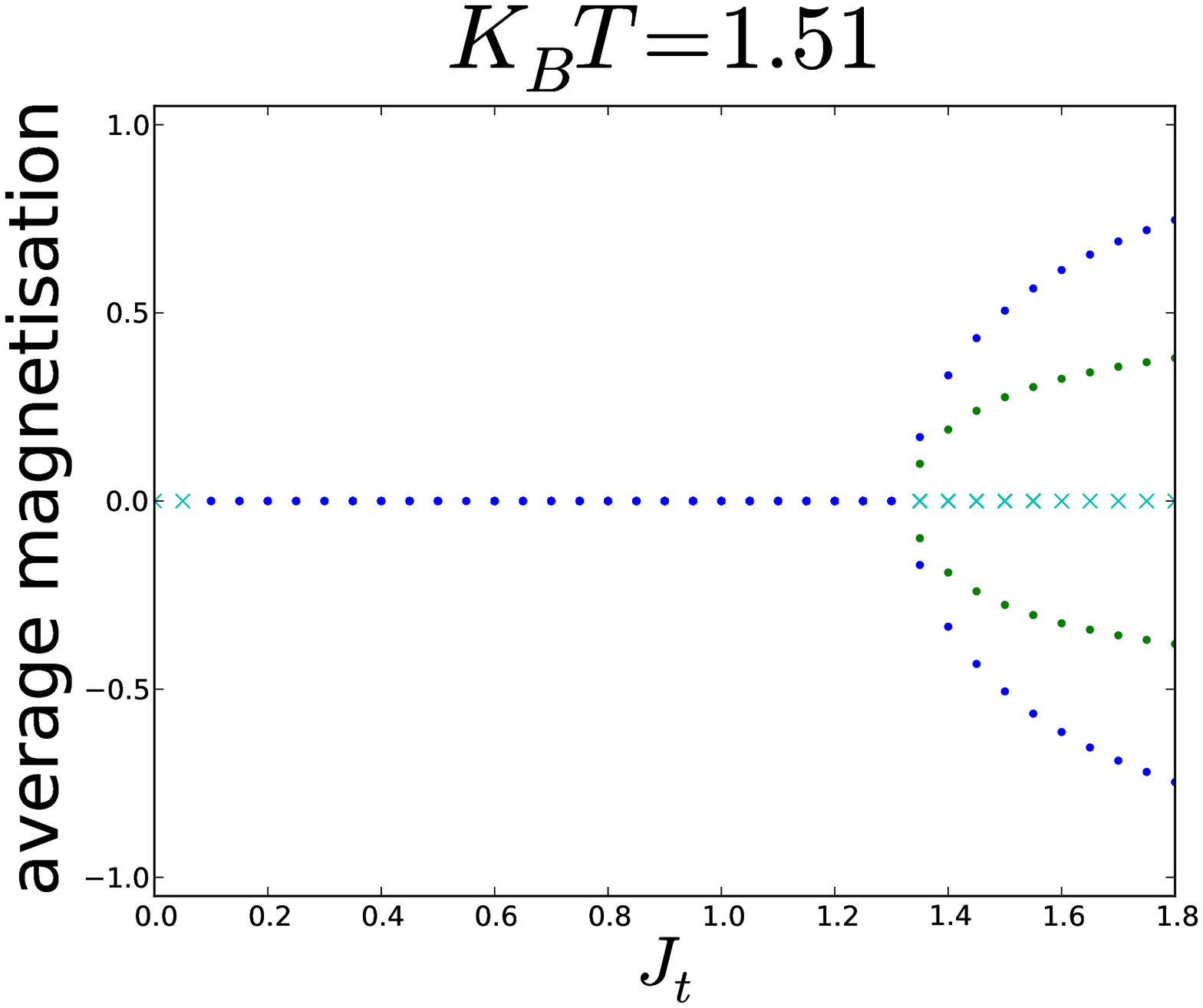}}
\subfloat[]{\includegraphics[width=0.33\textwidth]{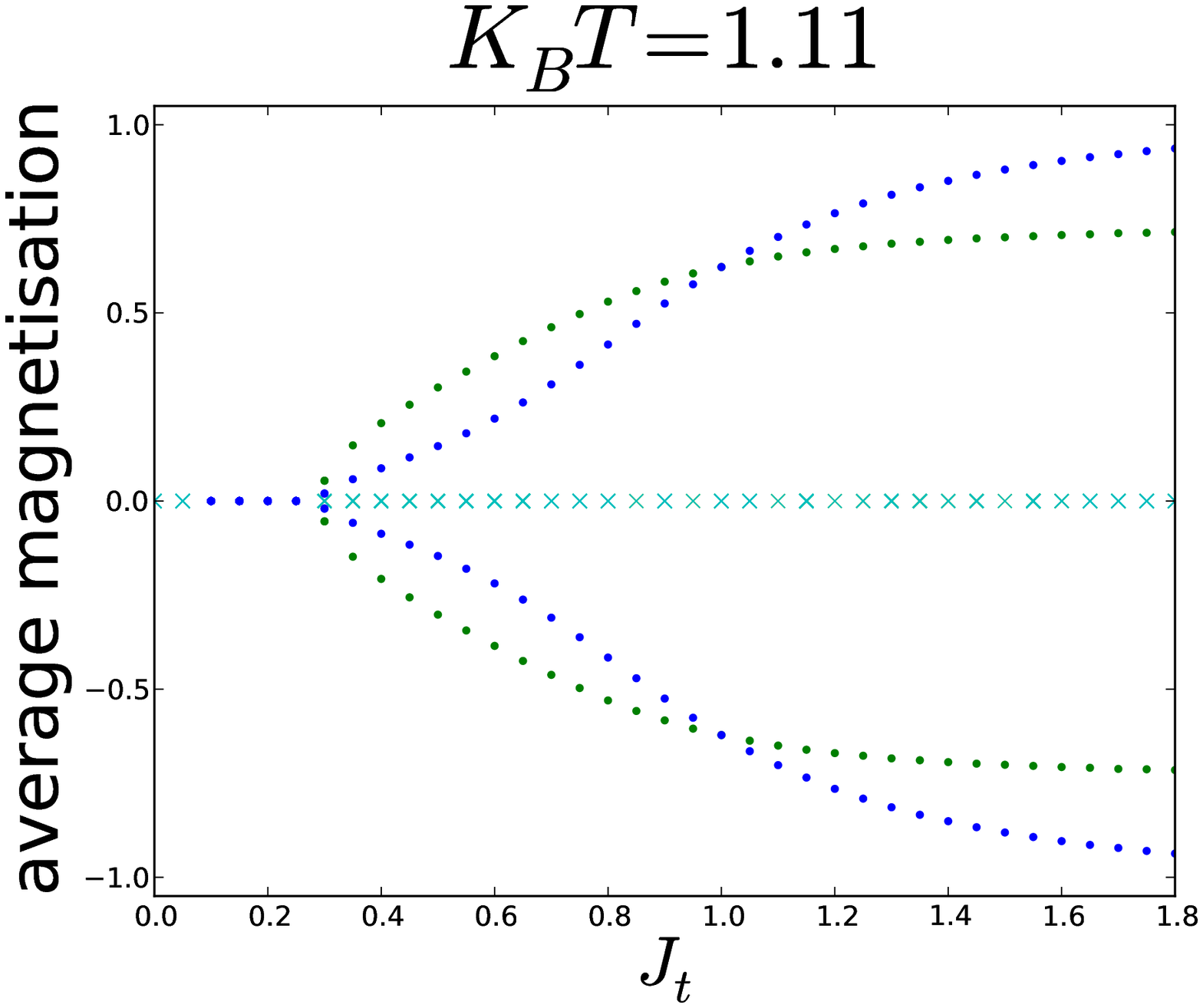}}
\subfloat[]{\includegraphics[width=0.33\textwidth]{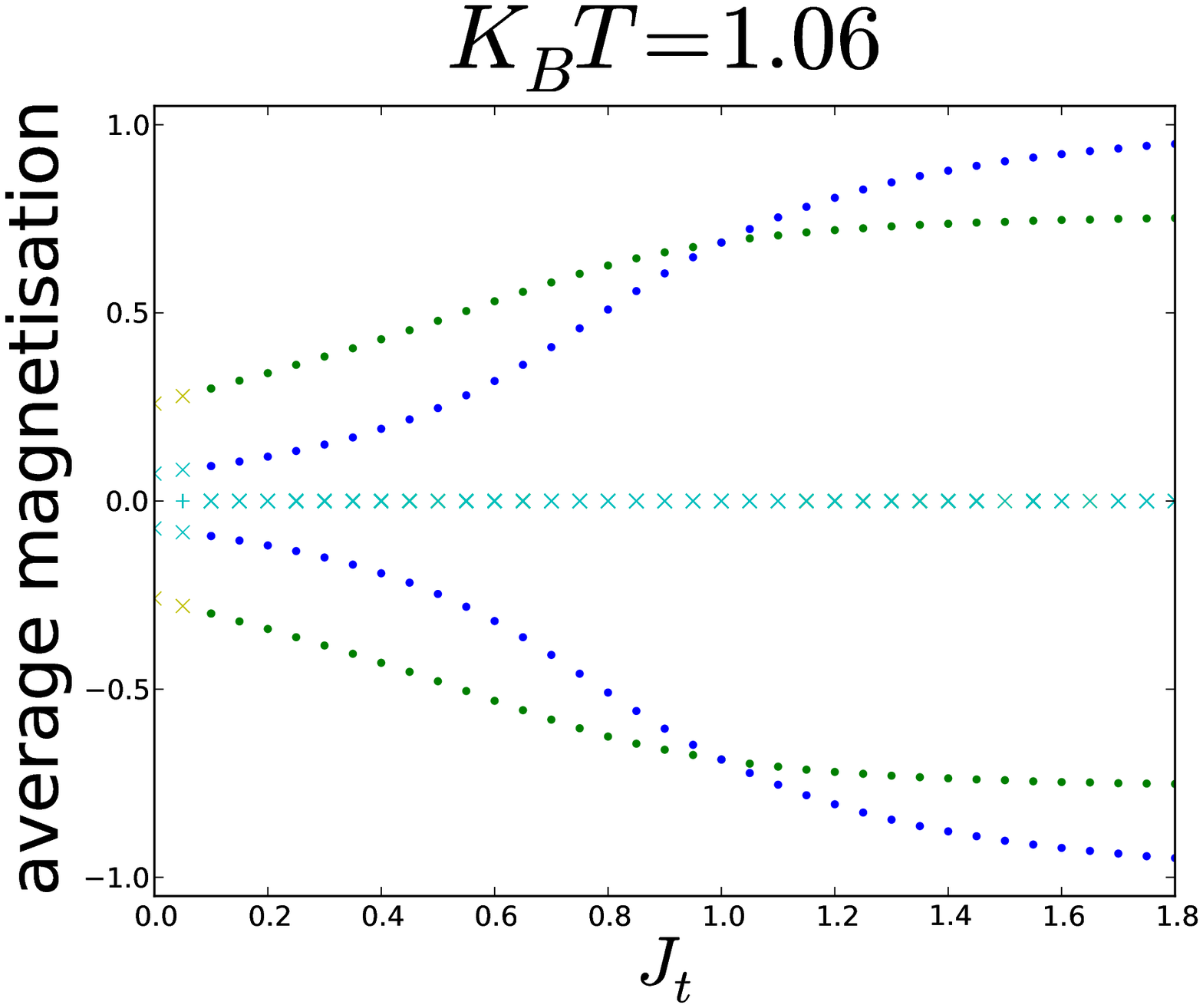}}\\
\subfloat[]{\includegraphics[width=0.33\textwidth]{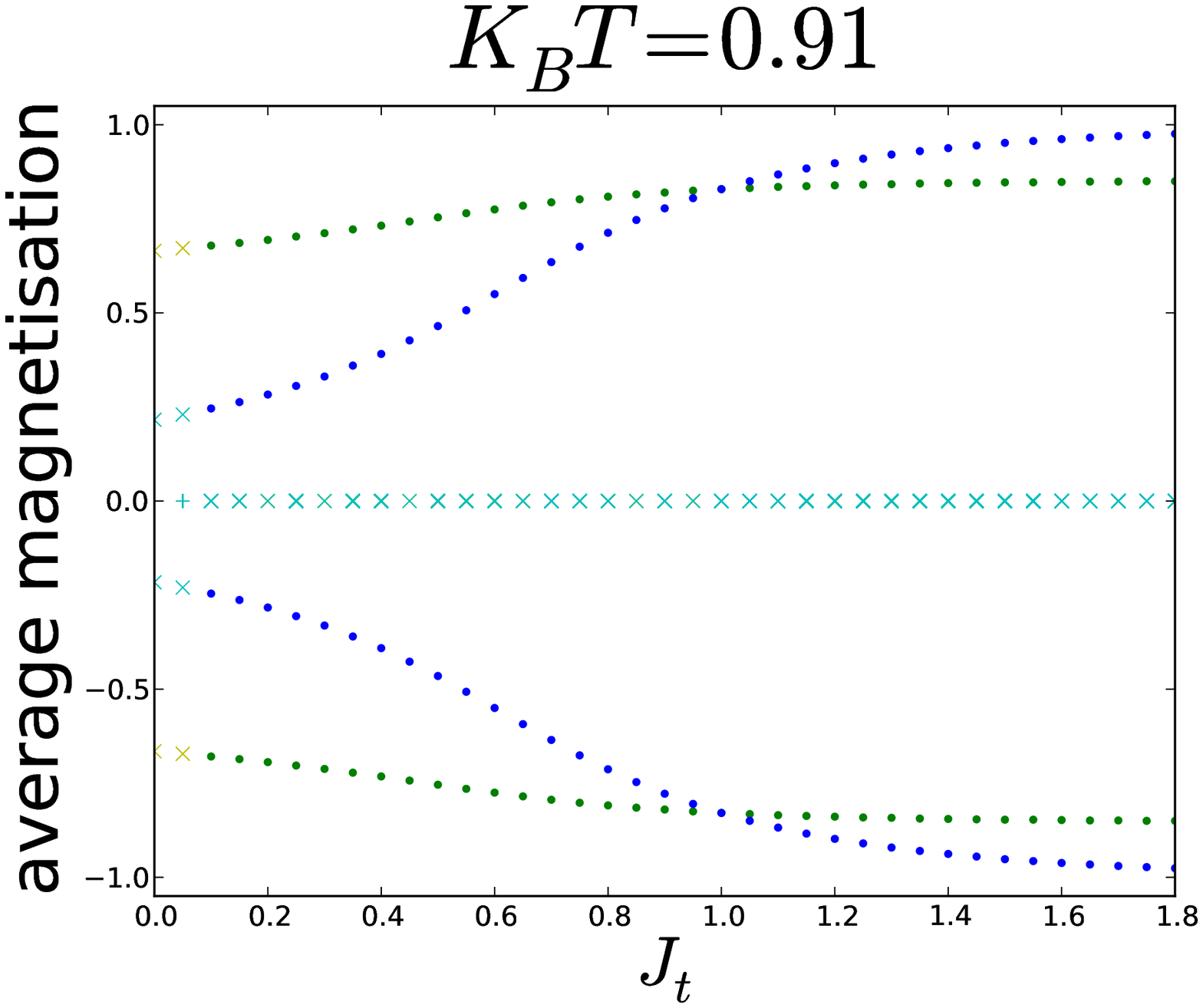}}
\subfloat[]{\includegraphics[width=0.33\textwidth]{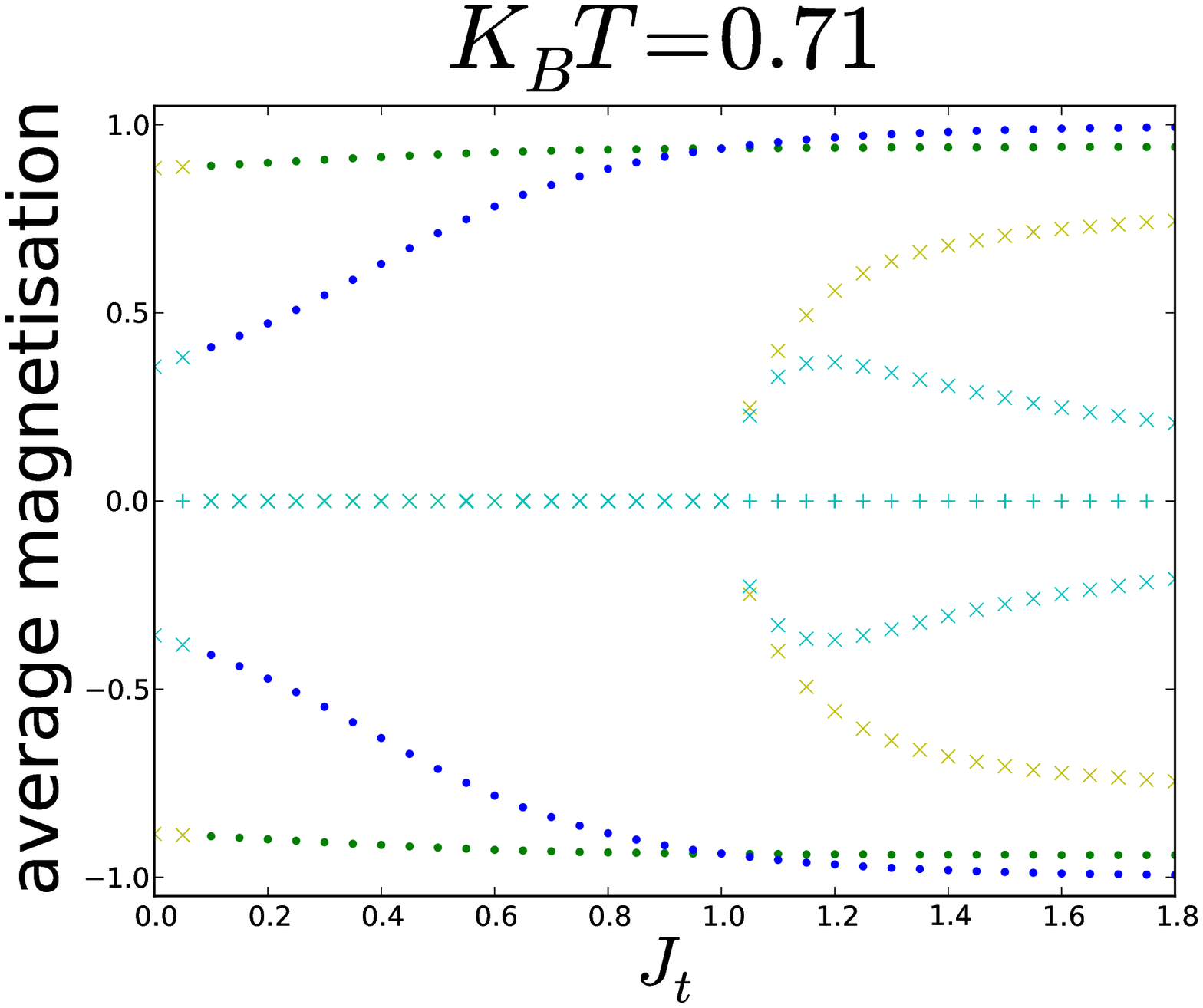}}
\subfloat[]{\includegraphics[width=0.33\textwidth]{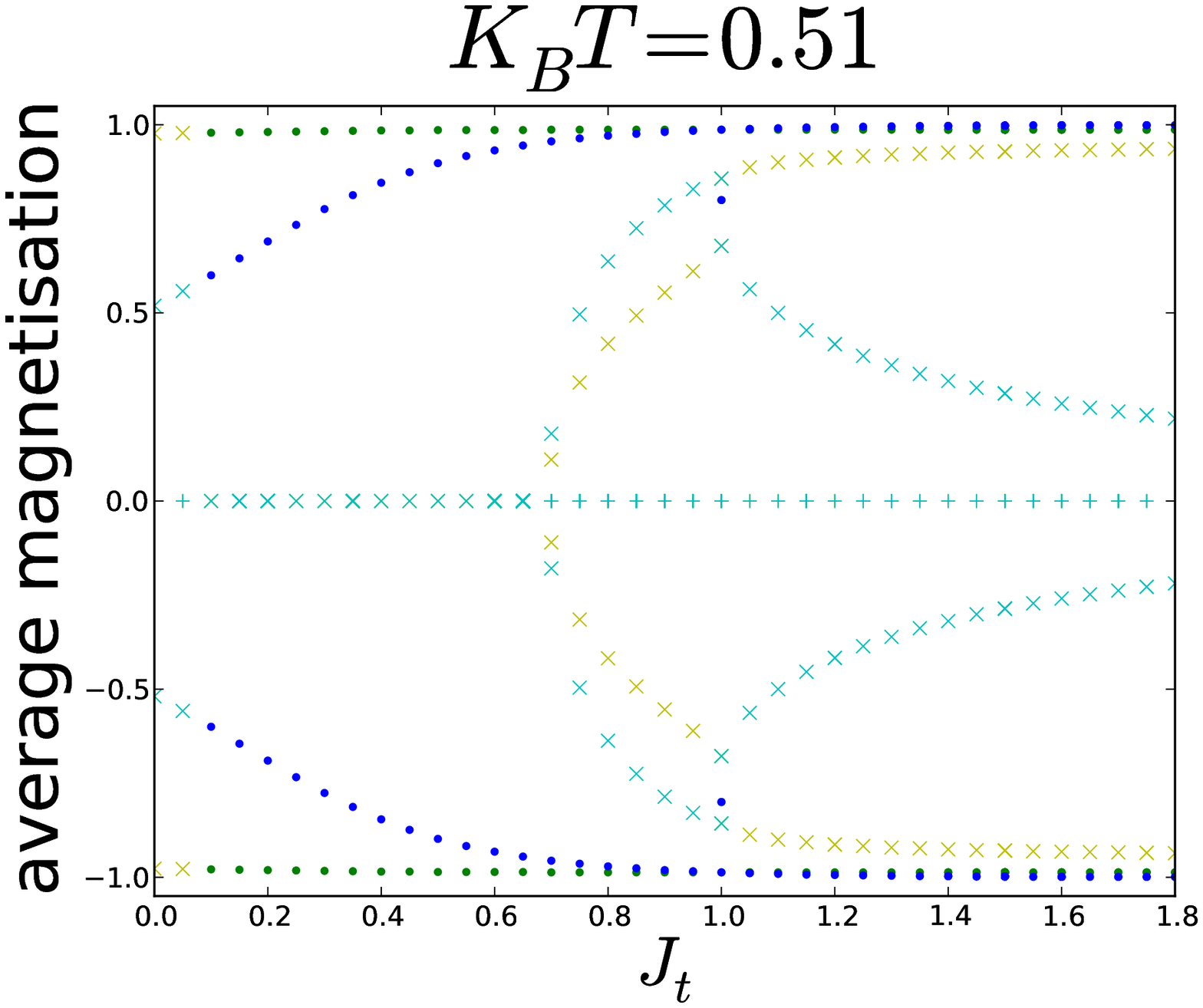}}\\
\subfloat[]{\includegraphics[width=0.33\textwidth]{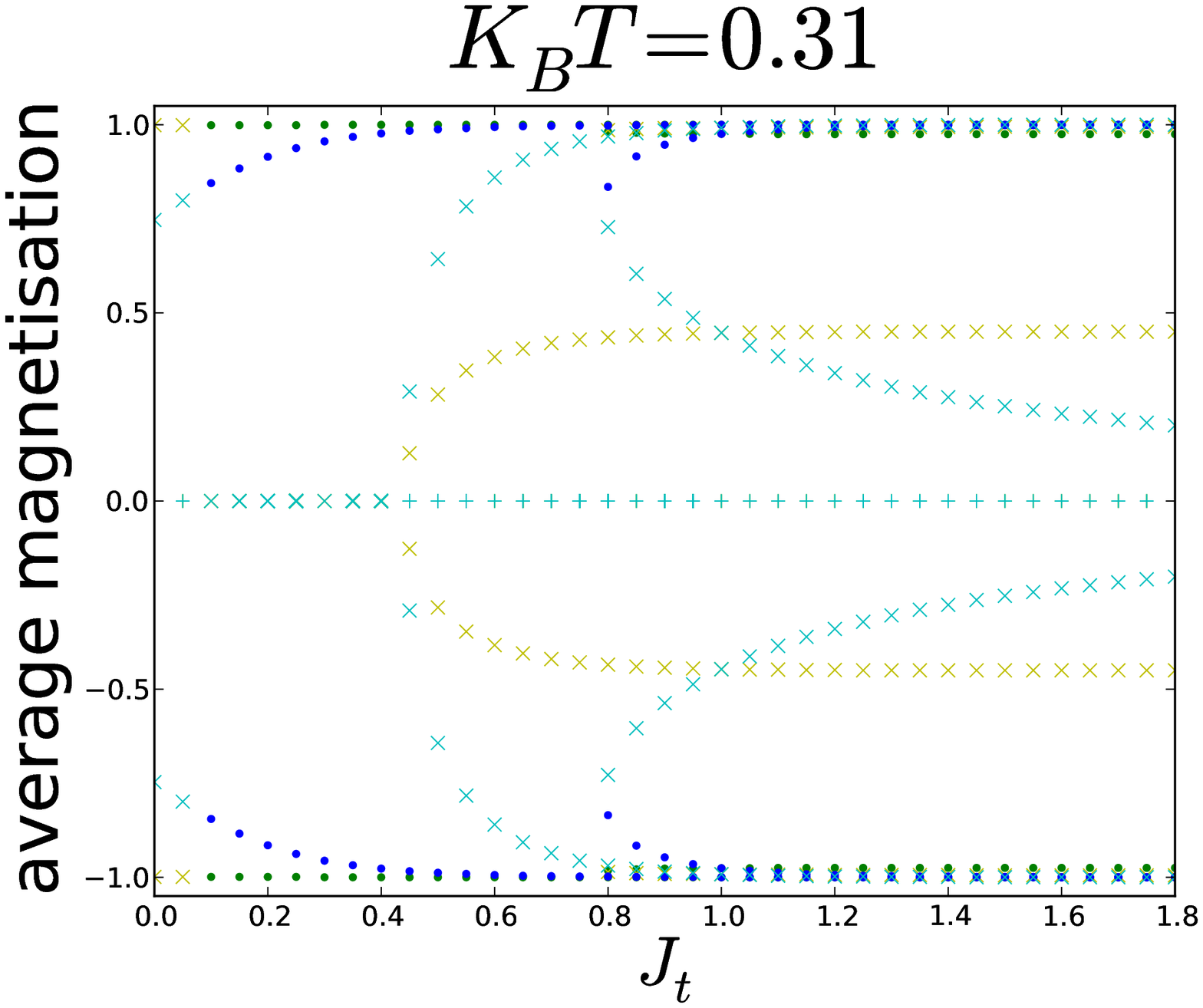}}
\subfloat[]{\includegraphics[width=0.33\textwidth]{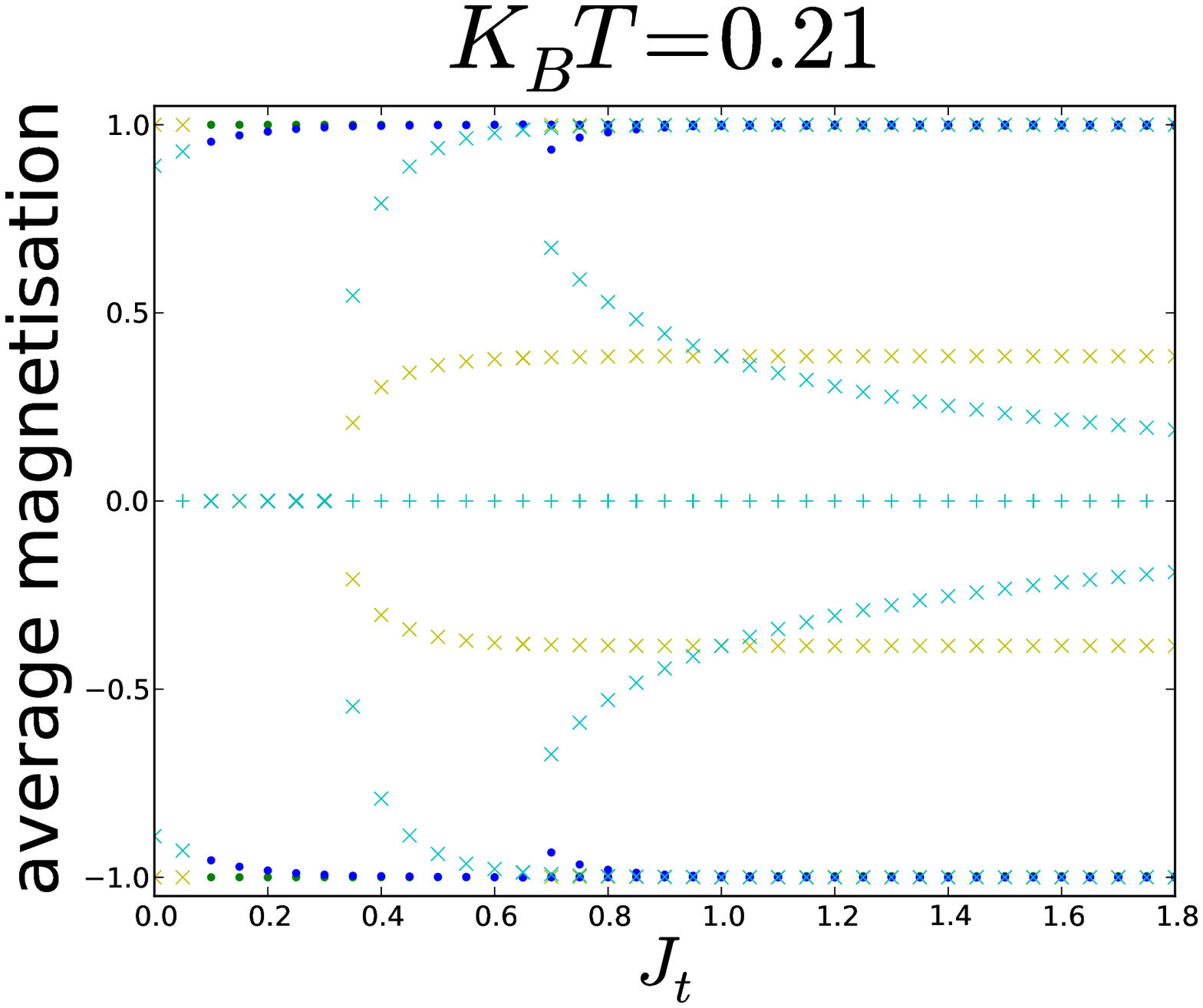}}
\subfloat[]{\includegraphics[width=0.33\textwidth]{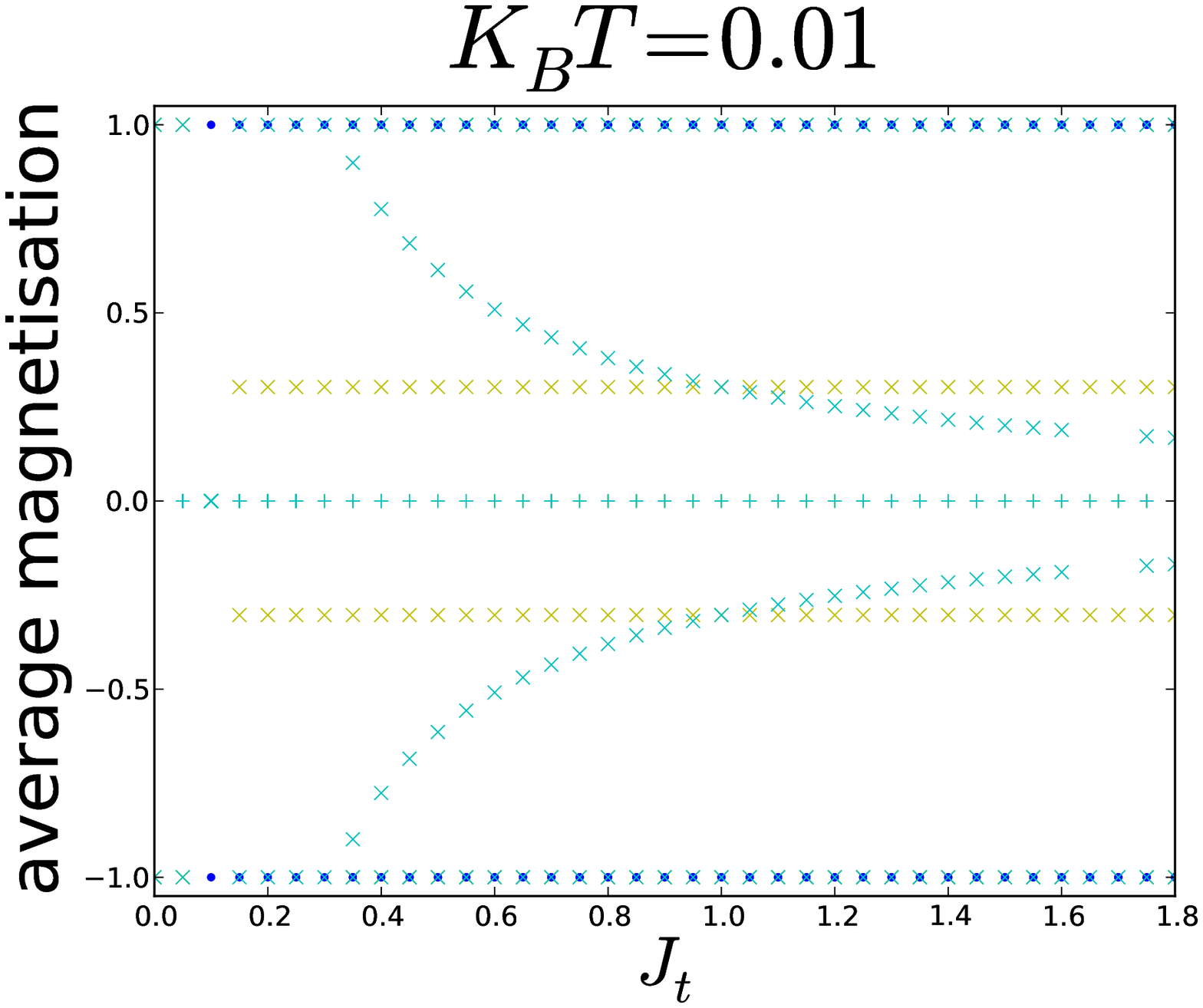}}
\caption{Dependence on intra-coupling $J_{t}$ of the numerically calculated average magnetisations $(s,t)$ for different values of the temperature $K_{B}T$. $J_{s}=1$ and $k=\pm 0.3$  for all plots. (a) $K_{B}T=1.51$, (b) $K_{B}T=1.11$, (c) $K_{B}T=1.06$, (d) $K_{B}T=0.91$, (e) $K_{B}T=0.71$, (f) $K_{B}T=0.51$, (g) $K_{B}T=0.31$, (h) $K_{B}T=0.21$ and (i) $K_{B}T=0.01$. In all cases, different solutions are plotted for intra-coupling $J_{t}$ between 0.01 and 1.8 every 0.05. Magnetisations are plotted in green for $s$ and blue for $t$. Dark points are used for stable solutions and lighter asp ($\times$, for saddle points) or cross ($+$, for maxima) for non stable solutions.}
\label{fig:nloanaTJ}
\end{figure}

Figure \ref{fig:nloanaTJ} shows how the dependence on $J_{t}$ varies as we change the temperature for fixed $k=\pm0.3$. Note for all figures there is always a small fixed region (for $J_{t}<J_{t}^{d}=0.09$) where there are no stable solutions (strong coupling regime). For high enough temperatures (figure \ref{fig:nloanaTJ} a), we are in the qualitative situation described described above with $J_{t}^{d}=0.09$ and by the critical intra-coupling $J_{t}^{c}$ at which there is a second order phase transition from paramagnetic to ferromagnetic states. As we lower the temperature $J_{t}^{c}$ moves to lower values, and when $J_{t}^{c}<J_{s}=1$, we begin to have ferromagnetic solutions with $|t|>|s|$ (for $J_{t}<1$) as well as with $|s|>|t|$ (for $J_{t}>1$, figure \ref{fig:nloanaTJ} b). At some point (when $J_{t}^{c}<J_{t}^{d}=0.09$), the paramagnetic phase will no longer be stable regardless of the intra-coupling strength (figure \ref{fig:nloanaTJ} c), and so there is no longer a phase transition. Both average magnetisations will grow in absolute value as we continue to lower the temperature (figure \ref{fig:nloanaTJ} d). At a low enough temperature, new saddle point solutions appear at the reappeared $J_{t}^{c}$, which is not a critical point anymore(figure \ref{fig:nloanaTJ} e). At even lower temperatures, metastable states appear for $J>J_{t}^{a}$ (figure \ref{fig:nloanaTJ} f)\footnote{There is in fact a small range of temperatures which differs depending on the specific value of fixed $k$ for which there seem to be metastable solutions for a small region $J_{a}<J<J_{a'}$ with $J_{a}\approx J_{a'}$. This is the case for figure \ref{fig:nloanaTJ} f where the numerical solution sampling yields a metastable state only for one value ($K_{B}T=0.51$, $J_{t}=1$). These may be however numerical effects on the spinodal border with no physical or sociological interpretation. }. As the temperature continues to drop, $J_{t}^{c}$ moves to lower values (never reaching zero), and both main and metastable ferromagnetic solutions will get closer to one in absolute value for smaller and smaller values of $J_{t}$ (figure \ref{fig:nloanaTJ} g, h, i).
 
\begin{figure}
\centering
\subfloat[]{\includegraphics[width=0.33\textwidth]{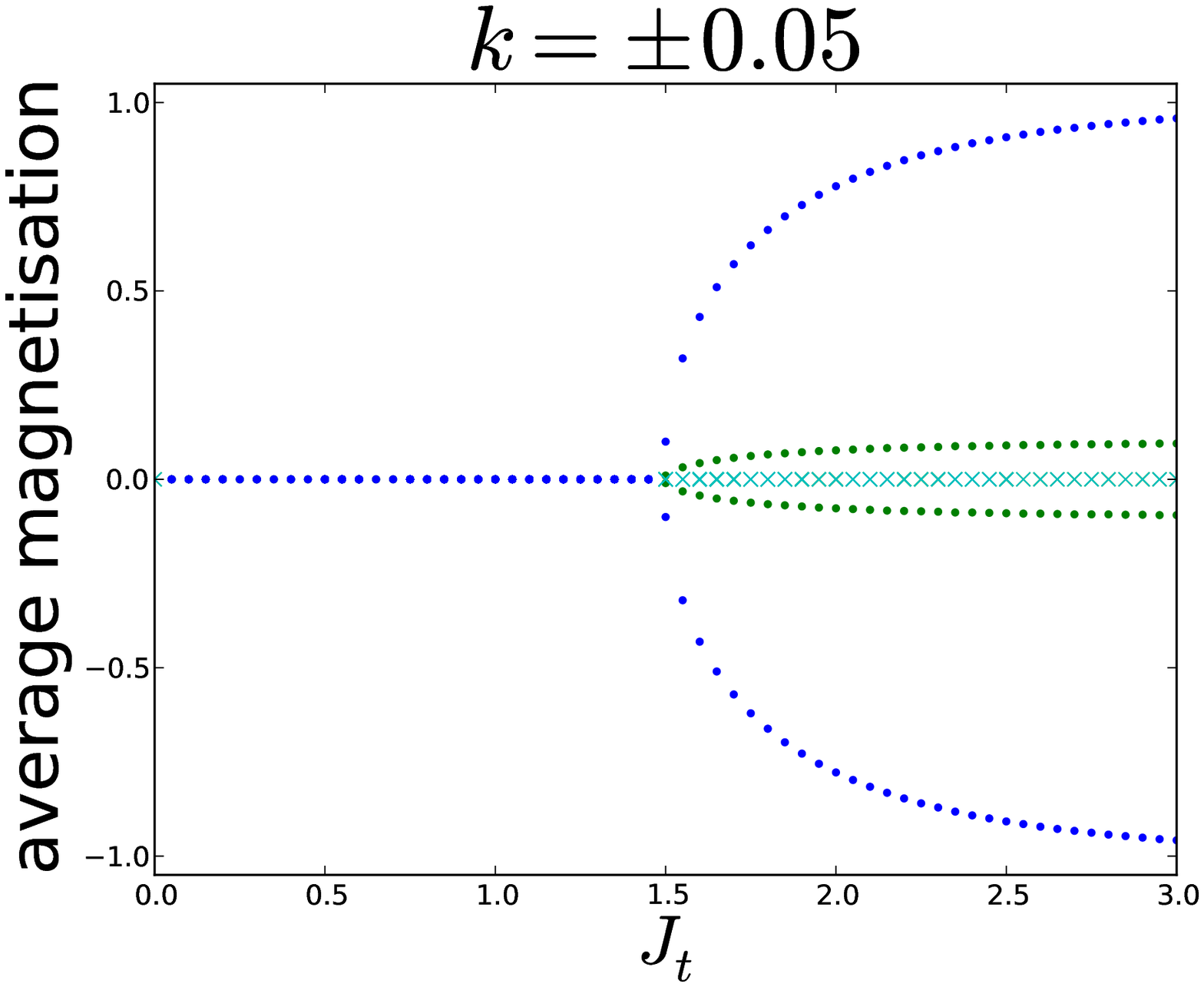}}
\subfloat[]{\includegraphics[width=0.33\textwidth]{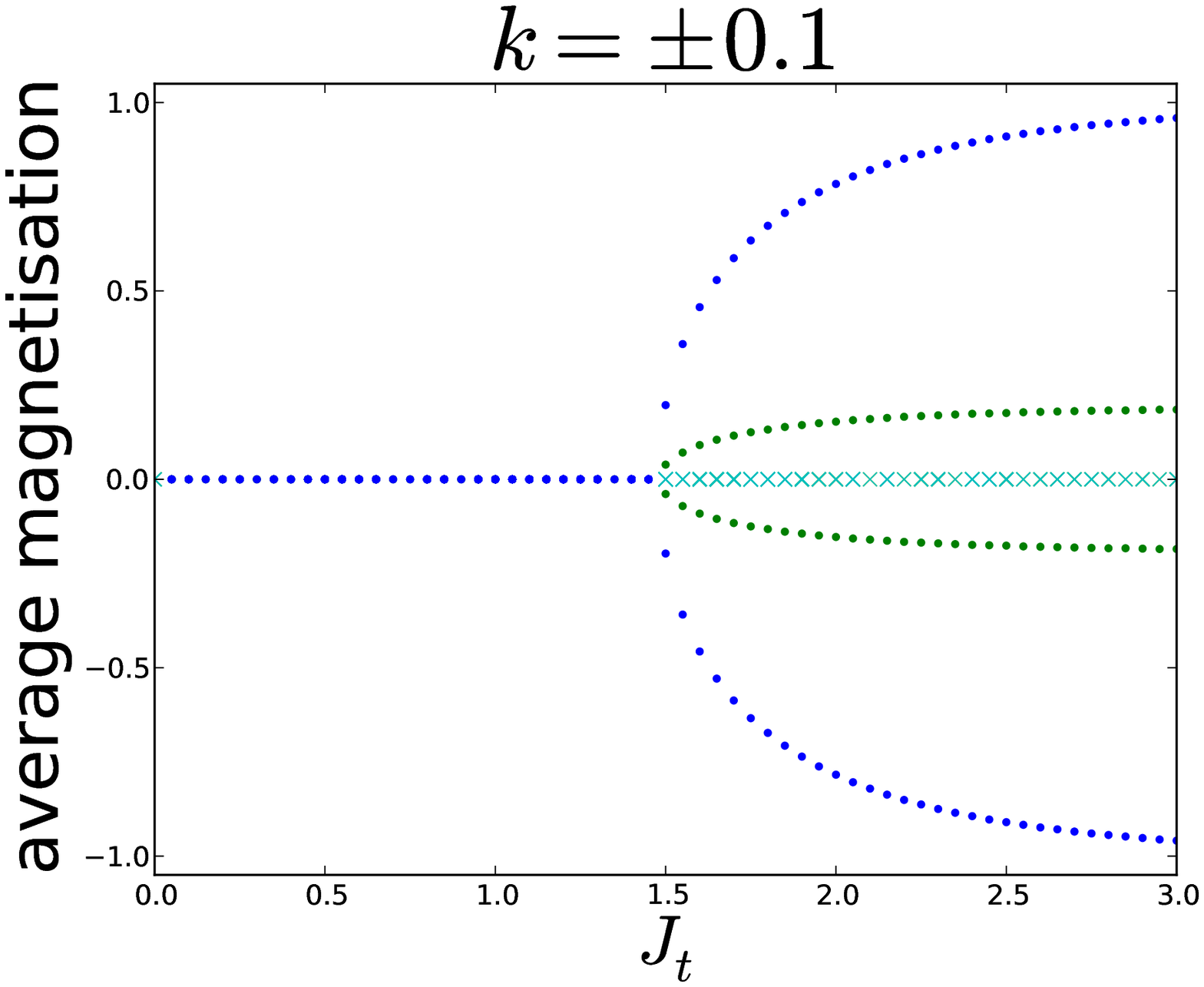}}
\subfloat[]{\includegraphics[width=0.33\textwidth]{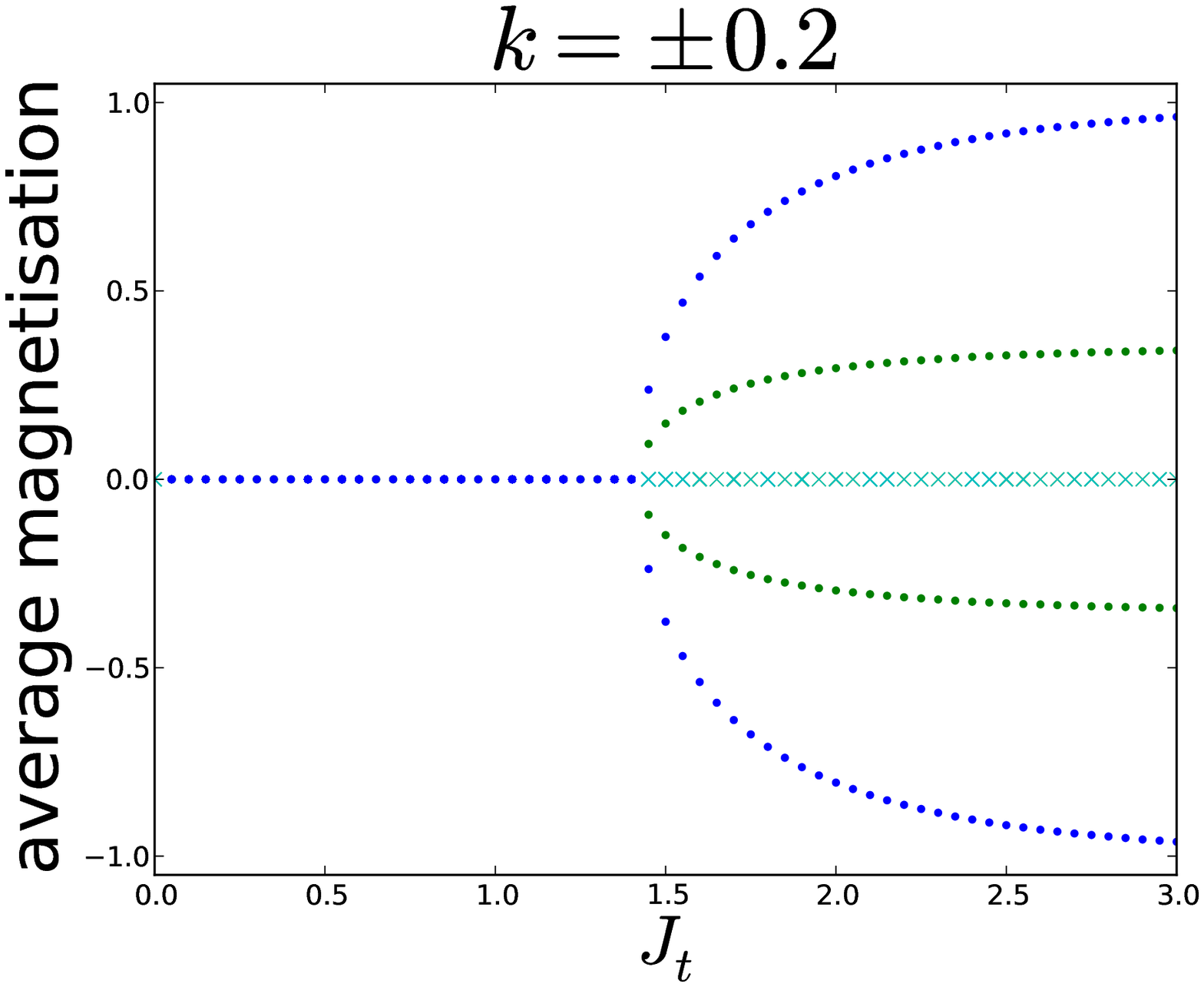}}\\
\subfloat[]{\includegraphics[width=0.33\textwidth]{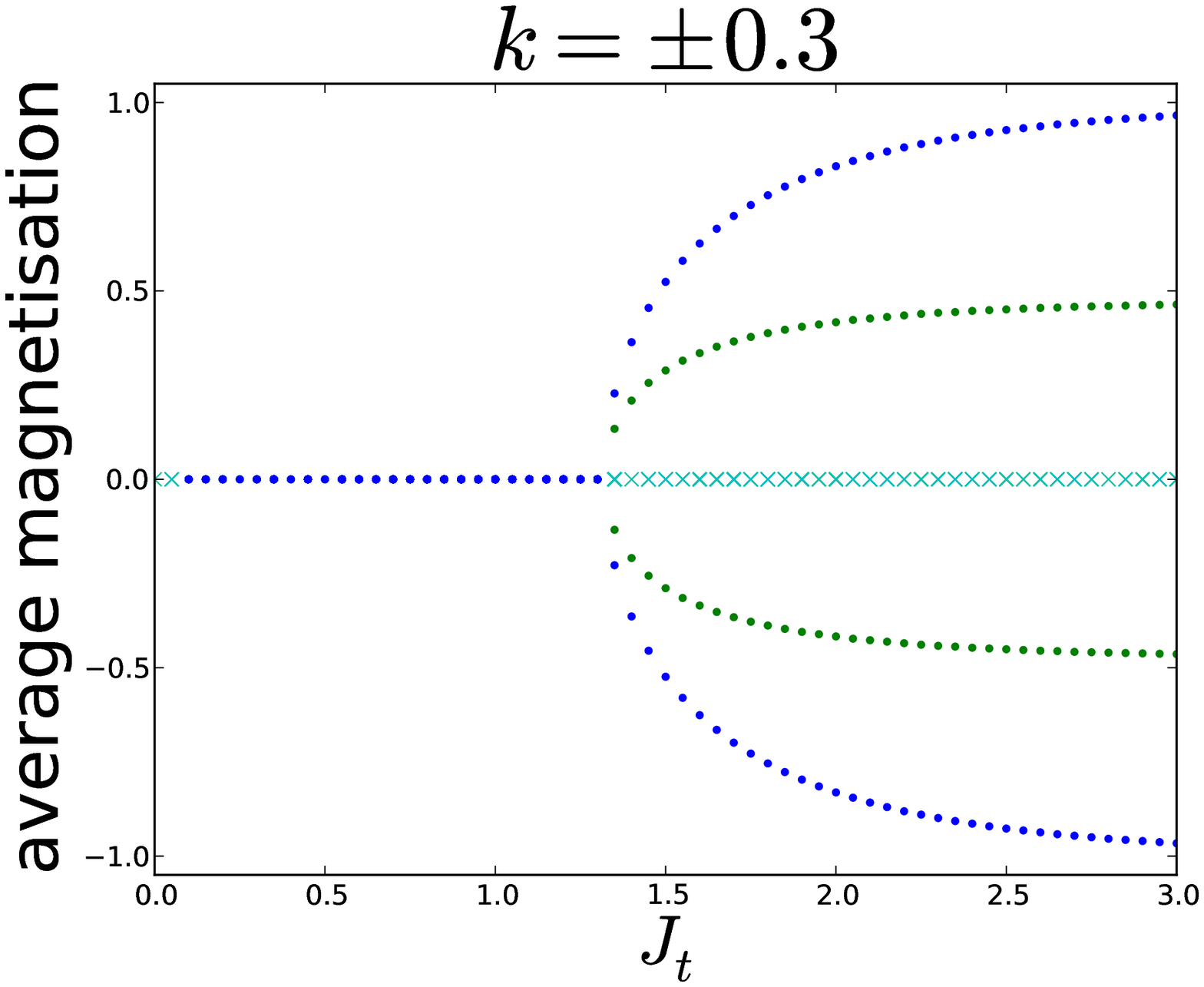}}
\subfloat[]{\includegraphics[width=0.33\textwidth]{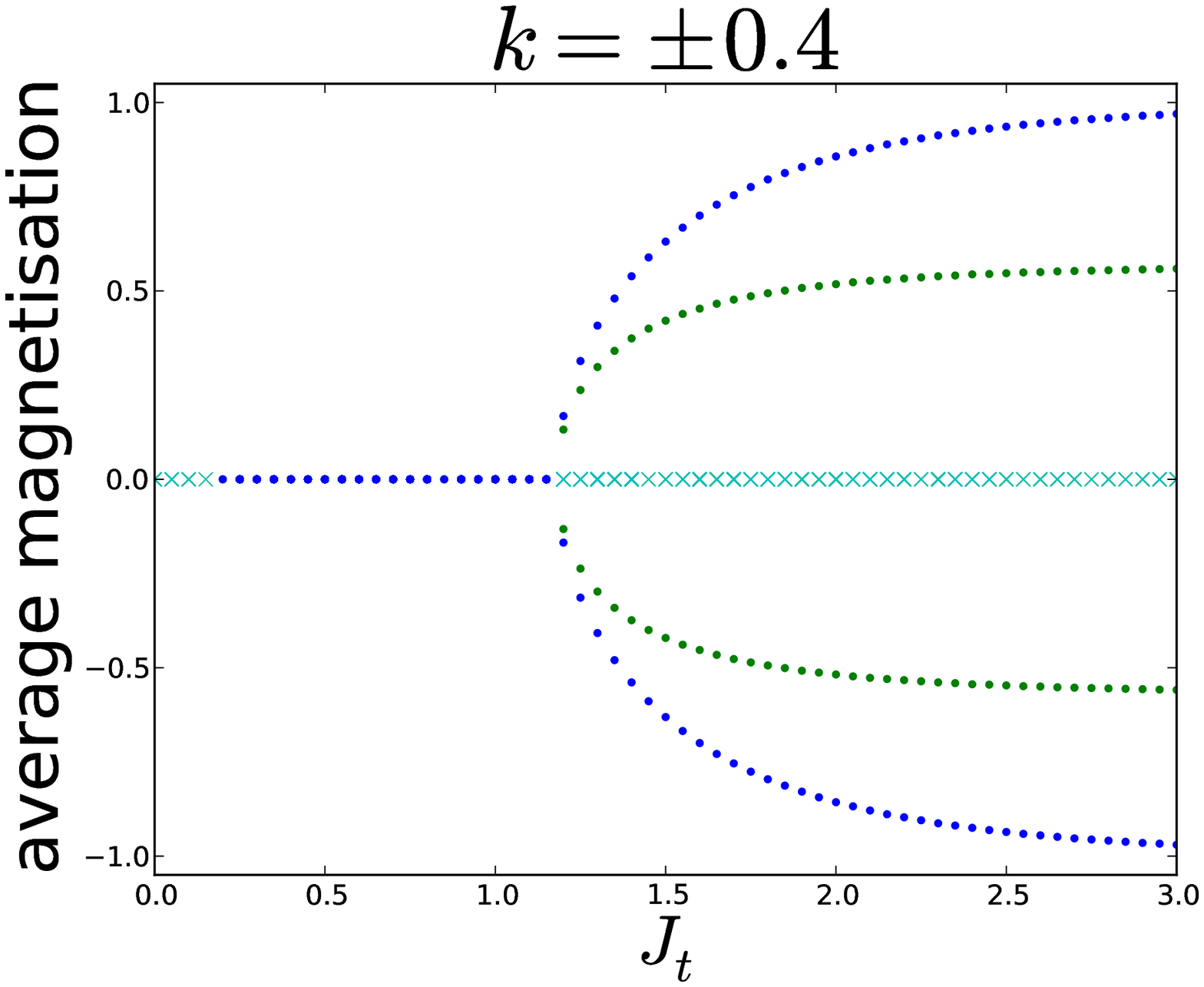}}
\subfloat[]{\includegraphics[width=0.33\textwidth]{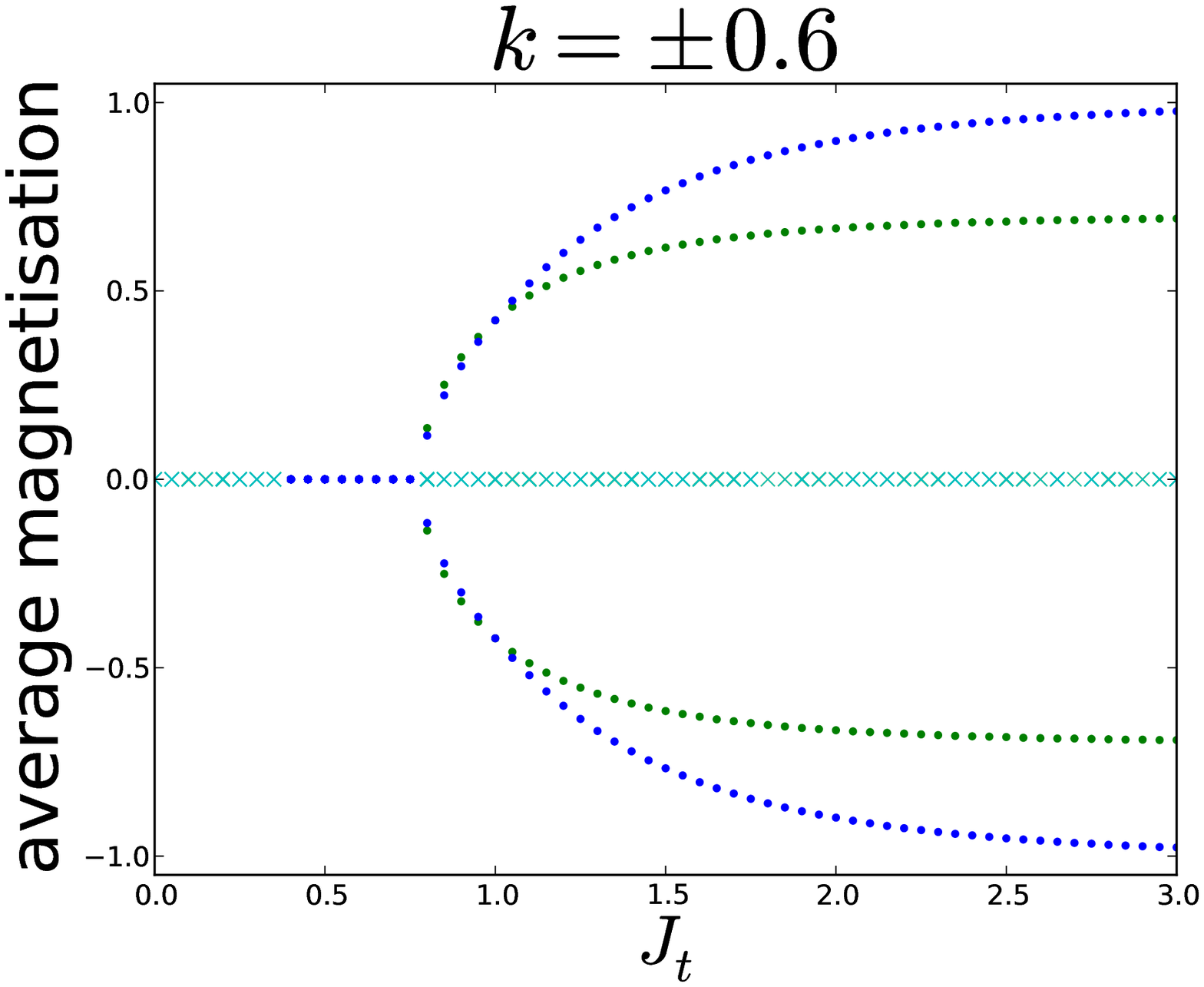}}\\
\subfloat[]{\includegraphics[width=0.33\textwidth]{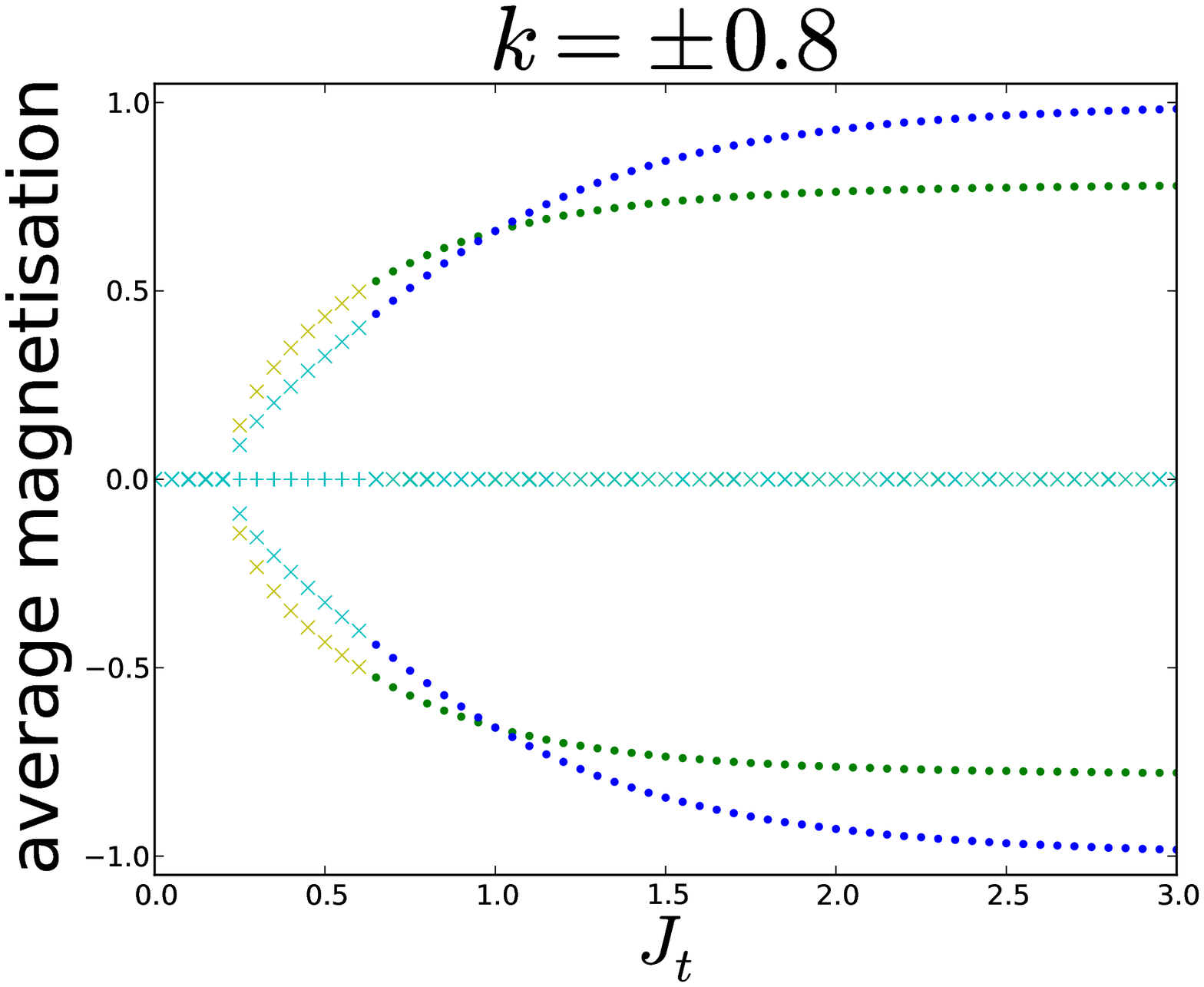}}
\subfloat[]{\includegraphics[width=0.33\textwidth]{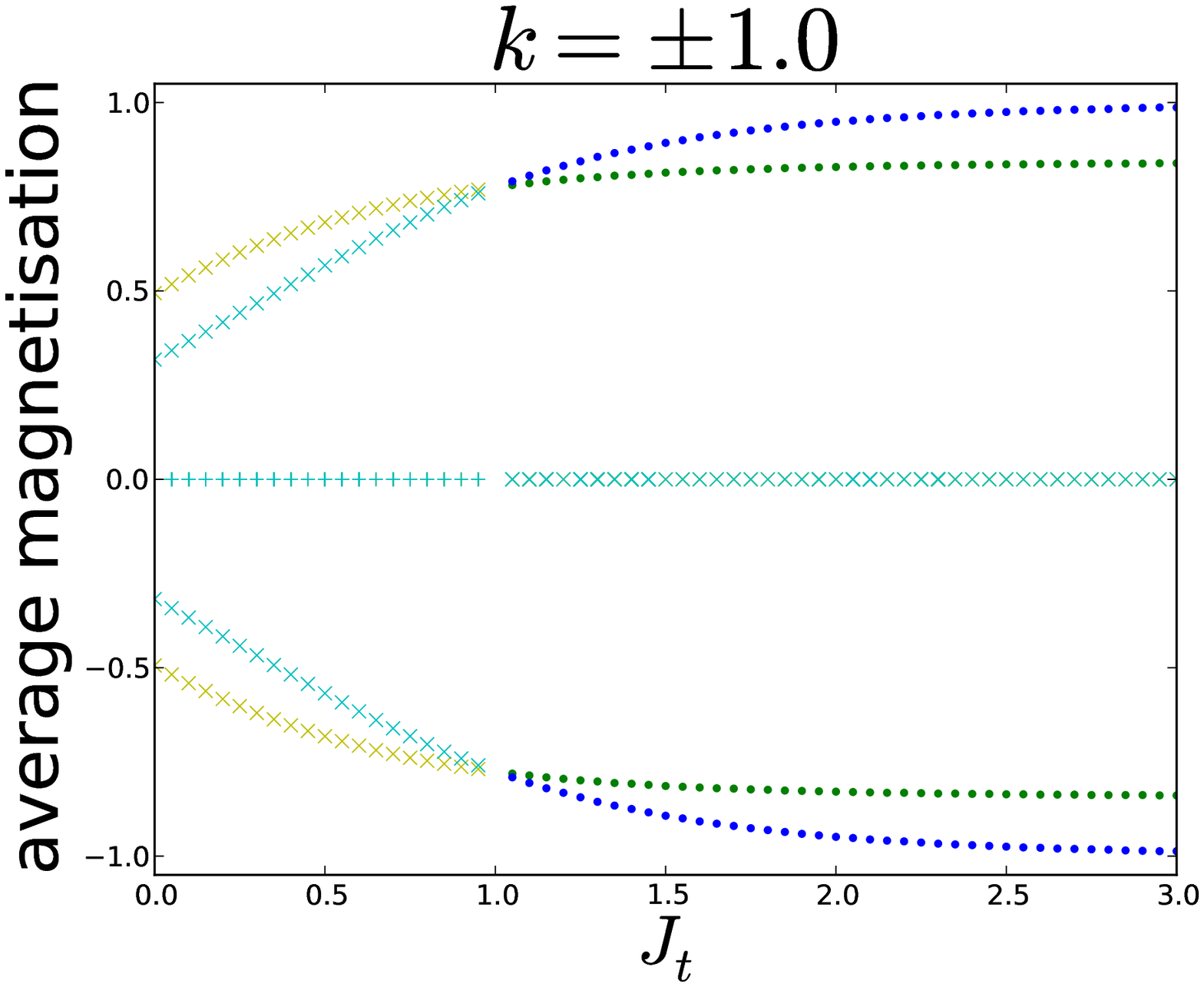}}
\subfloat[]{\includegraphics[width=0.33\textwidth]{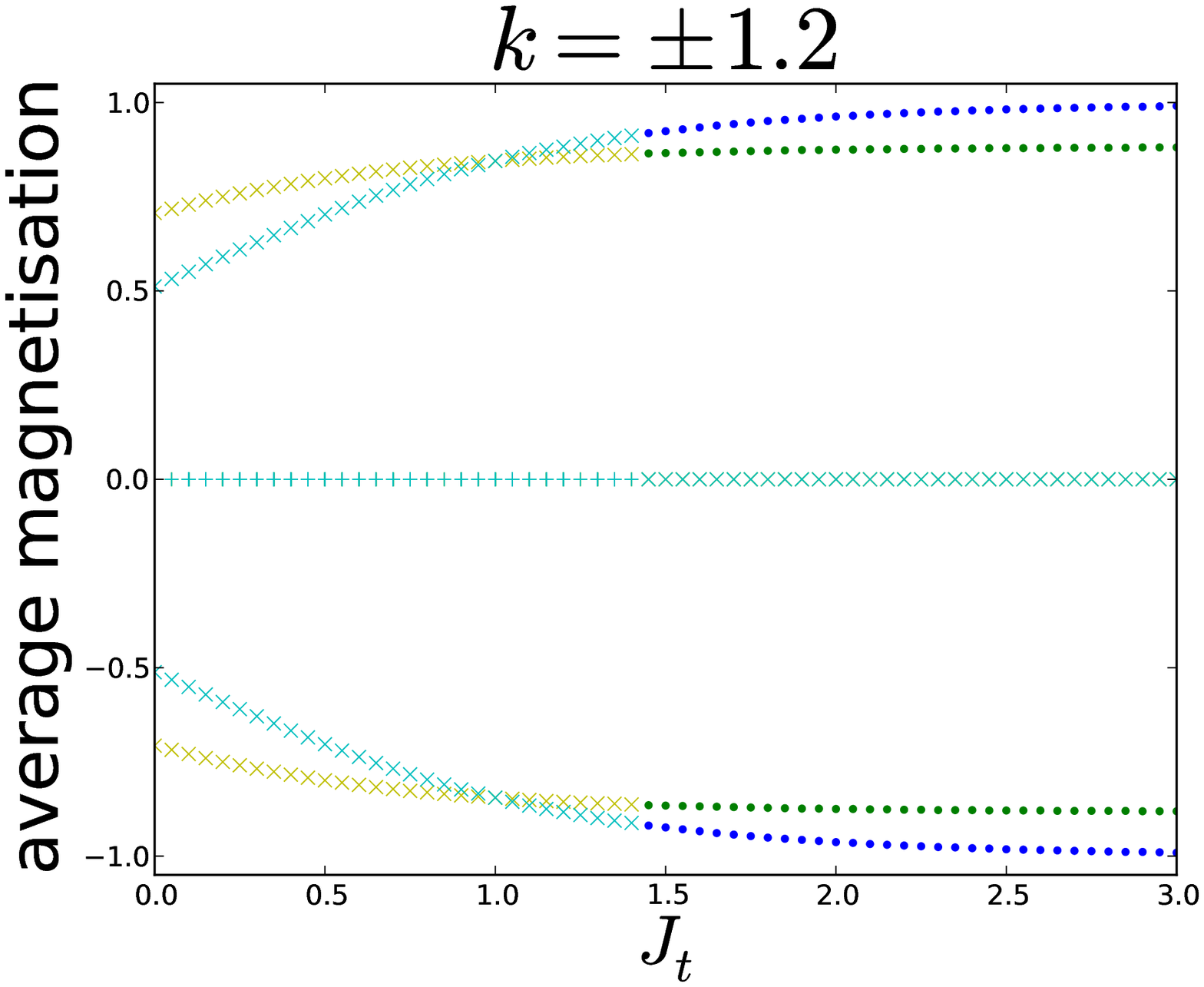}}
\caption{Dependence on intra-coupling $J_{t}$ of the numerically calculated average magnetisations $(s,t)$ for different values of the inter-coupling $k$ at high temperatures. $J_{s}=1$ and $K_{B}T=1.5$  for all plots. (a) $k=0.05$, (b) $k=0.1$, (c) $k=0.2$, (d) $k=0.3$, (e) $k=0.4$, (f) $k=0.6$, (g) $k=0.8$, (h) $k=1$ and (i) $k=1.2$. In all cases, different solutions are plotted for intra-coupling $J_{t}$ between 0.01 and 3 every 0.05. Magnetisations are plotted in green for $s$ and blue for $t$. Dark points are used for stable solutions and lighter asp ($\times$, for saddle points) or cross ($+$, for maxima) for non stable solutions.}
\label{fig:nloanakJ1}
\end{figure}

\begin{figure}
\centering
\subfloat[]{\includegraphics[width=0.33\textwidth]{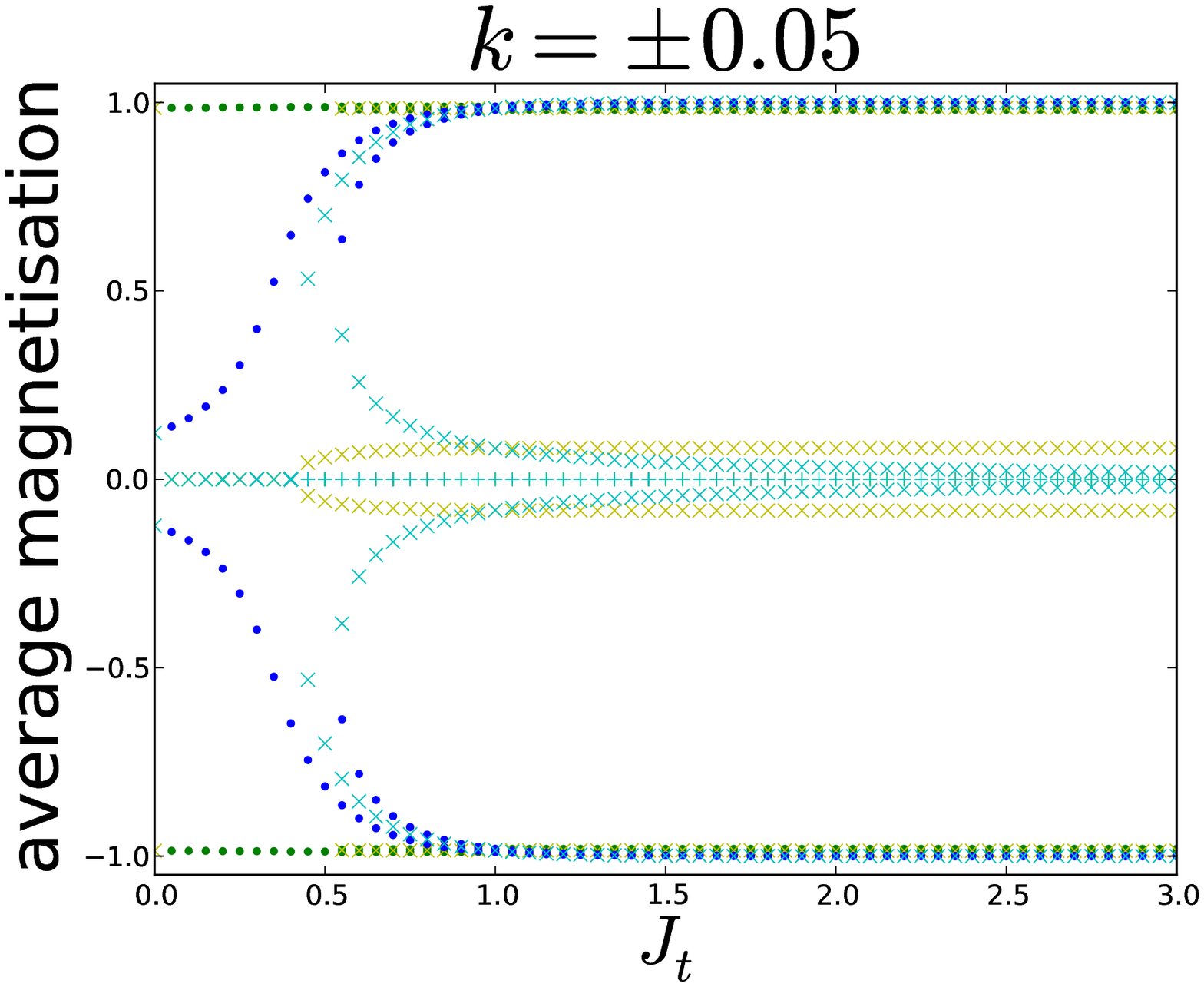}}
\subfloat[]{\includegraphics[width=0.33\textwidth]{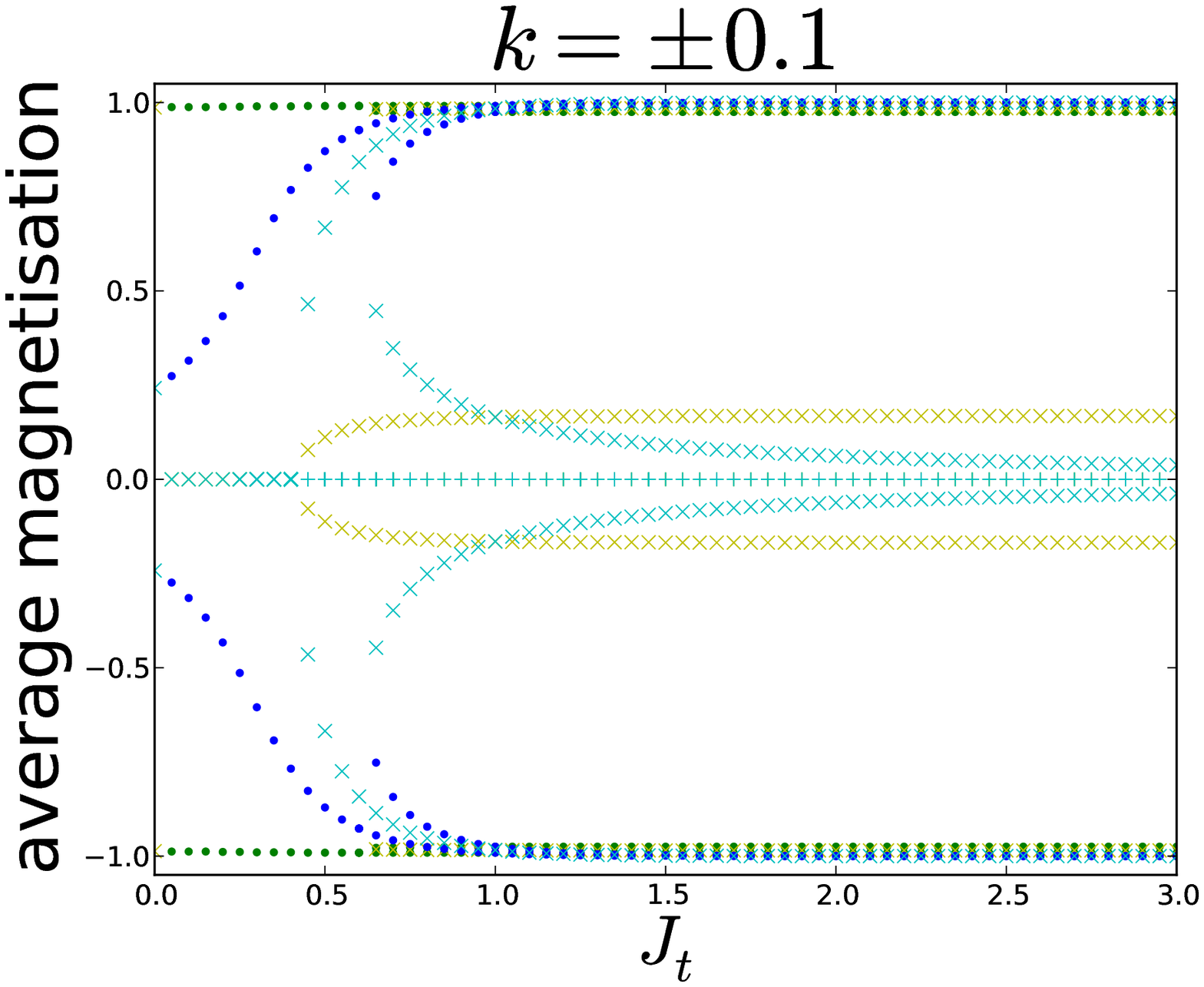}}
\subfloat[]{\includegraphics[width=0.33\textwidth]{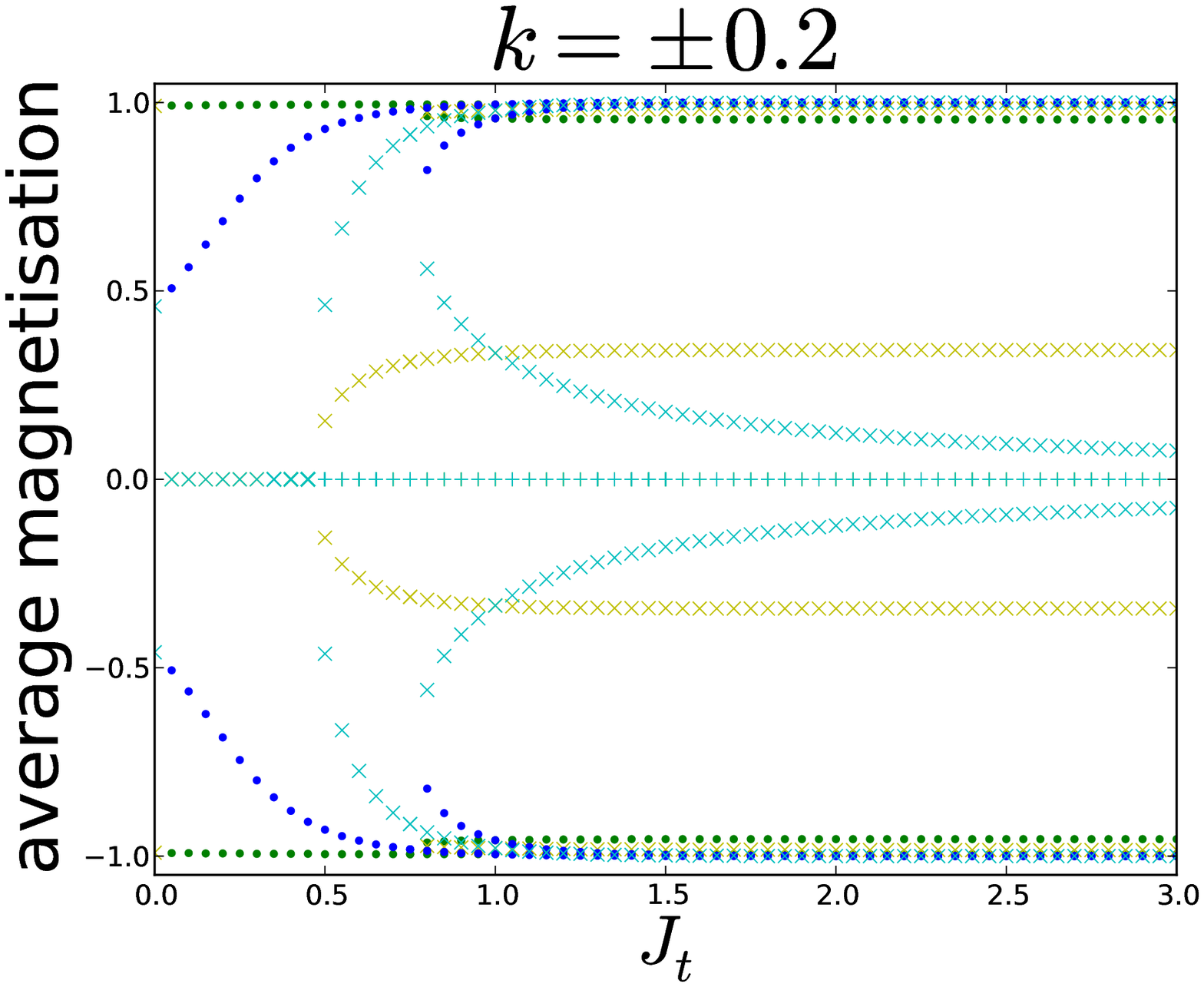}}\\
\subfloat[]{\includegraphics[width=0.33\textwidth]{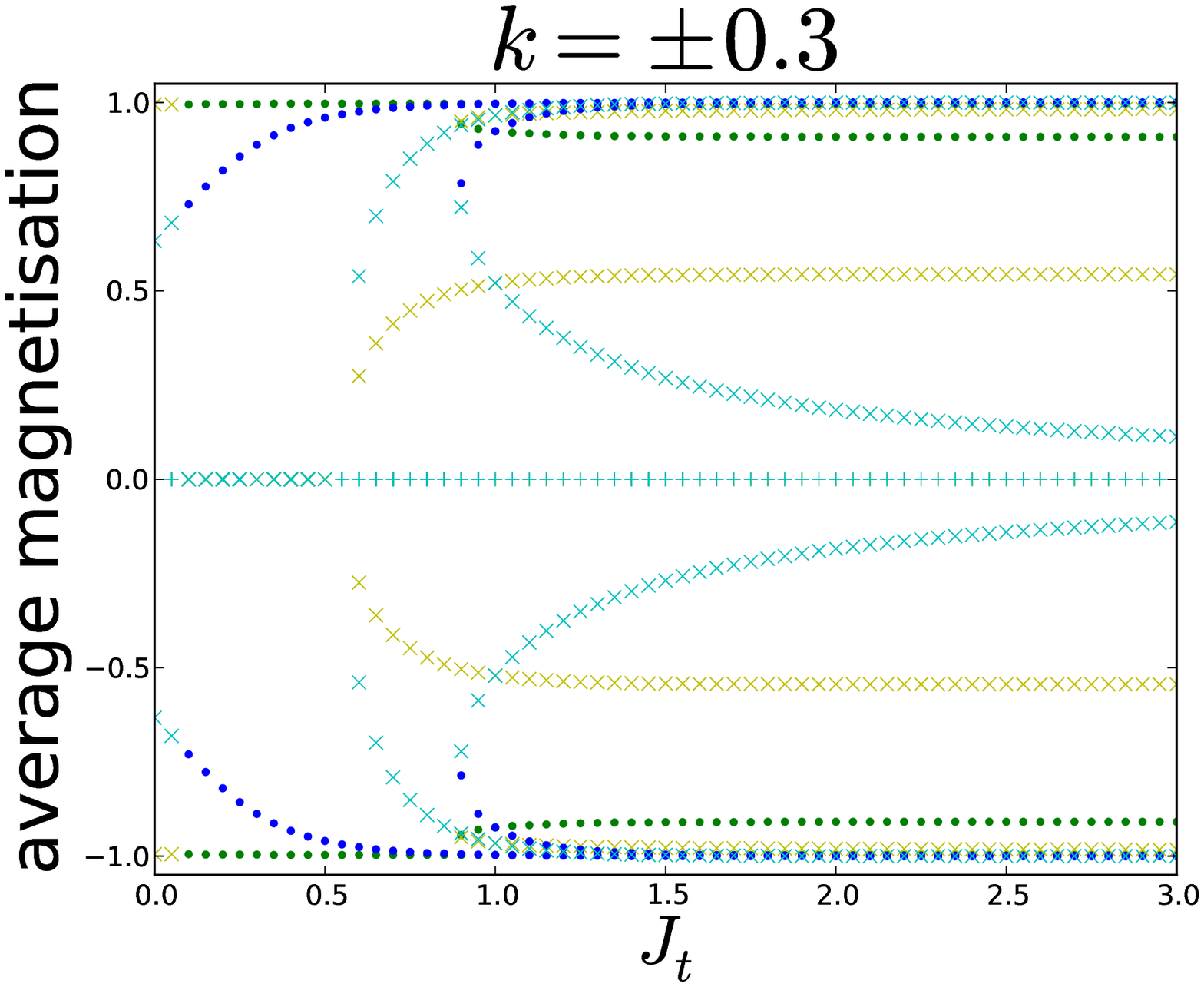}}
\subfloat[]{\includegraphics[width=0.33\textwidth]{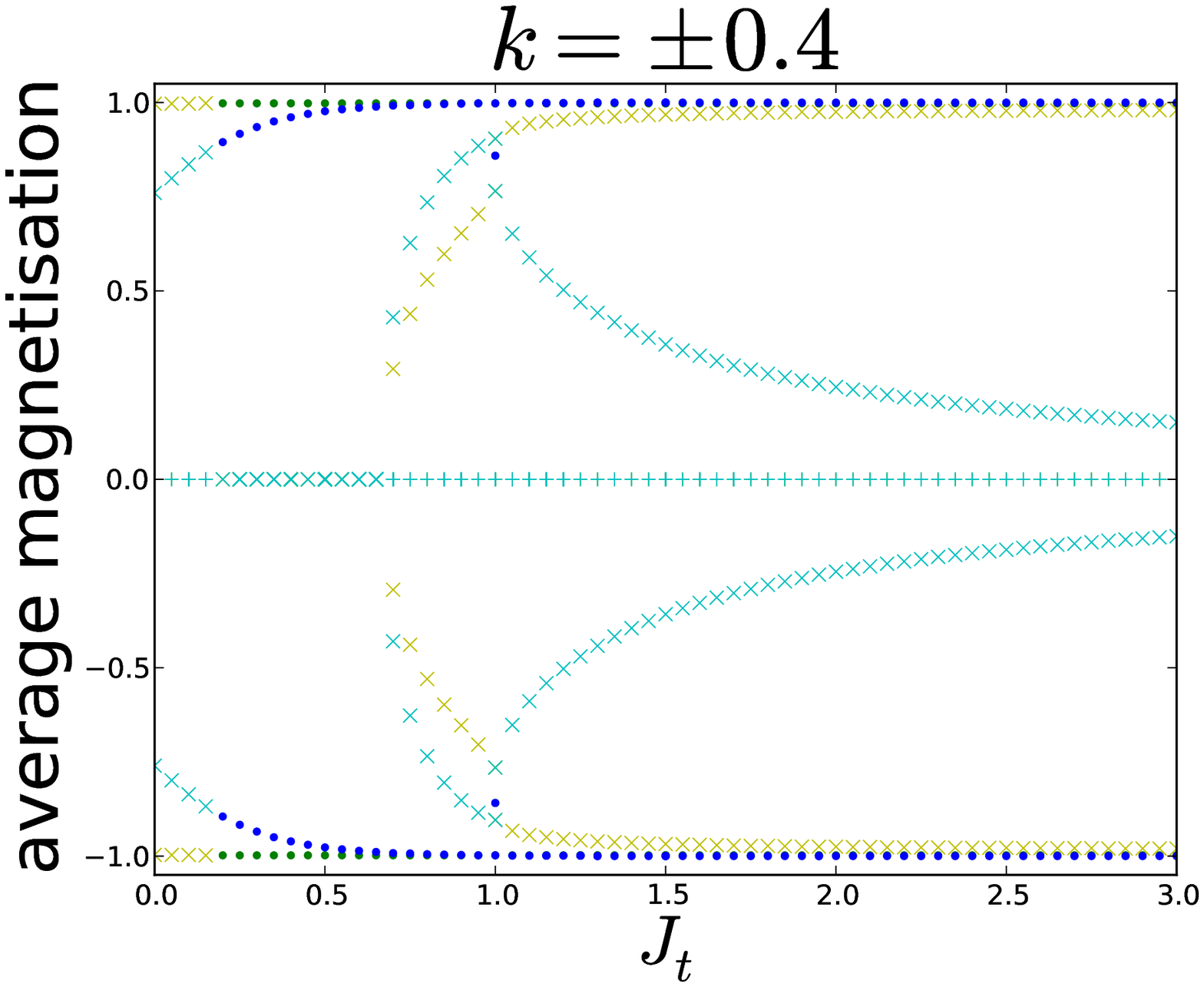}}
\subfloat[]{\includegraphics[width=0.33\textwidth]{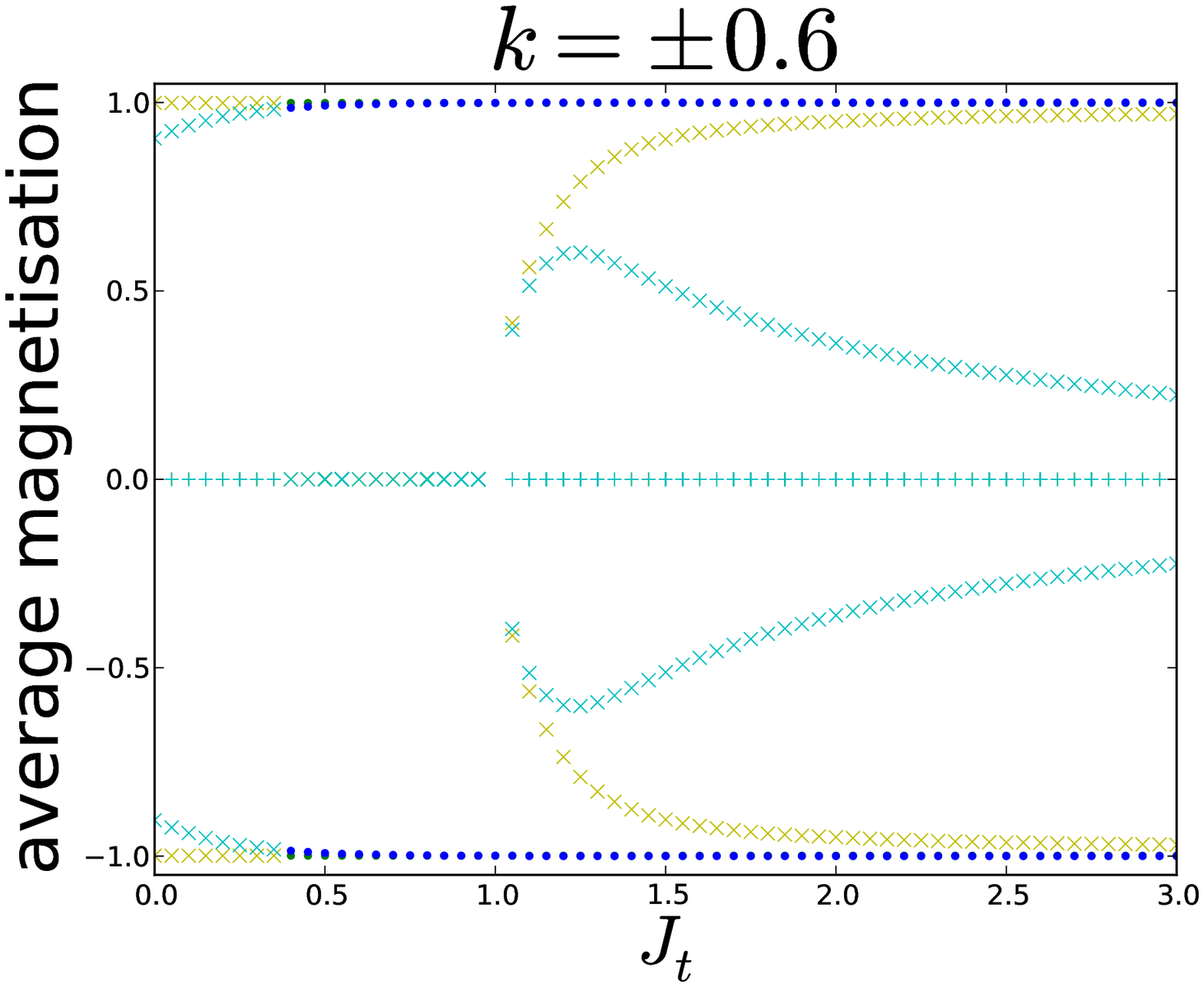}}\\
\subfloat[]{\includegraphics[width=0.33\textwidth]{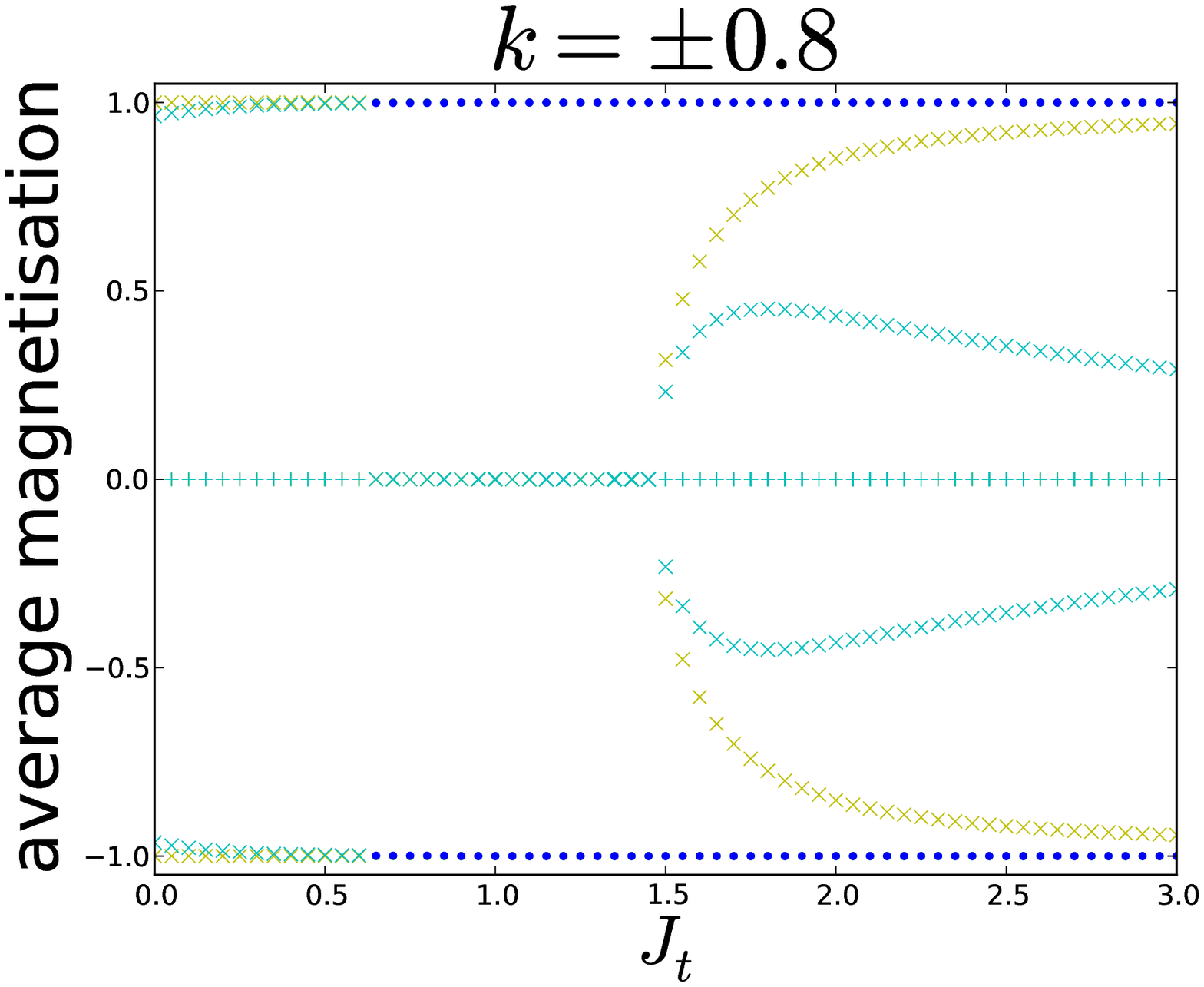}}
\subfloat[]{\includegraphics[width=0.33\textwidth]{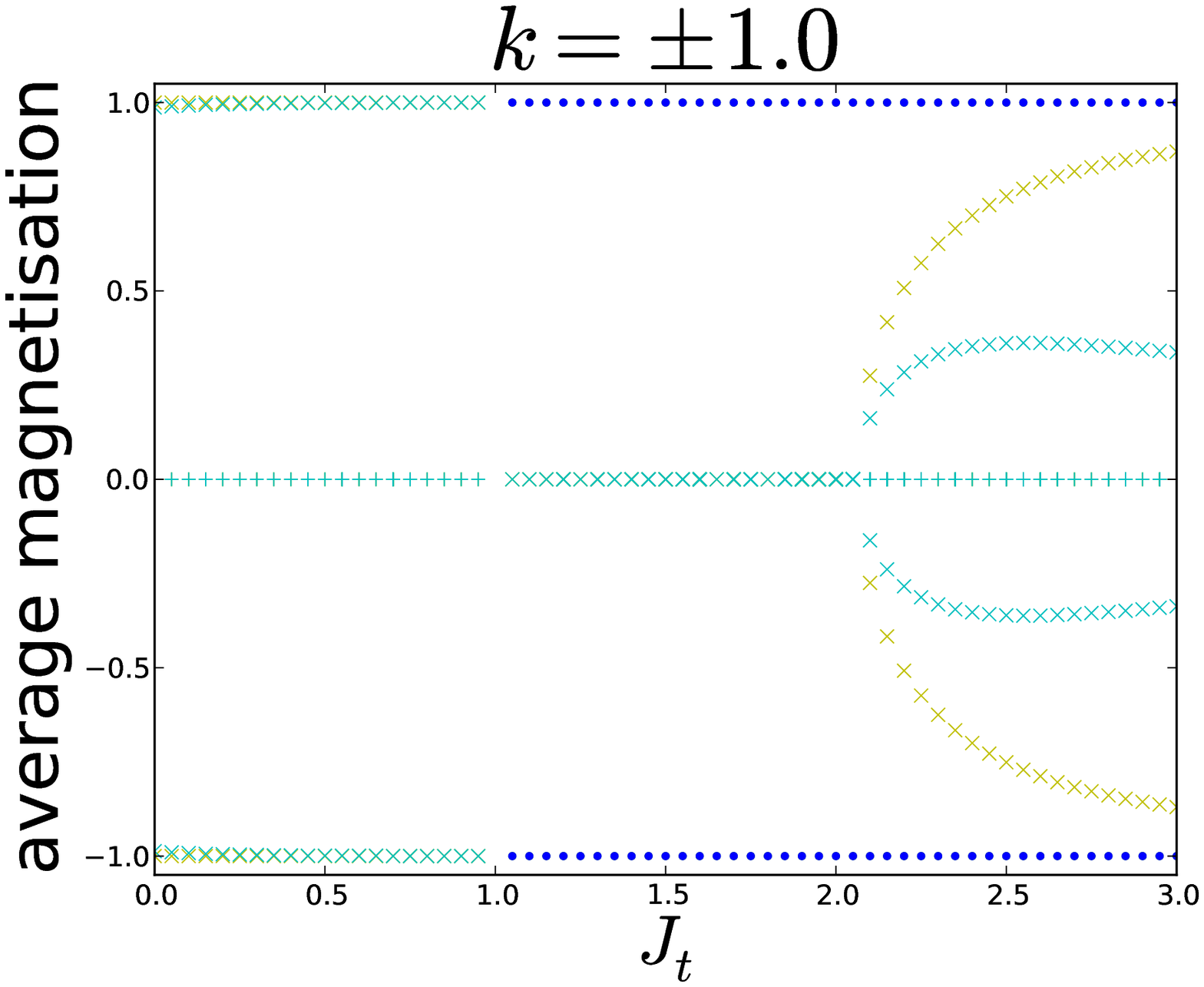}}
\subfloat[]{\includegraphics[width=0.33\textwidth]{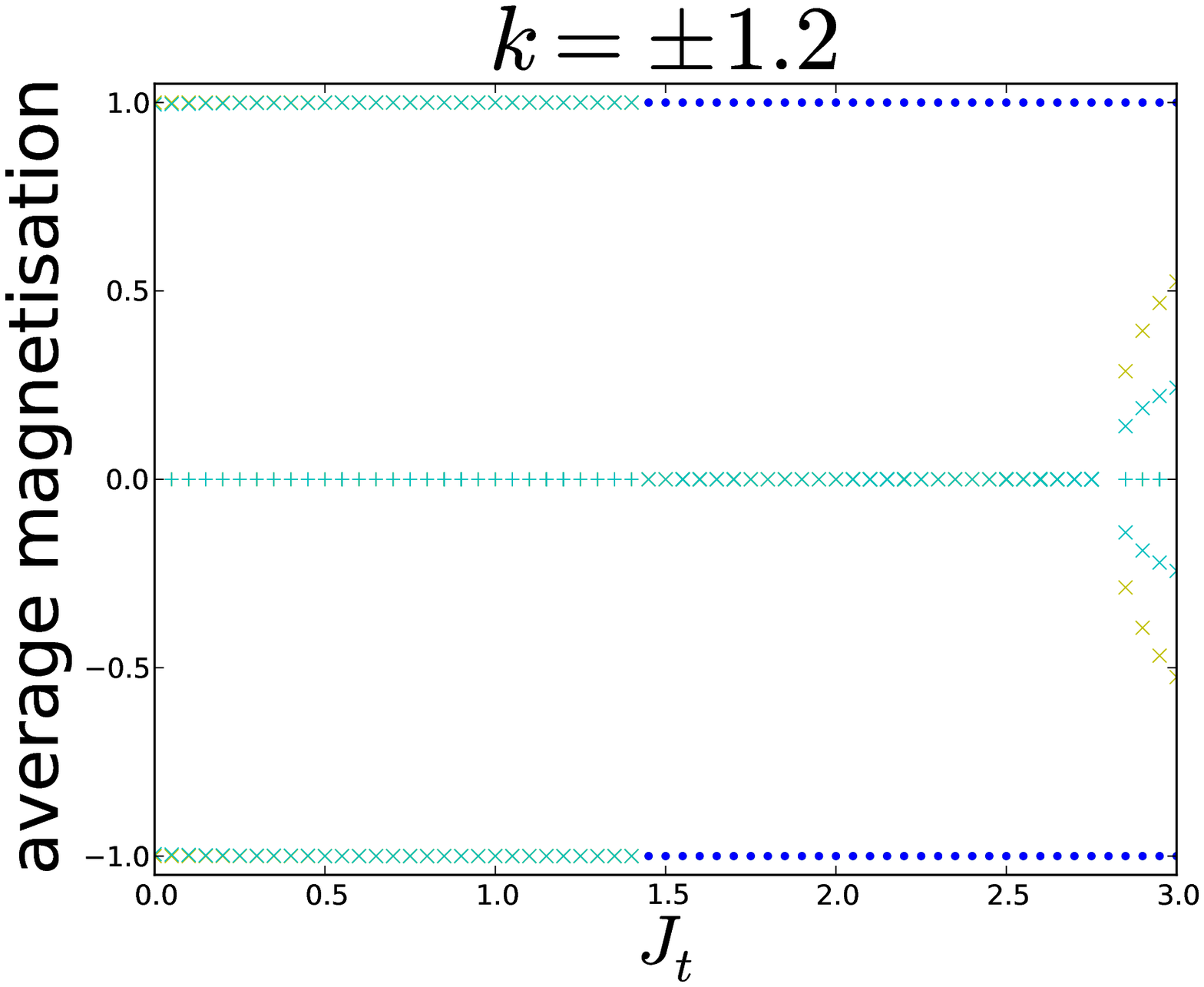}}
\caption{Dependence on intra-coupling $J_{t}$ of the numerically calculated average magnetisations $(s,t)$ for different values of the inter-coupling $k$ at low temperatures. $J_{s}=1$ and $K_{B}T=0.4$  for all plots. (a) $k=0.05$, (b) $k=0.1$, (c) $k=0.2$, (d) $k=0.3$, (e) $k=0.4$, (f) $k=0.6$, (g) $k=0.8$, (h) $k=1$ and (i) $k=1.2$. In all cases, different solutions are plotted for intra-coupling $J_{t}$ between 0.01 and 3 every 0.05. Magnetisations are plotted in green for $s$ and blue for $t$. Dark points are used for stable solutions and lighter asp ($\times$, for saddle points) or cross ($+$, for maxima) for non stable solutions.}
\label{fig:nloanakJ2}
\end{figure}

Figures \ref{fig:nloanakJ1} and \ref{fig:nloanakJ2} show how the dependence on $J_{t}$ varies as we change $|k|$ for constant high ($K_{B}T=1.5$) and low ($K_{B}T=0.4$) temperature respectively. In both cases, at low $k$ there are stable solutions for most values of $J_{t}$. The region of instability gets larger as $k$ increases (stable solutions are found only in the weak coupling regime $J_{t}>\frac{k^{2}}{J_{s}}$).

As shown in figure \ref{fig:nloanakJ1}, starting from low values of $J_{t}$, there are stable solutions for most values of $J_{t}$ . At $J_{t}^{c}$ there is a second order phase transition, and for $J>J_{t}^{c}$ the paramagnetic solution will be a saddle point and there are two ferromagnetic stable branches (figure \ref{fig:nloanakJ1} a). In this case, the ferromagnetic phase is only stable for values of $J_{t}$ significantly bigger than $J_{s}=1$, and so $t$ will be much bigger than $s$ in absolute value. As we increase the value of $|k|$ (figures \ref{fig:nloanakJ1}  b to e), no real qualitative change takes place for a while. The critical value of $J_{t}$ slowly moves towards lower values and the strong coupling regime becomes larger, and so the paramagnetic phase will be stable for smaller and smaller ranges of $J_{t}$ and the absolute values of $s$ and $t$ get more similar. When the critical value falls bellow the value $J_{s}=1$, a new small region for which $|s|>|t|$ appears (figure \ref{fig:nloanakJ1} f). It will get larger for a while as we increase $|k|$ (figure \ref{fig:nloanakJ1} g) until $J_{s}=1$ falls into the strong coupling regime and stable ferromagnetic solutions will have $|t|>|s|$ for all $J_{t}$ (figure \ref{fig:nloanakJ1} h). At the intermediate point where $J_{t}^{c}$ becomes smaller than the degenerate limit $J_{t}<\frac{k^{2}}{J_{s}}$ (figure \ref{fig:nloanakJ1} g), the paramagnetic phase disappears completely as a stable state (and so does the second order phase transition). As we now continue to increase $|k|$, the instability will grow to larger values of $J_{t}$, and $s$ and $t$ become more similar in absolute value (growing more rapidly to one). No metastable solutions appear regardless the values of the inter- and intra-couplings. 

For low temperatures, an example is shown in figure \ref{fig:nloanakJ2}. Now only ferromagnetic solutions exist (except for $k=0$ under certain circumstances), and so there is no phase transition. A {\itshape critical} value $J_{t}^{c}$ does always exist, but in this case it will represent a point where saddle type ferromagnetic solutions appear. It will move towards higher values of $J_{t}$ as we increase the value of $|k|$ and will never disappear. At low inter-couplings (figure \ref{fig:nloanakJ2} a), the behaviour is as that described for figure \ref{fig:locanaJ2} but with the system being in strong coupling regime only for very small values of $J_{t}$. As we increase $|k|$ (figures \ref{fig:nloanakJ2} b to e), the strong coupling regime gets larger, and $J_{t}^{c}$ and $J_{t}^{a}$ move to larger values\footnote{As in the previous example studying dependence on $J_{t}$ for different values of $T$ at fixed $k$, here again there seems to be a small region in $J_{t}-k$ for which metastable solutions stop existing for all $J_{t}>J_{t}^{a}$ and do so only for $J_{t}^{a'}>J_{t}>J_{t}^{a}$ with $J_{t}^{a} \approx J_{t}^{a'}$. In (figure \ref{fig:nloanakJ2} e) for example in this case, for $k=\pm 0.4$ the metastable solutions calculated exist only for $J_{t}=1$. This may be an artificial effect due to numerical calculations.}. Beyond some value they disappear (figure \ref{fig:nloanakJ2} f). From then on, increasing $|k|$ will only continue to enlarge the region in strong coupling regime, move $J_{t}^{c}$ to higher values, and make both magnetisations more and more similar to one in absolute value.\\

\section{Phase diagram for the zero field case}
\label{sec:nlophadia}
We have basically already described the phase diagram through the numerical solutions in the last section. In this section we will be summarising these results, focusing on stable solutions, and actually show some figures of how the phase diagram cross sections look like.

For completeness, we include here the expansion of the free energy density \eqref{eq:freeE} for $|s| \ll 1$ and $|t| \ll 1$ (and finite nonzero temperature):

\begin{equation}
\begin{array}{l}
f = -\frac{2}{\beta}\log2+\frac{1}{2}\left(J_{s}-\beta J_{s}^{2}-\beta k^{2}\right)s^{2}+\frac{1}{2}\left(J_{t}-\beta J_{t}^{2}-\beta k^{2}\right)t^{2}+\\
\qquad +k\left(1-\beta(J_{s}+J_{t})\right)st-\frac{3\beta^{3}}{8}\left(J_{s}^{4}+k^{4}\right)s^{4}-\frac{3\beta^{3}}{8}\left(J_{t}^{4}+k^{4}\right)t^{4}+\\
\qquad -\frac{7}{8}\beta^{3}k^{2}(J_{s}^{2}+J_{t}^{2})s^{2}t^{2}-\frac{5}{4}\beta^{3}k(J_{s}k^{2}+J_{t}^{3})st^{3}+\\
\qquad -\frac{5}{4}\beta^{3}k(J_{s}^{3}+J_{t}k^{2})s^{3}t
\end{array}
\end{equation}

In the next subsections we will be analysing our results (both numerical and the basic analytical derivations carried out) for the different cross sections of the phase space. We will be using the same parameter choice as for the examples in last section (figures showing how the dependence on one of the parameters varies when we change another of the parameters for the other two fixed).

\subsection{Two dimensional $k - T$ sections}

\begin{figure}
\centering
\includegraphics[width=\textwidth]{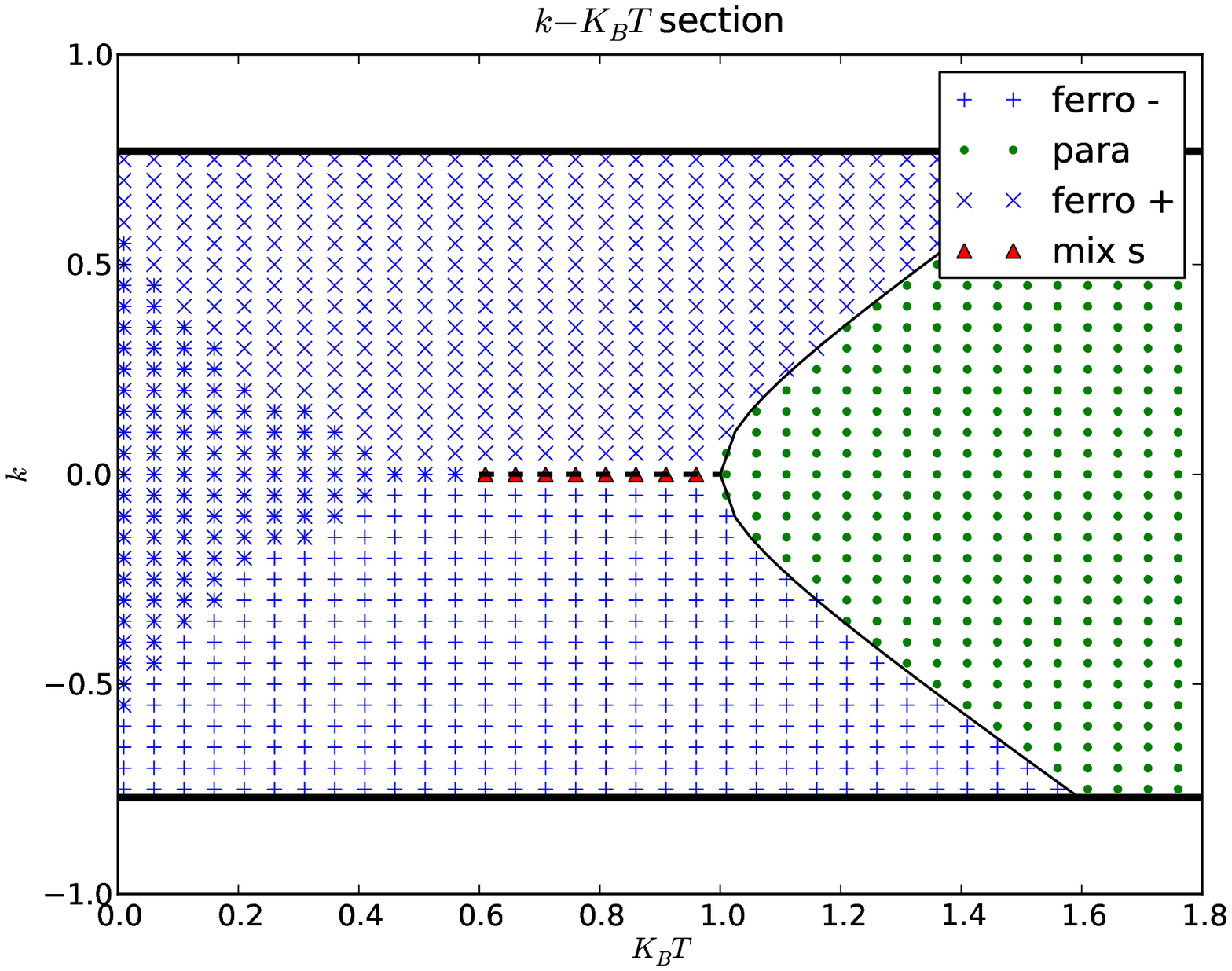}
\caption{Two dimensional $k-K_{B}T$ section of the phase diagram as given by the numerical average magnetisations used in the analysis in section \ref{sec:nlnumsol} ($J_{s}=1$ and $J_{t}=0.6$). Only stable solutions are shown. The phase associated to each solution is plotted for $K_{B}T$ between 0.05 and 1.8 and for $k$ between -1 and 1 and solutions are shown at intervals of 0.05. Green points are used for the paramagnetic phase , blue asps ({$\times$}) for ferromagnetic phases with $s$ and $t$ of the same sign and blue crosses ($+$) for ferromagnetic phases with $s$ and $t$ of opposite signs. Red triangles are used for the mixed phase where $s\neq 0$. Solid thick black lines are used for $k=\pm k_{d}$, solid thin black lines for the critical branches $k_{c}^{+}=\sqrt{J_{s}J_{t}-K_{B}T(J_{s}+J_{t})+(K_{B}T)^{2}}=\sqrt{0.6-1.6K_{B}T+(K_{B}T)^{2}}$ and $k_{c}^{-}=-\sqrt{J_{s}J_{t}-K_{B}T(J_{s}+J_{t})+(K_{B}T)^{2}}=-\sqrt{0.6-1.6K_{B}T+(K_{B}T)^{2}}$ when $K_{B}T>K_{B}T_{uc}^{s}=J_{s}=1$ and a dashed black line for the mixed phase segment $J_{t}<K_{B}T< J_{s}$.}
\label{fig:nlphadiakT}
\end{figure}

Figure \ref{fig:nlphadiakT} shows the $k-K_{B}T$ section for $J_{s}=1$ and $J_{t}=0.6$ (only stable solutions considered). Numerical solutions are represented using  green points, blue asps ({$\times$}), blue crosses ($+$) and red triangles for points of the phase diagram cross section with stable numerical solutions in the paramagnetic, ferromagnetic with same relative signs, ferromagnetic with opposite relative signs and mixed phase respectively. Besides, analytical results for the curves delimiting different regions are also plotted. Strong coupling regime is shown in white (no stable solutions found). The stable region (weak coupling regime) is centred in $k=0$ and is delimited by the line $k=\pm\sqrt{J_{s}J_{t}}=\pm 0.77=\pm k_{d}$ (drawn as a thick solid black lines in the figure).

For large enough values of $T$ only the paramagnetic phase is stable. As we move to lower temperatures, ferromagnetic phases appear at decreasing values of the temperature ($s$ and $t$ same sign for $k>0$ and $s$ and $t$ opposite sign for $k<0$). The critical curve separating ferromagnetic from paramagnetic phases is given by $k_{c}^{2}=J_{s}J_{t}-K_{B}T(J_{s}+J_{t})+(K_{B}T)^{2}=0.6-1.6K_{B}T+(K_{B}T)^{2}$  when $K_{B}T>K_{B}T_{uc}^{s}=J_{s}=1$\footnote{If we consider the whole range of temperatures, first both branches melt into the mixed phase segment between $K_{B}T=K_{B}T_{uc}^{t}=J_{t}$ and $K_{B}T=K_{B}T_{uc}^{s}=J_{s}$. Bellow $T_{uc}^{t}$,there are another two branches, symmetric to the ones drawn with respect to the $K_{B}T=\frac{K_{B}(T_{uc}^{t}+T_{uc}^{s})}{2}=0.8$ axis. These are the points where the $(0,0)$ critical point of the free energy changes from a saddle point to a maximum and new ferromagnetic saddle critical points appear.}. It is  drawn in solid black in the figure.  Temperature beyond which only the paramagnetic phase exists will be given by the intersection between these and $k=\pm k_{d}$ ($K_{B}T=J_{s}+J_{t}=1.6$). 

For $K_{B}T<K_{B}T_{uc}^{s}=J_{s}=1$ (or the analogous condition in $t$ if $J_{t}>J_{s}$,) the paramagnetic phase is no longer stable. At the $k=0$ axis, for temperatures bellow the latter but above the other uncoupled critical temperature $K_{B}T_{uc}^{t}=J_{t}=0.6$, stable solutions are in the mixed phase with $s\neq0$ ($J_{s}>J_{t}$). This segment is drawn as a dashed black line in the figure. Bellow this temperature, the region where there are metastable states begins, extending to higher values of $k$ as the temperature decreases.

For lower values of $J_{t}$ the mixed phase $k=0$ line  extends to lower temperatures as $T_{uc}^{t}$ decreases and the metastable region shrinks. For low enough values of $J_{t}$, the region with metastable states no longer exists at all. The overall region of stability gets smaller as we decrease $J_{t}$.

For larger values of $J_{t}$, the mixed phase $k=0$ lines extends this time to higher temperatures as $T_{uc}^{s}$ increases, the metastable region grows and the paramagnetic region is shifted towards higher temperatures. For $J_{s}=J_{t}$ the mixed phase line  disappears,  and for $J_{t}>J_{s}$  mixed phases present will be those  with $t=0$ (with the line extending between $\frac{J_{s}}{K_{B}}=T_{uc}^{s}<T<T_{uc}^{t}=\frac{J_{t}}{K_{B}}$). The metastable region continues to grow and the paramagnetic phase to move to higher and higher temperatures. The larger $J_{t}$, the broader the stability region (weak coupling regime).

\subsection{Two dimensional $J_{t} - T$ sections}

\begin{figure}
\centering
\includegraphics[width=\textwidth]{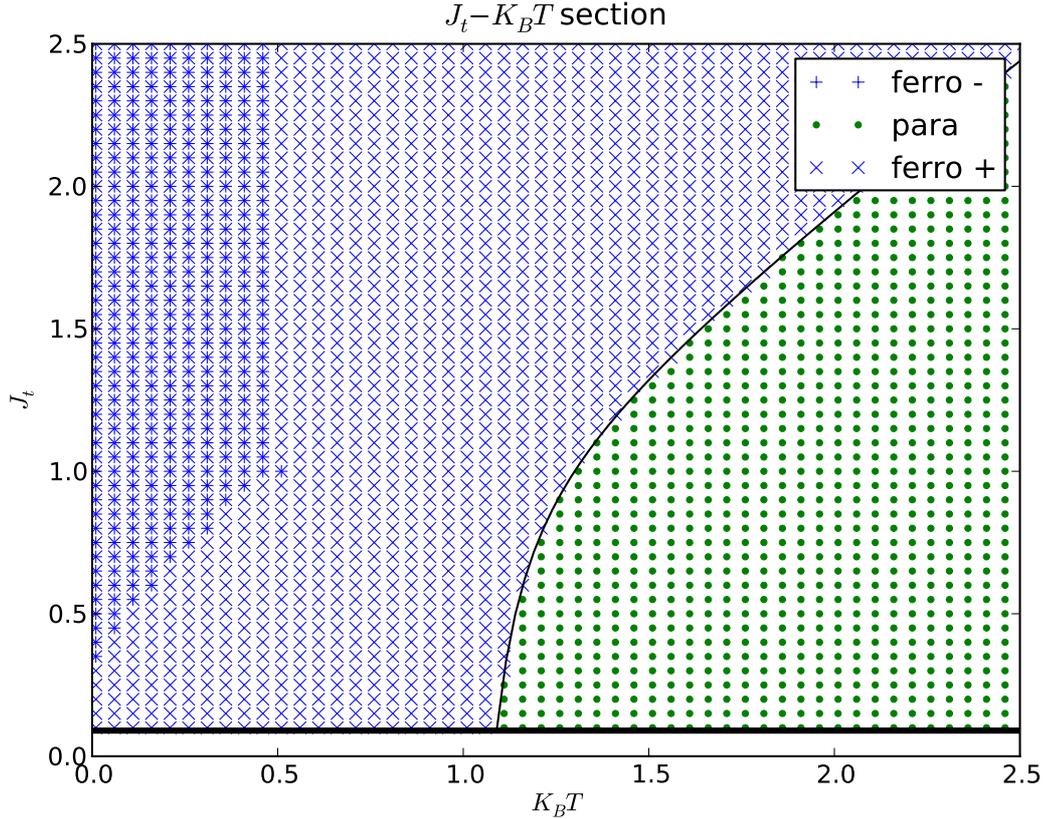}
\caption{Two dimensional $J_{t}-K_{B}T$ section of the phase diagram as given by the numerical average magnetisations used in the analysis in section \ref{sec:nlnumsol} ($J_{s}=1$ and $k=0.3$). Only stable solutions are shown. The phase associated to each solution is plotted for $K_{B}T$ and $J_{t}$ between 0.05 and 2.5 and solutions are shown at intervals of 0.05. Green points are used for the paramagnetic phase , blue asps ({$\times$}) for ferromagnetic phases with $s$ and $t$ of the same sign and blue crosses ($+$) for ferromagnetic phases with $s$ and $t$ of opposite signs. Red triangles are used for the mixed phase where $s\neq 0$. Solid thick black lines are used for $k=k_{d}$ and  solid thin black lines for the critical curve $J_{t}^{c}=\frac{k^{2}+J_{s}K_{B}T-(K_{B}T)^{2}}{J_{s}-K_{B}T}=\frac{0.09+K_{B}T-(K_{B}T)^{2}}{1-K_{B}T}$ for $K_{B}T>K_{B}T_{cs0}=J_{s}=1$ ($J_{s}>J_{t}$).}
\label{fig:nlphadiaJT}
\end{figure}

Figure \ref{fig:nlphadiaJT} shows the $J_{t}-K_{B}T$ section for $J_{s}=1$ and $k=0.3$ (only stable solutions). The same representation is used as for the $k-K_{B}T$ figure (figure \ref{fig:nlphadiakT}) discussed in last subsection both for numerical solutions and curves and lines delimiting regions. The region where stable solutions exist is $J_{t}>\frac{k^{2}}{J_{s}}=0.09=J_{t}^{d}$.

The critical curve separating ferromagnetic from paramagnetic phases is given by \begin{equation}
J_{t}^{c}=\frac{k^{2}+J_{s}K_{B}T-(K_{B}T)^{2}}{J_{s}-K_{B}T}=\frac{0.09+K_{B}T-(K_{B}T)^{2}}{1-K_{B}T}
\end{equation}
\noindent for $K_{B}T>K_{B}T_{uc}^{s}=J_{s}=1$ ($J_{s}>J_{t}$)\footnote{The other branch of the function (for $K_{B}T<1$) is symmetric to the one shown with respect to the centre (1,1).  This branch delimits the region where saddle ferromagnetic solutions appear and the paramagnetic phase changes from saddle point to maximum}. It intersects $J_{t}=J_{t}^{d}=0.09$ at $J_{t}=1.09$. So the paramagnetic phase is the only stable solution for large enough temperatures, that decrease as $J_{t}$ decreases, and for $K_{B}T<1.09$ only ferromagnetic phases exist.

 Metastable estates appear above a certain value of $J_{t}$, extending to higher values of the temperature as we raise $J_{t}$, until $J_{t}=J_{s}=1$. From $J_{t}>J_{s}$, the paramagnetic region has a constant width. Just before this region becomes constant, for $K_{B}T=1$ there is only one value of $J_{t}=1$ where there are metastable states. This can be a problem associated to the calculation of numerical solutions and should be investigated in further detail\footnote{The extent of this {\itshape diffuse} region both in $T$ and $J_{t}$ depends on the particular choice of parameter $k$.}. 

There are no mixed phases ($k\neq0$). For $k=-0.3$ the situation will be identical, but then ferromagnetic states occupying most of the section are of the type with $s$ and $t$ with opposite relative signs, and states with both of them of the same sign are only present in the metastable region.

For lower (positive) values of $k$, the metastable region grows, expanding towards lower values of $J_{t}$ and becoming broader in $K_{B}T$. The overall region of stability gets larger and the paramagnetic phase expands towards lower temperatures. 

For larger (positive) values of $k$, the metastable region shrinks as it retreats towards larger values of $J_{t}$ and narrows in $K_{B}T$. For high enough values, the region with metastable states no longer exists at all. The paramagnetic region is shifted towards higher temperatures. The larger $|k|$ is, the larger the instability region (strong coupling regime) at low $J_{t}$.

\subsection{Two dimensional $J_{t} - k$ sections}

\begin{figure}
\centering
\includegraphics[width=\textwidth]{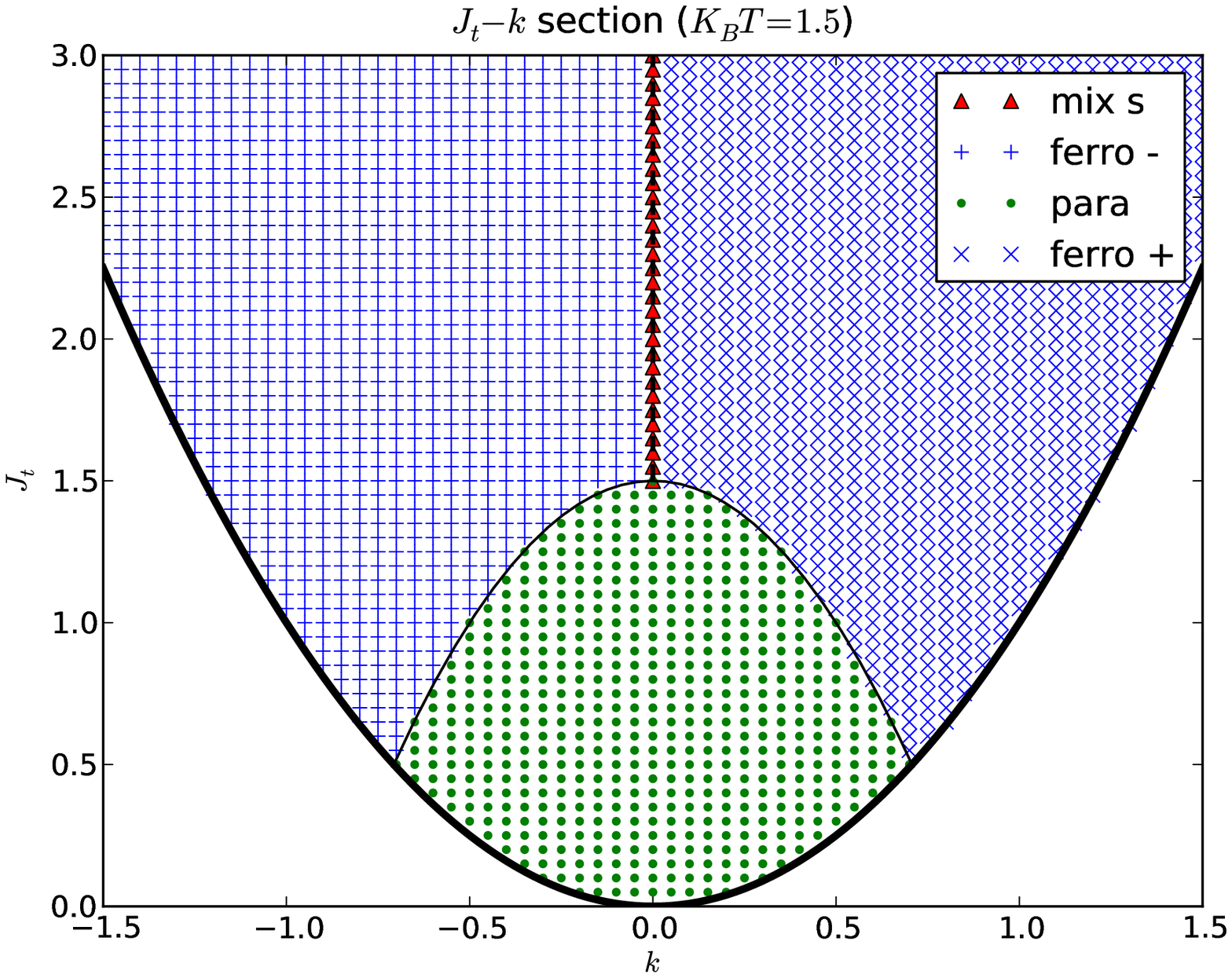}
\caption{Two dimensional $J_{t}-k$ sections of the phase diagram as given by the numerical average magnetisations for $J_{s}=1$ and $K_{B}T=1.5$. Only stable solutions are shown. The phase associated to each solution is plotted for $k$ between -1.5 and 1.5 and $J_{t}$ between 0.01 and 3  at intervals of 0.05. Green points are used for the paramagnetic phase, blue asps ({$\times$}) for ferromagnetic phases with $s$ and $t$ of the same sign and blue crosses ($+$) for ferromagnetic phases with $s$ and $t$ of opposite signs. Red triangles are used for the mixed phase where $s\neq 0$. Solid thick black lines are used for $J_{t}=J_{t}^{d}=\frac{k^{2}}{J_{s}}=k^{2}$, solid thin black lines for the critical curve $J_{t}=J_{t}^{c}=\frac{k^{2}+J_{s}K_{B}T-(K_{B}T)^{2}}{J_{s}-K_{B}T}=-2k^{2}+1.5$ and a dashed black line for the mixed phase segment $J_{t}>J_{uc}^{t}=K_{B}T=1.5$.}
\label{fig:nlophadiaJk1}
\end{figure}

\begin{figure}
\centering
\includegraphics[width=\textwidth]{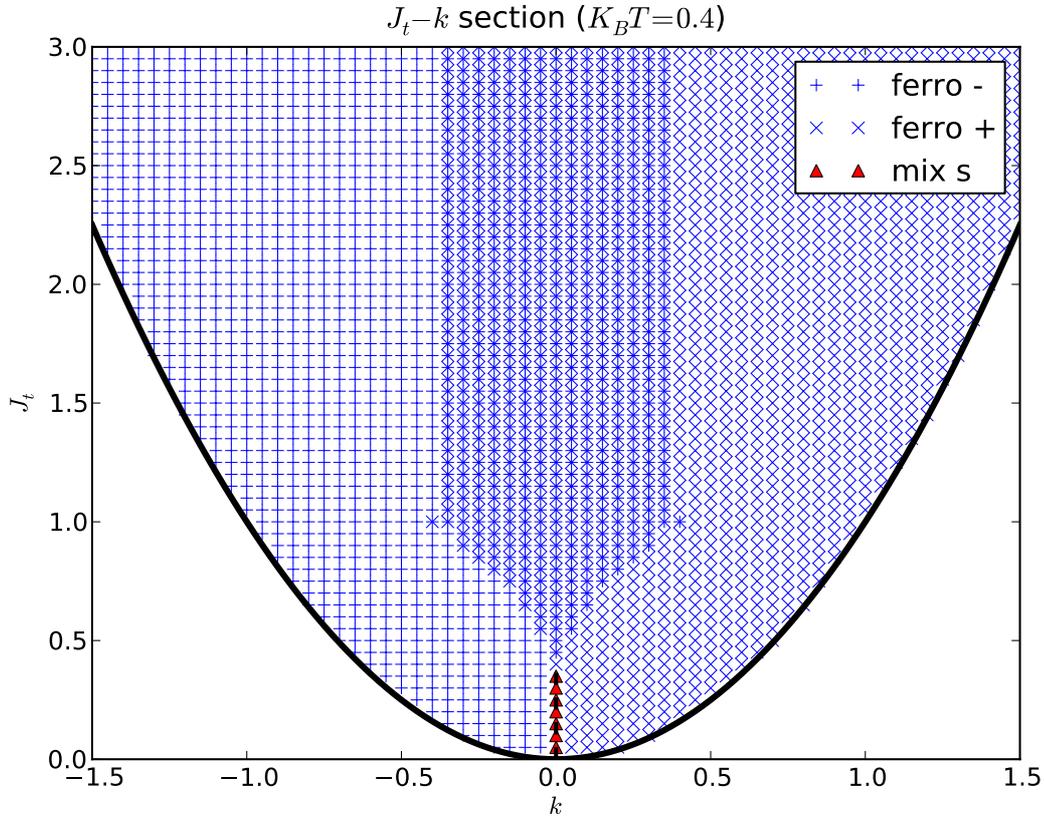}
\caption{Two dimensional $J_{t}-k$ sections of the phase diagram as given by the numerical average magnetisations for $J_{s}=1$ and  $K_{B}T=0.4$. Only stable solutions are shown. The phase associated to each solution is plotted for $k$ between -1.5 and 1.5 and $J_{t}$ between 0.01 and 3  at intervals of 0.05. Green points are used for the paramagnetic phase, blue asps ({$\times$}) for ferromagnetic phases with $s$ and $t$ of the same sign and blue crosses ($+$) for ferromagnetic phases with $s$ and $t$ of opposite signs. Red triangles are used for the mixed phase where $s\neq 0$. Solid thick black lines are used for $J_{t}=J_{t}^{d}=\frac{k^{2}}{J_{s}}=k^{2}$ and a dashed black line for the mixed phase segment $J_{t}<J_{uc}^{t}=K_{B}T=0.4$.}
\label{fig:nlophadiaJk2}
\end{figure}

Figures \ref{fig:nlophadiaJk1} and \ref{fig:nlophadiaJk1} show $J_{t}-k$ sections for $J_{s}=1$ and $K_{B}T=1.5$ (figure \ref{fig:nlophadiaJk1}) and for $J_{s}=1$ and $K_{B}T=0.4$ (figure \ref{fig:nlophadiaJk2}). We again use the same graphical conventions as for the previous cross sections studied. In both of them, the region where stable solutions exists is $J_{t}>\frac{k^{2}}{J_{s}}=k^{2}$.

For $K_{B}T=1.5$ (figure \ref{fig:nlophadiaJk1}), the critical curve separating ferromagnetic from paramagnetic phases is given by $J_{t}=J_{t}^{c}=\frac{k^{2}+J_{s}K_{B}T-(K_{B}T)^{2}}{J_{s}-K_{B}T}=-2k^{2}+1.5$. There are mixed phases ($s\neq0$) for $k=0$ and $J_{t}>J_{uc}^{t}=K_{B}T=1.5$ and there are no metastable solutions. The critical and degenerate curves intersect at $k=\pm 0.71$ ($J_{t}=0.50$). Therefore there are only ferromagnetic stable solutions for $|k|<0.71$ and only paramagnetic stable solutions for $J_{t}<0.50$. 

For $K_{B}T=0.4$ (figure \ref{fig:nlophadiaJk2}), the critical curve separating ferromagnetic from paramagnetic phases is given by $J_{t}=J_{t}^{c}=\frac{k^{2}+J_{s}K_{B}T-(K_{B}T)^{2}}{J_{s}-K_{B}T}=1.67k^{2}+0.4$. There are mixed phases ($s\neq0$) for $k=0$ and $J_{t}<J_{uc}^{t}=K_{B}T=0.4$. The metastable regions exists for $J_{t}>0.4$, broadens around $k=0$ as $J_{t}$ gets higher for $J_{t}<J_{s}=1$, and is constant for $J_{t}>J_{s}=1$. As in the $k-K_{B}T$ section case, just before this region becomes constant, for $|k|=0.4$ there is only one value $J_{t}=1$ where there are metastable states. As discussed before, this can be a problem associated to the calculation of numerical solutions\footnote{As for the $k-K_{BT}$ section, the extent of this {\itshape diffuse} region both in $k$ and $J_{t}$ depends on the particular choice of the temperature $K_{B}T$.}. 

For temperatures higher than the ones considered above, the paramagnetic phase will grow, extending to higher values of $J_{t}$. Note that the general stability region (weak coupling regime) does not vary when we change the temperature.

For $K_{B}T<1.5$, the paramagnetic phase shrinks to lower values of $J_{t}$ until it disappears, and we will then have the two possible ferromagnetic phases separated by mixed phases for $k=0$. At low enough temperatures,  the metastable region appears and we recover the situation depicted in figure \ref{fig:nlophadiaJk2} for $K_{B}T=0.4$. If we move to even lower temperatures the metastable region grows towards lower values of $J_{t}$ and the mixed segment shrinks.

\section{Socioeconomic interpretation}
\label{sec:nlosoc}

How can we translate this phase diagram to something more comprehensive in the context of discrete choice problems? It is useful to study what changes the coupling introduces with respect to the uncoupled model from one of the spin type's (group) perspective. That is, what are the differences that we can expect in the average decision made by group A when we add interaction with group B? Probably the most natural use of this model is for the coupling representing some social influence interaction and so we will be using this interpretation (we can consider utility terms arising from the group to group interaction as part of the social deterministic utility). We will be thus considering what happens to the group of choice moment of type $t$ as compared to the Ising model described in \ref{sec:sinisi}. Recall that we have studied in detail the $h=0$ case, so it is to the zero field Ising model that we should compare our results.

We will only be referring to the weak coupling regime as no stable states exist outside it. It is interesting to note that in our case, this accounts to assuming that social influence is stronger within both groups than that of one group over the other, which seems pretty natural if the groups are chosen appropriately\footnote{As noted before, this is not, strictly speaking, a coupled model. It is thus the definition of the $J_{ij}$ for the aggregate problem that causes the instability regime. The physical term is \emph{frustration}, a term that for once seems to fit more or less naturally in the socioeconomic context.}. The model does give  then a description of the problem when it is ({\itshape sociologically speaking}) well defined.

Note that this comparison is implicit in the cross sections of the phase diagram described in last section. For figures \ref{fig:nlphadiakT}, \ref{fig:nlophadiaJk1} and \ref{fig:nlophadiaJk2} the uncoupled case is described by the $k=0$ axis. Note here when there are four possible equilibria, these are all on equal grounds (no metastability). If we are only concerned with one of the groups, they are equivalent to only two different solutions (there will always be two solutions with equal $t$, one associated to $s$ and one to $-s$), the two symmetric solutions in the ferromagnetic phase discussed in \ref{sec:sinisi}. Figure \ref{fig:nlphadiaJT} shows the only cross section both the coupled and uncoupled model have in common. In the case of the uncoupled model, the critical curve would be the line $J_{t}=K_{B}T$ and the region of metastability would extend to all the ferromagnetic region and is in fact not metastable anymore (they are the same equilibria described above for the $k$ axis of the other figures).

In general, we can say that the group is shifted to higher consensus (whether in one direction or the other) by the introduction of the interdependence with another group. The paramagnetic phase (completely random binary choice) is shifted to higher values of the temperature (statistical fluctuations). There are no mixed phases so either there is some trend on both groups or there is none. From the point of view of only one group, both ferromagnetic phases are equivalent (two equilibria with $t=\pm m$). In the region of metastability, there will be two additional, though less probable, states with $t=\pm m'$ (of opposite sign to $m$). The existence of regions where four equilibria are possible (as opposed to two in the single Ising model) is, together with the shift of the paramagnetic phase, the qualitative features introduced in the model by the coupling. Note that now there can be hysteresis even when $h=0$. In the metastability region, there will be two possible equilibria, one of low and one of high demand or acceptance.

It is in general a similar picture to that of the nonzero field Ising model. Consensus is favoured and first order transitions and multiple regime exists when the symmetry is {\itshape not completely broken} (low values of the coupling at low temperatures in this case). The additional unbroken symmetry ($h=0$), makes the solutions appear in pairs, and so the sign of the magnetisation is not determined. Further more, in this picture, there is still margin for the paramagnetic phase at high temperatures. The metastability region and associated possibility of hysteresis in the low coupling region do give rise to interesting interpretations when two groups are {\itshape building confidence or influence on each other} (or loosing it with opposite reasoning). \emph{Previous states of confrontation, dislike or negative influence between two groups can prevent them from aligning in their decision making even when they have more to gain from it}. At least in the absence of deterministic private utilities. Again there is reason to be optimistic: if both groups persevere in improving their relations or image, there will be a drastic change. Groups that tended to have opposed average opinions will suddenly shift to the opposite tendency. Groups that tended to align their decisions moved by the {\itshape good old past} will {\itshape realise} the current situation has changed and abruptly change to a tendency to disagree with the other group.

How can we expect this picture to be modified by the introduction of private utilities (external fields in a model of magnetism)? In general, this will determine the sign of the average choice (magnetisation) and break the equilibria degeneracy, shifting the system to a more polarised (magnetised) state. The paramagnetic phase will be transformed to one where there is only one equilibrium (with signs determined by those of the groups' private utilities or external fields). Ferromagnetic phases will, in absence of metastability, also loose their degeneracy, and each group {\itshape chooses} sign according to that of their IWA (or aligns its spin with the external field). Critical curves will be now separating regions where social utility is important from those where it is not (or regions where there is spontaneous magnetisation from those where there is not). What sort of metastable behaviour will come out when the two first order phase transitions are present is more difficult to predict, and definitely worth studying in detail. It will allow us to answer qualitative questions such as: Will the statement made in the paragraph above about previously confronted groups hold even when their private utilities favour the same decision?  This case will be studied as a particular example in \cite{R'io2010}.

%% file: secs/1p2cho.tex
\chapter{Local coupling: two coupled choices in one population}
\label{cha:1p2cho}
We now turn to the case of a single group of individuals all of which are simultaneously making two interdependent choices subject to social influence. The strength of social influence may be different for the different choices but must be equal for all individuals. Agents are affected in their choice making by their perception of the degree of acceptance of each choice in the group, which is given by the average magnetisation or opinion (given by the fraction of adopters) for all individuals.

Each individual's utility function will have an extra term which bonuses or penalises if he chooses to make both choices positively. This means both choices are interrelated in that each agent's outcome is affected by each of his choices, but also by whether these are aligned. This additional benefit or cost will be constant for all individuals. Note that in this case the inter-coupling is local and has no possible interpretation as social interaction. It must refer to actual gains or losses of individuals depending on their choice making.

Besides social influence and decision interdependence, individuals have and {\itshape idiosyncratic willingness to adopt} (IWA) which will be considered to be constant across the population.   

This is equivalent to considering $N$ particles, each having two spin type properties with Weiss mean field Ising (with constant external field) dynamics, plus an additional constant local coupling between both spins for each particle. In the thermodynamic limit ($N\to \infty$) this is equivalent to considering infinite range intra-interactions for which Hamiltonian \eqref{eq:hamgen} takes the form:       

\begin{equation}
\label{eq:ham1p2cho1}
H = \sum_{(i, j)}\left(-\frac{J_{s}}{N}s_{i}s_{j}-\frac{J_{t}}{N}t_{i}t_{j}\right)-\sum_{i}\left(ks_{i}t_{i}+h_{s}s_{i}+h_{t}t_{i}\right)
\end{equation} 

\noindent where summations over $(i,j)$  are $1\leq i<j \leq N$ and summations over $i$ are $1 \leq i \leq N$.  

For $h_{s}=h_{t}=0$, this is a particular case of the  model considered in \cite{Galam1995} when the intra-coupling probability distribution (they consider random couplings) is constant $p(\eta_{i})=\delta(\eta_{i}-k)$. The problem is that, although some interesting general expressions are derived, the paper focuses on the use of symmetric distributions (relevant in the context of plastic phases as they do not destroy the symmetry of the ordered phases). In our context, this means that both choices must be coupled positively for half of the population and negatively for the other half, and thus seems unrealistic for most examples of social or economic interest. The study of the general random system would be however very interesting from the social sciences perspective as another way of encoding the population's heterogeneity. We will be using some of the results and general approach of \cite{Galam1995}.

In what remains of the section we will be carrying out a similar analysis to that of the previous section for the system of Hamiltonian \eqref{eq:ham1p2cho1}.

\section{The model}
Hamiltonian \eqref{eq:ham1p2cho1} can be rewritten:

\begin{equation}
\label{eq:ham1p2cho2}
H = \frac{J_{s}+J_{t}}{2}-\frac{J_{s}}{2N}\left(\sum_{i}s_{i}\right)^{2}-\frac{J_{t}}{2N}\left(\sum_{i}t_{i}\right)^{2}-\sum_{i}\left(ks_{i}t_{i}+h_{s}s_{i}+h_{t}t_{i}\right)
\end{equation}

Following \cite{Galam1995}, the partition function of the representative canonical ensemble can be computed exactly in the thermodynamic limit yielding

\begin{equation}
Z=\frac{\beta N}{2\pi}(J_{s}J_{t})^{\frac{1}{2}}e^{-\frac{\beta}{2}(J_{s}+J_{t})}\int_{-\infty}^{\infty}\int_{-\infty}^{\infty}\,ds\,dt e^{-\beta Ng(s,t)}
\end{equation} 

\noindent where $g(s,t)$ is the free energy functional such that the free energy density $f= \lim_{N\to \infty}\frac{F}{N}=$ \\$\int_{-\infty}^{\infty}\int_{-\infty}^{\infty}\,ds\,dt\, g(s,t)$ and is given by the expression:

\begin{equation}
\label{eq:g}
g = \frac{1}{2}J_{s}s^{2}+\frac{1}{2}J_{t}t^{2} -\frac{1}{\beta}\ln\lbrack 2e^{\beta k}\cosh\left(\beta\left(J_{s}s+J_{t}t+h_{s}+h_{t}\right)\right)
+2e^{-\beta k}\cosh\left(\beta\left(J_{s}s-J_{t}t+h_{s}-h_{t}\right)\right)\rbrack
\end{equation}
 
In statistical equilibrium, stable states are determined by the average magnetisations $(s,t)$ minimising the free energy $f$ and thus the functional $g$.

\section{Equations of state: solutions and stability}

Derivatives of \eqref{eq:g} are given by:

\begin{align}
\frac{ \partial g}{ \partial s} & = J_{s}s 
 -\frac{J_{s}e^{\beta k}\sinh\left(\beta(J_{s}s+J_{t}t+h_{s}+h_{t})\right)+J_{s}e^{-\beta k}\sinh\left(\beta(J_{s}s-J_{t}t+h_{s}-h_{t})\right)}{e^{\beta k}\cosh\left(\beta(J_{s}s+J_{t}t+h_{s}+h_{t})\right)+e^{-\beta k}\cosh\left(\beta(J_{s}s-J_{t}t+h_{s}-h_{t})\right)} \\
\frac{ \partial g}{ \partial t} & = J_{t}t 
-\frac{J_{t}e^{\beta k}\sinh\left(\beta(J_{s}s+J_{t}t+h_{s}+h_{t})\right)+J_{t}e^{-\beta k}\sinh\left(\beta(J_{t}t-J_{s}s+h_{t}-h_{s})\right)}{e^{\beta k}\cosh\left(\beta(J_{s}s+J_{t}t+h_{s}+h_{t})\right)+e^{-\beta k}\cosh\left(\beta(J_{t}t-J_{s}s+h_{t}-h_{s})\right)}
\end{align}

The system's equations of state giving $g$'s critical points will be then given by 

\begin{equation}
\label{eq:1p2choeqsta}
\begin{array}{l}
s = \displaystyle{\frac{\tanh\left(\beta(J_{s}s+h_{s})\right)+\tanh\left(\beta(J_{t}t+h_{t})\right)\tanh\left(\beta k\right)}{1+\tanh\left(\beta(J_{s}s+h_{s})\right)\tanh\left(\beta(J_{t}t+h_{t})\right)\tanh\left(\beta k\right)}} \\
t = \displaystyle{\frac{\tanh\left(\beta(J_{t}t+h_{t})\right)+\tanh\left(\beta(J_{s}s+h_{s})\right)\tanh\left(\beta k\right)}{1+\tanh\left(\beta(J_{s}s+h_{s})\right)\tanh\left(\beta(J_{t}t+h_{t})\right)\tanh\left(\beta k\right)}}
\end{array}
\end{equation}

\noindent introducing the notation $\alpha_{s}=\tanh\left(\beta(J_{s}s+h_{s})\right)$, $\alpha_{t}=\tanh\left(\beta(J_{t}t+h_{t})\right)$ and $\alpha_{k}=\tanh\left(\beta k\right)$ these become:

\begin{equation}
\label{eq:1p2choeqstasim}
\begin{array}{l}
s = \displaystyle{\frac{\alpha_{s}+\alpha_{t}\alpha_{k}}{1+\alpha_{s}\alpha_{t}\alpha_{k}} }\\
t = \displaystyle{\frac{\alpha_{t}+\alpha_{s}\alpha_{k}}{1+\alpha_{s}\alpha_{t}\alpha_{k}} }
\end{array}
\end{equation}

Second derivatives of $g$ determining stability of solutions to \eqref{eq:1p2choeqsta} are given by

\begin{align}
\frac{ \partial^{2} g}{ \partial s^{2}} & = J_{s}-\frac{\beta J_{s}^{2}\gamma_{s}\left(1-\alpha_{t}^{2}\alpha_{k}^{2}\right)}{(1+\alpha_{s}\alpha_{t}\alpha_{k})^{2}} \label{eq:locgss}\\
\frac{ \partial^{2} g}{ \partial t^{2}} & = J_{t}-\frac{\beta J_{t}^{2}\gamma_{t}\left(1-\alpha_{s}^{2}\alpha_{k}^{2}\right)}{(1+\alpha_{s}\alpha_{t}\alpha_{k})^{2}}\label{eq:locgtt}\\
\frac{ \partial^{2} g}{\partial s \partial t } & = - \frac{\beta J_{s}J_{t}\gamma_{s}\gamma_{t}\alpha_{k}}{(1+\alpha_{s}\alpha_{t}\alpha_{k})^{2}}
\end{align}

\noindent with $\gamma_{s}=\frac{1}{\cosh^{2}\left(\beta(J_{s}s+h_{s})\right)}$ and $\gamma_{t}=\frac{1}{\cosh^{2}\left(\beta(J_{t}t+h_{t})\right)}$.

 In this case the Hessian's determinant takes the form:

\begin{equation}
\label{eq:locdethess}
det(\mathcal{H})=J_{s}J_{t}\lbrack 1- \frac{\beta \left( J_{t}\gamma_{t}(1-\alpha_{s}^{2}\alpha_{k}^{2})+J_{s}\gamma_{s}(1-\alpha_{t}^{2}\alpha_{k}^{2})\right)}{(1+\alpha_{s}\alpha_{t}\alpha_{k})^{2}}
 +\frac{\beta^{2}J_{s}J_{t}\gamma_{s}\gamma_{t}\left((1-\alpha_{s}^{2}\alpha_{k}^{2})(1-\alpha_{t}^{2}\alpha_{k}^{2})-\gamma_{s}\gamma_{t}\alpha_{k}^{2}\right)}{(1+\alpha_{s}\alpha_{t}\alpha_{k})^{4}}\rbrack
\end{equation}

Signs of \eqref{eq:locdethess} and \eqref{eq:locgss}, when evaluated at a critical point $(m_{s},m_{t})$, will determine its stability: saddle points for negative determinant \eqref{eq:locdethess} and for positive one, minima for \eqref{eq:locgss} positive and maxima for \eqref{eq:locgss} negative\footnote{Or alternatively using \eqref{eq:locgtt} or the trace as discussed for the nonlocal case.}.

Note that both the local model just described and the nonlocal one discussed in last chapter are the same uncoupled model in the $k \to 0$ limit. The behaviour of both models must be therefore similar at least at low $k$.

\subsection{Zero field case}

As we did for the nonlocal case, let us begin by drawing some general conclusions upon simple inspection of the equations derived when $h_{s}=h_{t}=0$.

As for the nonlocal coupling, the paramagnetic solution $(s,t)=(0,0)$ is always present at finite temperature (and is the only one at high enough temperatures). The completely ordered ferromagnetic phase $(s,t)=(\pm 1, \pm 1)$ is the only solution for $T=0$ ($\beta \to \infty$), but in this case, $(s,t)=(1,1)$ and $(s,t)=(-1,-1)$ are the only solutions in the positive $k$ scenario, and $(s,t)=(-1,1)$ and $(s,t)=(1,-1)$ in the negative $k$ one. There are no metastable solutions for $T=0$. Note this suggests a different behaviour of the nonlocal and local models concerning the metastable states at least at low enough temperature, even at low $k$.

As in the nonlocal case, mixed phases (with non zero magnetisation associated to the highest intra-coupling constant) can only be critical points when $k=0$. In fact the $k=0$ phase diagram hyperplane of both the local and nonlocal model are identical, and so is the behaviour of these mixed phases.  

The equations of state also have the symmetries $(s,t)\to (-s,-t)$, $(s,t)\to (-s,t)$ and $k\to -k$. All ferromagnetic or mixed solutions will be therefore appearing in pairs, and critical points for $k$ and $-k$ are the same in absolute value (and have opposite relative sign between both average magnetisations).

Linearization of \eqref{eq:1p2choeqstasim} for $|s| \ll 1$ and $|t| \ll 1$ (at finite nonzero temperature) yields

\begin{equation}
\begin{array}{l}
s = \beta J_{s}s+\beta J_{t}\tanh(\beta k)t+ O(s^{3},t^{3}, s^{2}t, st^{2})\\
t = \beta J_{t}t+\beta J_{s}\tanh(\beta k)s+ O(s^{3},t^{3}, s^{2}t, st^{2})
\end{array}
\end{equation}

\noindent which can be simplified to

\begin{equation}
\label{eq:locTc}
l(\beta)=J_{s}J_{t}\left(1-\tanh^{2}(\beta k)\right)\beta^{2}-(J_{t}+J_{s})\beta+1 = 0
\end{equation}

\begin{figure}
\centering
\subfloat[]{\includegraphics[width=0.33\textwidth]{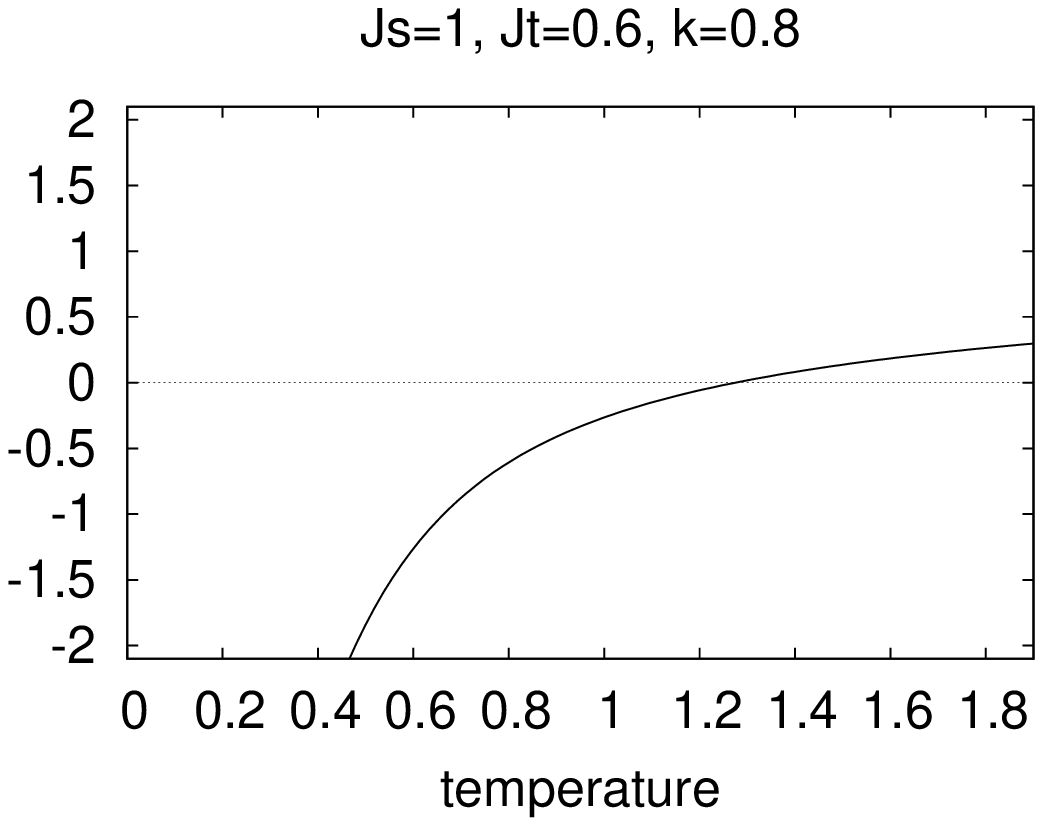}}
\subfloat[]{\includegraphics[width=0.33\textwidth]{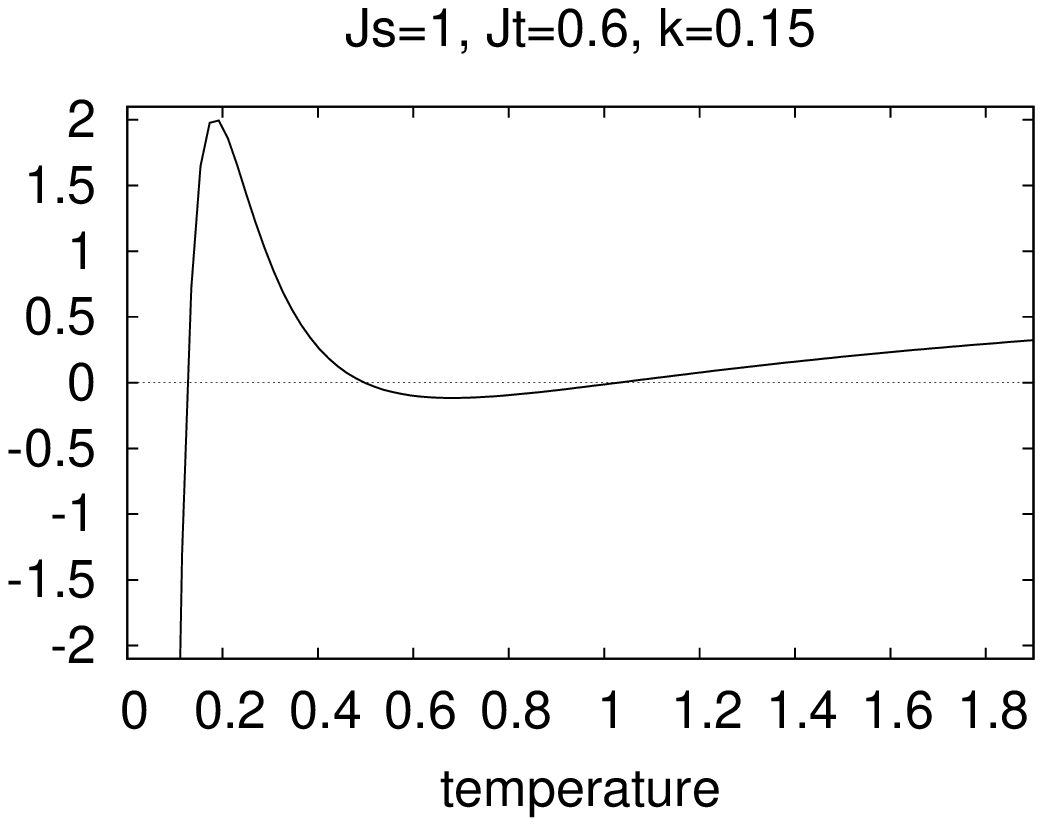}}
\subfloat[]{\includegraphics[width=0.33\textwidth]{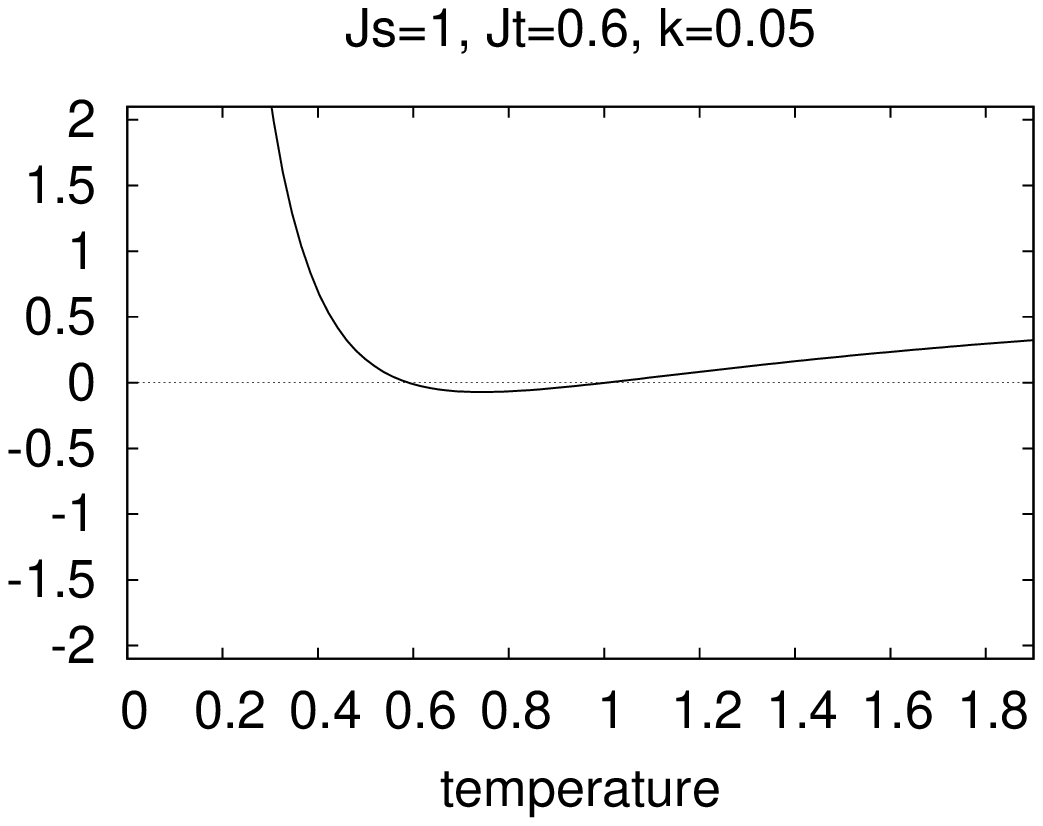}}
\caption{Plots of the function $l(K_{B}T)=J_{s}J_{t}\left(1-\tanh^{2}(\frac{k}{K_{B}T})\right)\frac{1}{(K_{B}T)^{2}}-(J_{s}+J_{t})\frac{1}{K_{B}T}+1$ against temperature ($K_{B}T$). $J_{s}=1$, $J_{t}=0.6$ and different values of inter-coupling are considered. In (a) $k = 0.$ and $l$ has one root. In (b) $k = 0.15$ and $l$ has two roots. In (c) $k = 0.05$ and $l$ has three roots.}
\label{fig:locTc}
\end{figure}

Here the first significant differences with the nonlocal case appear. As shown in figure \ref{fig:locTc}, depending on the values of the (intra- and inter-) couplings, there can be three qualitatively different scenarios when analysing equation \eqref{eq:locTc} in $\beta$, with either one, two or three different solutions (all positive and thus physically relevant).

Expressions concerning stability of solutions \eqref{eq:locdethess} and \eqref{eq:locgss} are more obscure than in the nonlocal case. We can however draw some simple conclusions upon inspection. First of all, the Hessian's determinant and the second derivative still retain the equations of motion symmetries, and thus solutions related by the transformations $(s,t)\to(-s,-t)$ and $(s,t)\to(-s,t)\; k\to-k$ or $(s,t)\to(s,-t)\; k\to-k$ will share the same stability condition.

As $\alpha_{s}$, $\alpha_{t}$ and $\alpha_{k}$ are bounded, for ferromagnetic solutions ($\gamma_{s}\neq 1$, $\gamma_{t}\neq 1$, $\alpha_{s}\neq 0$, $\alpha_{t}\neq 0$), behaviour at $T=0$ ($\beta \to \infty$) of equations \eqref{eq:locdethess} and \eqref{eq:locgss} will depend on the explicit dependence on $\beta$ and $\gamma_{s}$, $\gamma_{t}$, which go to zero. Thus ferromagnetic solutions are stable at $T=0$. 

As for mixed phases ($\gamma_{s}\neq 1$, $\gamma_{t} = 1$, $\alpha_{s} \neq 0$, $\alpha_{t} = 0$, $\alpha_{k}= 0$ or $\gamma_{s}= 1$, $\gamma_{t} \neq 1$, $\alpha_{s} = 0$, $\alpha_{t} \neq 0$, $\alpha_{k}= 0$), little can be said about their stability from this simple inspection, as the region where they exist is out of its scope. 

Let us study the paramagnetic phase in more detail. When $\gamma_{s}=\gamma_{t}=1$ and $\alpha_{s}=\alpha_{t}=0$ equations \eqref{eq:locdethess} and \eqref{eq:locgss} can be rewritten:

\begin{align}
det(\mathcal{H}) & =J_{s}J_{t}\lbrack 1- \beta \left( J_{s}+J_{t}\right)+\beta^{2}J_{s}J_{t}\left(1-\alpha_{k}^{2}\right)\rbrack = l(\beta)\label{eq:locdethesspara}\\
\frac{ \partial^{2} g}{ \partial s^{2}} & = J_{s}-\beta J_{s}^{2} \label{eq:locgsspara}
\end{align}

At very high temperatures ($\beta \to 0$), the paramagnetic is the only stable solution. At $T=0$ ($\beta \to \infty$), it is a maximum. When analysing where the sign of \eqref{eq:locdethesspara}  changes, we recover equation \eqref{eq:locTc}, and so the sign of the Hessian's determinant (and the stability of the paramagnetic phase) changes once, twice or three times depending on the coupling values (see figure \ref{fig:locTc}). The number of roots of function $l(\beta)$ defined in \eqref{eq:locTc} give the points where the stability of the $(0,0)$ solutions changes from saddle to maximum/minimum. For $K_{B}T>J_{s}$ equation \eqref{eq:locgsspara} is positive (and negative for $K_{B}T<J_{s}$).

Let us use a notation analogous to the nonlocal case where $T_{c}$  is the largest (or only) solution to equation \eqref{eq:locTc}, and $T_{b}$, $T_{b'}$ the second and third largest (when present). We have three possible scenarios. When all three {\itshape critical} temperatures exist, this means that for $T<T_{b'}$, the paramagnetic phase is a saddle point,  at $T=T_{b}$ it is becomes a maximum, for $T_{b}<T<T_{c}$ again a saddle point, and for $T>T_{c}$ a minimum. When only $T_{b}$ and $T_{c}$ exist, the paramagnetic critical point will be a maximum for $T<T_{b}$, a saddle point for $T_{b}<T<T_{c}$ and a minimum for $T>T_{c}$. When only $T_{c}$ exists, $(0,0)$ is a saddle point for temperatures lower than the critical, and  a minimum above it. As in the nonlocal case, there is a second order phase transition at the critical point $T_{c}$, but now for all values of the parameters. 

In this case, global minima are always at local critical points. When  $l(\beta)$ has only one root (high $k$), there are either two ferromagnetic minima (and the paramagnetic phase is a saddle point), or only the paramagnetic minimum, depending on the temperature. No metastable states appear. It resembles the behaviour of the nonlocal strong coupling regime, but saddle points are now minima and maxima saddle points. 

At low enough values of $k$, the free energy surface is very much like that of the nonlocal weak coupling case, specially in what concerns minima. At low temperatures, there are two main ferromagnetic minima and two additional metastable states. Above some temperature, the metastable states disappear and finally the two main ferromagnetic ones do too as $(0,0)$ becomes the only minimum. This behaviour is the same for $l(\beta)$ having two or three roots. In the last case however, at low enough temperatures, the paramagnetic solution is a saddle point instead of a maximum.

At $k=0$, the situation is that described in last chapter for the nonlocal case. As already discussed, metastable states are associated to a  first order transition at $k=0$.

\section{Numerical analysis for the zero field case}
\label{sec:locnumsol}
The same algorithm, methodology, convergence tolerance, iterations and initial values described in section \ref{sec:nlnumsol} are used. Python libraries and code can also be found in {\text https://github.com/anafrio/}. The same graphical conventions are used as in the previous chapter, and the same symmetry considerations must be also taken into account when interpreting the graphs. The same parameter choice cases as for the nonlocal case are also those studied in more detail in this section. This allows for a quantitative as well as qualitative comparison between both models.

\subsection{Dependence on temperature}

Although our previous analysis on the stability of the paramagnetic phase (figure \ref{fig:locTc}) would suggest the convenience of studying the dependence on temperature in three different cases, concerning our interests, it is in fact enough to consider, as in the nonlocal case, only two cases (which we can label high and low inter-coupling with respect to the intra-coupling, without going into further details for the time being). The reason is that the appearance of a third temperature $T_{b'}$ where the stability of the $(0,0)$ phase changes, has no effect on the minima of the free energy functional, and so we can consider the qualitative behaviour described in figure \ref{fig:locTc} (c) as a particular case of \ref{fig:locTc} (b) where $T_{b'} \to 0$. In many ways, the system's behaviour resembles that of the nonlocal case, with the remarkable difference that in the local coupling case we find stable solutions throughout all the parameter space.

In this case, it is not possible to analytically find the roots of equation \eqref{eq:locTc}. Whenever we give explicit values for $T_{c}$, $T_{b}$ or $T_{b'}$ we will be referring to numerically calculated values (again using the Newton-Raphson algorithm). 

\begin{figure}
\centering
\includegraphics[width=\textwidth]{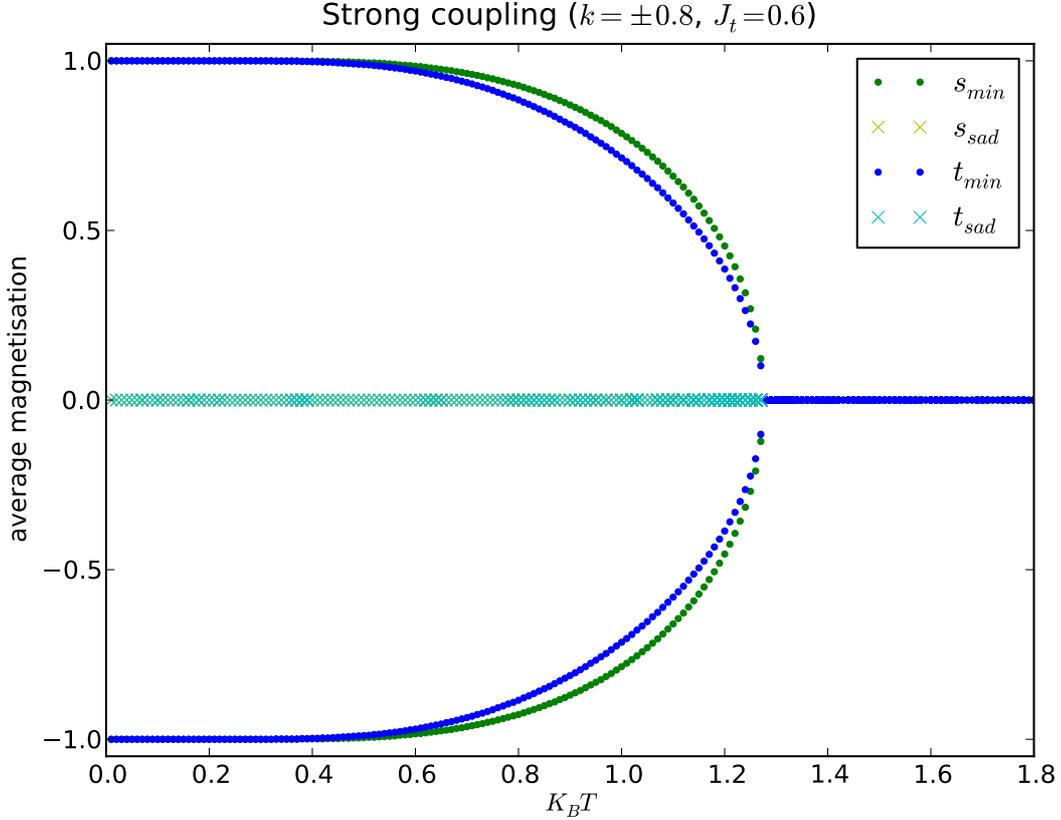}
\caption{Dependence on temperature of the numerically calculated average magnetisations for $J_{s}=1$, $J_{t}= 0.6$ ,$k = \pm 0.8$ ($K_{B}T_{c}=1.28$). Different solutions are plotted for temperatures  between 0.01 and 1.8 every 0.01 ($K_{B}T$). Magnetisations are plotted in green for $s$ and blue for $t$. Dark points are used for stable solutions and lighter asps ($\times$) for saddle point, non stable solutions.}
\label{fig:locanaT1}
\end{figure}

Figure \ref{fig:locanaT1} shows an example of what we will call high coupling behaviour ($l(\beta)$ only one root), for $J_{s}=1$, $J_{t}=0.6$ and $k=\pm 0.8$. In this case, $K_{B}T_{c}=1.28$ is the only root of $l(\beta)$ and the critical value of the temperature, where there is a second order phase transition from the ferromagnetic to paramagnetic phase (as in the nonlocal phase, note the smooth change in the average magnetisations to zero as the temperature approaches the critical value is indicating a second order phase transition). This behaviour is roughly similar to the strong coupling regime of the nonlocal case, with the already mentioned significant difference that saddle points are now minima and maxima saddle points, and so there are stable solutions for all values of the temperature. There are however no metastable solutions. Note that the value of $T_{c}$ is lower than for the nonlocal case (for which $K_{B}T_{c}=1.62$).

\begin{figure}
\centering
\includegraphics[width=\textwidth]{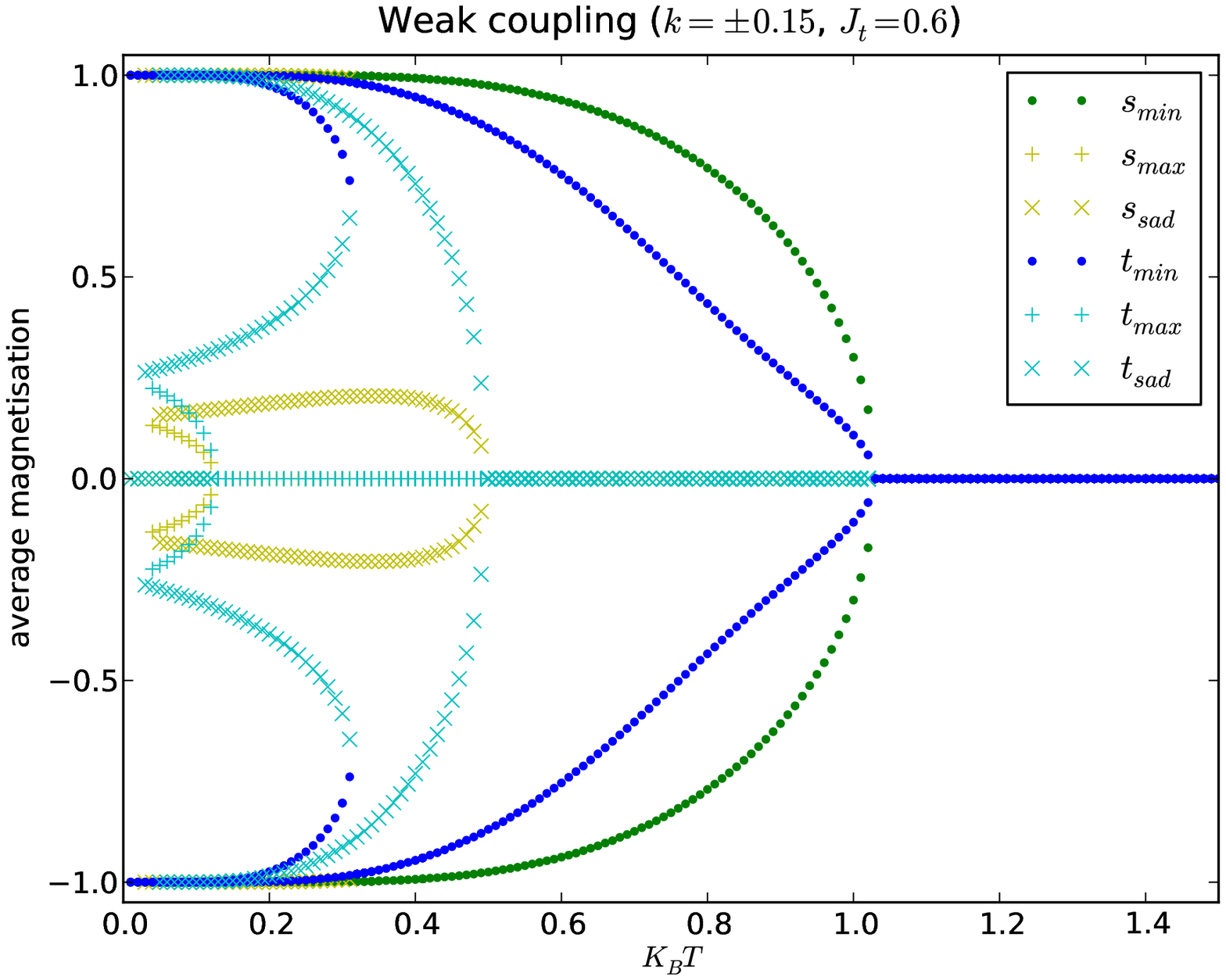}
\caption{Dependence on temperature of the numerically calculated average magnetisations for $J_{s}=1$, $J_{t}= 0.6$ ,$k = \pm 0.15$ ($K_{B}T_{c}=1.03$, $K_{B}T_{b}=0.50$ and $K_{B}T_{b'}=0.13$). Different solutions are plotted for temperatures  between 0.01 and 1.5 every 0.01 ($K_{B}T$). Magnetisations are plotted in green for $s$ and blue for $t$. Dark points are used for stable solutions and lighter asp ($\times$, for saddle points) or cross ($+$, for maxima) for non stable solutions. }
\label{fig:locanaT2}
\end{figure}

In figure \ref{fig:locanaT2}, the low coupling behaviour is depicted for $J_{s}=1$, $J_{t}=0.6$ and $k=\pm 0.15$. In this case there are three roots of $l(\beta)$: $K_{B}T_{c}=1.03$, $K_{B}T_{b}=0.50$ and $K_{B}T_{b'}=0.13$. This figure reminds very much of the nonlocal weak coupling regime dependence on the temperature with two basic differences (and only one of them significant to the stable solutions of the system which are our real interest). The basic difference concerning non stable solutions is that for $T<T_{b'}$, $(0,0)$ becomes a saddle point, and an additional pair of ferromagnetic maxima appear. These, as well as the saddle type and metastable ferromagnetic critical points, disappear for temperatures bellow another characteristic temperature of the system, to which we will refer to as $T_{a'}$. This is the basic qualitative difference with the nonlocal weak coupling regime: metastable solutions only exist for $T_{a'}<T<T_{a}$, and not for all temperatures bellow $T_{a}$ as in the nonlocal scenario, as was indicated by the behaviour at $T=0$ discussed in the previous section. Note that the values of both $T_{c}$ and $T_{b}$ are lower in the local case, and so the region where stable ferromagnetic solutions, as well as that where there are metastable ferromagnetic solutions, is smaller.

\begin{figure}
\centering
\subfloat[]{\includegraphics[width=0.33\textwidth]{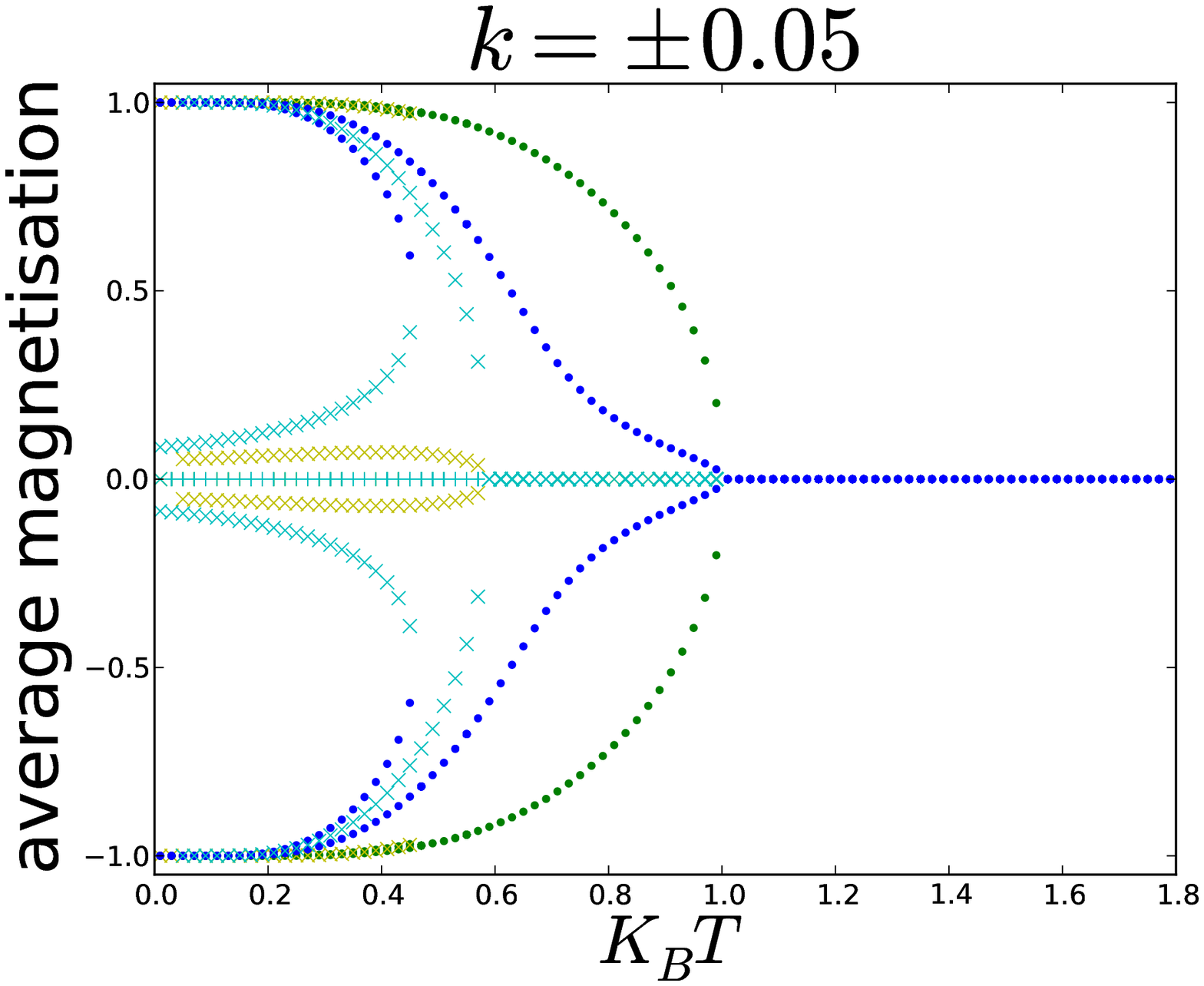}}
\subfloat[]{\includegraphics[width=0.33\textwidth]{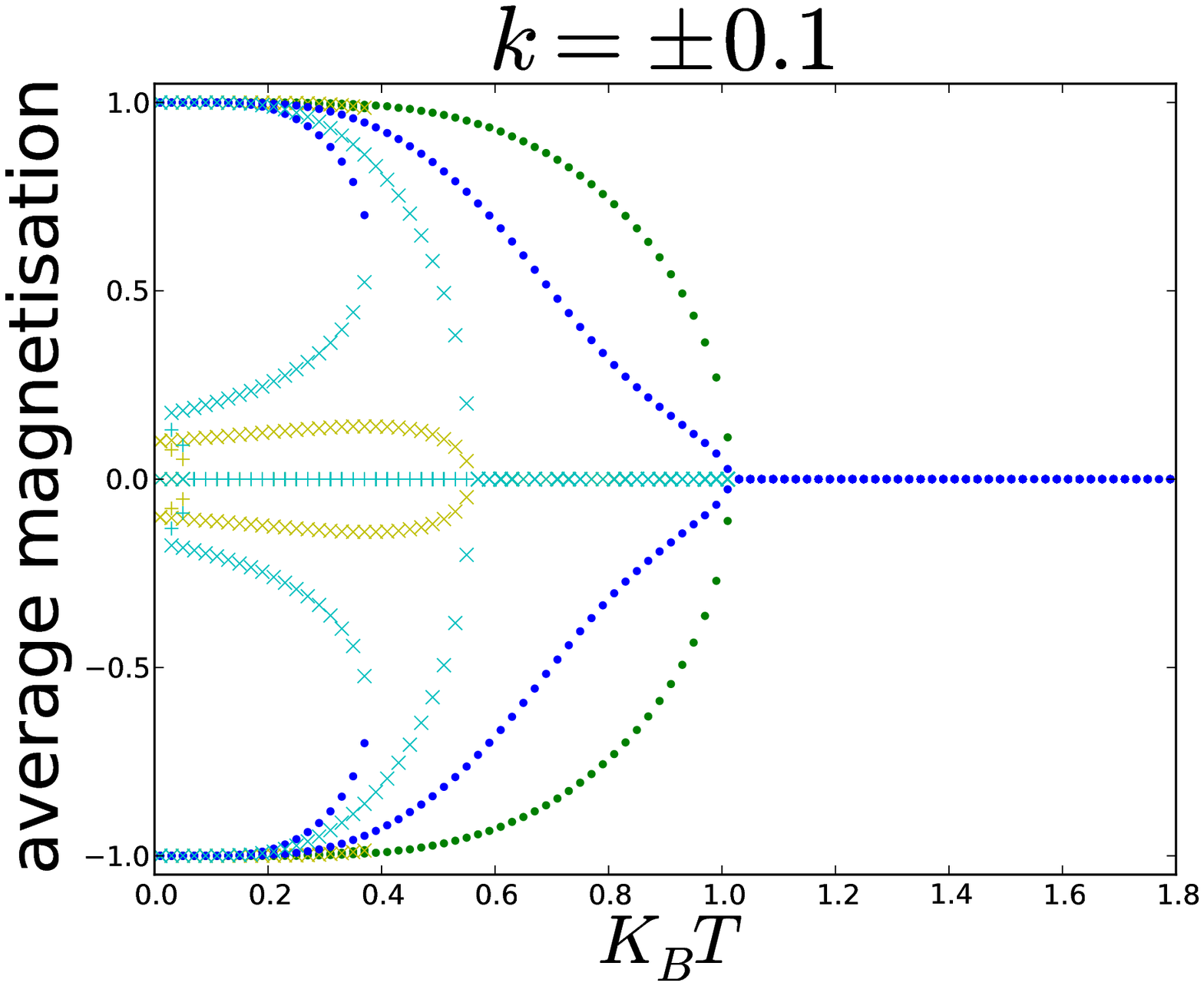}}
\subfloat[]{\includegraphics[width=0.33\textwidth]{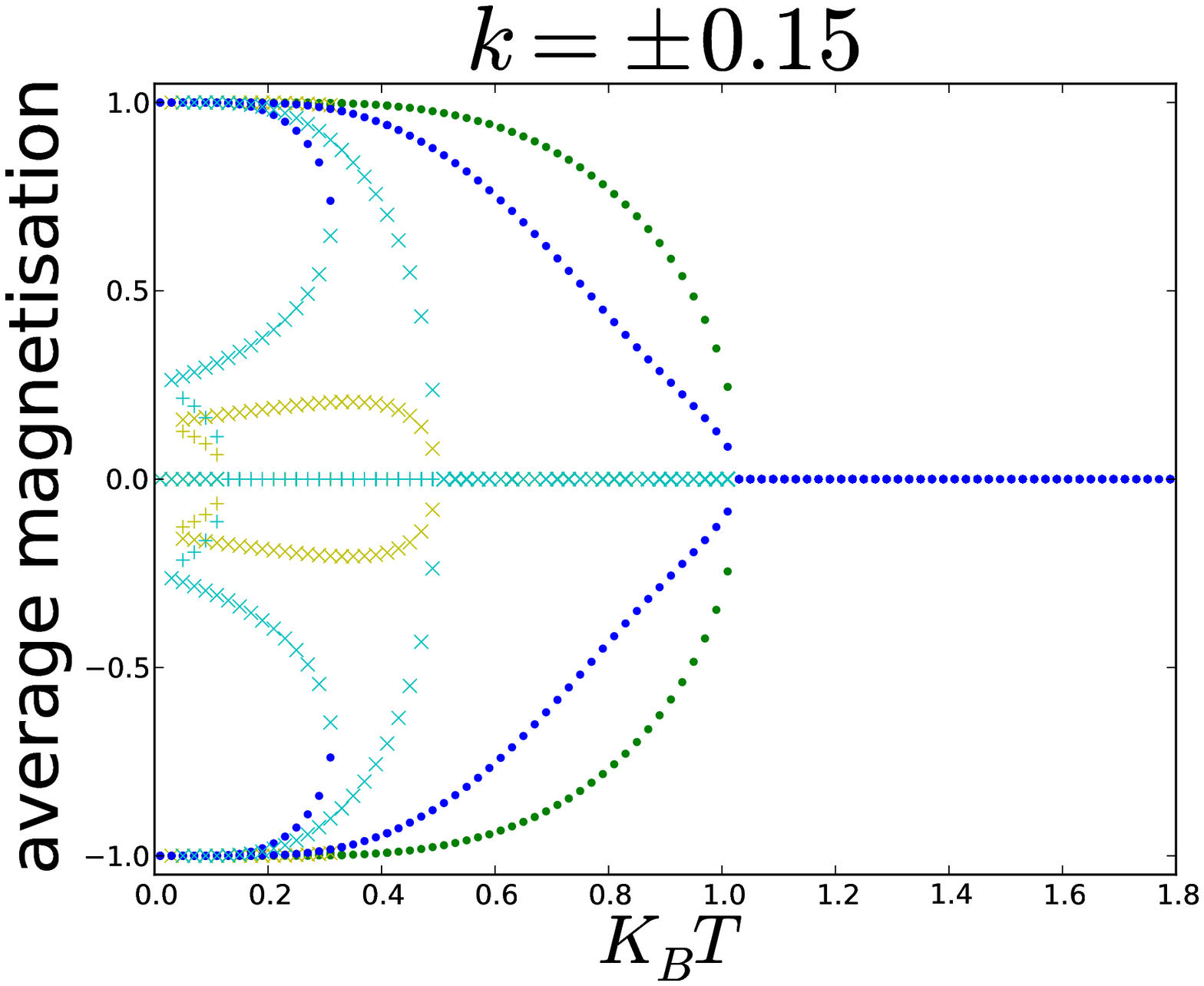}}\\
\subfloat[]{\includegraphics[width=0.33\textwidth]{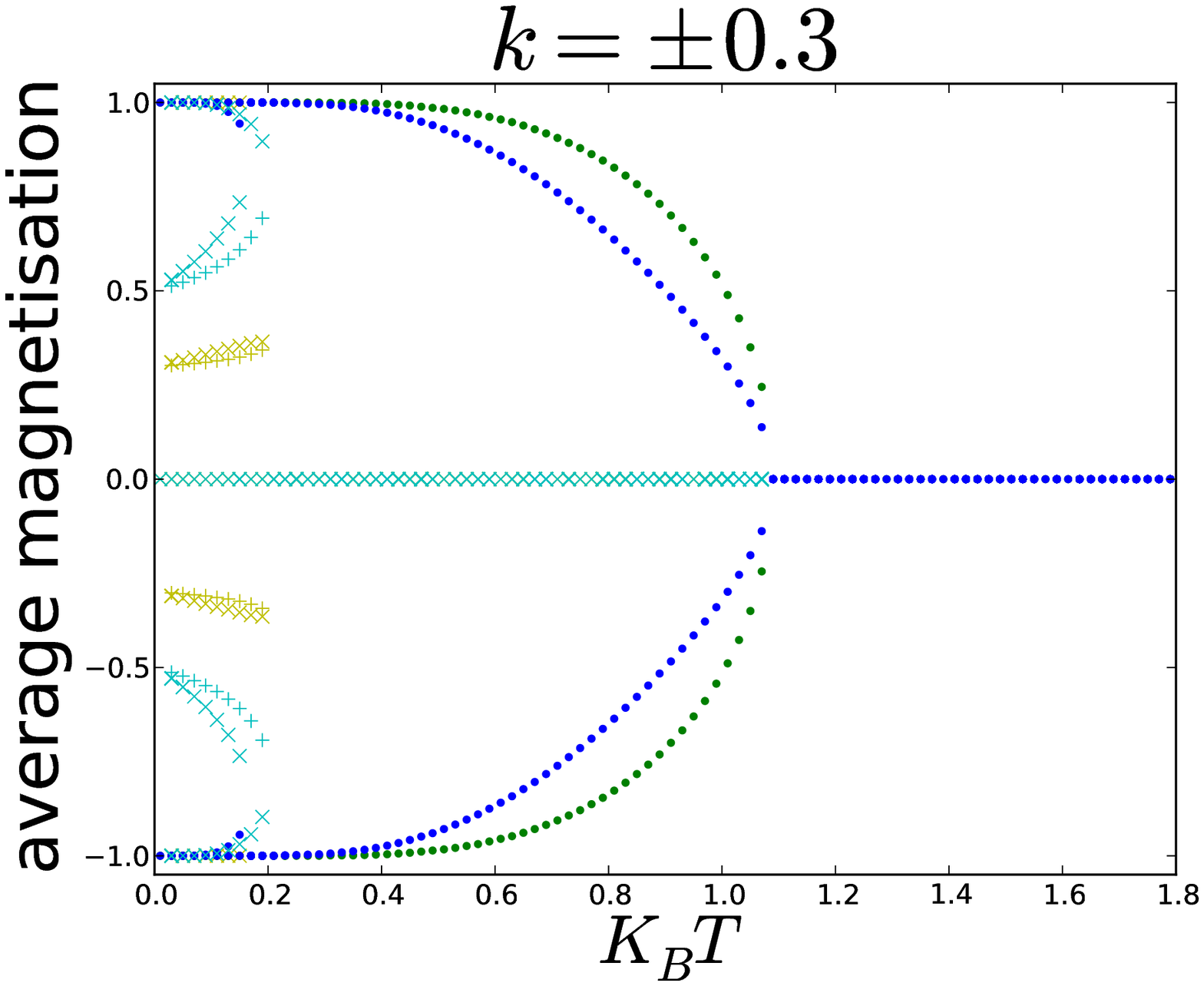}}
\subfloat[]{\includegraphics[width=0.33\textwidth]{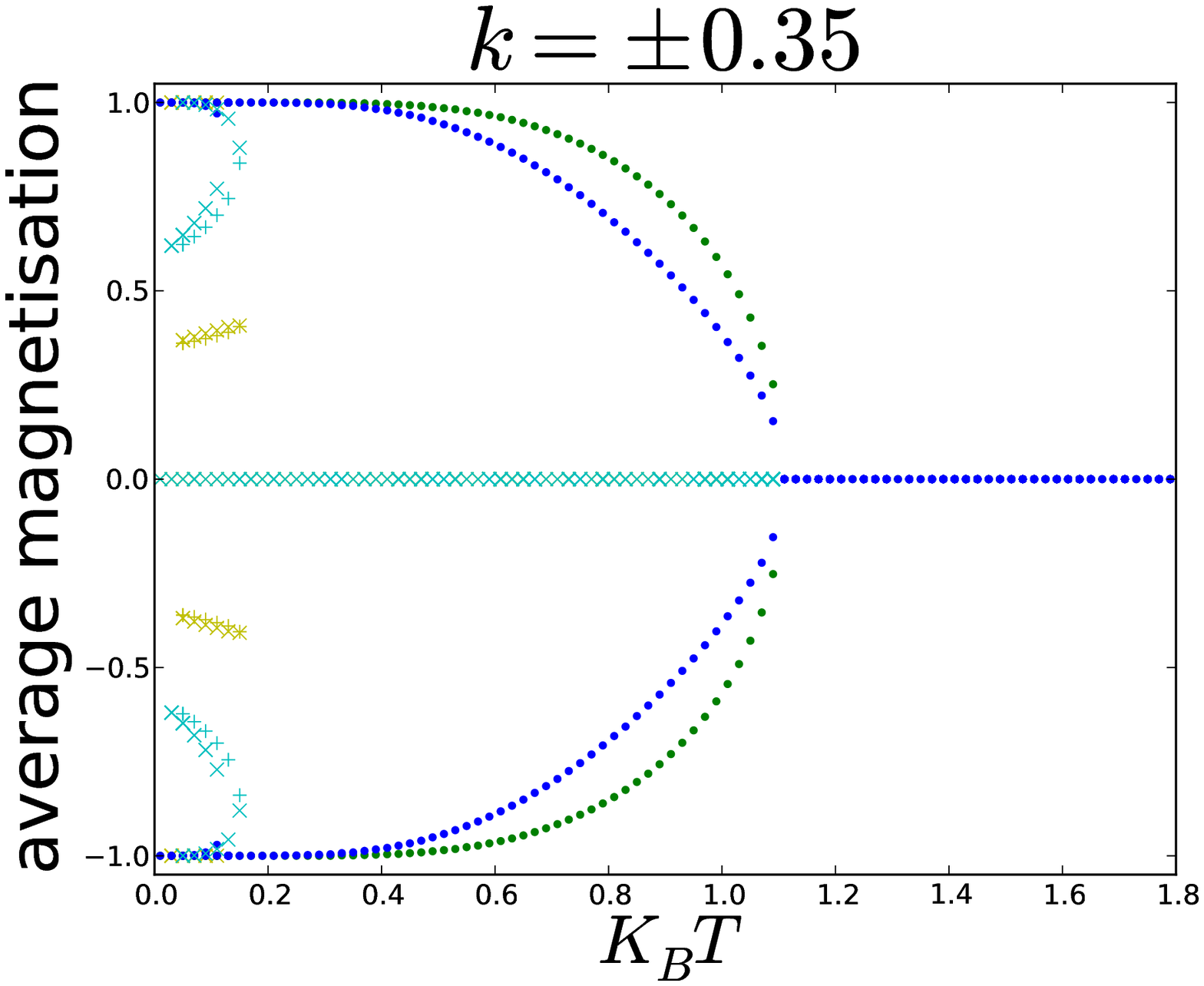}}
\subfloat[]{\includegraphics[width=0.33\textwidth]{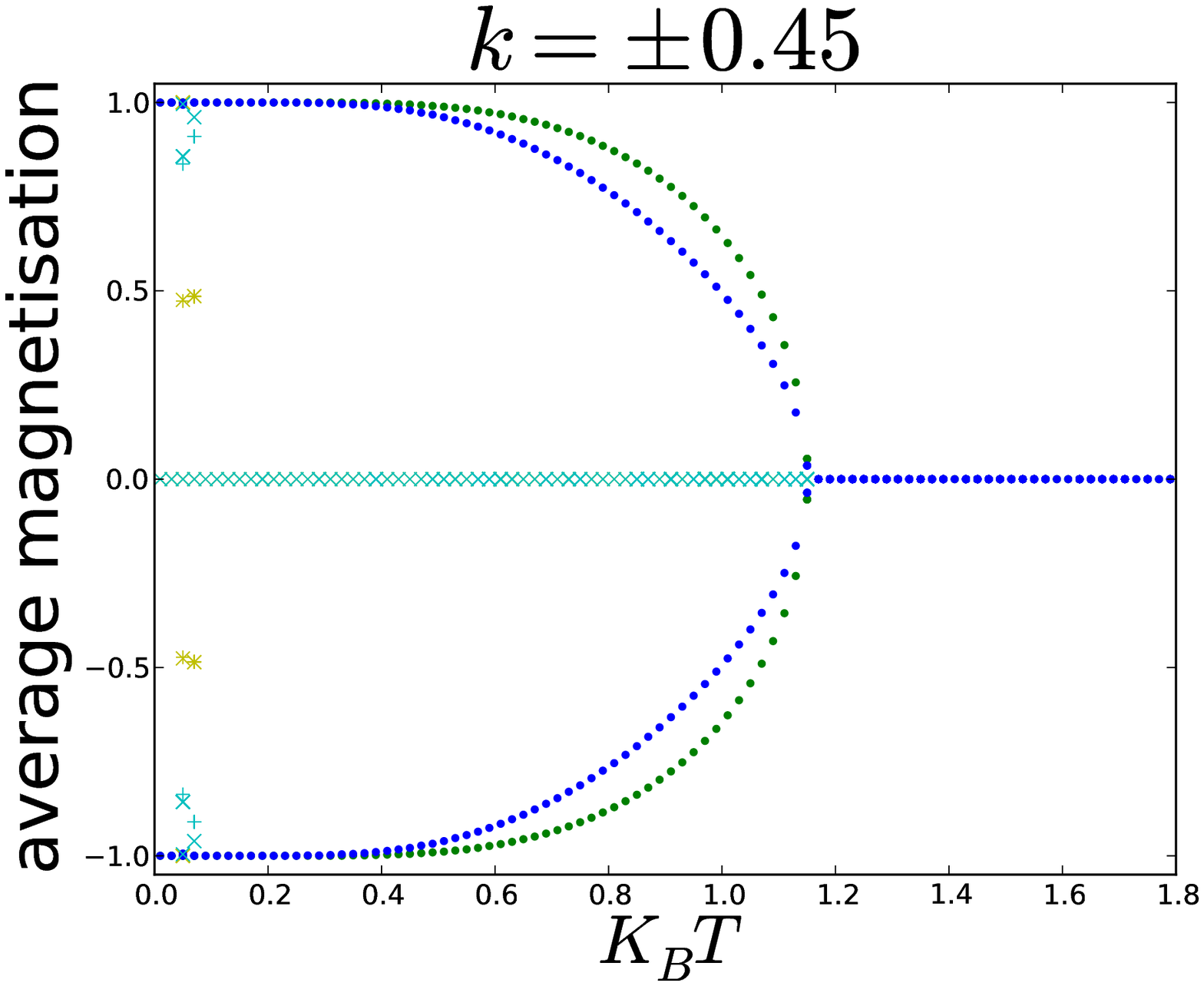}}\\
\subfloat[]{\includegraphics[width=0.33\textwidth]{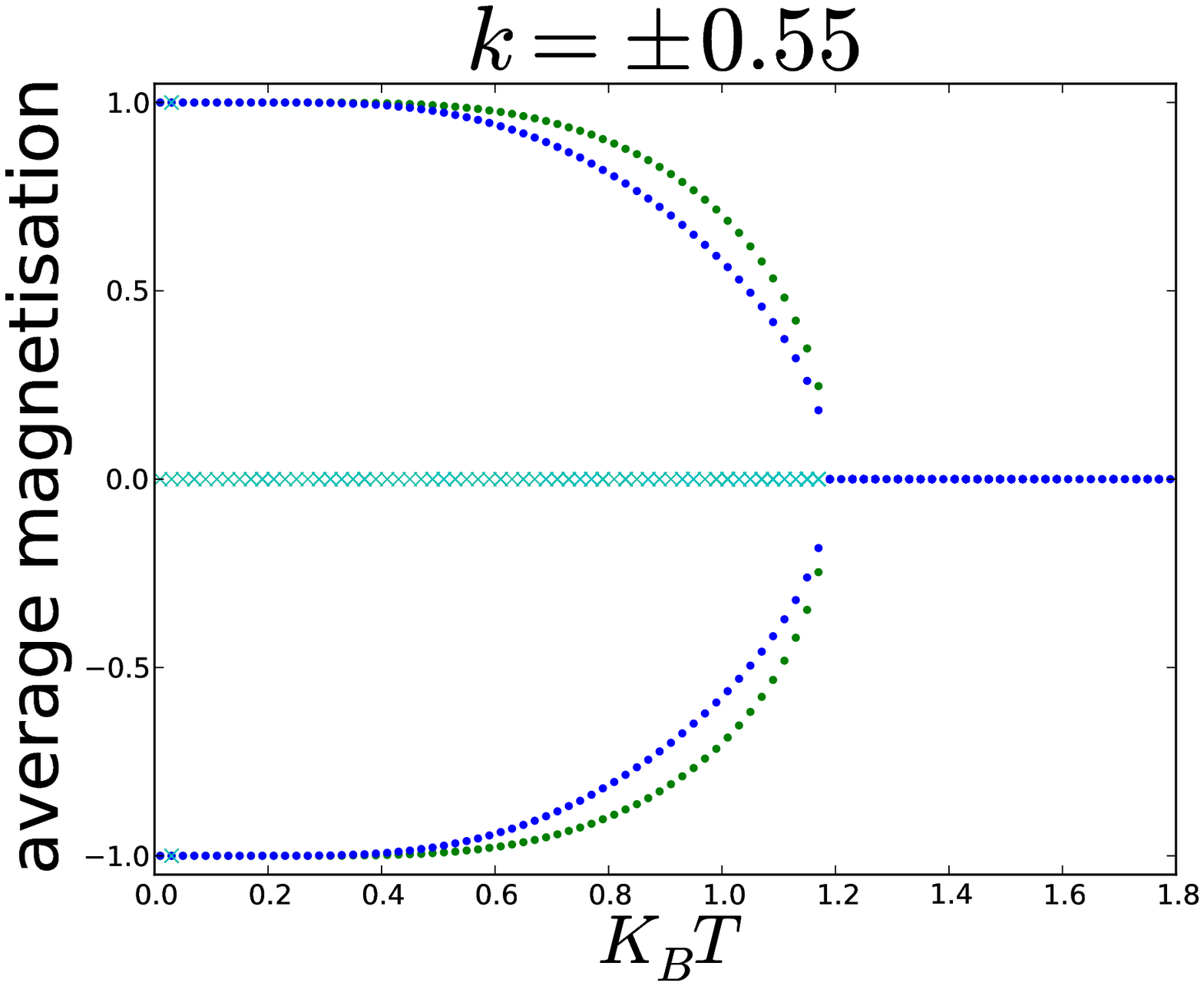}}
\subfloat[]{\includegraphics[width=0.33\textwidth]{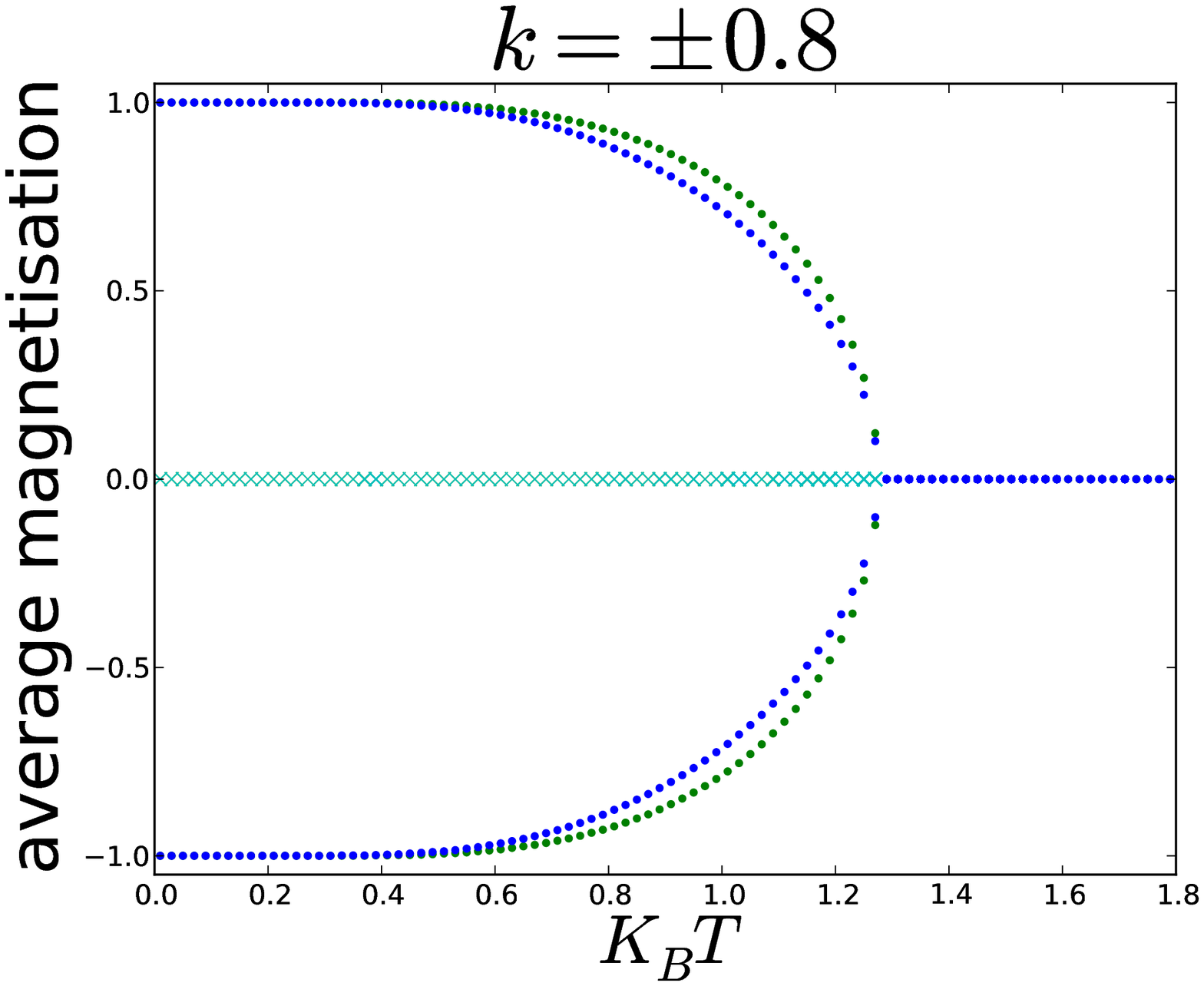}}
\subfloat[]{\includegraphics[width=0.33\textwidth]{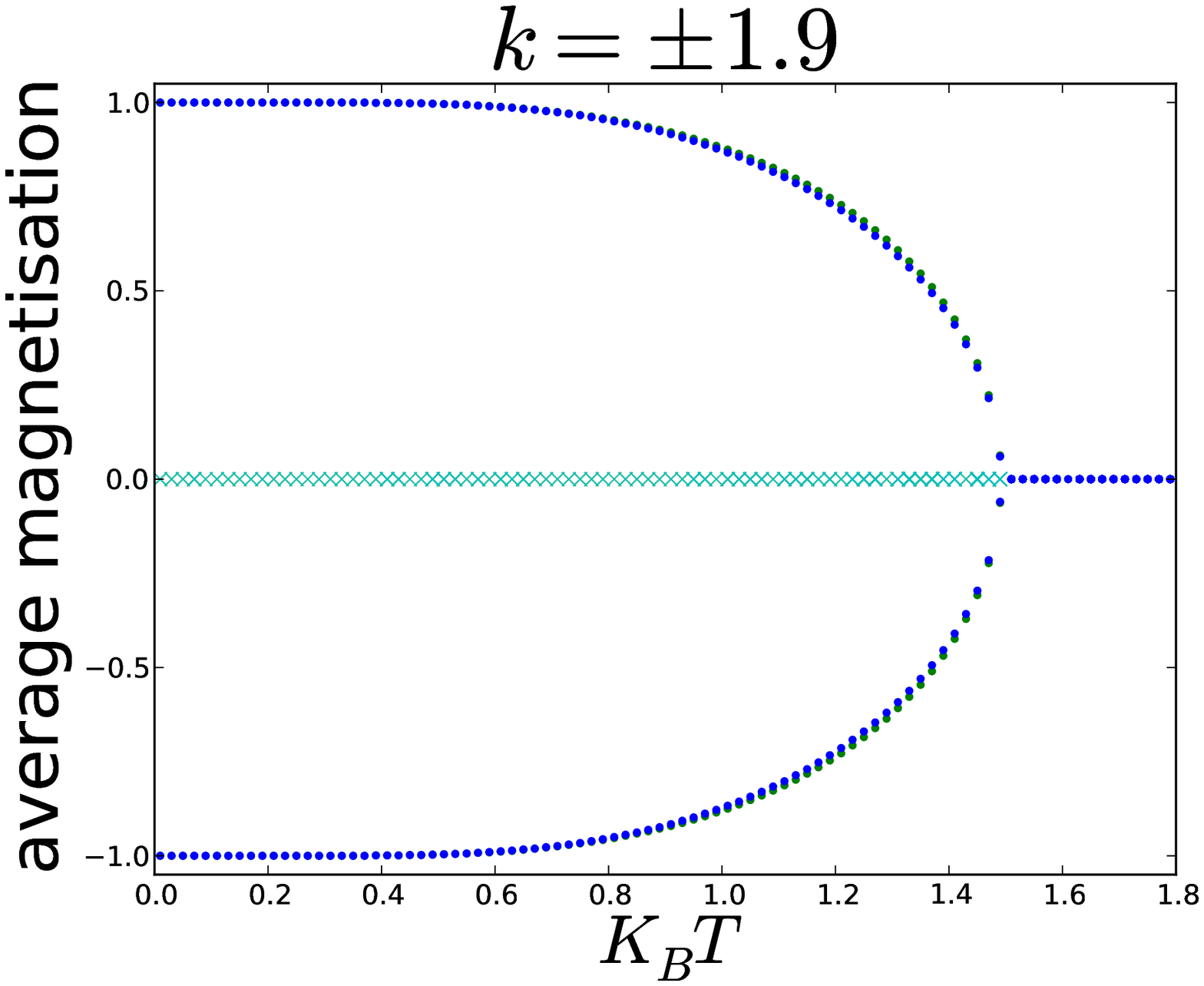}}
\caption{Dependence on temperature of the numerically calculated average magnetisations $(s,t)$ for different values of the inter-coupling $k$. $J_{s}=1$ and $J_{t}= 0.6$  for all plots. (a) $k =\pm  0.05$, (b) $k =\pm  0.1$, (c) $k =\pm  0.15$, (d) $k =\pm  0.3$, (e) $k =\pm  0.35$, (f) $k =\pm  0.45$, (g) $k =\pm  0.55$, (h) $k =\pm 0.8$ and (i) $k =\pm  1.9$. In all cases, different solutions are plotted for temperatures  between 0.01 and 1.8 every 0.02 ($K_{B}T$). Magnetisations are plotted in green for $s$ and blue for $t$. Dark points are used for stable solutions and lighter asp ($\times$, for saddle points) or cross ($+$, for maxima) for non stable solutions.}
\label{fig:locanakT}
\end{figure}

\begin{figure}
\centering
\subfloat[]{\includegraphics[width=0.33\textwidth]{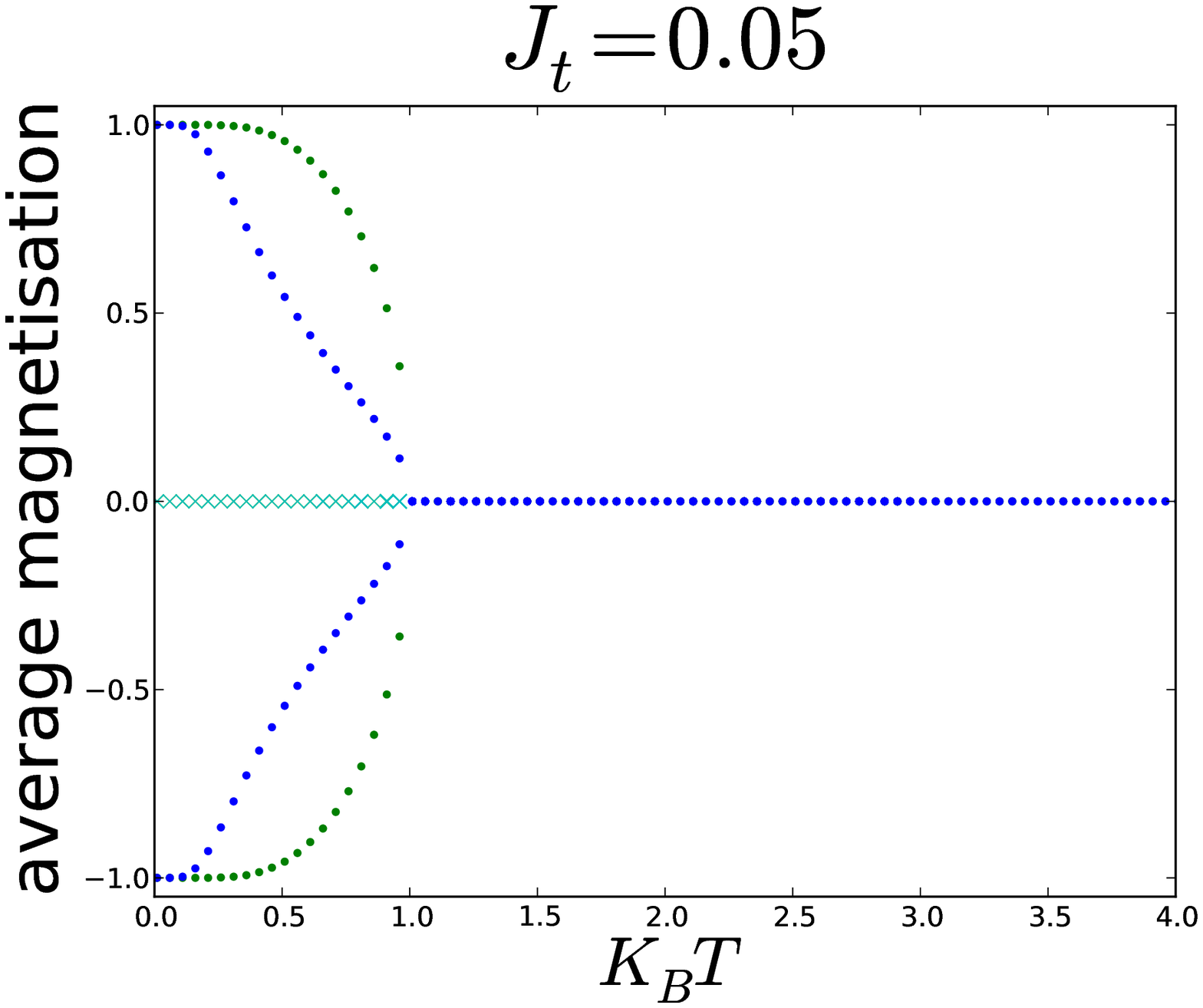}}
\subfloat[]{\includegraphics[width=0.33\textwidth]{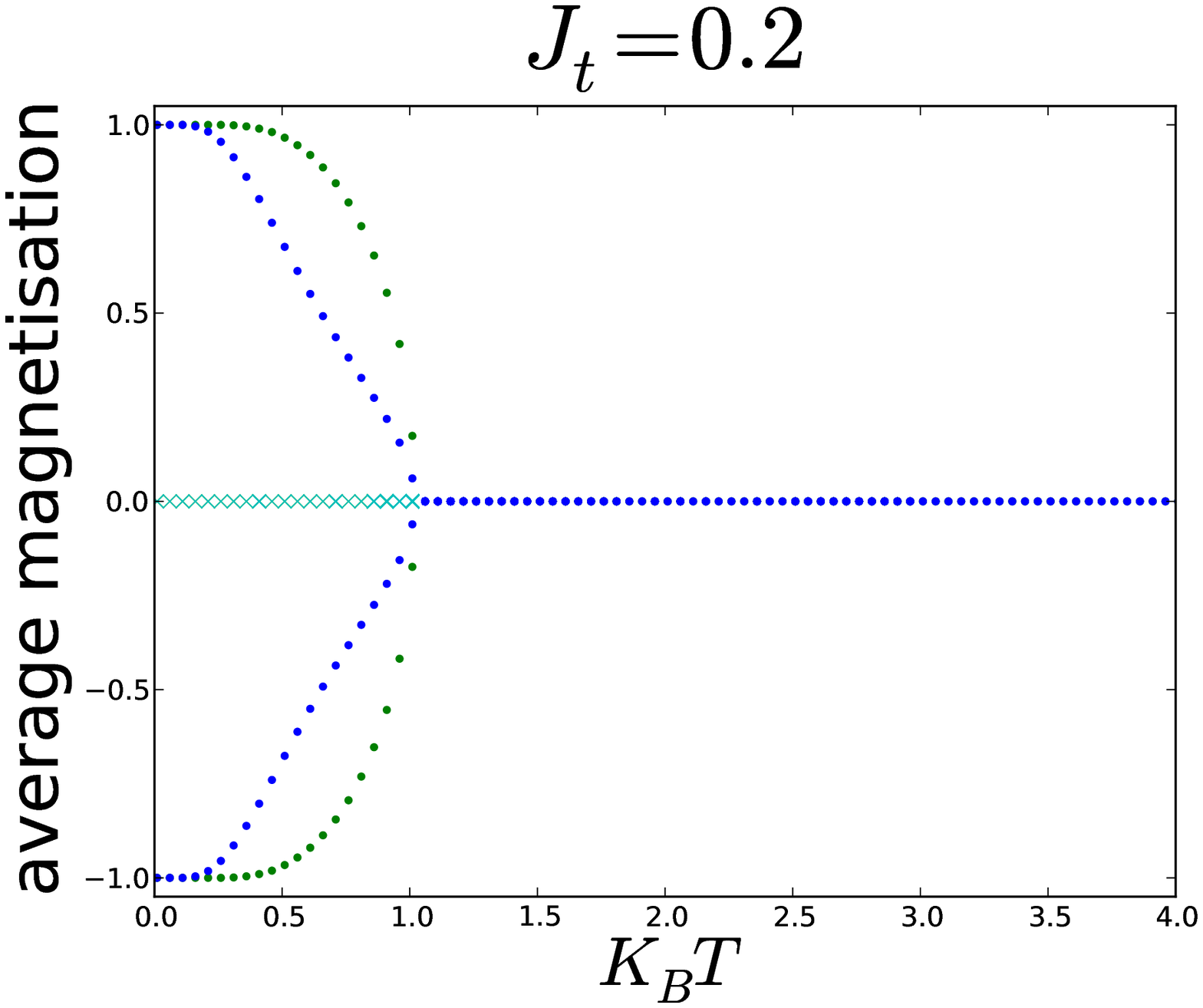}}
\subfloat[]{\includegraphics[width=0.33\textwidth]{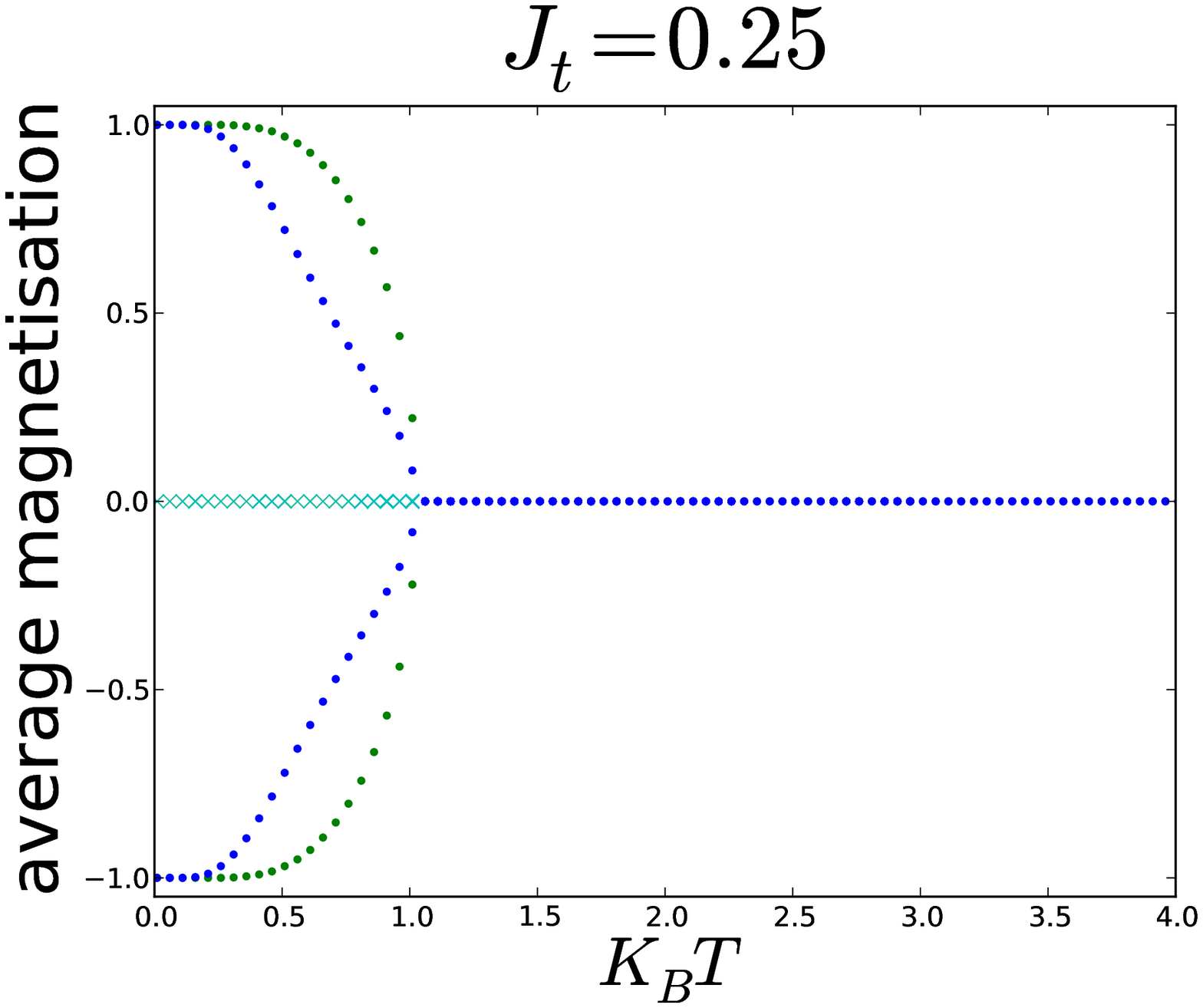}}\\
\subfloat[]{\includegraphics[width=0.33\textwidth]{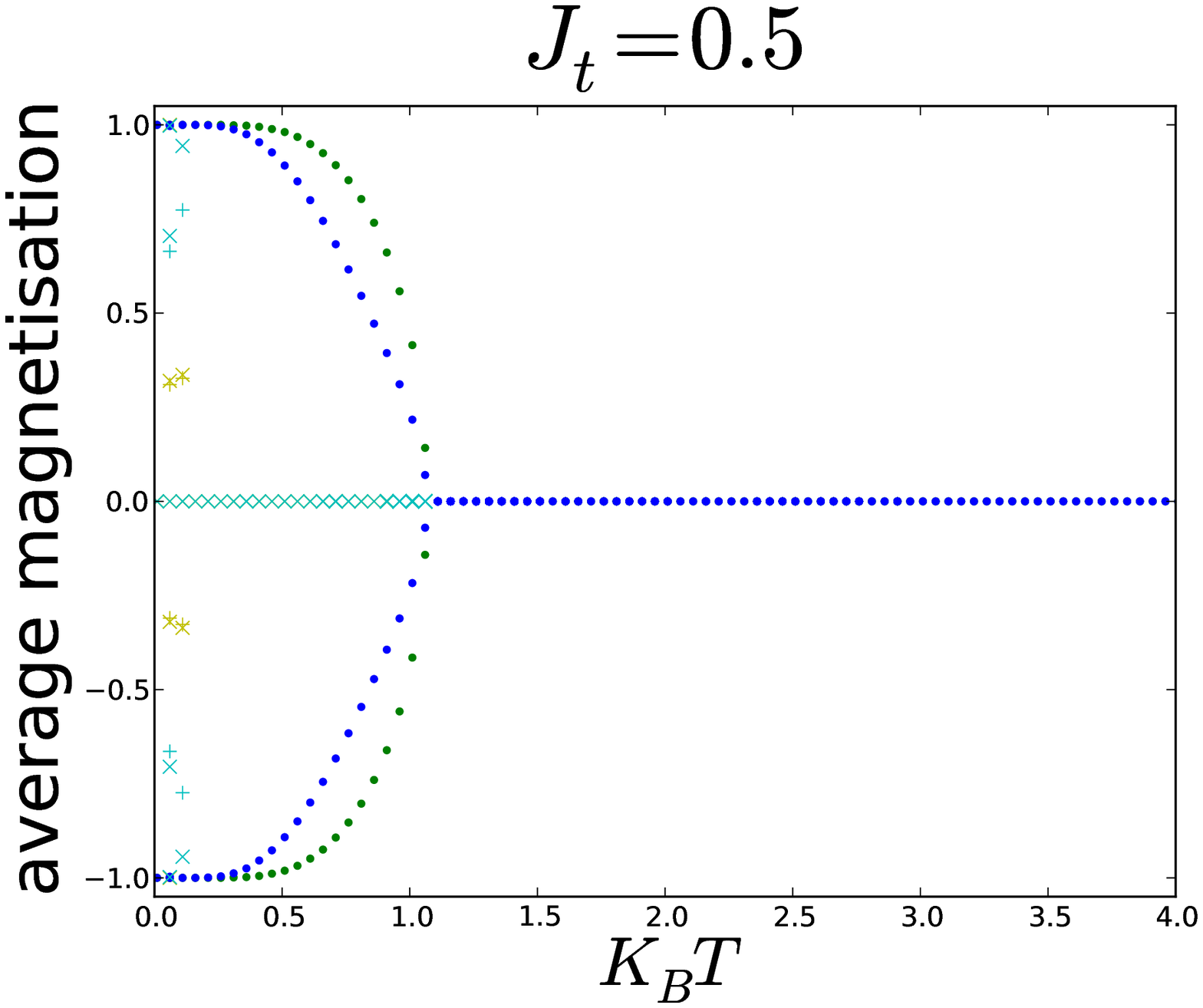}}
\subfloat[]{\includegraphics[width=0.33\textwidth]{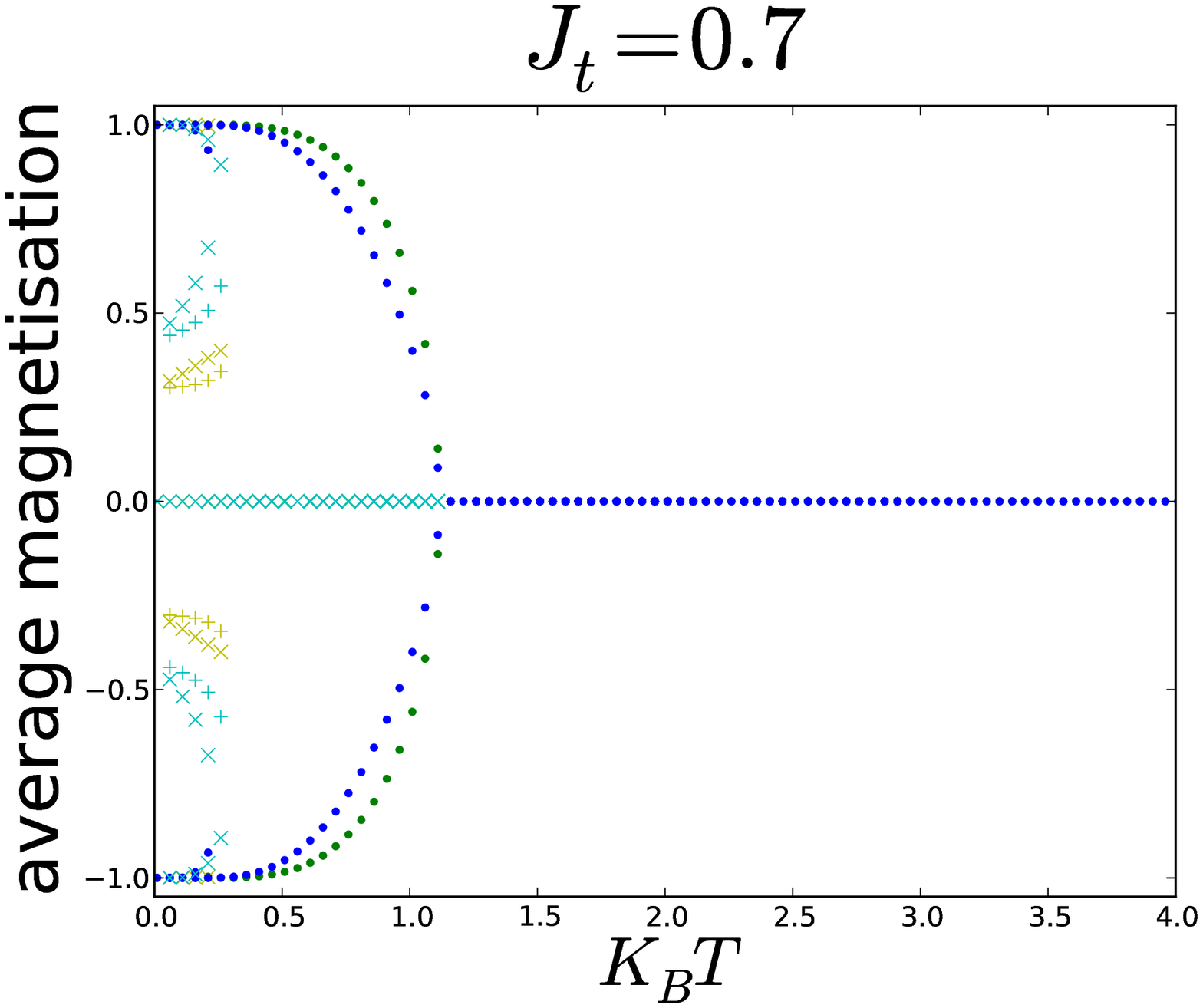}}
\subfloat[]{\includegraphics[width=0.33\textwidth]{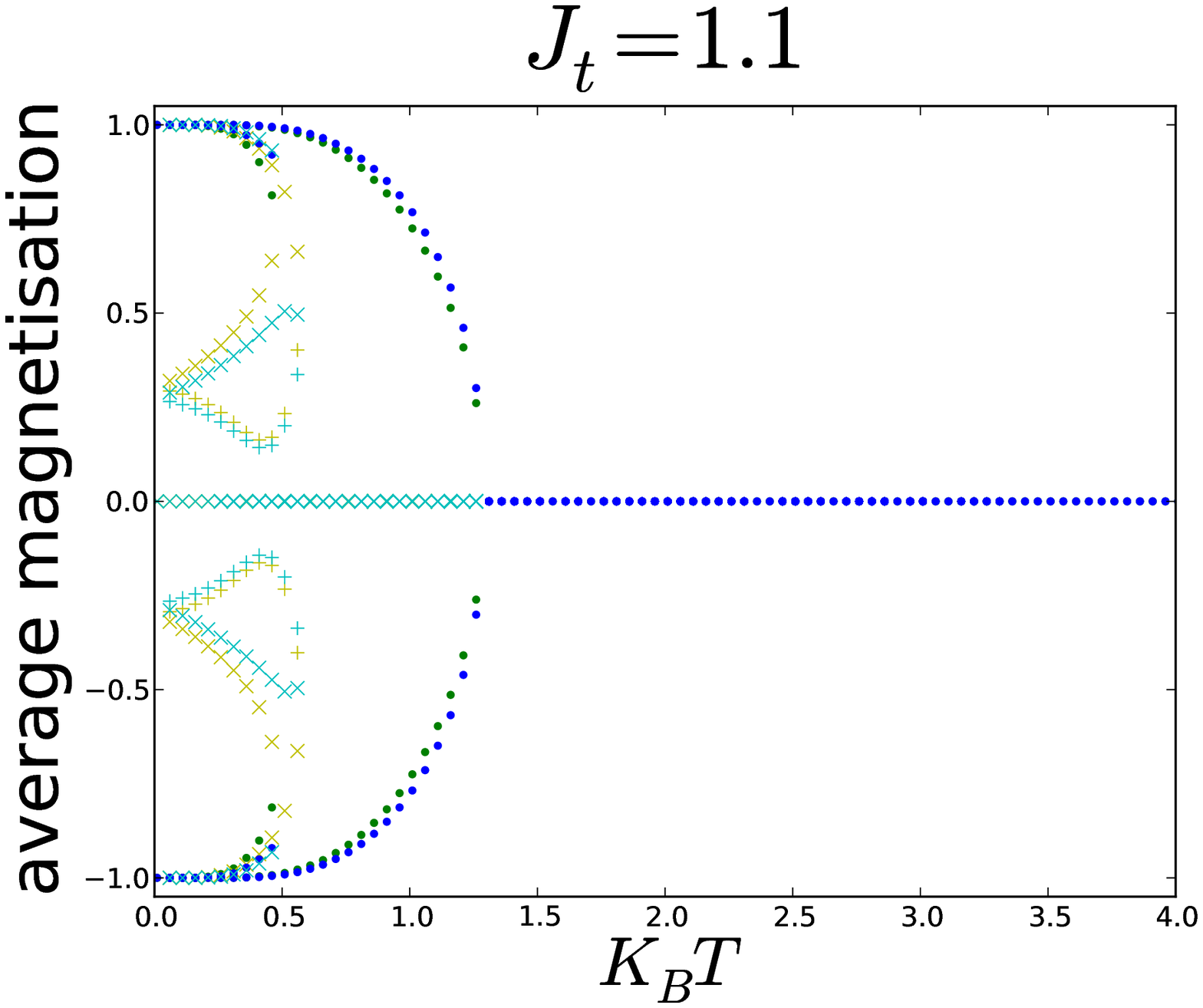}}\\
\subfloat[]{\includegraphics[width=0.33\textwidth]{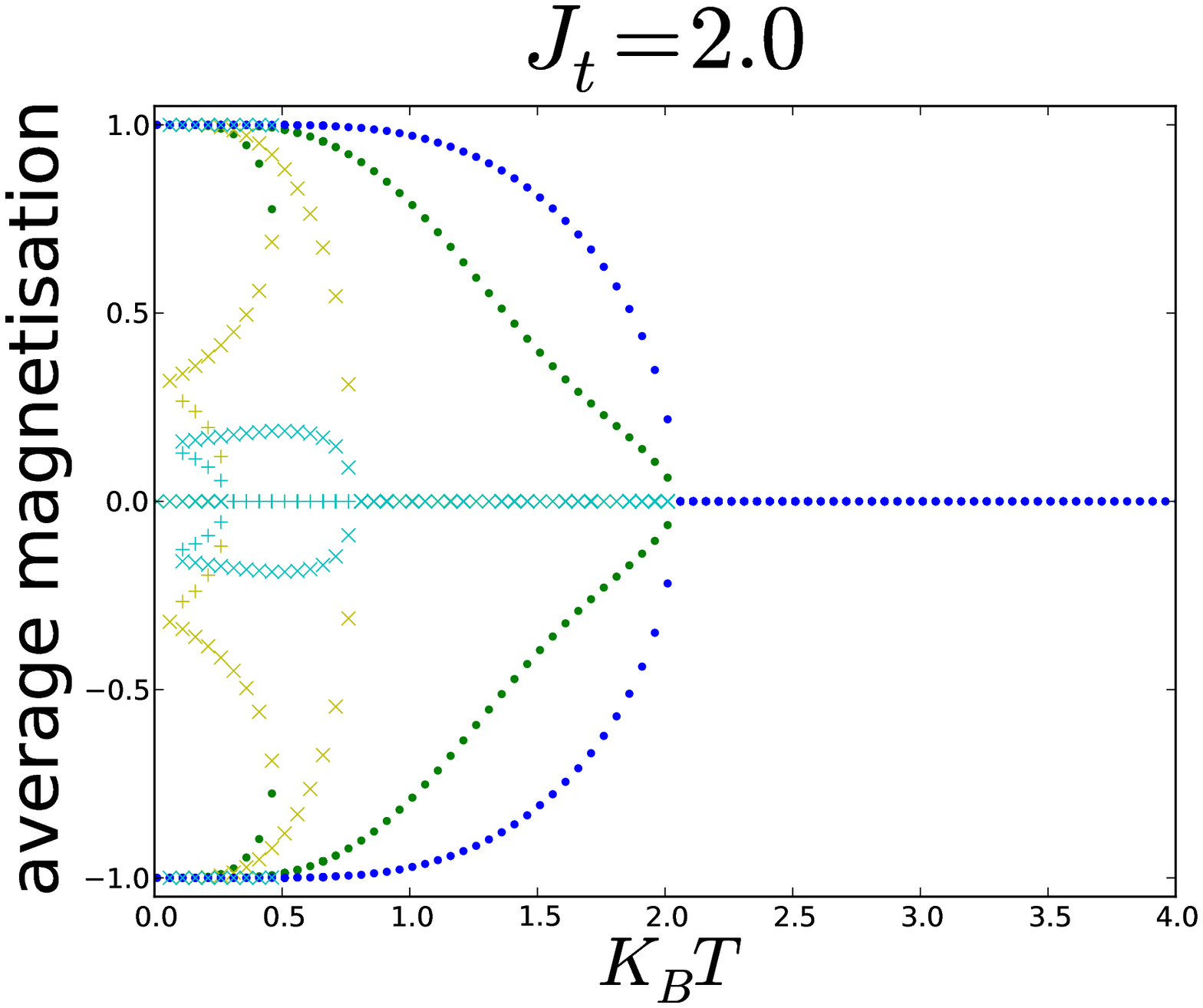}}
\subfloat[]{\includegraphics[width=0.33\textwidth]{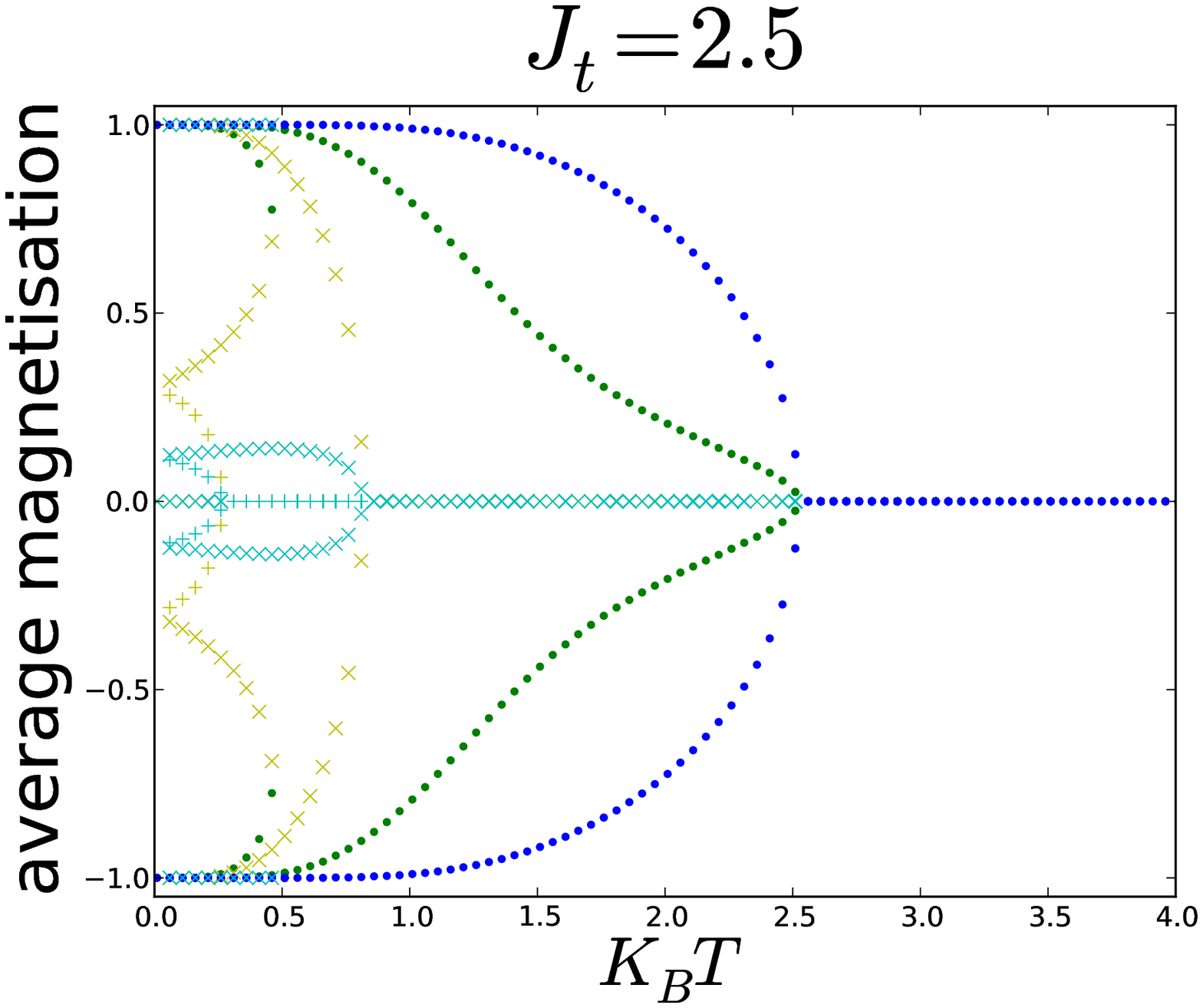}}
\subfloat[]{\includegraphics[width=0.33\textwidth]{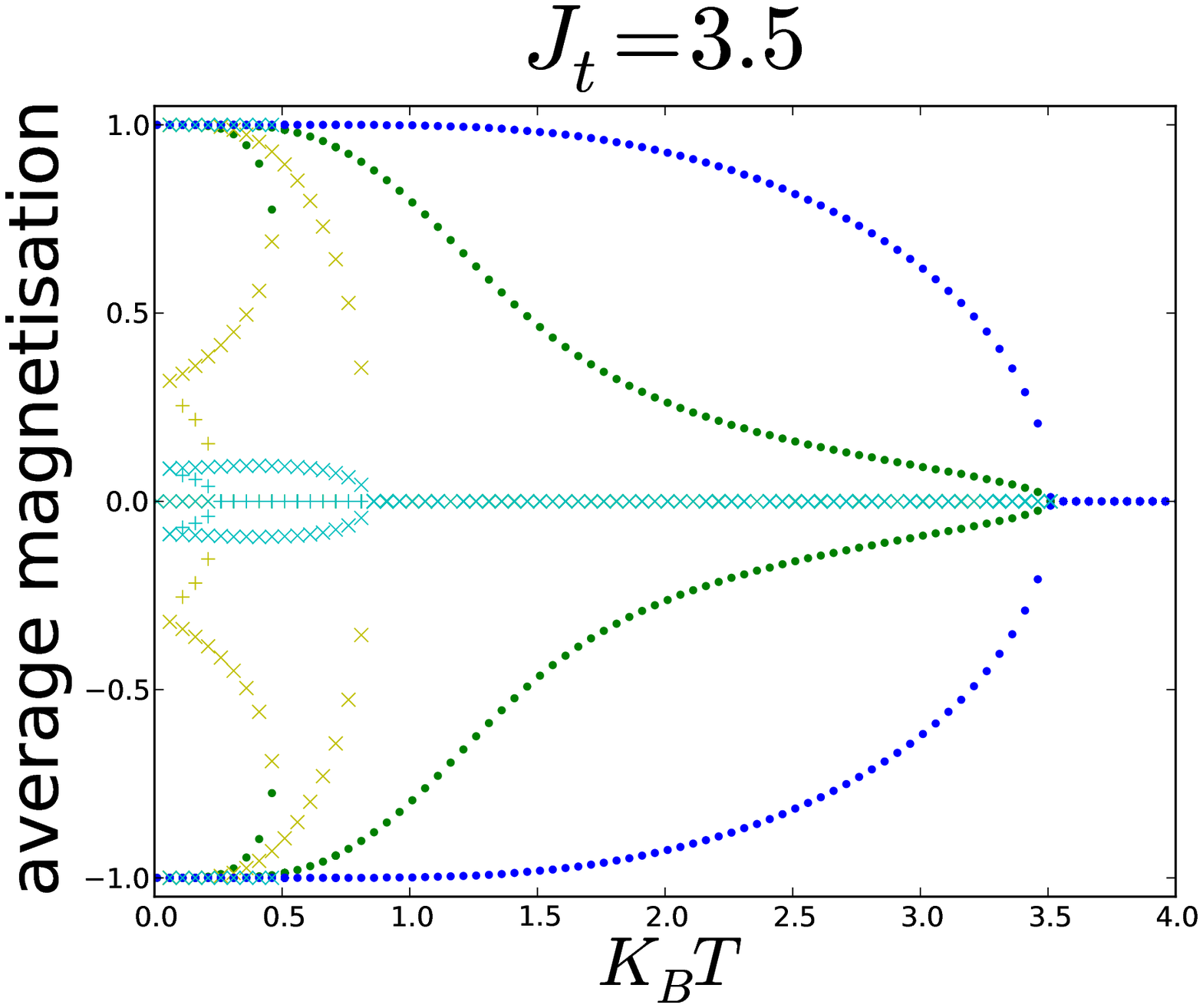}}
\caption{Dependence on temperature of the numerically calculated average magnetisations $(s,t)$ for different values of the intra-coupling $J_{t}$. $J_{s}=1$ and $k= 0.15$  for all plots. (a) $J_{t}=0.05$, (b) $J_{t}=0.2$, (c) $J_{t}=0.25$, (d) $J_{t}=0.5$, (e) $J_{t}=0.7$, (f) $J_{t}=1.1$, (g) $J_{t}=2$, (h) $J_{t}=2.5$ and (i) $J_{t}=3.5$. In all cases, different solutions are plotted for temperatures  between 0.01 and 4 every 0.05 ($K_{B}T$). Magnetisations are plotted in green for $s$ and blue for $t$. Dark points are used for stable solutions and lighter asp ($\times$, for saddle points) or cross ($+$, for maxima) for non stable solutions.}
\label{fig:locanaJT}
\end{figure}

As for how this dependence on $T$ varies as we change either $k$ or $J_{t}$ for fixed values of the rest of the parameters, these are shown in figures \ref{fig:locanakT} and \ref{fig:locanaJT} respectively. The situation is again qualitatively very similar to that of the nonlocal case with the differences already discussed. These are: that there are always stable solutions for all temperatures; that for some range of values the parameters an additional $T_{b'}$ appears at which the type of instability of the paramagnetic phase changes and new ferromagnetic pair of maxima appear; and that there is a temperature $T_{a'}$ bellow which only the main ferromagnetic branch persists (regardless of the stability). Another interesting difference is that there can be saddle point ferromagnetic solutions for some values of the temperatures without there existing any $T_{b}$ or $T_{b'}$ at which the paramagnetic phase changes its stability (figure \ref{fig:locanakT} d to f and figure \ref{fig:locanaJT} d to f).

\subsection{Dependence on inter-coupling}

Considering the dependence of the function $l$ defined in equation \eqref{eq:locTc} on the inter-coupling $k$, we can rewrite the function's roots as given by

\begin{equation}
k_{c}=\pm K_{B}T\tanh^{-1}\left(\sqrt{1-\frac{K_{B}T(J_{s}+J_{t})}{J_{s}J_{t}}+\frac{(K_{B}T)^{2}}{J_{s}J_{t}}}\right)
\end{equation}

This means there are two (equal absolute valued and with opposite signs) values of $k$ at which the stability of the paramagnetic phase changes from saddle to maximum/minimum, unless (for $J_{s}>J_{t}$) $K_{B}T_{uc}^{t}=J_{t}<K_{B}T<J_{s}=K_{B}T_{uc}^{s}$ (for which $1-\frac{K_{B}T(J_{s}+J_{t})}{J_{s}J_{t}}+\frac{(K_{B}T)^{2}}{J_{s}J_{t}}<0$), or $K_{B}T>J_{s}+J_{t}$ (for which the argument of the hyperbolic arctangent becomes lager than one in absolute value). The sign of the second derivative for the paramagnetic phase \eqref{eq:locgsspara} will be positive for $K_{B}T>K_{B}T_{uc}^{s}=J_{s}$ ($>J_{t}$). 

We thus recover the exact same interpretation for the uncoupled critical values of the temperature ($T_{uc}^{s}$, $T_{uc}^{t}$) as in the nonlocal case: for temperatures bellow both, the stability of the paramagnetic solution will change once for positive $k$, but is never a minimum (with $k_{c}$ also associated to the appearance of new saddle point ferromagnetic solutions, but not a real critical point); for temperatures between both the paramagnetic phase will be a saddle point for all values of $k$, and only for temperatures above both, does $k_{c}$ become a real critical point at which a second order phase transition takes place. Furthermore, the additional condition imposed by the hyperbolic arctangent marks another significant difference between the nonlocal and locally coupled models, as in the latter for $K_{B}T>J_{s}+J_{t}$, the stability of the paramagnetic phase will change for no value of $k$, but now it is always a minima. In these cases, only the paramagnetic phase is stable regardless of the value of $k$, which was never the case in the nonlocal case.   

\begin{figure}
\centering
\includegraphics[width=\textwidth]{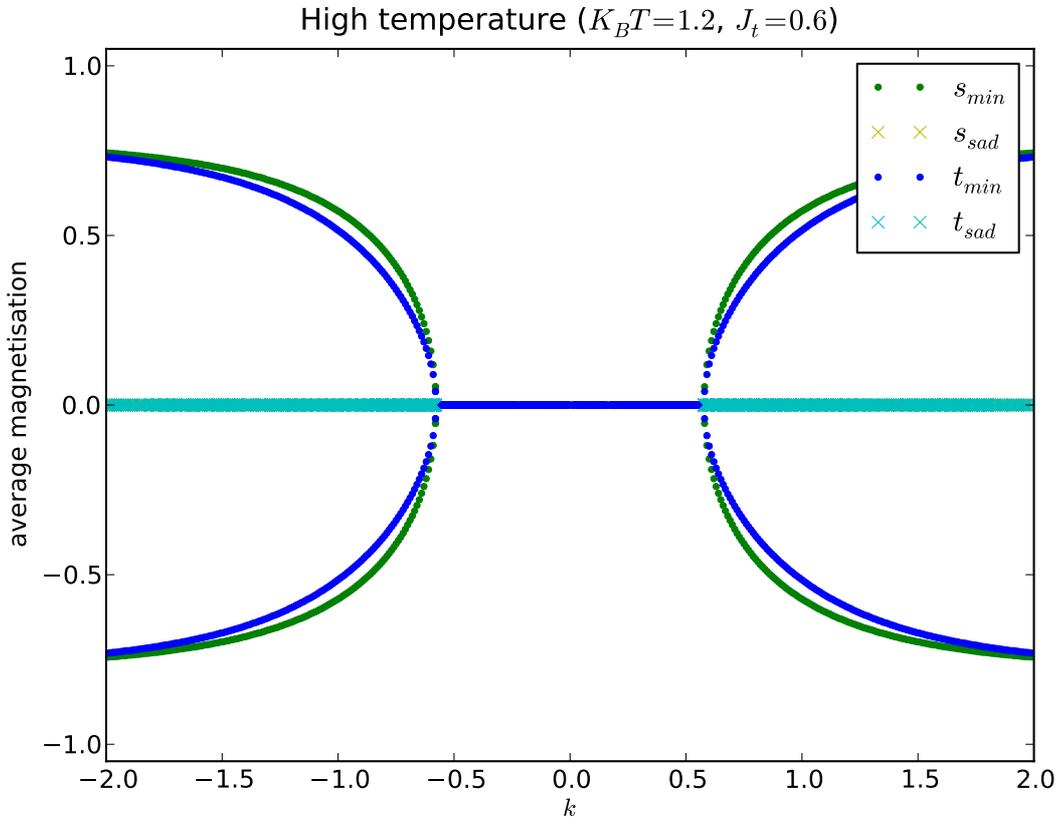}
\caption{Dependence on the inter-coupling $k$ of the numerically calculated average magnetisations at high temperature for $J_{s}=1$, $J_{t}= 0.6$ and  $K_{B}T = 1.2$ ($k_{c}=\pm 0.58$). Different solutions are plotted for $k$  between -2 and 2 every 0.01. Magnetisations are plotted in green for $s$ and blue for $t$. Dark points are used for stable solutions and lighter asps ($\times$) for saddle point, non stable solutions.}
\label{fig:locanak1}
\end{figure}

Figure \ref{fig:locanak1} shows what we have called {\itshape high temperature} case (although not as high as to make the paramagnetic phase the only possible stable solution regardless of $k$), i.e., $J_{t}<J_{s}<K_{B}T<J_{s}+J_{t}$. In particular, $J_{s}=1$, $J_{t}= 0.6$ , $K_{B}T = 1.2$ and so $k_{c}=\pm 0.58$. Note qualitatively the situation is identical to the nonlocal case, besides the fact that solutions will not loose stability at high values of $k$ (recall the strong coupling regime is always unstable in the nonlocal case studied in last chapter). Critical values $k_{c}$ represent points where a second order phase transition takes place. Note that the actual absolute value in this case is larger than that of the nonlocal case (for which $k_{c}=\pm 0.35$).

\begin{figure}
\centering
\includegraphics[width=\textwidth]{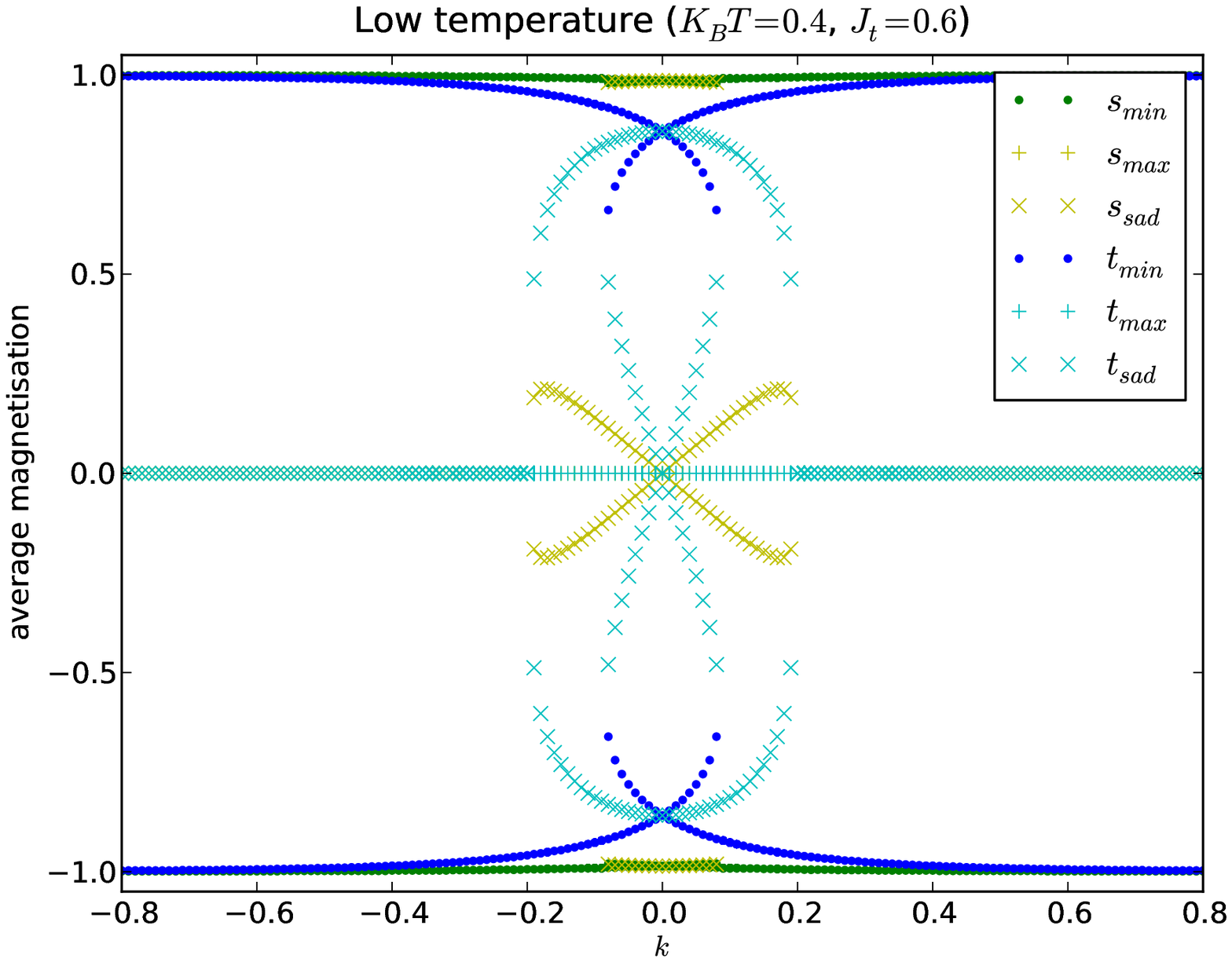}
\caption{Dependence on the inter-coupling $k$ of the numerically calculated average magnetisations at low temperature $J_{s}=1$, $J_{t}= 0.6$ and  $K_{B}T = 0.4$ ($k_{c}=\pm0.19$). Different solutions are plotted for $k$  between -0.8 and 0.8 every 0.01. Magnetisations are plotted in green for $s$ and blue for $t$. Dark points are used for stable solutions and lighter asp ($\times$, for saddle points) or cross ($+$, for maxima) for non stable solutions. }
\label{fig:locanak2} 
\end{figure}

Figure \ref{fig:locanak2} shows what we have called {\itshape low temperature} case ($K_{B}T<J_{t}$) for $J_{s}=1$, $J_{t}= 0.6$ , $K_{B}T = 0.4$ ($k_{c}=\pm0.19$). Qualitatively, this is again the same situation depicted in the nonlocal case (with stable solutions for all values of $k$). The absolute value of $k_{c}$ is however significantly lower in the local case ($k_{c}=\pm0.35$ in the nonlocal case). 

\begin{figure}
\centering
\subfloat[]{\includegraphics[width=0.33\textwidth]{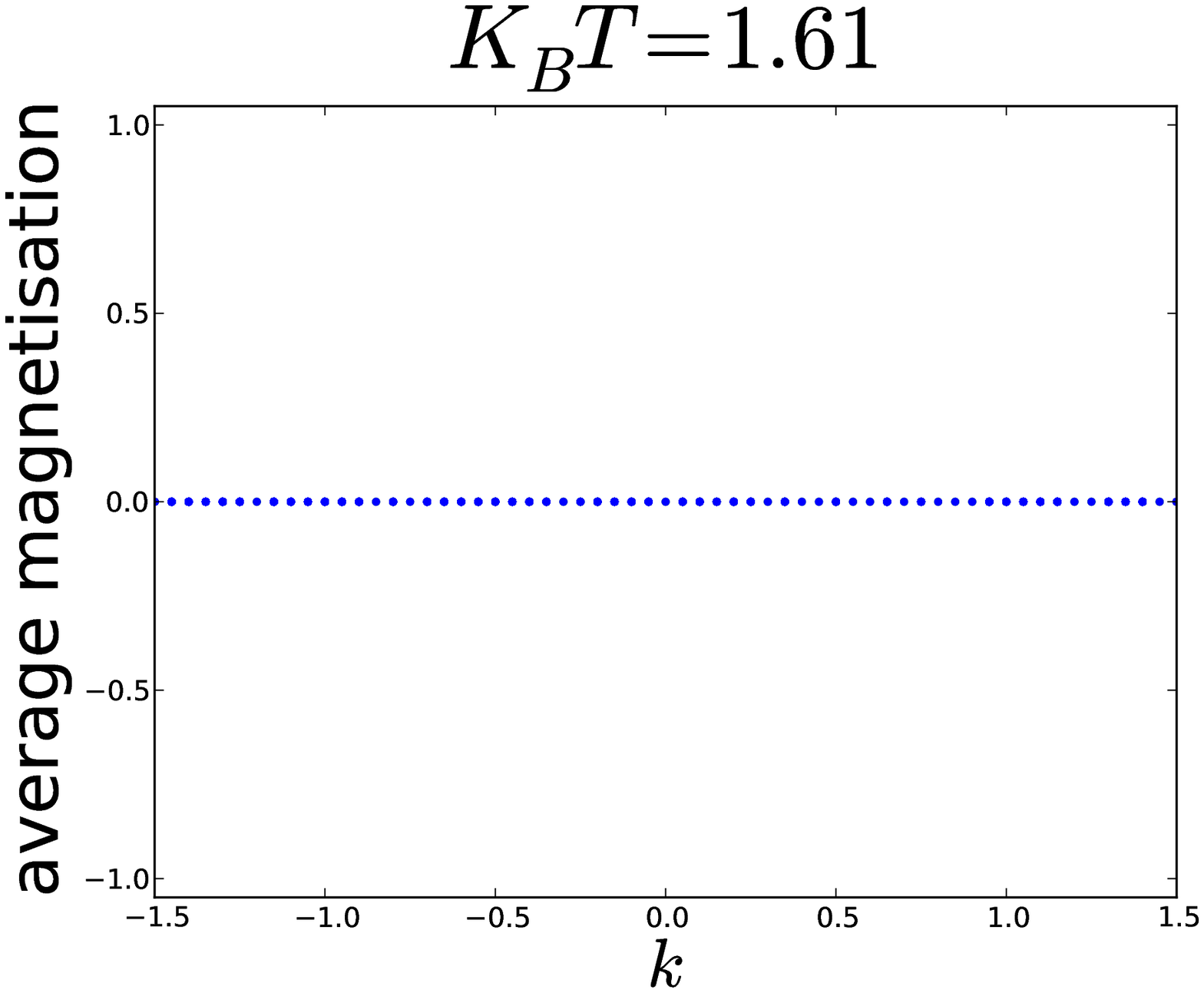}}
\subfloat[]{\includegraphics[width=0.33\textwidth]{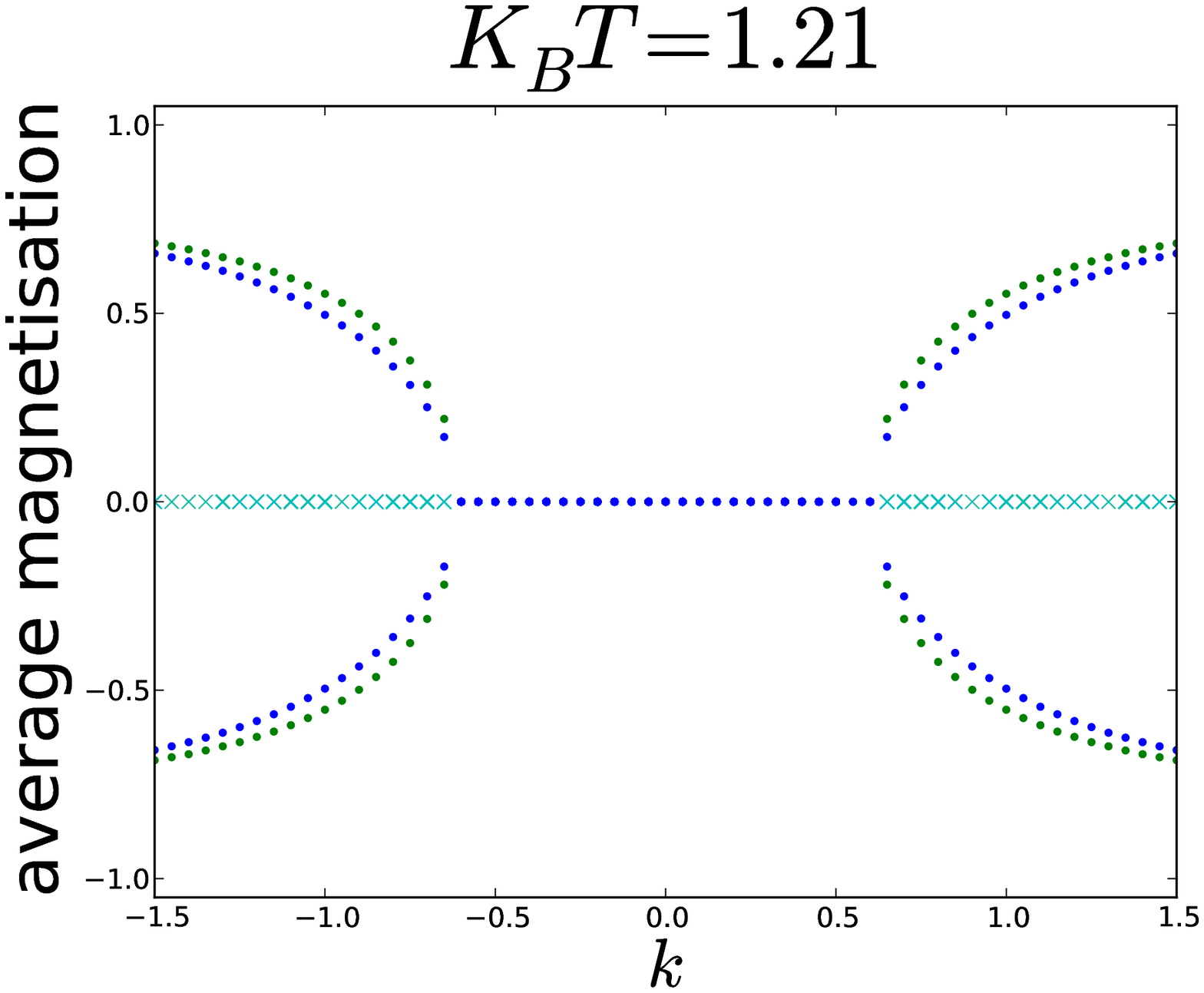}}
\subfloat[]{\includegraphics[width=0.33\textwidth]{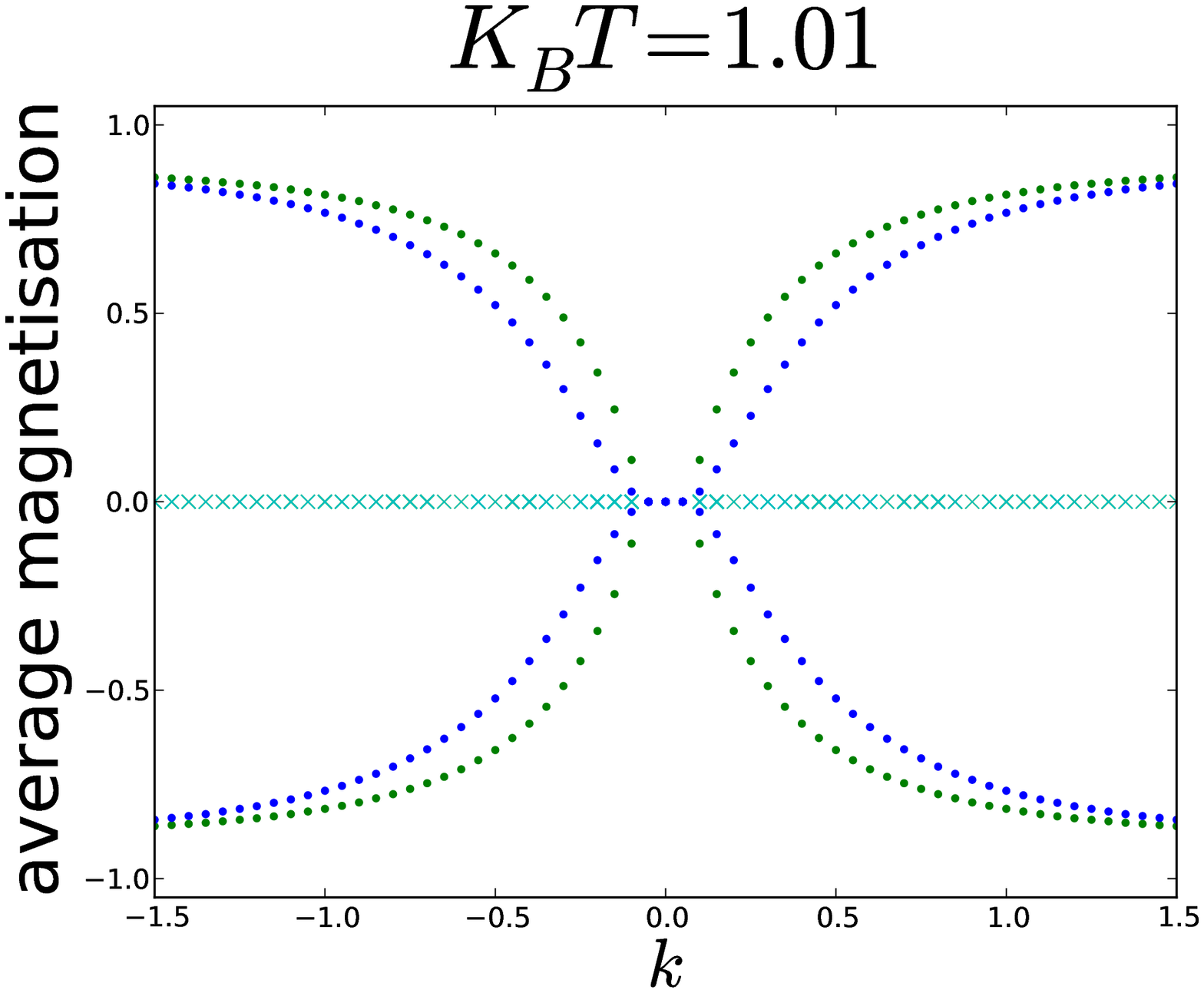}}\\
\subfloat[]{\includegraphics[width=0.33\textwidth]{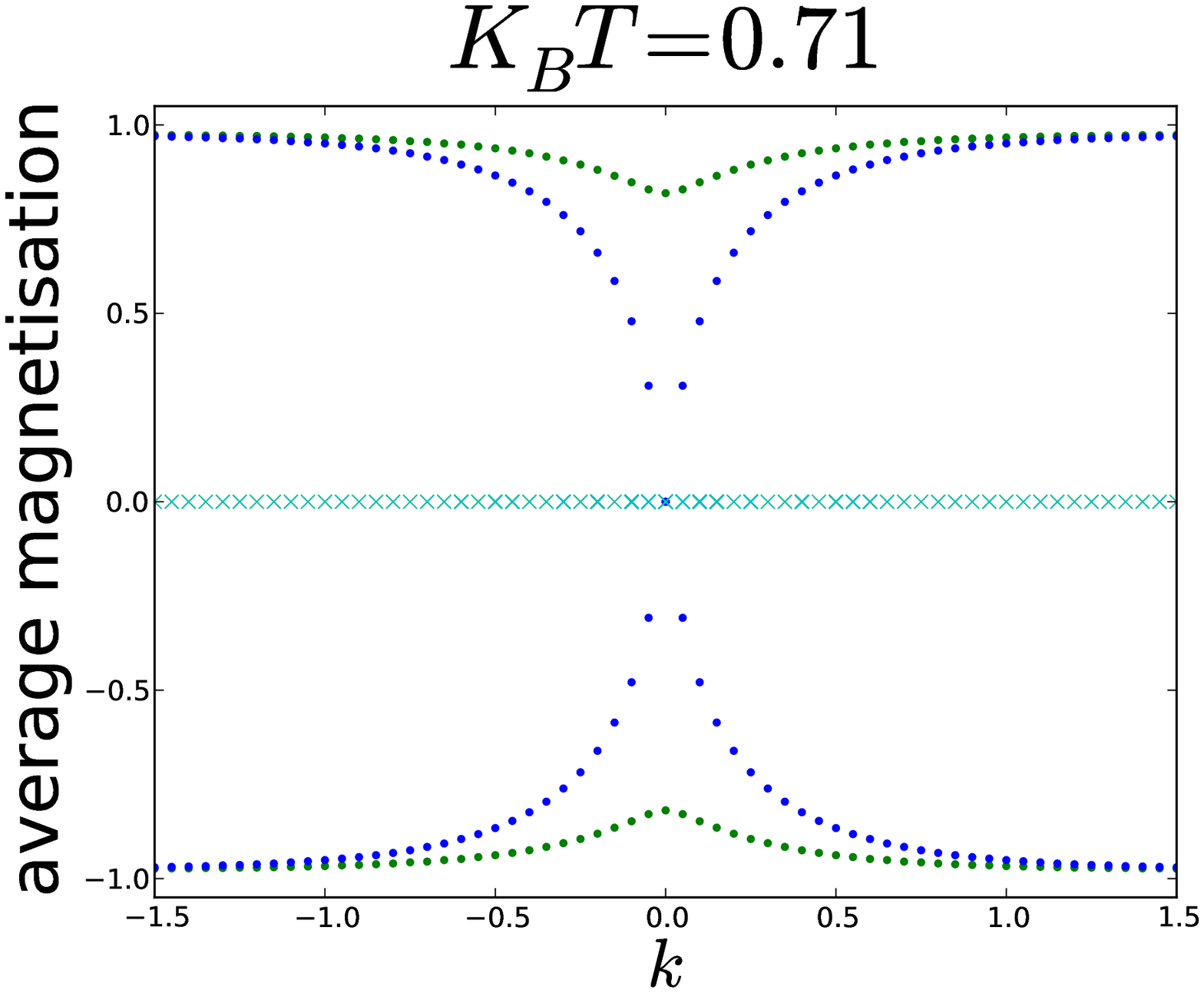}}
\subfloat[]{\includegraphics[width=0.33\textwidth]{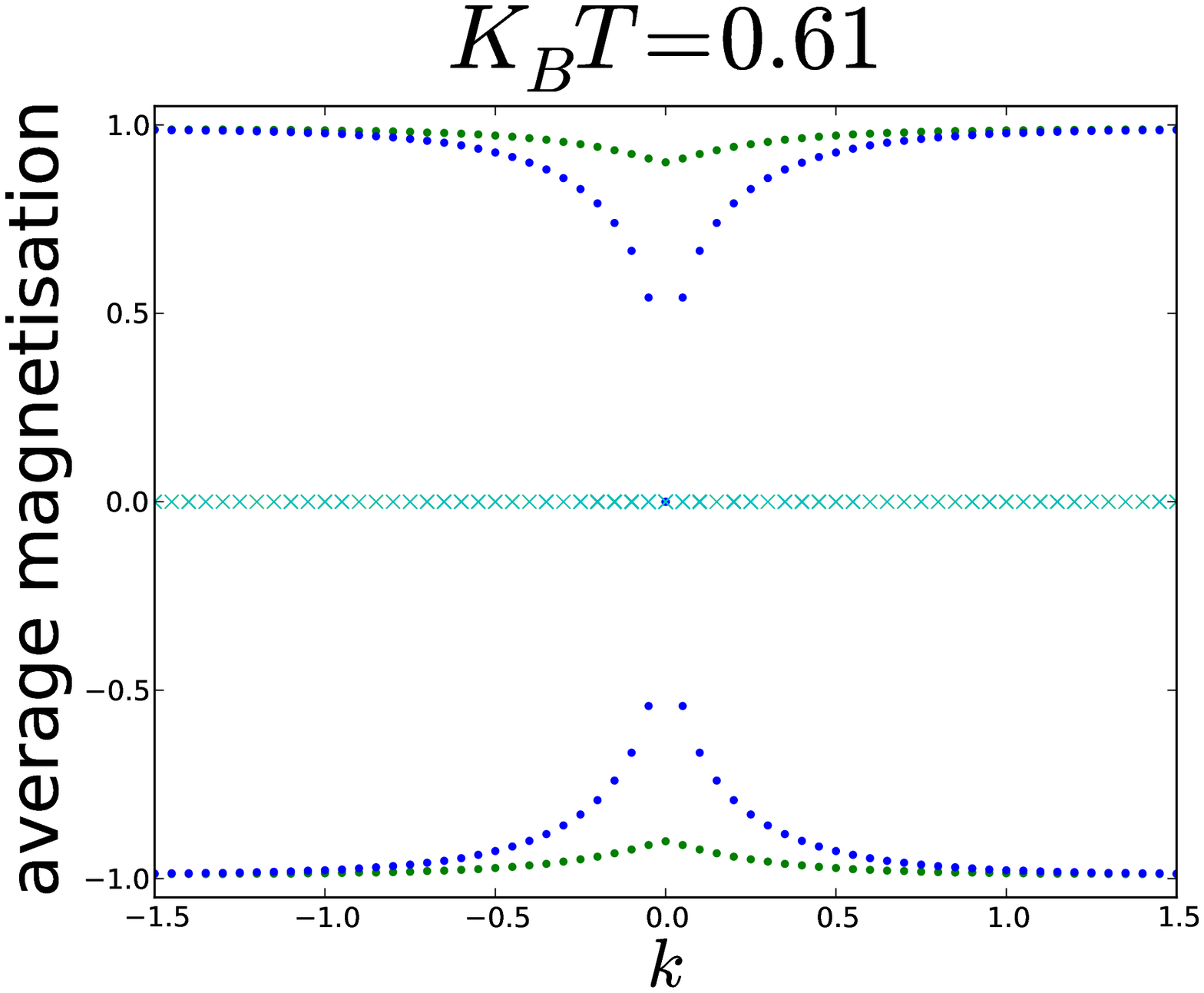}}
\subfloat[]{\includegraphics[width=0.33\textwidth]{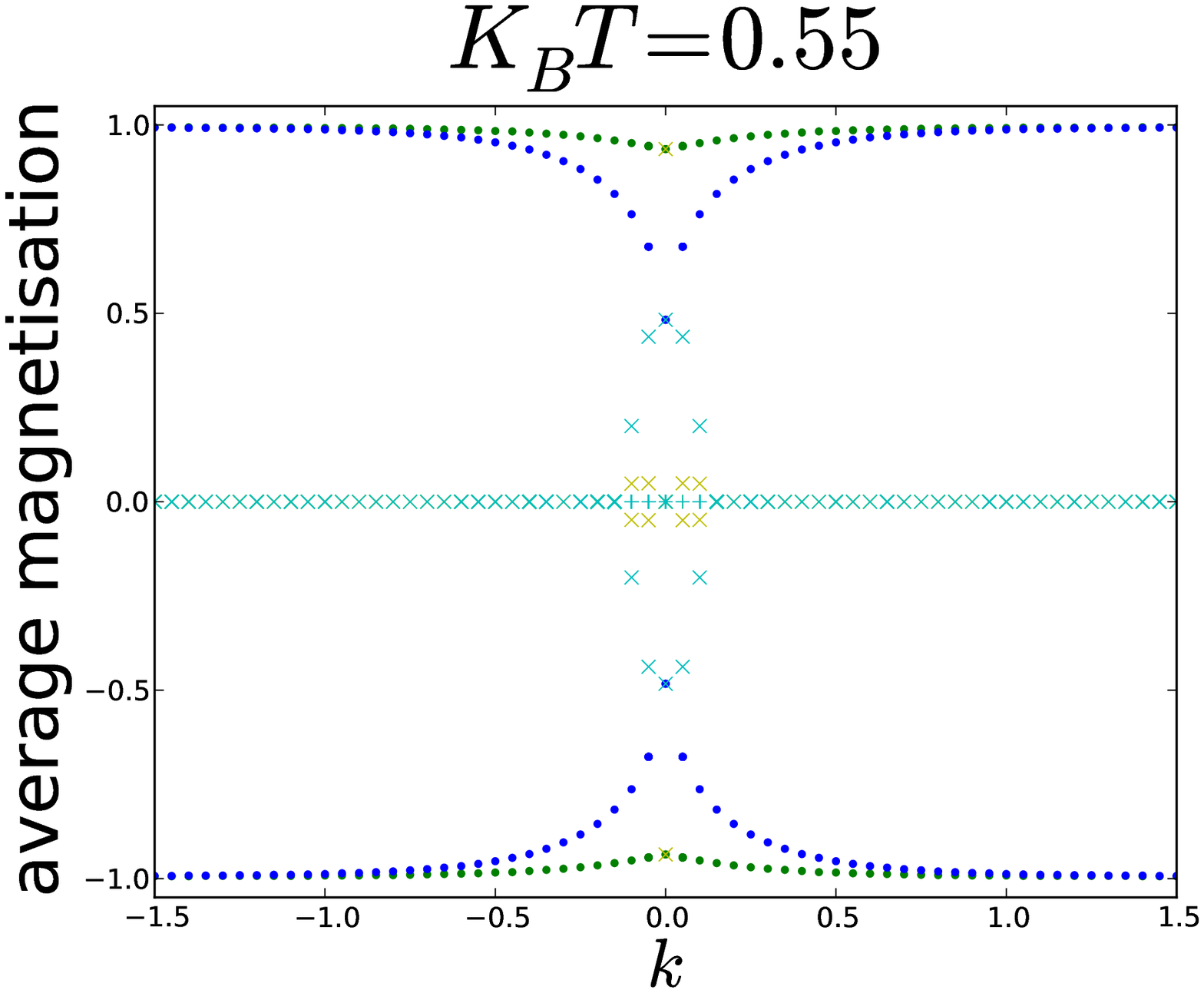}}\\
\subfloat[]{\includegraphics[width=0.33\textwidth]{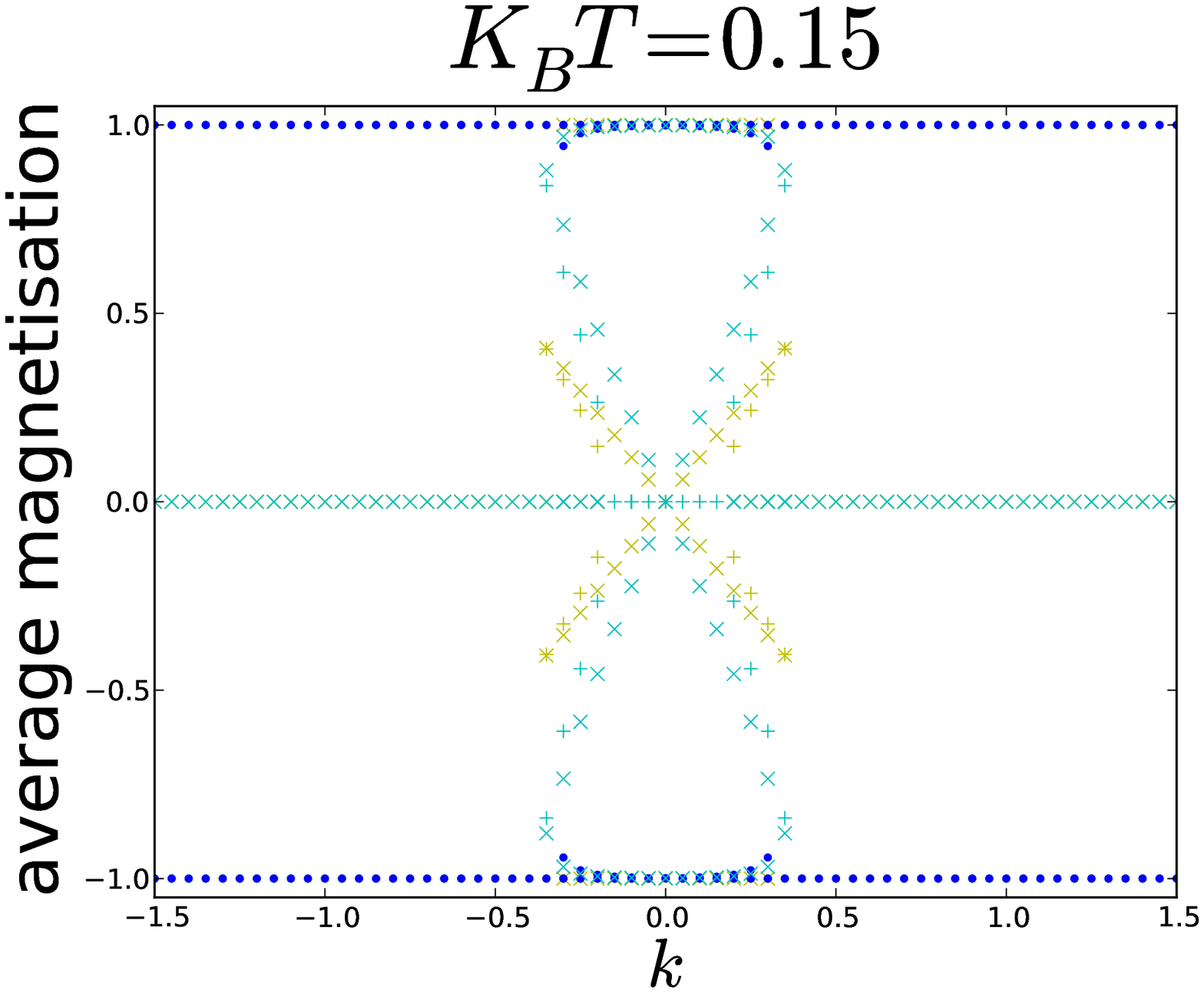}}
\subfloat[]{\includegraphics[width=0.33\textwidth]{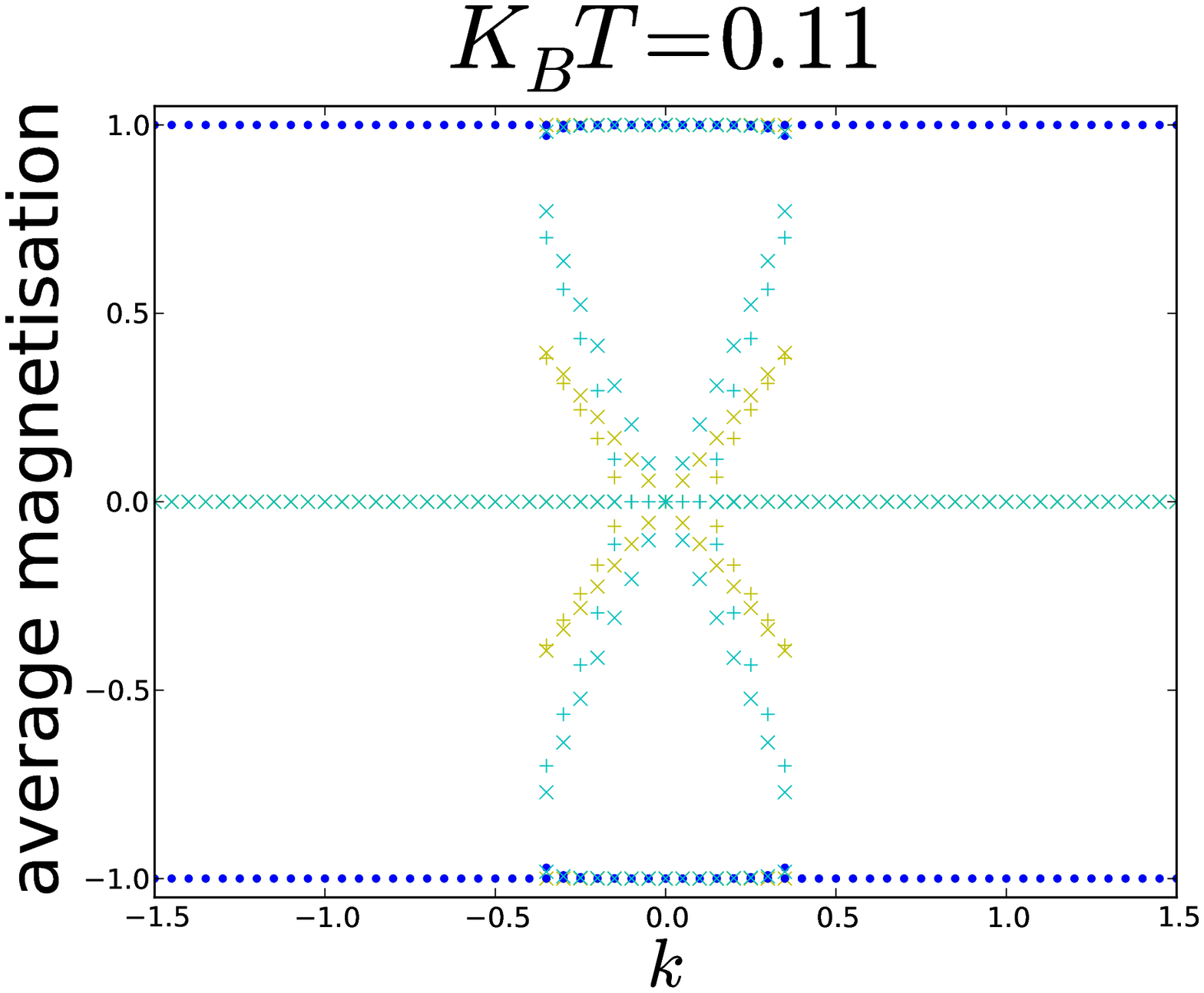}}
\subfloat[]{\includegraphics[width=0.33\textwidth]{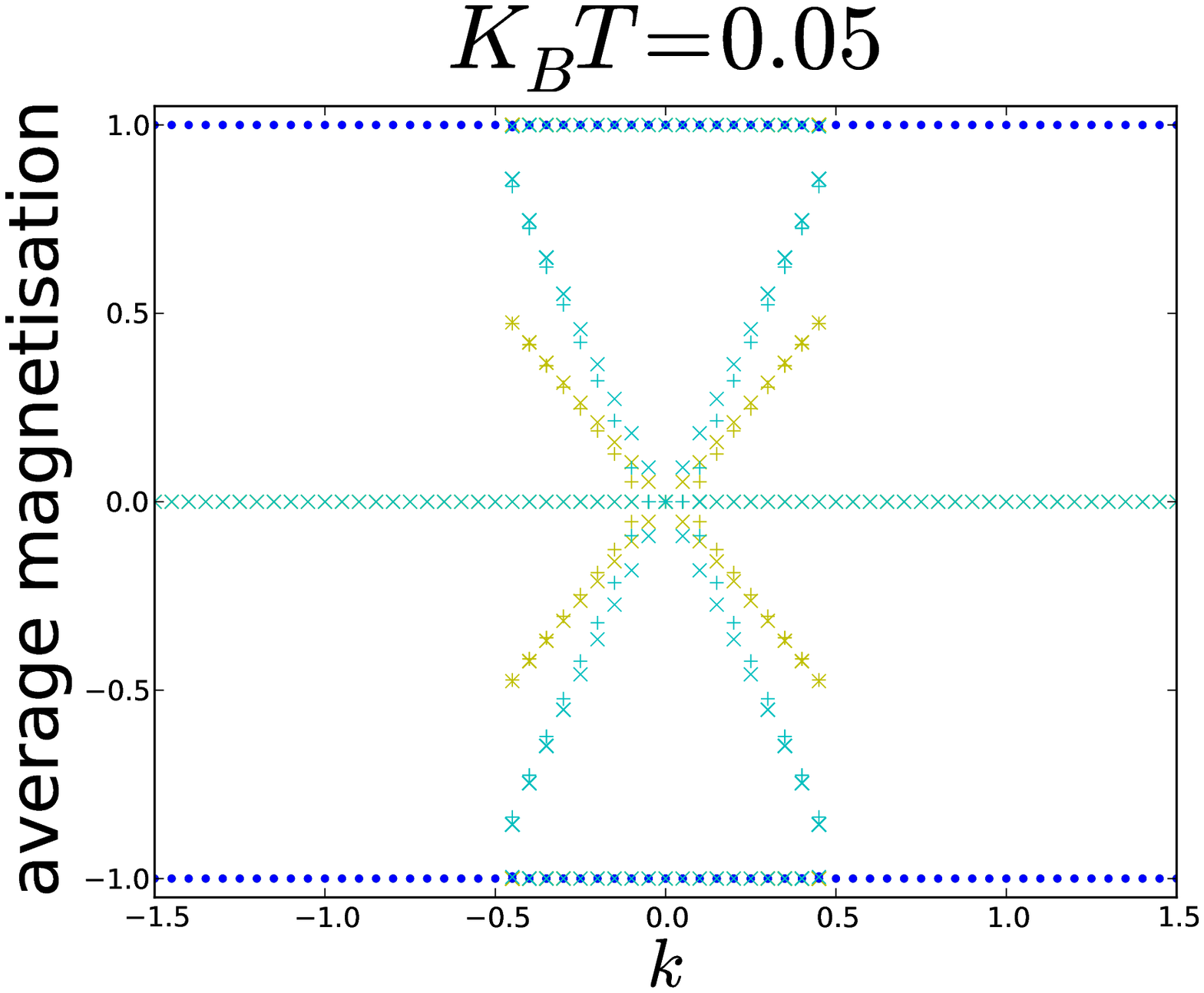}}
\caption{Dependence on inter-coupling $k$ of the numerically calculated average magnetisations $(s,t)$ for different values of the temperature $K_{B}T$. $J_{s}=1$ and $k=0.15$  for all plots. (a) $K_{B}T=1.61$, (b) $K_{B}T=1.21$, (c) $K_{B}T=1.01$, (d) $K_{B}T=0.71$, (e) $K_{B}T=0.61$, (f) $K_{B}T=0.55$, (g) $K_{B}T=0.15$, (h) $K_{B}T=0.11$ and (i) $K_{B}T=0.05$. In all cases, different solutions are plotted for inter-coupling  between -1.5 and 1.5 every 0.05 ($K_{B}T$). Magnetisations are plotted in green for $s$ and blue for $t$. Dark points are used for stable solutions and lighter asp ($\times$, for saddle points) or cross ($+$, for maxima) for non stable solutions.}
\label{fig:locanaTk}
\end{figure}

As expected from these two cases studied, the variation of the dependence in $k$ with $T$ for fixed values of the rest of the parameters, will be qualitatively similar to the nonlocal case, as shown in figure \ref{fig:locanaTk}. There are, however, some differences. As we have already noted, there is no change in stability at high values of $k$ in this case. At a high enough temperature, only the paramagnetic phase is stable for all values of $k$ (figure \ref{fig:locanaTk} a), which as we have seen was never the case in the nonlocal case. Another difference appears at very low temperatures, where additional ferromagnetic maxima appear around $k=0$ (figure \ref{fig:locanaTk} g, h and i), which never exist in the nonlocal case. At this point, it is no longer true that non stable ferromagnetic solutions exist only for $|k|<|k_{c}|$ either.

\begin{figure}
\centering
\subfloat[]{\includegraphics[width=0.33\textwidth]{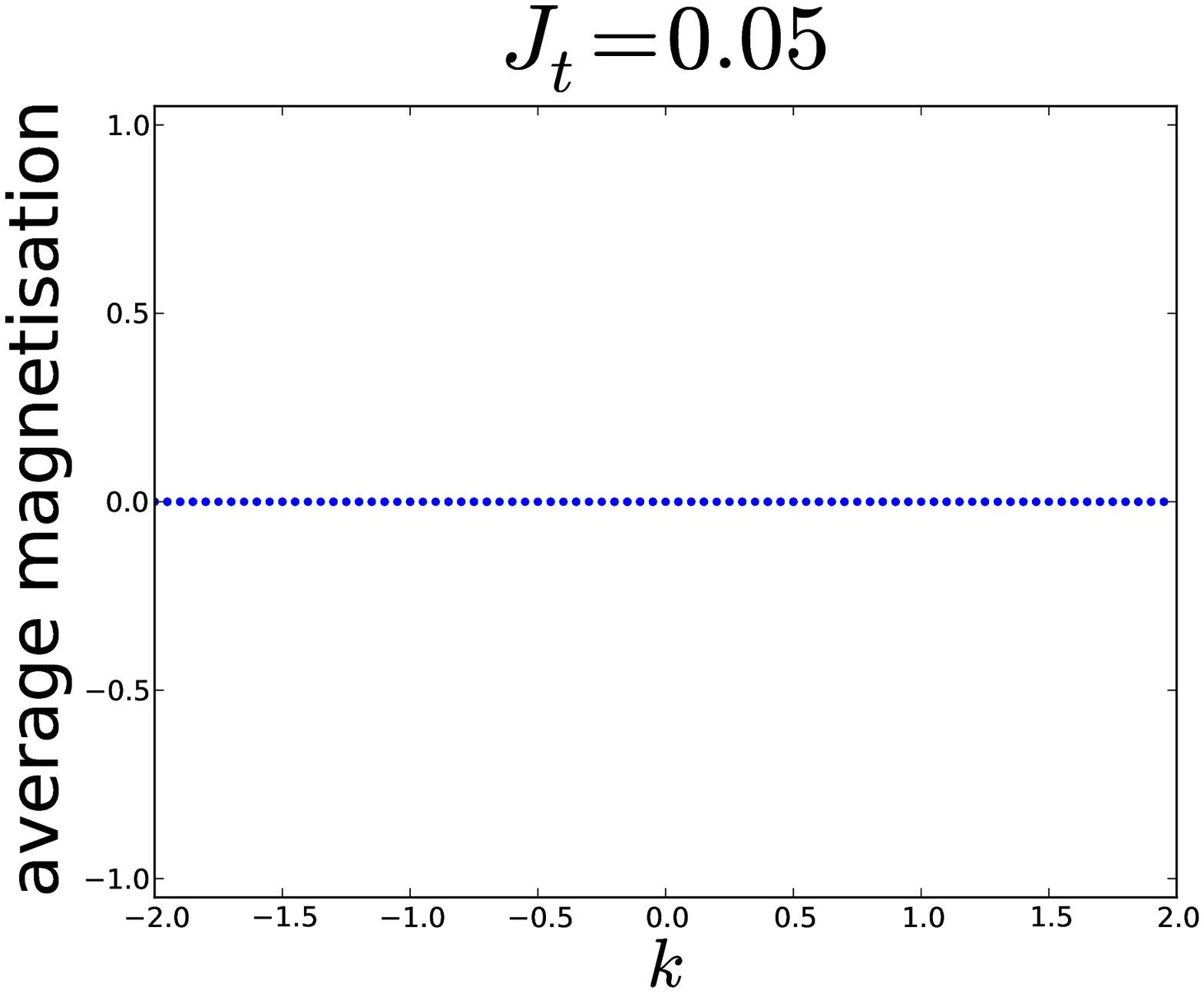}}
\subfloat[]{\includegraphics[width=0.33\textwidth]{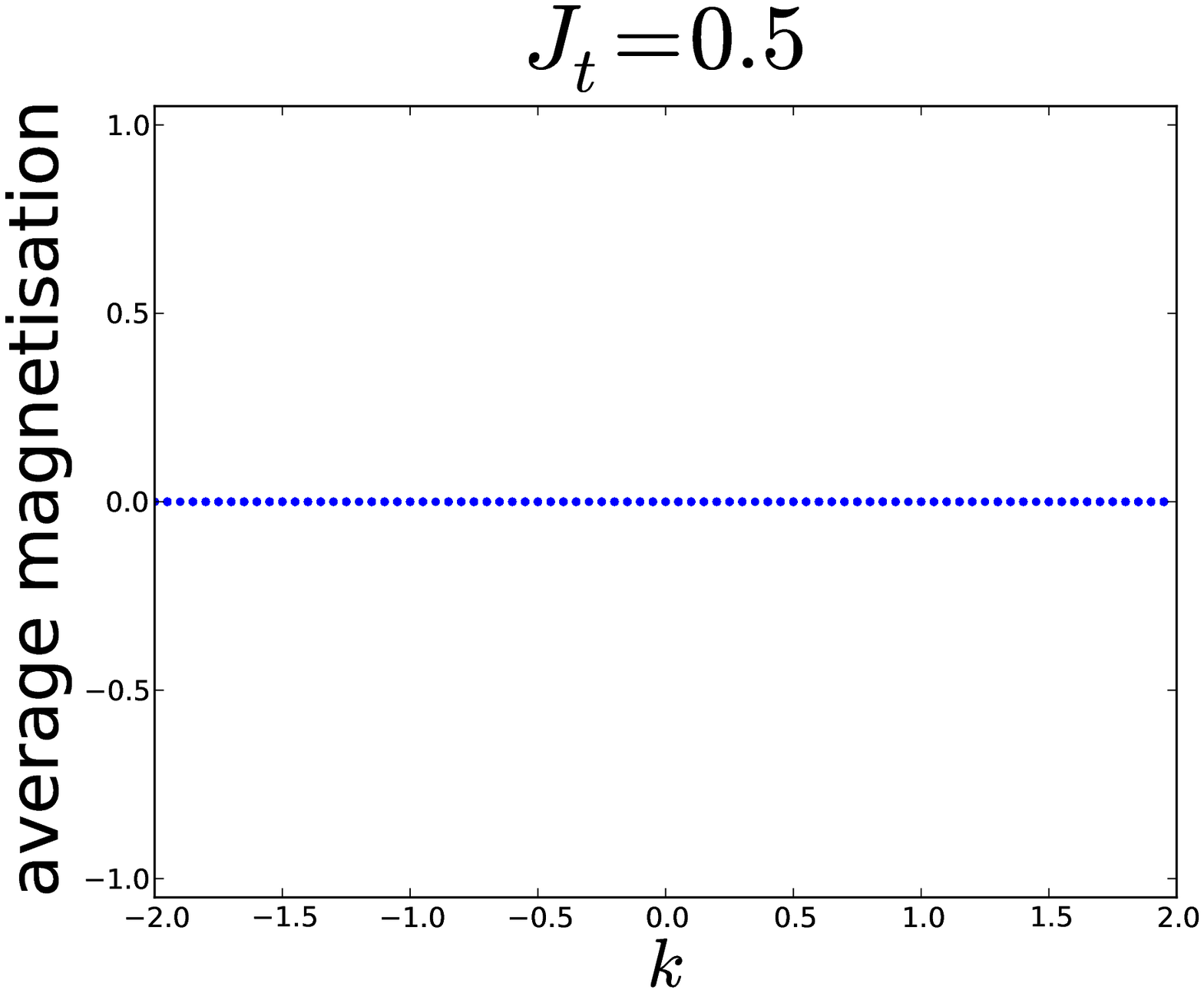}}
\subfloat[]{\includegraphics[width=0.33\textwidth]{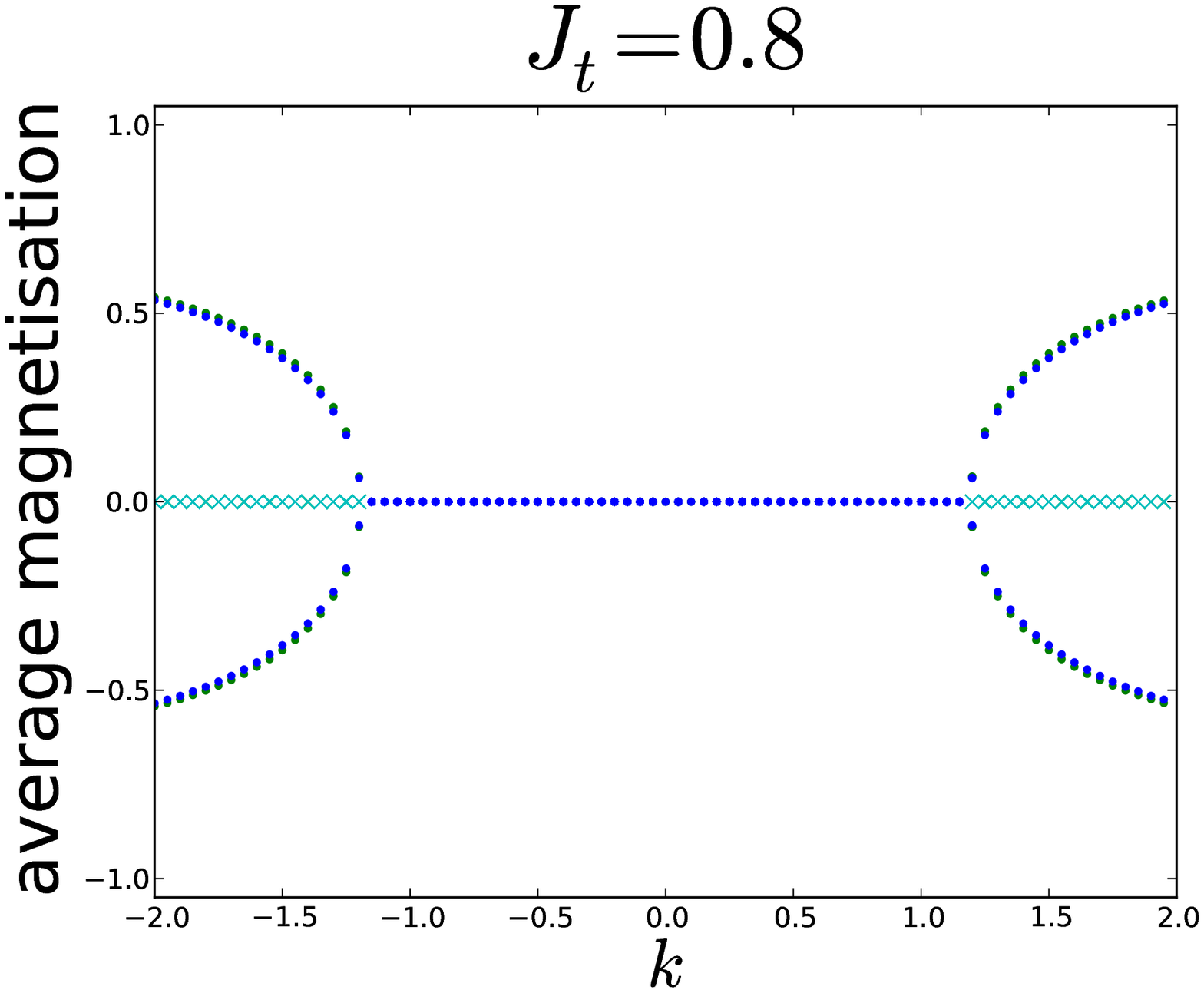}}\\
\subfloat[]{\includegraphics[width=0.33\textwidth]{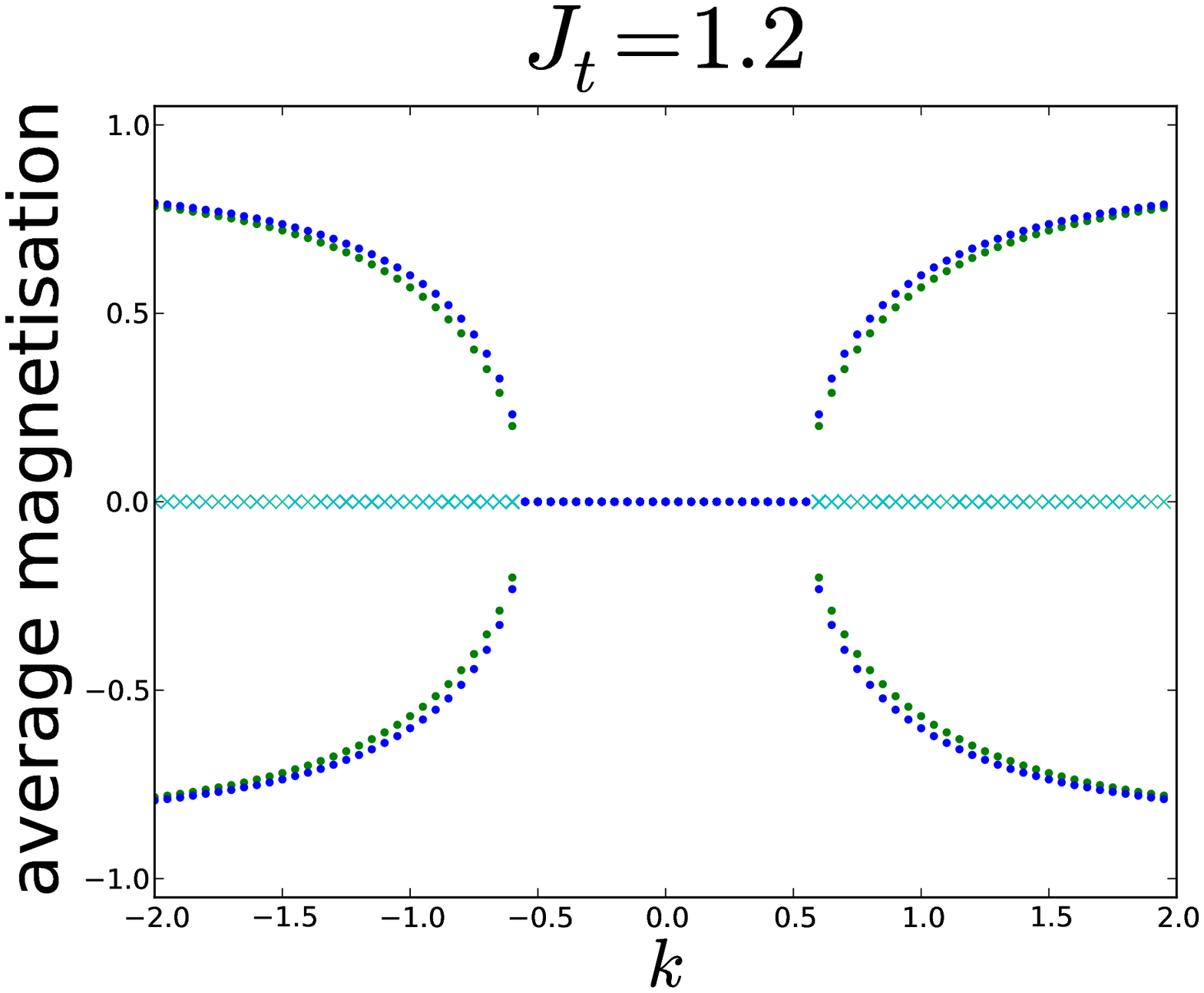}}
\subfloat[]{\includegraphics[width=0.33\textwidth]{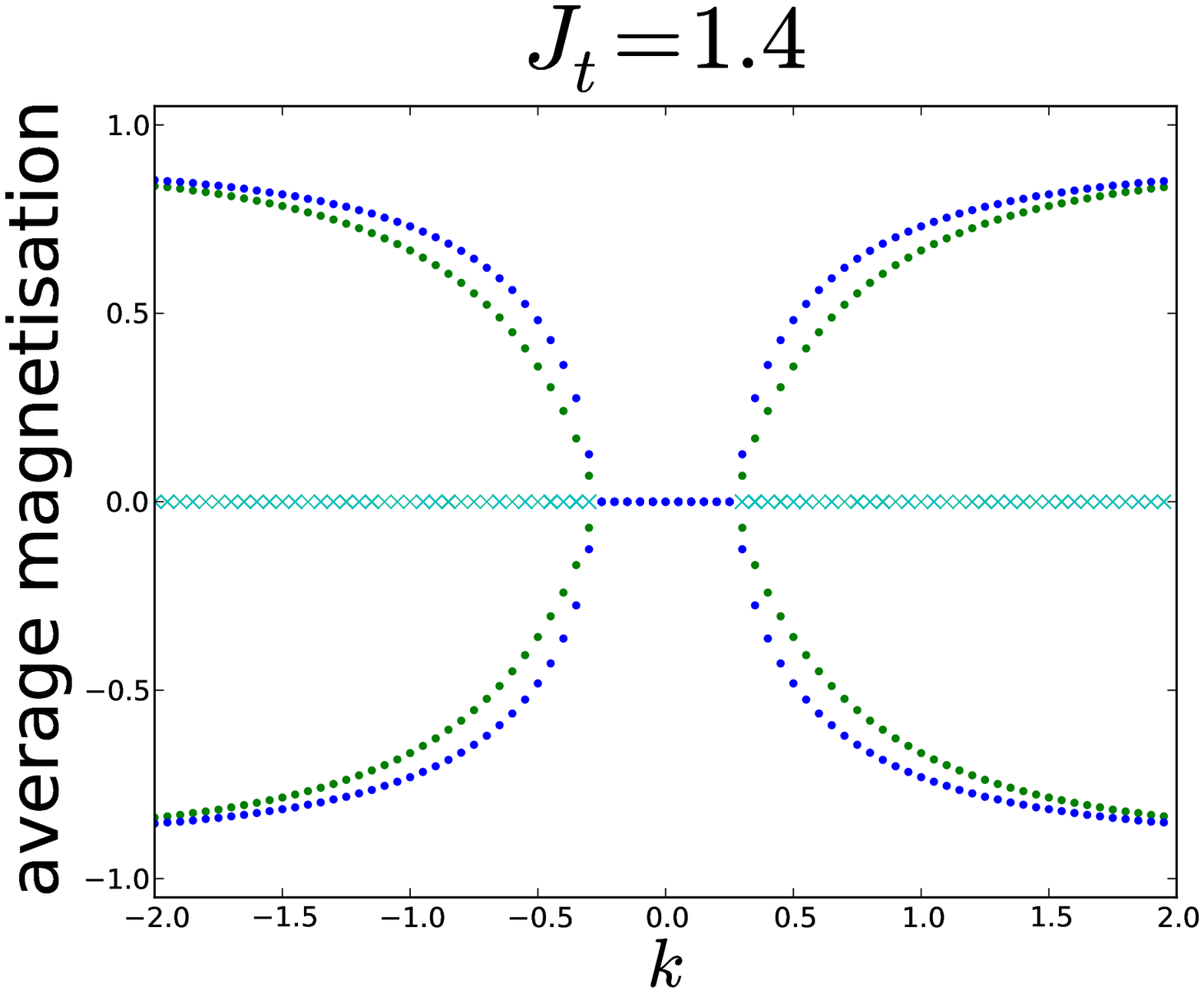}}
\subfloat[]{\includegraphics[width=0.33\textwidth]{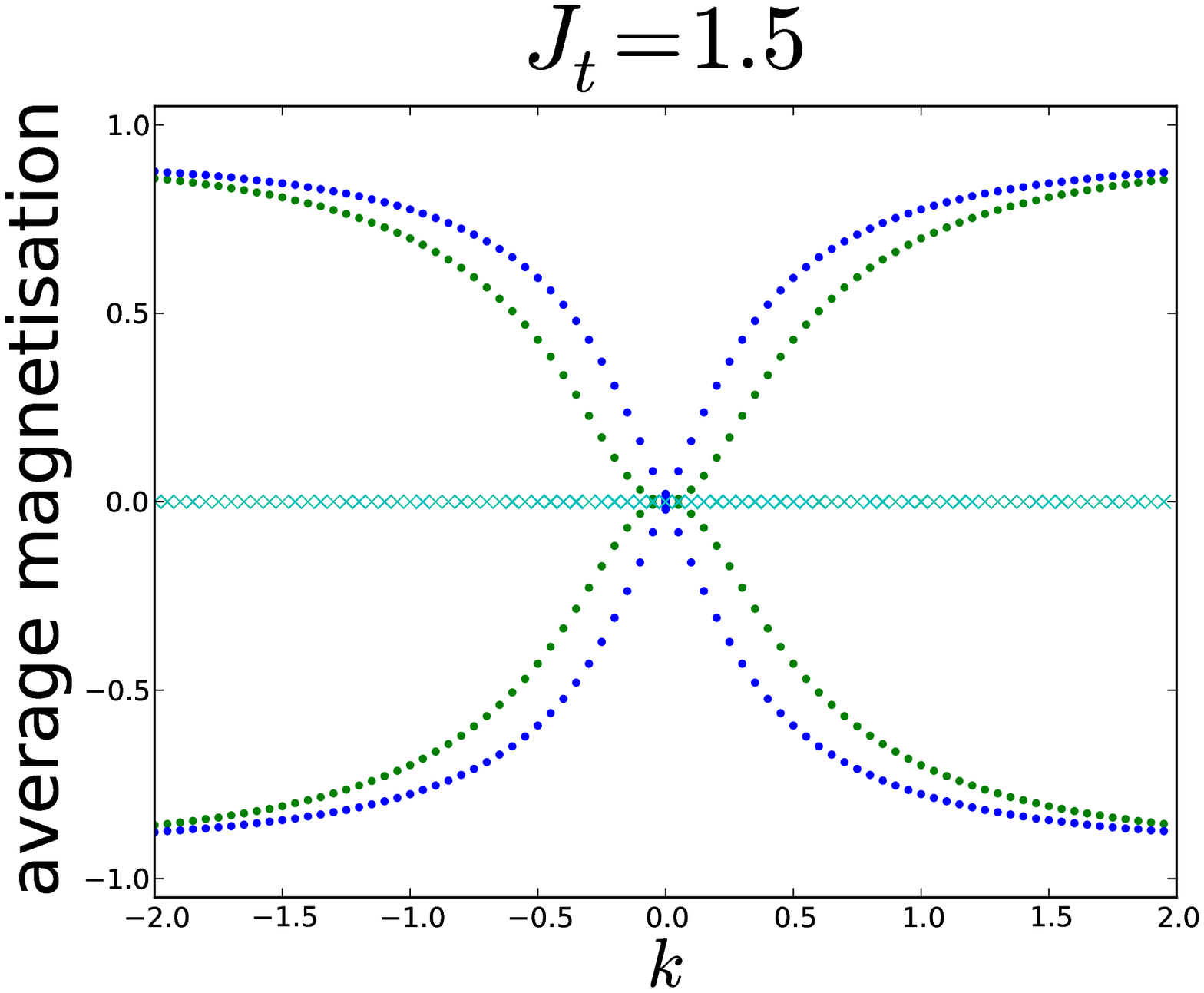}}\\
\subfloat[]{\includegraphics[width=0.33\textwidth]{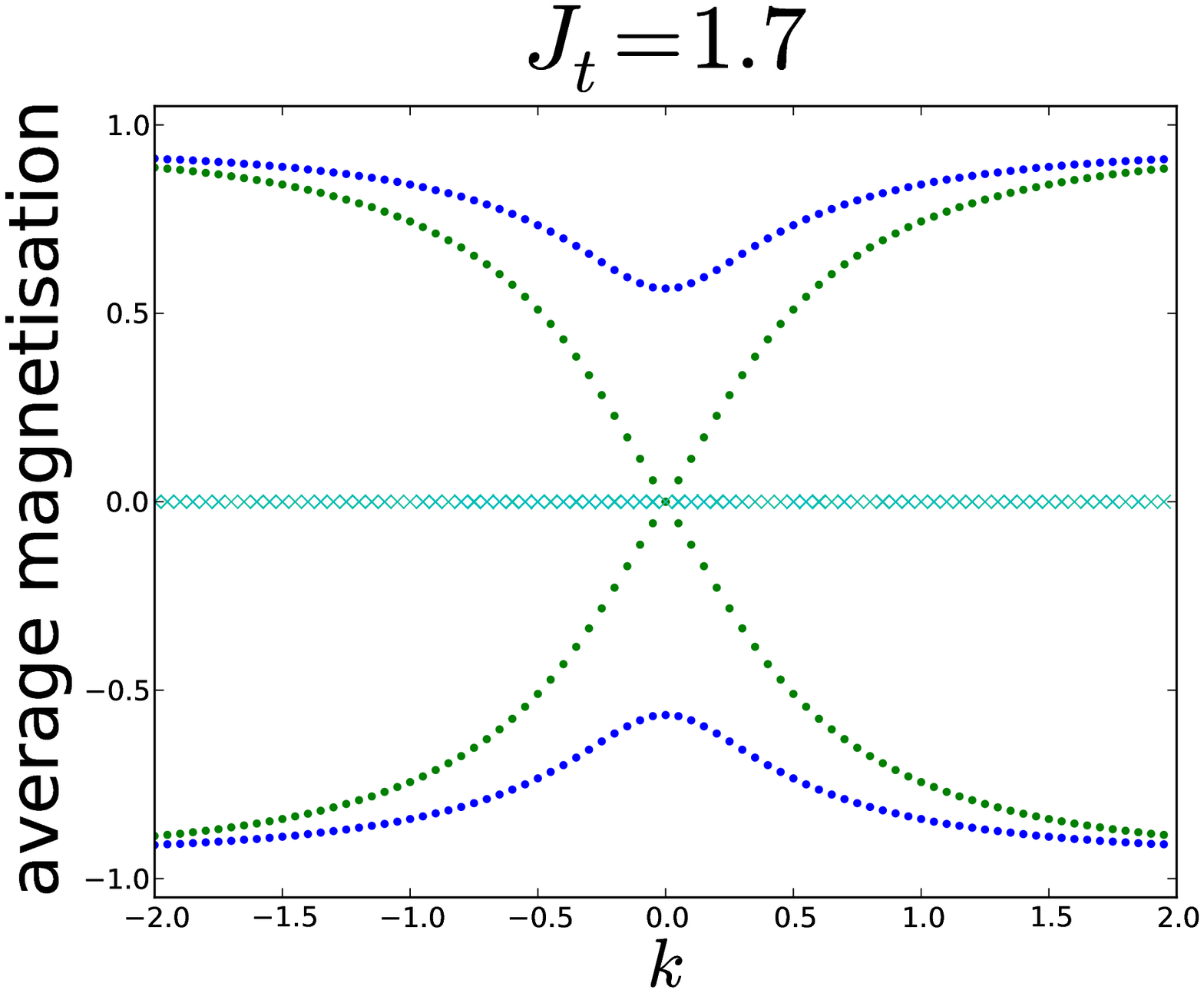}}
\subfloat[]{\includegraphics[width=0.33\textwidth]{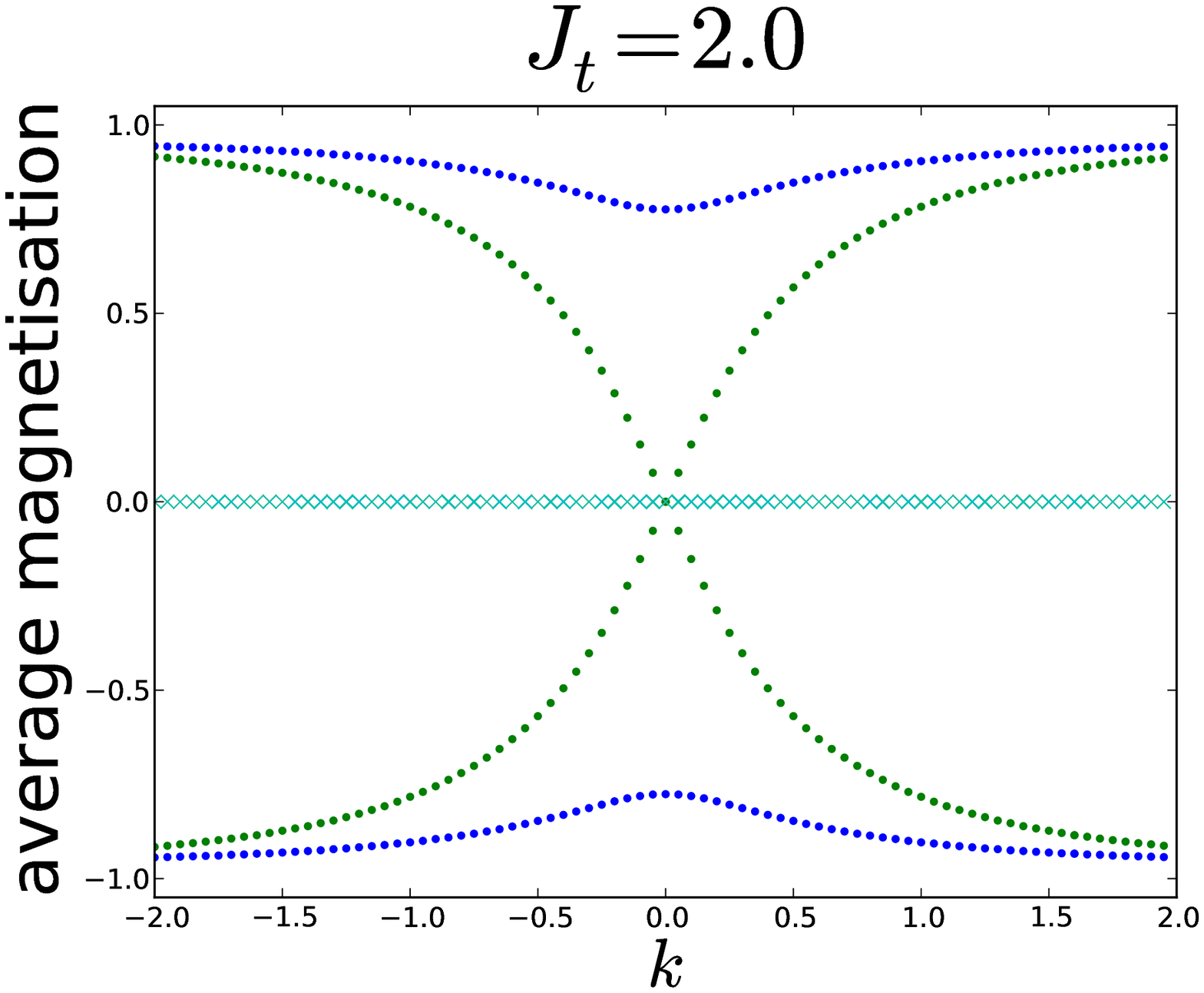}}
\subfloat[]{\includegraphics[width=0.33\textwidth]{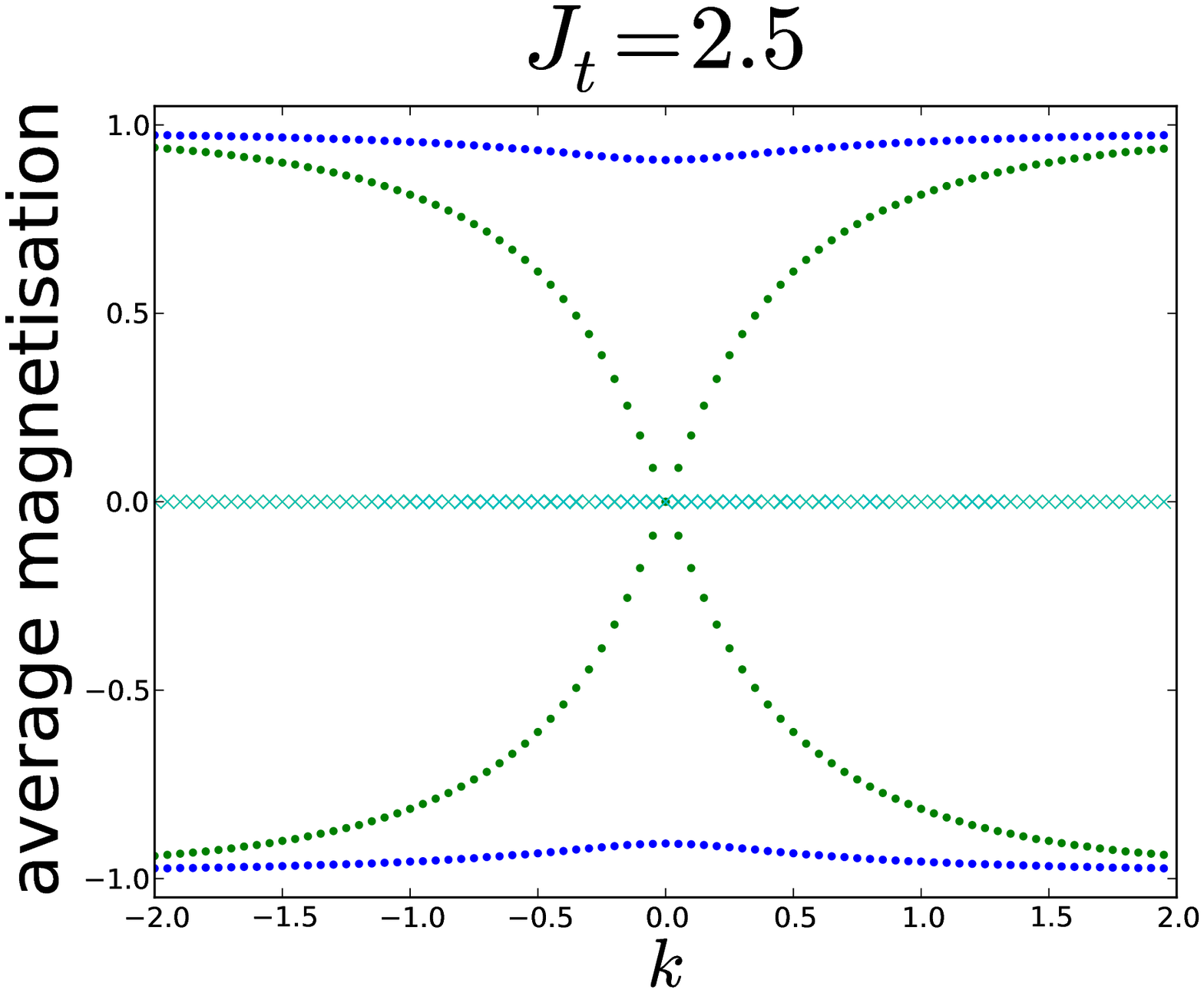}}
\caption{Dependence on inter-coupling $k$ of the numerically calculated average magnetisations $(s,t)$ for different values of the intra-coupling $J_{t}$ at high temperatures. $J_{s}=1$ and $K_{B}T=1.5$  for all plots. (a) $J_{t}=0.05$, (b) $J_{t}=0.5$, (c) $J_{t}=0.8$, (d) $J_{t}=1.2$, (e) $J_{t}=1.4$, (f) $J_{t}=1.5$, (g) $J_{t}=1.7$, (h) $J_{t}=2$ and (i) $J_{t}=2.5$. In all cases, different solutions are plotted for inter-coupling $k$ between -2 and 2 every 0.05. Magnetisations are plotted in green for $s$ and blue for $t$. Dark points are used for stable solutions and lighter asp ($\times$, for saddle points) or cross ($+$, for maxima) for non stable solutions.}
\label{fig:locanaJk1}
\end{figure}

\begin{figure}
\centering
\subfloat[]{\includegraphics[width=0.33\textwidth]{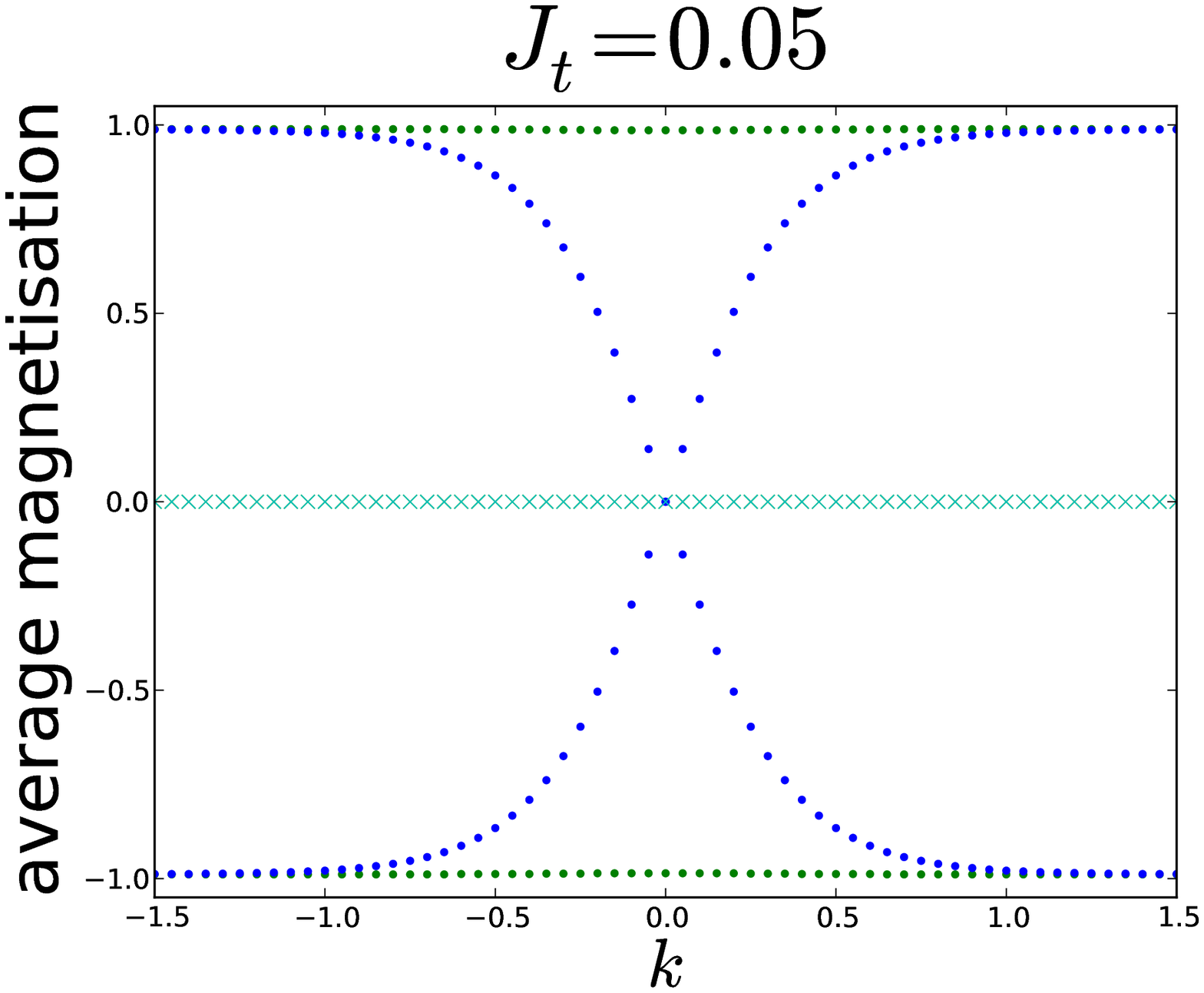}}
\subfloat[]{\includegraphics[width=0.33\textwidth]{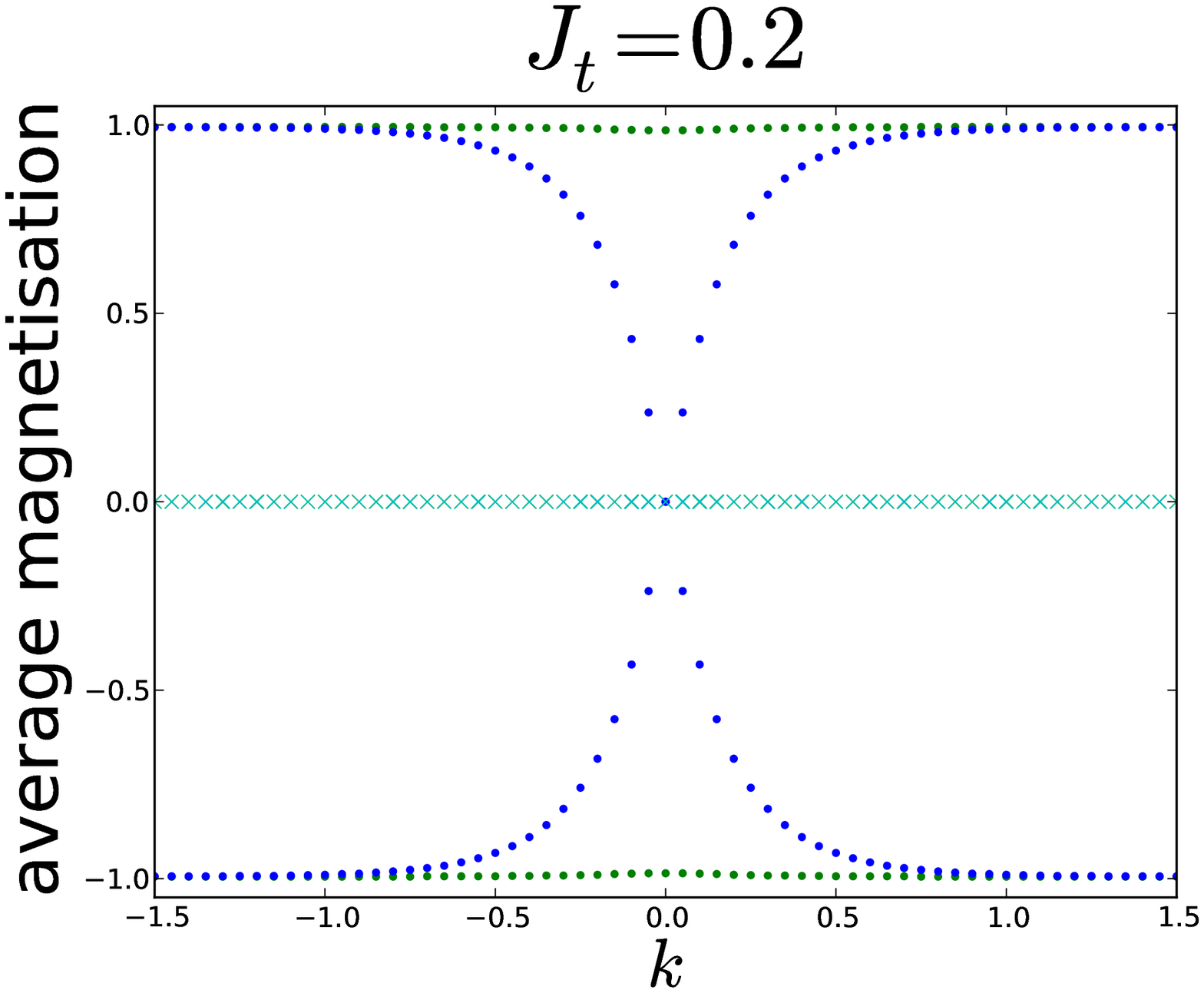}}
\subfloat[]{\includegraphics[width=0.33\textwidth]{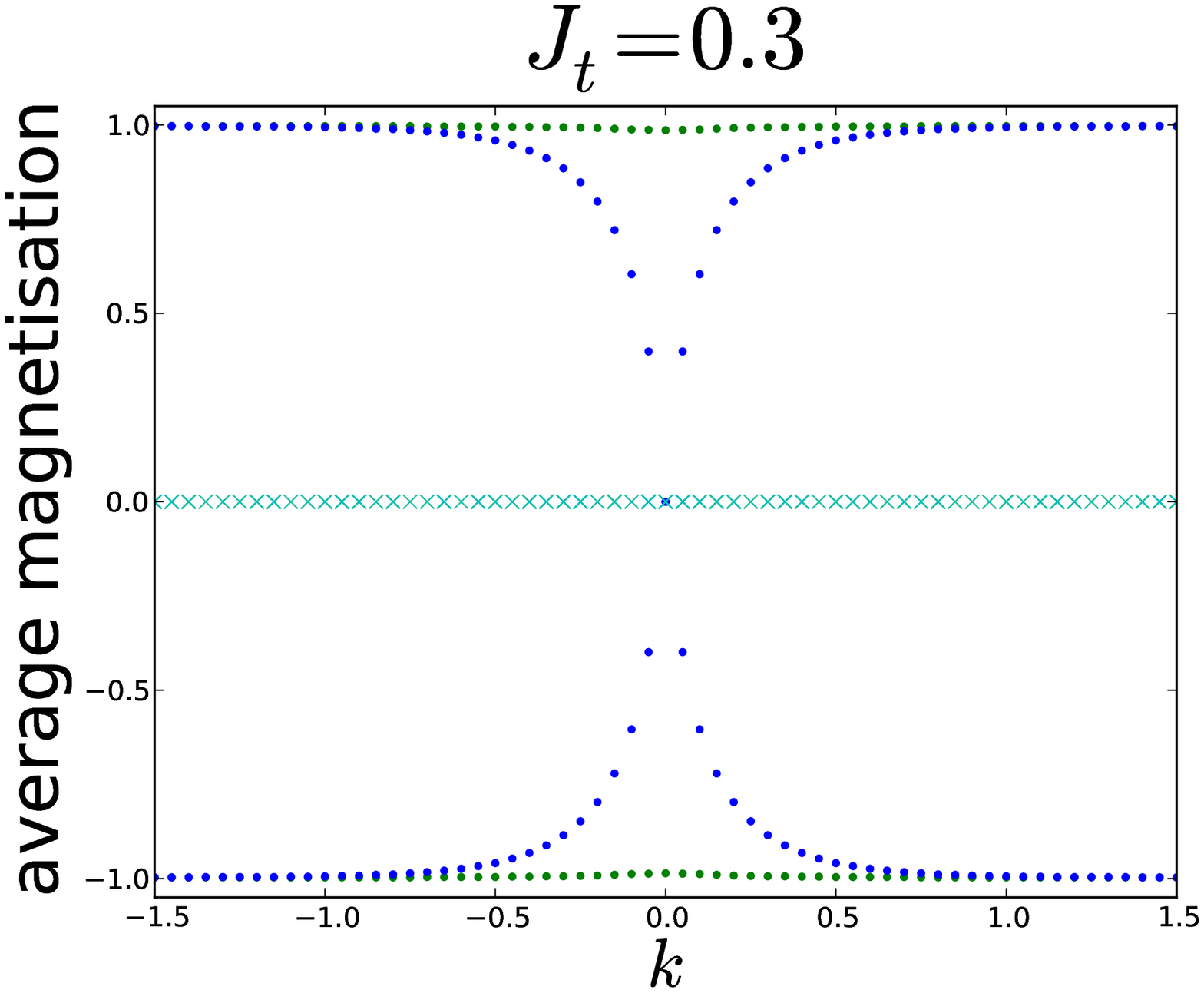}}\\
\subfloat[]{\includegraphics[width=0.33\textwidth]{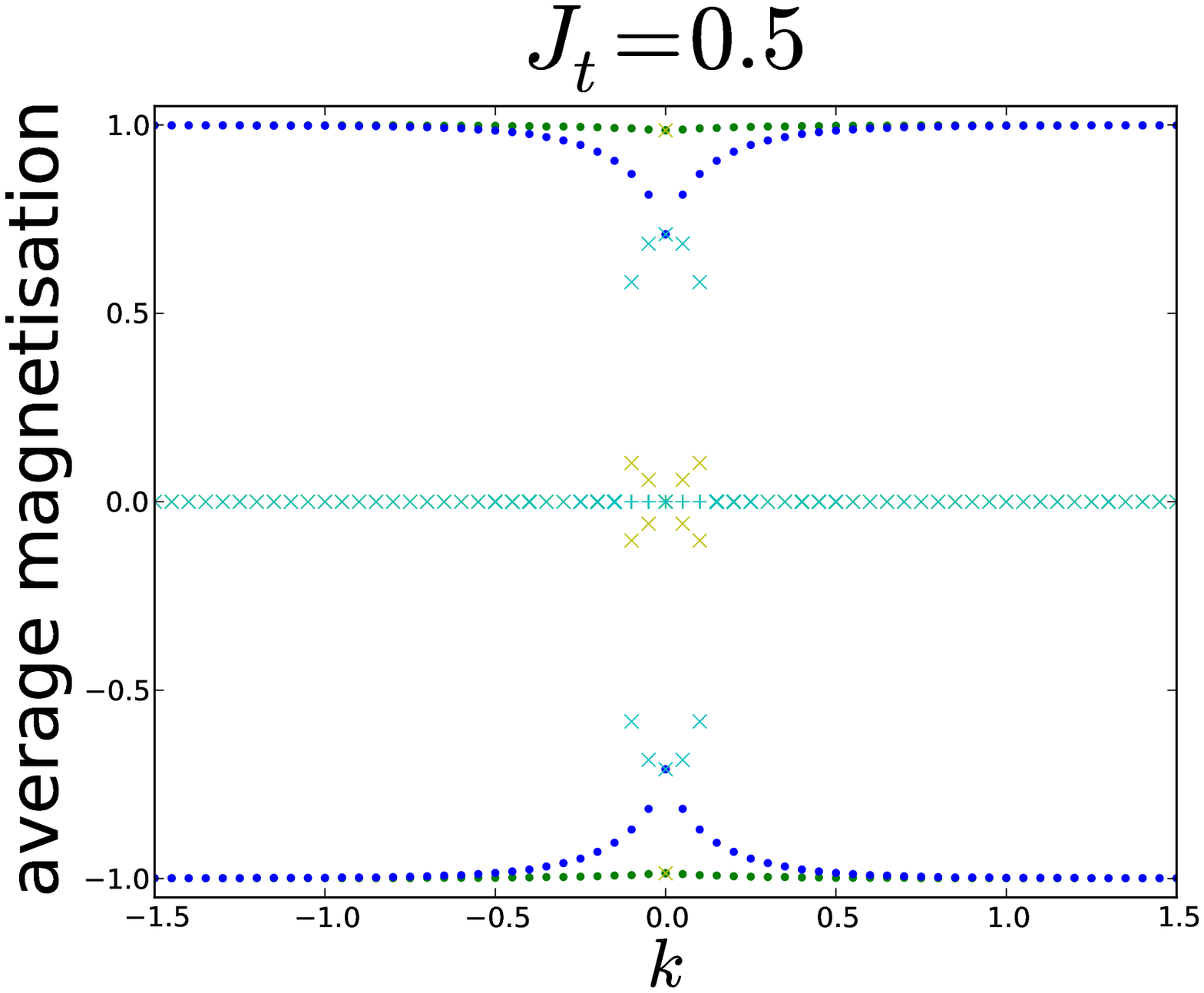}}
\subfloat[]{\includegraphics[width=0.33\textwidth]{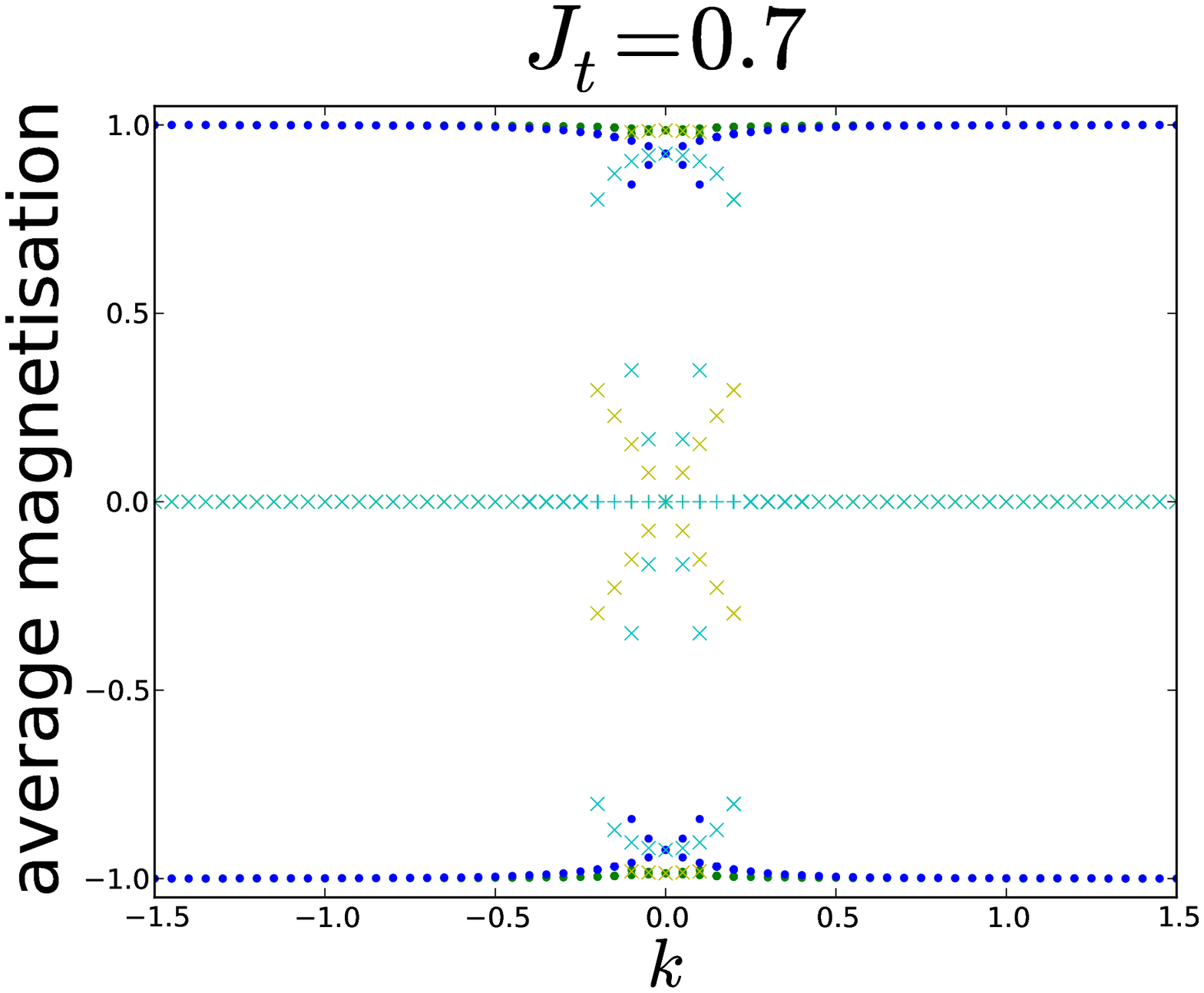}}
\subfloat[]{\includegraphics[width=0.33\textwidth]{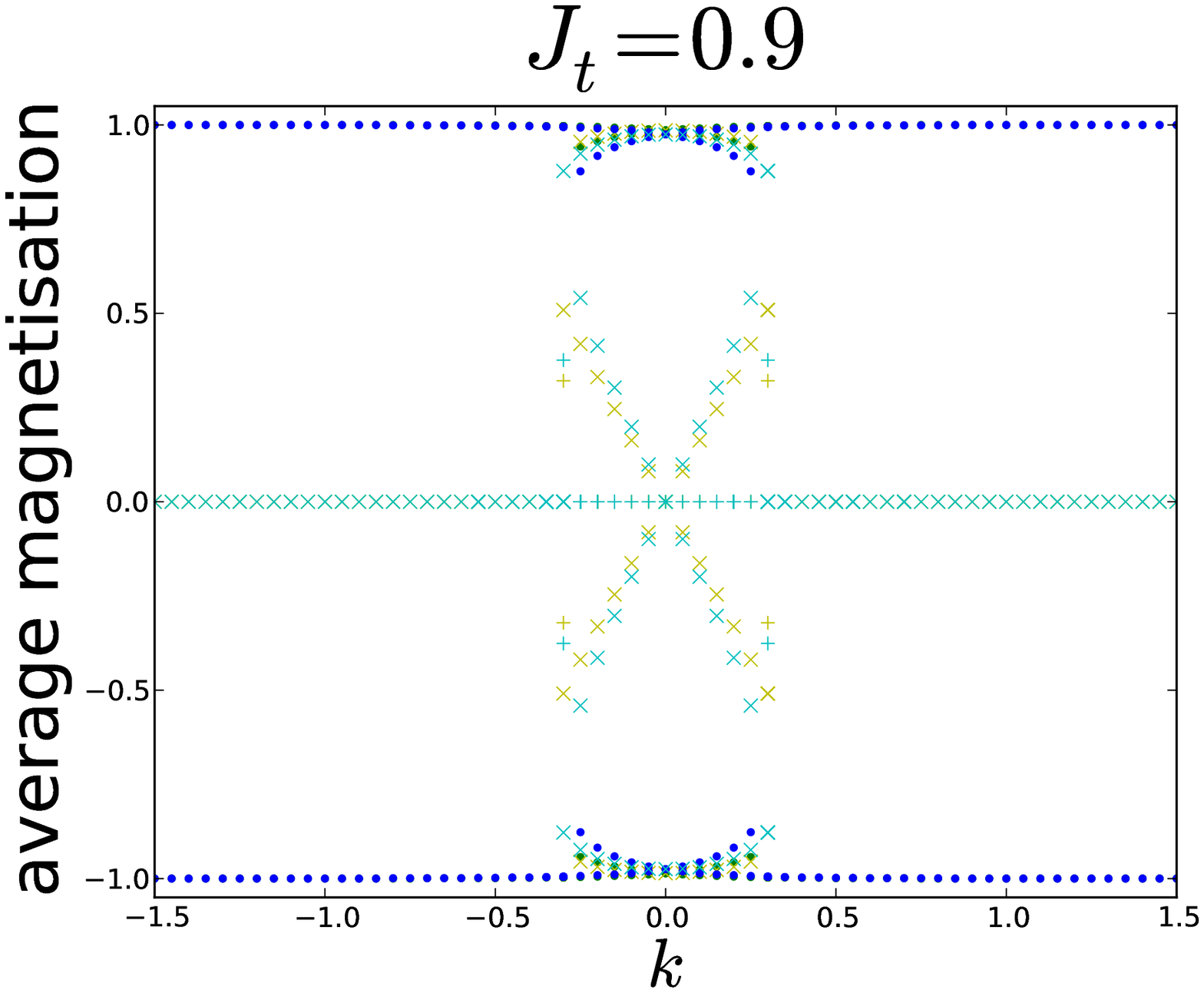}}\\
\subfloat[]{\includegraphics[width=0.33\textwidth]{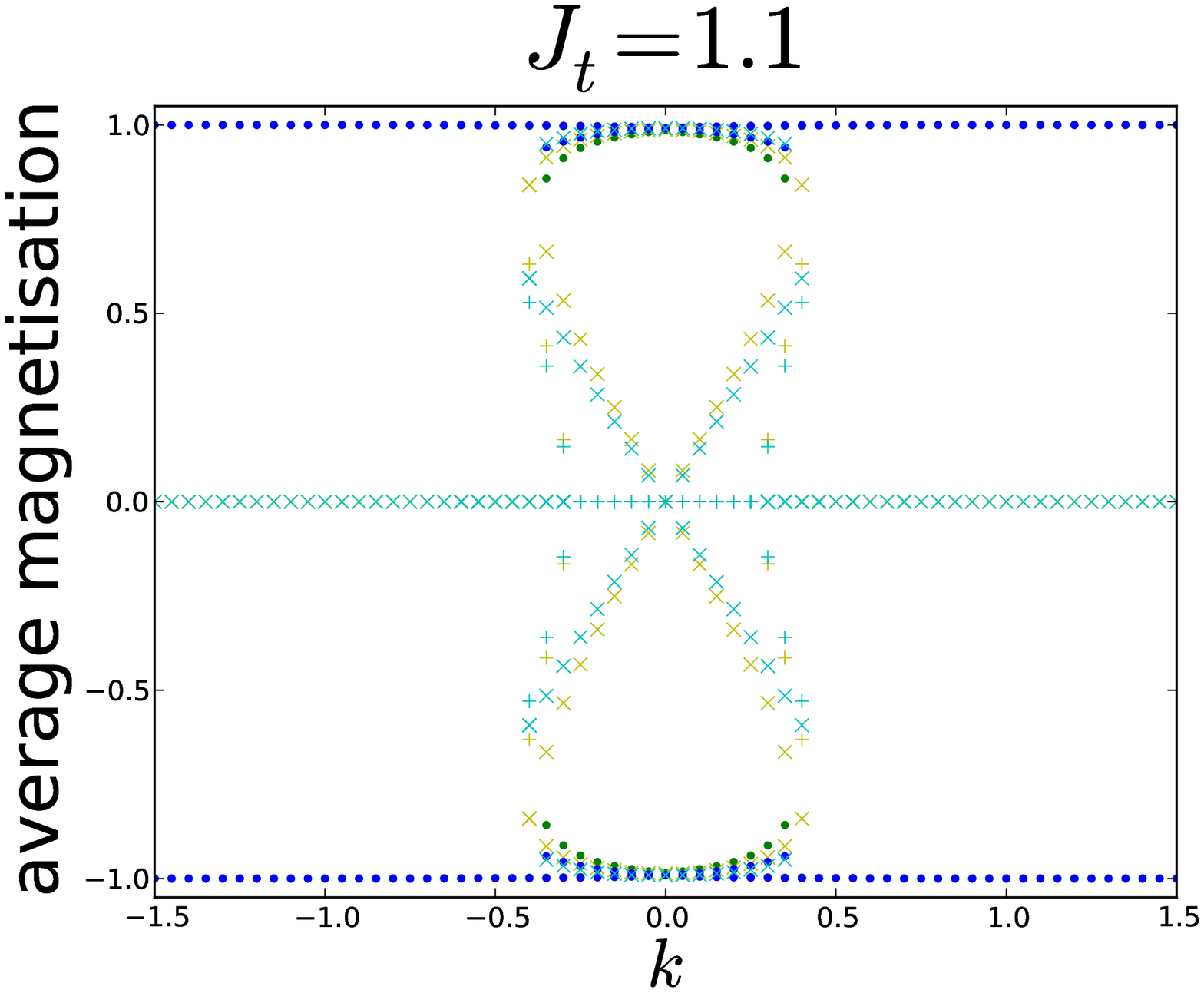}}
\subfloat[]{\includegraphics[width=0.33\textwidth]{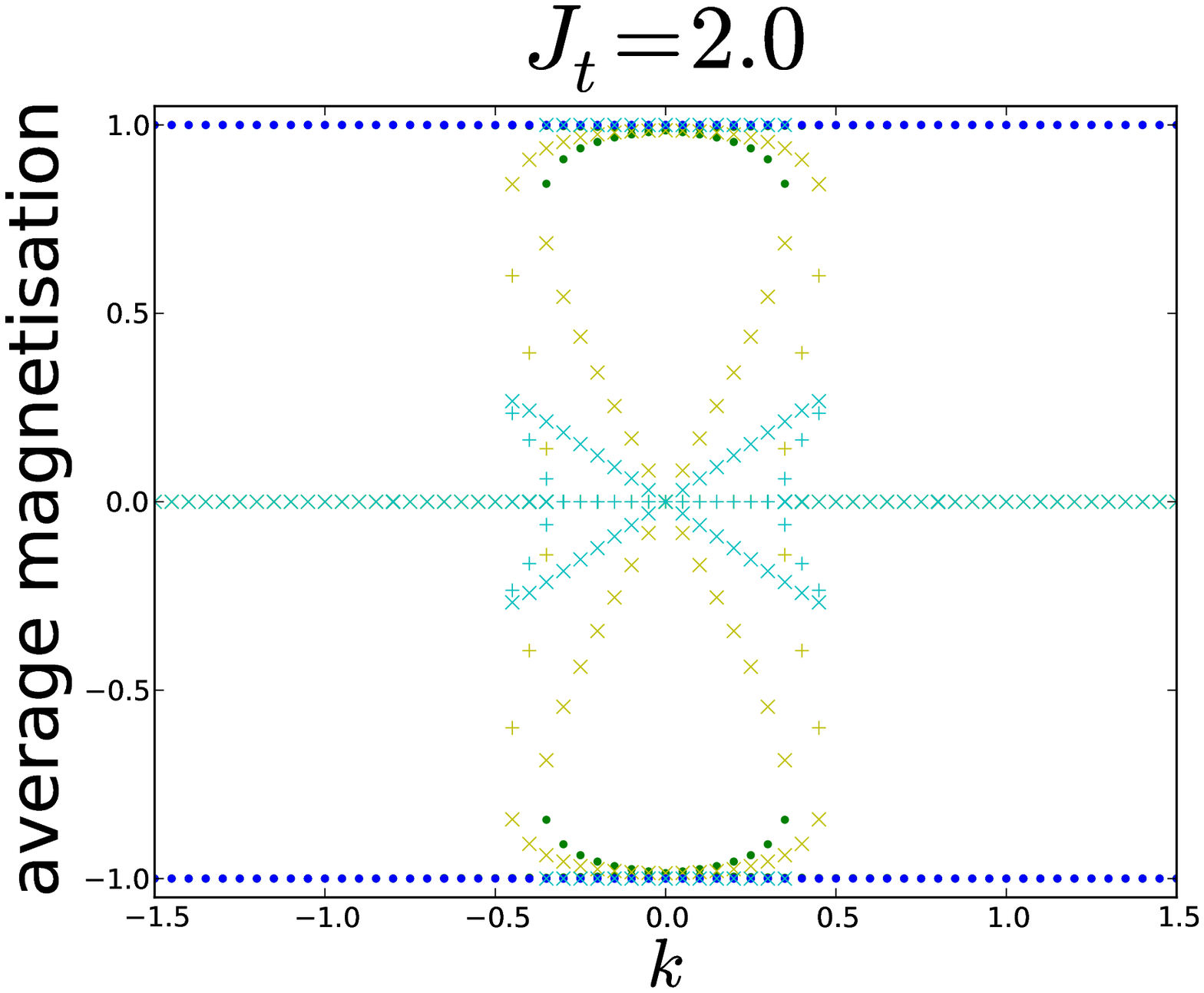}}
\subfloat[]{\includegraphics[width=0.33\textwidth]{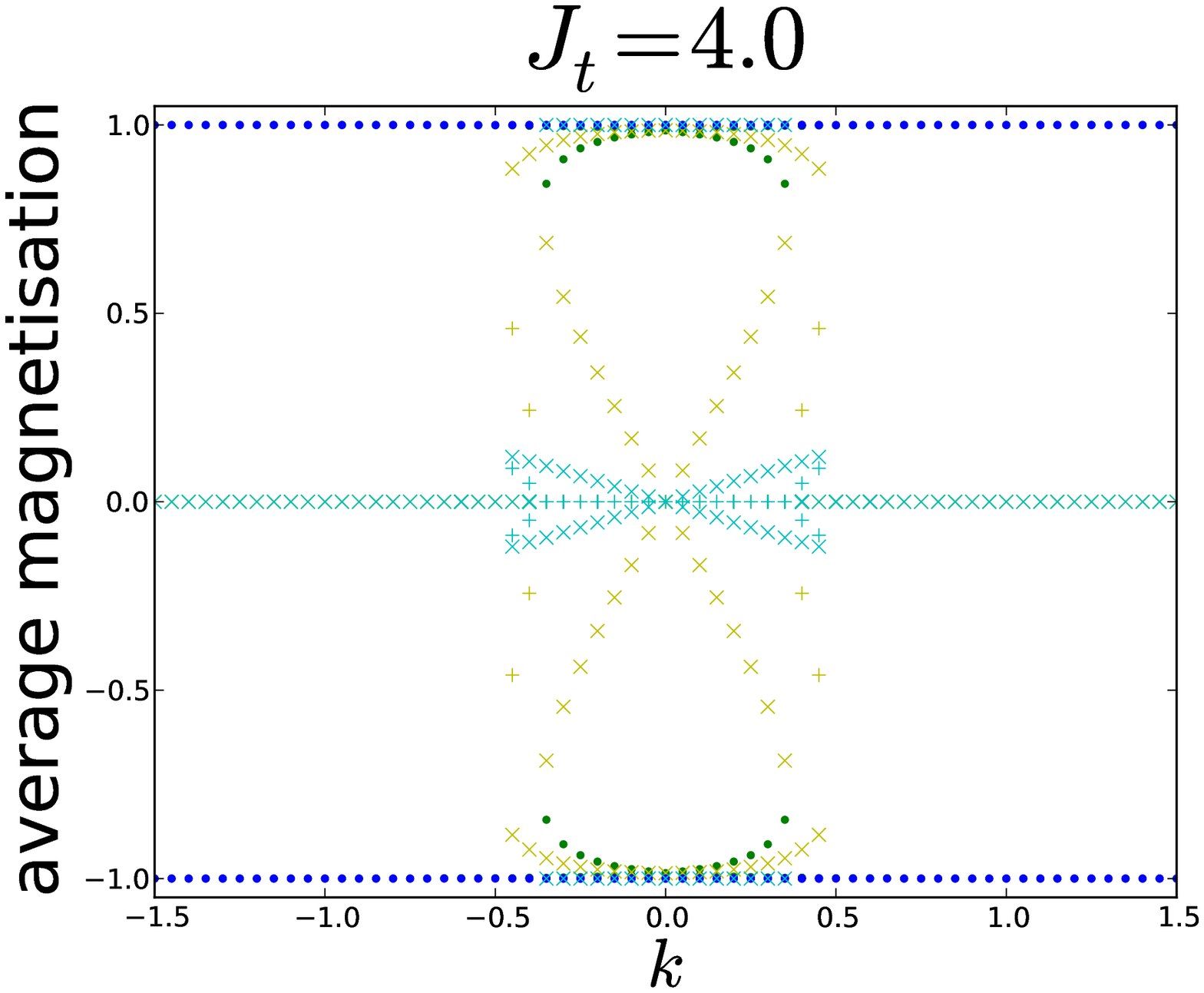}}
\caption{Dependence on inter-coupling $k$ of the numerically calculated average magnetisations $(s,t)$ for different values of the intra-coupling $J_{t}$ at low temperatures. $J_{s}=1$ and $K_{B}T=0.4$  for all plots. (a) $J_{t}=0.05$, (b) $J_{t}=0.2$, (c) $J_{t}=0.3$, (d) $J_{t}=0.5$, (e) $J_{t}=0.7$, (f) $J_{t}=0.9$, (g) $J_{t}=1.1$, (h) $J_{t}=2$ and (i) $J_{t}=4$. In all cases, different solutions are plotted for inter-coupling $k$ between -1.5 and 1.5 every 0.05. Magnetisations are plotted in green for $s$ and blue for $t$. Dark points are used for stable solutions and lighter asp ($\times$, for saddle points) or cross ($+$, for maxima) for non stable solutions.}
\label{fig:locanaJk2}
\end{figure}

When considering the variation of the dependence on $k$ with $J_{t}$ for fixed values of the rest of the parameters, it is again convenient to distinguish between the high temperature (figure \ref{fig:locanaJk1}) and the low temperature (figure \ref{fig:locanaJk2}) situation. As expected, the situation is qualitatively similar to the nonlocal case with some differences, the obvious one being that in the local case there is no change in the stability of the solutions as we move into the high coupling regime. Another important difference in the high temperature case is that at low enough intra-coupling $J_{t}$, in the local case only the paramagnetic solution exists and it is stable (figure \ref{fig:locanaJk1} a and b). In the nonlocal case, while only the paramagnetic solution exists for the range of $k$ in the weak coupling regime, there are additional saddle point ferromagnetic solutions at high enough values of $k$ for any $J_{t}$). In the low temperature case, an additional difference is related to the existence of ferromagnetic maxima at high enough values of $J_{t}$ (figure \ref{fig:locanaJk1} f to i), as was the case for the dependence on $k$. Again, at this point, it is no longer true that non stable ferromagnetic solutions exist only for $|k|<|k_{c}|$ either.

\subsection{Dependence on intra-couplings}

If we now turn to the dependence of $l$ (equation \eqref{eq:locTc}) on the intra-coupling $J_{t}$, we can rewrite the function\'{}s roots as given by

\begin{equation}
J_{t}^{c}=\frac{K_{B}T(J_{s}-K_{B}T)}{J_{s}(1-\alpha_{k}^{2})-K_{B}T}
\end{equation}

The numerator will be negative for $J_{s}<K_{B}T$ and positive for $J_{s}>K_{B}T$. The denominator is negative for $J_{s}<\frac{K_{B}T}{1-\alpha_{k}^{2}}$ and positive elsewhere (for positive $J_{s}$). As $K_{B}T \leq \frac{K_{B}T}{1-\alpha_{k}^{2}}$, this means there will exist a positive value of $J_{t}^{c}$, whenever $J_{s}$ is not in the range $(K_{B}T,\frac{K_{B}T}{1-\alpha_{k}^{2}})$, for which there is no physically relevant value of $J_{t}^{c}$. The sign of the second derivative for the paramagnetic phase \eqref{eq:locgsspara} will be positive for $K_{B}T>J_{s}$.

\begin{figure}
\centering
\includegraphics[width=\textwidth]{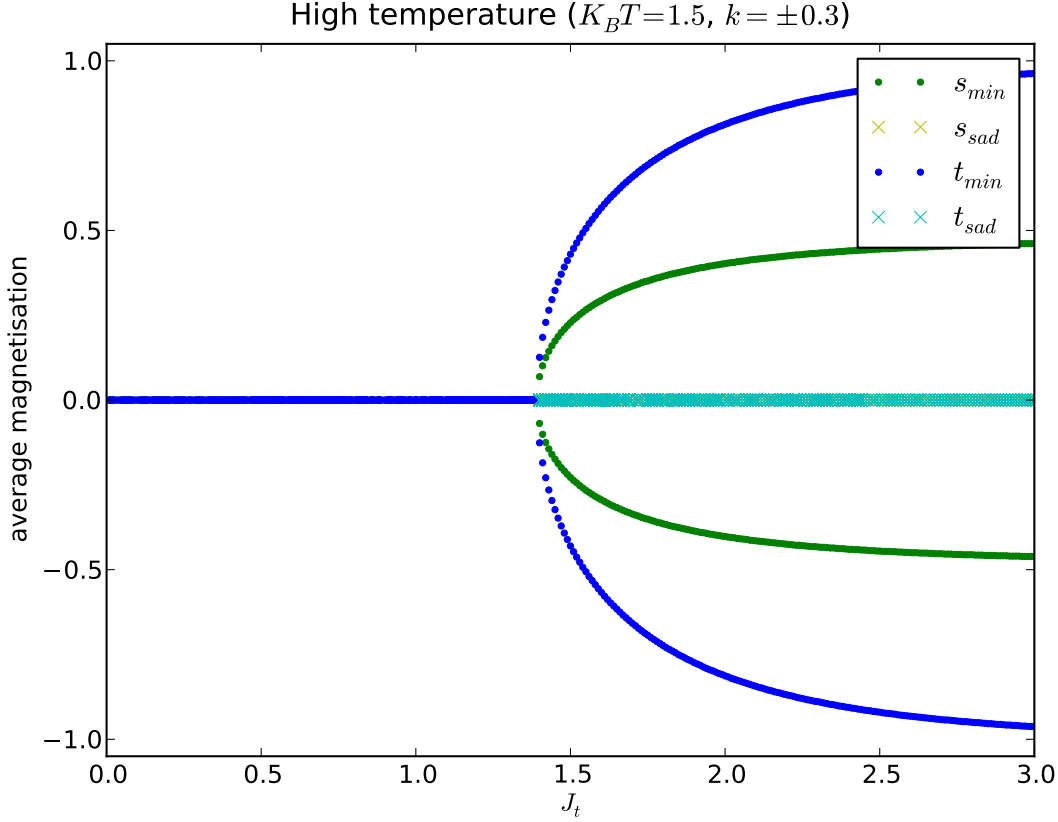}
\caption{Dependence on the intra-coupling $J_{t}$ of the numerically calculated average magnetisations for $J_{s}=1$, $k= 0.3$ , $K_{B}T = 1.5$ ($J_{t}^{c}=1.39$). Different solutions are plotted for $J_{t}$  between 0 and 3 every 0.01. Magnetisations are plotted in green for $s$ and blue for $t$. Dark points are used for stable solutions and lighter asps ($\times$) for saddle point, non stable solutions.}
\label{fig:locanaJ1}
\end{figure}

There are three different behaviours depending on the specific choice of parameters. When $J_{s}<K_{B}T$, there is value of $J_{t}^{c}$ at which there is a second order phase transition, and bellow which the paramagnetic solution is stable (above it is a saddle point). This is the case described in figure \ref{fig:locanaJ1}, where $J_{s}=1$, $k= 0.3$ and  $K_{B}T = 1.5$ ($J_{t}^{c}=1.39$). The behaviour is qualitatively identical to the nonlocal case (we the exception of the stability issue in the strong coupling regime), with a slightly higher value of $J_{t}^{c}$ (which is 1.32 in the nonlocal case).

\begin{figure}
\centering
\includegraphics[width=\textwidth]{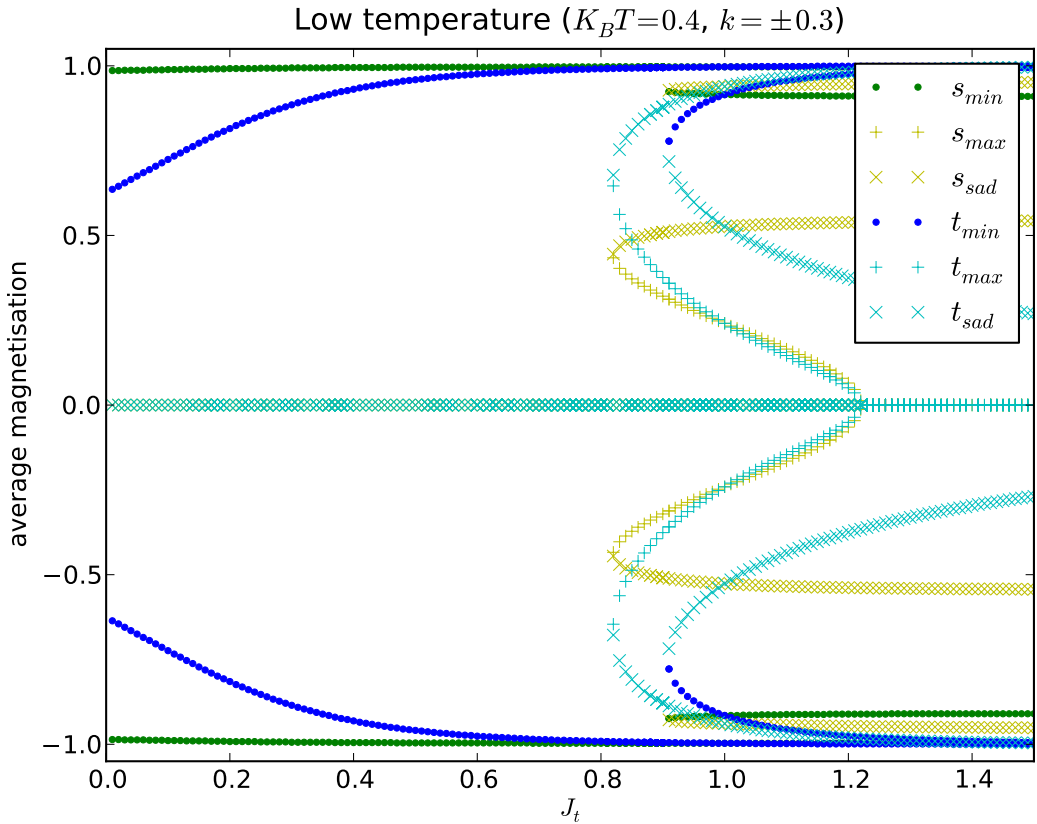}
\caption{Dependence on the intra-coupling $J_{t}$ of the numerically calculated average magnetisations for $J_{s}=1$, $k= 0.3$ and $K_{B}T = 0.4$ ($J_{t}^{c}=1.22$). Different solutions are plotted for $J_{t}$  between 0 and 1.5 every 0.01. Magnetisations are plotted in green for $s$ and blue for $t$. Dark points are used for stable solutions and lighter asp ($\times$, for saddle points) or cross ($+$, for maxima) for non stable solutions. }
\label{fig:locanaJ2}
\end{figure}

When $J_{s}>\frac{K_{B}T}{1-\alpha_{k}^{2}}$, there still is a value of $J_{t}^{c}$, but now it is not a real critical point as the paramagnetic phase changes from saddle to maximum (associated to the existence of saddle point ferromagnetic branches related to metastable states). Figure \ref{fig:locanaJ2} shows an example for $J_{s}=1$, $k= 0.3$ and $K_{B}T = 0.4$ ($J_{t}^{c}=1.22$). Behaviour of stable solutions is very similar to that of the nonlocal case (besides there been stable ferromagnetic solutions for all values of $J_{t}$), but there is an evident difference between the non stable solutions: there are ferromagnetic saddle points and maxima for  $J_{t}<J_{t}^{c}$ (the branches convexity changes in this case). According to numerical solutions, there are metastable states for $J_{t}>J_{t}^{a}$. Note that while the value of $J_{t}^{c}$ is significantly higher in the local case (it was 0.55 in the nonlocal case), the spinodal value remains very similar in both (due to the change in the behaviour of the ferromagnetic non stable branches).

\begin{figure}
\centering
\subfloat[]{\includegraphics[width=0.33\textwidth]{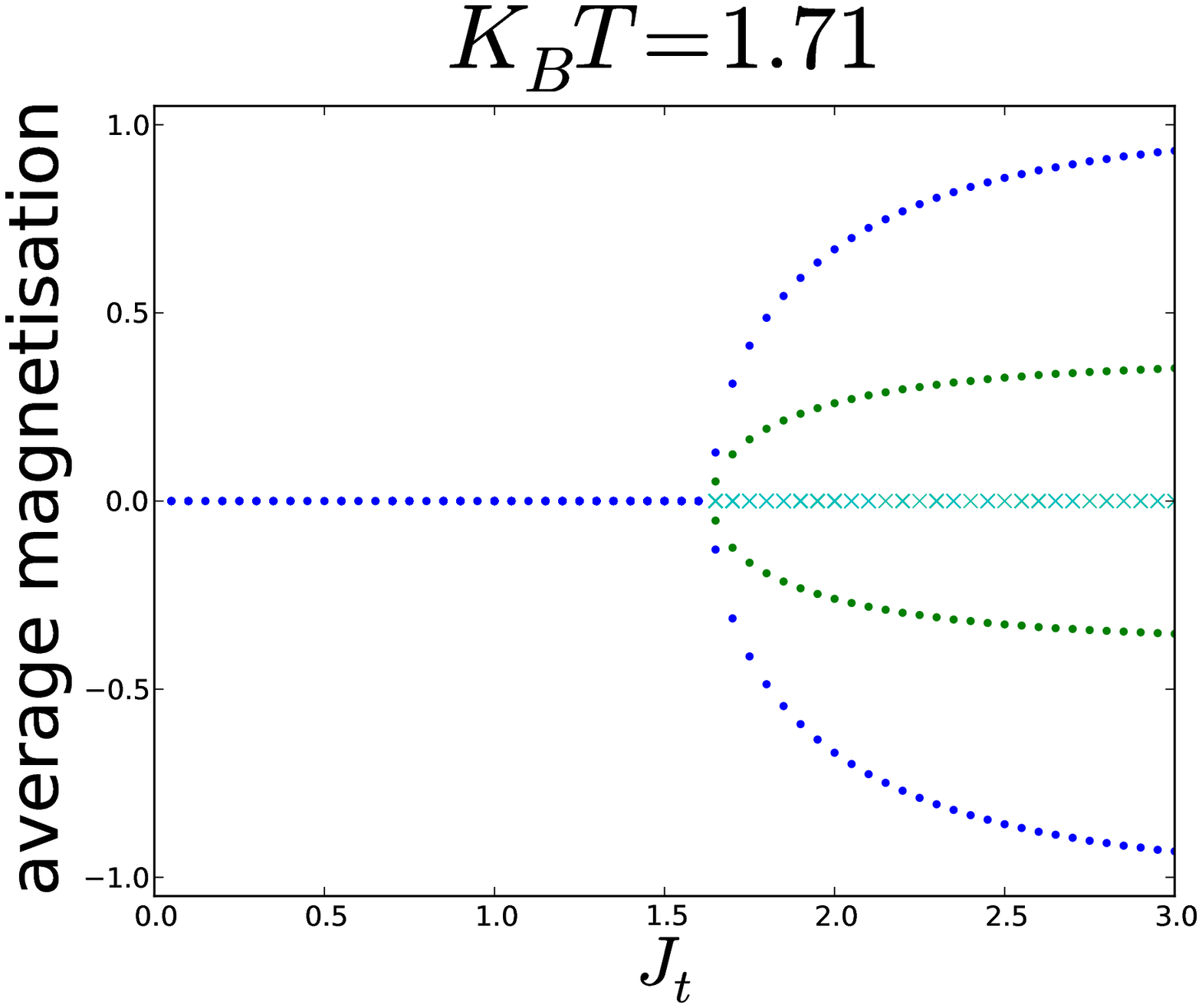}}
\subfloat[]{\includegraphics[width=0.33\textwidth]{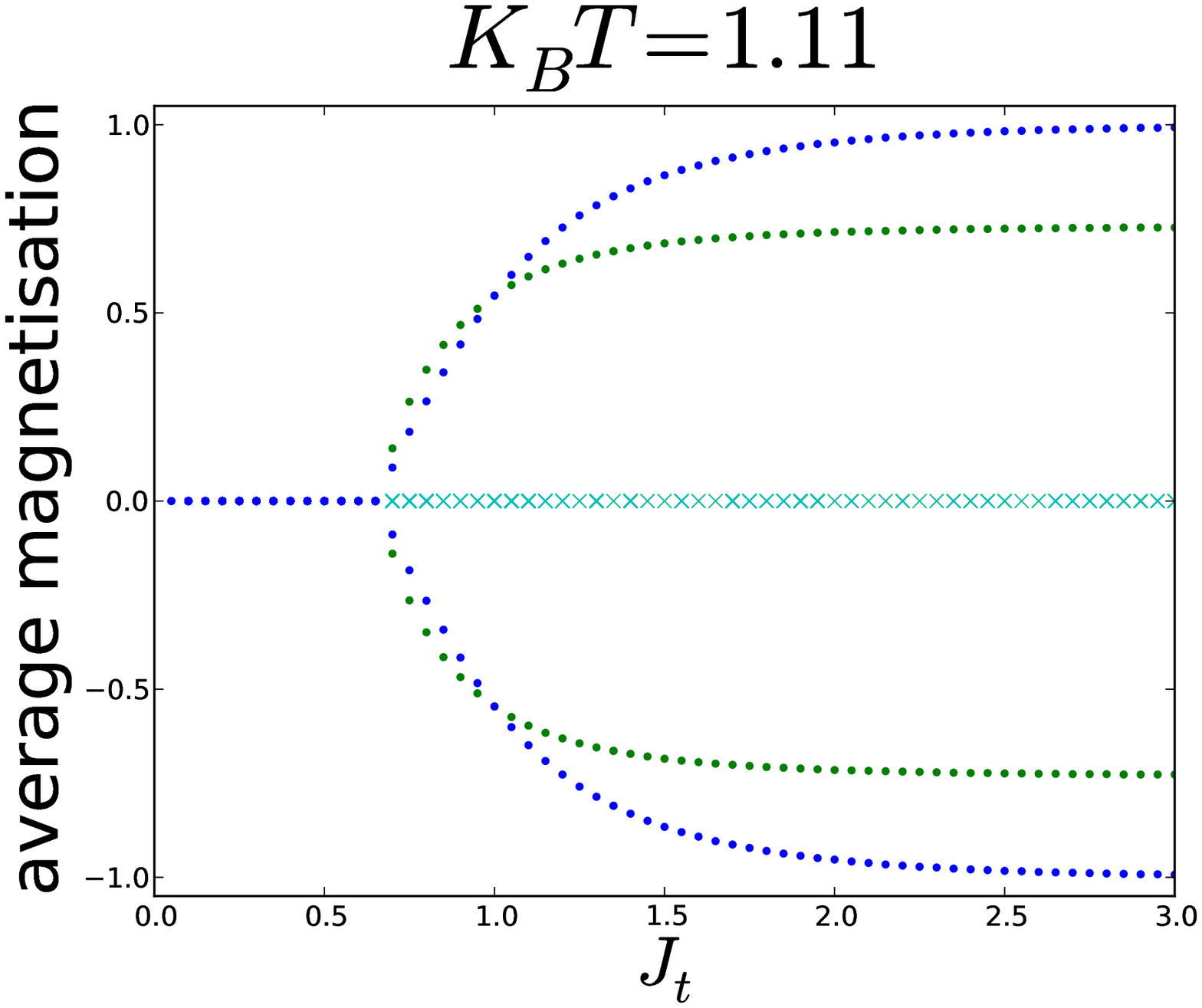}}
\subfloat[]{\includegraphics[width=0.33\textwidth]{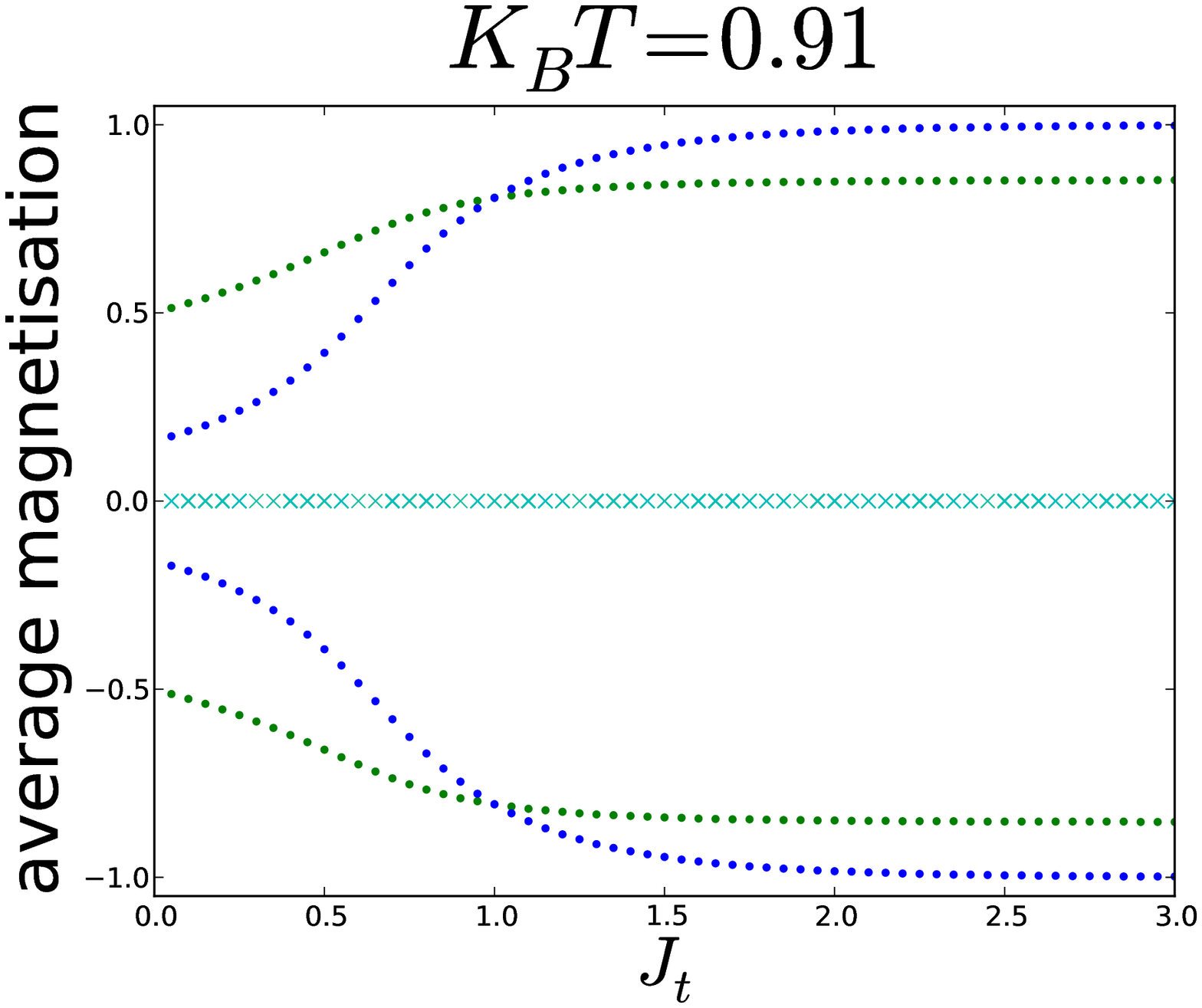}}\\
\subfloat[]{\includegraphics[width=0.33\textwidth]{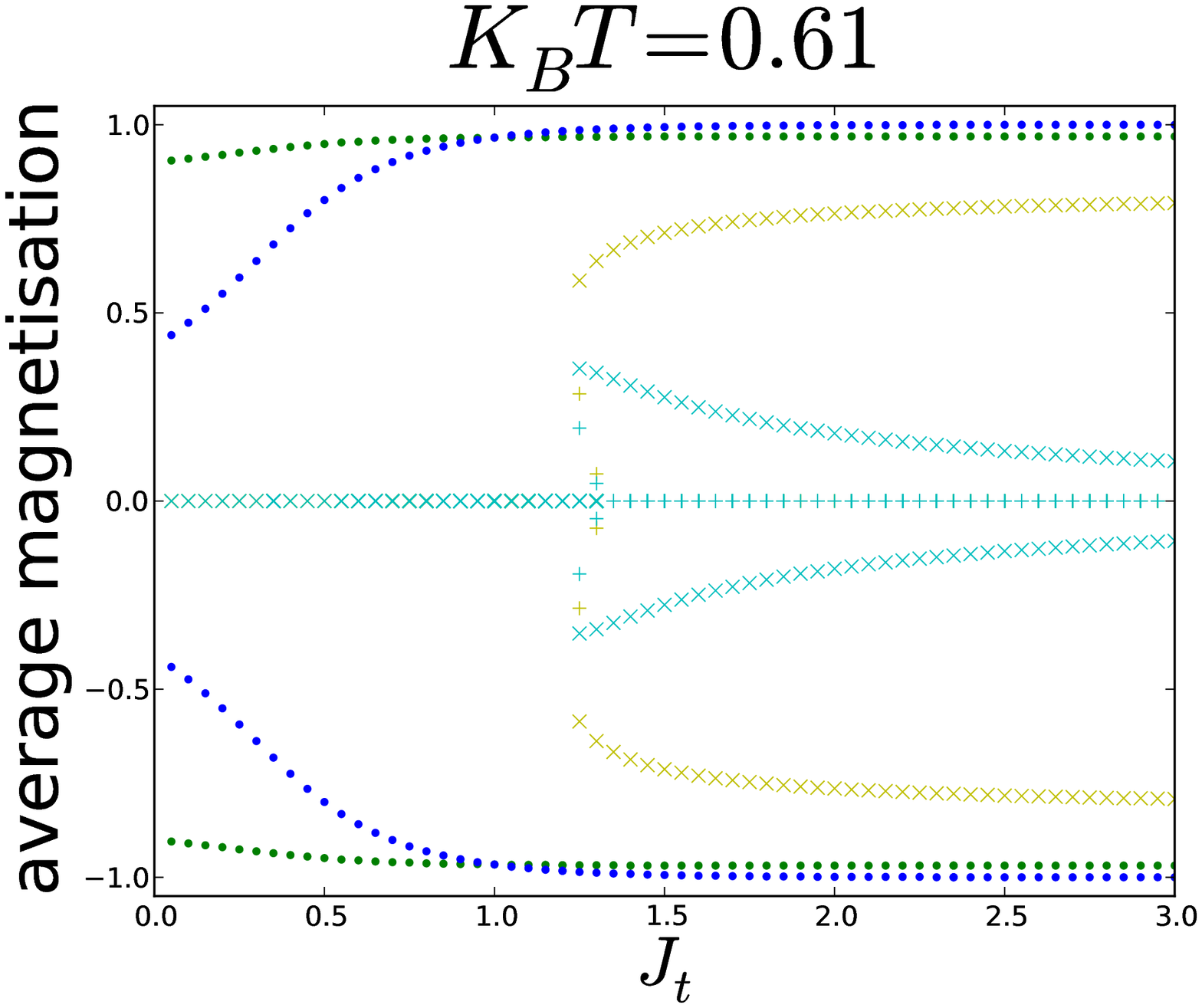}}
\subfloat[]{\includegraphics[width=0.33\textwidth]{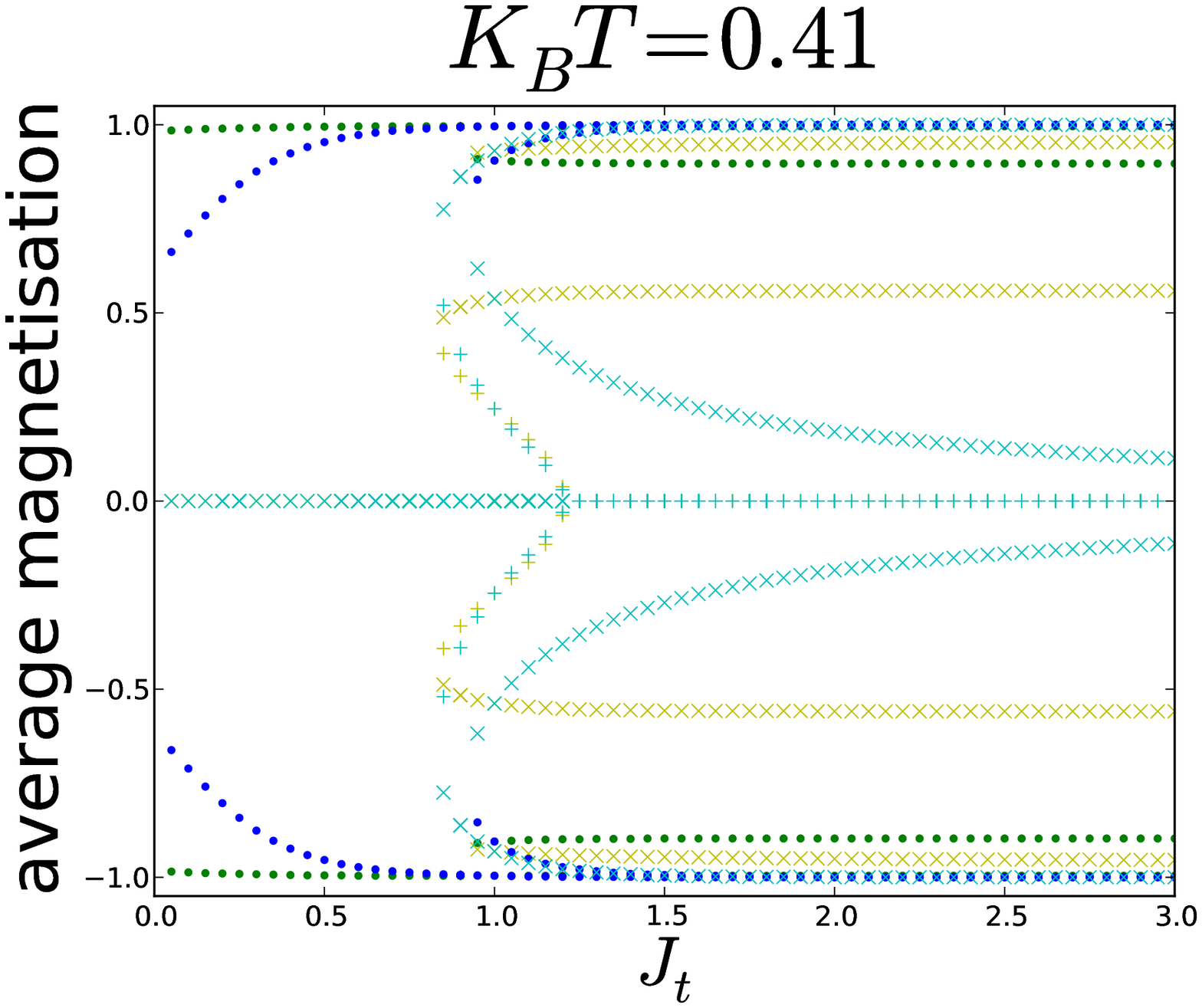}}
\subfloat[]{\includegraphics[width=0.33\textwidth]{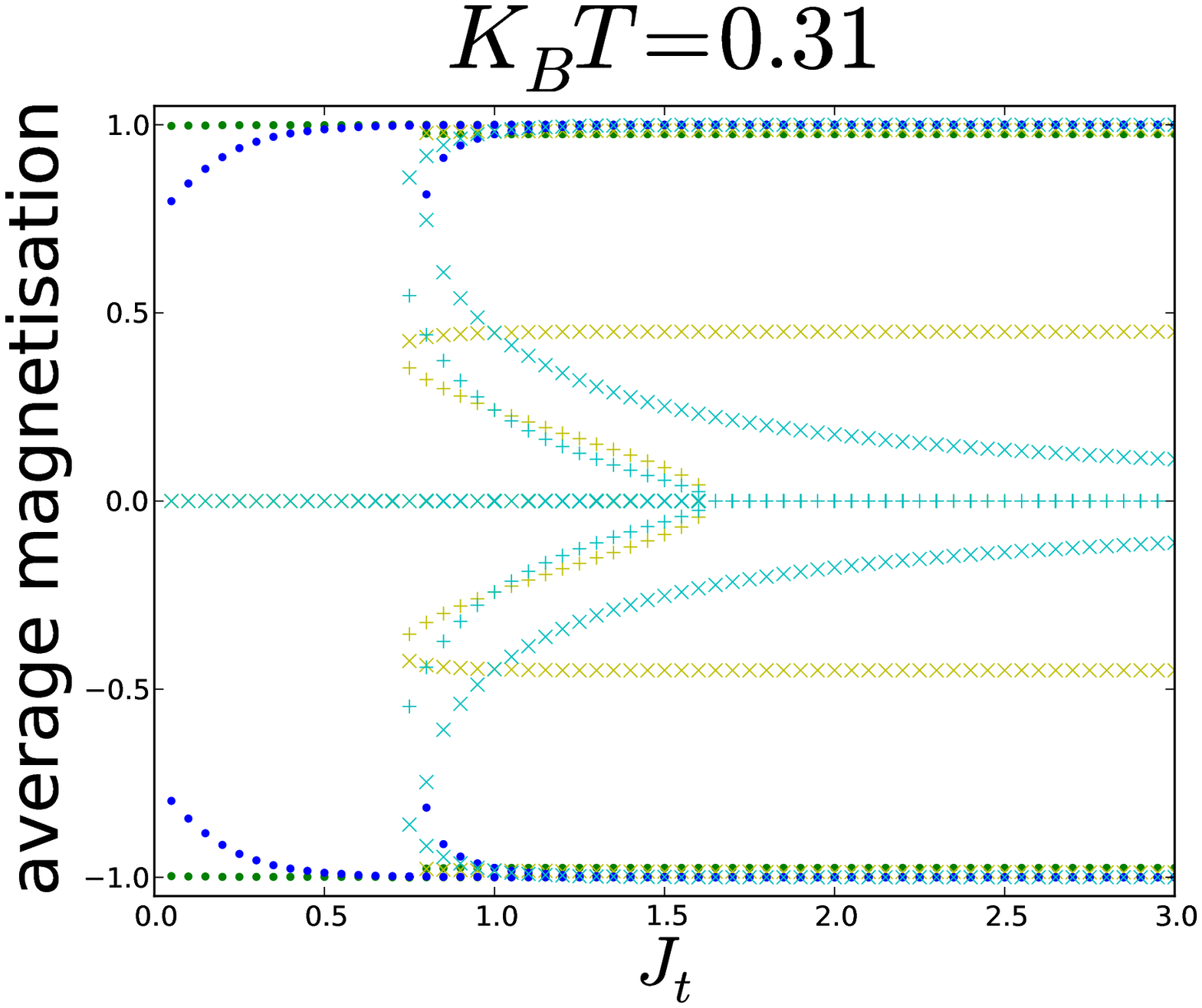}}\\
\subfloat[]{\includegraphics[width=0.33\textwidth]{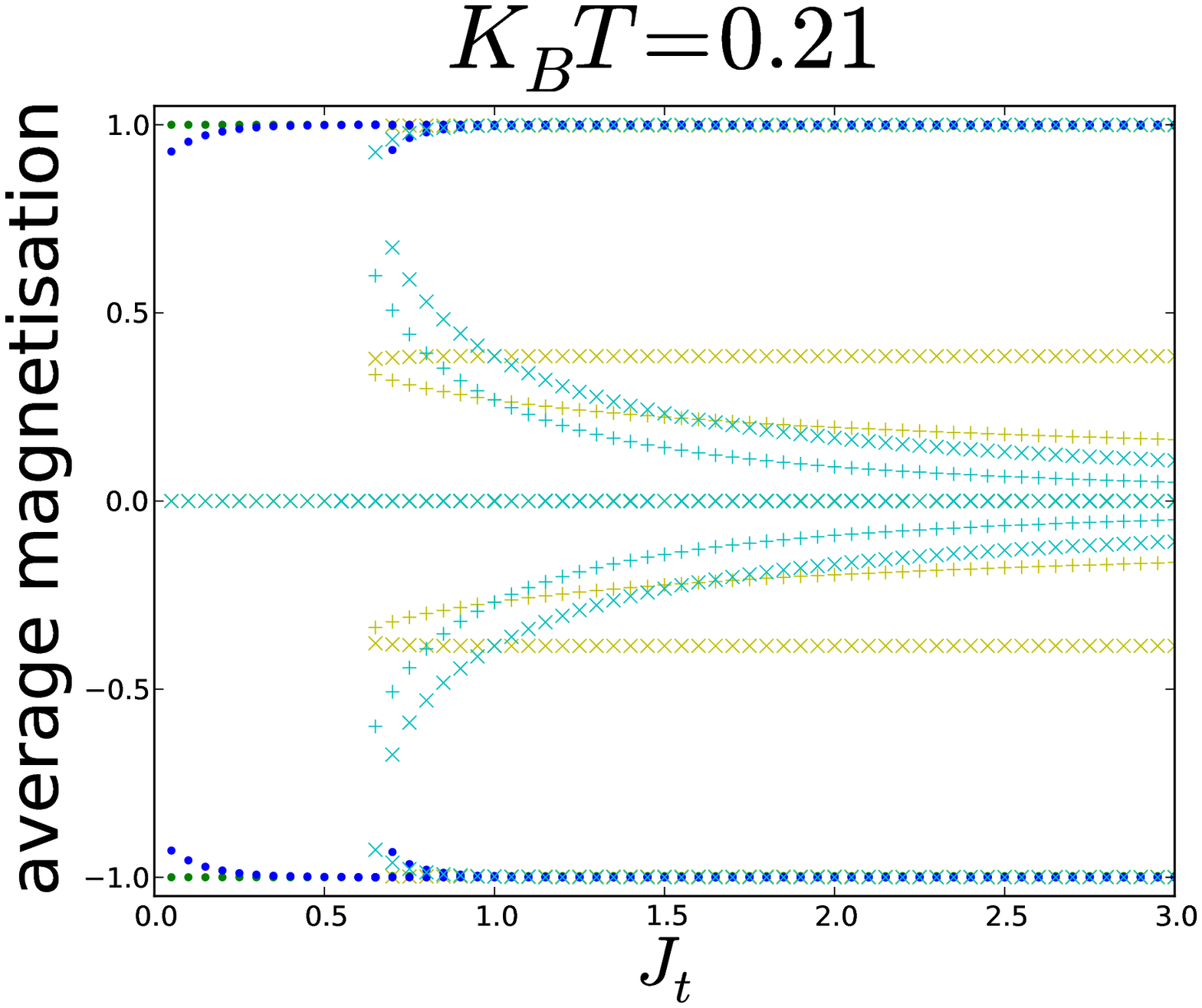}}
\subfloat[]{\includegraphics[width=0.33\textwidth]{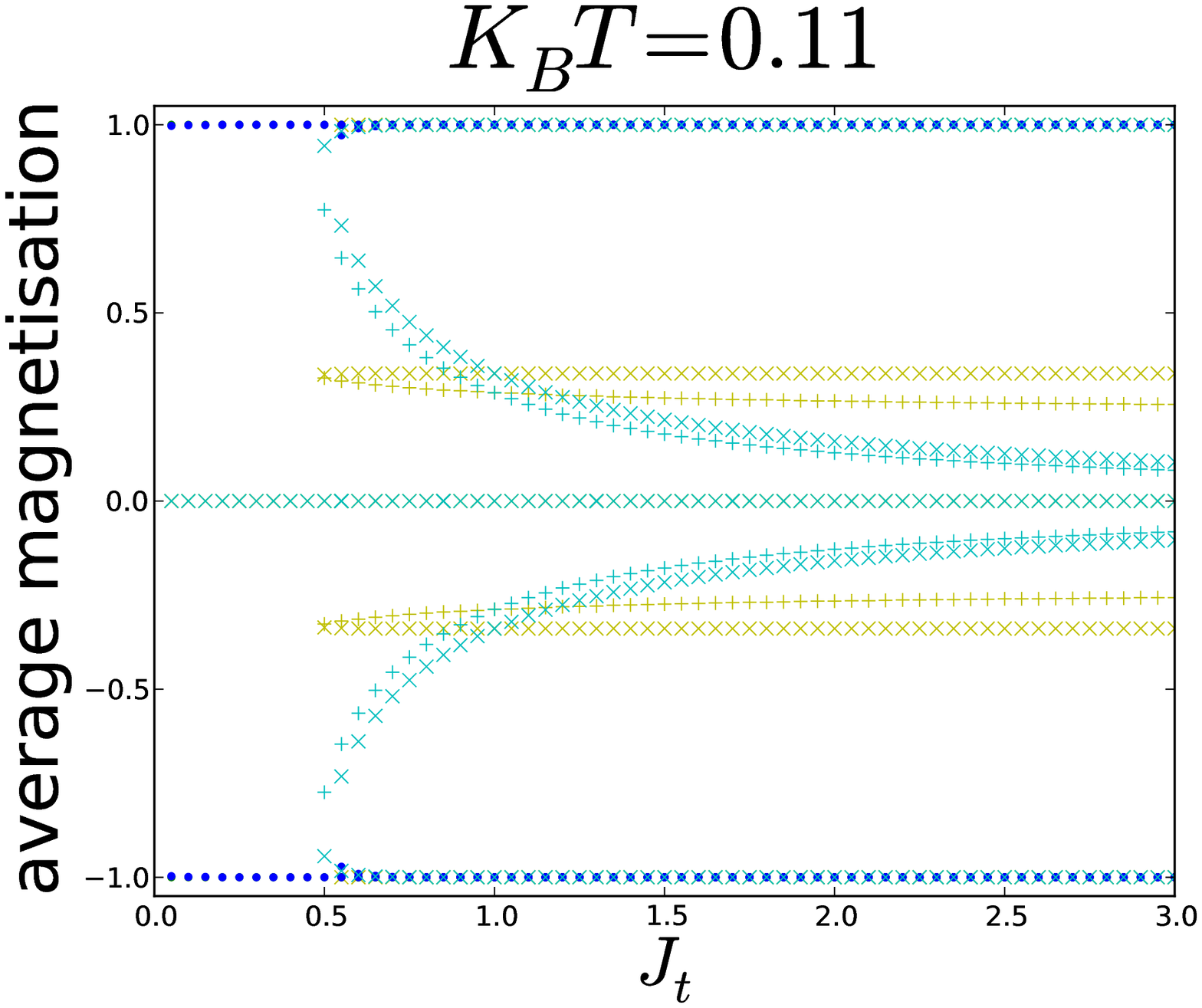}}
\subfloat[]{\includegraphics[width=0.33\textwidth]{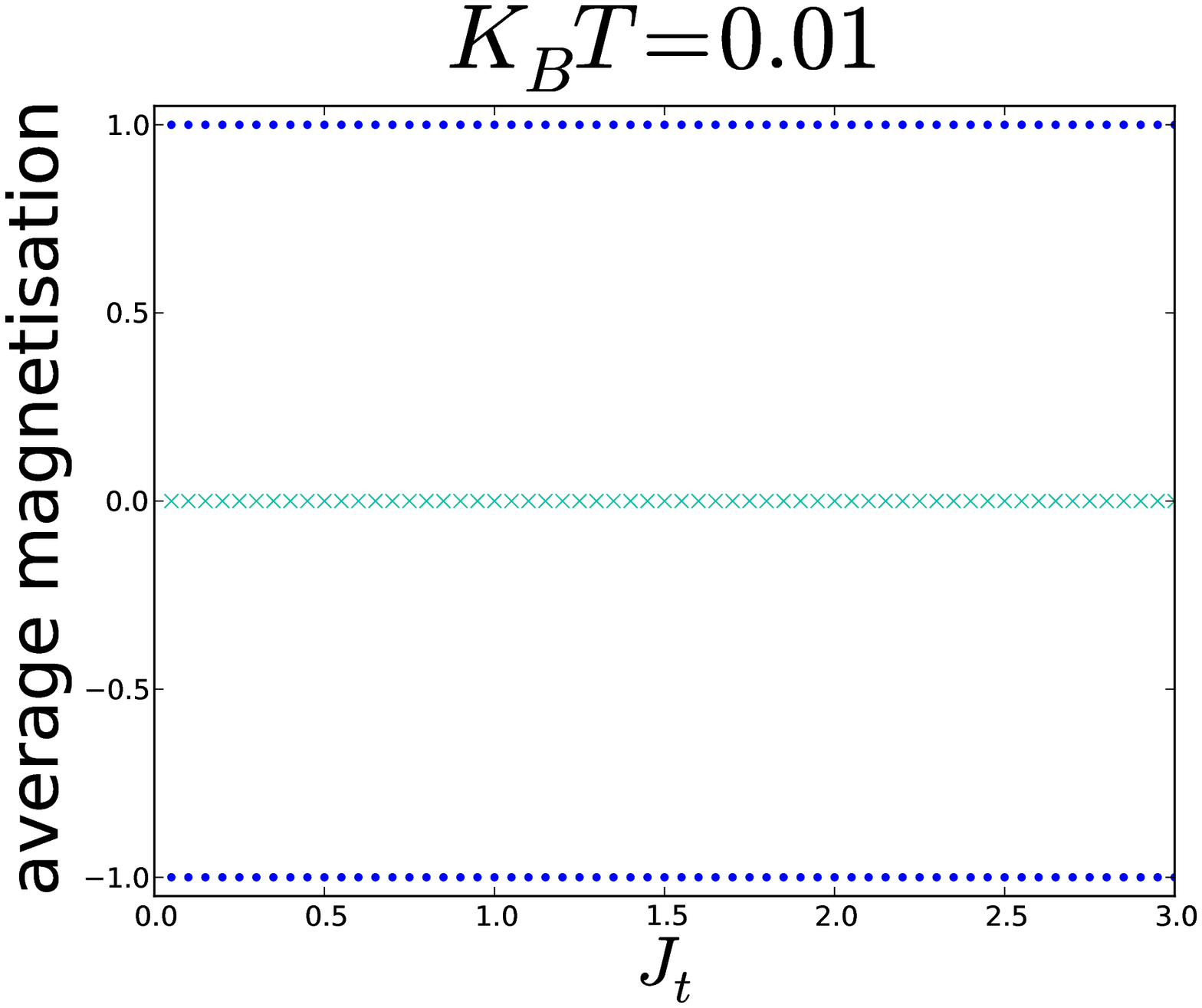}}
\caption{Dependence on intra-coupling $J_{t}$ of the numerically calculated average magnetisations $(s,t)$ for different values of the temperature $K_{B}T$. $J_{s}=1$ and $k=0.15$  for all plots. (a) $K_{B}T=1.71$, (b) $K_{B}T=1.11$, (c) $K_{B}T=0.91$, (d) $K_{B}T=0.61$, (e) $K_{B}T=0.41$, (f) $K_{B}T=0.31$, (g) $K_{B}T=0.21$, (h) $K_{B}T=0.11$ and (i) $K_{B}T=0.01$. In all cases, different solutions are plotted for intra-coupling $J_{t}$ between 0.01 and 3 every 0.05. Magnetisations are plotted in green for $s$ and blue for $t$. Dark points are used for stable solutions and lighter asp ($\times$, for saddle points) or cross ($+$, for maxima) for non stable solutions.}
\label{fig:locanaTJ}
\end{figure}

For the region in between ($K_{B}T<J_{s}<\frac{K_{B}T}{1-\alpha_{k}^{2}}$), which will be larger the larger $k$ is, there is no relevant $J_{t}^{c}$. The paramagnetic phase is always a saddle point, and the only stable solutions are the main ferromagnetic branches for all values of $J_{t}$. We will not study this case in more detail as it can be considered the limiting case of that described in figure \ref{fig:locanaJ2} when there are no ferromagnetic non-stable (or metastable) branches (figures \ref{fig:locanaTJ} c and i are examples). 

Figure \ref{fig:locanaTJ} shows how the dependence on $J_{t}$ varies as we move to lower values of the temperature for fixed values of the rest of the parameters. Besides the general features that have already been commented upon (stable solutions for all values of $J_{t}$ and behaviour of the non stable ferromagnetic branches), there is another qualitative difference with the nonlocal case: for low enough values of $T$, metastable states (and non stable ferromagnetic branches) disappear (figure \ref{fig:locanaTJ} i).

\begin{figure}
\centering
\subfloat[]{\includegraphics[width=0.33\textwidth]{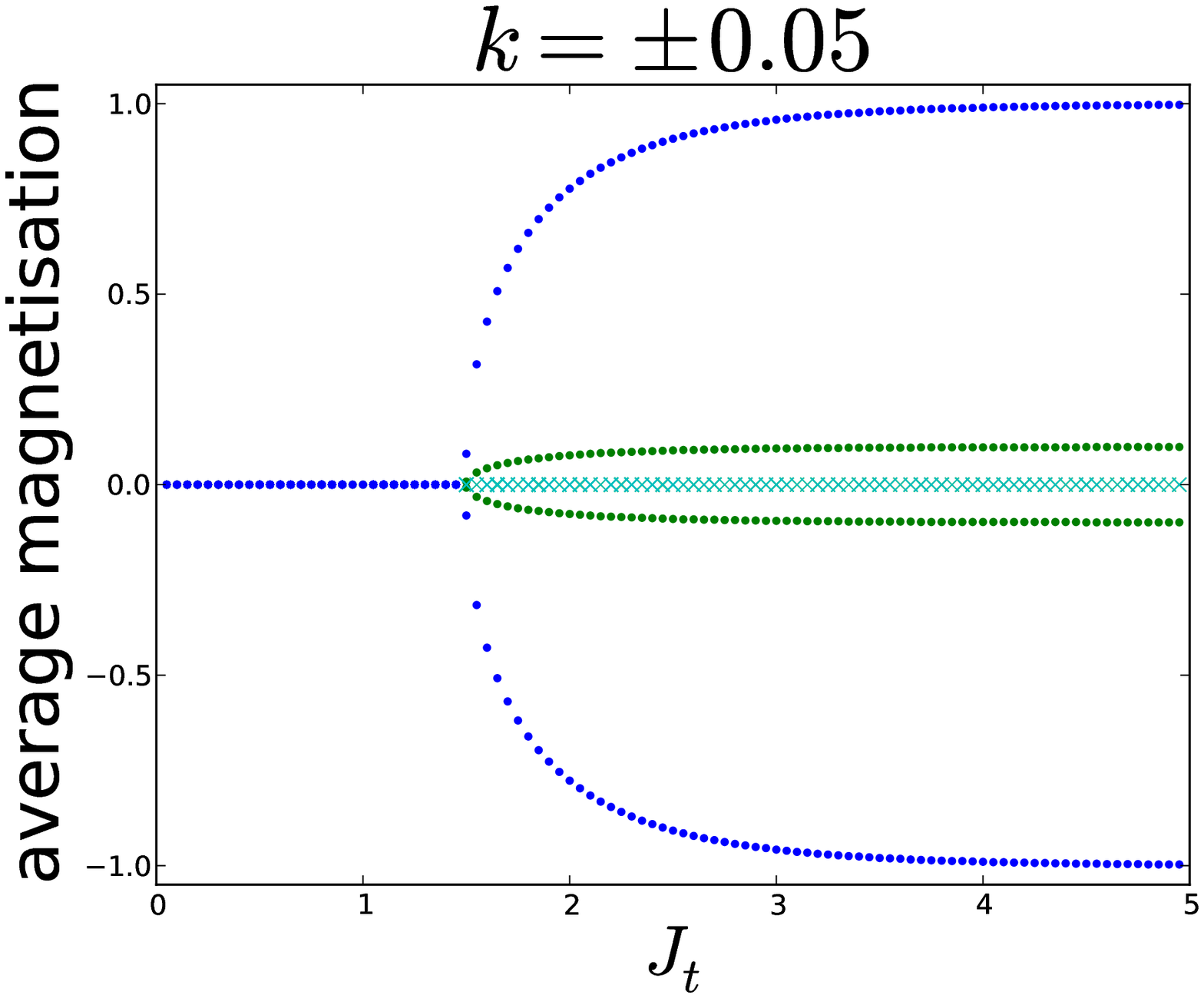}}
\subfloat[]{\includegraphics[width=0.33\textwidth]{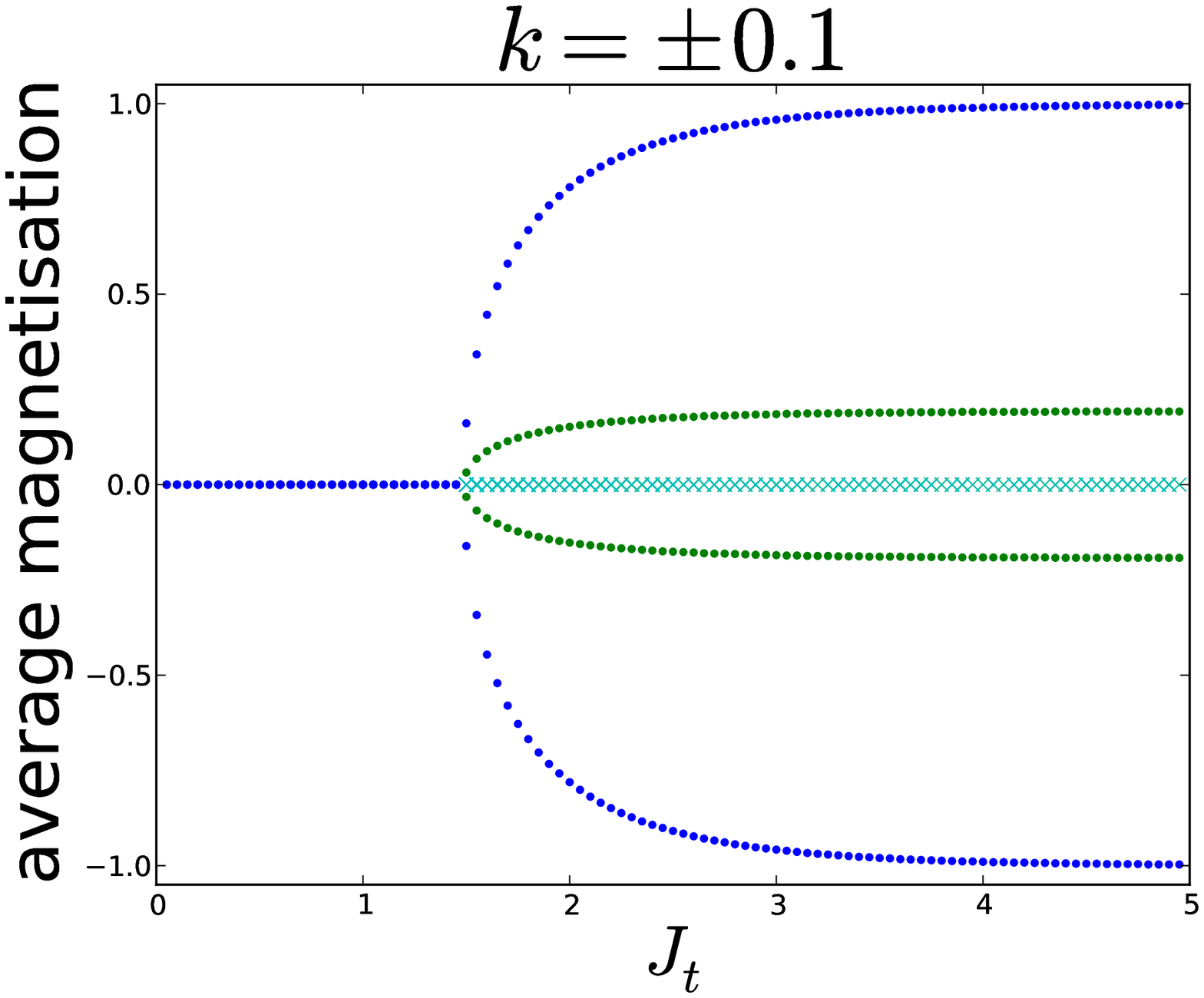}}
\subfloat[]{\includegraphics[width=0.33\textwidth]{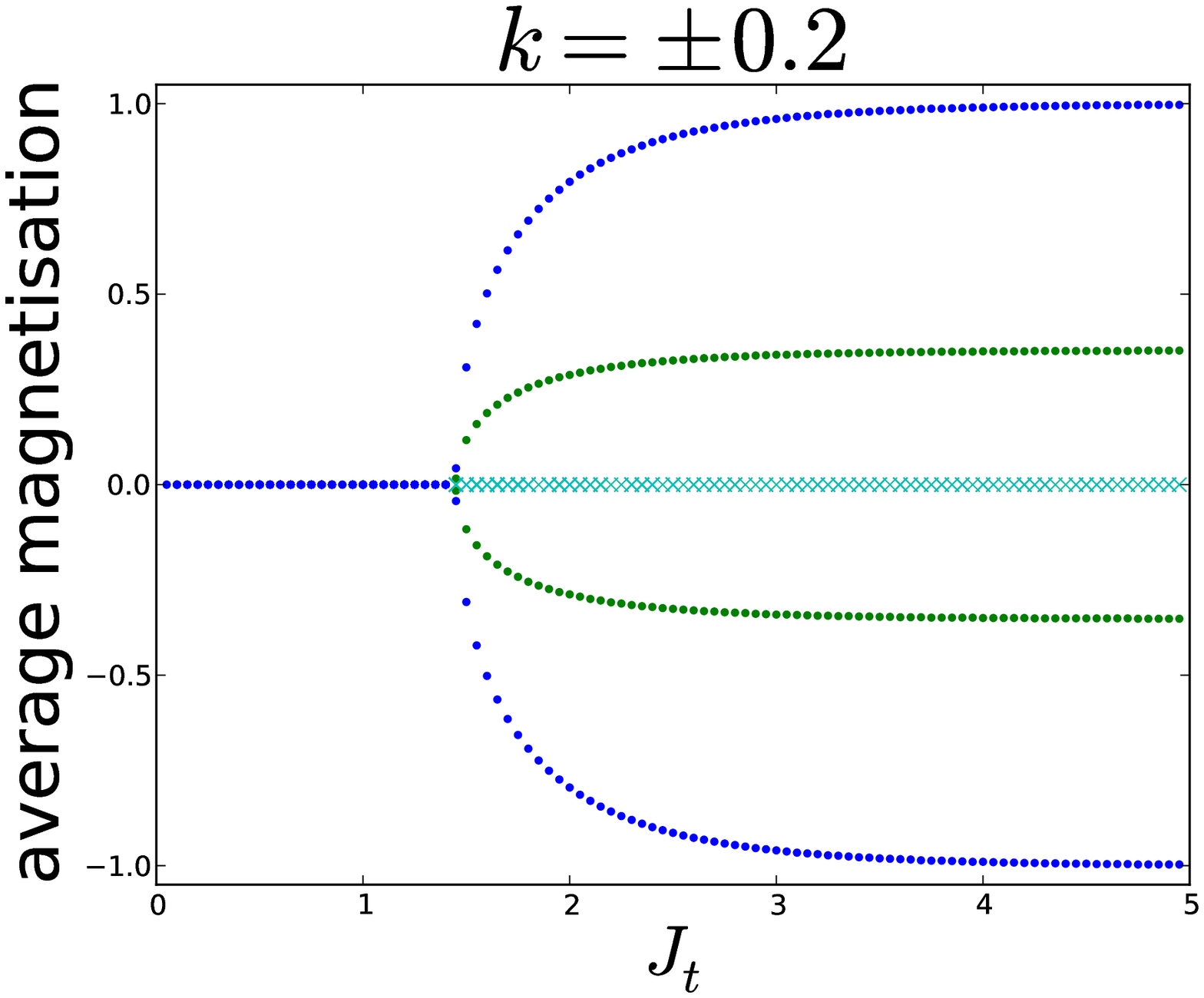}}\\
\subfloat[]{\includegraphics[width=0.33\textwidth]{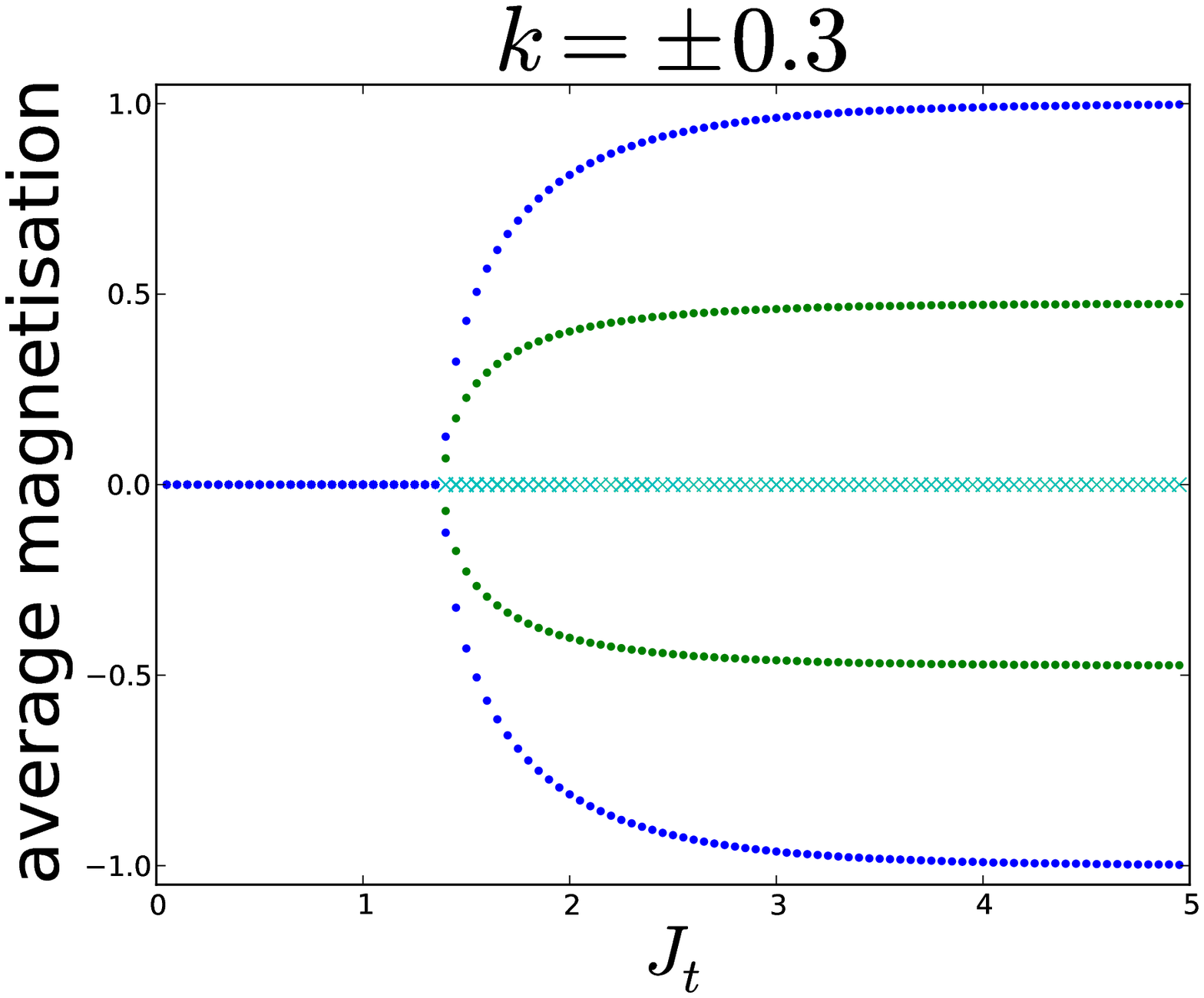}}
\subfloat[]{\includegraphics[width=0.33\textwidth]{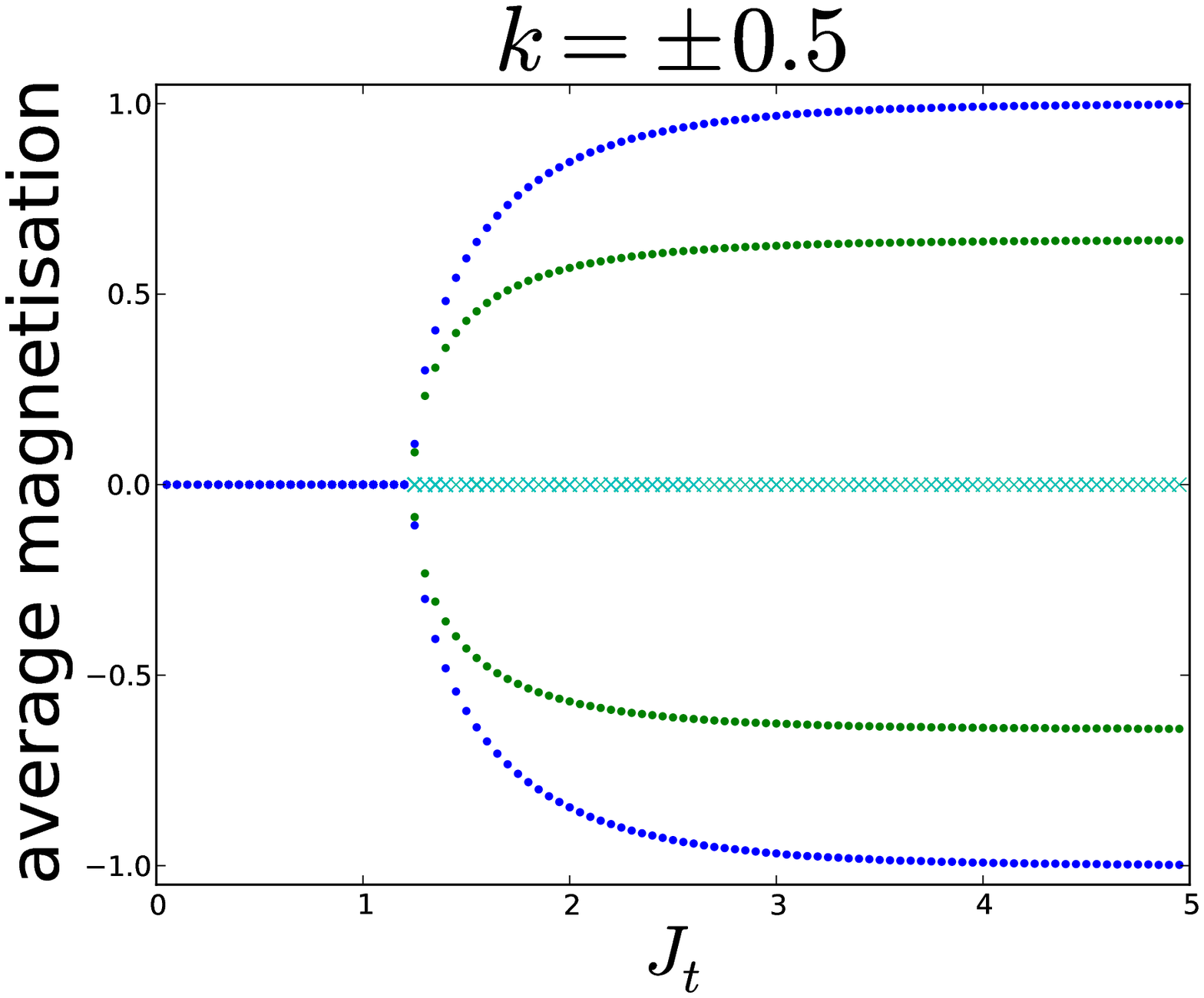}}
\subfloat[]{\includegraphics[width=0.33\textwidth]{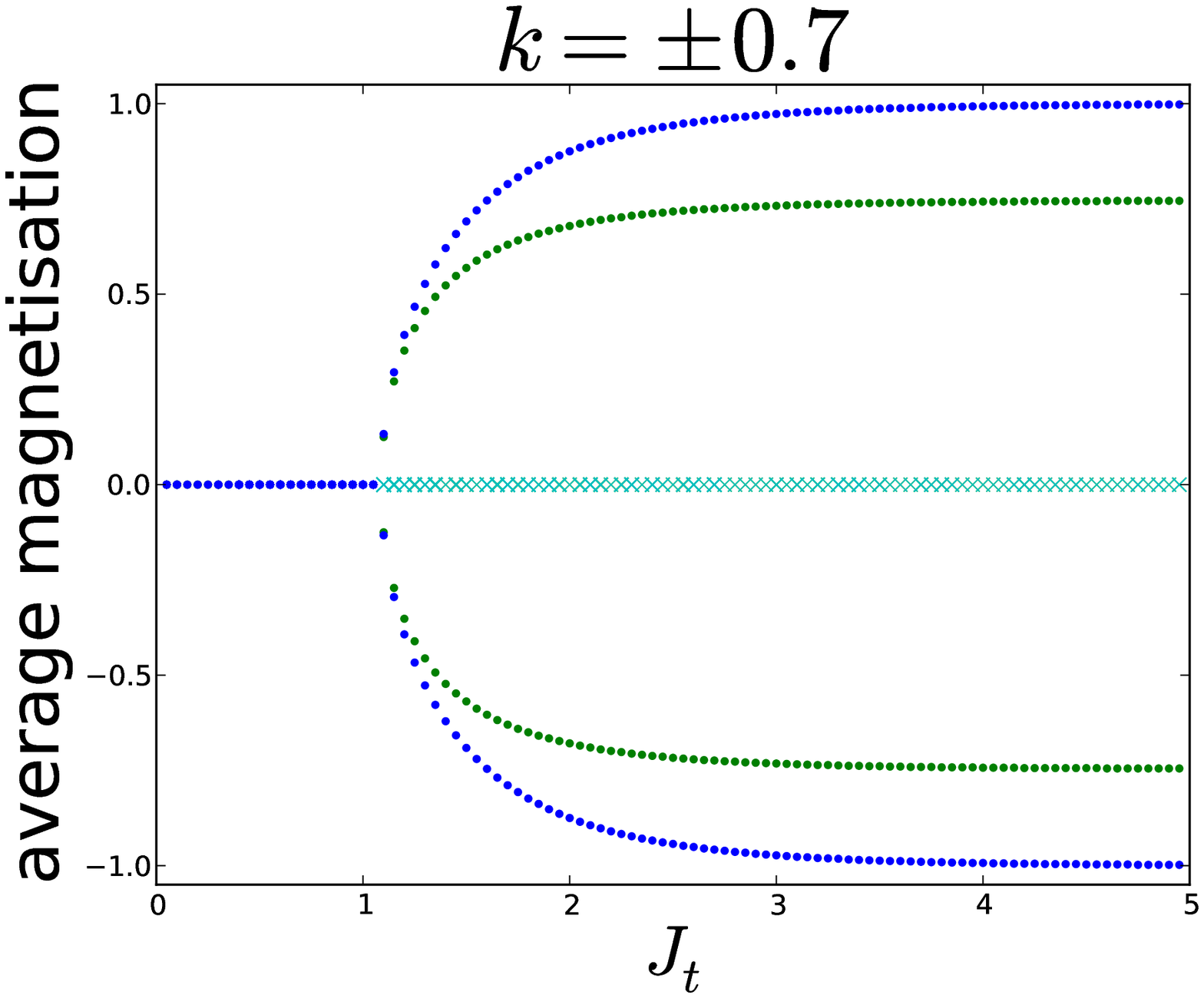}}\\
\subfloat[]{\includegraphics[width=0.33\textwidth]{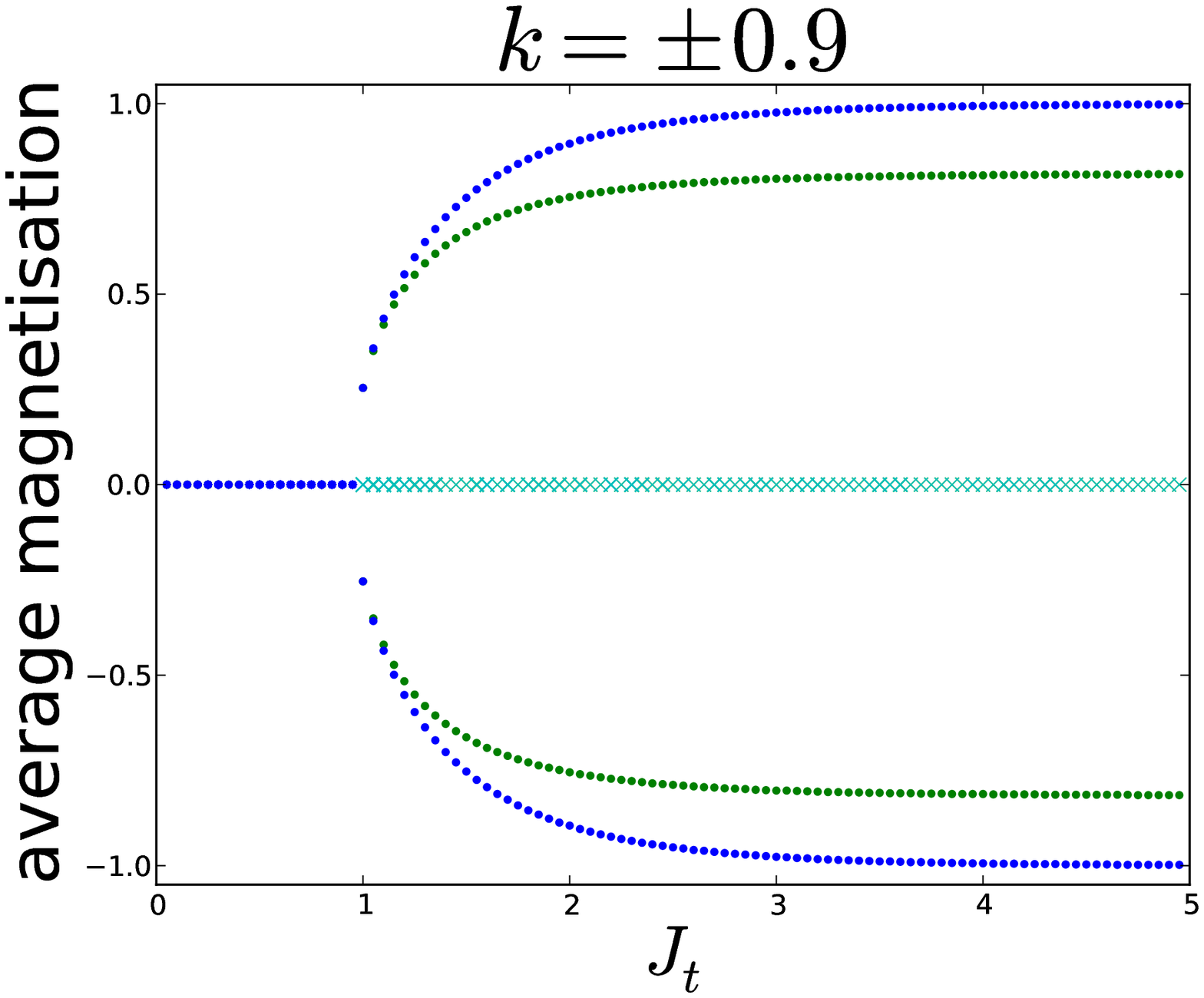}}
\subfloat[]{\includegraphics[width=0.33\textwidth]{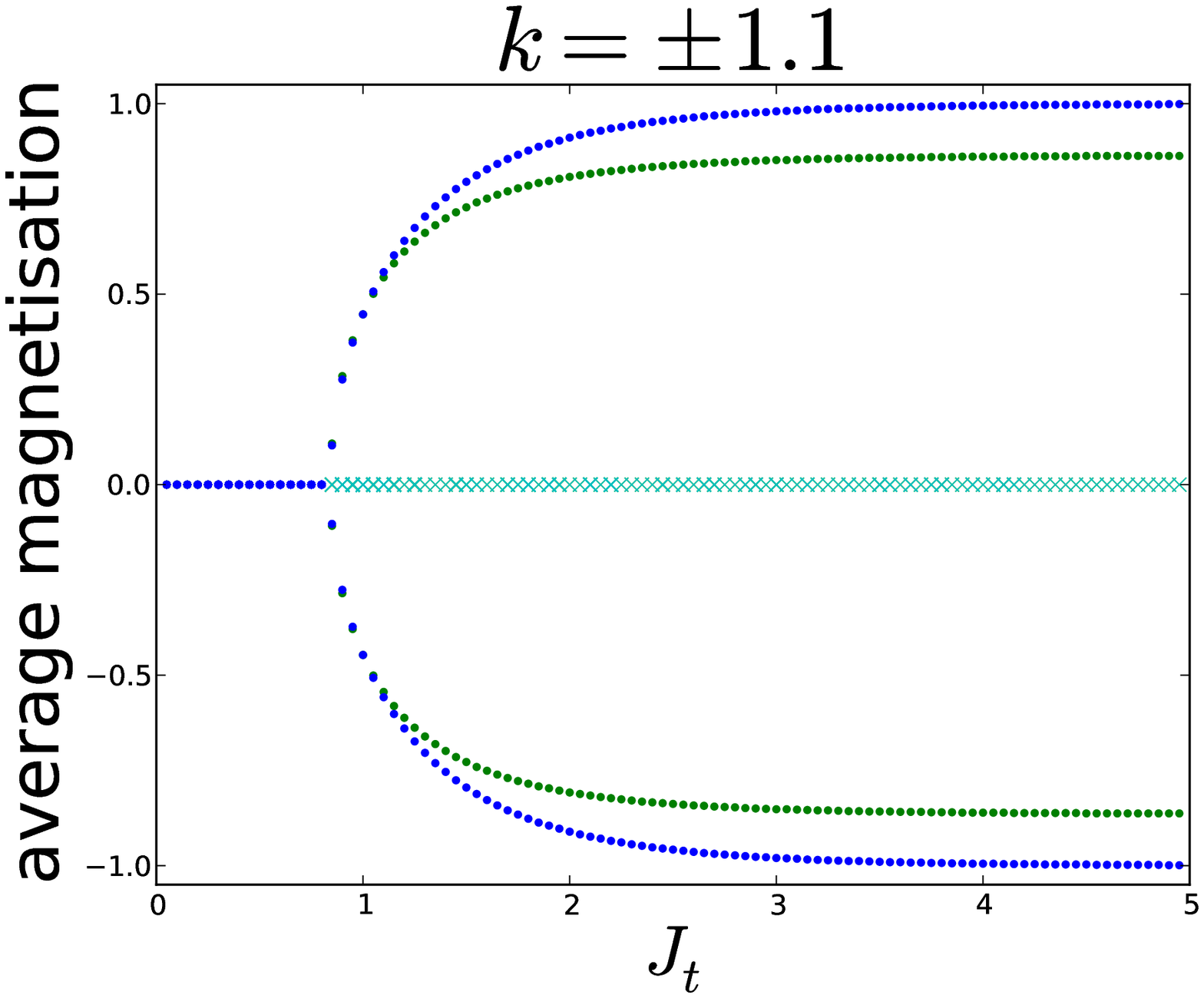}}
\subfloat[]{\includegraphics[width=0.33\textwidth]{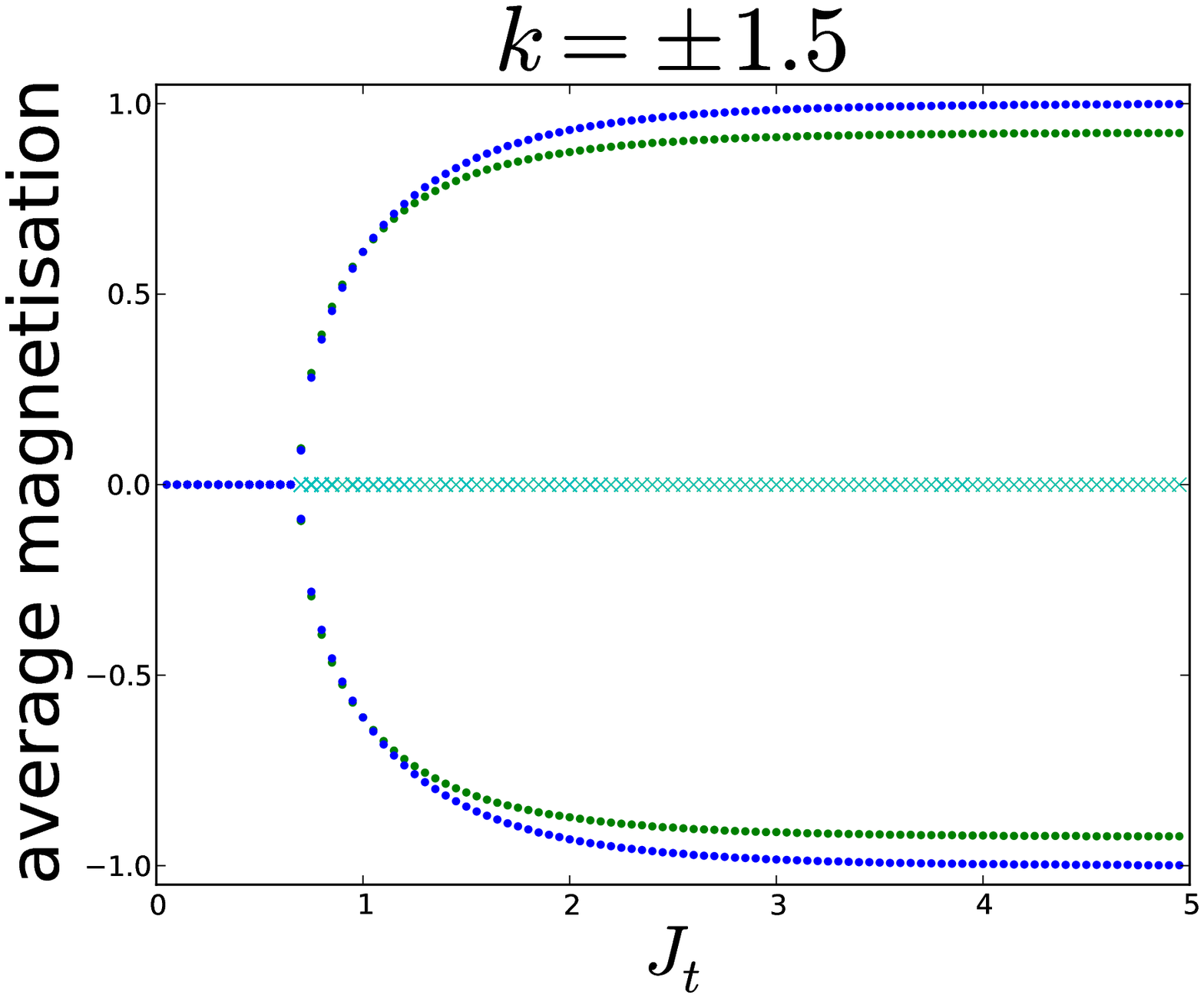}}
\caption{Dependence on intra-coupling $J_{t}$ of the numerically calculated average magnetisations $(s,t)$ for different values of the inter-coupling $k$ at high temperatures. $J_{s}=1$ and $K_{B}T=1.5$  for all plots. (a) $k=0.05$, (b) $k=0.1$, (c) $k=0.2$, (d) $k=0.3$, (e) $k=0.5$, (f) $k=0.7$, (g) $k=0.9$, (h) $k=1.1$ and (i) $k=1.5$. In all cases, different solutions are plotted for intra-coupling $J_{t}$ between 0.01 and 5 every 0.05. Magnetisations are plotted in green for $s$ and blue for $t$. Dark points are used for stable solutions and lighter asp ($\times$, for saddle points) or cross ($+$, for maxima) for non stable solutions.}
\label{fig:locanakJ1}
\end{figure}

\begin{figure}
\centering
\subfloat[]{\includegraphics[width=0.33\textwidth]{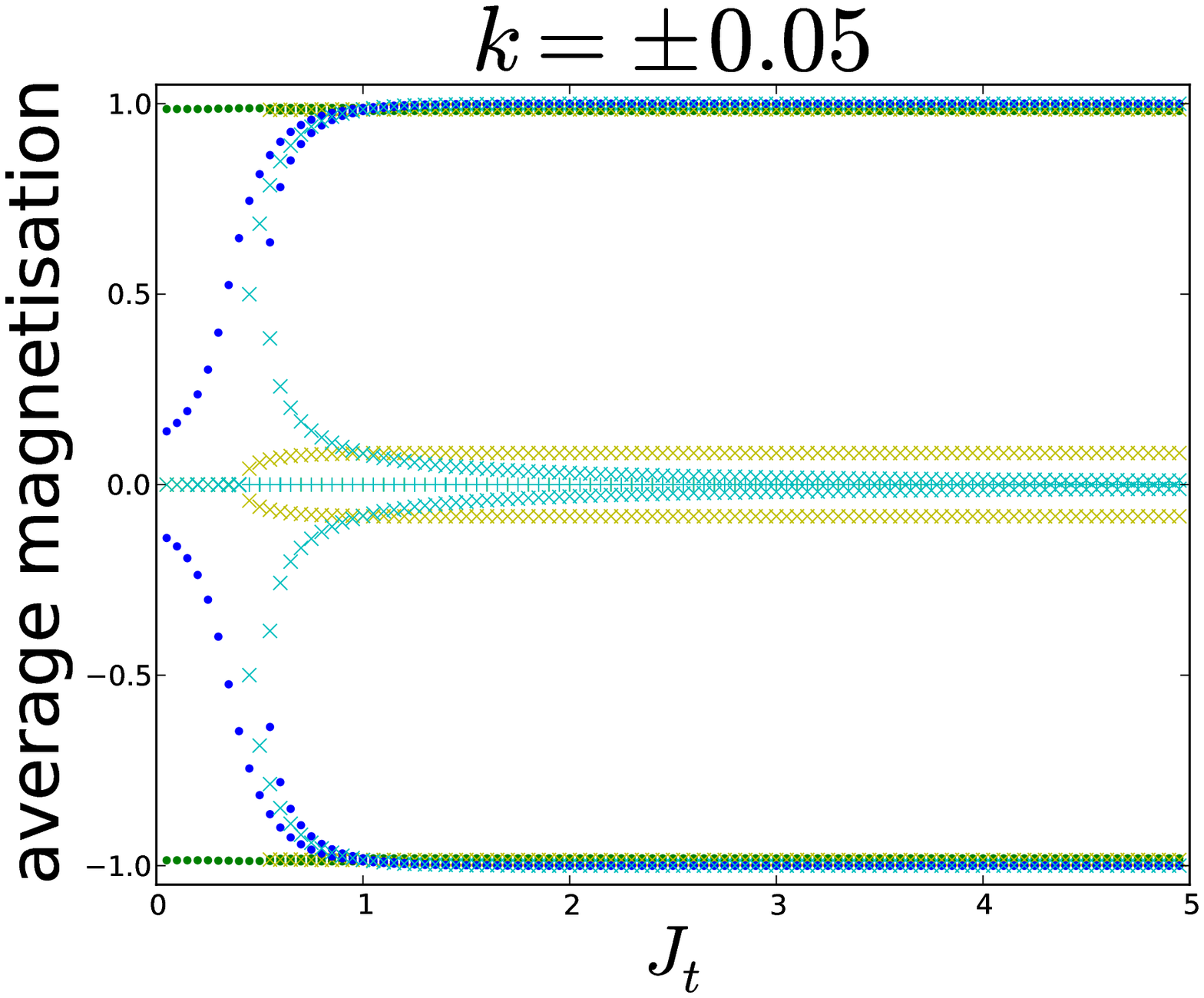}}
\subfloat[]{\includegraphics[width=0.33\textwidth]{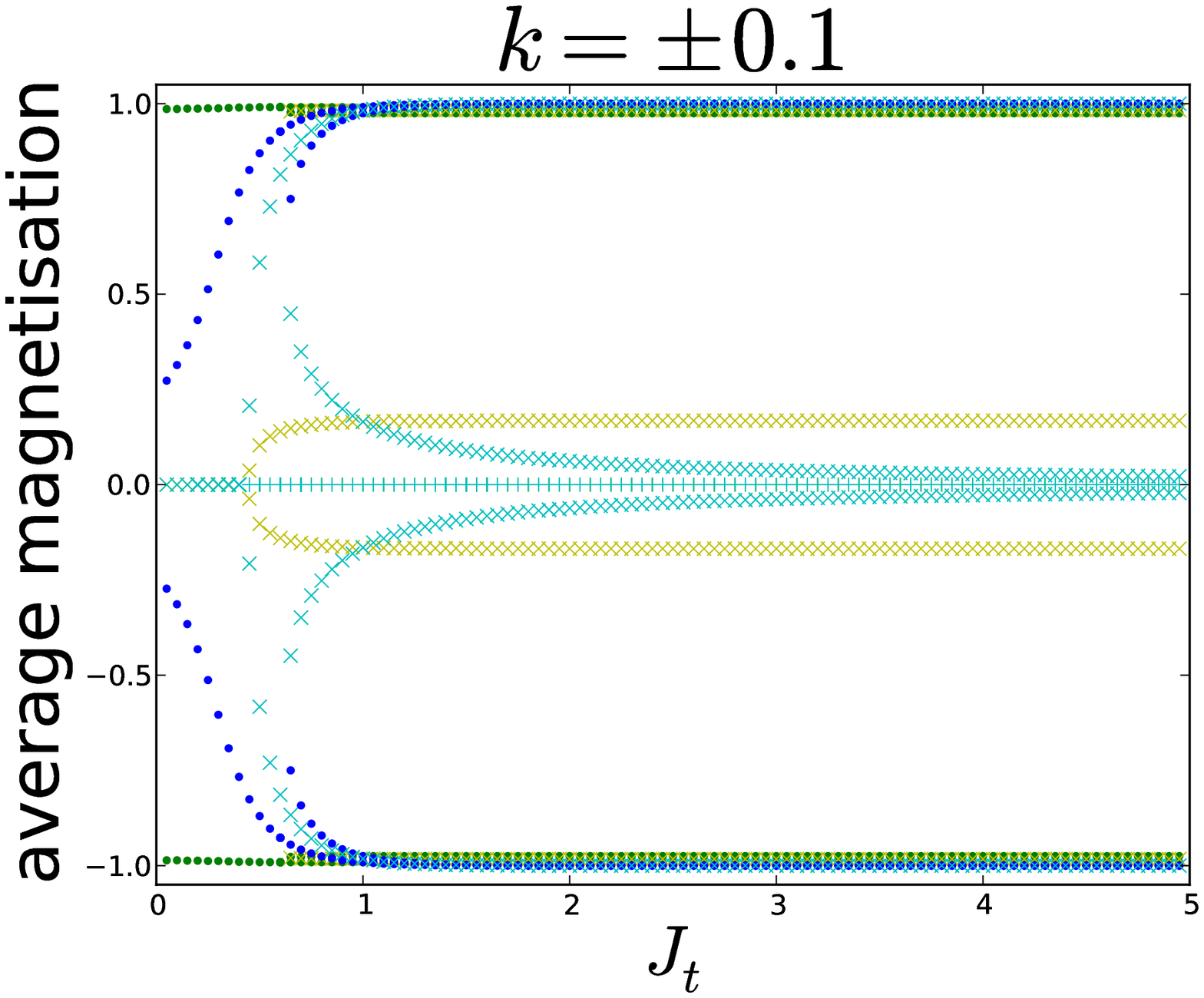}}
\subfloat[]{\includegraphics[width=0.33\textwidth]{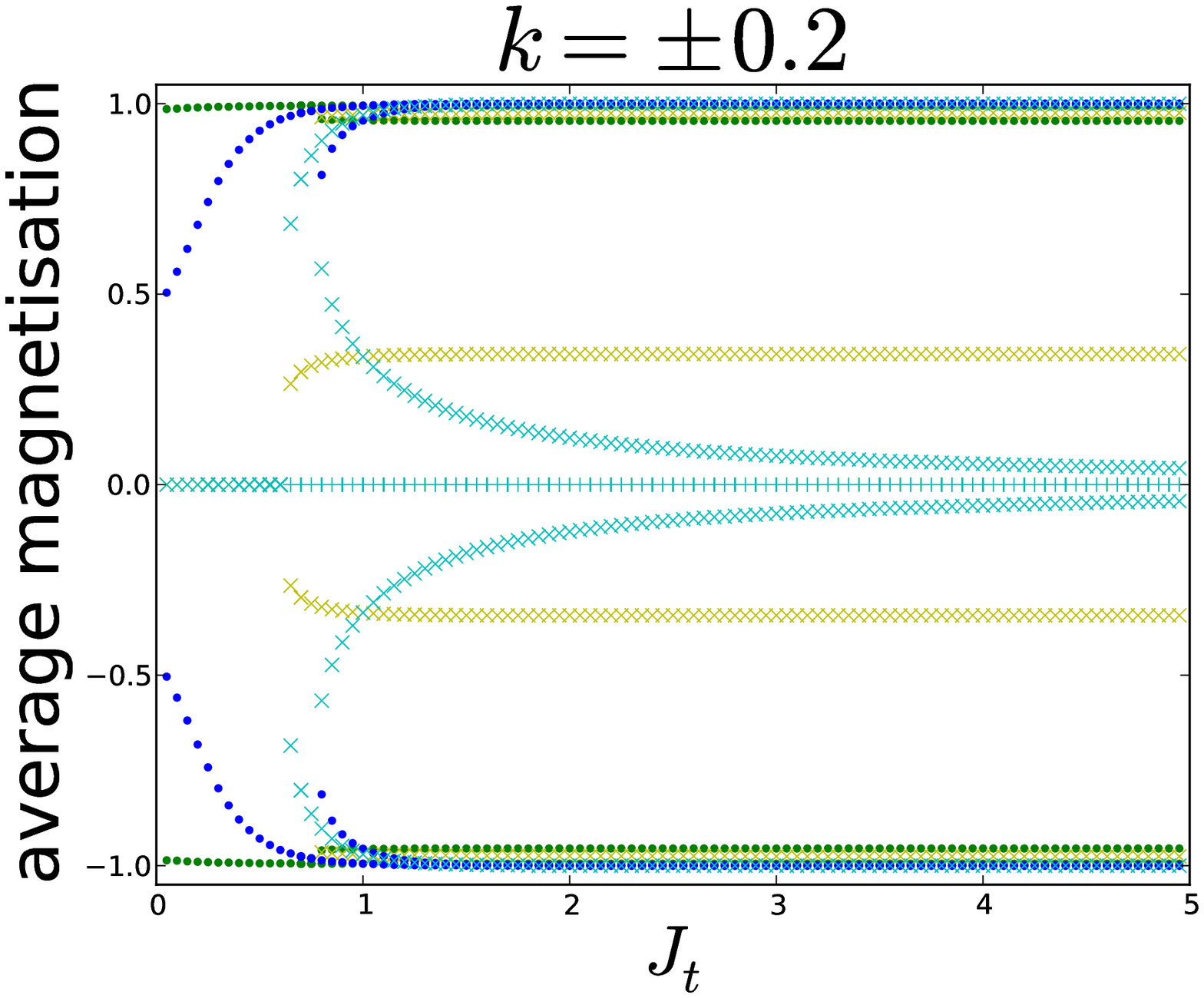}}\\
\subfloat[]{\includegraphics[width=0.33\textwidth]{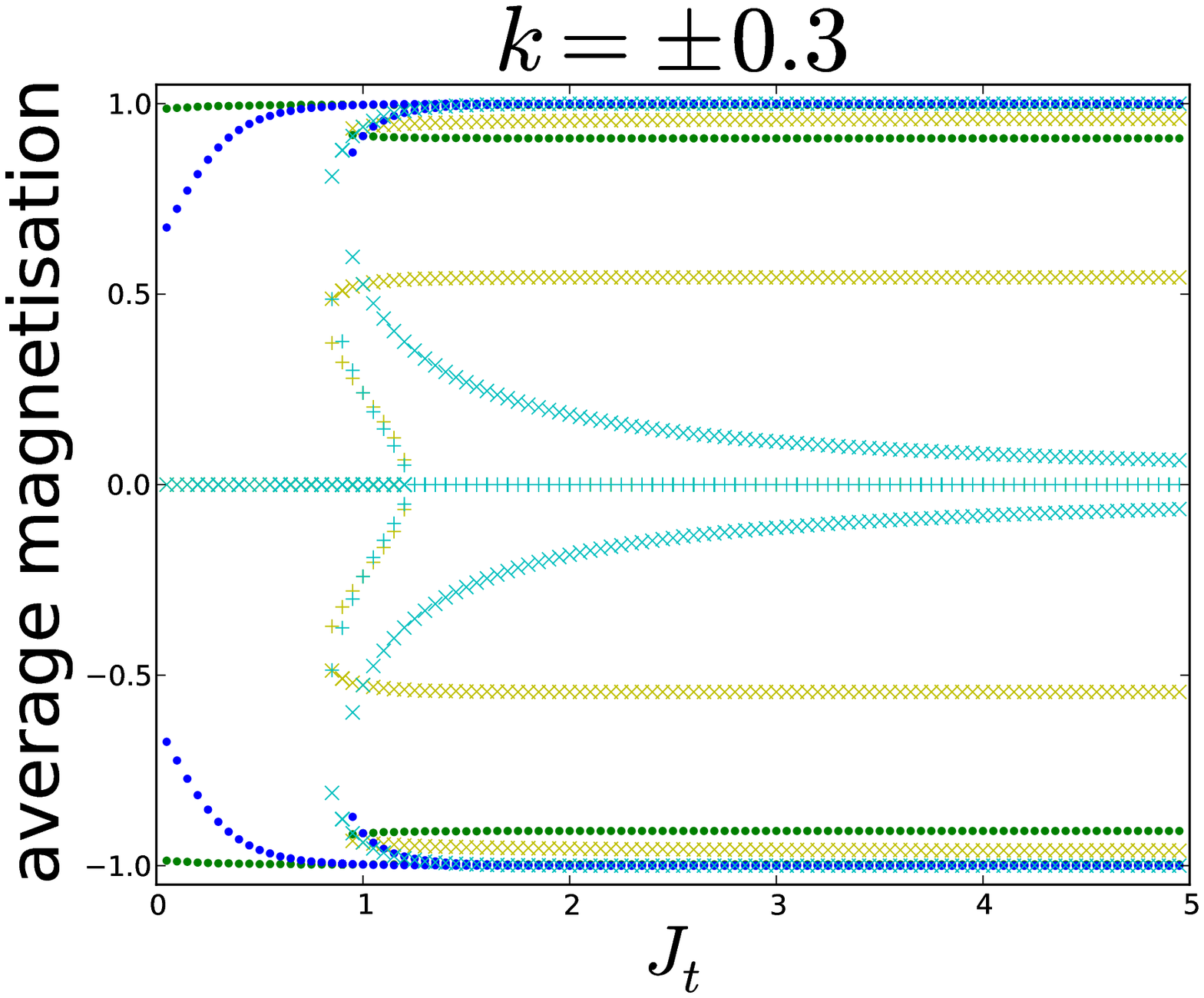}}
\subfloat[]{\includegraphics[width=0.33\textwidth]{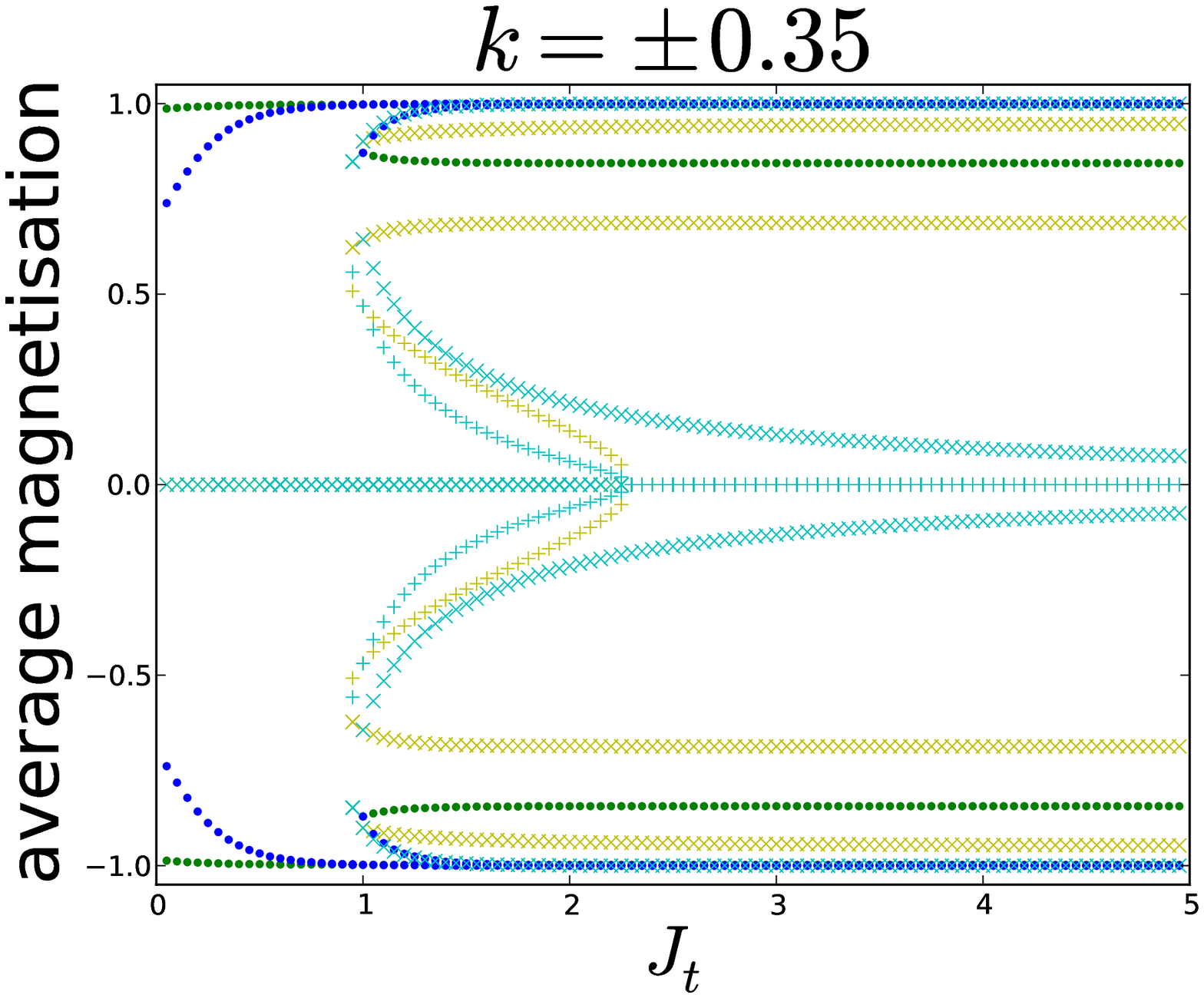}}
\subfloat[]{\includegraphics[width=0.33\textwidth]{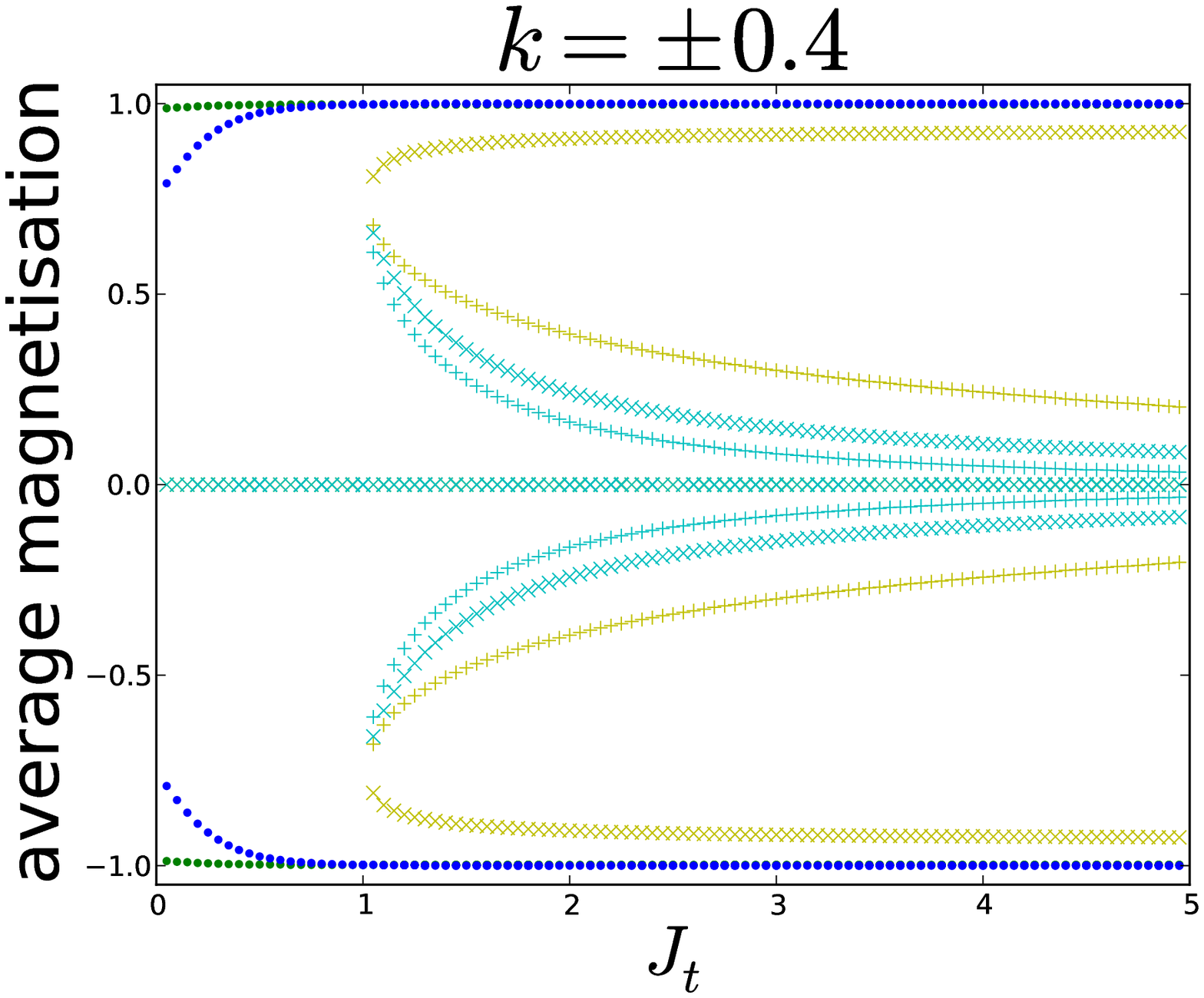}}\\
\subfloat[]{\includegraphics[width=0.33\textwidth]{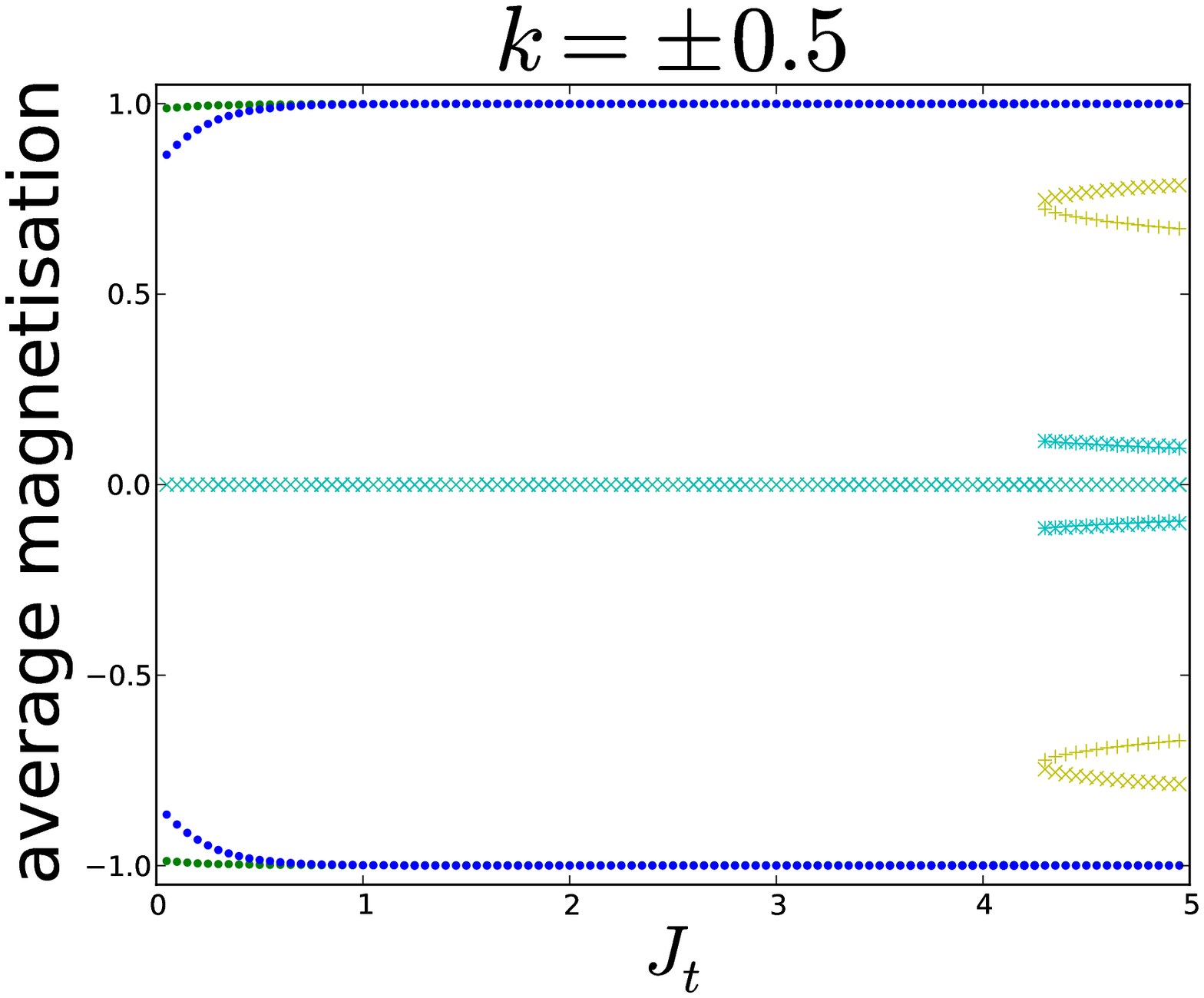}}
\subfloat[]{\includegraphics[width=0.33\textwidth]{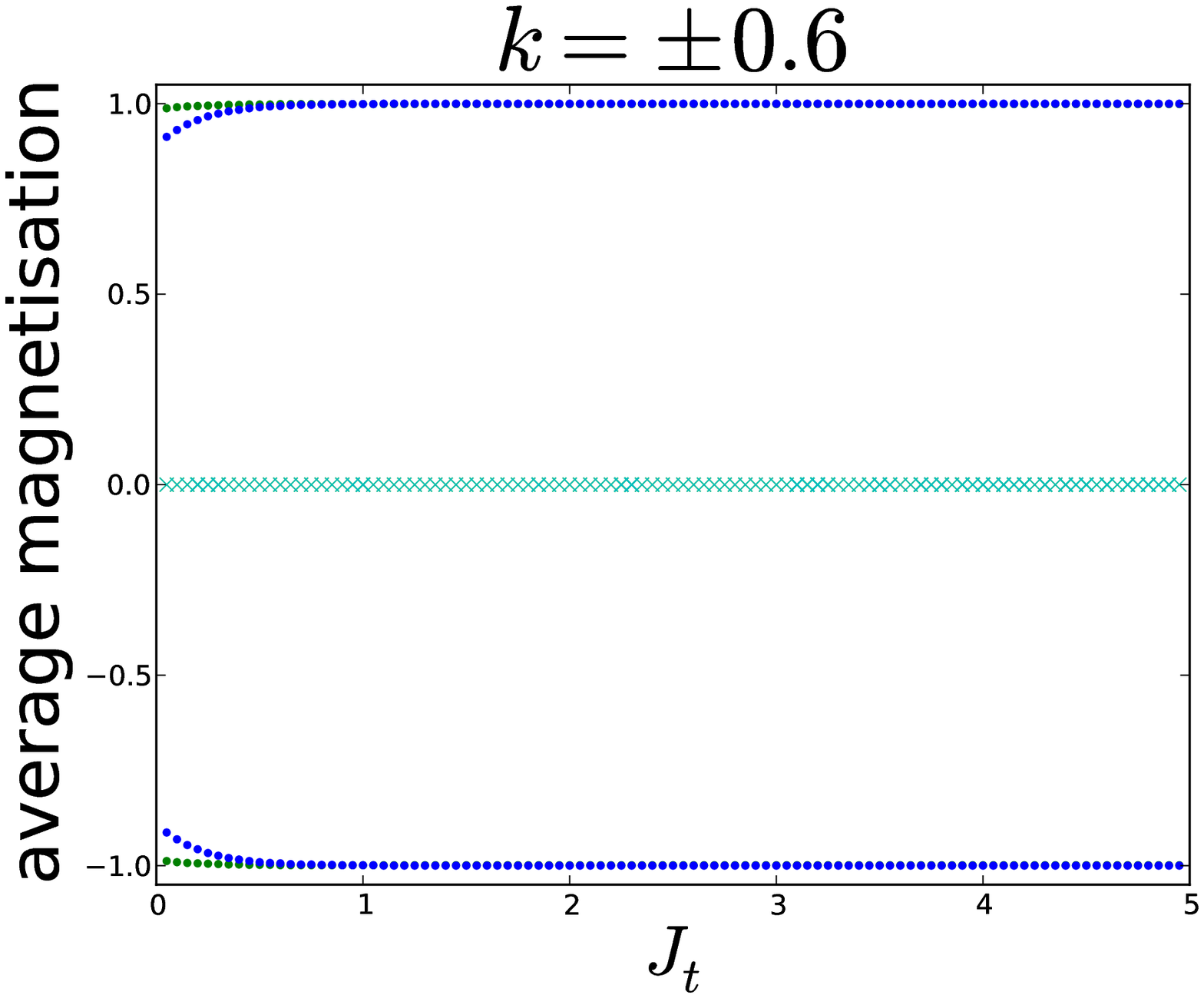}}
\subfloat[]{\includegraphics[width=0.33\textwidth]{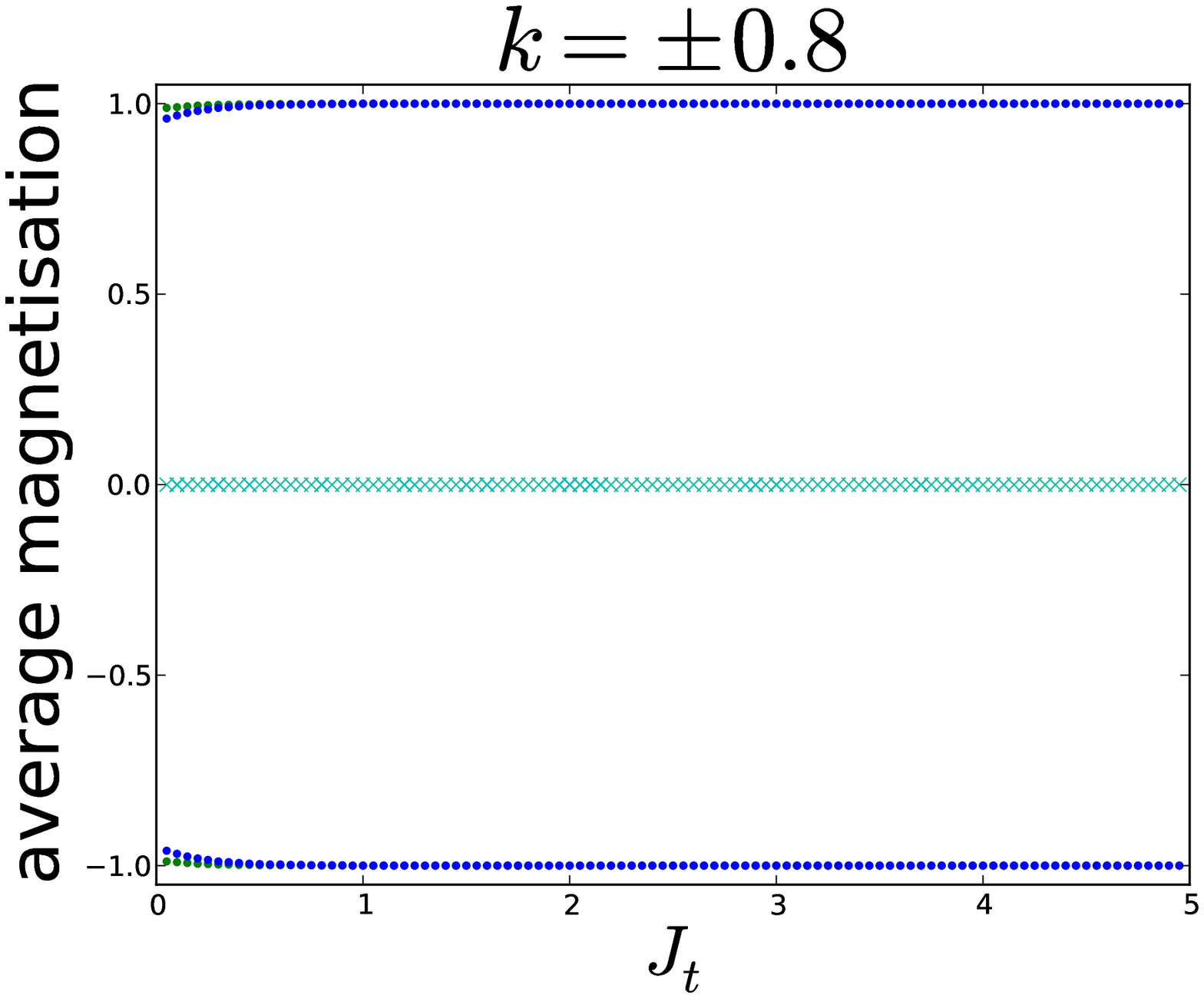}}
\caption{Dependence on intra-coupling $J_{t}$ of the numerically calculated average magnetisations $(s,t)$ for different values of the inter-coupling $k$ at low temperatures. $J_{s}=1$ and $K_{B}T=0.4$  for all plots. (a) $k=0.05$, (b) $k=0.1$, (c) $k=0.2$, (d) $k=0.3$, (e) $k=0.35$, (f) $k=0.4$, (g) $k=0.5$, (h) $k=0.6$ and (i) $k=0.8$. In all cases, different solutions are plotted for intra-coupling $J_{t}$ between 0.01 and 5 every 0.05. Magnetisations are plotted in green for $s$ and blue for $t$. Dark points are used for stable solutions and lighter asp ($\times$, for saddle points) or cross ($+$, for maxima) for non stable solutions.}
\label{fig:locanakJ2}
\end{figure}

Figures \ref{fig:locanakJ1} and \ref{fig:locanakJ2} show how the dependence on $J_{t}$ varies as we raise the (absolute) value of  $k$ for fixed values of the rest of the parameters, for the high and low temperature scenarios studied in detail for the nonlocal case. Besides the already noted generic differences, we can see that in the high temperature case, $J_{t}^{c}$ always exists, and so there is always a region of low intra-coupling at which the paramagnetic phase is stable. In the low temperature case, at high enough values of $|k|$, $J_{t}^{c}$ disappears (becomes negative in fact), and so do all ferromagnetic saddle points and maxima (figure \ref{fig:locanakJ2} h and i), which was not the case when studying the nonlocal model.\\

\section{Phase diagram for the zero field case}
\label{sec:locphadia}
We will be using the same methodology and conventions as in section \ref{sec:nlophadia}. 

In this case, the expansion of the free energy density functional \eqref{eq:g} for $|s| \ll 1$ and $|t| \ll 1$ (and finite nonzero temperature) is

\begin{equation}
\begin{array}{l}
f = -\frac{2}{\beta}\log2-\frac{1}{\beta}\log\lbrack\cosh(\beta k) \rbrack 
+\frac{1}{2}J_{s}\left(1-\beta J_{s}\right)s^{2}+\frac{1}{2}J_{t}\left(1-\beta J_{t}\right)t^{2}+\\
\qquad +\tanh(\beta k)\beta J_{s}J_{t}st+\frac{1}{12}\beta^{3}J_{s}^{4}s^{4}+\frac{1}{12}\beta^{3}J_{t}^{4}t^{4}+\frac{1}{2}\tanh^{2}(\beta k)\beta^{3}J_{s}^{2}J_{t}^{2}s^{2}t^{2}+\\
\qquad +\frac{1}{3}\tanh(\beta k)\beta^{3}J_{s}J_{t}^{3}st^{3}+\frac{1}{3}\tanh(\beta k)\beta^{3}J_{s}^{3}J_{t}s^{3}t
\end{array}
\end{equation}

Let us now review how the different cross sections of the phase diagram look throughout the next subsections. Recall that only stable solutions will be considered.

\subsection{Two dimensional $k - T$ sections}

\begin{figure}
\centering
\includegraphics[width=\textwidth]{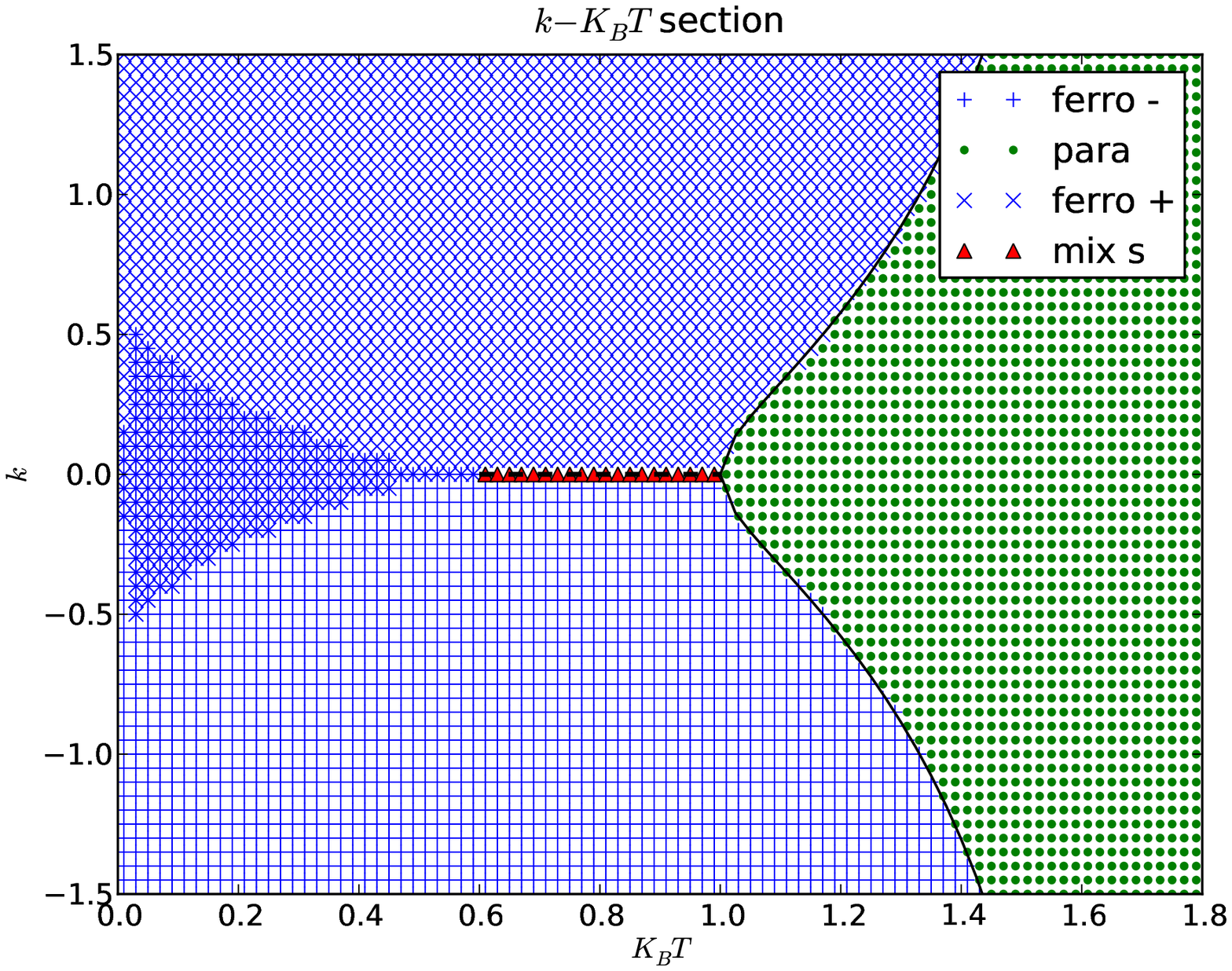}
\caption{Two dimensional $k-K_{B}T$ section of the phase diagram as given by the numerical average magnetisations used in the analysis in section \ref{sec:locnumsol} ($J_{s}=1$ and $J_{t}=0.6$). Only stable solutions are shown. The phase associated to each solution is plotted for $K_{B}T$ between 0.05 and 1.8 and for $k$ between -1 and 1 and solutions are shown at intervals of 0.02 for $K_{B}T$ and 0.05 for $k$. Green points are used for the paramagnetic phase , blue asps ({$\times$}) for ferromagnetic phases with $s$ and $t$ of the same sign and blue crosses ($+$) for ferromagnetic phases with $s$ and $t$ of opposite signs. Red triangles are used for the mixed phase where $s\neq 0$. Solid black lines are used for the critical branches $k_{c}^{+}=K_{B}T\tanh^{-1}\left(\sqrt{1-\frac{K_{B}T(J_{s}+J_{t})}{J_{s}J_{t}}+\frac{(K_{B}T)^{2}}{J_{s}J_{t}}}\right)= K_{B}T\tanh^{-1}\left(\sqrt{1-2.67K_{B}T+1.67(K_{B}T)^{2}}\right)$ and $k_{c}^{-}=-K_{B}T\tanh^{-1}\left(\sqrt{1-\frac{K_{B}T(J_{s}+J_{t})}{J_{s}J_{t}}+\frac{(K_{B}T)^{2}}{J_{s}J_{t}}}\right)=-K_{B}T\tanh^{-1}\left(\sqrt{1-2.67K_{B}T+1.67(K_{B}T)^{2}}\right)$ and a dashed line for the mixed segment $J_{t}<K_{B}T<J_{s}$ ($k=0$). }
\label{fig:locphadiakT}
\end{figure}

Figure \ref{fig:locphadiakT} shows the $k-K_{B}T$ section for $J_{s}=1$ and $J_{t}=0.6$. Besides numerical results, analytical derivations for the curves delimiting the different phase regions are plotted. In this case these are $k_{c}=\pm K_{B}T\tanh^{-1}\left(\sqrt{1-\frac{K_{B}T(J_{s}+J_{t})}{J_{s}J_{t}}+\frac{(K_{B}T)^{2}}{J_{s}J_{t}}}\right)=\pm K_{B}T\tanh^{-1}\left(\sqrt{1-2.67K_{B}T+1.67(K_{B}T)^{2}}\right)$ for $K_{B}T>K_{B}T_{c}^{s0}=J_{s}=1$\footnote{For $0.6=J_{t}=K_{B}T_{uc}^{t}<K_{B}T<K_{B}T_{uc}^{s}=J_{s}=1$, the critical curve does not exist. For $K_{B}T<J_{t}$ (not shown in figure \ref{fig:locphadiakT}), it is an ellipse shaped curved with one of its axis along the $K_{B}T$ axis and with one of its diameters lying between $K_{B}T=0$ and $K_{B}T=J_{t}$. These are points where the paramagnetic phase changes its stability from saddle to maxima, and the fact that it goes to zero (instead of infinity as in the nonlocal case) for $T\to0$, is what makes the non stability type of the paramagnetic phase have an additional change in the local case, related to the fact that metastable states do not exist at sufficiently low temperatures. Note that the critical curves for the two models have important qualitative differences for $K_{B}T<J_{t}$, precisely where its points are not related to real phase transitions.} (solid black line) separating ferromagnetic from paramagnetic regions, and the mixed phase segment (identical to the nonlocal case) $k=0$ and $J_{t}<K_{B}T<J_{s}$ (dashed line).

When compared to the nonlocal model's cross section, we note some qualitative differences. First of all, the always present region of instability for the nonlocal model (strong coupling regime) which is never the case for this cross section. We can now find solutions for the whole of the parameter space. The other two differences that were already noted when analysing the numerical results for the average magnetisation vector are, that in the present case, at low enough temperatures there are no metastable states regardless of the value of $k$, and at high enough temperatures only the paramagnetic phase is stable for all values of $k$. This is due to the fact that the critical curve has an horizontal asymptote $K_{B}T= J_{s}+J_{t}=1.6$. For values of the temperature above it, the system will be in the paramagnetic phase no matter how high the absolute value of $|k|$ gets. 

Besides these interesting features, both phase diagram sections look remarkably similar, even in its actual values of critical points, specially for small $|k|$.

\subsection{Two dimensional $J_{t} - T$ sections}

\begin{figure}
\centering
\includegraphics[width=\textwidth]{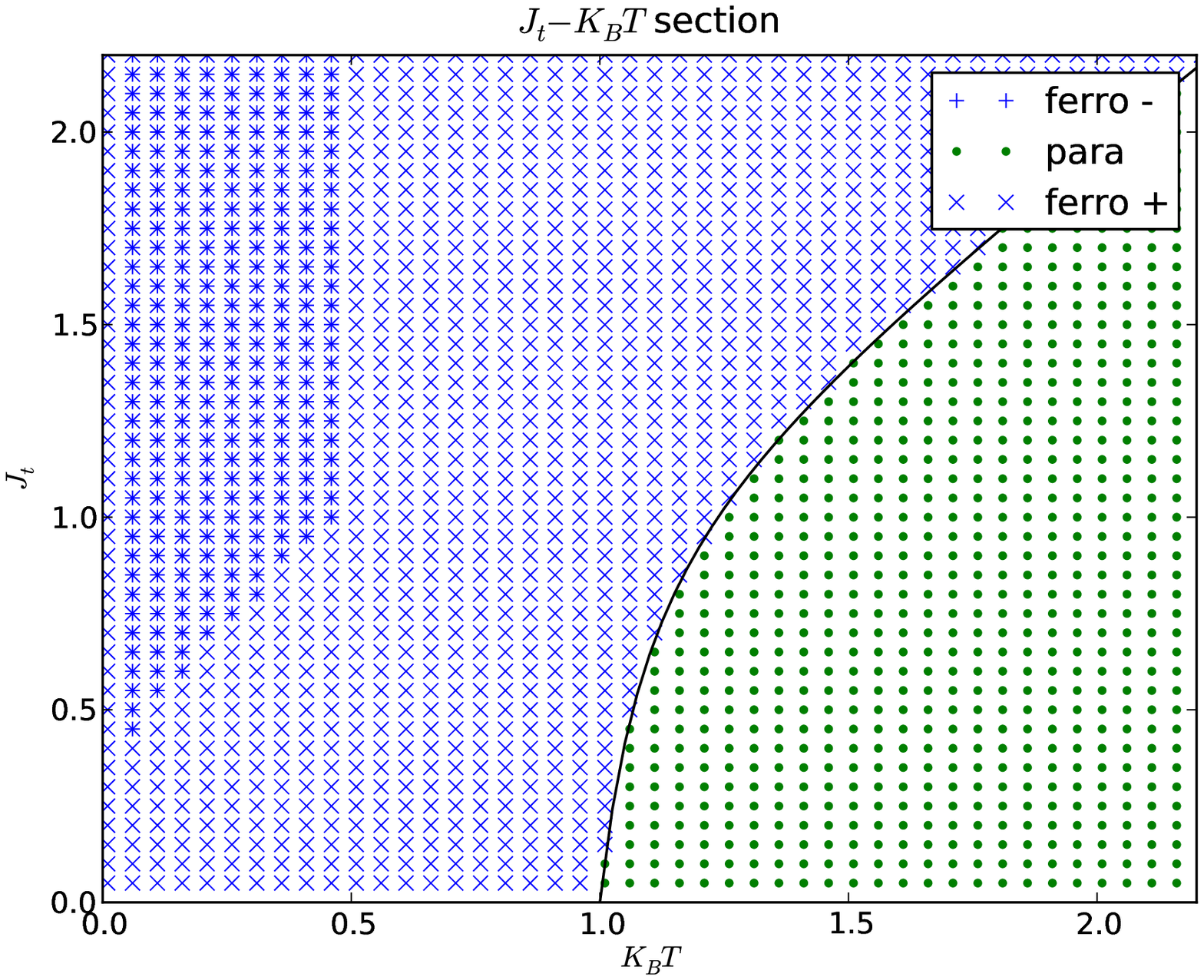}
\caption{Two dimensional $J_{t}-K_{B}T$ section of the phase diagram as given by the numerical average magnetisations used in the analysis in section \ref{sec:locnumsol} ($J_{s}=1$ and $k=0.3$). Only stable solutions are shown. The phase associated to each solution is plotted for $K_{B}T$ and $J_{t}$ between 0.01 and 2.2  and solutions are shown at intervals of 0.05 for both $K_{B}T$ and $J_{t}$. Green points are used for the paramagnetic phase , blue asps ({$\times$}) for ferromagnetic phases with $s$ and $t$ of the same sign and blue crosses ($+$) for ferromagnetic phases with $s$ and $t$ of opposite signs. A solid black line is used for the critical curve $J_{t}^{c}=\frac{K_{B}T(J_{s}-K_{B}T)}{J_{s}(1-\alpha_{k}^{2})-K_{B}T}=\frac{K_{B}T(1-K_{B}T)}{1-\tanh^{2}(\frac{0.3}{K_{B}T})-K_{B}T}$ when $K_{B}T>J_{s}=1$.}
\label{fig:locphadiaJT}
\end{figure}

Figure \ref{fig:locphadiaJT} shows  the $J_{t} - T$ section for $J_{s}=1$ and $k=0.3$. Same graphical conventions as in last section and previous chapter are used. The critical curve (solid black line) is now given by $J_{t}^{c}=\frac{K_{B}T(J_{s}-K_{B}T)}{J_{s}(1-\alpha_{k}^{2})-K_{B}T}=\frac{K_{B}T(1-K_{B}T)}{1-\tanh^{2}(\frac{0.3}{K_{B}T})-K_{B}T}$ when $K_{B}T>J_{s}=1$\footnote{When $K_{B}T<J_{s}$, the function (although no longer represents critical points) is not symmetric and is bell shaped with its branches pointing upwards, giving the additional points where the non stability type of the paramagnetic changes. Although not drawn in figure \ref{fig:locphadiaJT}, it would look strange, as it goes over regions with and without metastable states. We must remember in the local case, the relation between points where the paramagnetic solution changes from saddle to maximum are not directly related to the onset of non stable ferromagnetic solutions (and so their connection to metastable states is even more subtle than in the nonlocal case).}.  

Leaving aside the behaviour of non-stable solutions, we again find that cross sections for both nonlocal and local models are very similar, even quantitatively. The only qualitative difference is that there are stable solutions for all of the plane and no metastable states for low enough temperatures in the local case\footnote{Although for this particular choice of parameters in the local case the {\itshape diffuse} effect (around the temperature at which the width of the metastable region becomes fixed) observed in the nonlocal case is not present, for other choices of the parameters this behaviour also exists for this model and so it is not something characteristic of non-locality. }.

\subsection{Two dimensional $J_{t} - k$ sections}

\begin{figure}
\centering
\includegraphics[width=\textwidth]{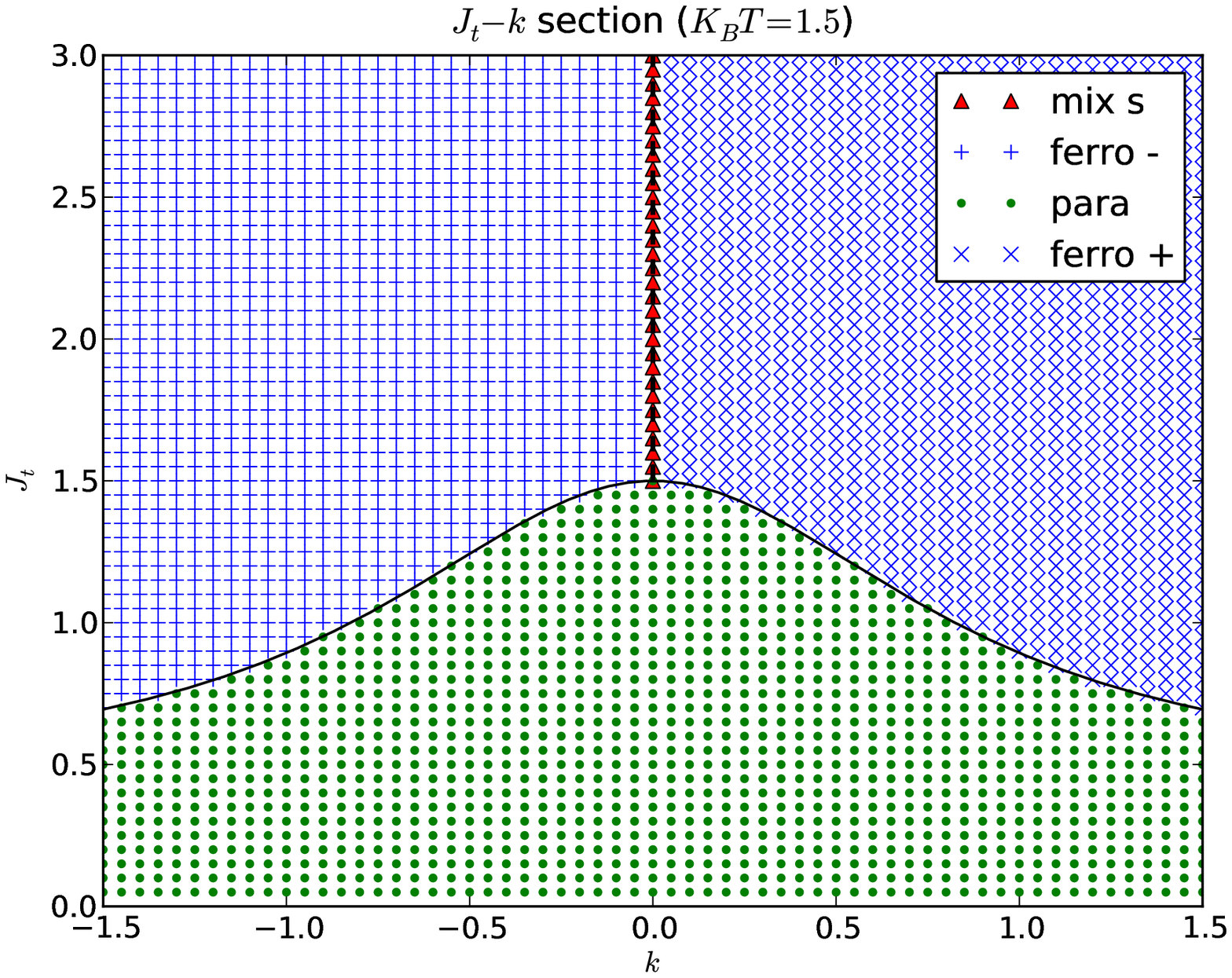}
\caption{Two dimensional $J_{t}-k$ sections of the phase diagram as given by the numerical average magnetisations for $J_{s}=1$ and  $K_{B}T=1.5$. Only stable solutions are shown. The phase associated to each solution is plotted for $k$ between -1.5 and 1.5 and $J_{t}$ between 0.01 and 3  at intervals of 0.05. Green points are used for the paramagnetic phase, blue asps ({$\times$}) for ferromagnetic phases with $s$ and $t$ of the same sign and blue crosses ($+$) for ferromagnetic phases with $s$ and $t$ of opposite signs. Red triangles are used for the mixed phase where $s\neq 0$. The critical curve (solid black line) is given by  $J_{t}^{c}=\frac{K_{B}T(J_{s}-K_{B}T)}{J_{s}(1-\alpha_{k}^{2})-K_{B}T}=\frac{0.75}{\tanh^{2}(\frac{2k}{3})+0.5}$ and the mixed segment by $k=0$ and $J_{t}>J_{t}^{c0}=K_{B}T=1.5$ (dotted line). }
\label{fig:locphadiaJk1}
\end{figure}

\begin{figure}
\centering
\includegraphics[width=\textwidth]{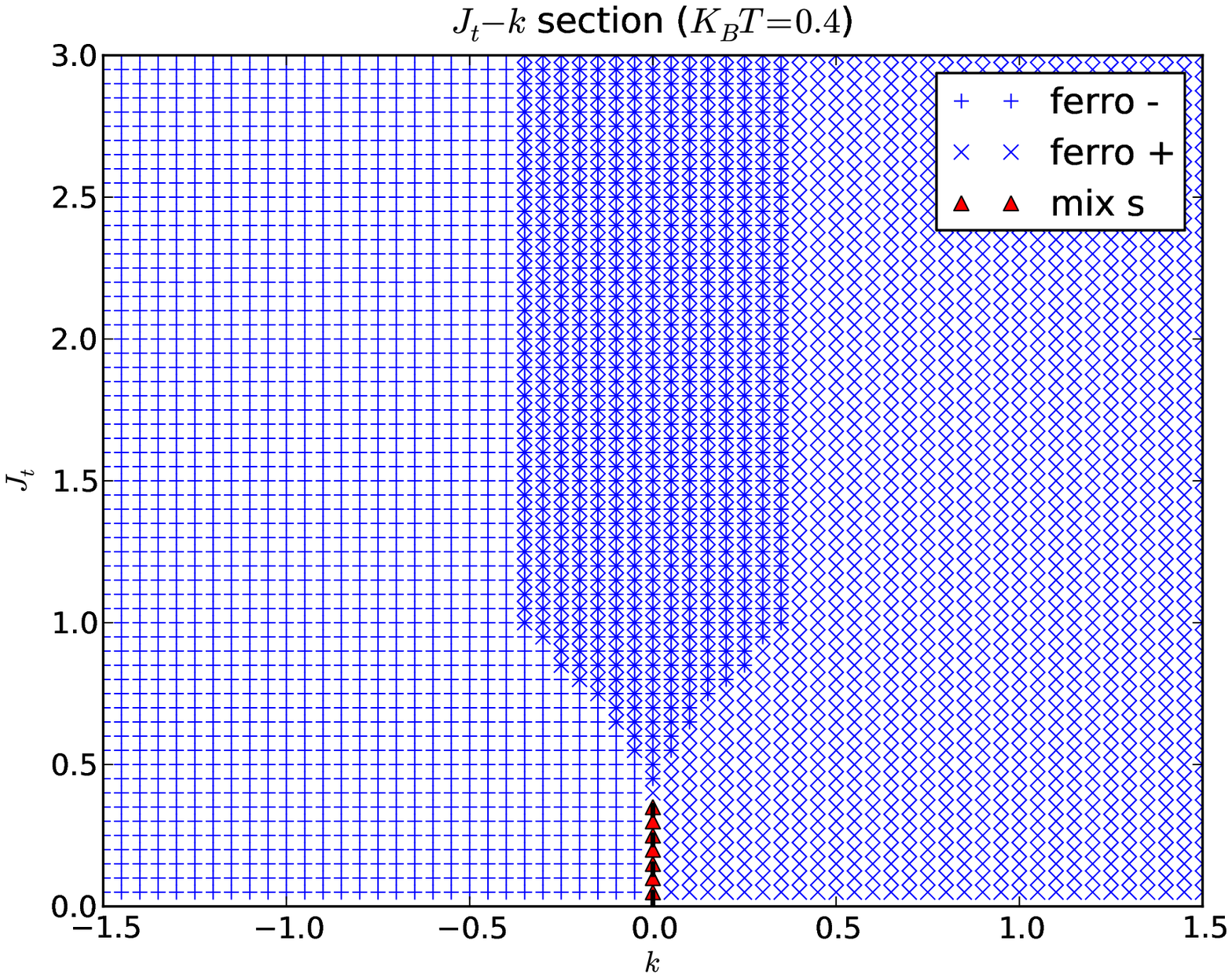}
\caption{Two dimensional $J_{t}-k$ sections of the phase diagram as given by the numerical average magnetisations for $J_{s}=1$ and  $K_{B}T=0.4$. Only stable solutions are shown. The phase associated to each solution is plotted for $k$ between -1.5 and 1.5 and $J_{t}$ between 0.01 and 3  at intervals of 0.05. Green points are used for the paramagnetic phase, blue asps ({$\times$}) for ferromagnetic phases with $s$ and $t$ of the same sign and blue crosses ($+$) for ferromagnetic phases with $s$ and $t$ of opposite signs. Red triangles are used for the mixed phase where $s\neq 0$. The mixed segment ($s\neq0$) is given by  $k=0$ and $J_{t}<J_{t}^{c0}=K_{B}T=0.4$ (dotted line).}
\label{fig:locphadiaJk2}
\end{figure}

Figures \ref{fig:locphadiaJk1} and \ref{fig:locphadiaJk2} show $J_{t}-k$ sections for $J_{s}=1$ at high $K_{B}T=1.5$ (\ref{fig:locphadiaJk1}) and low $K_{B}T=0.4$ (\ref{fig:locphadiaJk2}). For the former, the critical curve is given by $J_{t}^{c}=\frac{K_{B}T(J_{s}-K_{B}T)}{J_{s}(1-\alpha_{k}^{2})-K_{B}T}=\frac{0.75}{\tanh^{2}(\frac{2k}{3})+0.5}$ and the mixed segment, as in the local case, by $k=0$ and $J_{t}>J_{uc}^{t}=K_{B}T=1.5$. For the latter, there are mixed phases ($s\neq0$) for $k=0$ and $J_{t}<J_{t}^{c0}=K_{B}T=0.4$.

The low temperature case is particularly similar to the nonlocal cross section (even quantitatively)\footnote{Although in this case the metastable region does not seem to have {\itshape diffuse} edges, as was already mentioned for the $J_{t}-K_{B}T$ sections, for other specific choices of the parameters this behaviour appears, as did in the nonlocal cross sections.} except for the fact that there are stable solutions for all values of the parameters. For low temperatures, besides this difference, the critical curve has an asymptotic behaviour such that $\lim_{|k|\to\infty}J_{t}^{c}= K_{B}T-J_{s}=0.5$. This means for $J_{t}<0.5$, only the paramagnetic solution is stable regardless the value of $|k|$.

\section{Socioeconomic interpretation}
\label{sec:locsoc}

As has been repeatedly noted throughout the chapter, the local case is very similar to the nonlocal one in its basic qualitative characteristics. Most of what was said in section \ref{sec:nlosoc} can be applied directly with the only need to translate from two groups to two different choices in one group. In particular, the single choice (single group in the nonlocal case) interpretation of the phases are the same.

In this case there is no change in the stability of the solutions when the interdependence gets too strong. Note in the local setup it does make sense to have problems in which the interrelation between each individual's choices is more important than social influence. For example, consider the study of a group where social or cultural traits make women strongly motivated to marry and have children. Even if they are willing to have children and all women around are doing so, probably few will be tempted to have them out of marriage. In the same way, they probably will not be willing to marry a man who has no intention of becoming a father. And there is certainly a very small probability of someone buying some additional i-pod accessory if he doesn't decide to buy an i-pod, no matter how many people are doing so. In fact, in this case inter-choice coupling effects must be considered as part of the deterministic private utility. It is a {\itshape nice} property of this framework that it seems to provide well defined answers when the problem makes sense socioeconomicly and that it {\itshape breaks down} when it does not.

In the case of no {\itshape pure} deterministic private utilities (terms associated only to $h_{s}$, $h_{t}$), trends will be reinforced in both options by the introduction of the coupling, although the sign of the choices is not defined. There is still room for completely random choice making (both social and choice interaction effects are negligible) at high enough temperatures. There is a multiple equilibria regime (associated to the $h$ symmetry) with up to four possible stable states. The latter appear only in a condition of metastability, and there is the possibility of hysteresis for weak enough interdependence between the choices. In the absence of IWAs (zero external fields $h_{s}$, $h_{t}$), \emph{when the interrelation between two choices individuals have to make is changing from negative to positive (or from positive to negative), social utility may prevent the group from changing its trends even when their private utility dictates so}. At some point though, when the intensity of the interdependence becomes strong enough, the system will rapidly collapse to its {\itshape natural} new configuration.

There are additional differences between the nonlocal and the local models regarding their stable behaviour as we have observed in last section. The fact that metastable states disappear at low enough temperatures does also make sense in the light of social applications. Imagine the case of zero temperature (completely deterministic behaviour or all relevant information encoded in private utilities, interaction between both choices in this case). In the absence of statistical fluctuations, in the local case, all of the rational agents know exactly if they are better off aligning or not both of their choices (although not in which direction), and it is completely up to them as the interaction is through their deterministic private utility. In the nonlocal case however, the interaction is part of the social deterministic utility. When there are no statistical fluctuations, there will still be a globally preferred condition of alignment or not between both groups, but the choice is not up to every agent, as the coupling depends on what the other group is doing, so there is still some probability of the system ending in a state where the groups are not aligning or not between them according to the sign of $k$.

The other qualitative difference between both couplings is the asymptotic behaviour of the critical curves in the local case. Recall that critical curves will still be separating regions where social utility is relevant from those where it is not even when we turn on the IWAs. This seems to suggest there is {\itshape more place} for disorder in the local model (in that the paramagnetic phase encompasses a larger region). These will become well behaved, single equilibrium regions when constant IWAs are turned on.  At high enough temperatures or low enough social bounding for the individual choice, we can make the interdependence of both choices as strong as we want without making social influence important, which makes sense. In the nonlocal case, the strength of interdependence between both groups is bounded by the strength of social interaction within each group and so we are not concerned with  high $k$\footnote{In this case though, as the interdependence is of social kind, there would be place for regions where social interactions are relevant even at very high temperatures or low intra group social interaction effects. The framework however simply {\itshape falls apart} when trying to consider such cases.} cases.

As discussed for the nonlocal case, the study in detail of the nonzero IWAs case is of a lot of interest specially in what concerns effects on the metastability region. It will also be considered as a particular example in \cite{R'io2010}.   

%% file: secs/conc.tex
\chapter{Conclusions}
\label{cha:conc}

The interest in studying socioeconomic discrete choice problems in a context involving direct interactions between individuals (of social norm, role model, or peer pressure type) can be motivated both empirically and based on common sense intuition arguments for a wide variety of problems. As some economists began to notice, specially since the 1990s, this type of formulation allows for the use of statistical mechanical tools in many such cases, and even for a direct translation between conventional models from condensed matter into specific discrete choice settings.

The simplest possible model is the infinite range Ising model with constant field (deterministic private utility or idiosyncratic willingness to adopt in the socioeconomic literature) and coupling. The direct use of this model in social sciences may be limited specially because all characteristics of the individuals of a group must be identical (homogeneous populations). It presumes that direct interaction is through the correctly perceived fraction of adopters, which makes sense for many of the problems of interest. It can be used to reflect a desire to conform to the social norm on choices on which there is good information available for all individuals. It could also be a good approximation to peer pressure (or role model influence) when applied to the study of smaller  groups (only formed by peers). Or in general, whenever we can consider that each agent's estimation of reality through the real subset of his interactions deviates little from the global average (i.e., the choice is distributed quite homogeneously). 

This simple system does however already give rise to interesting socioeconomic interpretations that will still be present in more sophisticated models. The more remarkable of them is the existence of multiple equilibria for some range of values of the parameters, for which the microeconomic specification of the model is not enough to uniquely determine its macroeconomic properties. Even in the absence of private utilities (trends or traditions), for temperatures (statistical fluctuations) bellow a certain critical temperature (which depends on the value of the coupling and where a second order phase transition takes place), social influence is enough for a tendency to consensus to emerge.  In this regime there are two possible equilibria, one with a fraction of adopters under $1/2$, the other above $1/2$. For non zero pure private deterministic utilities, the symmetry of the problem is broken and a state of average choice of the same sign as $h$ is privileged. For weak enough IWA $h$ though, there is an additional (metastable) equilibrium of average choice of opposite sign as that dictated by private utilities. This means that even when private utilities favour a certain option, social utility can make the opposite option persist in the population and allow for the possibility of hysteresis. It provides a simple framework to jointly understand and conciliate the lack of incentives and negative social culture approaches to explaining persistence of social pathologies.  These metastable states are associated to a first order phase transition at $h=0$. Even when $h\neq 0$, the critical temperature described for the zero field case still separates values of the temperature for which the interactive effect is relevant (spontaneous magnetisation) from those for which it is not. 

We have studied the case of two coupled Ising models in the context of socioeconomic interactions. This problem arises naturally in many socioeconomic contexts both in trying to assess how two groups affect each other in their decisions and  in studying decisions that are strongly interdependent of each other for each individual. The goal has been to understand and compare the two model's zero field phase diagrams in a socioeconomic context. 

In the first case, we have a nonlocal coupling, with an infinite range (mean field)  term in the Hamiltonian containing the interaction of all the individuals of the first group with all of the second. This is equivalent to considering individuals in any group to be affected in their choice making by their (accurate) perception of the average behaviour of both his own group and the other. Examples of interest can be the study of public opinion on a given subject in two neighbouring countries (regions, cities, neighbourhoods\ldots), companies in two related business sectors and their production technology option\ldots 

In the second case, coupling is completely local, only through each individual, and there is a single group considered. Agents will be thus bonused (or penalised) if both of his spin choices align. This is interesting in many contexts, for example, the interaction between different social pathologies (dropping from school vs teenage pregnancy, drug abuse vs unemployment, \ldots  ), other social traits (joining the labour force vs having a child, living in the suburbs vs having a child, \ldots), opinion dynamics and political science (voting yes to two different propositions coming from the same party, opinion on two related subjects\ldots) and indeed in economical contexts  (buying different products of the same/competitor brand, interrelated technology elections \ldots).

Both models have remarkable similarities, specially qualitatively. There is however a major difference: in the nonlocal case, for strong enough inter choice coupling, there are no stable solutions whatsoever. In the local case, there is at least a stable solution for all values of all the parameters. As has been discussed, this can be naturally explained in a socioeconomic context. In the nonlocal case the additional interaction between groups can be considered as an additional social effect, while in the local model it will imply an additional term in the private deterministic utility. This makes problems in the strong coupling only well defined for the local case.

For both models, mixed phases can only exist at $k=0$, i.e., if social interactions start to count for one of the decisions they automatically will tend to make the other choice align too through the interdependence between them. Both models have an unbroken symmetry ($h=0$) and so whenever there is spontaneous magnetisation (prevalence of the social utility) there will be two physically identical equilibria. If the interdependence between the choices/groups is positive (they tend to reinforce each other), these will be ones in which both average decisions have the same sign (both have a low or both have a high demand or fraction of adopters). If it is negative, average choices will be opposite signed in both equilibria.  Critical curves where a second order phase transition takes place have been analytically calculated. Note that in the nonzero field case, these will still separate regions where social utility matters from regions where it does not.  

Another interesting feature common to both models is that even in the absence of private deterministic utilities there is a first order phase transition at $k=0$ (and metastable states and hysteresis). For small enough inter couplings (and temperatures), the symmetry of the problem is not completely broken and  besides the two stable equilibria two additional (of opposite relative sign as the previous) metastable ones appear. This means that if the coupling between both groups or choices is not too big, there can still be a situation where social/choice interaction gives an outcome which opposes private utility.

When compared to the uncoupled case (or to the single Ising model from one of the spin type variables {\itshape point of view}), the interdependence (inter coupling $k$) introduces a higher trend to consensus (although not favouring any specific direction). It also gives rise to interesting considerations concerning metastability and hysteresis in the light of interacting groups whose perception is changing from negative to positive (or positive to negative) in the nonlocal case or of choices for which interdependence is reversing in the local case. This can be of interest, for example,  when studying opinion formation in social or political groups when their traditional inter-influence is changing (for example, party excisions). When considering the local model, it can be used to study situations such as strong government action to compensate for {\itshape natural} reinforcement between undesired trends (for example, well designed aid plans for teenage mothers to continue with their studies or improving available resources provided by the state towards family conciliation to facilitate mother's incorporation to the labour market and incentive a sustainable birth rate). In both cases the metastability regions involve situations in which the groups/options may be prevented from aligning (or disaligning) even when they would be better off doing so.   

There are some additional differences between both models besides the regions of instability existing for the nonlocal case. In the nonlocal case, these metastable equilibria exist for low temperatures (provided $k$ is small enough) up to $T=0$. For the local case however, at a small enough temperature, these disappear and are not stable at $T=0$. This means that even if we suppose we have an exact model for behaviour and that it is completely deterministic (no statistical fluctuations), the nonlocal model allows for four possibly equilibria while the local one only for two.

Another difference is that in the local case, when considering the dependence in $k$, for low enough intra couplings or large enough temperatures, there will be no second order phase transition. That is, for the local model, if social pressure to conform to the norm is small enough or statistical fluctuations large enough, no order will emerge regardless how much we increase $k$. In the nonlocal case, ferromagnetic solutions would appear always at high enough inter coupling, however, if they lie in the strong coupling regime they too will be unstable.

As we have reviewed, both of these differences can also be explained in relation to whether the interdependence between both variables will add an effective term to the social (nonlocal case) or to the private (local case) deterministic utilities. This difference makes it possible for the agents in a local setting to be able to completely determine whether they should align or not their options in the absence of statistical fluctuations (no metastable states at $T=0$). It also makes sense that in this scenario we can make interdependence very strong and if statistical fluctuations are strong enough or social pressure is low enough, this will not make the social utility term relevant (related to the asymptotic behaviour of critical curves in the local case). 

When introducing nonzero fields, the paramagnetic phase will disappear and critical curves will be now dividing regions where spontaneous choice self organisation appears from where it is completely dominated by deterministic pure private utilities. Symmetry on $k$ will be broken and so solutions will no longer appear in pairs (of opposite signed magnetisations). We can expect however, as happened for the single Ising model case, that this will give rise to additional metastable equilibria for small enough private utilities. How these will interact with those associated to the first order transition in $k$ and so what the metastability picture emerging will look like, is of great interest as it will provide more insight on how the behaviour for groups/ choices modifying their inter-influence relates to their specific alignment or not of their IWAs. 

These models need to be tested for their suitability to explain real data. They are probably too simple to provide an accurate fit for any real dataset of interest, although they can help us understand general features of these type of behaviours and challenge our preconceptions. The best way to put these models to the test with real data is also one of our main concerns. Testing if the performance of the coupled model is better than the uncoupled one, for problems where we expect this to be the case, should start teaching us things about how sensible this approach is. It can also help us understand in more detail some problems, providing us with hypothesis about average values of parameters that will be useful when introducing probability distributions to account for population heterogeneity (in an attempt to {\itshape really} explain the data).

The introduction of disorder, of individual deterministic private utilities that really encode the group's particularities, seems to be the next obvious step if we wish to really put these models to work on real phenomena (and not only as supporting evidence for general, abstract considerations). We can expect the introduction of probability distributions to enrich the equilibria landscape much more, which makes the picture given by these models radically different form non interactive ones. We can expect them to provide both a rich theoretical framework on which to work \emph{deductively} and also a reasonable applied method for explaining real data for convenient choices of distributions. In the theory side, it will for example allow us to make considerations on how many equilibria or how they look like depending on the type of distribution codifying socioeconomic factors of relevance for the problem. In the experiment side, much work has still to be done, but it will hopefully provide us with a way of satisfactorily explaining real sets of data of different types.